\documentclass[12pt]{thloria}
\usepackage[dvips]{color,graphicx,epsfig}
\usepackage[french]{minitoc}
\usepackage{color}
\usepackage{amssymb}
\usepackage{latexsym}

\newcommand\UnivLogo{\vtop
{\hbox{}\hbox{\includegraphics[height=2cm]{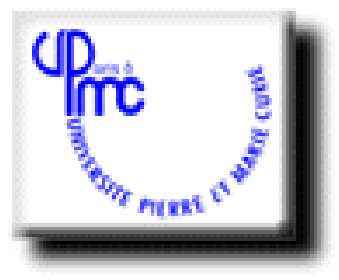}}\vss}}
\newcommand\EDLogo{\vtop
{\hbox{}\hbox{\includegraphics[height=2cm]{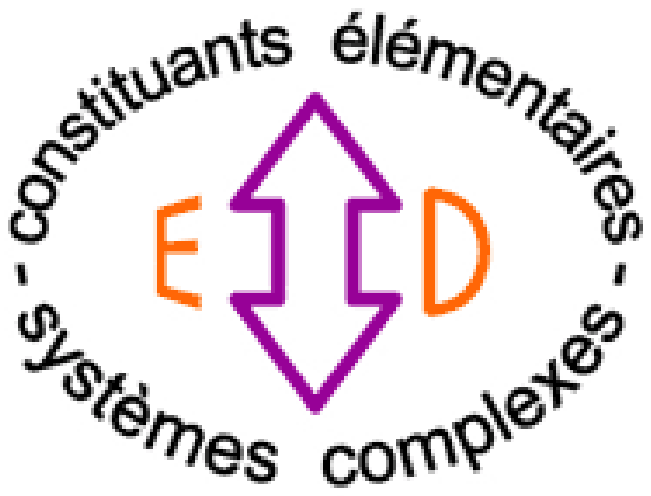}}\vss}}
\makeatletter
\renewcommand{\@NancyIhe@d}{{\UseEntryFont{ThesisFirstPageHead}\noindent
	\indent{$\raise2cm\hbox{\UnivLogo}$ \hspace{10.7cm}
	$\raise2cm\hbox{\EDLogo}$}\par
	\centerline{\'Ecole
	Doctorale ``Constituants \'El\'ementaires
	$\longleftrightarrow$ Syst\`emes Complexes''}%
	\par\nointerlineskip \vskip1mm \noindent\hrulefill } }

\newcommand\ThesisJussieu{\renewcommand{\@ThesisFirstPageHead}{\@NancyIhe@d}%
                         \ThesisDiploma{{\UseEntryFont{ThesisDiploma}%
             Docteur de l'Universit\'e Pierre et Marie Curie -- Paris VI \\[3mm]
            {\UseEntryFont{ThesisSpecialty}(Sp\'ecialit\'e  ``Champs,
    Particules, Mati\`eres'')}}}}
\makeatother

\renewcommand\TheBanner{\textsl{Version 1.1}}

\ShowLogoUHP
\ResetFootnotesAtChapters

\newcommand\itheadings[1]{\textit{#1}}
\FormatHeadingsWith{\itheadings}

\NoChapterNumberInRef
\NoChapterPrefix

\ptcrule

\newcommand{\journal}[1]{\textit{#1}}
\newcommand{\editeur}[1]{\textnormal{#1}}
\newcommand{\numero}[1]{\textbf{#1}}
\newcommand{\page}[1]{\textsl{#1}}
\newcommand{\annee}[1]{\textnormal{(#1)}}
\newcommand{\eprint}[1]{\texttt{#1}}

\newcommand{\kc}{\mathsf{k}}
\newcommand{\uc}{{\mathrm{c}}}

\newcommand{\ud}{{\mathrm{d}}}
\newcommand{\ue}{{\mathrm{e}}}
\newcommand{\mcroix}{m_{\times}}
\newcommand{\ug}{{\mathrm{g}}}
\newcommand{\uem}{{\mathrm{em}}}

\newcommand{\urad}{{\mathrm{rad}}}
\newcommand{\umat}{{\mathrm{mat}}}
\newcommand{\udec}{{\mathrm{dec}}}
\newcommand{\urec}{{\mathrm{rec}}}
\newcommand{\ueq}{{\mathrm{eq}}}
\newcommand{\uq}{{\mathrm{q}}}
\newcommand{\Hc}{{\mathcal{H}}}
\newcommand{\uL}{{\mathrm{L}}}
\newcommand{\Vc}{{\mathcal{V}}}
\newcommand{\Lc}{{\mathcal{L}}}
\newcommand{\Lcb}{\overline{\mathcal{L}}}
\newcommand{\gt}{\widetilde{g}}
\newcommand{\Gc}{{\mathcal{G}}}

\newcommand{\Ec}{{\mathcal{E}}}
\newcommand{\Fc}{{\mathcal{F}}}
\newcommand{\Sc}{{\mathcal{S}}}
\newcommand{\Scb}{\overline{\mathcal{S}}}
\newcommand{\Cc}{{\mathcal{C}}}
\newcommand{\Dc}{{\mathcal{D}}}
\newcommand{\Oc}{{\mathcal{O}}}
\newcommand{\meV}{\,{\mathrm{meV}}}
\newcommand{\eV}{\,{\mathrm{eV}}}
\newcommand{\KeV}{\,{\mathrm{KeV}}}
\newcommand{\MeV}{\,{\mathrm{MeV}}}
\newcommand{\GeV}{\,{\mathrm{GeV}}}
\newcommand{\TeV}{\,{\mathrm{TeV}}}
\newcommand{\Kel}{\,{\mathrm{K}}}
\newcommand{\Pl}{{\mathrm{Pl}}}
\newcommand{\utot}{{\mathrm{tot}}}
\newcommand{\zero}{{_{0}}}
\newcommand{\uup}{{\mathrm{p}}}
\newcommand{\uZ}{{\mathrm{Z}}}
\newcommand{\uH}{{\mathrm{H}}}
\newcommand{\uS}{{\mathrm{S}}}
\newcommand{\uf}{{\mathrm{f}}}
\newcommand{\ui}{{\mathrm{i}}}
\newcommand{\ub}{{\mathrm{b}}}
\newcommand{\uG}{{\mathrm{G}}}
\newcommand{\Psib}{\overline{\Psi}}
\newcommand{\psib}{\overline{\psi}}
\newcommand{\chib}{\overline{\chi}}
\newcommand{\Uc}{{\mathcal{U}}}
\newcommand{\Jc}{{\mathcal{J}}}
\newcommand{\Rc}{{\mathcal{R}}}
\newcommand{\Tp}{\widehat{{\mathcal{T}}}}
\newcommand{\Dt}{\widetilde{D}}
\newcommand{\Vt}{\widetilde{V}}
\newcommand{\ueff}{{\mathrm{eff}}}
\newcommand{\uR}{{\mathrm{R}}}
\newcommand{\ur}{{\mathrm{r}}}
\newcommand{\us}{{\mathrm{s}}}
\newcommand{\uT}{{\mathrm{T}}}
\newcommand{\ucm}{{\mathrm{cm}}}
\newcommand{\uw}{{\mathrm{w}}}
\newcommand{\uew}{{\mathrm{ew}}}
\newcommand{\ugrav}{{\mathrm{grav}}}
\newcommand{\uh}{{\mathrm{h}}}
\newcommand{\bx}{{\mathbf{x}}}
\newcommand{\bp}{{\mathbf{p}}}

\newcommand{\bk}{{\mathbf{k}}}
\newcommand{\uPl}{{\mathrm{Pl}}}
\newcommand{\uinf}{{\mathrm{inf}}}
\newcommand{\uint}{{\mathrm{int}}}
\newcommand{\unue}{{\mathrm{nue}}}
\newcommand{\Qb}{\overline{Q}}
\newcommand{\xib}{\overline{\xi}}
\newcommand{\uwz}{{\mathrm{wz}}}
\newcommand{\ulib}{{\mathrm{lib}}}
\newcommand{\uv}{{\mathrm{v}}}
\newcommand{\ugn}{{\mathrm{gn}}}
\newcommand{\nablab}{{\overline{\nabla}}}
\newcommand{\first}{{\P}}
\newcommand{\Tb}{{\overline{T}}}
\newcommand{\fb}{{\overline{f}}}
\newcommand{\jb}{{\overline{j}}}
\newcommand{\nub}{{\overline{\nu}}}
\newcommand{\mub}{{\overline{\mu}}}
\newcommand{\Lambdab}{{\overline{\Lambda}}}
\newcommand{\cb}{{\overline{c}}}
\newcommand{\pb}{{\overline{p}}}
\newcommand{\Lambdatd}{{\widetilde{\Lambda}}}
\newcommand{\lambdat}{\widetilde{\lambda}}

\newcommand{\varpib}{\varpi}
\newcommand{\varpitd}{\widetilde{\varpi}}
\newcommand{\ctd}{{\widetilde{c}}}
\newcommand{\ptd}{{\widetilde{p}}}
\newcommand{\ct}{c_{_{\mathrm{T}}}}
\newcommand{\cl}{c_{_{\mathrm{L}}}}
\newcommand{\gammal}{\gamma_\ell}
\newcommand{\telque}{{\,\diagup\,}}
\newcommand{\privede}{{\,\setminus\,}}
\newcommand{\dotvec}[2]
{\dot{\vec{#1}}^{\,\raisebox{-1pt}[0pt][0pt]{\tiny #2}}}
\newcommand{\primevec}[2]
{\vec{#1}^{\,' \raisebox{2.2pt}[0pt][0pt]{\tiny #2}}}
\newcommand{\ddotvec}[2]
{\ddot{\vec{#1}}^{\,\raisebox{-1pt}[0pt][0pt]{\tiny #2}}}
\newcommand{\pprimevec}[2]
{\vec{#1}^{\,'' \raisebox{2.2pt}[0pt][0pt]{\tiny #2}}}
\newcommand{\ETAL}{{et al.}}
\newcommand{\ET}{et~}
\newcommand{\uV}{{\mathrm{V}}}
\newcommand{\uA}{{\mathrm{A}}}
\newcommand{\Rho}{{\mathcal{R}}}
\newcommand{\Rhop}{{\widehat{\mathcal{R}}}}
\newcommand{\Chi}{{\mathcal{X}}}
\newcommand{\cpsir}{c_{\psi_\uR}}
\newcommand{\cpsil}{c_{\psi_\uL}}
\newcommand{\cchir}{c_{\chi_\uR}}
\newcommand{\cchil}{c_{\chi_\uL}}
\newcommand{\cphi}{c_{\phi}}
\newcommand{\jbt}{\tilde{\overline{j}}}
\newcommand{\I}{I}
\newcommand{\Ferm}{{\mathcal{F}}}
\newcommand{\norm}{{\mathcal{N}}}
\newcommand{\M}{{\mathcal{M}}}
\newcommand{\Pstate}{{\mathcal{P}}}
\newcommand{\Eferm}{{\mathcal{E}}}
\newcommand{\chip}{{\chi_{\mathcal{P}}}}
\newcommand{\psip}{{\psi_{\mathcal{P}}}}
\newcommand{\cfl}{c_{\Ferm_{\mathrm{L}}}}
\newcommand{\cfr}{c_{\Ferm_{\mathrm{R}}}}
\newcommand{\miv}{\overline{m}}

\newcommand{\rhot}{\widetilde{\rho}}

\newcommand{\atd}{\widetilde{\alpha}}
\newcommand{\ab}{\overline{\alpha}}
\newcommand{\ah}{\widehat{\alpha}}
\newcommand{\bt}{\widetilde{\beta}}
\newcommand{\nut}{\widetilde{\nu}}
\newcommand{\Sigmab}{\overline{\Sigma}}
\newcommand{\ft}{\widetilde{f}}

\newcommand{\vht}{\widehat{v}}
\newcommand{\uht}{\widehat{u}}
\newcommand{\xpar}{x_{\parallel}}
\newcommand{\xper}{x_{\perp}}
\newcommand{\syse}{\left({\mathcal{S}}_{\varepsilon}\right)}
\newcommand{\sysp}{\left({\mathcal{S}}_+\right)}
\newcommand{\sysm}{\left({\mathcal{S}}_-\right)}
\newcommand{\vsep}{\vspace{4pt}}
\newcommand{\kappabu}{\kappa_{_5}}
\newcommand{\mbu}{m_{_5}}
\newcommand{\Gbu}{\Gc_{_5}}

\newcommand{\hb}{\overline{h}}

\newcommand{\dd}{\ud}
\newcommand{\de}{\ue}

\newcommand{\DD}{{\mathcal{D}}}
\newcommand{\U}{{\mathcal{U}}}

\newcommand{\Pb}{{\mathcal{P}}}
\newcommand{\Tt}{\widetilde{\mathcal{T}}}
\newcommand{\Th}{\widehat{\mathcal{T}}}
\newcommand{\Ut}{\widetilde{\mathcal{U}}}
\newcommand{\Uh}{\widehat{\mathcal{U}}}
\newcommand{\mt}{\mu}
\newcommand{\Mt}{\gamma_{_{\mathrm{F}}}}
\newcommand{\mut}{\ell}

\newcommand{\gsim}{\gtrsim}

\begin{document}

\pagestyle{ThesisHeadingsI}

\dominitoc
\doparttoc

\ThesisTitle{\'Etude des courants fermioniques sur les objets \'etendus}
\ThesisDate{5 juillet 2002}
\ThesisAuthor{Christophe Ringeval}
\ThesisInOrderToGet{pour l'obtention du grade de}
\ThesisFirstPageFoot{}
\ThesisJussieu

\NewJuryCategory{Directeur}{\textit{Directeur :}}{\textit{Directeurs :}}

\Rapporteurs= {M. Tom KIBBLE \\
	M. Bernard LINET}
\Invites= {M. Fran\c cois BOUCHET}
\Examinateurs= {M. Pierre BIN\'ETRUY \\ M. Richard KERNER \\
	M. Patrick PETER \\ M. Jean-Philippe UZAN}

\SetBinding{0mm}	
\MakeThesisTitlePage

\AlignTitlesRight

\WriteThisInToc
\begin{ThesisAcknowledgments}
Mes remerciements se portent naturellement vers mes parents qui m'ont
toujours soutenu et aid\'e dans mes choix passionn\'es. Je suis
\'egalement redevable \`a Raoul Talon, Pierre Bin\'etruy et Yves
Charon pour m'avoir donn\'e le privil\`ege de suivre des enseignements
de qualit\'e.

Je remercie tous les membres du D\'epartement d'Astrophysique
Relativiste et de Cosmologie de l'Observatoire de Meudon dans lequel
j'ai commenc\'e ma th\`ese, ainsi que ceux de l'Institut
d'Astrophysique de Paris dans lequel je l'ai termin\'ee. Mes pens\'ees
concernent particuli\`erement les membres du laboratoire de
Gravitation Relativiste et de Cosmologie qui ont toujours pr\^et\'e
attention \`a mes interrogations. Merci \`a David Langlois et
Jean-Philippe Uzan pour m'avoir donn\'e mon embryon de connaissance
sur les dimensions suppl\'ementaires, \`a Alain Riazuelo pour m'avoir
\'eclair\'e sur de nombreux sujets, ainsi qu'\`a G\"unter Sigl pour ses
reponses \`a de si na\"\i ves questions sur les rayons cosmiques. Je
voudrais \'egalement adresser un remerciement particulier \`a Martin
Lemoine, J\'er\^ome Martin et Olivier Poujade qui m'ont donn\'e, en
plus de leur savoir, et de leurs connaissances pratiques en
balistique, leur amiti\'e.

Mes connaissances en informatique me viennent de Fr\'ed\'eric
Magnard\footnote{derF.}, j'ai encore honte du temps que je lui ai fait
passer \`a m'expliquer ses tours de magie. Je suis aussi redevable \`a
St\'ephane Colombi\footnote{Totor.} de m'avoir initi\'e au calcul
num\'erique en parall\`ele, ainsi qu'\`a Emmanuel
Bertin\footnote{Pr\'esident des Nerds.} pour ses connaissances sur le
meilleur mat\'eriel et logiciel du moment. Merci \'egalement \`a
Philippe Canitrot et Reinhardt Prix pour m'avoir converti \`a
\textsc{Linux} et \LaTeX. Je dois aussi remercier la tour de
connexions informatiques et ses $65 \, \mathrm{dB}$ pour m'avoir
appris la concentration en milieu bruyant.

Je voudrais adresser un remerciement particulier \`a Pierre
Bin\'etruy, Fran\c{c}ois Bouchet, Nathalie Deruelle, Ruth Durrer et
Jean-Philippe Uzan pour l'int\'er\^et et le soutien qu'ils ont
port\'e sur mes travaux scientifiques. Je remercie \'egalement, \`a ce
titre, les membres du jury de m'honorer de leur pr\'esence, et
particuli\`erement les rapporteurs, Tom Kibble et Bernard Linet.

Enfin, j'ai eu l'immense chance et privil\`ege d'avoir effectu\'e
cette th\`ese sous la direction de Patrick Peter. Il a su m'indiquer
les chemins \`a suivre tout en me laissant la libert\'e, et la
satisfaction, de m'y frayer un passage. Sa comp\'etence et son
attention dans cette t\^ache m'ont enseign\'e le m\'etier de
chercheur, ce dont je lui en serai toujours reconnaissant.

\end{ThesisAcknowledgments}

\begin{ThesisDedication}
{\Large
\`A La\"etitia
}
\end{ThesisDedication}

\WriteThisInToc
\tableofcontents

\mainmatter

\SpecialSection{Introduction}

Depuis la d\'ecouverte de l'expansion de l'univers par Hubble en
1925~\cite{hubble25,hubble26}, et celle du rayonnement fossile par
Penzias et Wilson en 1965~\cite{penzias65}, l'id\'ee que notre univers
a une histoire s'est impos\'ee. \`A partir d'un \'etat de
temp\'erature et de densit\'e extraordinairement \'elev\'ees,
l'expansion l'a refroidi jusqu'\`a lui donner les propri\'et\'es que
l'on observe aujourd'hui. Les succ\`es de la th\'eorie de la
gravitation d'Einstein pour d\'ecrire cette \'evolution font de la
relativit\'e g\'en\'erale la pierre angulaire de la cosmologie
moderne\footnote{Inversement, l'univers n'\'etant pas vide de
mati\`ere, il est un moyen de tester la relativit\'e g\'en\'erale en
pr\'esence de sources.}. Par ailleurs, lorsque l'on s'int\'eresse aux
premiers instants, quand l'univers \'etait dense et chaud, la
description correcte de la mati\`ere fait appel \`a la physique des
particules, au travers de la th\'eorie quantique des champs. Celle-ci
constitue notre meilleure compr\'ehension des trois autres
interactions de la Nature, l'\'electromagn\'etisme, l'interaction
faible et la force nucl\'eaire forte, telles qu'elles sont observ\'ees
dans les acc\'el\'erateurs de particules. La cosmologie primordiale
est l'\'etude de l'univers dans ses premiers instants et se trouve
ainsi au croisement des ces diverses th\'eories. L'unification des
interactions \'electromagn\'etique et faible donne l'\'echelle
d'\'energie de pr\'edilection au del\`a de laquelle le terme
``primordial'' est de rigueur, c'est-\`a-dire $10^3 \GeV$, soit un
univers \^ag\'e de moins de $10^{-11} \, \us$. C'est dans ce cadre que
T.~Kibble a montr\'e, en 1976~\cite{kibble76}, que des transitions de
phases induites par des brisures de sym\'etrie, telles qu'on les
rencontre en physique des particules pour unifier les interactions,
peuvent g\'en\'erer des d\'efauts topologiques du vide. Ces objets
\'etendus sont stables par topologie et, bien que pour l'instant non
d\'etect\'es, ils pourraient \^etre encore pr\'esent
aujourd'hui. L'\'evolution cosmologique de telles reliques
primordiales peut cependant influer sur la dynamique de l'univers. La
compatibilit\'e de leur existence avec les observations cosmologiques
donne alors des contraintes sur la physique des particules qui est \`a
l'origine de leur formation, c'est-\`a-dire \`a des \'echelles
d'\'energie qui sont, et seront dans un futur pr\'evisible,
inaccessibles aux acc\'el\'erateurs. Dans cette th\`ese, nous
\'etudions du point de vue cosmologique, et particulaire, des
d\'efauts topologiques filiformes, plus commun\'ement appel\'es des
cordes cosmiques, qui ont la particularit\'e de pouvoir \^etre
parcourues de courants de particules dont les cons\'equences ne sont
que partiellement comprises.

La premi\`ere partie de ce m\'emoire introduit le cadre g\'en\'eral
pr\'ec\'edemment \'evoqu\'e. Le premier chapitre est consacr\'e au
mod\`ele standard de cosmologie, d\'ecoulant de la relativit\'e
g\'en\'erale, et le deuxi\`eme chapitre est son analogue pour la
physique des particules, d\'ecrit par la th\'eorie quantique des
champs. Dans le troisi\`eme chapitre nous verrons comment ces deux
mod\`eles se compl\`etent et s'\'eclairent mutuellement, et comment
leurs extensions habituelles peuvent conduire \`a la formation de
cordes cosmiques. L'int\'er\^et particulier port\'e sur ce type de
d\'efauts sera \'egalement discut\'e, ainsi que quelques
propri\'et\'es pouvant permettre leur \'eventuelle d\'etection.

La deuxi\`eme partie est d\'edi\'ee \`a la dynamique cosmologique des
cordes, c'est-\`a-dire \`a leur \'evolution temporelle au cours de
l'expansion de l'univers. Le quatri\`eme chapitre pr\'esente un outil
math\'ematique privil\'egi\'e pour arriver \`a cette fin: le
formalisme covariant. Il a \'et\'e d\'evelopp\'e par B.~Carter depuis
1989~\cite{carter89,carter89b,carter94b,carter97} et mod\'elise une
corde comme une $2$-surface de l'espace-temps plong\'ee dans une
vari\'et\'e quadri-dimensionnelle. Seules quelques quantit\'es
macroscopiques s'av\`erent finalement n\'ecessaire pour d\'ecrire la
dynamique et la stabilit\'e des cordes: l'\'energie par unit\'e de
longueur, la tension, et un param\`etre additionnel tenant compte de
l'existence d'un courant interne. \\ Le cinqui\`eme chapitre
pr\'esente l'application directe des r\'esultats de ce formalisme \`a
l'\'etude cosmologique des cordes, dans le cas le plus simple o\`u
celles-ci ne transportent pas de courant de particules. Le recours \`a
des simulations num\'eriques est indispensables et permet de donner
des contraintes sur les grandeurs macroscopiques, associ\'ees \`a ce
type de corde, qui sont compatibles avec les observations
cosmologiques. Dans cette optique, le meilleur code existant,
d\'evelopp\'e par F.~Bouchet et D.~Bennett en
1988~\cite{bennett88,bennett90}, a \'et\'e repris et modernis\'e dans
le but d'\'etudier leurs signatures observationnelles dans le
rayonnement fossile.

E.~Witten a montr\'e en 1985~\cite{witten}, que l'existence de courants
de particules le long des cordes semble en \^etre une propri\'et\'e
g\'en\'erique. Pour cela il propose deux m\'ecanismes, du point de vue
de la th\'eorie des champs, permettant de pi\'eger des bosons et des
fermions, respectivement, sur une corde cosmique. La dynamique et la
stabilit\'e de ces cordes, dites \emph{supraconductrices}, a \'et\'e
d\'ecrite, dans les ann\'ees 1990~\cite{bps,neutral,enon0}, pour des
courants de particules scalaires. Ces travaux ont d\'emontr\'e la
validit\'e du formalisme covariant de B.~Carter lorsque le param\`etre
interne est identifi\'e \`a l'intensit\'e du courant, et ont montr\'e
que la dynamique de ces cordes est g\'en\'eralement de type
\emph{supersonique}. Cette approche pr\'esente \'egalement l'avantage
de relier les quantit\'es macroscopiques r\'egissant la dynamique des
cordes aux param\`etres microscopiques du mod\`ele de Witten
directement issus de la physique des particules. Ces propri\'et\'es
sont pr\'esent\'ees dans le sixi\`eme chapitre.\\ Le principal
int\'er\^et cosmologique des cordes conductrices est dans la formation
potentielle de boucles stables, appel\'ees \emph{vortons}. En effet,
l'existence d'une structure interne, par un courant, brise
l'invariance de Lorentz longitudinale, et il existe une configuration
d'\'equilibre o\`u la force centrifuge compense exactement la force de
tension. La pr\'esence dans l'univers de ces vortons conduit
g\'en\'eralement \`a une catastrophe cosmologique: ils finissent
toujours par dominer les autres formes d'\'energie. Cette
caract\'eristique permet donc d'invalider les th\'eorie de physique
des particules \`a l'origine de leur formation. Le formalisme
covariant permet \'egalement d'\'etablir des crit\`eres de stabilit\'e
de ces vortons vis-\`a-vis de leur dynamique, et il appara\^\i t que
les r\'egimes supersoniques, comme ceux induits par les courants de
scalaires, conduisent \`a des vortons g\'en\'eriquement
instables~\cite{carter93,martinpeter,martin94}.

La troisi\`eme partie concerne l'\'etude de la dynamique des cordes
lorsqu'elles sont parcourues par des courants de fermions. Ce travail
s'ins\`ere ainsi dans l'\'etude de la structure interne des cordes
cosmiques, et prolonge les travaux sur les cordes parcourues par des
courants de bosons.\\ Le chapitre~\ref{chapitrezero} pr\'esente une
approche semi-analytique permettant de calculer l'\'equation d'\'etat
dans le mod\`ele de Witten fermionique, c'est-\`a-dire l'\'energie par
unit\'e de longueur et la tension. Pour pouvoir tenir compte des
effets d'exclusion, propres aux fermions, une quantification des
champs a \'et\'e effectu\'ee le long de la corde, alors que les
\'equations de champs ont \'et\'e r\'esolues num\'eriquement dans les
dimensions transverses~\cite{ringeval}. Cette approche s'est
initialement limit\'ee aux cas des fermions de masse nulle, tels ceux
introduits originellement par Witten. Les r\'esultats diff\`erent
essentiellement de l'id\'ee en vogue selon laquelle la dynamique de
telles cordes aurait d\^u ressembler \`a celle des cordes poss\'edant
un courant de bosons. D'une part, l'\'equation d'\'etat obtenue
d\'epend g\'en\'eralement de plusieurs param\`etres internes qui
s'identifient aux nombres d'occupation des fermions pr\'esents sur la
corde. D'autres parts, elle est de type \emph{subsonique}, et conduit
\`a des vortons classiquement stables. Au vue de leurs cons\'equences
cosmologiques, certains de ces r\'esultats ont \'et\'e confirm\'es,
dans le huiti\`eme chapitre, par une approche purement macroscopique
bas\'ee sur le formalisme covariant, dans la mesure o\`u celui-ci peut
s'appliquer, ce qui n'est pas toujours possible avec des courants de
fermions~\cite{prep}. De plus, les effets de retroaction par
les courants ont \'et\'e \'egalement \'etudi\'es et montrent que les
modes z\'eros n\'ecessaires \`a leur g\'en\'eration peuvent acqu\'erir
une faible masse.\\ Le neuvi\`eme chapitre est une \'etude
d\'etaill\'ee des diff\'erents modes de propagation des fermions le
long d'une corde, et de leurs influences sur l'\'equation
d'\'etat~\cite{ringeval2}. Comme la retroaction le
sugg\'erait, les modes z\'eros ne sont pas des solutions
g\'en\'eriques. Les fermions dans une corde pr\'esentent
n\'ecessairement un spectre de masse discret, o\`u chaque masse
contribue \`a la g\'en\'eration du courant. La quantification
initi\'ee pour les modes de masse nulle a pu \^etre \'elargie pour
englober tous les modes massifs, et utilis\'ee dans l'\'etablissement
de l'\'equation d'\'etat g\'en\'erale. Finalement, il appara\^\i t que
les modes massifs favorisent les r\'egimes supersoniques, alors que
les modes z\'eros et ultrarelativistes privil\'egient les r\'egimes
subsoniques. L'\'equation d'\'etat pr\'esente ainsi des transitions
entre ces deux r\'egimes lorsque la densit\'e de fermions pi\'eg\'es
varie.

Les r\'esultats concernant le spectre de masse des fermions sur les
cordes sont tout \`a fait g\'en\'eriques d\`es que ces derniers sont
coupl\'es \`a un champ de Higgs formant un d\'efaut topologique. Dans
la derni\`ere partie, en collaboration avec P.~Peter et J.-P.~Uzan,
nous avons utilis\'e ces propri\'et\'es pour confiner des fermions,
vivant dans un espace-temps \`a cinq dimensions, sur un mur de domaine
quadri-dimensionnel repr\'esentant notre univers~\cite{rpu}. Ce
mod\`ele, pr\'esent\'e dans le chapitre~\ref{chapitremur}, s'ins\`ere
dans un domaine de la cosmologie discutant de la g\'eom\'etrie de
l'univers au travers de l'existence de dimensions
suppl\'ementaires. Notre approche concerne le mod\`ele de Randall et
Sundrum, pr\'esent\'e dans le dixi\`eme chapitre, o\`u la cinqui\`eme
dimension est non-compacte, mais courb\'ee de telle sorte que la
gravit\'e effective sur la membrane formant notre univers s'identifie,
au premier ordre, \`a la gravit\'e usuelle d'Einstein. Comme dans le
cas des cordes, un spectre de masse discret pour les fermions est
obtenue, donnant une explication \`a la quantification de la masse en
g\'en\'eral.

Les cons\'equences cosmologiques de ces nouveaux r\'esultats sont
discut\'ees en conclusion, ainsi que leurs extensions \`a venir.

\part{Cadre g\'en\'eral}
\label{partiegene}
\chapter{Le mod\`ele cosmologique standard}
\label{chapitrecosmo}
\minitoc
\section{Introduction}

La gravitation, la plus faible des interactions de la Nature, est
aujourd'hui la force dominante \`a grande distance. Par son
caract\`ere strictement attractif et sa port\'ee
infinie\footnote{Contrairement
\`a l'\'electrodynamique, il n'existe pas de masse n\'egative pouvant
\'ecranter l'interaction.} elle fa\c conne le
monde qui nous entoure. La cosmologie, ou l'\'etude de l'univers dans
son ensemble, consid\'er\'e comme un syst\`eme physique, repose
essentiellement sur notre compr\'ehension de cette interaction. Elle
suppose ainsi que la gravitation est effectivement d\'ecrite par la
th\'eorie de la relativit\'e g\'en\'erale
d'Einstein~\cite{einstein15,hilbert15} reliant la g\'eom\'etrie de
l'espace temps \`a son contenu
\'energ\'etique. Les \'equations locales de la relativit\'e
g\'en\'erale permettent de conna\^{\i}tre la dynamique et la
g\'eom\'etrie globale de l'univers moyennant deux autres hypoth\`eses
simplificatrices: l'univers est suppos\'e homog\`ene et isotrope aux
grandes \'echelles de distance, c'est le \emph{principe
cosmologique}~\cite{einstein17}; et il est simplement
connexe\footnote{Une autre hypoth\`ese implicite, car incluse dans le
principe d'\'equivalence faible, est que les lois de la physique sont
valables dans tout l'univers.}. Le principe cosmologique est
aujourd'hui v\'erifi\'e par diff\'erentes observations (comme les
comptage de galaxies~\cite{strauss92}, les mesures du
CMBR\footnote{Cosmic Microwave Background Radiation ou fond cosmique
micro onde.}~\cite{mather90,smoot92}, le fond de rayons
X~\cite{peebles} ou la r\'epartition des radiosources~\cite{peebles}),
alors que des tests sur la topologie de notre univers sont encore \`a
venir\footnote{En ce qui concerne ce m\'emoire, la topologie de
l'univers ne jouera pas un r\^ole crucial.}~\cite{lachieze95,
uzan00}. Ce sont Friedmann et Lema\^{\i}tre qui, historiquement, ont
d\'eriv\'e la m\'etrique d'un tel univers, les menant \`a construire
les premiers sc\'enarios cosmologiques non
stationnaires~\cite{friedmann22, lemaitre27} formant le mod\`ele
standard baptis\'e de Friedmann-Lema\^{\i}tre-Robertson-Walker
(FLRW)~\cite{robertson28, walker36} aux pr\'edictions largement
confirm\'ees.

\section{Le mod\`ele de FLRW}

\subsection{La m\'etrique}

Dans un syst\`eme de coordonn\'ees $x^\mu$, le tenseur m\'etrique $g_{\mu
\nu}$ permet de d\'efinir l'intervalle quadri-dimensionnel
\begin{equation}
\label{intervalle}
\ud s^2 = g_{\mu \nu} \, \ud x^{\mu} \ud x^{\nu}.
\end{equation}
Cependant, l'hypoth\`ese d'isotropie du principe cosmologique assure
l'existence d'observateurs isotropes, c'est-\`a-dire de courbes de
genre temps sur lesquelles il est impossible de d\'eterminer une
direction spatiale privil\'egi\'ee. Si $t$ d\'esigne le temps propre
associ\'e \`a ces observateurs, leurs lignes d'univers sont
caract\'eris\'ees par les quadrivecteurs vitesse
\begin{equation}
\label{quadrivitesse}
u^\mu=\frac{\ud x^\mu}{\ud t},
\end{equation}
en fonction desquels la m\'etrique (\ref{intervalle}) se re\'ecrit
\begin{equation}
\label{intervalleiso}
\ud s^2 = u_\mu u_\nu \ud x^{\mu} \ud x^{\nu} + \left(g_{\mu \nu} - u_\mu
u_\nu\right) \ud x^{\mu} \ud x^{\nu}.
\end{equation}
L'homog\'en\'eit\'e du principe cosmologique impose de plus
l'existence d'hypersurfaces de genre espace, \`a chaque instant, dans
lesquelles il est impossible de choisir un point privil\'egi\'e. Afin
de ne pas violer l'isotropie, ces hypersurfaces doivent \^etre
orthogonales aux lignes d'univers des observateurs isotropes. Le terme
en $g_{\mu \nu} - u_\mu u_\nu$ dans (\ref{intervalleiso}) appara\^\i t
donc comme un projecteur sur les sections spatiales permettant, \`a
l'aide de (\ref{quadrivitesse}), de simplifier la m\'etrique
en\footnote{Dans toute la suite les indices latins varieront de $1$
\`a $3$ et les grecs de $0$ \`a $3$}
\begin{equation}
\label{intervallehomiso}
\ud s^2 = c^2 \ud t^2 + g_{i j} \, \ud x^{i} \ud x^{j},
\end{equation}
o\`u les $x^{i}$ sont des coordonn\'ees spatiales, dites
\emph{comobiles}, sur les hypersurfaces, et $c$ la vitesse de la
lumi\`ere. Les sections spatiales \'etant homog\`enes et isotropes,
elles sont \'egalement \`a sym\'etrie maximale. On peut montrer dans
ce cas que l'\'el\'ement infinit\'esimal de longueur se met sous la
forme~\cite{waldbook,weinbergbook}
\begin{equation}
\ud \ell^2= -g_{i j} \, \ud x^{i} \ud x^{j}= a^2(t)\left[\frac{\ud
r^2}{1-\kc r^2} + r^2 \left(\ud\theta^2 + \sin^2\theta \, \ud
\phi^2\right) \right],
\end{equation}
avec $(r,\theta,\phi)$ les coordonn\'ees sph\'eriques comobiles
adimensionn\'ees obtenues \`a partir des $x^i$, et $\kc$ le
param\`etre de courbure. Celui-ci se r\'eduit, apr\`es un changement
d'unit\'e, \`a $-1$, $0$, ou $1$, suivant que les sections spatiales
sont respectivement hyperboliques, plates, ou elliptiques
localement. La d\'ependance en $t$, le \emph{temps cosmique}, est pour
sa part d\'ecrite au travers du facteur d'\'echelle\footnote{Il a ici
les dimensions d'une longueur.} $a(t)$. La m\'etrique de FLRW est
finalement,
\begin{eqnarray}
\label{metriqueFLRW}
\ud s^2 & = &\ud t^2 - a^2(t)\left[\frac{\ud
r^2}{1-\kc r^2} + r^2 \left(\ud\theta^2 + \sin^2\theta \, \ud
\phi^2\right) \right],
\end{eqnarray}
o\`u la vitesse de la lumi\`ere $c=1$. Afin de simplifier les
notations, dans toute la suite, les unit\'es seront choisies telles
que la constante de Planck $\hbar$ et la constante de Boltzmann
$k_\ub$ soient \'egalement \'egales \`a l'unit\'e $\hbar=k_\ub=1$. Il
est commode d'introduire le temps conforme $\eta$ reli\'e au temps
cosmique $t$ par
\begin{equation}
\ud t = a(\eta) \ud \eta,
\end{equation}
pour exprimer la m\'etrique sous sa forme conforme
\begin{eqnarray}
\label{metriqueFLRWconf}
\ud s^2 & = & a^2(\eta) \left(\ud \eta^2 - \gamma_{ij}\ud x^i \ud x^j
\right),
\end{eqnarray}
avec
\begin{equation}
\gamma_{i j} \ud x^i \ud x^j = \frac{\ud
r^2}{1-\kc r^2} + r^2 \left(\ud\theta^2 + \sin^2\theta \, \ud
\phi^2\right).
\end{equation}
La g\'eom\'etrie de l'univers est donc compl\`etement d\'etermin\'ee
par la donn\'ee de la m\'etrique (\ref{metriqueFLRWconf}). Afin d'en
d\'eduire son \'evolution par les \'equations d'Einstein il faut se
donner le contenu \'energ\'etique. Conform\'ement au principe
cosmologique, l'univers \`a grande \'echelle peut \^etre mod\'elis\'e
par un fluide parfait de densit\'e d'\'energie $\rho(\eta)$ et de
pression $P(\eta)$, dont le tenseur d'\'energie impulsion est le seul
compatible avec les hypoth\`eses d'homog\'en\'eit\'e et d'isotropie
\begin{equation}
\label{tmunutout}
T^{\mu \nu}=\left(\rho + P \right)u^{\mu} u^{\nu} - P g^{\mu \nu}.
\end{equation}

\subsection{Les \'equations de Friedmann}

Il est commode d'introduire le param\`etre de Hubble $H(t)$ et son
analogue conforme $\Hc(\eta)$ pour d\'ecrire la dynamique de l'univers
\begin{eqnarray}
\label{hubbleparam}
H & = & \frac{\dot{a}(t)}{a(t)},\\
\Hc & = & \frac{a'(\eta)}{a(\eta)}=a H,
\end{eqnarray}
o\`u le ``point'' d\'esigne la d\'eriv\'ee par rapport au temps
cosmique $t$ et le ``prime'' par rapport au temps conforme $\eta$. La
dynamique de l'univers est donn\'ee par les \'equations
d'Einstein
\begin{equation}
\label{einsteineq}
G_{\mu \nu} + \Lambda g_{\mu \nu}= \kappa ^2 T_{\mu\nu},
\end{equation}
avec $\Lambda$ la constante cosmologique, $G_{\mu\nu}$ le tenseur
d'Einstein et
\begin{equation}
\kappa^2=8 \pi {\mathcal{G}},
\end{equation}
${\mathcal{G}}$ \'etant la constante de gravitation universelle. Ces
\'equations se r\'eduisent pour la m\'etrique
(\ref{metriqueFLRWconf}), et le tenseur \'energie impulsion
(\ref{tmunutout}), aux \'equations de Friedmann
\begin{eqnarray}
\label{friedmann0}
\Hc^2 + \kc - \Lambda \frac{a^2}{3}& = &  \kappa^2 \frac{a^2}{3} \rho,\\
\label{friedmann1}
\Hc' - \Lambda \frac{a^2}{3} & = & -\kappa^2 \frac{a^2}{6} \left( \rho +
3 P \right).
\end{eqnarray}
En d\'erivant la premi\`ere \'equation de Friedmann (\ref{friedmann0})
par rapport au temps conforme, la d\'eriv\'ee du param\`etre de Hubble
peut \^etre exprim\'ee en fonction de $\rho$ et sa
d\'eriv\'ee. L'\'equation (\ref{friedmann1}) se r\'eduit donc \`a
\begin{equation}
\rho' = -3 \Hc \left(\rho + P \right).
\end{equation}
Exprim\'ee en fonction du facteur d'\'echelle, on retrouve alors
l'\'equation de conservation de l'\'energie d'un fluide
parfait\footnote{Elle est en effet implicitement contenue dans les
\'equation d'Einstein puisque $\nabla_\mu T^{\mu \nu}=0$.}
\begin{equation}
\label{bianchi3}
\ud \!\!\left(a^3 \rho \right)=-P \ud \!\!\left(a^3\right).
\end{equation}
La dynamique de l'univers est donc compl\`etement d\'etermin\'ee par
les trois param\`etres $\Hc$, $\rho$ et $P$ v\'erifiant
(\ref{friedmann0}) et (\ref{bianchi3}) pourvu que l'on se donne son
\'equation d'\'etat, c'est-\`a-dire une relation entre la densit\'e
d'\'energie et la pression. Il est d'usage de se restreindre aux
\'equations d'\'etat de type barotropique avec
\begin{equation}
\label{etatunivers}
P=w(\eta) \rho.
\end{equation}
Physiquement, elle introduit dans le mod\`ele la nature du fluide
cosmologique, et donc $w$ d\'epend \emph{a priori} de l'\^age de
l'univers. Ainsi, actuellement, l'univers \'etant domin\'e par la
mati\`ere, le terme de pression est compl\`etement n\'egligeable
devant la densit\'e d'\'energie menant \`a $w_{\mathrm{mat}}=0$. La
densit\'e d'\'energie \'evolue donc en
\begin{equation}
\label{densmat}
\rho_{\mathrm{mat}} \propto a^{-3},
\end{equation}
caract\'eristique d'une dilution par dilatation des distances en $a$.

Les param\`etres couramment utilis\'es pour d\'ecrire les diff\'erents
mod\`eles cosmologiques, i.e. les diverses solutions des \'equations
(\ref{friedmann0}), (\ref{bianchi3}) et (\ref{etatunivers}) sont la
densit\'e critique $\rho_\uc$ et le param\`etre de densit\'e $\Omega$
d\'efinis par
\begin{eqnarray}
\label{rhocrit}
\rho_\uc & = & \frac{3 H^2}{\kappa^2}=\frac{3 \Hc^2}{a^2
\kappa^2},
\end{eqnarray}
et
\begin{eqnarray}
\label{omegafluide}
\Omega & = & \frac{\rho}{\rho_\uc}.
\end{eqnarray}
D'apr\`es l'\'equation (\ref{friedmann0}), la densit\'e critique est
donc la densit\'e d'\'energie du fluide cosmologique tel que les
sections spatiales soient toujours plates, i.e. $\kc=0$, pour une
constante cosmologique nulle\footnote{Aujourd'hui $\rho_\uc \simeq
10^{-29} \ug/\ucm$.} $\Lambda=0$. La premi\`ere \'equation de
Friedmann (\ref{friedmann0}) se r\'e\'ecrit donc
\begin{equation}
\Omega + \frac{\Lambda/\kappa^2}{\rho_\uc} = 1 + \frac{\kc}{\Hc ^2}.
\end{equation}
La constante cosmologique peut alors \^etre vue comme un fluide
parfait d'\'equation d'\'etat
\begin{equation}
\label{rholambda}
P_{\Lambda}=- \rho_{\Lambda} = - \frac{\Lambda}{\kappa^2},
\end{equation}
donnant la relation
\begin{equation}
\label{omegafctH}
\Omega + \Omega_{\Lambda} - 1 = \frac{\kc}{\Hc^2}.
\end{equation}
La mesure a un instant donn\'e, par exemple aujourd'hui, de ces trois
param\`etres, $\Omega$, $\Omega_{\Lambda}$ et $\rho_\uc$ d\'etermine
donc le mod\`ele cosmologique \`a tous les temps, une fois donn\'e sa
nature physique, i.e. son \'equation d'\'etat. Notons, que le
param\`etre de courbure $\kc$ est lui aussi parfaitement d\'etermin\'e
par la mesure \`a un seul instant. Si aujourd'hui le param\`etre de
densit\'e total, incluant la constante cosmologique
\begin{equation}
\Omega_\utot= \Omega + \Omega_\Lambda
\end{equation}
est inf\'erieur \`a l'unit\'e $\Omega_{\utot_\zero} < 1$ alors
$\kc=-1$ et les sections spatiales sont de type
hyperboliques\footnote{On choisit d'indexer par ``0'' les quantit\'es
pr\'esentes.}, de m\^eme qu'une valeur sup\'erieure \`a l'unit\'e est
caract\'eristique d'un univers ferm\'e $\kc=1$, alors que le cas
euclidien est obtenu pour $\Omega_{\utot_\zero}=1$.

\subsection{L'expansion de l'univers}
\label{sectionexpansion}
La caract\'eristique majeure du mod\`ele de FLRW est certainement la
non staticit\'e de l'univers introduite au travers du facteur
d'\'echelle $a$. En effet, la m\'etrique (\ref{metriqueFLRW}), ainsi
que les solutions g\'en\'eriques aux \'equations de Friedmann
(\ref{friedmann0}) et (\ref{friedmann1}), apparaissent comme des
fonctions du temps. C'est \`a partir de la compr\'ehension de la
nature cosmologique de la d\'ecouverte du d\'ecalage syst\'ematique
vers le rouge des raies spectrales d'objets distants~\cite{slipher15},
par Hubble~\cite{hubble25,hubble26}, que l'expansion de l'univers a
acquis son statut de fait observationnel.

Pour des photons se propageant le long des g\'eod\'esiques radiales de
genre lumi\`ere $\ud s^2=0$, et \`a l'aide de la m\'etrique
(\ref{metriqueFLRW}), il vient
\begin{equation}
\frac{\ud t}{a} = \pm \frac{\ud r}{\sqrt{1-\kc r^2}}.
\end{equation}
Ainsi, une onde lumineuse ayant \'et\'e \'emise au temps cosmique
$t_\uem$ en $r_\uem$, et re\c cue aujourd'hui en
($t_\zero$,$r_\zero$) suivra la relation
\begin{equation}
\label{propphoton}
\int_{t_\uem}^{t_\zero}{\frac{\ud t}{a}} =
\int_{r_\uem}^{r_\zero}{\frac{\ud r}{\sqrt{1-\kc r^2}}}.
\end{equation}
De la m\^eme mani\`ere, un paquet d'ondes \'emis \`a l'instant suivant
$t_\uem + \delta t_\uem$ par la m\^eme source en
$r_\uem$ sera d\'etect\'e \`a l'instant $t_\zero + \delta t_\zero$ et
aura les m\^emes propri\'et\'es. Le membre de droite de
(\ref{propphoton}) ne d\'ependant que des coordonn\'ees comobiles
(fixes), on en d\'eduit
\begin{equation}
\int_{t_\uem}^{t_\zero}{\frac{\ud t}{a}}=\int_{t_\uem +
\delta t_\uem}^{t_\zero+\delta 
t_\zero}{\frac{\ud t}{a}},
\end{equation}
soit apr\`es quelques manipulations, et pour $\delta t \ll t$,
\begin{equation}
\label{retard}
\frac{\delta t_\uem}{a(t_\uem)}=\frac{\delta t_\zero}{a(t_\zero)}.
\end{equation}
La forme de la m\'etrique (\ref{metriqueFLRW}) introduit ainsi un ``effet
D\"oppler'' apparent dans la propagation des ondes lumineuses dans
l'espace-temps par le biais du facteur d'\'echelle
\begin{equation}
\frac{\lambda_\uem}{\lambda_\zero}=\frac{a(t_\uem)}{a(t_\zero)}.
\end{equation}
La quantit\'e physique couramment utilis\'ee en astronomie est le
d\'ecalage spectral vers le rouge $z$, ou \emph{redshift}, d\'efini
par
\begin{equation}
\label{redshift}
z=\frac{\lambda_\zero}{\lambda_\uem}-1 =
\frac{a(t_\zero)}{a(t_\uem)}-1.
\end{equation}
L'observation d'un d\'ecalage des raies spectrales des objets
cosmologiques vers les grandes longueurs d'onde est donc, dans le
cadre du mod\`ele FLRW, le r\'esultat de l'expansion de l'univers: $a$
est une fonction croissante du temps. D'un point de vue newtonnien, le
facteur d'expansion dilatant la distance physique $\ell \propto a $
entre deux points de coordonn\'ees comobiles fixes, ceux-ci semblent
s'\'eloigner l'un de l'autre \`a une vitesse $v \propto \dot{a}$ telle
que
\begin{equation}
\label{loihubble}
v = \frac{\dot{a}}{a} \ell = H \ell,
\end{equation}
induisant un ``effet D\"oppler'' dilatant les longueurs d'onde.
\begin{figure}
\begin{center}
\epsfig{file=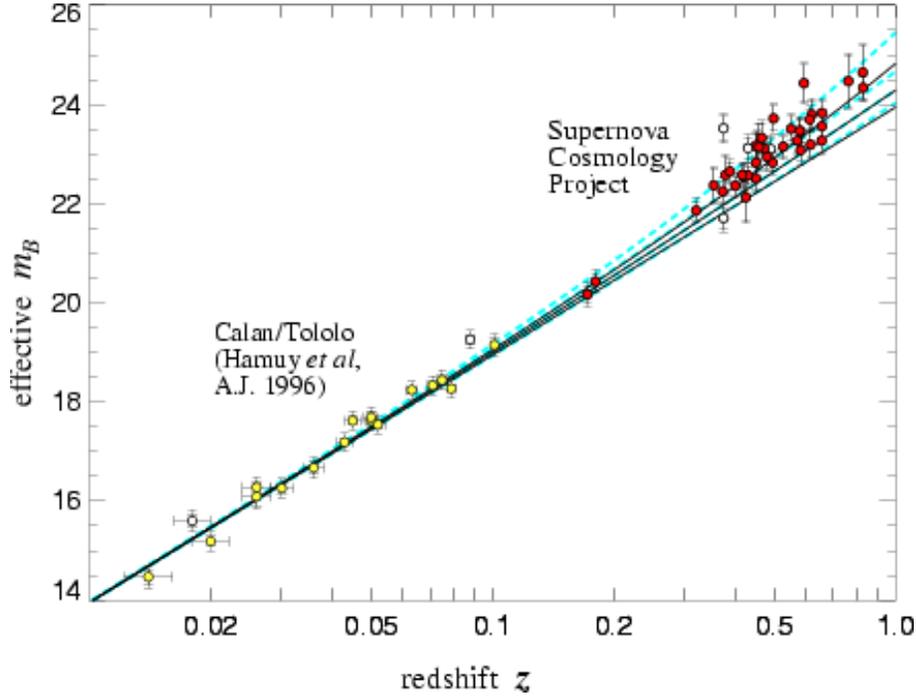,width=12cm}
\caption[Diagramme de Hubble repr\'esentant la distance luminosit\'e
en fonction du redshift.]{Diagramme de Hubble
repr\'esentant le logarithme de $H_\zero d_\uL$ en fonction du
redshift $z$. Ces donn\'ees sont en faveur d'une acc\'el\'eration de
l'expansion de l'univers~\cite{perlmutter99}.}
\label{figloihubble}
\end{center}
\end{figure}
La loi de Hubble (\ref{loihubble}) n'est cependant correcte que pour
des objets proches pour lesquels la notion de distance n'est que peu
affect\'ee par la g\'eom\'etrie de l'univers. Ce n'est plus le cas
pour des objets \'eloign\'es o\`u diff\'erentes distances peuvent
\^etre d\'efinies en fonction des diverses observables comme le
diam\`etre angulaire ou la luminosit\'e apparente. Cette derni\`ere
permet de d\'efinir la \emph{distance luminosit\'e} $d_\uL$ par
\begin{equation}
d_\uL^2 = \frac{\Lc}{4 \pi \Fc},
\end{equation}
$\Lc$ \'etant la luminosit\'e absolue, i.e. la puissance totale
intrins\`eque de la source, et $\Fc$ le flux mesur\'e, i.e. la
puissance par unit\'e de surface re\c cue. Dans le cas d'une
m\'etrique de type Minkowski, $d_\uL$ est simplement la distance
physique \`a la source, alors que dans le cadre du mod\`ele FLRW, on
s'attend \`a ce qu'elle d\'epende de l'expansion de
l'univers. Intuitivement, pour une source situ\'ee \`a la coordonn\'ee
comobile $r_\uem$, la dilution de l'\'energie par l'expansion
introduit un facteur $(1+z)$ par dilatation des longueurs d'onde et un
autre d\^u \`a la dilatation des temps d'\'emission (\ref{retard}). La
surface actuelle de la sph\`ere sur laquelle se dilue l'\'energie
\'emise par la source\footnote{On fait l'hypoth\`ese que l'observateur
est en $r_\zero$=0.} est, d'apr\`es (\ref{metriqueFLRW}), $\Sc = 4\pi
a^2_\zero r_\uem^2$. Le flux re\c cu est \'egal \`a $\Fc =
\Lc/\left[\Sc (1+z)^2\right]$ et on en d\'eduit la distance
luminosit\'e
\begin{equation}
d_\uL = a_\zero r_\uem (1+z).
\end{equation}
La coordonn\'ee comobile $r_\uem$ de la source peut \^etre exprim\'ee
en fonction des temps d'\'emission et de r\'eception par
(\ref{propphoton}). De plus, en d\'eveloppant le redshift
(\ref{redshift}) en $t_\uem-t_\zero$ pour des sources proches,
\begin{equation}
z \simeq H_\zero(t_\zero-t_\uem)+
\left(1+\frac{q_\zero}{2}\right)(t_\zero-t_\uem)^2,
\end{equation}
et en inversant cette relation, la distance luminosit\'e se
d\'eveloppe en fonction du redshift
\begin{equation}
\label{loihubblelum}
d_\uL \simeq \frac{z}{H_\zero} + \frac{1-q_\zero}{2 H_\zero} z^2.
\end{equation}
La quantit\'e $q_\zero=-\ddot{a}_\zero a_\zero/\dot{a_\zero}^2$ est le
param\`etre de d\'ec\'el\'eration, et s'exprime en fonction du
param\`etre de Hubble
\begin{equation}
q_\zero=-\frac{\Hc_\zero'}{\Hc_\zero^2}.
\end{equation}
\`A l'aide de l'\'equation de Friedmann (\ref{friedmann1}), avec
$P_\zero=0$ dans l'\`ere de mati\`ere, il vient
\begin{equation}
\label{paramdec}
q_\zero=\frac{\Omega_\zero}{2}-\Omega_{\Lambda_\zero}.
\end{equation}
La relation (\ref{loihubblelum}) est la version ``correcte'' de la loi
de Hubble (\ref{loihubble}), et au premier ordre en redshift, on
retrouve la proportionnalit\'e $z \simeq H_0 d_\uL$ (voir
Fig.~\ref{figloihubble}). Le param\`etre de d\'ec\'el\'eration
quantifie donc les \'ecarts \`a la loi de Hubble. Sa d\'etermination
est de plus un moyen de mesure des param\`etres cosmologiques dont la
densit\'e d'\'energie associ\'ee \`a la constante cosmologique [voir
Eq.~(\ref{paramdec})].

\begin{figure}
\begin{center}
\epsfig{file=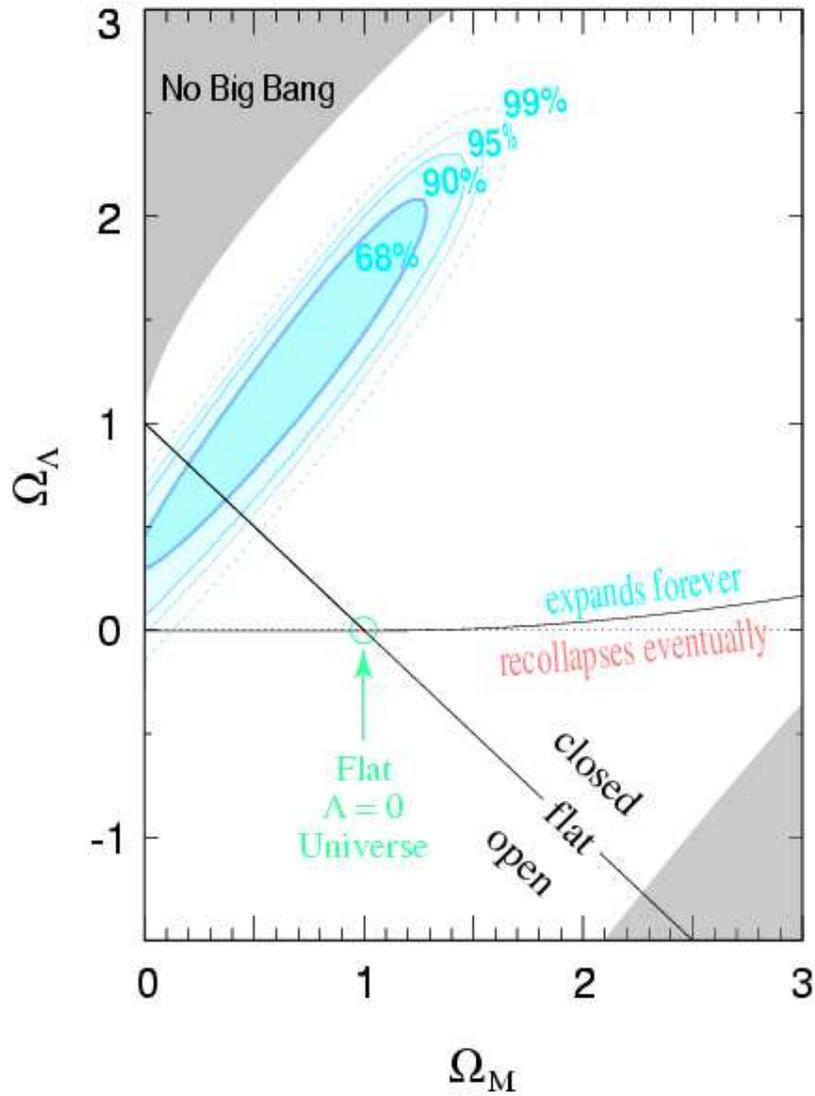,width=11cm}
\caption[Les param\`etres de densit\'e actuels.]{D\'etermination des
param\`etres de densit\'e \`a partir de l'\'evolution de la distance
luminosit\'e pour les supernovae de type Ia. Dans le cadre du mod\`ele
standard, l'acc\'el\'eration de l'expansion de l'univers r\'esulte de
la pr\'edominance actuelle de la constante
cosmologique~\cite{perlmutter99}.}
\label{figparamcosmo}
\end{center}
\end{figure}

La d\'etermination de $d_\uL$ comme une fonction du redshift a fait
l'objet de multiples m\'ethodes astrophysiques exploitant les
propri\'et\'es intrins\`eques des
sources~\cite{leavitt,baade,faber79,dressler,collins}
(c\'eph\'e\"{\i}des, relations masse-luminosit\'e des galaxies,
supernovae de type Ia\footnote{SNIa.}
\dots). La figure~\ref{figloihubble} repr\'esente des r\'esultats
relativement r\'ecents dans la d\'etermination de $d_\uL(z)$ par les
courbes de luminosit\'es des SNIa~\cite{perlmutter99}. Les valeurs
estim\'ees des param\`etres cosmologiques correspondants sont
repr\'esent\'es sur la figure~\ref{figparamcosmo}.  La
d\'eg\'en\'erescence sur la d\'etermination de la densit\'e
d'\'energie associ\'ee \`a la mati\`ere peut \^etre lev\'ee par
l'utilisation d'autres donn\'ees
astrophysiques~\cite{boomerang,maxima,maxima2}. Ces r\'esultats
confirment de fa\c con \'eclatante le mod\`ele standard et semble
indiquer, aujourd'hui, que la nature du fluide cosmologique fait
intervenir une proportion importante de constante cosmologique
\begin{equation}
\Omega_{\Lambda_\zero} \simeq 0.7, \qquad \Omega_\zero \simeq 0.3,
\end{equation}
indiquant que nous sommes actuellement dans une phase d'expansion
acc\'el\'er\'ee d'un univers \`a faible courbure $\Omega_\utot
\simeq 1$.

\subsection{Histoire thermique}

La cons\'equence majeure de l'expansion de l'univers est que le
facteur d'\'echelle d\'ecro\^{\i}t lorsqu'on remonte dans le
temps. Cette constatation implique, d'apr\`es les \'equations
(\ref{densmat}) et (\ref{redshift}), que l'univers \'etait plus dense
et plus chaud par le pass\'e, menant ainsi au concept de ``big bang''
chaud pour d\'esigner l'instant initial o\`u le facteur d'\'echelle
pourrait s'annuler. La domination de la mati\`ere n'est donc
certainement plus une hypoth\`ese acceptable lorsqu'on remonte
suffisamment loin dans le temps.

Si l'on consid\`ere, en effet, un fluide de radiation (ou de
particules ultrarelativistes pour lesquelles la masse est
n\'egligeable devant l'impulsion), le terme de pression n'est plus
n\'egligeable
\begin{equation}
P_\urad=\frac{1}{3} \rho_\urad,
\end{equation}
et l'\'equation de conservation (\ref{bianchi3}) donne une \'evolution
de la densit\'e d'\'energie en
\begin{equation}
\label{rhorad}
\rho_\urad \propto a^{-4}.
\end{equation}
De mani\`ere g\'en\'erale, pour une \'equation d'\'etat
(\ref{etatunivers}) avec $w$ constant, (\ref{bianchi3}) s'int\`egre et
la densit\'e d'\'energie varie en
\begin{equation}
\label{rhow}
\rho \propto a^{-3(1+w)}.
\end{equation}
Il existe donc un instant $t_\ueq$ o\`u les densit\'es d'\'energie
associ\'e \`a la mati\`ere et au rayonnement \'etaient comparables,
s\'eparant l'\`ere de radiation pour $t<t_\ueq$ o\`u $\rho \simeq
\rho_\urad$, de l'\`ere de mati\`ere o\`u $\rho \simeq
\rho_\umat$\footnote{Le redshift d'\'equivalence est estim\'e \`a
$z_\ueq \simeq 10^4$.}. Il est possible d'introduire une temp\'erature
associ\'ee \`a l'\'etat thermodynamique du gaz de particule alors en
interaction. Si l'on se place dans des conditions de pression telles
que l'\'equilibre thermodynamique local soit v\'erifi\'e, la densit\'e
d'\'energie peut \^etre obtenue \`a partir des statistiques de
Fermi-Dirac ou de Bose-Einstein selon le spin des particules, et la
temp\'erature $\Theta$ d\'efinie au travers de la fonction de
distribution dans l'espace des phases. Dans ce cas, la densit\'e
d'\'energie est donn\'ee par
\begin{equation}
\label{fdistrib}
\rho = \varsigma \int{\ud^3 \vec{p}\,\,E f\left(\vec{p}\right)} =
\varsigma \int{\ud^3 \vec{p}\,\, 
\delta_\ue\frac{E}{\ue^{(E-\chi)/\Theta} +
\varepsilon}},
\end{equation}
avec $E^2=p^2+m^2$ l'\'energie totale de chaque particule, $\varsigma$
le facteur de d\'eg\'en\'erescence associ\'ee \`a
$\delta_\ue=1/(2\pi)^3$ la densit\'e d'\'etat, $\chi$ le potentiel
chimique\footnote{$\chi=0$ \`a l'\'equilibre chimique.}, et
$\varepsilon=\pm 1$ pour des fermions et bosons respectivement. Lors
de l'\`ere domin\'ee par la radiation, pour des particules
ultra-relativistes telles que $E \simeq p$, en \'equilibre chimique,
il vient
\begin{equation}
\rho_{\mathrm{B}}=\varsigma_{\mathrm{B}} \frac{\pi^2}{30} \Theta^4 \qquad
\rho_{\mathrm{F}} = \frac{7}{8} \varsigma_{\mathrm{F}}
\frac{\pi^2}{30} \Theta^4,
\end{equation}
pour les bosons et les fermions respectivement\footnote{Le facteur $7/8$
vient du principe d'exclusion de Pauli agissant au travers de
$\varepsilon=1$.}. Ainsi, pour le fluide cosmologique constitu\'e de
mati\`ere baryonique usuelle et de rayonnement, la densit\'e totale
d'\'energie est li\'ee \`a la temp\'erature par
\begin{equation}
\label{rhothermal}
\rho = \varsigma \frac{\pi^2}{30} \Theta^4,
\end{equation}
avec
\begin{equation}
\varsigma = \sum_{\mathrm{B}} \varsigma_{\mathrm{B}} +
\frac{7}{8} \sum_{\mathrm{F}} \varsigma_{\mathrm{F}}.
\end{equation}
La somme ne porte que sur les esp\`eces effectivement relativistes \`a
la temp\'erature $\Theta$, et il est raisonnable de n\'egliger
l'\'energie associ\'ee aux esp\`eces non relativistes. Les \'equations
(\ref{rhorad}) et (\ref{rhothermal}) montrent que l'univers en
expansion se refroidit comme on pouvait s'y attendre. Ainsi, au fur et
\`a mesure du refroidissement de l'univers, les esp\`eces vont de moins en
moins interagir et vont finir par se d\'ecoupler de l'\'equilibre
thermodynamique, essentiellement lorsque leur libre parcours moyen
$\Gamma^{-1} \simeq H^{-1}$, avec $H$ le param\`etre de
Hubble\footnote{$H^{-1}$ est un ordre de grandeur de la distance au
del\`a de laquelle les r\'egions sont causalement disjointes, et
$\Gamma$ est le taux d'interactions.}. D\'ependant de la nature des
esp\`eces pr\'esentes, le facteur de d\'eg\'en\'erescence va donc
subir de brusques d\'ecroissances au cour de l'expansion jusqu'au
dernier d\'ecouplage, celui des photons.

\begin{figure}
\begin{center}
\epsfig{file=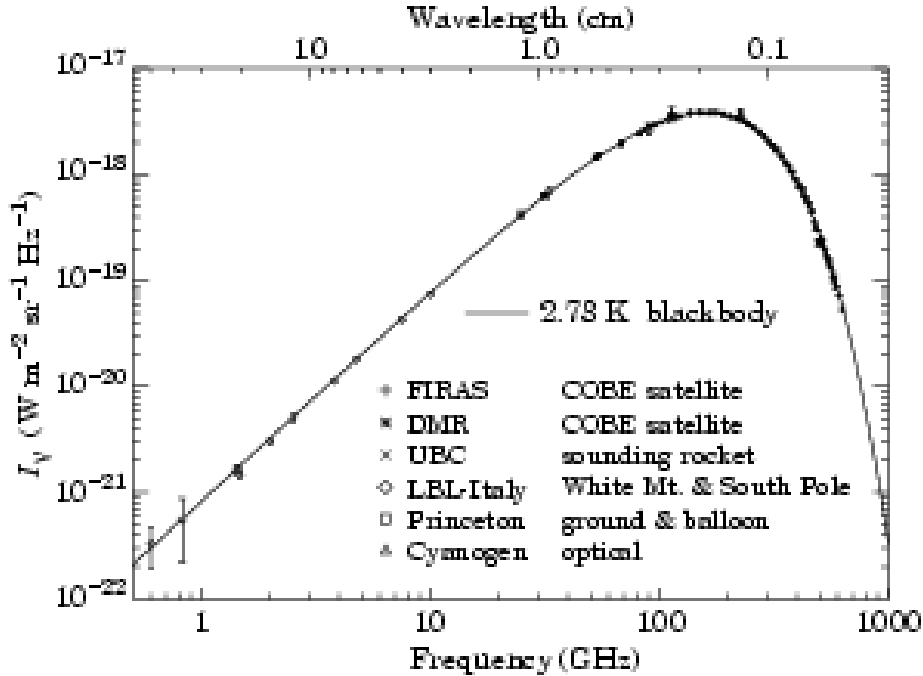,height=9cm}
\caption[Le spectre de corp noir du rayonnement fossile.]{Le spectre
du CMBR mesur\'e par diff\'erentes exp\'eriences~\cite{groom00}.}
\label{figfiras}
\end{center}
\end{figure}

Si l'on s'int\'eresse au devenir des esp\`eces ultra-relativistes
d\'ecoupl\'ees de l'\'equilibre, la relation (\ref{rhothermal}) n'est
\emph{a priori} plus valable, mais du fait de l'expansion, leur
\'energie est att\'enu\'ee d'un facteur $a_\udec/a$ [voir
Eq.~(\ref{redshift})] pendant que la densit\'e de particules $n$
d\'ecro\^{\i}t en $1/a^3$ par dilatation des longueurs. La fonction de
distribution $f\left(\vec{p}\right) \propto \ud^3n/\ud \vec{p}^{\, 3}$
reste donc invariante et \'egale \`a $f_\udec$. Ceci implique,
d'apr\`es (\ref{fdistrib}), que $E/\Theta$ est constant, soit
\begin{equation}
\label{tempdec}
\Theta = \Theta_\udec \frac{a_\udec}{a}.
\end{equation}
De la m\^eme mani\`ere, pour des esp\`eces non relativistes,
l'impulsion subissant toujours le redshift (\ref{redshift}), la
fonction de distribution reste invariante apr\`es d\'ecouplage, et $E$
d\'ecro\^{\i}t d'un facteur $a^2$ menant \`a une variation de la
temp\'erature en $\Theta_{\mathrm{m}} \propto a^{-2}$. La
cons\'equence fondamentale de ce sc\'enario est l'existence actuelle
d'un fond de rayonnement diffus de photons dans tout l'univers
v\'erifiant une distribution de corps noir \`a la temp\'erature
(\ref{tempdec}). Ce fond de rayonnement d'origine
cosmologique\footnote{CMBR} a \'et\'e effectivement d\'etect\'e par
Penzias et Wilson~\cite{penzias65}, et v\'erifie une loi de corps noir
\`a la temp\'erature actuelle\footnote{C'est le meilleur corps noir
actuellement connu.} de $2.728 \pm 0004 \Kel$ (voir Fig.~\ref{figfiras}).
L'origine physique du rayonnement fossile est dans le d\'ecouplage
entre la mati\`ere et les photons essentiellement d\^u \`a la
recombinaison entre les \'electrons et les protons. Le redshift de la
surface de derni\`ere diffusion, i.e. le redshift \`a partir duquel
l'univers est devenu transparent aux photons, peut ainsi \^etre
estim\'e par celui correspondant \`a la recombinaison, soit $z_\urec
\simeq 1300$, et $\Theta_\urec \simeq 3500 \Kel$ ou une
\'energie\footnote{$1 \eV \simeq 11604 \Kel$.} voisine de $0.3
\eV$~\cite{kolbturner}.
 
\begin{figure}
\begin{center}
\epsfig{file=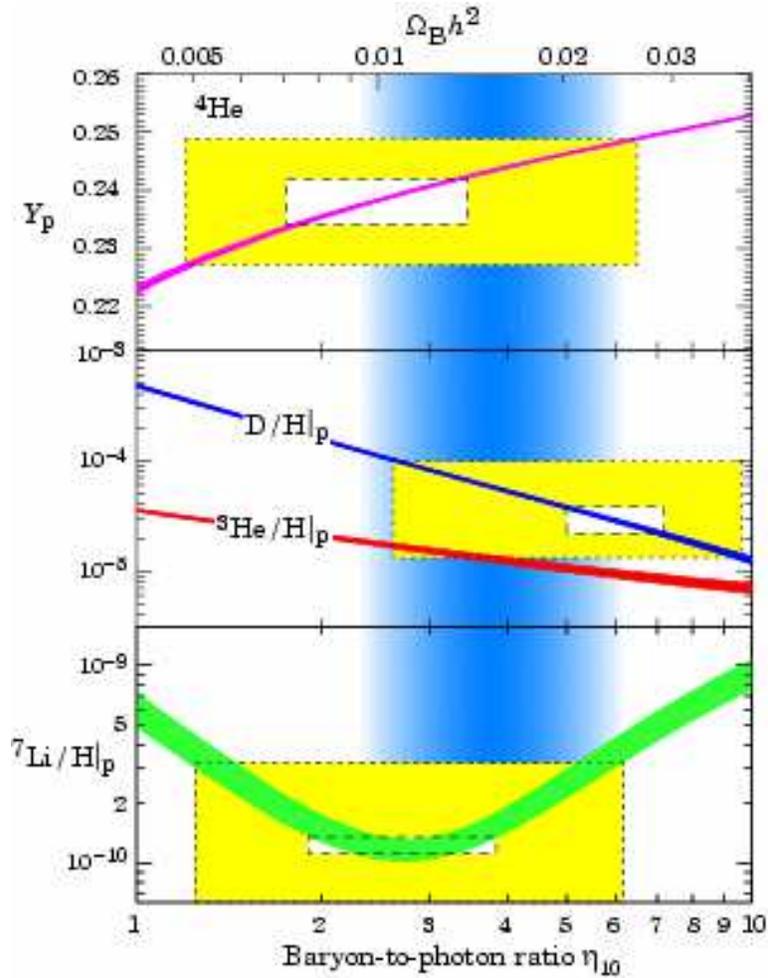,height=13cm}
\caption[Abondance des \'el\'ements l\'egers issus de la
nucl\'eosynth\`ese primordiale.]{Pr\'ediction \`a $95 \% $ de
confiance de l'abondance des \'el\'ements l\'egers issus de la
nucl\'eosynth\`ese primordiale en fonction du rapport de la densit\'e
baryonique \`a la densit\'e de rayonnement ($\eta_{10}$). Les diverses
observables sont repr\'esent\'es par des bo\^{\i}tes rectangulaires et
sont clairement en accord avec les
pr\'edictions~\cite{nollett00,groom00}.}
\label{fignucleo}
\end{center}
\end{figure}

L'histoire thermique de l'univers peut ainsi \^etre comprise en
s'approchant de la singularit\'e initiale \`a partir de la
connaissance de la microphysique dominante aux \'echelles d'\'energie
en question. \`A des \'echelles d'\'energie de l'ordre de $\Theta
\simeq 0.1-10 \MeV$, les r\'eactions nucl\'eaires prennent place dans
le plasma et sont \`a l'origine de la nucl\'eosynth\`ese
primordiale. Tester les cons\'equences de celle-ci revient \`a
v\'erifier la validit\'e du sc\'enario du big bang chaud \`a
l'\'epoque la plus recul\'ee qui nous soit actuellement accessible. En
particulier, les abondances relatives des diff\'erents \'el\'ements
produit par ces r\'eactions nucl\'eaires peuvent \^etre calcul\'ees et
sont en accord avec les observations (voir Fig.~\ref{fignucleo}).  Pour
des temp\'eratures de l'ordre de $\Theta \simeq 1 \MeV$ le mod\`ele
standard pr\'evoit \'egalement le d\'ecouplage des neutrinos du
plasma, leur temp\'erature d\'ecro\^{\i}t alors en $a^{-1}$, comme les
photons alors encore en \'equilibre thermique [cf.~Eqs.~(\ref{rhorad})
et (\ref{rhothermal})]. Cependant, le rayonnement de photons subit un
r\'echauffement par annihilation \'electrons positrons. Il est
possible de montrer\footnote{La conservation de l'\'energie
(\ref{bianchi3}) impose \`a l'entropie par unit\'e de volume comobile,
associ\'e \`a l'\'equilibre thermique des photons
$s_{\mathrm{ph}}=(\rho_{\mathrm{ph}} + P_{\mathrm{ph}})/\Theta
\propto \varsigma (a \Theta)^3$, d'\^etre conserv\'e au cour de
l'expansion. Celle-ci \'etant proportionnelle au facteur de
d\'eg\'en\'erescence, qui passe de $\varsigma=2 + 2
\times 2 \times 7/8$ (photons, \'electrons et positrons), \`a
$\varsigma=2$ (photons), la temp\'erature des photons augmente donc
d'un facteur $(11/4)^{1/3}$, le facteur d'expansion $a$ \'etant
continu lors de la transition.} que la conservation d'entropie
g\'en\`ere sur le fluide de photons un accroissement de la
temp\'erature d'un facteur $(11/4)^{1/3}$. Puisque les neutrinos ne
subissent pas cet effet c'est donc \'egalement le rapport de
temp\'erature entre les deux fluides apr\`es la recombinaison et il
vient, pour l'\'epoque actuelle, $\Theta_{\nu_\zero} \simeq 1.95 \Kel$.

Aux \'echelles d'\'energie au del\`a de $100 \MeV$, la description de
l'univers n\'ecessite l'emploi de la physique des hautes
\'energies que nous verrons dans le chapitre suivant. Quoiqu'il en
soit, il existe une limite absolue au mod\`ele standard qui est
l'\'epoque de Planck $\Theta_\Pl \simeq 10^{19} \GeV$ au del\`a de
laquelle il est peu raisonnable\footnote{Mais non exclu.} de supposer
que la relativit\'e g\'en\'erale et les autres interactions ne
constituent plus une description correcte de la Nature.

\section{Les faiblesses du sc\'enario standard}

La validit\'e du mod\`ele standard de cosmologie ainsi bri\`evement
d\'ecrit repose donc essentiellement sur l'observation du d\'ecalage
vers le rouge des objets distants, de l'existence du fond diffus
cosmologique et des abondances des \'el\'ements l\'egers pr\'edites par
la nucl\'eosynth\`ese primordiale. Bien qu'il soit possible de
construire des th\'eories alternatives pour chacune de ces
observations s\'epar\'ement~\cite{bondi48,dicke70,hoyle48,milne34}, le
mod\`ele du big bang chaud reste le plus simple et complet pour
d\'ecrire l'univers jusqu'\`a des
\'energies de l'ordre de la dizaine de$\MeV$.

Cependant, quelques observations ne trouvent pas d'explications
raisonnables dans le cadre du mod\`ele standard et doivent \^etre
d\'ecrites par l'interm\'ediaire de param\`etres \emph{ad hoc}. Comme
nous le verrons dans le chapitre~\ref{chapitreaudela}, elles ne
peuvent \^etre correctement interpr\'et\'ees que dans le cadre de
l'univers primordial, \`a des \'echelles d'\'energie bien au del\`a de
la nucl\'eosynth\`ese primordiale, n\'ecessitant l'extrapolation de la
physique des hautes
\'energies.

\subsection{L'univers plat}
\label{sectionplat}
Dans la section~\ref{sectionexpansion}, nous avons vu que les valeurs
mesur\'ees des param\`etres cosmologiques montrent que l'univers est
actuellement de faible courbure $\Omega_\utot \simeq 1$
(cf. Fig.~\ref{figparamcosmo}). Pour une \'equation d'\'etat avec $w$
constant, \`a partir des \'equations (\ref{friedmann1}) et
(\ref{omegafctH}), il vient
\begin{equation}
\label{omegaevol}
\Omega_\utot'=\Hc \left[(1+3w) \Omega - 2 \Omega_\Lambda
\right] \left (\Omega_\utot - 1 \right).
\end{equation} 
La solution $\Omega_\utot = 1$ est soit instable, soit un attracteur,
selon le signe de $(1+3w) \Omega - 2 \Omega_\Lambda$. Dans un univers
en expansion $\Hc > 0$, et actuellement avec $w=0$ et
$\Omega_{\Lambda_\zero} \simeq 0.7$, $\Omega_\zero \simeq 0.3$, cette
solution est un attracteur. Cependant, il est facile de voir \`a
partir des \'equations (\ref{rhocrit}) \`a (\ref{rholambda}),
(\ref{redshift}) et (\ref{rhow}) que
\begin{equation}
\label{damplambda}
\frac{\Omega_\Lambda}{\Omega}=\frac{\Omega_{\Lambda_\zero}}{\Omega_\zero}
\,  (1+z)^{-3(1+w)},
\end{equation}
et la domination du terme de constante cosmologique s'efface pour $z
\simeq 0.7$. Il en r\'esulte que pour des redshifts $z \gtrsim 1$, la
platitude de l'univers est instable~\cite{uzan01,goliath99}. Ceci
reste valable dans l'\`ere de radiation avec $w=1/3$. Ainsi, pour que
nous observions aujourd'hui $\Omega_{\utot_\zero} \simeq 1$, il a
fallu qu'au moment de l'\'equivalence, pour $z_\ueq \simeq 10^4$, le
param\`etre de densit\'e soit \'egal \`a l'unit\'e \`a $10^{-3}$
pr\`es, et \`a $10^{-60}$ pr\`es \`a l'\'epoque de
Planck\footnote{L'univers \'etait donc extr\^emement proche du cas
euclidien $\kc=0$ dans l'univers primordial.}.

Un tel ajustement des conditions initiales\footnote{\emph{Fine
tunning}.} nuit \`a la g\'en\'ericit\'e du mod\`ele et refl\`ete
certainement la pr\'esence d'autres ph\'enom\`enes physiques dans
l'univers primordial que nous
\'evoquerons au chapitre~\ref{chapitreaudela}.

\subsection{L'homog\'en\'eit\'e}

L'hypoth\`ese d'homog\'en\'eit\'e de l'univers est contenue dans le
principe cosmologique qui nous a permis d'\'ecrire la m\'etrique FLRW
(\ref{metriqueFLRW}). Elle est donc \`a la base du mod\`ele standard
et confirm\'ee par les observations. On peut n\'eanmoins
s'interroger sur les m\'ecanismes physiques \`a l'origine de cette
homog\'en\'eit\'e. La fraction d'univers observable actuellement peut
\^etre quantifi\'ee \`a partir de la m\'etrique
(\ref{metriqueFLRW}). Pour un observateur actuel situ\'e au rayon
comobile $r_\zero=0$, les photons \'emis \`a $\eta_\uem=0$, en la
position $r_\uH$ l'atteindront en un temps conforme $\eta$ tel que
\begin{equation}
\label{poshorizon}
\int_{\eta_\uem}^{\eta} \ud \eta' = \pm \int_{r_\uH}^{r_\zero}
\frac{\ud r}{\sqrt{1-\kc r^2}}. 
\end{equation}
Autrement dit, au temps conforme $\eta$, \emph{l'horizon des
particules} est pr\'ecis\'ement \`a la coordonn\'ee comobile $r_\uH$,
les photons situ\'es au del\`a de $r_\uH$ n'ont donc pas encore,
\`a $\eta$, atteint l'observateur. La distance propre \`a l'horizon
est alors, d'apr\`es (\ref{metriqueFLRW}) et (\ref{poshorizon})
\begin{equation}
\label{disthorizonconf}
d_\uH(\eta)= a(\eta) \int_0^{r_\uH} \frac{\ud r}{\sqrt{1-\kc
r^2}}=a(\eta)\eta,
\end{equation}
ou, exprim\'ee en fonction du temps cosmique
\begin{equation}
\label{disthorizon}
d_\uH(t)=a(t)\int_0^{t}\frac{\ud t'}{a(t')}.
\end{equation}
Cette derni\`ere \'equation montre plus clairement que la distance \`a
l'horizon peut \^etre infinie. Dans le cadre du mod\`ele standard,
pour $w$ constant et $w>-1$, l'\'equation de Friedmann
(\ref{friedmann0}) s'int\`egre \`a l'aide de (\ref{rhow}) pour donner
l'\'evolution du facteur d'\'echelle\footnote{En n\'egligeant la
constante cosmologique pour $z>1$ [cf. Eq.~(\ref{damplambda})].}
\begin{eqnarray}
\label{evolechelle}
a(t) \propto t^{\frac{2}{3(1+w)}} & \Leftrightarrow & a(\eta) \propto
\eta^{\frac{2}{1+3w}}.
\end{eqnarray}
Si $w<-1/3$, l'\'equation (\ref{disthorizon}) diverge en z\'ero et la
distance \`a l'horizon est infinie. Inversement, pour $w>-1/3$, elle
s'int\`egre et $d_\uH$ est une fonction du temps\footnote{On a
\'egalement dans ce cas $d_\uH \propto H^{-1}$.}
\begin{eqnarray}
\label{evolhorizon}
d_\uH(t) \propto t  & \Leftrightarrow & d_\uH(\eta) \propto
\eta^{\frac{3+3w}{1+3w}}.
\end{eqnarray}
Il en r\'esulte que l'univers observable aujourd'hui contient des
r\'egions causalement disjointes du pass\'e, la distance \`a l'horizon
s'accroissant avec le temps cosmique. Ainsi, la distance propre
actuelle \`a la surface de derni\`ere diffusion\footnote{Elle
correspond au d\'ecouplage des photons \`a $\eta_\udec$.} est donn\'ee
par (\ref{poshorizon}) et (\ref{disthorizonconf}), avec
$\eta_\uem=\eta_\udec$, i.e.
\begin{equation}
d_{\udec}(\eta_\zero)=a_\zero \left(\eta_\zero-\eta_\udec\right).
\end{equation}
et donc au moment du d\'ecouplage, elle correspondait \`a une distance
caract\'eristique de l'ordre de
$d_{\udec}(\eta_\udec)=(1+z_\udec)^{-1}d_{\udec}(\eta_\zero)$. Il est
possible de la comparer \`a la distance \`a l'horizon \`a cette
\'epoque, i.e. la distance propre au del\`a de laquelle les r\'egions
de l'univers \`a cette \'epoque n'\'etaient plus causalement
connect\'ees. D'apr\`es (\ref{disthorizonconf}), celle-ci se r\'eduit
\`a $d_{\uH_\udec}=a_\udec \eta_\udec$. Le nombre de cellules
projet\'ees sur le ciel aujourd'hui et causalement d\'econnect\'es
lors de la recombinaison est donc approximativement
\begin{equation}
n_\udec \simeq \left[
\frac{d_{\udec}(\eta_\udec)}{d_{\uH_\udec}}
\right]^{3} = \left(
\frac{\eta_\zero-\eta_\udec}{\eta_\udec} \right)^{3},
\end{equation}
soit $n_\udec \simeq 10^5$ pour $z_\udec \simeq z_\urec \simeq
1300$. Inversement, la distance \`a l'horizon \emph{actuelle} pour une
cellule causale de la surface de derni\`ere diffusion est
$d_{\uH_\udec}(\eta_\zero) = (1+z_\udec)d_{\uH_\udec}$, elle est donc
actuellement vue sous un angle
\begin{equation}
\delta \theta \simeq \frac{d_{\uH_{\udec}}(\eta_\zero)}
{d_{\udec}(\eta_\zero)} = \frac{\eta_\udec}{\eta_\zero - \eta_\udec}
\simeq 0.8^{\circ}.
\end{equation}
L'homog\'en\'eit\'e entre ces r\'egions causalement d\'econnect\'ees
lors de la recombinaison est observ\'ee aujourd'hui dans l'isotropie
du spectre du CMBR (cf. Fig.~\ref{figfiras}) \`a $10^{-4}$ pr\`es. Le
mod\`ele standard ne permet pas de d'expliquer de telles
corr\'elations entre ces diff\'erentes r\'egions du ciel.

\subsection{La formation des structures}
\label{sectionstruct}
La pr\'esence de structures gravitationnelles non lin\'eaires aux
petites \'echelles telles que les galaxies et les amas est
interpr\'et\'ee comme le r\'esultat de la croissance
d'inhomog\'en\'eit\'es de densit\'e initiales par instabilit\'e de
Jeans (voir annexe~\ref{annexeaa}). Dans le cadre du mod\`ele de FLRW,
en l'absence de constante cosmologique (voir Sect.~\ref{sectionplat}),
l'\'equation (\ref{omegaevol}) s'int\`egre en fonction du facteur
d'\'echelle
\begin{equation}
\frac{\Omega -1}{\Omega} \propto a^{1+3w},
\end{equation}
qui peut se r\'e\'ecrire
\begin{equation}
\frac{\delta \Omega}{\Omega} \propto
a^{1+3w},
\end{equation}
avec $\delta \Omega= \Omega - 1$ repr\'esentant l'\'ecart du param\`etre
de densit\'e \`a sa valeur critique dans un univers plat. Autrement
dit, le contraste de densit\'e dans un univers plat d'une r\'egion
finie ayant une densit\'e sup\'erieure \`a la densit\'e critique,
i.e. $\Omega \gtrsim 1$, s'amplifie avec l'expansion de l'univers
jusqu'\`a entrer dans le r\'egime non-lin\'eaire menant aux structures
gravitationnelles observ\'ees. Dans le r\'egime lin\'eaire, il est
toujours possible de d\'evelopper le contraste de densit\'e $\delta
\rho / \rho$ en s\'erie de Fourier permettant de d\'efinir les
longueurs d'onde $\lambda$ associ\'ees \`a ces fluctuations. Dans le
mod\`ele de FLRW, celles-ci sont dilat\'ees par l'expansion, et les
structures observ\'ees actuellement de longueur caract\'eristique
$\lambda_\zero$ correspondent \`a des fluctuations ant\'erieures de
longueur d'onde
\begin{equation}
\label{evolonde}
\lambda = \frac{a}{a_\zero} \lambda_\zero.
\end{equation}
\`A partir des \'equations (\ref{evolechelle}) et (\ref{evolhorizon}),
la distance \`a l'horizon \'evolue \'egalement avec le facteur
d'\'echelle en
\begin{equation}
\label{evolhorizonechelle}
d_\uH \propto a^{\frac{3(1+w)}{2}}.
\end{equation}
En comparant (\ref{evolonde}) et (\ref{evolhorizonechelle}), on voit
donc qu'il existe un redshift pour lequel $\lambda > d_\uH$ quelque
soit $\lambda_\zero$. Il est alors impossible dans le cadre du
mod\`ele de FLRW de trouver un m\'ecanisme physique permettant de
fixer de telles conditions initiales sur des distances correspondant
\`a des r\'egions causalement disjointes, ou \emph{superhorizons}.
Ce probl\`eme, comme celui de l'homog\'en\'eit\'e, r\'esulte
directement de l'existence d'un horizon des particules et de la
d\'ec\'el\'eration de l'univers. Nous verrons dans le
chapitre~\ref{chapitreaudela} qu'il est pleinement r\'esolu par
l'inflation.

\section{Conclusion}

Le mod\`ele standard de la cosmologie fond\'e sur la gravitation
d'Einstein est donc remarquablement confirm\'e par les observations
pour des \'echelles d'\'energie allant de $10^{-3} \eV$ aujourd'hui \`a
quelques dizaines de$\MeV$ lors de la nucl\'eosynth\`ese
primordiale. Il attribue n\'eanmoins des propri\'et\'es inexpliqu\'ees
\`a notre univers, telle la platitude, l'acc\'el\'eration r\'ecente de
l'expansion, une homog\'en\'eit\'e ``magique''{\dots}
Ces probl\`emes
laissent entrevoir l'existence d'autres ph\'enom\`enes physiques \`a
l'\oe uvre dans l'univers primordial. L'\'echelle de Planck, qui
marque l'insuffisance suppos\'ee d'une description purement
g\'eom\'etrique de l'espace-temps, se trouvant \`a des \'energies de
l'ordre de $10^{19} \GeV$, il est raisonnable d'esp\'erer d\'ecrire
ces ph\'enom\`enes physiques, en de\c{c}a de cette \'energie, par la
physique des hautes \'energies couramment utilis\'ee dans les
exp\'eriences de physique des particules. Dans le prochain chapitre
nous pr\'esenterons donc bri\`evement le mod\`ele standard de la
physique des particules, avant de nous int\'eresser \`a ses
cons\'equences sur la cosmologie.

\chapter{Le mod\`ele standard de physique des particules}
\label{chapitrepp}
\minitoc
\section{Introduction}

Aux petites \'echelles de distances, par sa tr\`es faible intensit\'e,
la gravitation peut \^etre n\'eglig\'ee. Le monde microscopique
sond\'e dans les acc\'el\'erateurs de particules est ainsi r\'egi par
les trois autres interactions fondamentales de la Nature:
l'\'electromagn\'etisme, l'interaction faible et l'interaction
nucl\'eaire forte. Le mod\`ele standard de physique des particules
donne un cadre th\'eorique unifi\'e d\'ecrivant l'interaction de la
mati\`ere avec ces trois forces fondamentales. Pour cela, il s'appuie
sur la th\'eorie quantique des champs, issue de la relativit\'e
restreinte et de la m\'ecanique quantique, o\`u les objets
fondamentaux sont des particules, \'ev\'enements de l'espace-temps,
interagissant en respectant certaines sym\'etries et dont les
propri\'et\'es d\'ependent de l'\'echelle d'\'energie \`a laquelle ils
sont observ\'es. Le mod\`ele standard aujourd'hui comprend le mod\`ele
de Glashow, Weinberg et Salam~\cite{glashow61,weinberg67,salam64} qui
ont unifi\'e l'\'electrodynamique
quantique~\cite{feynmann49,dyson49,schwinger51} et l'interaction
faible en une th\'eorie de jauge fond\'ee sur la sym\'etrie $SU(2)
\times U(1)$, \`a l'aide du m\'ecanisme de
Higgs~\cite{englert64,higgs64,guralnik64,kibble67}. Le m\'ecanisme de
Higgs\footnote{Dans toute la suite, conform\'ement \`a l'usage, ce
m\'ecanisme d\'ecouvert simultan\'ement par Englert et
Brout~\cite{englert64}, Higgs~\cite{higgs64}, et Guralnik, Hagen et
Kibble~\cite{guralnik64}, sera ainsi d\'enom\'e.} est ainsi utilis\'e
en physique des particules pour unifier les interactions et donner une
explication \`a la masse des particules. Son existence est motiv\'ee
par l'observation de ses cons\'equences\footnote{Et peut
\^etre m\^eme directement par la d\'etection de sa particule associ\'e
\`a $115 \GeV$~\cite{L301}.}, et les pr\'edictions largement
confirm\'ees faites par ce mod\`ele. Les g\'en\'eralisations de ce
m\'ecanisme sont intens\'ement utilis\'ees en cosmologie lors des
transitions de phase, il est en outre \`a l'origine de la formation
des cordes cosmiques par le m\'ecanisme de
Kibble~\cite{kibble76}. L'interaction forte pouvant elle aussi \^etre
d\'ecrite par une th\'eorie de jauge s'appuyant sur la sym\'etrie
$SU(3)$, la chromodynamique quantique, on parle alors de mod\`ele
standard $SU(3) \times SU(2) \times U(1)$.

\section{Les th\'eories de jauge}

La repr\'esentation des invariances impos\'ees par la relativit\'e
restreinte au travers du groupe de Poincar\'e se trouve r\'ealis\'ee
en m\'ecanique quantique par la diff\'erenciation des particules selon
leur masse et leur spin. Ces sym\'etries de base de l'espace-temps
repr\'esent\'ees dans l'espace de Hilbert de la m\'ecanique quantique
sont effectivement r\'ealis\'ees dans la Nature par les bosons,
particules de spin entier, et les fermions de spin demi-entier. Les
particules observ\'ees se r\'eduisent ainsi aux scalaires $\Phi$ de
spin nul, aux fermions $\Psi$ de spin $1/2$ et aux champs vectoriels
$V^{\mu}$ de spin unit\'e\footnote{Le graviton de spin $2$ pourrait
\^etre \'egalement introduit, mais il n'existe pas \`a
l'heure actuelle de th\'eorie quantique pr\'edictive de la
gravitation.}. En plus des sym\'etries de base de l'espace-temps \`a
l'origine des propri\'et\'es intrins\`eques de la mati\`ere, la
confirmation exp\'erimentale du mod\`ele standard sugg\`ere que les
interactions entre ces particules sont elles aussi le produit de
l'existence de sym\'etries locales: les interactions sont en effet
invariantes sous certaines transformations de coordonn\'ees et des
champs. Des charges peuvent \^etre d\'efinies pour toutes les
particules quantifiant leur comportement face \`a ces sym\'etries
locales, on construit alors une th\'eorie, dite ``de jauge'',
quantifi\'ee et
pr\'edictive~\cite{weyl18,yang54,feynman63,dewitt67,fadeev67}.

\subsection{L'\'electrodynamique}

La th\'eorie r\'egissant les interactions \'electromagn\'etiques est
bas\'ee sur la sym\'etrie de jauge locale $U(1)$, i.e. l'invariance de
l'interaction par la multiplication des champs par une phase complexe
d\'ependant de l'\'ev\'enement. Cette sym\'etrie est une des plus
simple continue, car ab\'elienne, permettant de construire une
th\'eorie de jauge et servira de mod\`ele de pr\'edilection pour
l'\'etude des cordes cosmiques dans les chapitres suivants.

\`A l'aide du tenseur de Faraday $F^{\mu \nu}$, les \'equations de
Maxwell~\cite{maxwell} peuvent \^etre r\'eexprim\'ees sous leur forme
covariante\footnote{Le tenseur de Faraday est reli\'e aux champs
\'electriques et magn\'etiques par $E^{i}=F^{0i}$ et $2 B^i =
\varepsilon^{ijk}F_{jk}$.}
\begin{equation}
\label{mvtem}
\partial_\mu F^{\mu \nu}=j^\nu, \qquad \partial_\mu
\widetilde{F}^{\mu \nu}=0,
\end{equation}
avec
\begin{equation}
\widetilde{F}^{\mu \nu}=\frac{1}{2} \varepsilon^{\mu \nu \rho \sigma}
F_{\rho \sigma},
\end{equation}
le tenseur dual de $F$ dans l'espace de Minkowski, $j^{\mu}$ le
quadrivecteur courant source du champ, et $\varepsilon^{\mu \nu \rho
\sigma}$ le tenseur de Levi-Civita compl\`etement antisym\'etrique
dans ses indices et tel que $\varepsilon^{0123}=1$. Le tenseur
$F^{\mu \nu}$ antisym\'etrique de rang $2$ v\'erifiant les identit\'es
de Bianchi
\begin{equation}
\partial_\mu F_{\nu \rho} + \partial_\nu F_{\rho \mu} + \partial_\rho
F_{\mu \nu} \equiv \partial_{(\mu}F_{\nu \rho)}=0,
\end{equation}
il s'exprime \'egalement comme un champ de gradient
\begin{equation}
\label{potentielem}
F_{\mu \nu}= \partial_{\mu} A_\nu - \partial_\nu A_\mu \equiv
\partial_{[\mu} A_{\nu]},
\end{equation}
o\`u $A^\mu$ est le potentiel vecteur associ\'e. L'\'equation
(\ref{potentielem}) est d\'efinie \`a un gradient pr\`es,
c'est-\`a-dire qu'il existe une \emph{invariance de jauge} des
\'equations de Maxwell par rapport \`a n'importe quelle transformation
du potentiel vecteur $A^\mu$ du type
\begin{equation}
\label{potjauge}
A^{\mu} \rightarrow A^{\mu} - \partial^{\mu} \Uc(x^\mu),
\end{equation}
o\`u $\Uc(x^\mu)$ est une fonction scalaire. L'\'evolution des
champs \'electromagn\'etiques libres (\ref{mvtem}) peut
\'egalement \^etre obtenue par les \'equations d'Euler-Lagrange \`a
partir de la densit\'e lagrangienne\footnote{L'invariance de cette
densit\'e lagrangienne sous les transformations de jauge
(\ref{potjauge}) est assur\'ee par la conservation du quadrivecteur
courant $j^\mu$ par (\ref{mvtem}), apr\`es int\'egration par partie de
(\ref{lagem}).}
\begin{equation}
\label{lagem}
\Lc_\uem = -\frac{1}{4} F^{\mu \nu}F_{\mu \nu} -j_\mu A^\mu.
\end{equation}

Si l'on s'int\'eresse maintenant aux fermions $\Psi(x^\mu)$ de masse
$m$, leur propagation libre est r\'egie par le
lagrangien\footnote{Dans toute la suite, nous d\'esignerons par
\emph{lagrangien}, la densit\'e lagrangienne correspondante.} de
Dirac~\cite{dirac27}
\begin{equation}
\label{lagdirac}
\Lc_\uf = \Psib \left(i \gamma^\mu \partial_\mu - m\right)\Psi,
\end{equation}
qui est trivialement invariant sous un changement de phase global
$\Psi \rightarrow \Psi \ue^{i \alpha}$. L'id\'ee g\'en\'erale des
th\'eories de jauge est alors d'imposer une invariance locale,
autrement dit on souhaiterait rendre le lagrangien de Dirac
(\ref{lagdirac}), ind\'ependant des transformations $U(1)$ locales,
i.e. $\Psi \rightarrow \Psi \ue^{i q \Uc(x^\mu)}$. Il est facile
de voir, \`a partir de (\ref{lagdirac}) que ceci n'est possible que
par l'introduction d'un champ quadrivectoriel $A^\mu$ tel que
\begin{equation}
\label{invem}
\Psi \rightarrow \Psi \ue^{i \Uc(x^\mu)} \quad \Rightarrow \quad
A^{\mu} \rightarrow A^{\mu} - \partial^{\mu} \Uc(x^\mu),
\end{equation}
\`a condition de remplacer le terme cin\'etique par une d\'eriv\'ee
covariante sous ces transformations
\begin{equation}
D_\mu = \partial_\mu + i q A_\mu.
\end{equation}
Le lagrangien des fermions invariant s'\'ecrit alors
\begin{equation}
\label{lagjmuamu}
\Lc_\uf = \Psib \left(i \gamma^\mu D_\mu - m \right) \Psi = \Psib
\left(i \gamma^\mu \partial_\mu - m\right) \Psi -q \Psib \gamma^\mu
\Psi A_\mu.
\end{equation}
La dynamique propre du champ vectoriel $A_\mu$ v\'erifiant
(\ref{invem}) ne peut \^etre donn\'ee que par les \'equations
(\ref{mvtem}) et (\ref{potentielem}). Le dernier terme du lagrangien
de Dirac invariant sous $U(1)$ s'interpr\`ete comme le terme source du
champ \'electromagn\'etique avec $j^\mu =
\Psib \gamma^\mu \Psi$. Il est conserv\'e car c'est \'egalement le
courant de N{\oe}ther associ\'e \`a l'invariance $U(1)$ de la
th\'eorie. Le champ bosonique $A^{\mu}$ auquel il est coupl\'e [voir
Eq.~(\ref{lagjmuamu})] est la repr\'esentation de la particule
m\'ediatrice de l'interaction: le photon.

Ainsi, le simple fait d'imposer aux fermions libres l'invariance
locale sous les transformations de jauge $U(1)$ permet de construire
une th\'eorie d\'ecrivant leurs interactions \'electromagn\'etiques
par le biais d'un champ vectoriel qu'on interpr\`ete comme \'etant le
photon. Le nouveau param\`etre $q$ introduit est simplement la
charge repr\'esentant la sensibilit\'e des particules \`a ce nouveau
couplage. Le lagrangien d\'ecrivant l'\'electrodynamique des fermions
est donc
\begin{equation}
\Lc_{\uf\uem}=-\frac{1}{4}F_{\mu \nu}F^{\mu \nu} + \Psib\left(i
\gamma^\mu D_\mu -m \right) \Psi.
\end{equation}
De la m\^eme mani\`ere, \`a partir du lagrangien de
Klein-Gordon~\cite{klein26} d\'ecrivant l'\'evolution libre des particules
scalaires complexes,
\begin{equation}
\label{lagkg}
\Lc_\ub = \frac{1}{2} \left(\partial_\mu \Phi\right)^{\dag}
\left(\partial^\mu \Phi \right) -\frac{1}{2} m^2 \Phi^\dag \Phi,
\end{equation}
l'invariance par la sym\'etrie de jauge locale $U(1)$ permet d'imposer
leur couplage au champ \'electromagn\'etique. Le lagrangien de
l'\'electrodynamique scalaire est alors donn\'e par
\begin{equation}
\label{emscalaire}
\Lc_{\ub\uem}=-\frac{1}{4}F_{\mu \nu}F^{\mu \nu} + \frac{1}{2}
\left(D_\mu \Phi\right)^{\dag} \left(D^\mu \Phi \right) - \frac{1}{2}
m^2 \Phi^\dag \Phi.
\end{equation}
La th\'eorie $U(1)$ \'electromagn\'etique est aujourd'hui
remarquablement v\'erifi\'ee par l'exp\'erience, ses pr\'edictions
sont en effet v\'erifi\'ees \`a $10^{-13}$ pr\`es dans les
acc\'el\'erateurs~\cite{groom00}.

\subsection{La chromodynamique}

L'interaction nucl\'eaire forte, responsable de la coh\'esion des
noyaux atomiques, peut \'egalement \^etre d\'ecrite par une th\'eorie
de jauge fond\'ee sur le groupe de sym\'etrie non ab\'elien $SU(3)$
rendant compte du confinement de cette interaction. Contrairement \`a
l'\'electrodynamique, de port\'ee infinie, cette derni\`ere n'agit
qu'aux \'echelles des nucl\'eons, de l'ordre de $10^{-15}\,
\mathrm{m}$.

Par analogie avec ce qui pr\'ec\`ede, on d\'ecrit l'interaction forte
des quarks \`a partir du lagrangien de Dirac (\ref{lagdirac}), en
imposant son invariance sous les transformations du groupe $SU(3)$ des
matrices $\Uc$ unitaires de rang $3$ et de d\'eterminant unit\'e,
i.e. sous la transformation
\begin{equation}
\Psi^i \rightarrow \Uc^i_j \, \Psi^j,
\end{equation}
$\Psi^i$ \'etant cette fois un triplet de spineurs. Par soucis de
simplicit\'e, nous omettrons \`a partir d'ici ses indices. En suivant
la m\^eme d\'emarche que pour la sym\'etrie $U(1)$, l'invariance
locale est obtenue par l'introduction d'une matrice de champ vectoriel
$\left(A^\mu\right)^i_j$ se transformant sous $SU(3)$ par
\begin{equation}
A^\mu \rightarrow \Uc A^\mu \Uc^\dag - \frac{i}{g}\, \Uc \partial_\mu
\Uc^\dag,
\end{equation}
permettant de d\'efinir la matrice de d\'eriv\'ees covariantes
\begin{equation}
D_\mu = \partial_\mu + i g A_\mu.
\end{equation}
Par propri\'et\'e de $SU(3)$, il y a huits\footnote{Pour des matrices
unitaires de rang $N$ et de d\'eterminant unit\'e, il y a $N^2-1$
composantes ind\'ependantes.} champs vectoriels $A^{\mu}$
ind\'ependants qui s'identifient aux g\'en\'erateurs du groupe et qui
correspondent aux huit particules scalaires portant l'interaction et
appel\'ees les gluons. Leur dynamique propre peut \'egalement
\^etre obtenue par sym\'etrie, en imposant l'invariance de jauge. La
g\'en\'eralisation de (\ref{potentielem}) pour une sym\'etrie non
ab\'elienne est
\begin{equation}
\label{forcenonab}
\Gc_{\mu \nu} = \partial_{[\mu} A_{\nu]} + i g \left[A_\mu,A_\nu\right],
\end{equation}
et on montre que le lagrangien associ\'e, invariant de Lorentz et
conservant ls sym\'etrie de parit\'e, peut toujours se ramener \`a
\begin{equation}
\Lc_\uc =-\frac{1}{2} \mathrm{Tr} \left(\Gc_{\mu \nu} \Gc^{\mu \nu}\right),
\end{equation}
la trace portant sur les composantes
matricielles~\cite{weinbergbook2}. La chromodynamique d\'ecrivant
l'interaction des quarks et des gluons peut par cons\'equent \^etre
d\'ecrite par le lagrangien
\begin{equation}
\Lc_{\uq\uc} = \sum_{\uq} \Psib_\uq \left(i \gamma^\mu D_\mu - m_\uq
\right) \Psi_\uq  + \Lc_\uc.
\end{equation}
Il est alors possible de montrer que la charge port\'ee par les
quarks, appel\'ee la \emph{couleur}, est inobservable \`a nos
\'echelles d'\'energie d\^u aux propri\'et\'es de confinement de cette
int\'eraction: son intensit\'e est en effet d'autant plus forte que
les \'echelles de distances observ\'ees sont
grandes~\cite{coleman73,feynman69,bjorken69,drell71,zimmermann73}. Les
particules sensibles \`a l'int\'eraction forte, \`a nos \'echelles
d'\'energie, sont alors un assemblage de quarks et gluons de charge
totale de couleur nulle, ou
\emph{blanche}, int\'eragissant par \'echange de mesons, des
particules scalaires form\'es d'une paire
quark-antiquark~\cite{gellmann63,zweig63,groom00}.

\subsection{L'interaction \'electrofaible}

La force faible est \`a l'origine des d\'esint\'egrations $\beta^\pm$
violant la parit\'e. La premi\`ere th\'eorie effective propos\'ee
\'etait bas\'ee sur une interaction de type
``courant-courant''~\cite{feynman58,sudarshan58,sakurai58} de
lagrangien
\begin{equation}
\Lc_\ueff = -\frac{1}{\sqrt{2}} G_\uf \Jc^\mu \Jc^\dag_\mu,
\end{equation}
o\`u les courants charg\'es $\Jc^\mu$ font intervenir les diff\'erents
fermions coupl\'es par l'interaction, i.e. les quarks $q$ et leptons
$\ell$
\begin{equation}
\label{lagfaibleeff}
\Jc^\mu =
\sum_{\ell} \Psib_\ell \gamma^\mu \frac{1-\gamma^5}{2}
\Psi_{\nu_\ell} + \sum_{q} \Psib_q \gamma^\mu \frac{1 - \gamma^5}{2}
\Psi_{q'}.
\end{equation}
L'op\'erateur $\gamma^5$ introduit artificiellement, permet la
violation de la parit\'e observ\'ee dans les int\'eractions, et les
couplages s'effectuent entre des doublets de particules g\'en\'erant
une sym\'etrie dite
\emph{d'isospin faible}. Ils sont form\'es d'un lepton et de son
neutrino associ\'e $(\ell,\nu_\ell)$, et du doublet de quarks \'etats
propres de l'interaction $(q,q')$. Cette th\'eorie effective pose de
nombreux probl\`emes, le plus important d'entre eux \'etant que la
constante de couplage $G_\uf$ est dimensionn\'ee\footnote{C'est
\'egalement le cas de la gravitation.}, rendant la th\'eorie non
renormalisable apr\`es quantification, i.e. non pr\'edictive. Elle
n'est pas une th\'eorie de jauge et n'explique pas les interactions
par courant neutre telles que les diffusions de neutrinos sur des
\'electrons.

En constatant une sym\'etrie globale d'isospin faible, on pourrait
penser naturellement, suivant les exemples de l'\'electromagn\'etisme
et des int\'eractions fortes, construire une th\'eorie de jauge
d\'ecrivant l'interaction faible se basant sur le groupe de Lie
$SU(2)$. Comme pour la chromodynamique ou l'\'electrodynamique, une
telle th\'eorie introduirait des bosons vecteurs additionnels, au
nombre de trois\footnote{$2^2-1=3$ c'est le nombre de g\'en\'erateurs
du $SU(2)$.}. La violation de la parit\'e peut, quant \`a elle, \^etre
introduite au travers des repr\'esentations irr\'eductibles de
$SU(2)$, les particules de chiralit\'e ``droite'' faisant partie d'un
singulet de charge d'isospin nulle, et les particules ``gauches'',
sensibles \`a l'interaction, d'un doublet de charge $1/2$, comme
sugg\'er\'e par (\ref{lagfaibleeff}). En notant $\Psi_\uL$ les
doublets et $\Psi_\uR$ les singulets, le lagrangien d'une telle
th\'eorie s'\'ecrirait alors comme une somme sur les diff\'erentes
familles de particules, les ``saveurs'', des termes invariants de
jauge
\begin{equation}
\label{lagw}
\Lc_\uw = \sum_\us \left[ \, \Psib^\us_\uL \left(i \gamma^\mu D_\mu -
m_\us \right)
\Psi^\us_\uL + \Psib^\us_\uR \left(i \gamma^\mu \partial_\mu - m_\us
\right) \Psi^\us_\uR \right] - 
\frac{1}{2} \mathrm{Tr}\left(W_{\mu \nu} W^{\mu \nu} \right),
\end{equation}
o\`u $W^\mu$ est la matrice de $SU(2)$ des bosons vecteurs, $W_{\mu
\nu}$ son tenseur de type Faraday associ\'e [cf. Eq.~(\ref{forcenonab})]
et la d\'eriv\'ee covariante pour les doublets
\begin{equation}
D_\mu = \partial_\mu + i g W_\mu,
\end{equation}
$g$ \'etant la constante de couplage de l'interaction. En
d\'ecomposant les bosons vecteurs $W^\mu$ sur les g\'en\'erateurs de
$SU(2)$, i.e. les trois matrices de Pauli $\sigma^{i}$,
\begin{equation}
W = \sum_i W^i \frac{\sigma^i}{2},
\end{equation}
il est possible de retrouver les courants charg\'es de la th\'eorie
effective, \`a ceci pr\`es que la constante de couplage $G_\uf$ est
maintenant remplac\'ee par un terme faisant intervenir les
propagateurs des bosons vectoriels charg\'es $W^\pm$ avec $W^\pm =
(1/\sqrt{2}) \left(W^1 \pm W^2 \right)$. De m\^eme, les int\'eractions
de type courant neutre s'expliquent maintenant par l'\'echange du
boson $W^3$.

Ces interactions ont d'autres parts la propri\'et\'e de faire
intervenir les charges \'electromagn\'etiques au sein m\^eme des
bosons vecteurs $W^\pm$. Dans les interactions observ\'ees
exp\'erimentalement, on constate \'egalement la stricte conservation
d'un nombre scalaire, \emph{l'hypercharge}
\begin{equation}
Y=Q-T^3,
\end{equation}
o\`u $Q$ est la charge \'electromagn\'etique et $T^3$ la projection
d'isospin associ\'ee aux particules de $SU(2)$, i.e. $0$ pour les
singulets et $\pm 1/2$ pour chaque composante des doublets,
respectivement. Cette sym\'etrie peut \^etre jaug\'ee par une
sym\'etrie locale $U(1)$ additionnelle unifiant
l'\'electromagn\'etisme et les courants neutres~\cite{glashow61}. Le
lagrangien de la th\'eorie s'\'ecrit alors
\begin{equation}
\Lc_\uew = \Lc_\uw - \frac{1}{4}B_{\mu \nu}B^{\mu\nu},
\end{equation}
o\`u les nouvelles d\'eriv\'ees covariantes apparaissant dans
(\ref{lagw}) pour les composantes gauches et droites,
respectivement, sont donn\'ees par
\begin{eqnarray}
D_\mu & = & \partial_\mu + i g W_\mu + i g' Y^\us_\uL B_\mu, \\ D_\mu
& = & \partial_\mu + i g' Y^\us_\uR B_\mu.
\end{eqnarray}

Une telle th\'eorie ne d\'ecrit absolument pas les interactions
qu'elle se propose de mod\'eliser. En effet, les bosons vecteurs $W^i$
y sont pr\'edits de masse nulle, sugg\'erant une port\'ee infinie de
l'interaction faible, contrairement \`a ce qui est observ\'e. Il
faudrait en r\'ealit\'e qu'ils soient tr\`es massifs afin que
l'influence de l'impulsion $k^\mu$ de ceux-ci soit n\'egligeable
devant leur masse pour que les termes en $g W^i$ dans le lagrangien
donnent des propagateurs constants \`a basse \'energie compatibles
avec l'interaction ``courant-courant''. Or des termes de masse en $m^2
W^\mu W_\mu$ brisent l'invariance de jauge de la th\'eorie. Le
probl\`eme trouve sa solution dans le m\'ecanisme de
Higgs~\cite{englert64,higgs64,guralnik64}, qui introduit une particule
scalaire suppl\'ementaire \`a l'origine de la masse des particules
auxquelles elle est coupl\'ee.

\section{Le m\'ecanisme de Higgs}
\label{sectionhiggs}
Le m\'ecanisme de Higgs~\cite{englert64, higgs64} met en jeu la
brisure d'une sym\'etrie, i.e. le non respect de la sym\'etrie d'une
th\'eorie par l'\'etat fondamental: l'id\'ee en est simplement que si
une \'equation admet une certaine sym\'etrie et plusieurs solutions,
celles-ci peuvent ne pas la respecter\footnote{Le principe de Curie
assure cependant que \emph{l'ensemble} des solutions doit respecter la
sym\'etrie.}. Pour illustrer ce m\'ecanisme, il est d'usage de
consid\'erer une sym\'etrie $U(1)$ locale.

\begin{figure}
\begin{center}
\epsfig{file=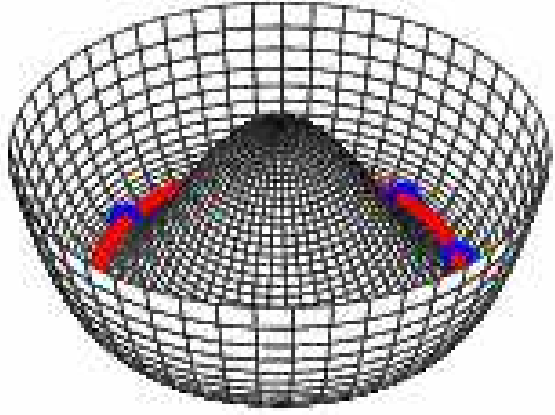,height=8cm}
\caption[Le potentiel de Higgs dans le mod\`ele $U(1)$ ab\'elien.]{Le
potentiel de Higgs $V(\Phi)$ trac\'e dans le plan complexe du champ
scalaire $\left[\Re(\Phi),\Im(\Phi)\right]$. L'\'etat d'\'energie
minimale est repr\'esent\'e en rouge et correspond \`a $|\Phi|=\eta$,
diff\'erentes valeurs de phase apparaissent en bleu.}
\label{figvacua}
\end{center}
\end{figure}

Soit le champ scalaire complexe $\Phi$ de Higgs, invariant sous les
transformation de jauge $U(1)$, en auto-interaction dans un potentiel
$V(\Phi)$. D'apr\`es (\ref{emscalaire}), sa dynamique est r\'egie par
le lagrangien
\begin{equation}
\label{laghiggs}
\Lc_\uh=\frac{1}{2} \left(D_\mu \Phi\right)^\dag \left(D^\mu \Phi
\right) - \frac{1}{4} H_{\mu \nu} H^{\mu \nu} - V(\Phi),
\end{equation}
o\`u $H_{\mu \nu}$ est le tenseur de type Faraday associ\'e au boson
vectoriel $B_\mu$ tel que
\begin{equation}
D_\mu = \partial_\mu + i g B_\mu.
\end{equation}
La th\'eorie est alors effectivement invariante sous les
transformations de jauge
\begin{equation}
\label{higgsjauge}
\Phi \rightarrow \Phi \ue^{ig \Uc(x^\mu)} \quad \textrm{et} \quad
A^\mu \rightarrow A^\mu - \partial_\mu\Uc.
\end{equation}
Pour pouvoir d\'efinir une th\'eorie quantique renormalisable, et donc
pr\'edictive, le potentiel $V$ ne peut faire intervenir que des termes
de masse et d'auto-interaction en $\Phi^4$, au plus. La brisure de
sym\'etrie intervient lorsque le potentiel $V$ est de la forme
\begin{equation}
\label{pothiggs}
V(\Phi) = \frac{\lambda}{8} \left(|\Phi|^2 - \eta^2 \right)^2,
\end{equation}
avec $\lambda$ la constante d'autocouplage du champ scalaire et $\eta$
sa valeur moyenne dans le vide\footnote{\emph{Vacuum expectation value} ou
\emph{vev}.}. Dans ce cas, l'\'etat fondamental stationnaire du champ est
alors obtenu en imposant une \'energie minimale, i.e. une valeur du
potentiel minimale qui, d'apr\`es (\ref{pothiggs}), est obtenue pour
$|\Phi|^2 = \eta$ (cf. Fig.~\ref{figvacua}).

L'ensemble des valeurs moyennes du champ de Higgs dans son \'etat
fondamental est donc de la forme
\begin{equation}
\langle \Phi \rangle = \Phi_\zero = \eta \, \ue^{i \alpha},
\end{equation}
avec $\alpha$ une phase complexe arbitraire
(cf. Fig.~\ref{figvacua}). Chacun de ces \'etats n'est clairement plus
invariant sous les transformation de jauge $U(1)$, brisant ainsi la
sym\'etrie en question. Notons que le lagrangien \`a l'origine de la
brisure et la th\'eorie sous-jacente sont toujours invariants sous
$U(1)$, ainsi que \emph{l'ensemble} des \'etats vides.

Pour explorer les cons\'equences physiques du ph\'enom\`ene il est
commode de d\'evelopper le champ de Higgs autour de son \'etat
fondamental afin d'en extraire la forme de ses excitations,
\begin{equation}
\Phi = \left[\eta + \rho(x^\mu) \right] \ue^{i \varphi(x^\mu)/\eta}.
\end{equation}
Pour de petites fluctuations autour de $\Phi_\zero$, et en choisissant
$\alpha=0$, ce qui localement est toujours possible, cette expression
se r\'eduit au premier ordre en
\begin{equation}
\Phi \simeq \eta + \rho + i \varphi,
\end{equation}
dans la mesure o\`u $\rho \ll \eta$ et $\varphi \ll \eta$.  Le lagrangien
(\ref{laghiggs}) se d\'eveloppe alors, au deuxi\`eme ordre en ces
champs, en
\begin{eqnarray}
\Lc_\uh  & \rightarrow & \frac{1}{2} \partial_\mu \rho \partial^\mu \rho +
\frac{1}{2} \partial_\mu \varphi \partial^\mu \varphi - g B^\mu \left(
\varphi \partial_\mu \rho - \rho \partial_\mu \varphi -\eta
\partial_\mu \varphi \right) \\ \nonumber
& & + g^2 \eta^2 B_\mu B^\mu + g^2 B_\mu B^\mu
\left( \rho^2 + \varphi^2 + 2 \eta \rho \right) -\frac{\lambda}{2}
\eta^2 \rho^2.
\end{eqnarray}
La brisure de sym\'etrie a donc donn\'e naissance, \emph{a priori},
\`a deux champs scalaires r\'eels $\rho$ et $\varphi$. Le champ $\rho$
est un champ scalaire r\'eel en interaction de masse $\sqrt{\lambda}
\eta$, alors que le champ $\varphi$, quoiqu'\'egalement en
int\'eraction, ne comporte pas de terme de masse
[cf. Eq.~(\ref{lagkg})]; il est couramment appel\'e boson de
Goldstone~\cite{goldstone62,goldstone61,nambu61}. Physiquement les
bosons de Goldstone ne peuvent exister que pour la brisure d'une
sym\'etrie globale, i.e. sans champ de jauge associ\'e. En effet, dans
le cas pr\'esent, le champ $\varphi$ peut \^etre absorb\'e dans les
transformations de jauge $U(1)$ de l'\'equation (\ref{higgsjauge}) en
choisissant
\begin{equation}
\Uc(x^\mu)= -\frac{\varphi(x^\mu)}{g \eta},
\end{equation}
il ne constitue donc pas, pour une sym\'etrie locale, un degr\'e de
libert\'e physique. Le r\'esultat essentiel de ce m\'ecanisme et que
les bosons de jauge $B^\mu$ poss\`edent apr\`es brisure une masse
physique $g \eta$ proportionnelle \`a la \emph{vev} du champ de
Higgs. Il est clair sur l'\'equation (\ref{laghiggs}) que le
lagrangien est effectivement invariant de jauge, et met en jeu
maintenant, par le biais du champ de Higgs, des bosons vecteurs
massifs.

Pour revenir \`a l'interaction \'electrofaible, le m\^eme m\'ecanisme
est g\'en\'eralis\'e \`a la sym\'etrie $SU(2) \times U(1)$. La
th\'eorie d\'ecrivant cette interaction est effectivement celle
r\'esum\'ee au paragraphe pr\'ec\'edent, mais du fait du m\'ecanisme
de brisure de sym\'etrie, seul la jauge $U(1)$ reste encore observable
\`a basse \'energie au travers de l'\'electrodynamique. Le m\'ecanisme
de Higgs appliqu\'e \`a $SU(2)$ permet donc de donner une masse aux
bosons vecteurs $W$, tout en conservant une th\'eorie
renormalisable~\cite{thooft71, thooft71b,thooft72}; il est par
cons\'equent pr\'edictif. En effet, consid\`erons un doublet de champs
scalaires complexes d'hypercharge $Y_\uh=1/2$
\begin{equation}
\Phi= \left(
\begin{array}{c}
\phi^+ \\
\phi_\zero
\end{array}
\right),
\end{equation}
de lagrangien (\ref{laghiggs}) dans lequel la d\'eriv\'ee covariante
l'est maintenant par rapport aux transformations de $SU(2) \times
U(1)$
\begin{equation}
D_\mu= \partial_\mu + igW_\mu + ig'Y_\uh B_\mu.
\end{equation}
En choisissant la valeur moyenne dans le vide du doublet de Higgs le
long de la composante non charg\'ee\footnote{C'est un choix de jauge
sans effet physique a priori.}
\begin{equation}
\langle \Phi \rangle= \left(
\begin{array}{c}
0 \\
\eta
\end{array}
\right),
\end{equation}
il est possible d'\'etudier, comme pour la brisure de $U(1)$, les
fluctuations des champs dans leur nouvel \'etat de vide en les
d\'ecomposant sur les g\'en\'erateurs de $SU(2)$
\begin{equation}
\Phi=\left(
\begin{array}{c}
0 \\
\eta + \rho
\end{array}
\right) \exp{\left(i \frac{\varphi_j}{\eta} \frac{\sigma^j}{2} \right)}.
\end{equation}
Une transformation de jauge permet, comme dans le cas ab\'elien, de
r\'eabsorber les phases $\varphi_j$ dans les masses, et de
s'int\'eresser seulement au champ $\rho$. Le lagrangien se r\'eduit
donc, au deuxi\`eme ordre en ces nouveaux champs, \`a
\begin{eqnarray}
\label{lagewbrok}
\Lc_\uh  & \rightarrow & \frac{1}{2} \partial_\mu \rho \partial^\mu
\rho -\frac{\lambda}{2} \eta^2 \rho^2 + \frac{1}{2} g^2 \eta^2 W^-_\mu
W^{\mu^+} + \frac{g^2}{2} \left(\rho^2 + 2 \eta \rho \right) W^-_\mu
W^{\mu^+} \\ \nonumber
& & + \left(\eta + \rho \right)^2 \left[\left(g' Y_\uH B_\mu -
\frac{g}{2} W_\mu^3 \right)^2 \right].
\end{eqnarray}
Comme pour la brisure de $U(1)$, il appara\^\i t un champ
scalaire neutre $\rho$ de masse $\sqrt{\lambda} \eta$, le boson de
Higgs, et les bosons vecteurs $W^\pm$ acqui\`erent une masse $ (1/2) g
\eta$ comme $W^3$ et $B_\mu$ qui se trouvent maintenant
coupl\'es. Le couplage entre les bosons vecteurs neutres peut \^etre
mieux compris physiquement en introduisant deux nouveaux champs
\begin{eqnarray}
Z_\mu & = & \cos(\theta_\uw) W_\mu^3 - \sin(\theta_\uw) B_\mu \\
A_\mu & = & \sin(\theta_\uw) W_\mu^3 + \cos(\theta_\uw) B_\mu,
\end{eqnarray}
avec $\theta_\uw$ l'angle faible\footnote{\emph{Weak angle}.} d\'efini
par
\begin{equation}
\cos(\theta_\uw)=\frac{g}{\sqrt{g^2+g'^2}} \qquad
\sin(\theta_\uw)=\frac{g'}{\sqrt{g^2+g'^2}}.
\end{equation}
Il est facile de voir \`a partir de (\ref{lagewbrok}) que les termes
de masse et de couplage entre $W_\mu^3$ et $B_\mu$ se r\'eduisent
uniquement \`a un terme de couplage pour le boson $Z$ en
\begin{equation}
\Lc_\uZ = \frac{1}{4}(\eta + \rho)^2 \left(g^2 + g'^2 \right) Z_\mu Z^\mu,
\end{equation}
et que, autrement dit, le boson $A_\mu$ reste de masse nulle: il peut
\^etre identifi\'e au photon g\'en\'erant la sym\'etrie $U(1)$
\'electromagn\'etique, alors que le boson vecteur $Z_\mu$, \`a
l'origine des courants neutre, est de masse $(1/2) \eta
\sqrt{g^2+g'^2}$. \`A partir du couplage entre neutrinos, de charge
\'electromagn\'etique nulle, donn\'e par (\ref{lagw}), il est possible
de relier les constantes de couplage introduites, i.e. $g$, $g'$ et
$\theta_\uw$, \`a la constante de couplage \'electromagn\'etique $e$
par
\begin{equation}
e = g'\cos(\theta_\uw) = g \sin(\theta_\uw).
\end{equation}

\begin{figure}
\begin{center}
\epsfig{file=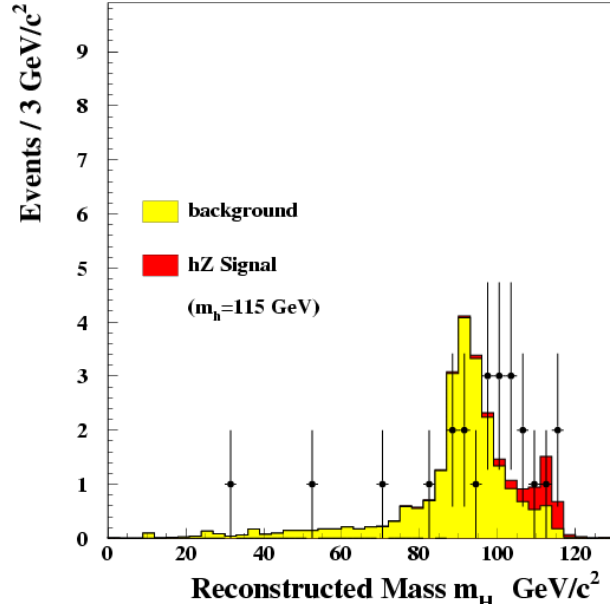,height=8cm}
\caption[Observation d'une \'eventuelle r\'esonance du boson de Higgs
\`a $115 \GeV$.]{Distribution en masse du nombre d'\'ev\`enements
observ\'es au LEP durant l'ann\'ee 2000 et qui pourrait \^etre
interpr\'et\'ee comme une r\'esonance du boson de Higgs \`a $115
\GeV$~\cite{L301}.}
\label{fighiggsevent}
\end{center}
\end{figure}

Exp\'erimentalement, connaissant la constante de couplage effective de
l'interaction faible $G_\uf$, celle de l'\'electromagn\'etisme $e$, et
l'angle faible $\theta_\uw$ d\'etermin\'e \`a partir des couplages $Z$
quarks et leptons~\cite{groom00}, la masse des $W^\pm$ pr\'edite est
de $80.4
\GeV$, et celle du $Z$, $91.6 \GeV$. Ces particules ont effectivement
\'et\'e observ\'ees dans les acc\'el\'erateurs avec des masses
extr\^emement voisines des valeurs pr\'edites~\cite{groom00},
confirmant ainsi le mod\`ele de Glashow-Weinberg-Salam. Actuellement,
la particule associ\'e au Higgs $\rho$ aurait peut \^etre \'et\'e
directement d\'etect\'ee (cf. Fig.~\ref{fighiggsevent}), la mesure de
sa masse fixerait alors la constante d'autocouplage\footnote{$\eta$
est d\'ej\`a d\'etermin\'e par les mesures des masses des bosons
vecteurs massifs, $\eta \simeq 175\GeV$.} $\lambda$ et confirmerait le
m\'ecanisme de Higgs.

La masse des autres particules est obtenue par leur couplage au champ
de Higgs. Ainsi pour des fermions, un couplage de type
Yukawa~\cite{yukawa35} en
\begin{equation}
\lambda_\uf \Psib \Phi \Psi,
\end{equation}
permet d'obtenir apr\`es brisure de sym\'etrie des fermions $\Psi$
massifs de masse $\lambda_\uf  \eta$. Les diff\'erentes masses des
diverses particules sont alors introduites par le biais de leur
constante de couplage au champ de Higgs.

\section{Pr\'edictions et limites du mod\`ele standard}

Dans les sections pr\'ec\'edentes, nous avons bri\`evement d\'ecrit
les bases classiques du mod\`ele standard. La th\'eorie compl\`ete
d\'ecrivant les interactions fondamentales n\'ecessite cependant
d'\^etre quantifi\'ee, donnant ainsi tout son sens \`a la notion de
particule. Sans entrer dans le d\'etail et les difficult\'es
inh\'erentes \`a la quantification\footnote{La m\'ethode de
``quantification canonique'' sera n\'eanmoins d\'etaill\'ee dans la
partie~\ref{partieferm}.}, celle-ci permet essentiellement de
pr\'edire les sections efficaces de toutes les interactions entre les
particules, telles qu'on peut les mesurer dans les acc\'el\'erateurs,
mais \'egalement de les extrapoler \`a des
\'echelles d'\'energie bien plus \'elev\'ees, telles celles de
l'univers primordial.

\subsection{La quantification et la renormalisation}
\label{soussectionrenorm}
L'existence physique d'une configuration de particules peut \^etre
identifi\'ee, dans la th\'eorie quantique des champs, \`a un \'etat
donn\'e d'un espace de Fock, i.e. une supperposition infinie d'espaces
de Hilbert dans lesquels \'evolue chaque particule. L'op\'erateur
d'\'evolution $\widehat{U}(t_i,t_f)$ agissant dans cet espace permet
de conna\^{\i}tre l'amplitude de probabilit\'e $S_{\ui\uf}$ de passage
d'une configuration de particules $|\Cc_\ui \rangle$ en $t_\ui$, \`a
la configuration finale $|\Cc_\uf \rangle$ en $t_\uf$. Il est possible
de montrer que ces amplitudes de probabilit\'es se r\'eduisent aux
calculs de la valeurs moyenne dans le vide initial du produit
ordonn\'e des champs en interaction~\cite{lehmann55,schwinger59}
\begin{equation}
S_{\ui \uf} \propto \langle 0 | \Tp \left[\Fc_1 \dots \Fc_n
\exp \left(i \int{\ud^4 \bx} \, \Lc_{\mathrm{int}}\right) \right] | 0
\rangle,
\end{equation}
o\`u les $\Fc_i$ sont les champs consid\'er\'es dans les \'etats de
Fock de transition et $\Tp$ l'op\'erateur produit ordonnant les champs
selon les temps d\'ecroissants. Dans la repr\'esentation
d'interaction, le terme $\Lc_{\mathrm{int}}$ se r\'eduit aux parties
du lagrangien de la th\'eorie classique autre que celles d\'ecrivant
la propagation libre des champs. Le membre de droite est aussi not\'e
$G^{(n)}$ comme fonction de Green en interaction \`a $n$ points. Dans
l'approche perturbative on d\'eveloppe l'op\'erateur d'\'evolution (le
terme en exponentielle) en puissances successives de la constante de
couplage et des champs libres. La fonction de Green de l'interaction
se r\'eduit en une somme infinie de termes faisant intervenir les
fonctions de Green libres de la th\'eorie \`a deux, trois, \dots, $n$
points qui s'identifient \`a
\begin{equation}
\label{greenlibre}
G^{(n)}_\ulib =\langle 0 | \Tp \left[\Fc_1 \dots \Fc_n \right] | 0
\rangle
\end{equation}
Celles-ci s'interpr\`etent comme des corrections dues aux fluctuations
quantiques des champs $\Fc$ en interactions, et sont souvent les
seules qu'il est possible de calculer analytiquement.

Le calcul effectif des corrections quantiques peut s'effectuer au
moyen de la quantification par l'int\'egrale de
chemin~\cite{dirac33,feynman48}. En effet, il est possible de montrer
que la fonctionnelle\footnote{C'est
\'egalement l'amplitude de persistance du vide en pr\'esence des
sources externes $\langle 0|0
\rangle_{\{\Jc\}}$.}
\begin{equation}
\label{genegreen}
W\left[\left\{\Jc \right\}\right] \propto \int \left\{\Dc \Fc \right\}
\exp \left[\int{\ud^4 \bx} \left(\Lc + \Jc \Fc \right) \right],
\end{equation}
avec $\{\Jc\}$ l'ensemble des densit\'es de courant externe, est
g\'en\'eratrice des fonctions de Green en interaction
\begin{equation}
G^{(n)} = \frac{1}{i^n} \left. \frac{\delta^n W\left[\left\{\Jc \right\}
\right]}{\delta \Jc^n}\right \vert_{\{\Jc\}=0}.
\end{equation}

Il est ainsi possible de tenir compte des fluctuations quantiques des
champs dans les interactions. Cependant, leur nature quantique ne
s'accommode pas du tout de leur d\'efinition locale, et les termes
apparaissant dans ce d\'eveloppement sont en g\'en\'eral infinis. Pour
rendre la th\'eorie pr\'edictive il faut alors proc\'eder \`a la
\emph{renormalisation}.

\begin{figure}
\begin{center}
\epsfig{file=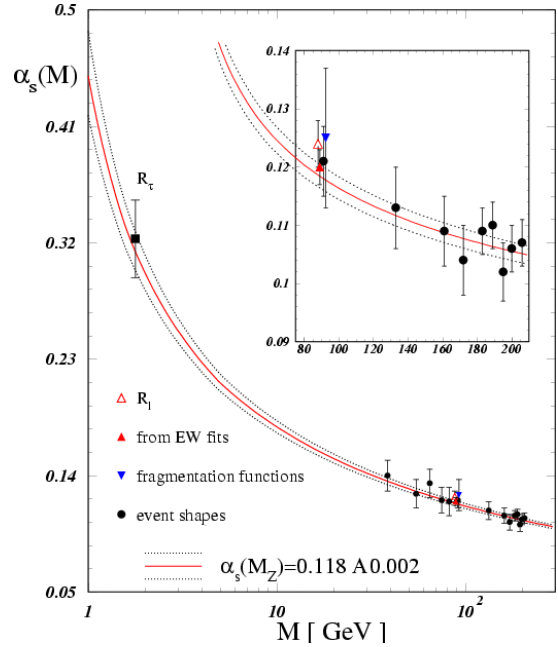,height=8.5cm}
\caption[\'Evolution de la constante de couplage de l'interaction
forte avec l'\'echelle d'\'energie.]{\'Evolution de la constante de
couplage de l'interaction forte avec l'\'echelle d'\'energie telle
qu'elle est pr\'edite par renormalisation, compar\'ee aux valeurs
mesur\'ees~\cite{polonsky95}. Sa d\'ecroissance avec l'\'energie est
caract\'eristique de la libert\'e asymptotique de cette interaction.}
\label{figrunstrong}
\end{center}
\end{figure}

Le processus de renormalisation consiste \`a ajouter un nombre fini de
contre-termes, \`a un ordre donn\'e, dans le lagrangien initial de
mani\`ere \`a annuler ces divergences. Ces contre-termes peuvent alors
\^etre absorb\'es dans une red\'efinition des champs et des constantes
de la th\'eorie, celles-ci devenant alors des fonctions de l'\'echelle
d'\'energie consid\'er\'ee. Les pr\'edictions de la th\'eorie sont
finalement assur\'ees par la connaissance de ces fonctions dont les
variations avec l'\'energie sont compar\'ees et v\'erifi\'ees par
l'exp\'erience (voir figure~\ref{figrunstrong}). C'est 't
Hooft~\cite{thooft71,thooft71b,thooft72} qui a montr\'e que les
th\'eories de jauge avec brisure de sym\'etrie
\'etaient renormalisables sous certaines conditions fixant de mani\`ere
unique la forme m\^eme des termes d'interaction \`a celles que nous
avons utilis\'ees dans les sections pr\'ec\'edentes. Le mod\`ele
standard de la physique des particules ainsi construit d\'ecrit de
mani\`ere satisfaisante la physique sond\'ee dans les
acc\'el\'erateurs pour des \'energies allant jusqu'\`a la centaine de
$\GeV$. Il laisse n\'eanmoins certaines questions en suspens dont les
tentatives de r\'eponse sugg\`erent que ce mod\`ele ne constitue
qu'une th\'eorie effective d'une th\'eorie physique plus compl\`ete.

\subsection{L'unification}
\label{sectionunif}
\begin{figure}
\begin{center}
\epsfig{file=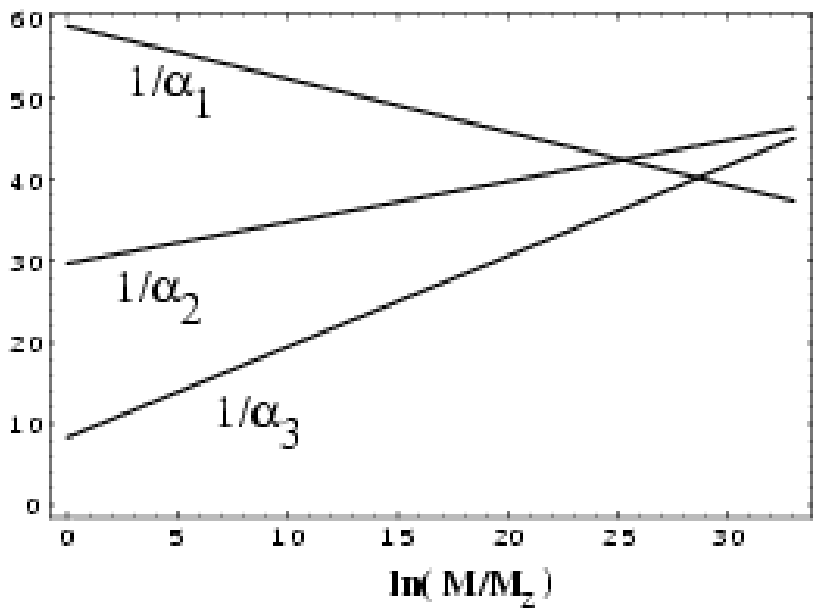,height=8cm}
\caption[\'Evolution des constantes de couplage des trois
interactions fondamentales avec l'\'echelle d'\'energie.]{\'Evolution
des constantes de couplage des trois interactions fondamentales avec
l'\'echelle d'\'energie. $\alpha_1$ est la constante associ\'ee \`a la
sym\'etrie d'hypercharge, i.e. $\alpha_1=(5/3)\alpha_\ue /
\cos^2(\theta_\uw)$ et $\alpha_2$ celle associ\'ee \`a $SU(2)$,
i.e. $\alpha_2=\alpha_\ue/\sin^2(\theta_\uw)$ o\`u $\alpha_\ue =
\ue^2/(4\pi)$. $\alpha_3$ est la constante de couplage de
l'interaction forte. Elles semblent s'unifier autour de $10^{15}
\GeV$~\cite{lopez96}.}
\label{figunifsm}
\end{center}
\end{figure}

La forme unifi\'ee de l'interaction \'electrofaible du mod\`ele
standard pose naturellement le probl\`eme de l'unification des autres
interactions. Pourquoi deux interactions seulement seraient la
manifestation d'un m\^eme ph\'enom\`ene ? Par la renormalisation des
constantes de couplage, il est possible d'en calculer leur \'evolution
avec l'\'echelle d'\'energie (cf. Fig.~\ref{figrunstrong}), et les
mesures actuelles semblent montrer que ces constantes sont du m\^eme
ordre de grandeur lorsque les \'energies \'echang\'ees sont de l'ordre
de $10^{15} \GeV$ (cf. Fig.~\ref{figunifsm}), sugg\'erant que la
chromodynamique devait \'egalement \^etre unifi\'ee avec l'interaction
\'electrofaible~\cite{georgi74,cline78}.

Un probl\`eme d'envergure r\'eside \'egalement dans l'inclusion de la
gravitation. Celle-ci peut \'egalement \^etre exprim\'ee sous forme
d'une th\'eorie de jauge sous les transformations locales de
coordonn\'ees dont l'action s'\'ecrit~\cite{hilbert15}
\begin{equation}
\Sc_\ug = \int{\frac{1}{2 \kappa^2} \left(R - 2 \Lambda \right)
\sqrt{-g}\, \ud^4 \bx},
\end{equation}
o\`u $R$ est le scalaire de Ricci et $g$ le d\'eterminant de la
m\'etrique. La constante de couplage $\kappa^2$ \'etant
dimensionn\'ee, la gravitation d'Einstein n'est pas renormalisable, et
donc non pr\'edictive apr\`es quantification. Du point de vue
th\'eorique, ceci renforce l'id\'ee selon laquelle le mod\`ele
standard incluant la gravit\'e cette fois-ci, n'est qu'une th\'eorie
effective \`a basse \'energie. N\'eanmoins, l'\'echelle d'\'energie
\`a laquelle les effets quantiques sont susceptibles de modifier
notablement la gravit\'e est celle de
Planck\footnote{$\ell_{\mathrm{Pl}}=\sqrt{\Gc \hbar/c^3}$ est la seule
longueur que l'on puisse construire \`a partir des ces trois
constantes fondamentales.}, i.e. $10^{19}\GeV$ et donc dans la
pratique, largement au del\`a des \'echelles d'unification des autres
interactions. La relativit\'e g\'en\'erale est encore, \`a ces
\'energies, une tr\`es bonne description de la gravitation.

\subsection{Les param\`etres libres}

Bien que v\'erifi\'e au pourcent pr\`es, le mod\`ele standard met en
jeu $19$ param\`etres ind\'ependants qui sont les constantes de
couplage, les angles de m\'elange et les masses (ou constantes de
couplage au Higgs) des diverses particules, ce qui n'est pas
compl\`etement satisfaisant du point de vue th\'eorique.

Le mod\`ele standard ne propose aucune explication relative \`a la
quantification de la charge, i.e. au fait que les charges des quarks
sont en $e/3$ exactement, $e$ \'etant la charge de l'\'electron. Cette
sym\'etrie entre leptons et hadrons sugg\`ere \'egalement que ces
param\`etres \emph{a priori} ind\'ependants pourraient \^etre reli\'es
dans une th\'eorie plus g\'en\'erale.

Le mod\`ele standard ne donne pas non plus d'explication au probl\`eme
de hi\'erarchie des masses. La brisure de sym\'etrie \'electrofaible
donne une \'echelle d'\'energie naturelle \`a la th\'eorie autour de
$\eta \simeq 175 \GeV$. Bien que le quark top poss\`ede une masse de
cet ordre, les autres particules s'\'etalent sur un spectre de masse
bien plus large, ainsi, pour d\'ecrire des \'electrons de masse $511
\KeV$, il faut que leur constante de couplage soit extraordinairement
faible, ce qui est encore une forme de \emph{fine tunning}.

Enfin, la renormalisation de la th\'eorie introduit un nouveau
param\`etre $M$ qui est l'\'echelle d'\'energie consid\'er\'ee, et
dont d\'ependent les autres param\`etres du mod\`ele. Pour un champ
scalaire, \`a l'ordre d'une boucle, il est possible de montrer que les
contre-termes ajout\'es au lagrangien initial (nu) renormalisent sa
masse en
\begin{equation}
m^2_\zero \rightarrow m^2=m^2_\zero + \Oc(M^2).
\end{equation}
La masse physique observ\'ee est $m$ alors que $m_\zero$ est le
param\`etre fondamental de la th\'eorie. Si la th\'eorie dont on
mesure les effets aujourd'hui est valable jusqu'\`a l'\'echelle de
Planck, i.e. $M\simeq 10^{19} \GeV$, alors pour observer un boson de
Higgs de $115 \GeV$ il faudrait que $m_\zero/M$ soit ajust\'e \`a
$10^{-17}$ pr\`es dans la th\'eorie fondamentale.

\subsection{L'\'energie du vide}
\label{enervide}
Le vide tel qu'il est interpr\'et\'e en th\'eorie quantique des champs
est sujet aux fluc\-tuations quantiques des multiples champs en
interactions. Bien que de valeurs moyennes nulles, celles-ci
contribuent n\'eanmoins \`a l'\'energie du vide telle qu'elle
appara\^\i t dans la constante cosmologique des \'equations de
Friedmann (\ref{friedmann0}) si on suppose que les th\'eories de
physique des particules sont coupl\'ees \`a la gravitation, ne
serait-ce que par les termes cin\'etiques\footnote{Par exemple, pour
le terme cin\'etique d'un champ scalaire on a $g_{\mu
\nu} \partial^\mu \Phi \partial^\nu \Phi$.}. Ainsi pour un champ
scalaire libre de lagrangien (\ref{lagkg}), la densit\'e d'\'energie
du champ dans l'\'etat de vide s'obtient \`a partir du tenseur
\'energie-impulsion
\begin{equation}
\langle 0 | T_{tt} |0 \rangle = \langle 0 | \frac{1}{2} \left[ \sum_\mu
\left(\partial_\mu \Phi \right)^2 + m^2 \Phi^2 \right] | 0 \rangle.
\end{equation}
Avec $\bp$ le quadrivecteur impulsion du champ et apr\`es
quantification du champ scalaire\footnote{Un calcul similaire sera
d\'etaill\'e au chapitre~\ref{chapitrezero}.}, il vient
\begin{equation}
\label{rhovac}
\langle 0 | T_{tt} | 0 \rangle =  \int{\frac{\ud^3 \vec{p}}{(2\pi)^3
2 p^0}} \, \left(|\vec{p}\,|^2 + m^2\right).
\end{equation}
Dans le cas particulier d'un champ de masse nulle, et en int\'egrant
jusqu'\`a l'\'energie de Planck, toujours en supposant que le mod\`ele
standard y est encore valable, l'\'equation (\ref{rhovac}) se
simplifie en
\begin{equation}
\label{rhovide}
\langle 0 | T_{tt} | 0 \rangle \simeq \frac{1}{16 \pi^2} M_\uPl^4.
\end{equation}
La densit\'e critique actuelle est de l'ordre de $10^{-29} \ug/\ucm^3$,
soit dans les unit\'es favorites $1 \meV^4$. La contribution d'un
tel champ \`a l'\'energie du vide serait donc de l'ordre de $
\Omega_{\mathrm{vac}} \simeq 10^{124}$, \`a comparer \`a la valeur
actuelle $\Omega_{\Lambda_\zero} \simeq 0.7$ mesur\'ee par l'expansion
de l'univers. Ce terme ne pose pas de probl\`eme s'il n'y a pas de
couplage gravitationnel ($\Gc=0$) car il peut \^etre renormalis\'e
\`a z\'ero, mais ce n'est plus le cas lorsque la gravit\'e est
consid\'er\'ee.

\section{Conclusion}

Le mod\`ele standard de physique des particules est une excellente
description des interactions \'electrofaible et nucl\'eaire forte au
moins jusqu'\`a des \'energies de l'ordre de quelques centaines de
$\GeV$. Ses pr\'edictions v\'erifi\'ees au pourcent pr\`es constituent
une base solide des m\'ecanismes qu'il invoque, tel que la brisure de
sym\'etrie par le biais d'un champ de Higgs. La d\'ecouverte de sa
particule associ\'ee en sera certainement la meilleure
confirmation. N\'eanmoins, il laisse ouvert de nombreuses questions
montrant qu'il n'est que la partie visible \`a nos \'echelles d'une
th\'eorie plus fondamentale. Le probl\`eme est que l'\'energie
potentiellement accessible dans des acc\'el\'erateurs humains
n'atteindra jamais, dans un futur pr\'evisible, les \'echelles o\`u la
physique semble \^etre radicalement diff\'erente. Cependant, comme
nous l'avons \'evoqu\'e au chapitre~\ref{chapitrecosmo}, ces
\'energies ont d\'ej\`a \'et\'e atteintes dans l'histoire de l'univers
du fait de l'expansion. Ainsi, l'\'etude de l'univers primordial \`a
des \'echelles d'\'energies sup\'erieures \`a $10^3 \GeV$, i.e. un
univers \^ag\'e de moins de $10^{-11} \us$, permet de sonder les
th\'eories de physique des particules alors \`a l'\oe
uvre. Inversement, la description quantique des champs est
certainement celle qu'il convient d'utiliser dans l'univers
primordial. La symbiose existant entre ces deux branches de la
physique est effectivement fructueuse, et, comme nous le verrons dans
les chapitres suivants, permet de donner des \'el\'ements de r\'eponse
aux faiblesses des deux mod\`eles standards, mais
\'egalement de poser de nouvelles questions.

\chapter{Au del\`a des mod\`eles standards}
\label{chapitreaudela}
\minitoc
\section{Introduction}

Les mod\`eles standard de cosmologie et de physique des particules
d\'ecrivant correctement la physique ``\`a basse \'energie'', les
quelques extensions \'evoqu\'ees dans cette section vont donc
concerner l'univers primordial, i.e des \'energies sup\'erieures \`a
l'\'echelle de brisure \'electrofaible o\`u l'univers \'etait \^ag\'e
de moins de $10^{-11} \us$. Le r\^ole essentiel du champ de Higgs dans
cette transition de phase est certainement un exemple de l'importance
qu'ont pu avoir les champs scalaires \`a ces \'epoques. Ainsi, \`a
l'approche hydrodynamique usuelle d\'ecrivant le contenu mat\'eriel de
l'univers, est souvent pr\'ef\'er\'ee l'approche th\'eorie des champs
en espace-temps courbe. La dynamique de l'univers primordial est alors
r\'egie par les lois d'\'evolution d'un ou plusieurs champs
quantiques.

\section{Les champs scalaires cosmologiques}

\subsection{L'inflation}

L'id\'ee qu'un champ scalaire ait pu dominer la dynamique de l'univers
primordial est \`a l'origine du m\'ecanisme d'inflation, i.e. une
phase d'acc\'el\'eration de l'expansion de l'univers, qui s'av\`ere
r\'esoudre les probl\`emes du mod\`ele de FLRW li\'es \`a l'existence
d'un horizon des particules~\cite{guth81}. L'id\'ee g\'en\'erale est que
l'observation actuelle de r\'egions causalement d\'econnect\'ees du
pass\'e r\'esulte de l'accroissement plus rapide de la distance \`a
l'horizon par rapport au facteur d'\'echelle, ce qui est directement
la cons\'equence du ralentissement de l'expansion dans l'\`ere de
radiation et de mati\`ere. Le plus simple de ces mod\`eles invoque le
lagrangien (\ref{lagkg}) d'un champ scalaire $\Phi$ dans un potentiel
$V(\Phi)$ en espace-temps courbe
\begin{equation}
\label{laginf}
\Lc_\uinf=\frac{1}{2} g_{\mu \nu} \partial^\mu \Phi
\partial^\nu \Phi - V(\Phi),
\end{equation}
dont l'\'evolution s'obtient \`a partir de l'\'equation de
Klein-Gordon
\begin{equation}
\frac{1}{\sqrt{-g}} \, \partial_\mu \left(\sqrt{-g} g^{\mu \nu}
\partial_\nu \Phi \right) + \frac{\ud V}{\ud \Phi} = 0,
\end{equation}
avec $g$ le d\'eterminant de la m\'etrique. Pour une
m\'etrique de FLRW (\ref{metriqueFLRW}) et un champ scalaire
homog\`ene, il vient
\begin{equation}
\label{mvtinf}
\ddot{\Phi} + 3 H \dot{\Phi} + \frac{\ud V}{\ud \Phi} = 0.
\end{equation}
La densit\'e d'\'energie et la pression associ\'ees \`a ce champ peuvent
s'obtenir \`a partir de son tenseur \'energie-impulsion
\begin{equation}
\label{tmunulag}
T_{\mu \nu} = \frac{2}{\sqrt{-g}} \frac{\delta \left(\sqrt{-g}
\Lc \right)}{\delta g^{\mu \nu}}=2 \frac{\delta \Lc}{\delta g^{\mu \nu}}
- g_{\mu \nu} \Lc.
\end{equation}
Par identification de (\ref{tmunulag}) avec le tenseur
\'energie-impulsion (\ref{tmunutout}) d'un fluide parfait,
l'\'equation (\ref{laginf}) donne
\begin{eqnarray}
\label{rhoscalar}
\rho_\phi & = & \frac{1}{2} \dot{\Phi}^2 + V(\Phi), \\
\label{pscalar}
P_\phi & = & \frac{1}{2} \dot{\Phi}^2 - V(\Phi).
\end{eqnarray}
La dynamique de l'univers est donn\'ee par les \'equations de
Friedmann (\ref{friedmann0}) et (\ref{friedmann1}) qui deviennent
\begin{eqnarray}
\label{inflation0}
H^2 & = & \frac{\kappa^2}{3}\left[\frac{1}{2} \dot{\Phi}+ V(\Phi)
\right], \\
\label{inflation1}
\dot{H} & = & -\frac{\kappa^2}{2} \dot{\Phi}^2,
\end{eqnarray}
o\`u les termes de courbure et de constante cosmologique ont \'et\'e
n\'eglig\'es\footnote{La constante cosmologique peut n\'eanmoins
\^etre consid\'er\'ee comme un terme constant du potentiel $V(\Phi)$.}
puisque l'on s'int\'eresse \`a l'univers primordial
(cf. Sect.~\ref{sectionplat}). Dans le r\'egime dit de ``roulement
lent'' o\`u l'\'energie cin\'etique du champ scalaire reste
n\'egligeable devant son \'energie potentielle, i.e.
\begin{equation}
\label{slowroll}
\frac{1}{2} \dot{\Phi}^2 \ll V(\Phi) \quad \textrm{et} \quad
\ddot{\Phi} \ll 3 H \dot{\Phi},
\end{equation}
il vient
\begin{equation}
\label{scalarcosmo}
\rho_\phi \simeq -P_\phi \simeq V(\Phi),
\end{equation}
et la dynamique de l'univers s'apparente alors \`a celle conduite par
un terme de constante cosmologique pure $P_\Lambda = -\rho_\Lambda$
[voir Eq.~(\ref{rholambda})]. Les \'equations de Friedmann
(\ref{inflation0}) et (\ref{inflation1}) se simplifient en
\begin{equation}
\label{hubbleinf}
H^2 \simeq \frac{\kappa^2}{3} V(\Phi), \qquad  |\dot{H}|
\ll H^2 ,
\end{equation}
et l'expansion de l'univers est acc\'el\'er\'ee garantissant la phase
d'inflation
\begin{equation}
\Hc'= a \ddot{a} = a^2 \left(\dot{H} + H^2 \right) > 0.
\end{equation}
\`A partir des Eqs.~(\ref{mvtinf}) et (\ref{slowroll}), les potentiels
satisfaisant les conditions de roulement lent doivent v\'erifier
\begin{equation}
\left| \frac{1}{V} \frac{\ud V}{\ud \Phi}\right|^2 \ll 6 \kappa^2.
\end{equation}

La pr\'esence d'une telle p\'eriode inflationnaire permet de
r\'esoudre le probl\`eme de l'horizon. Au premier ordre, d'apr\`es
(\ref{hubbleinf}), le param\`etre de Hubble est constant donnant lieu
\`a une croissance exponentielle du facteur d'\'echelle
\begin{equation}
a \propto \ue^{H t}.
\end{equation}
Pendant cette p\'eriode, la distance \`a l'horizon des particules
(\ref{disthorizon}) se r\'eduit \`a
\begin{equation}
\label{disthorizoninf}
d_{\uH_\uinf} = \frac{1}{H} \ue^{Ht}\left(1-\ue^{-Ht}\right) \simeq
\frac{1}{H} \ue^{H t},
\end{equation}
pourvu que $Ht$ soit suffisamment grand, ce qui sera v\'erifi\'e
\emph{a posteriori}. Ainsi, le rapport des distances \`a l'horizon
entre le d\'ebut d'une phase inflationnaire \`a $t_\ui$ et sa fin \`a
$t_\uf$ vaut
\begin{equation}
\frac{d_{\uH}(t_\uf)}{d_{\uH}(t_\ui)} = \ue^{H \Delta t},
\end{equation}
avec $\Delta t=t_\uf - t_\ui$. La distance \`a l'horizon au temps de
Planck est d'apr\`es (\ref{disthorizonconf}) $d_{\uH_\uPl} = a_\uPl\,
\eta_\uPl$, alors que la distance propre actuelle au mur de Planck est
$d_{\uPl}(\eta_\zero)=a_\zero \, (\eta_\zero-\eta_\uPl)$, qui
ramen\'ee au temps de Planck devient $d_{\uPl}(\eta_\uPl)=a_\uPl \,
(\eta_\zero-\eta_\uPl)$. Comme pour la surface de derni\`ere
diffusion, cette derni\`ere est de plusieurs ordres de grandeur plus
grande que l'horizon \`a cette \'epoque. Le probl\`eme de l'horizon
est donc r\'esolu par l'inflation si
\begin{equation}
d_{\uH_{\uPl_\uinf}}= \ue^{H \Delta t} d_{\uH_\uPl} > d_{\uPl}(\eta_\uPl),
\end{equation}
ou encore pour une phase inflationnaire telle que
\begin{equation}
\ue^{H \Delta t} > \frac{\eta_\zero - \eta_\uPl}{\eta_\uPl} \simeq 10^{29}.
\end{equation}
Un tel facteur d'expansion est obtenue pour $H \Delta t \simeq 67$ et
v\'erifie donc l'hypoth\`ese (\ref{disthorizoninf}). Physiquement, la
phase inflationnaire gonfle rapidement une r\'egion causale au temps
de Planck de sorte qu'elle englobe aujourd'hui tout l'univers
observable.

L'inflation r\'esout aussi naturellement le probl\`eme de la
platitude, en effet l'\'equation (\ref{omegaevol}) d'\'evolution du
param\`etre de densit\'e devient
\begin{equation}
\dot{\Omega}_\utot=-2 H \Omega_\utot \left(\Omega_\utot-1\right),
\end{equation}
qui s'int\`egre en
\begin{equation}
\frac{\Omega_{\utot}(t_\uf) - 1}{\Omega_{\utot}(t_\uf)} =
\frac{\Omega_{\utot}(t_\ui)-1}{\Omega_{\utot}(t_\ui)} \ue^{-2 H \Delta t}.
\end{equation}
Ainsi, pour un taux d'expansion de $H \Delta t \simeq 67$, le
param\`etre de densit\'e se retrouve $10^{60}$ fois plus proche de
l'unit\'e qu'il ne l'\'etait avant la phase d'inflation, ce qui
compense l'instabilit\'e de $\Omega_\utot=1$ par l'\'evolution
ult\'erieure du facteur d'\'echelle (cf. chapitre~\ref{sectionplat}).

Enfin, l'inflation r\'esout \'egalement le probl\`eme de la formation
des grandes structures. Comme pr\'esent\'e dans la
section~\ref{sectionstruct}, les \'equations (\ref{evolonde}) et
(\ref{evolhorizonechelle}) montrent que les longueurs d'onde
$\lambda_\zero$ caract\'eristiques des structures gravitationnelles
actuellement observ\'ees se retrouvent toujours, dans le mod\`ele
standard, en dehors de l'horizon initialement, ce qui pose le probl\`eme
de leur origine physique. Par l'inflation, la distance \`a l'horizon
\'etant exponentiellement dilat\'ee selon (\ref{disthorizoninf}), les
fluctuations originelles de longueur d'onde $\lambda$, donn\'ee par
(\ref{evolonde}), se retrouvent initialement \`a des
\'echelles \emph{subhorizons}~\cite{kolbturner}. L'inflation autorise
ainsi l'existence de m\'ecanismes physiques causals leurs donnant
naissance\footnote{Les fluctuations quantiques du champ $\Phi$ sont
g\'en\'eralement consid\'er\'ees comme la source de ces fluctuations
primordiales.}.

Bien que tr\`es attrayant, le m\'ecanisme d'inflation laisse
ind\'etermin\'e la nature du champ scalaire dominant la dynamique,
``l'inflaton''. De nombreux mod\`eles ont \'et\'e d\'evelopp\'es
permettant de le resituer dans un cadre plus
large~\cite{polarski92,linde94,la89,linde83}. En particulier, les
transitions de phase peuvent \^etre utilis\'ees pour lui donner
naissance~\cite{guth81,guth82,linde82}, la brisure de sym\'etrie
permettant de changer la configuration du champ de Higgs de $\Phi=0$
\`a $\Phi=\eta$ (voir chap.~\ref{chapitrepp}), et sous certaines
conditions d'obtenir un r\'egime de roulement lent. Ces mod\`eles
permettent alors de pr\'evoir le devenir de l'inflaton lorsque
celui-ci atteint son nouvel
\'etat d'\'equilibre\footnote{Il doit osciller autour de son \'etat
d'\'equilibre cr\'eant ainsi des particules, on parle alors de
\emph{reheating} pour d\'esigner l'accroissement d''entropie qui
s'ensuit.}, afin de mieux contraindre les classes de mod\`eles
admissibles~\cite{albrecht82,abbott82,dolgov82,kolbturner}.

\subsection{La quintessence}
\label{soussectionquint}
Comme cela a \'et\'e \'evoqu\'e au chapitre~\ref{chapitrepp}, le
mod\`ele standard de physique des particules pr\'evoit une
contribution gigantesque des champs quantiques \`a l'\'energie du vide
qui n'est absolument pas observ\'e en cosmologie. Ce
probl\`eme~\cite{weinberg89} est le r\'esultat de l'impossibilit\'e de
red\'efinir arbitrairement le vide lorsque l'on consid\`ere la
gravit\'e, et dont la cause peut
\^etre vue comme due \`a notre ignorance de la th\'eorie quantique de la
gravitation. Si elle existe, les \'equations d'Einstein
(\ref{einsteineq}) peuvent en effet se r\'e\'ecrire sous la forme
\begin{equation}
\label{gravquant}
G_{\mu \nu} = \kappa ^2 T_{\mu\nu} - \Lambda_\unue \, g_{\mu \nu}.
\end{equation}
Il est clair que, comme on l'a fait dans la section~\ref{enervide},
consid\'erer une valeur moyenne dans le vide pour l'op\'erateur
tenseur \'energie-impulsion $\langle 0 |T_{\mu \nu}|0
\rangle$, revient \`a bien d\'efinir la valeur moyenne de
l'op\'erateur $G_{\mu\nu}$, ce que l'on ne sait actuellement pas
faire. N\'eanmoins, il est toujours possible de consid\'erer que les
champs classiques bosoniques en jeu dans les deux membres de
(\ref{gravquant}) repr\'esentent la valeur moyenne des excitations de
ces hypoth\'etiques champs quantiques bien d\'efinis d'une th\'eorie
plus vaste, et l'on a ici not\'e $\Lambda_\unue$ le terme de constante
cosmologique ainsi obtenu. Il est raisonnable d'imaginer qu'une
sym\'etrie de cette th\'eorie puisse induire une compensation exacte
telle que
\begin{equation}
\label{lambdanue}
\kappa^2 \langle 0 |T_{\mu \nu}|0 \rangle = \Lambda_\unue g_{\mu \nu},
\end{equation}
annulant ainsi le terme d'\'energie du vide (\ref{rhovide}) dans la
dynamique cosmologique~\cite{witten00}. L'\'echelle naturelle de
$\Lambda_\unue$ serait alors de l'ordre la masse de Planck, ce qui
semble compatible avec le domaine de pr\'edominance de la gravit\'e
quantique.

Le probl\`eme est alors d'expliquer pourquoi la constante cosmologique
effective actuellement observ\'ee ($\Omega_{\Lambda_\zero} \simeq
0.7$) n'est pas rigoureusement nulle, sans introduire un \emph{fine
tunning} entre les deux termes de l'\'equation (\ref{lambdanue})
puisqu'alors la diff\'erence entre les deux termes serait ajust\'ee
\`a $10^{-124}$ pr\`es. Les mod\`eles de quintessence proposent que l'effet
observ\'e actuellement soit d\^u \`a un autre champ scalaire
cosmologique induisant une acc\'el\'eration de l'univers, la
compensation dans (\ref{lambdanue}) \'etant par ailleurs suppos\'ee
exacte~\cite{wang00,wetterich95}.

Un tel champ scalaire peut \'egalement \^etre d\'ecrit par
(\ref{laginf}) avec cette fois un potentiel de la
forme~\cite{peebles88,ratra88}
\begin{equation}
\label{ratrapeebpot}
V(\Phi)=\frac{M^{4+\alpha}}{\Phi^\alpha},
\end{equation}
o\`u $M$ et $\alpha$ sont des param\`etres libres. De tels potentiels
permettent l'existence d'un attracteur pour l'\'evolution cosmologique
du champ scalaire. Il est possible de montrer~\cite{binetruy98} que
ind\'ependamment des conditions initiales, la densit\'e d'\'energie du
champ de quintessence (\ref{rhoscalar}) \'evolue en
\begin{equation}
\rho_\phi \propto a^{-\frac{3\alpha(1+w)}{2+\alpha}}.
\end{equation}
Elle d\'ecro\^{\i}t plus lentement que celle associ\'ee \`a la
mati\`ere et la radiation pourvu que $\alpha>0$ [voir
Eq.~(\ref{rhow})], et donc il existe un moment, apr\`es la phase
domin\'ee par la mati\`ere, o\`u l'\'energie du champ va dominer
l'univers, et d'apr\`es les \'equations (\ref{scalarcosmo}), agir
comme un terme de constante cosmologique dans l'approximation du
roulement lent. Une fois fix\'e $\rho_{\phi_\zero}$ tel que
$\Omega_{\phi_\zero} \simeq 0.7$, la pr\'edominance uniquement
r\'ecente du champ scalaire sur la mati\`ere est \'egalement
v\'erifi\'ee, comme c'\'etait le cas pour une constante cosmologique
pure [voir Eq.~(\ref{damplambda})]. D'autres parts l'insensibilit\'e
aux conditions initiales est assur\'ee par le fait qu'il existe un
attracteur, et fixe les param\`etres du mod\`ele \`a des \'echelles
d'\'energie raisonnables, $M\simeq 10^{6} \GeV$ lorsque $\alpha=6$ par
exemple, rendant le mod\`ele acceptable du point de vue de la
physique des hautes \'energies.

La quintessence r\'esout donc le deuxi\`eme probl\`eme de fine tunning
de la constante cosmologique, elle est de plus pr\'edictive puisque la
domination actuelle du champ scalaire en question donne une \'equation
d'\'etat diff\'erente de celle d'une constante cosmologique
pure. D'apr\`es les \'equations (\ref{rhoscalar}) et (\ref{pscalar}),
lorsque le champ entre dans le r\'egime de roulement lent
\begin{equation}
w_\phi=\frac{P_\phi}{\rho_\phi} \simeq -1 + \frac{\dot{\Phi}^2}{V(\Phi)}
> -1.
\end{equation}
La mesure du param\`etre $w$, pr\'evue dans les ann\'ees \`a venir, de
l'\'equation d'\'etat du fluide cosmologique qui domine maintenant,
permettra de tester les mod\`eles de quintessence~\cite{goliath01}.

Comme dans le cas de l'inflation, l'origine du champ de quintessence
est ind\'etermin\'ee. Il est possible de montrer que le potentiel
(\ref{ratrapeebpot}) m\`ene \`a une valeur physique du champ actuelle
$\Phi \simeq M_\uPl$ sugg\'erant un lien avec des th\'eories
quantiques de gravit\'e. L'existence physique de ces champs scalaires
cosmologiques est \'egalement motiv\'ee par les extensions du
mod\`eles standard de physique des particules, et principalement par
la supersym\'etrie.

\subsection{La supersym\'etrie}

Les repr\'esentations du groupe de Poincar\'e sont r\'ealis\'ees dans
la Nature par les fermions et des bosons. Cependant, le
mod\`ele standard de physique des particules $SU(3) \times SU(2)
\times U(1)$ introduit une dissym\'etrie entre ces deux familles: les
fermions constituent essentiellement la mati\`ere alors que les
bosons\footnote{\`A part le champ de Higgs.} n'apparaissent que comme
vecteurs de leurs interactions. La supersym\'etrie introduit une
nouvelle sym\'etrie entre fermions et bosons, qui de ce fait doit
s'ins\'erer dans le groupe de Poincar\'e. Il est possible de montrer
qu'une telle sym\'etrie est unique, et ne peut \^etre g\'en\'er\'ee
que par des op\'erateurs $Q_r$ de type fermionique\footnote{Le
th\'eor\`eme de Coleman-Mandula~\cite{coleman67} prouve en effet que
toute alg\`ebre de Lie additionnelle au groupe de Poincar\'e en est
n\'ecessairement disjointe, i.e. que ses g\'en\'erateurs commutent
avec les g\'en\'erateurs des translations $P^\mu$ et des
transformations de Lorentz $M^{\mu \nu}$. Ceci n'est plus valable dans
le cas d'une alg\`ebre de Lie gradu\'ee~\cite{golfand71}, dont les
g\'en\'erateurs v\'erifient des relations d'anticommutations. Une
sym\'etrie g\'en\'er\'ee par des op\'erateurs fermioniques peut donc
s'ins\'erer dans le groupe de Poincar\'e. Il est possible de montrer
que la seule th\'eorie non triviale ne peut mettre en jeu que des
g\'en\'erateurs de spin $1/2$, rendant en ce sens la supersym\'etrie
unique~\cite{haag75}.} v\'erifiant les relations
d'anticommutations~\cite{wess74,salam74}
\begin{eqnarray}
\label{anticomsusy}
\left\{Q_r,\Qb_s \right\} & = & 2 \gamma^\mu_{rs} P_\mu,\\
\label{transsusy}
\left[Q_r,P^\mu \right] & = & 0,\\
\label{comsusy}
\left[Q_r,M^{\mu \nu} \right] & = & i \, \sigma^{\mu \nu}_{rs} Q_s,
\end{eqnarray}
o\`u les $Q_s$ sont des op\'erateurs de spin $1/2$ sous les
transformations de Lorentz de chiralit\'e choisie, autrement dit des
spineurs de Majorana. Le choix de l'anticommutateur est impos\'e par
le th\'eor\`eme de Coleman-Mandula~\cite{coleman67} et les autres
relations de commutations sont fix\'ees par les conditions de
covariance et de fermeture de l'alg\`ebre ainsi
g\'en\'er\'ee. $\sigma^{\mu \nu}$ est le g\'en\'erateur des
transformations de Lorentz $M^{\mu \nu}$ dans la repr\'esentation de
spin $1/2$
\begin{equation}
\sigma^{\mu \nu} = \frac{1}{4} \left[ \gamma^\mu,\gamma^\nu \right].
\end{equation}

La r\'ealisation d'une telle sym\'etrie impose qu'\`a tout fermion
$|\uf \rangle$ soit associ\'e un boson de m\^eme masse $|\ub
\rangle$. La relation de commutation (\ref{transsusy}) donne en effet
\begin{equation}
m^2_\ub | \ub \rangle = P^\mu P_\mu Q_s |\uf \rangle = Q_s P^\mu
P_\mu |\uf \rangle = m_\uf^2 |\ub \rangle.
\end{equation}
Le fermion et son boson associ\'e peuvent \^etre vus comme des
partenaires supersym\'etriques d'un m\^eme objet, un
supermultiplet. Dans le cas d'un champ scalaire et d'un spineur, on
parle de supermultiplet chiral dont le lagrangien le plus simple doit
\^etre invariant sous les transformations globales de supersym\'etrie
$\delta_\xi$: les g\'en\'erateurs $Q$ \'etant spinoriels, il en est de
m\^eme pour les param\`etres infinit\'esimaux $\xi$ de ces
transformations. Le supermultiplet chiral se transforme
en~\cite{bailinlove}
\begin{eqnarray}
\label{psisusy}
\delta_\xi \Psi & = & i \gamma^\mu \partial_\mu \left( A + i\gamma^5 B
\right)\xi \\
\delta_\xi A  & = & \xib \Psi \\
\label{Bsusy}
\delta_\xi B & = & i \xib \gamma^5 \Psi,
\end{eqnarray}
o\`u les champs $A$ et $B$ sont les parties r\'eelles et imaginaires
du champ scalaire complexe $\Phi$. On peut alors v\'erifier que le
lagrangien de Wess-Zumino~\cite{wess74}
\begin{equation}
\label{lagwz}
\Lc_\uwz = \partial_\mu \Phi^\dag \partial^\mu \Phi + i \Psib
\gamma^\mu \partial_\mu \Psi,
\end{equation}
est effectivement invariant sous les transformations (\ref{psisusy})
\`a (\ref{Bsusy}), sur couche de masse, i.e. lorsque les champs
v\'erifient leurs \'equations du mouvement. Une telle sym\'etrie n'est
pas observ\'ee \`a nos \'echelles d'\'energie. Par cons\'equent, si
elle existe, elle a  n\'ecessairement \'et\'e bris\'ee dans l'univers
primordial. L'existence de la supersym\'etrie est n\'eanmoins
sugg\'er\'ee par certaines de ses cons\'equences.

La sym\'etrie fermions bosons g\'en\'er\'ee par la supersym\'etrie
r\'esout le probl\`eme des corrections radiatives
\'evoqu\'ees au chapitre~\ref{chapitrepp}. Le d\'eveloppement
perturbatif de la fonctionnelle g\'en\'eratrice des fonctions de Green
(\ref{genegreen}) comprend alors autant de termes invoquant des
fermions que des bosons, leur contribution \'etant de signe oppos\'ee,
les divergences \'eventuelles tendent ainsi \`a se compenser. Les
th\'eories supersym\'etriques sont donc plus facilement
renormalisables, quand elles ne sont pas simplement finies. Ces
compensations ont \'egalement la particularit\'e d'annuler l'\'energie
du vide quantique (\ref{rhovac}). Bien s\^ur, la brisure de
supersym\'etrie restaure ce probl\`eme, mais \`a une \'echelle
d'\'energie tr\`es inf\'erieure \`a celle de Planck\footnote{Le fine
tunning passe de $124$ ordres de grandeur \`a $60$.}.

\begin{figure}
\begin{center}
\epsfig{file=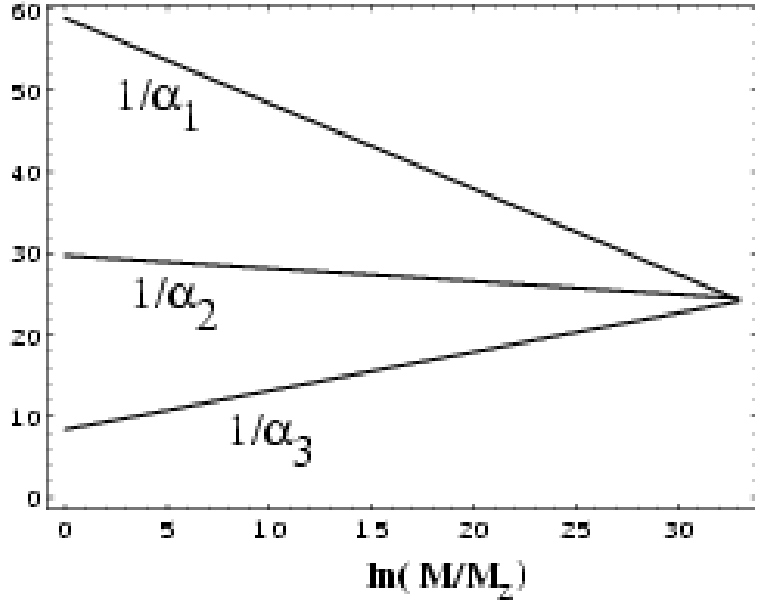,height=8cm}
\caption[Unification supersym\'etrique des constantes de
couplage.]{\'Evolution des constantes de couplage des trois
interactions fondamentales avec l'\'echelle d'\'energie, dans le cadre
des th\'eories de grande unification
supersym\'etriques~\cite{lopez96}. Les param\`etres $\alpha_i$ sont
les constantes de couplage associ\'ees aux trois interactions tels
qu'ils sont d\'efinis sur la figure~\ref{figunifsm}.}
\label{figunifsusy}
\end{center}
\end{figure}

Une autre pr\'ediction attrayante de l'existence de la supersym\'etrie
concerne les th\'eories d'unification des
interactions~\cite{nilles84}. Dans la section~\ref{sectionunif}, on a
vu que la variation des constantes de couplage avec l'\'echelle
d'\'energie tend \`a indiquer leur unification autour de $10^{15}
\GeV$. Les th\'eories de grande unification\footnote{GUT ou
\emph{Grand Unified Theory}.} essaient de trouver des groupes de jauge
redonnant ceux du mod\`ele standard apr\`es brisure d'une sym\'etrie
plus vaste [comme $SU(5)$ ou $SO(10)$] par le m\'ecanisme de
Higgs. L'adjonction de la supersym\'etrie dans ces th\'eories donne
une meilleure unification des constantes de couplages sugg\'erant sa
restauration \`a haute \'energie (cf. Figs.~\ref{figunifsm}
et~\ref{figunifsusy}).

Enfin il est possible de construire une th\'eorie invariante sous les
transformations \emph{locales} de supersym\'etrie. Les relations de
commutations (\ref{anticomsusy}) \`a (\ref{comsusy}) mettant en jeu
les g\'en\'erateurs du groupe de Poincar\'e, l'invariance sous les
transformations locales de coordonn\'ees sera donc naturellement
assur\'ee, d'o\`u sa d\'enomination de supergravit\'e. \`A partir du
lagrangien de Wess-Zumino (\ref{lagwz}), il est possible de
montrer~\cite{bailinlove,freedman76,vannieuw81} que la th\'eorie
correspondante invariante sous $\delta_{\xi(\bx)}$ introduit alors la
relativit\'e g\'en\'erale ainsi qu'un fermion de spin $3/2$
interpr\'et\'e comme le partenaire supersym\'etrique du graviton.
\begin{equation}
\delta_{\xi(\bx)} \Lc_\uwz = 0 \Rightarrow \Lc_\uwz = \partial_\mu
\Phi^\dag \partial^\mu \Phi + i \Psib \gamma^\mu \partial_\mu \Psi +
\frac{1}{2 \kappa^2} R + \frac{1}{2} \varepsilon^{\mu \nu \rho \sigma}
\Psib_\mu \gamma^5 \gamma_\nu \partial_\rho \Psi_\sigma,
\end{equation}
avec $R$ le scalaire de Ricci, et $\Psi^\mu$ le champ fermionique de
spin $3/2$ d\'ecrit par l'action de
Rarita-Schwinger~\cite{rarita41}. La supergravit\'e, en plus de donner
un cadre unifi\'e \`a toutes les interactions, offre des m\'ecanismes
de brisure de supersym\'etrie dans des secteurs cach\'es, i.e. pour
des champs coupl\'es aux particules usuelles uniquement par la
gravitation. Ces brisures douces\footnote{\emph{Soft supersymmetry
breaking}.} aboutissent \`a une s\'eparation de la masse des scalaires
et fermions \`a basse
\'energie compatible avec l'absence
d'observations~\cite{cremmer83,nilles84,barbieri82}, et
potentiellement d\'etectable. Par exemple, pour une \'echelle
d'\'energie de brisure de supersym\'etrie de l'ordre de $10^{10}
\GeV$, le gravitino aurait une masse de $100\GeV$~\cite{bailinlove}.
Du point de vue cosmologique, les potentiels de supergravit\'e
permettent de donner un cadre plus rigoureux \`a la quintessence. Il
est en effet possible de montrer que les potentiels en exponentielle
\begin{equation}
V(\Phi) \propto \ue^{-\alpha \Phi},
\end{equation}
naturellement pr\'esents en supergravit\'e, m\`enent \`a des solutions
attractives compatibles avec les observations~\cite{brax99,brax01},
tout en r\'esolvant le probl\`eme des corrections dues \`a la
gravitation (cf. Sect.~\ref{soussectionquint}).

La supersym\'etrie donne donc un cadre th\'eorique motivant
l'existence de champs scalaires dans l'univers primordial. De plus
toutes ces th\'eories mettent en jeu des m\'ecanismes de brisure de
sym\'etrie, faisant \'egalement intervenir d'autres champs scalaires,
permettant de retrouver le comportement du mod\`ele standard \`a basse
\'energie. La physique des particules actuelle donne donc la vision
d'un univers primordial \`a la dynamique r\'egie par l'\'evolution
d'un ou plusieurs de ces champs et ponctu\'ee de brisures de
sym\'etrie. Cependant, les brisures spontan\'ees de sym\'etrie ne sont
pas sans cons\'equences du point de vue cosmologique. T.~Kibble a en
effet montr\'e dans les ann\'ees 1970~\cite{kibble80,kibble76}
qu'elles pouvaient conduire \`a l'apparition de d\'efauts topologiques
du vide, c'est \`a dire des r\'egions de l'espace dans lesquelles les
champs restent pi\'eg\'es dans leur ancien \'etat de vide. L'\'energie
contenue dans ces objets d\'ependant explicitement de l'\'echelle
d'\'energie de la brisure de sym\'etrie, et pouvant \^etre assez
\'elev\'ee, leur existence peut effectivement modifier les
propri\'et\'es de l'univers.

\section{Les d\'efauts topologiques}
\label{sectiondefauts}
La transition de phase survenant lors de la brisure de sym\'etrie par
m\'ecanisme de Higgs (cf. Sect.~\ref{sectionhiggs}) s'ins\`ere dans le
cadre cosmologique lorsque l'on consid\`ere les effets de la
temp\'erature. Aux \'echelles d'\'energie du mod\`ele standard,
l'\'etat fondamental permettant de d\'efinir la fonctionnelle
g\'en\'eratrice des fonctions de Green (\ref{genegreen}) est le vide
\`a temp\'erature nulle (cf. Sect.~\ref{soussectionrenorm}). Cette
approximation n'est cependant plus valable dans l'univers primordial
puisque les interactions entre les divers champs s'effectuent dans un
bain thermique de temp\'erature $\Theta$ non nulle, et les fonctions
de Green libres (\ref{greenlibre}) doivent \^etre red\'efinies sur un
ensemble complet d'\'etats $|\Omega \rangle$ qui sont suppos\'es
appartenir \`a l'ensemble grand canonique:
\begin{equation}
G^{(n)}_\Theta \propto \sum_\Omega \ue^{-\beta E_\Omega} \langle
\Omega| \Tp \left[\Fc_1 \dots \Fc_n \right] | \Omega \rangle,
\end{equation}
o\`u $\beta = 1/\Theta$. Il est possible de montrer que le formalisme
de renormalisation, utilis\'e \`a temp\'erature nulle pour calculer les
corrections radiatives, s'applique \'egalement au calcul des
corrections due au bain
thermique~\cite{matsubara55,martin59,abrikosov59,brandi85} pourvu que
la variable temporelle soit born\'ee sur l'intervalle $[0,-i\beta]$.

Ainsi, dans le cadre du mod\`ele de Higgs ab\'elien, il est possible
de tenir compte des corrections au potentiel du champ scalaire
(\ref{pothiggs}) \`a la temp\'erature $\Theta$. Le potentiel effectif
\`a une boucle s'\'ecrit alors~\cite{weinberg74,dolan74,kirzhnits74}
\begin{equation}
\label{pothiggseff}
V_\ueff \left(\Phi,\Theta \right) = \frac{\lambda}{8} |\Phi|^4 +
\frac{\lambda}{24} \left(\Theta^2 - 6 \eta^2 \right) |\Phi|^2 +
\eta^4.
\end{equation}
Pour des temp\'eratures
\begin{equation}
\label{tempcrit}
\Theta > \Theta_\uc =\sqrt{6} \eta,
\end{equation}
ce potentiel ne poss\`ede qu'un minimum \`a $|\Phi|=0$. Lorsque la
temp\'erature devient inf\'erieure \`a cette valeur critique, on se
retrouve dans une situation de brisure spontan\'ee de sym\'etrie o\`u
l'\'etat fondamental devient d\'eg\'en\'er\'e $|\Phi| = \eta$ (voir
Fig.~\ref{figvacua}). Inversement, les sym\'etries bris\'ees des
mod\`eles de physique des particules sont donc restaur\'ees dans
l'univers primordial pourvu que la temp\'erature y soit suffisamment
\'elev\'ee. En cons\'equence, lors de son refroidissement, l'univers
est le si\`ege de multiples transitions de phase o\`u l'\'etat
d'\'energie du vide change de mani\`ere plus ou moins rapide selon la
nature de la transition\footnote{Dans les transitions de phase du
premier ordre, l'\'etat fondamental non d\'eg\'en\'er\'e doit
traverser une barri\`ere de potentiel pour rejoindre le nouvel \'etat
d'\'energie inf\'erieure, pouvant consid\'erablement ralentir la
cin\'etique de la transition.}.

\begin{figure}
\begin{center}
\input{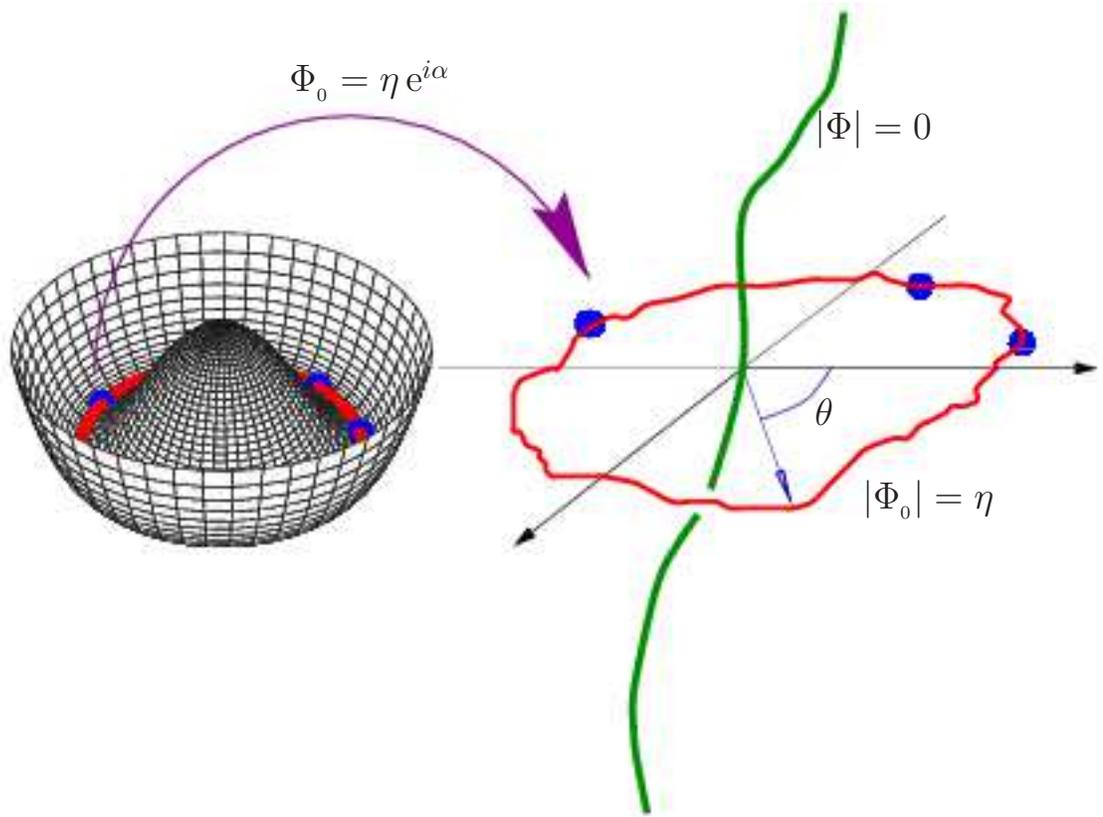}
\caption[Formation des cordes cosmiques par le m\'ecanisme de
Kibble.]{Le potentiel de Higgs du mod\`ele $U(1)$ ab\'elien dans le
plan complexe $\left[\Re\left( \Phi \right),\Im\left( \Phi \right)
\right]$ et la topologie des phases du nouvel \'etat de vide, dans
l'espace physique, menant \`a la formation d'une corde cosmique.}
\label{figcorde}
\end{center}
\end{figure}

Toujours pour le mod\`ele de Higgs ab\'elien, lorsque la temp\'erature
de l'univers passe en dessous de $\Theta_\uc$, le champ ``tombe'' en
chaque point de l'espace physique dans son nouvel \'etat de valeur
moyenne
\begin{equation}
\Phi_\zero = \eta \, \ue^{i \alpha}.
\end{equation}
La phase complexe \'etant \emph{a priori} quelconque\footnote{On la
rep\'esente sous la forme d'une variable al\'eatoire.}, sur des
distances physiques plus grandes que la longueur de corr\'elation
$\ell_\uc$ de la transition de phase, celle-ci va \^etre une fonction
continue de la position $\alpha\left(\vec{x}\right)$. L'existence
d'une longueur de corr\'elation est assur\'ee par l'existence d'une
distance finie \`a l'horizon \`a cette \'epoque. En pratique, un ordre
de grandeur de $\ell_\uc$ est plut\^ot donn\'e par la microphysique
\`a l'\oe uvre lors de la transition et peut \^etre reli\'e \`a
l'\'echelle de distance des fluctuations
thermiques~\cite{kibble80}. Une fois les phases d\'etermin\'ees par la
transition, on peut chercher un contour ferm\'e le long duquel
$\alpha\left(\vec{x}\right)$ varie contin\^ument de $0$ \`a un
multiple entier de $2 \pi$. Si une telle configuration existe, cela
implique n\'ecessairement, par continuit\'e, l'existence d'un point
\`a l'int\'erieur de cette boucle o\`u la phase $\alpha$ ne peut
\^etre correctement d\'efinie. Le seul \'etat du champ de Higgs
autorisant cette phase singuli\`ere est $\Phi_\zero=0$, soit l'\'etat
de vide avant la transition. Par translation dans les dimensions
transverses \`a la boucle, il se forme donc une structure lin\'eique
o\`u le champ de Higgs est nul et appel\'ee corde de Kibble ou corde
cosmique (voir Fig.~\ref{figcorde}). Plus intuitivement, l'ensemble
des choix de phases possibles en chaque point de l'espace survenant
lors de la transition va g\'en\'erer des configurations telles qu'en
certains points singuliers l'ancien vide ne peut choisir son nouvel
\'etat, et une fois les fluctuations de temp\'erature suffisamment
faibles pour ne plus modifier la configuration des
phases\footnote{C'est-\`a-dire pour des temp\'eratures inf\'erieures
\`a la temp\'erature de Ginzburg $\Theta_\uG$ pour laquelle
les fluctuations du champs sont du m\^eme ordre de grandeur que sa
valeur moyenne. Pour une transition de phase du premier ordre, on a
$\Theta_\uG \simeq (1-\lambda)
\Theta_\uc$~\cite{ginzburg60,kirzhnits76}.}, cet \'etat se retrouve
gel\'e en une corde cosmique.

\subsection{Cordes de Nielsen-Olesen}
\label{sectionnielsen}
\begin{figure}
\begin{center}
\epsfig{file=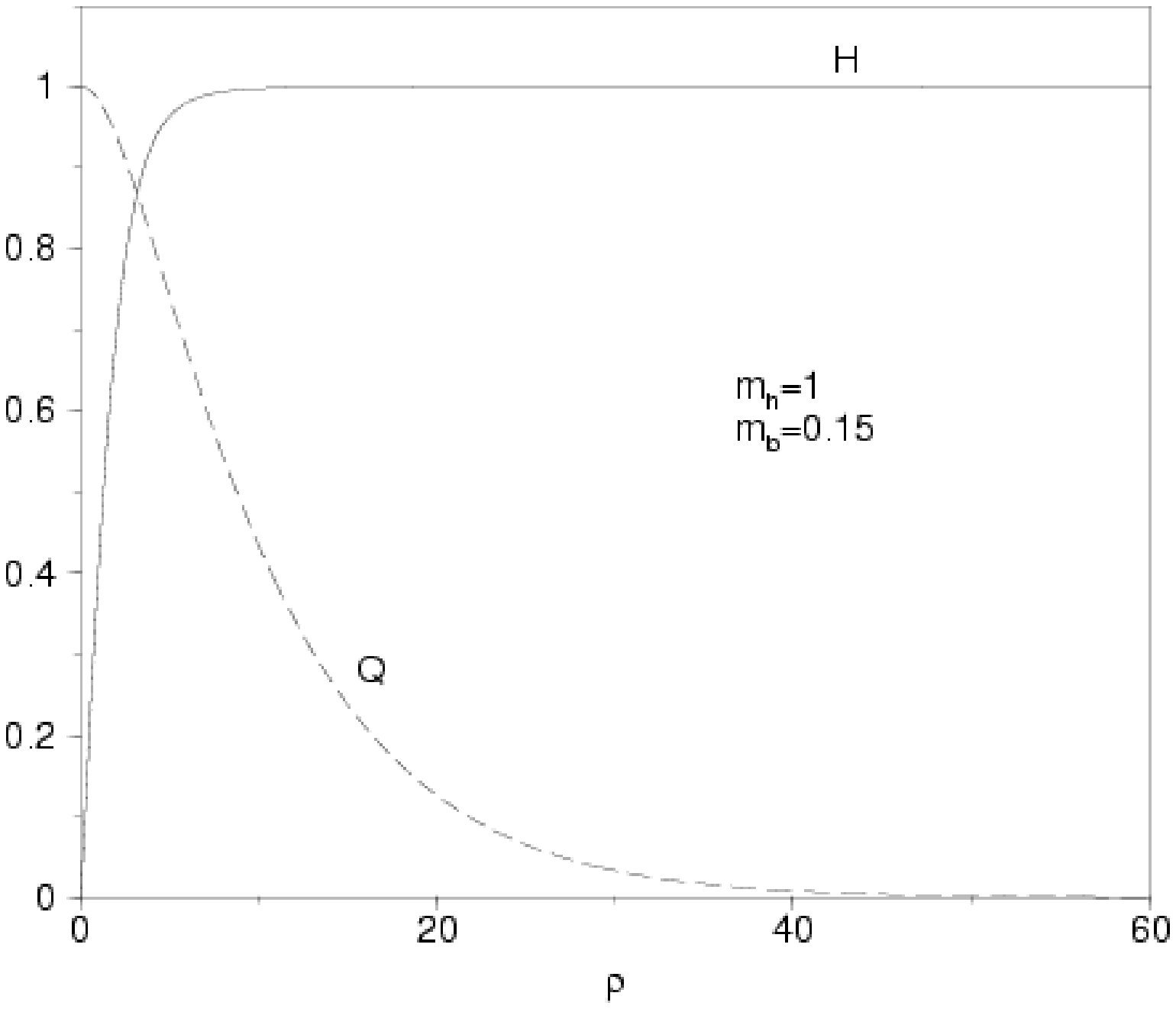,width=10cm}
\caption[Le champ de Higgs et le champ vectoriel d'une corde
cosmique.]{Les solutions de cordes statiques, \`a sym\'etrie
cylindrique, aux \'equations de champs d\'eduites du lagrangien de
Higgs ab\'elien (\ref{laghiggs}). Le champ de Higgs $H$ s'annule au
centre du vortex et rejoint son \'etat de vide usuel sur des distances
de l'ordre de $1/m_\uh$, alors que les bosons de jauge se condensent
sur la corde.}
\label{figback}
\end{center}
\end{figure}

\`A partir du lagrangien du mod\`ele de Higgs ab\'elien
(\ref{laghiggs}), des solutions de type cordes cosmiques peuvent
\^etre trouv\'ees en se restreignant \`a des solutions statiques \`a
sym\'etrie cylindrique, et il est toujours possible d'\'ecrire
localement les champs de Higgs et de jauge sous la
forme~\cite{abrikosov57, nielsen73}
\begin{equation}
\label{defvarphi}
\Phi = \varphi(r) \ue^{i n \theta}, \qquad B_\mu = B_\theta(r)
\delta_{\mu \theta},
\end{equation}
avec $\varphi$ et $B_\theta$ des champs scalaires r\'eels dans le plan
polaire $(r,\theta)$. Les \'equations du mouvement v\'erifi\'ees par
le champ de Higgs et le champ de jauge se simplifient alors en
\begin{eqnarray}
\label{tildehiggs}
\frac{\ud^2 H}{\ud \varrho^2}+\frac{1}{\varrho} \frac{\ud H}{\ud
\varrho} & = & \frac{H Q^2}{\varrho^2}+\frac{1}{2}H(H^2-1), \\
\label{tildegauge}
\frac{\ud^2 Q}{\ud \varrho^2} -\frac{1}{\varrho}\ \frac{\ud Q}{\ud
\varrho} & = & \frac{m_\ub^2}{m_\uh^2}H^2 Q,
\end{eqnarray}
avec $m_\ub=g \eta$ et $m_\uh = \sqrt{\lambda} \eta$ les masses du
boson vectoriel et du boson de Higgs, et les variables
adimensionn\'ees
\begin{eqnarray}
\label{defHQvarrho}
H = \frac{\varphi}{\eta},
\quad
Q = n + g B_\theta
\quad \textrm{et} \quad
\varrho = m_\uh r.
\end{eqnarray}
La solution de ces \'equations est repr\'esent\'ee sur la
figure~\ref{figback} pour un nombre d'enroulement $n=1$. Le champ de
Higgs s'annule au centre du vortex pour rejoindre sa valeur moyenne
$\eta$ dans le vide loin de la corde, alors que le champ vectoriel se
condense sur la corde o\`u la sym\'etrie n'est pas
bris\'ee~\cite{adler,bps,neutral,ringeval}. La largeur physique de la
corde est donn\'ee par l'\'echelle de distance sur laquelle varie le
champ de Higgs, et celle-ci est de l'ordre de sa longueur d'onde
Compton (cf. Fig~\ref{figback})
\begin{equation}
\varnothing_\uc \simeq 1/m_\uh.
\end{equation}
L'\'energie contenue dans cette configuration de corde peut \^etre
explicitement calcul\'ee \`a partir de ces solutions de champs et des
\'equations (\ref{laghiggs}) et (\ref{tmunulag}). Par sym\'etrie
cylindrique, seules les composantes temporelles et axiales sont non
nulles apr\`es int\'egration sur les variables
transverses~\cite{peter94}, il vient alors
\begin{equation}
\label{tmunuhiggs}
T^{tt} = - T^{zz} = - \Lc_\uh = \frac{1}{2} \left(\partial_r \varphi
\right)^2 + \frac{1}{2} \frac{\varphi^2 Q^2}{r^2} + V(\varphi) +
\frac{1}{2 g^2} \frac{\left(\partial_r Q \right)^2}{r^2},
\end{equation}
soit, en fonction des variables adimensionn\'ees
\begin{equation}
T^{tt} = \frac{\lambda \eta^4}{2} \left[ \left(\partial_\varrho H
\right)^2 + \frac{Q^2 H^2}{\varrho^2} + \frac{(H^2-1)^2}{4} +
\frac{1}{\lambda g^2} \frac{\left( \partial_\varrho Q
\right)^2}{\varrho^2} \right].
\end{equation}
L'\'energie par unit\'e de longueur $U$ s'obtient ensuite par
int\'egration sur les variables transverses
\begin{equation}
U = C\!\left(\lambda g^2 \right) \, \eta^2,
\end{equation}
avec $C$ une fonction des constantes de couplage de l'ordre de
l'unit\'e~\cite{book}. La densit\'e d'\'energie d'une corde cosmique est
donc directement donn\'ee par l'\'echelle d'\'energie de la brisure de
sym\'etrie. Ainsi, des cordes de grande unification form\'ees \`a
$10^{15} \GeV$ ont une masse lin\'eique voisine de $10^{15}$ tonnes
par cm et un diam\`etre $10^{-22}$ fois plus petit que celui de
l'atome d'hydrog\`ene. Pour des \'echelles d'\'energie autour de $ 10
\TeV$, la densit\'e d'\'energie est proche du gramme par
centim\`etre pour des rayons dix million de fois plus petits que
l'atome d'hydrog\`ene. Dans la vision hydrodynamique, par analogie
avec (\ref{tmunutout}), $-T^{zz}$ repr\'esente la tension de la corde
$T$. D'apr\`es l'\'equation (\ref{tmunuhiggs}) les cordes cosmiques
bas\'ees sur le mod\`ele de Higgs ab\'elien v\'erifient
donc\footnote{Il s'agit de corde de Goto-Nambu dans la limite
d'\'epaisseur nulle~\cite{goto,nambu}.}
\begin{equation}
\label{enereta}
U=T \simeq \eta^2.
\end{equation}

La brisure de la sym\'etrie $U(1)$ par le m\'ecanisme de Higgs donne
ainsi naissance \`a des d\'efauts dans la configuration spatiale du
vide. Comme \'evoqu\'e au paragraphe pr\'ec\'edent, la stabilit\'e de
ces objets est intimement li\'ee \`a la topologie du vide apr\`es la
transition de phase. D'une mani\`ere plus g\'en\'erale, il est
possible d'utiliser une approche math\'ematique bas\'ee sur les
groupes d'homotopie du vide pour statuer sur leur nature et existence
apr\`es une brisure de sym\'etrie.

\subsection{Topologie du vide}

De mani\`ere g\'en\'erale, lors d'une transition de phase par brisure
de sym\'etrie, l'\'etat d'\'energie minimale passe d'une invariance
sous les transformations d'un groupe de Lie compact $G$, \`a un sous
groupe $H$. La brisure de sym\'etrie est g\'en\'eralement not\'ee $G
\rightarrow H$. Avec les notations utilis\'ees dans la section
pr\'ec\'edente, l'\'etat de vide brisant la sym\'etrie est d\'efini
par
\begin{equation}
\Phi_\zero = \langle 0 | \Phi | 0 \rangle,
\end{equation}
et le sous groupe non bris\'e
\begin{equation}
H \equiv \left\{h \in G \telque \Rc(h) \Phi_\zero = \Phi_\zero
\right\},
\end{equation}
o\`u $\Rc(h)$ sont les matrices de la repr\'esentation du groupe de
sym\'etrie $G$ dans l'espace vectoriel o\`u \'evolue le champ
$\Phi$. Le vide obtenu peut alors \^etre repr\'esent\'e par une
vari\'et\'e $\Vc$ d\'efinie par
\begin{equation}
\Vc \equiv \left\{\Phi_\uv \telque \Phi_\uv = \Rc(v) \Phi_\zero; v \in
G \privede H \right\},
\end{equation}
i.e. l'ensemble des \'etats obtenus \`a partir d'un \'etat de vide
bris\'e $\Phi_\zero$ par les transformations ne le laissant pas
invariant. Autrement dit, chaque \'el\'ement $\Phi_\uv$ repr\'esente
le sous ensemble $v H$ du groupe $G$. La vari\'et\'e $\Vc$
repr\'esente donc l'ensemble des ensemble $vH$ tel que $v\in G
\privede H$, i.e. le groupe quotient de $H$ dans $G$
\begin{equation}
\Vc \sim G/H.
\end{equation}
R\'eciproquement, $G/H$ contient par d\'efinition l'ensemble des
\'el\'ements de la forme $g H$, et ceux-ci ne sont trivialement pas
identique \`a $H$ que si $g \in G \privede H$, c'est \`a dire $\Vc$.

La structure du vide peut donc \^etre d\'ecrite par la topologie du
groupe quotient $G/H$. En particulier, l'existence de cordes de Kibble
est directement reli\'ee \`a l'impossibilit\'e de r\'eduire un chemin
ferm\'e \`a un point, ce qui doit appara\^\i tre n\'ecessairement dans
la structure de $\Vc$. Intuitivement, dans le cas de la sym\'etrie
$U(1)$, si l'on repr\'esente $\Vc$ dans le plan complexe comme une
surface de $\mathbb{R}^2$, de telles boucles ne peuvent exister que si
la surface poss\`ede un trou. Du point de vue math\'ematique, il est
commode d'introduire les groupes d'homotopie pour d\'ecrire les
propri\'et\'es de tels espaces. Un chemin ferm\'e existant dans la
vari\'et\'e $\Vc$ et passant par un point\footnote{Dans toute la suite
la vari\'et\'e et sa correspondance dans l'espace euclidien de
r\'ef\'erence seront identifi\'es.} $\bx$ peut \^etre d\'ecrit par une
application continue $c(u)$ dans $\Vc$ et d\'efinie sur $[0,1]$ telle
que
\begin{equation}
c(0)=c(1)=\bx.
\end{equation}
Deux chemins ferm\'es $c(u)$ et $d(u)$ passant par $\bx$ sont dit
\emph{homotopes} si et seulement si il existe une application $h(u,v)$
d\'efinie sur $[0,1]\times[0,1]$ permettant de passer contin\^ument de
$c$ \`a $d$, i.e.
\begin{equation}
\begin{array}{ccc}
h(u,0) & = & c(u),\\
h(u,1) & = & d(u),
\end{array}
\end{equation}
le point $\bx$ restant fixe,
\begin{equation}
h(0,v)=h(1,v)=\bx.
\end{equation}
Afin de construire une structure de groupe, il est plus judicieux de
consid\'erer la classe d'\'equivalence associ\'ee \`a un chemin
ferm\'e par la relation d'homotopie. On adoptera ici la convention
usuelle dans laquelle $[c]$ d\'esigne l'ensemble de tous les chemins
passant par $\bx$ et homotopes \`a $c(u)$. Une loi de composition peut
\^etre d\'efinie en passant d'un chemin
\`a l'autre. Par d\'efinition $f=c*d$ est le chemin tel que
\begin{equation}
f(u)= \left\{
\begin{array}{lr}
c(2u) & \forall u \in \displaystyle
\left[0,\frac{1}{2}\right], \\
\\
\displaystyle
d(2u-1) & \forall u \in \displaystyle \left[\frac{1}{2},1\right].
\end{array}
\right.
\end{equation}
L'ensemble des classes d'\'equivalence $\left\{[c]\right\}$ munie de
la loi de composition $\circ$ d\'efinie par
\begin{equation}
[c] \circ [d] = [c*d] = [f],
\end{equation}
est un groupe appel\'e premier groupe d'homotopie, ou groupe
fondamental de $\Vc$. Si la vari\'et\'e $\Vc$ est connexe par arc, ce
groupe ne d\'epend pas pas du point\footnote{Plus rigoureusement les
groupes d'homotopie sont isomorphes.} $\bx$, et il est alors not\'e
$\pi_1 \left(\Vc \right)$. Si toutes les boucles de $\Vc$ peuvent
\^etre contin\^ument r\'eduites \`a un point, le premier groupe
d'homotopie se r\'eduit \`a l'identit\'e $\pi_1 \sim I$. Inversement,
dans le cas de la brisure de la sym\'etrie $U(1)$, il y aura autant
d'\'el\'ements dans $\pi_1$ que de familles de boucles homotopes,
i.e. de nombre de tours diff\'erents autour d'un trou. Le groupe
fondamental est donc isomorphe au groupe $\pi_1 \sim \mathbb{Z}$ des
entiers relatifs qui s'identifient aux nombres des enroulements
possibles du champ de Higgs autour de la corde. On voit par cet
exemple comment le groupe d'homotopie permet de classer les
diff\'erents type de cordes\footnote{Ce n'est rigoureusement vrai que
pour $\pi_1$ ab\'elien, dans le cas contraire deux \'el\'ements
diff\'erents de $\pi_1$ peuvent repr\'esenter la m\^eme structure de
vide. Dans ce cas on se sert des classes de conjugaison d\'efinies sur
la relation d'homotopie libre: deux chemins $c$ et $d$ sont
homotopiquement libres si il existe un chemin $f$ tel que $f^{-1} c f$
soit homotope \`a $d$.}  pouvant se former lors de la brisure
$G\rightarrow H$.

De la m\^eme mani\`ere, il est possible de d\'efinir le $n$-i\`eme
groupe d'homotopie de $\Vc$ en rempla\c cant les chemins ferm\'es par
des $n$-surfaces. Dans cette nomenclature, $\pi_0$ compte le nombre de
parties connexes disjointes et $\pi_2$ les diff\'erentes surfaces
homotopes enveloppant les points de la vari\'et\'e. Par analogie avec
la formation de cordes cosmiques, une brisure de sym\'etrie donnant
lieu \`a un vide $\Vc$ tel que $\pi_0(\Vc) \nsim I$ conduit \`a
l'apparition de
\emph{murs de domaine}. Une telle transition de phase est obtenue par
le lagrangien (\ref{laghiggs}) avec cette fois un champ scalaire
r\'eel $\Phi$, et donc pour $G \sim \{-1,1\}$ et $H \sim I$. Le
potentiel correspondant est alors la section r\'eelle du potentiel de
Higgs ab\'elien (cf. Fig.~\ref{figcorde}) et le champ de Higgs prend
cette fois un signe arbitraire apr\`es la brisure de sym\'etrie. Le
mur de domaine est la surface\footnote{La formation d'une telle
surface est tout \`a fait analogue aux surfaces d'aimantation nulle
dans les mat\'eriaux ferromagn\'etiques s\'eparant les domaines de
Weiss.} sur laquelle $\Phi=0$ s\'eparant la r\'egion o\`u $\Phi=\eta$
de celle o\`u $\Phi=-\eta$. La solution de champ
\'equivalente \`a la figure~\ref{figback} peut \^etre calcul\'e
analytiquement~\cite{book}:
\begin{equation}
\Phi(r)= \eta \tanh \left(\frac{1}{2}\sqrt{\lambda} \eta r \right),
\end{equation}
qui s'annule en $r=0$ et tend vers $\pm \eta$ \`a l'infini.  De la
m\^eme mani\`ere, la non trivialit\'e de $\pi_2$ conduit \`a
l'apparition de
\emph{monop\^oles} lors d'une transition de phase, i.e. des points de
l'espace o\`u le champ de Higgs s'annule.

La formation de d\'efauts topologiques survenant lors d'une transition
de phase est donc conditionn\'ee par la nature des groupes de Lie
bris\'es. Signalons deux th\'eor\`emes concernant les groupes
d'homotopie~\cite{nakahara90,wolf67} qui s'av\`erent utiles pour leur
d\'etermination. Si $G$ v\'erifie $\pi_n(G) \sim \pi_{n-1}(G) \sim I$
alors
\begin{equation}
\label{proptopie}
\pi_n(G/H) \sim \pi_{n-1}(H).
\end{equation}
Dans le cas des cordes cosmiques, si $G$ est simplement connexe, alors
$\pi_1(G/H) \sim \pi_0(H)$ et l'\'etude de la topologie du vide se
r\'eduit au d\'enombrement des parties disjointes du sous groupe $H$.
Enfin, dans le cas o\`u le vide s'exprime comme le produit de groupes
de Lie $Q=Q_1 \times Q_2$, son premier groupe d'homotopie
v\'erifie
\begin{equation}
\pi_1(Q) \sim \pi_1\left(Q_1\right) \oplus \pi_1 \left( Q_2 \right).
\end{equation}
Selon ces crit\`eres, il est int\'eressant de noter que la brisure de
sym\'etrie \'electrofaible $SU(2) \times U(1) \rightarrow U(1)$ ne
forme pas de d\'efaut topologique stables, $\pi_1\left[SU(2)\right]
\sim I$. Il est n\'eanmoins possible de former des d\'efauts
semi-topologiques dont la stabilit\'e d\'epend de la dynamique des
champs~\cite{vachaspati91,vachaspati93}. Dans le cas du mod\`ele
standard \'electrofaible, les valeurs des param\`etres, comme la masse
des bosons de jauge, sont tels que ces configurations sont
instables~\cite{hindmarsh92,achucarro92}.

\subsection{Effets gravitationnels}
\label{sectioneffetgrav}
\begin{figure}
\psfig{file=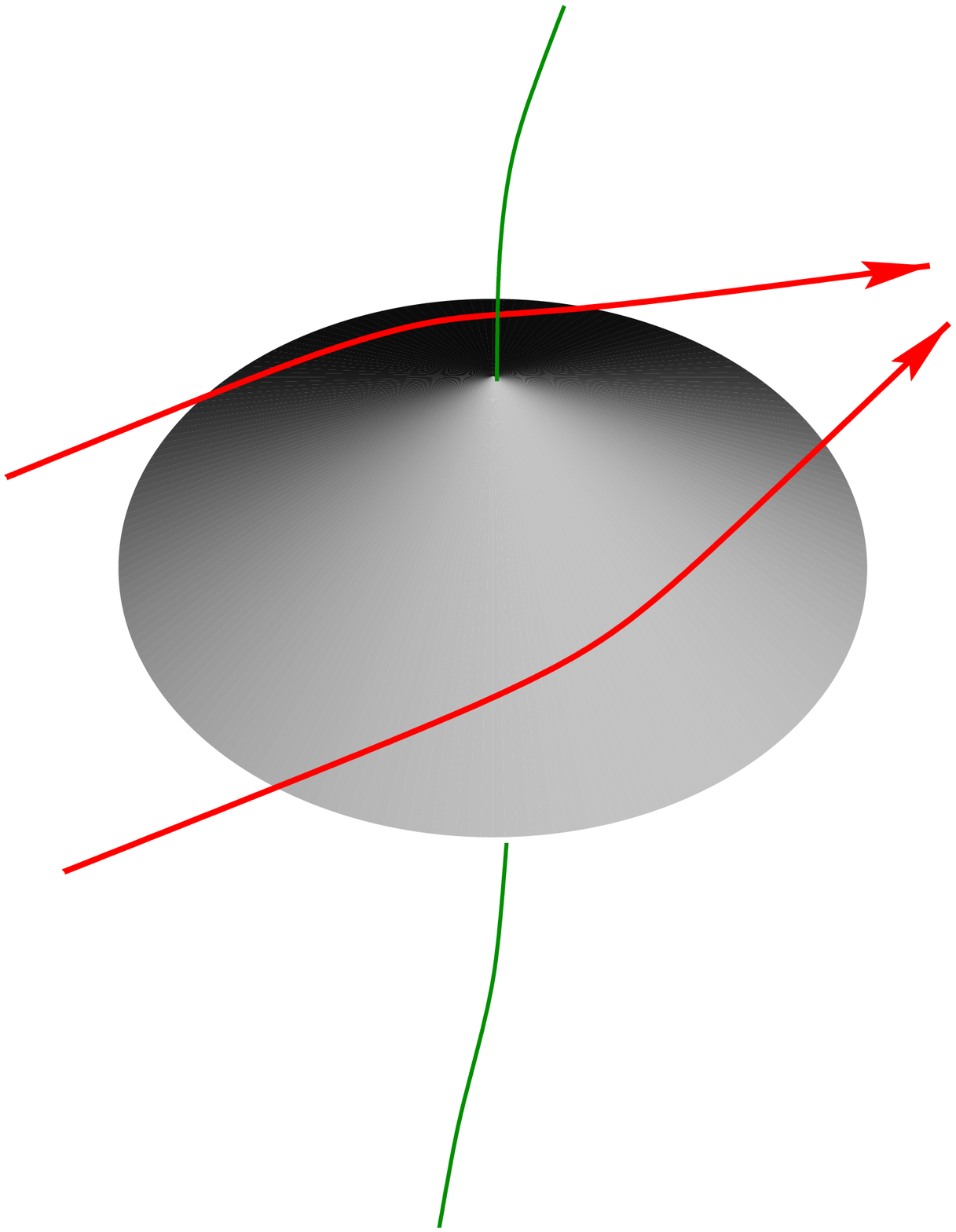,height=12cm}
\psfig{file=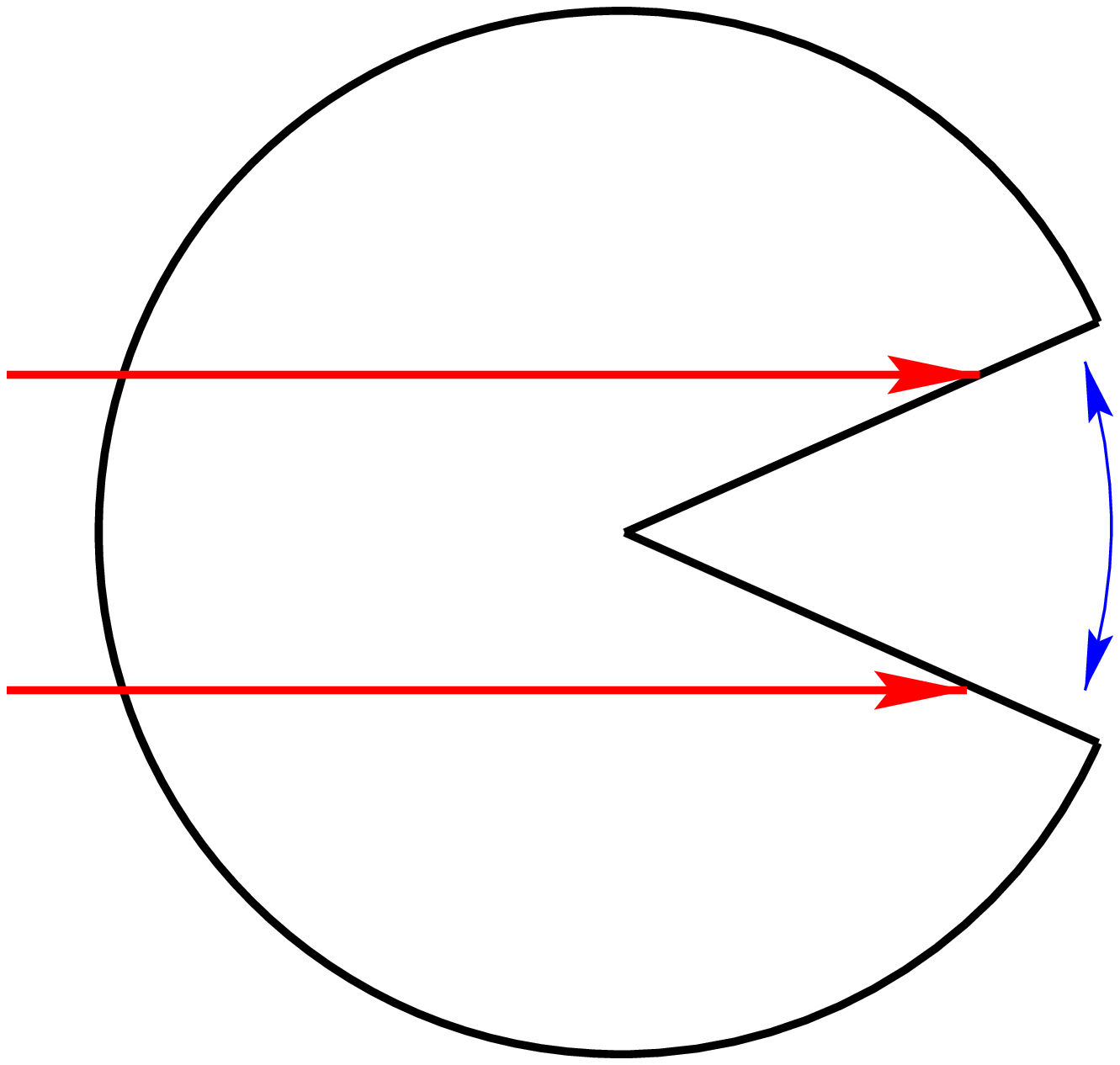,height=10cm}
\caption[M\'etrique d'une corde cosmique sous critique.]{Surface \`a
$t$ et $z$ constant de la m\'etrique autour d'une corde cosmique sous
critique ($\eta \ll M_\uPl$). L'angle manquant d\'efl\'echit les
rayons lumineux de $4 \pi \Gc U$ ind\'ependamment du param\`etre
d'impact.}
\label{figmetric}
\end{figure}

Malgr\'e leur tr\`es grande densit\'e d'\'energie, la structure
lin\'eique des cordes cosmiques ne g\'en\`ere que des effets
gravitationnels mod\'er\'es. Dans la limite newtonienne des
\'equations d'Einstein (\ref{einsteineq}), le potentiel gravitationnel
v\'erifie
\begin{equation}
\label{poissongrav}
\nabla^2 V_\ugrav = 4 \pi \Gc \,\mathrm{Tr}\left(T^{\mu \nu} \right),
\end{equation}
et pour une corde, les termes d'\'energie par unit\'e de longueur $U$
et de tension $T$ s'annulant dans le membre de droite, il n'y a pas
d'effet gravitationnel statique. Les effets gravitationnels dominants
sont directement donn\'es par la g\'eom\'etrie de l'espace-temps
autour de la corde. Celle-ci peut \^etre calcul\'ee approximativement
en supposant la corde infiniment fine et le champ de gravitation
g\'en\'er\'e suffisamment faible pour pouvoir lin\'eariser les
\'equations d'Einstein (\ref{einsteineq}). Pour une corde statique et \`a
sym\'etrie cylindrique, il vient la m\'etrique~\cite{vilenkin81b}
\begin{equation}
\ud s^2 = \ud t^2 - \ud z^2 -\ud r^2 - r^2 \ud \theta^2,
\end{equation}
avec
\begin{equation}
\theta \in \left[0,2 \pi - 8 \pi \Gc U \right].
\end{equation}
L'espace-temps autour d'une corde est plat mais poss\`ede un angle
manquant. Les surfaces \`a $t$ et $z$ constants sont donc des c\^ones
(cf. Fig.~\ref{figmetric}) obtenus en retirant cet angle et en
identifiant les bords. Quantitativement, cette approximation n'est
valable que lorsque le champ g\'en\'er\'e est faible, ou encore
lorsque l'angle manquant est inf\'erieur \`a $2 \pi$
\begin{equation}
\delta \theta = 8 \pi \Gc U \ll 2 \pi.
\end{equation}
D'apr\`es (\ref{enereta}), ceci est v\'erifi\'e pourvu que $\eta \ll
M_\uPl$, ce qui est satisfait pour la plupart des transitions de phase
consid\'er\'ees. Dans les cas o\`u $\eta \sim M_\uPl$, les sections
\`a $t$ et $z$ constants deviennent cylindriques pour $\delta \theta =
2\pi$ et pr\'esentent des singularit\'es\footnote{C'est un moyen
d'obtenir de ``l'inflation topologique''.} pour $\delta \theta > 2\pi$
\`a distance finie du c\oe ur~\cite{gott85,linet90}. Enfin, la
structure de la m\'etrique pr\`es de la corde d\'epend de sa nature
mais tend vers sa forme conique asymptotiquement~\cite{linet85}.

\begin{figure}
\begin{center}
\psfig{file=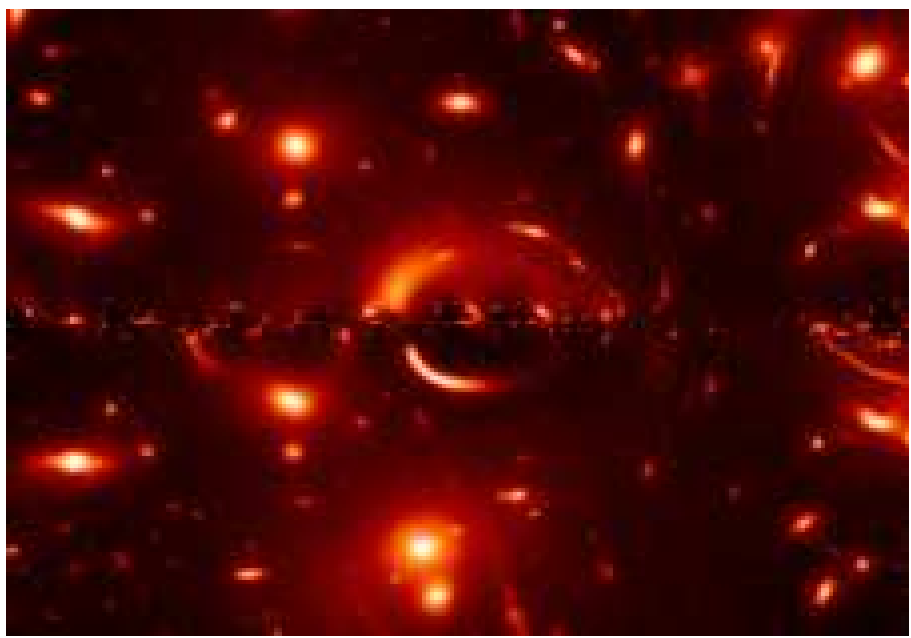,width=12cm}
\caption[Simulation de distorsions lumineuses induites par une corde
cosmique poss\'edant une microstructure.]{Effets simul\'es de lentille
gravitationnelle induits par une corde cosmique poss\'edant une allure
de marche al\'eatoire aux petites \'echelles de distance. L'existence
d'une telle micro-structrure est motiv\'ee par les simulations
num\'eriques d'\'evolution cosmologique de cordes (voir
Chap.~\ref{chapitreevol}). L'image de fond est un amas de galaxies de
redshift $z=1$ et la corde se trouve \`a $z=0.8$, pour une
r\'esolution angulaire inf\'erieure \`a $0.1''$~\cite{bernardeau00}.}
\label{figlensmap}
\end{center}
\end{figure}

\begin{figure}
\begin{center}
\epsfig{file=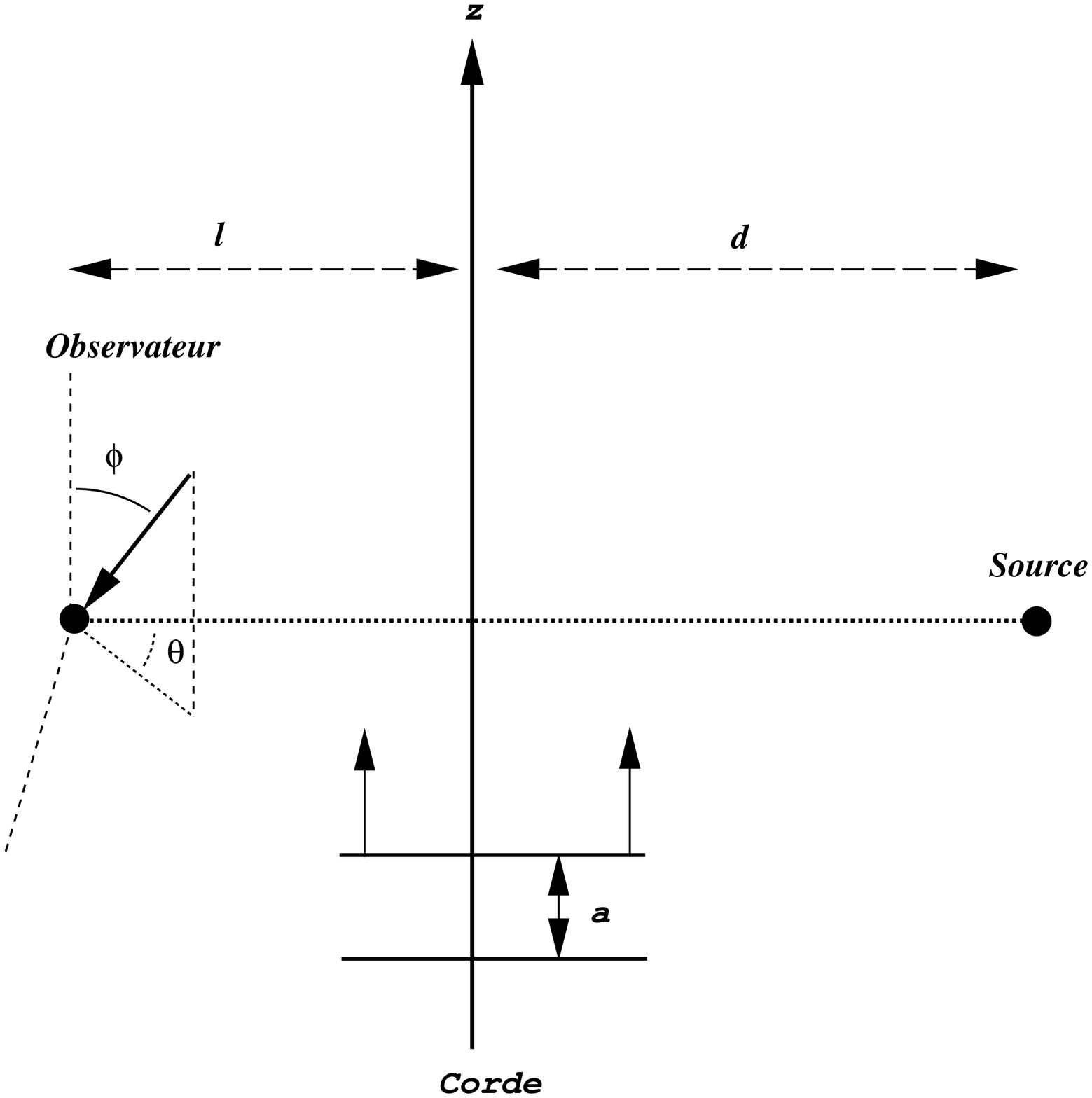,width=7cm}
\epsfig{file=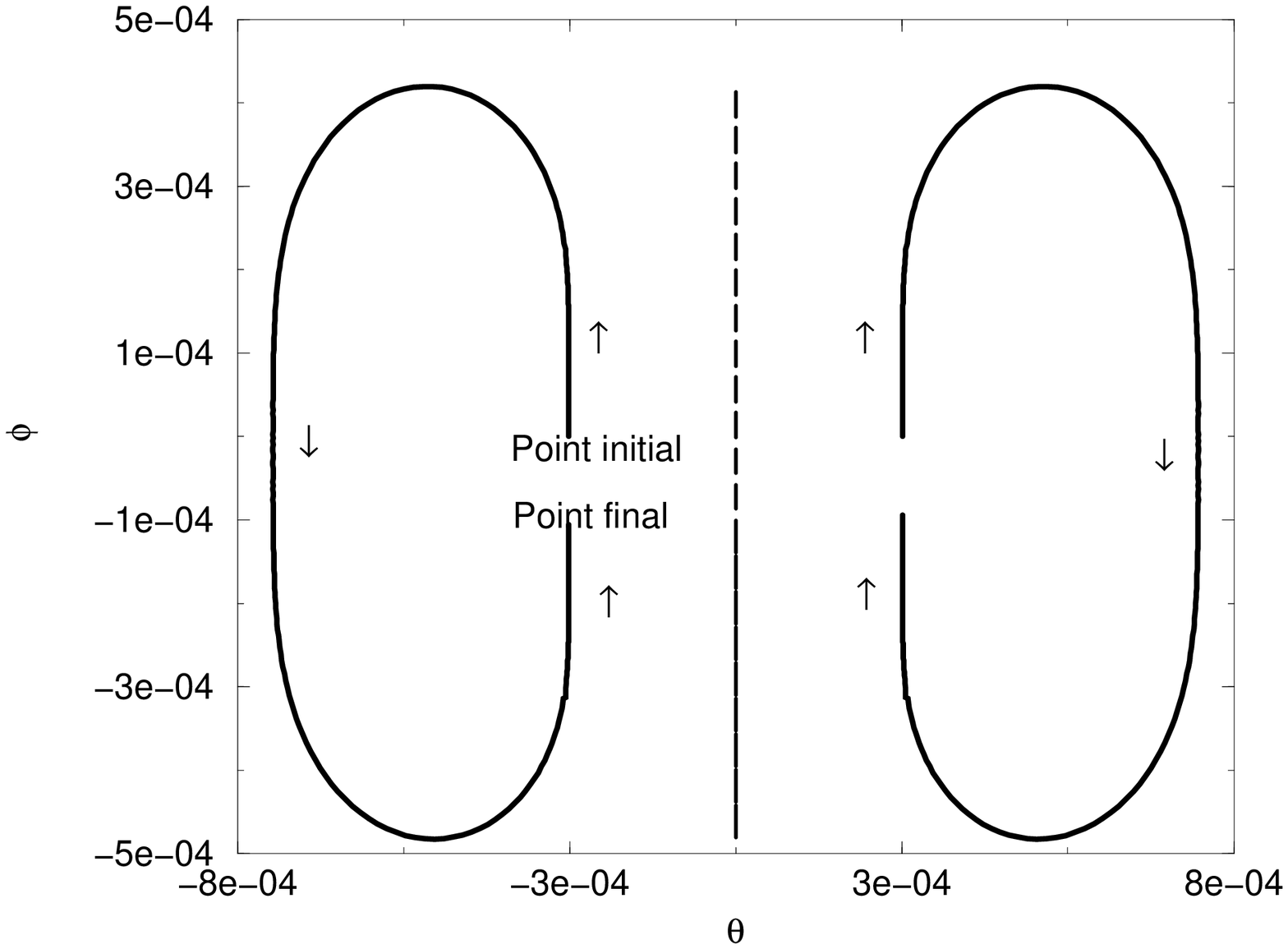,width=8cm}
\caption[D\'eflexion des rayons lumineux par une corde cosmique
parcourue par un courant de genre lumi\`ere.]{Mouvement apparent des
mirages gravitationnels g\'en\'er\'es par une corde cosmique suite au
passage d'un pulse de courant de genre lumi\`ere. Dans la
configuration d'observation (figure de gauche), le temps
caract\'eristique de d\'eplacement est $\ell \Gc U$ pour une source
situ\'ee \`a l'infini ($d \gg \ell$). Ainsi, pour une corde de grande
unification ($\Gc U \simeq 10^{-6}$) situ\'ee \`a quelques
m\'egaparsecs, le mouvement des images doubles de la source
s'effectuerait pendant quelques ann\'ees sur une ouverture angulaire
maximale de $0.1''$~\cite{garrigapeter}.}
\label{figpulse}
\end{center}
\end{figure}

Bien qu'il n'y ait pas d'effet statique, la g\'eom\'etrie de
l'espace-temps autour d'une corde induit tout de m\^eme des effets
gravitationnels potentiellement observables. Par exemple, les rayons
lumineux issus d'une source lointaine peuvent \^etre d\'efl\'echis par
la pr\'esence d'une corde entre la source et
l'observateur~\cite{vilenkin84, gott85}. Dans le r\'ef\'erentiel de
la corde, et dans son plan perpendiculaire, l'angle de d\'eflexion
s'identifie clairement \`a l'angle manquant $\delta D \sim 8 \pi \Gc
U$ (cf. Fig.~\ref{figmetric}). Dans le cas g\'en\'eral, avec $l$ et
$d$ les distances de la corde \`a la source et \`a l'observateur
respectivement, et $i$ l'angle d'inclinaison des rayons lumineux par
rapport au plan perpendiculaire, il vient~\cite{book}
\begin{equation}
\delta D = \delta \theta \sin(i) \frac{l}{l+d}.
\end{equation}
L'effet sur la mati\`ere est similaire. Deux particules de vitesse $v$
passant de part et d'autre de la corde vont acqu\'erir une vitesse
radiale l'une vers l'autre due \`a l'angle manquant de la m\'etrique
(cf. Fig.~\ref{figmetric}). Par projection sur la direction les
reliant, il vient
\begin{equation}
\label{deltav}
\delta v \sim \gamma v \delta \theta
\end{equation}
dans le r\'ef\'erentiel d'une particule\footnote{$\gamma$ est le
facteur de Lorentz.}. Cet effet de sillage laiss\'e par les cordes
dans la mati\`ere environnante a \'et\'e originellement consid\'er\'e
comme un autre m\'ecanisme permettant de g\'en\'erer les perturbations
gravitationnelles n\'ecessaires \`a la formation des grandes
structures (voir Chap.~\ref{chapitrecosmo}). Cependant, comme nous le
verrons dans la partie~\ref{partiedyn} les contraintes cosmologiques
actuelles semblent maintenant \'ecarter ce
sc\'enario~\cite{albrecht97,albrecht99,battye97,bouchet88,copeland99}.
Les effets gravitationnels pr\'ec\'edemment \'evoqu\'es ne concernent
cependant que les cordes rectilignes et sans structure interne. Or,
nous verrons dans le chapitre~\ref{chapitreevol} que leur
\'evolution cosmologique tend naturellement \`a leur donner une
allure fractale aux petites \'echelles de distance. L'existence de
cette sous-structure peut alors exacerber la d\'eflexion lumineuse
induite par la corde du fait de l'apparition de
caustiques~\cite{bernardeau00,uzan00b} (voir
Fig.~\ref{figlensmap}). Les mod\`eles plus r\'ealistes de cordes que
nous pr\'esenterons dans le chapitre~\ref{chapitrecour} invoquent de
plus l'existence de courants de particules pouvant \'egalement
g\'en\'erer des effets gravitationnels observables~\cite{garrigapeter}
(voir Fig.~\ref{figpulse}).

Les transitions de phase survenant dans l'univers primordial telles
qu'elles sont pr\'edites par le mod\`ele standard de physique des
particules peuvent donc g\'en\'erer des d\'efauts topologiques comme
les murs de domaine, les cordes cosmiques ou les
monop\^oles\footnote{Il est \'egalement possible de g\'en\'erer des
\emph{textures} lorsque le troisi\`eme groupe d'homotopie est non
trivial~\cite{kibble76}. Leur stabilit\'e n'est cependant pas
assur\'ee uniquement par la topologie, celle-ci ne pouvant agir que
sur les trois dimensions spatiales~\cite{skyrme61,kibble76}.}. Comme
nous le verrons dans les chapitres suivants, leur existence peut
modifier significativement la dynamique de l'univers, et les
contraintes cosmologiques s'imposant sur ces objets se trouvent
finalement \^etre des contraintes fortes sur les sym\'etries qu'il a
\'et\'e possible de briser dans l'univers primordial. L'\'etude des
d\'efauts topologiques permet donc de sonder la physique des
particules \`a des \'echelles d'\'energie qu'il est impossible
d'atteindre dans les acc\'el\'erateurs.

\section{Quelques contraintes}
\label{sectionqqcont}
Les th\'eories r\'esum\'ees dans ce chapitre permettent de r\'esoudre
certains probl\`emes des mod\`eles standard, mais en posent de
nouveaux, comme la formation des d\'efauts topologiques. Il est de ce
fait indispensable d'explorer ces cons\'equences afin de pouvoir mieux
contraindre la nouvelle physique en jeu. Certains faits observationnels
simples permettent d\'ej\`a de situer les probl\`emes que pose
l'existence de ces reliques de transition de phase.

La sym\'etrie \'electrofaible comportant le groupe $U(1)$, dans une
th\'eorie de grande unification ce groupe doit faire partie de la
cha\^\i ne de brisures de sym\'etrie d'un groupe compact
$G$. D'apr\`es (\ref{proptopie}), puisque $\pi_1 \left[ U(1)
\right] \sim
\mathbb{Z}$, il existe n\'ecessairement un vide dans la cha\^\i ne tel que
$\pi_2(\Vc) \nsim I$ menant \`a la formation de
monop\^oles~\cite{preskill79} (voir la section pr\'ec\'edente). Le
probl\`eme de ces objets est leur abondance: ils se comportent en
effet comme de la mati\`ere ordinaire et l'\'evolution de leur
densit\'e rapport\'ee \`a celle de l'entropie ne varie avec
l'expansion que par leur annihilation avec des antimonop\^oles,
\'egalement form\'es lors de la transition de phase. Or, ce
taux d'annihilation est tel~\cite{zeldov78} qu'on peut montrer que,
pour des monop\^ole de grande unification avec $\eta \sim 10^{15}
\GeV$, leur densit\'e actuelle dominerait la densit\'e critique de
plusieurs ordres de grandeur~\cite{preskill79}. L'existence des
monop\^oles n'est donc pas compatible avec le mod\`ele standard
cosmologique.

Si l'on s'int\'eresse aux murs de domaines, l'\'equation de Poisson du
potentiel gravitationnel (\ref{poissongrav}) se r\'eduit \`a
\begin{equation}
\nabla^2 V_\ugrav = - 4 \pi \Gc U,
\end{equation}
avec $U$ la densit\'e d'\'energie surfacique du mur. La pr\'esence
d'un tel d\'efaut conduit donc \`a une force gravitationnelle
r\'epulsive\footnote{Le mur ayant deux degr\'es de libert\'e
longitudinaux, par analogie avec les cordes (\ref{enereta}), son
tenseur \'energie-impulsion comprend deux termes spatiaux diagonaux
\'egaux \`a $-U$, et est donc de trace n\'egative.}, source
d'effets d\'etectables dans le CMBR \`a partir d'une \'echelle
d'\'energie de $\eta \sim 1 \MeV$, c'est-\`a-dire largement en
de\c{c}a de la transition
\'electrofaible~\cite{zeldov75,kibble76,stebbins89}. Inversement, un
mur de domaine form\'e lors de cette transition aurait une \'energie
$10^8$ fois sup\'erieure \`a celle de toute la mati\`ere connue, et
dominerait donc la dynamique de l'univers. La formation de ces
d\'efauts est donc \'egalement exclue par la cosmologie, et de ce fait
la brisure d'une sym\'etrie discr\`ete par m\'ecanisme de Higgs.

Une solution possible (et en fait privil\'egi\'ee \`a l'heure
actuelle) est celle de l'inflation. L'univers observable aujourd'hui
provenant d'une r\'egion bien plus petite que l'horizon avant
l'inflation, le facteur de dilution est consid\'erable et la
domination actuelle des reliques est \'evit\'ee~\cite{guth81,turner82}. Un
autre m\'ecanisme fait appel aux cordes cosmiques. Lors de transitions
de phases ult\'erieures, des cordes peuvent \^etre form\'ees et se
connecter aux autres d\'efauts alors pr\'esents. Elles catalysent
ensuite leur d\'esint\'egration \'evitant ainsi la catastrophe
cosmologique~\cite{langacker80,copeland86,holman92}.

\begin{figure}
\begin{center}
\epsfig{file=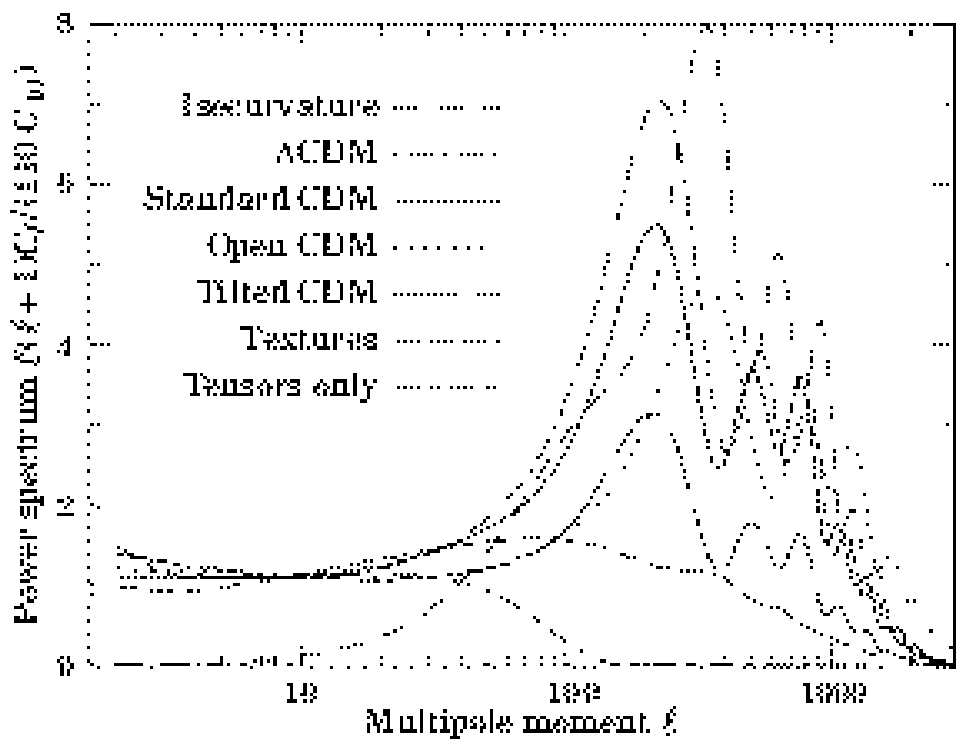,height=10cm}
\caption[Le spectre de puissance du rayonnement fossile pr\'edit par
diff\'erentes th\'eories d'inflation.]{Le spectre de puissance des
anisotropies de temp\'erature dans le CMBR pr\'edit par diff\'erentes
th\'eories d'inflation, et par des d\'efauts topologiques de type
textures~\cite{groom00}. La pr\'esence d'oscillations est observ\'ee
dans les donn\'ees actuelles (voir Fig.~\ref{figdata}).}
\label{figpredict}
\end{center}
\end{figure}

\begin{figure}
\begin{center}
\epsfig{file=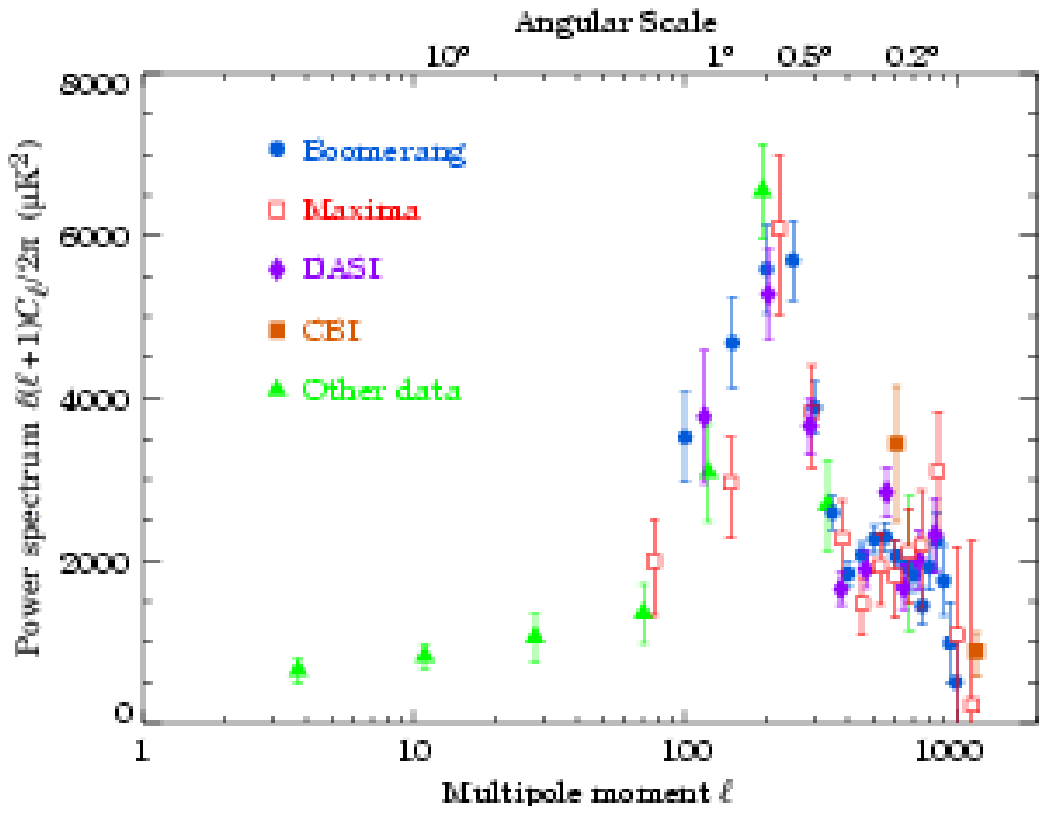,height=10cm}\\
\epsfig{file=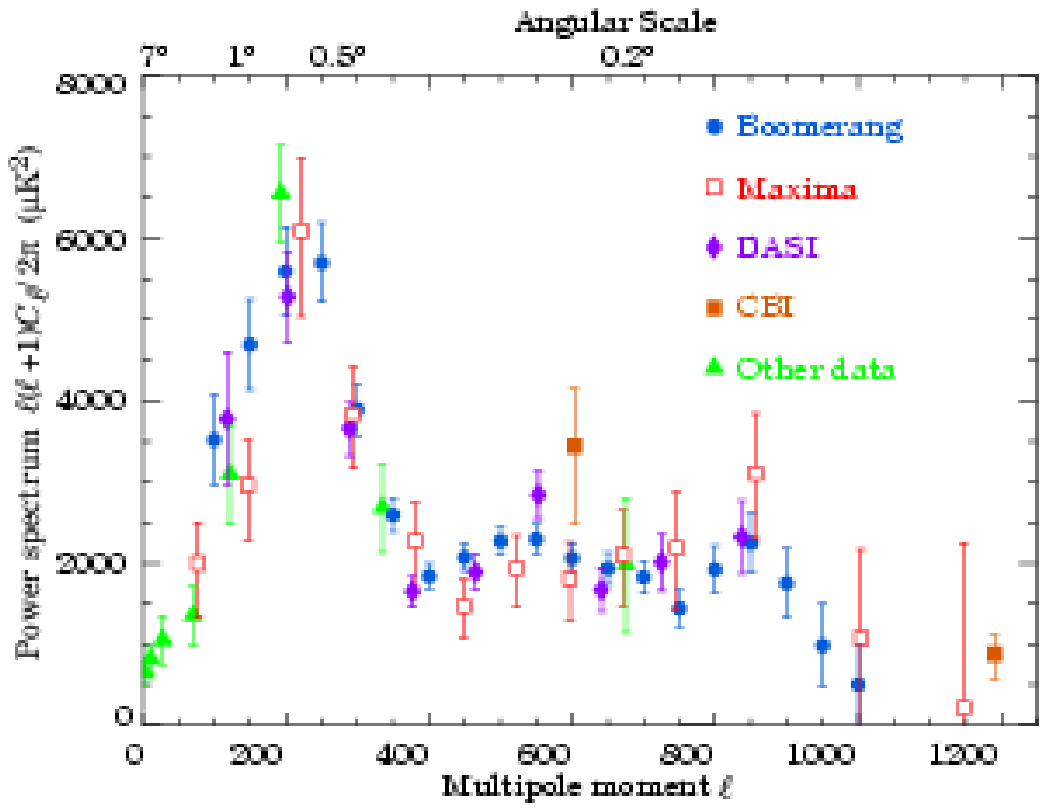,height=10cm}
\caption[Le spectre de puissance du rayonnement fossile observ\'e.]{Le
spectre de puissance des anisotropies de temp\'erature dans le CMBR
observ\'e actuellement~\cite{groom00}.}
\label{figdata}
\end{center}
\end{figure}

\begin{figure}
\begin{center}
\epsfig{file=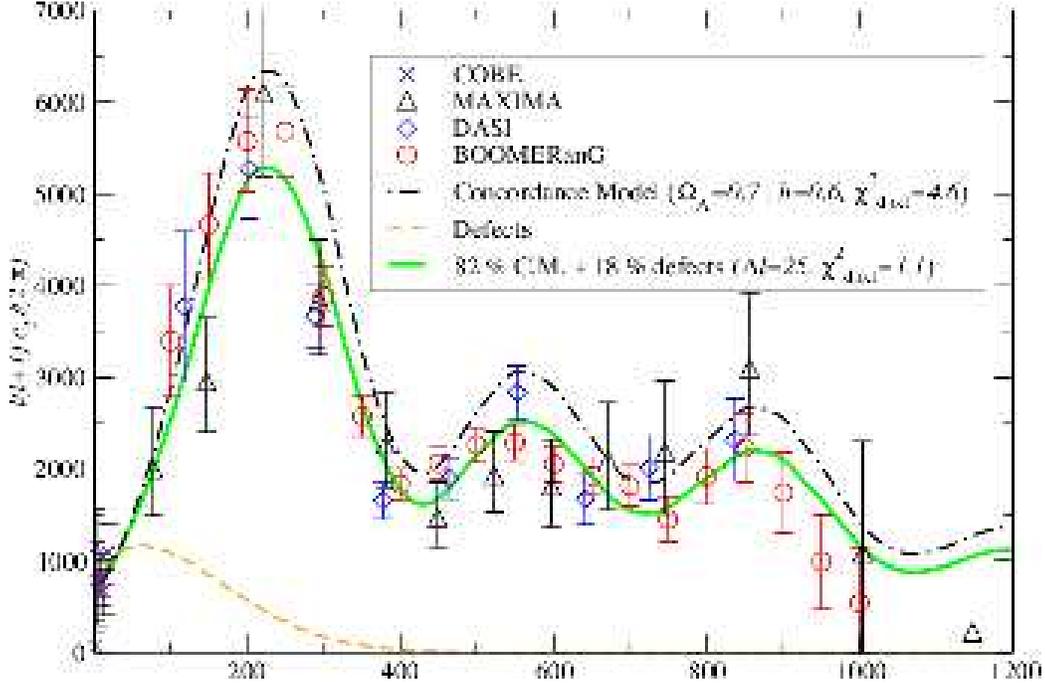,width=9cm,angle=270}
\caption[Le spectre de puissance du rayonnement fossile pr\'edit par
la coexistence d'une phase d'inflation avec des d\'efauts
topologiques.]{Spectre de puissance des fluctuations du CMBR pr\'edit
par la coexistence d'une phase d'inflation avec des d\'efauts
topologiques globaux. La proportion donnant le meilleur ajustement est
\'egalement indiqu\'ee~\cite{bouchet02}.}
\label{figmix}
\end{center}
\end{figure}

Les cordes de Kibble sont actuellement compatibles avec les
contraintes observationnelles, et ce n'est que dans des effets plus
fin que leur existence pourra \^etre, ou non,
infirm\'ee. L'observation de la surface de derni\`ere diffusion au
travers du spectre de puissance du CMBR est actuel\-lement un des
moyens les plus prometteurs pour contraindre ces mod\`eles. En effet,
l'observation du rayonnement de fond diffus dans toutes les directions
$\vec{u}$ du ciel permet d'en mesurer le contraste de temp\'erature
$\delta \Theta/\Theta$. La statistique de ces fluctuations est alors
donn\'ee par la connaissance de ses fonctions de corr\'elation. \`A
l'ordre dominant, la fonction de corr\'elation \`a deux points est
habituellement d\'evelopp\'ee sur la base des polyn\^omes de Legendre
\begin{equation}
\overline{\left(\left.\frac{\delta \Theta}{\Theta} \right|_{\vec{u}_1}
\left. \frac{\delta \Theta}{\Theta}\right|_{\vec{u}_2}\right)} = \frac{1}{4
\pi} \sum_{\ell=2}^\infty C_\ell P_\ell(\cos \theta),
\end{equation}
o\`u $\overline{()}$ d\'esigne la valeur moyenne sur toutes les paires
$(\vec{u}_1, \vec{u}_2)$ d'\'ecart angulaire $\theta$. Il est commode
de d\'ecomposer le contraste de temp\'erature sur les harmoniques
sph\'eriques
\begin{equation}
\left.\frac{\delta \Theta}{\Theta}\right|_{\vec{u}} =
\sum_{\ell=2}^\infty \sum_{m=-\ell}^{\ell} a_{\ell m} Y_\ell^m(\vec{u}),
\end{equation}
o\`u moments multipolaires $C_\ell$ se r\'eduisent \`a
\begin{equation}
C_\ell = \frac{1}{2 \ell + 1} \sum_m \left|a_{\ell m} \right|^2.
\end{equation}
Le spectre de puissance des fluctuations est finalement donn\'ee par
la d\'ependance en $\ell$ des moments $C_\ell$. Sur la
figure~\ref{figpredict} sont repr\'esent\'ees les courbes de puissance
pr\'edites par diff\'erents mod\`eles d'inflation, ainsi que celles
produites par des d\'efauts topologiques de type textures. La
pr\'esence d'oscillations dans les pr\'edictions des mod\`eles
inflationnaires, qui se retrouvent dans les observations (voir
Fig.~\ref{figdata}), semble \^etre une confirmation suppl\'ementaire
du m\'ecanisme d'inflation, initialement introduit pour r\'esoudre le
probl\`eme de l'horizon et de la platitude.
Les cordes cosmiques sont \'egalement des sources actives pouvant
g\'en\'erer des fluctuations de temp\'erature dans le CMBR. La
m\'etrique conique induisant des variations de vitesse relative des
particules dans le sillage de la corde [cf. Eq.~(\ref{deltav})], il
est en de m\^eme pour les photons\footnote{Effet Kaiser-Stebbins.} du
CMBR~\cite{kaiser84,gott85,vachaspati86,bouchet88}. La pr\'esence de
cordes cosmiques doit donc induire des fluctuations de temp\'erature
de l'ordre de
\begin{equation}
\overline{\left(\frac{\delta \Theta}{\Theta}\right)}_2 \simeq
\overline{ \left(\delta v\right)}_2 \simeq \Gc U,
\end{equation}
o\`u $\overline{\left( \right)}_2$ d\'esigne la moyenne quadratique.
Pour des cordes de grande unification $\Gc U \simeq 10^{-6}$ et
l'ordre de grandeur de ces fluctuations est aussi celui qui est
observ\'e. Bien que le spectre de puissance g\'en\'er\'e par un
r\'eseau de cordes cosmiques locales dans FLRW ne soit pas encore bien
connu, le m\'ecanisme de g\'en\'eration des fluctuations est par
nature incoh\'erent~\cite{magueijo96,vincent97,durrer97,durrer99} et
ne peut mener \`a la pr\'esence des pics tels qu'ils sont observ\'es
(voir Fig.~\ref{figdata}). Sur la figure~\ref{figpredict},
l'\'evolution des moments multipolaires des textures s'effondre
rapidement \`a grands $\ell$, et on peut s'attendre \`a un
comportement similaire pour des cordes cosmiques. La pr\'esence d'un
r\'eseau de cordes cosmiques sans phase d'inflation est donc exclue
par ces observations. N\'eanmoins, leur coexistence semble mieux
s'accorder \`a tout point de vue. D'une part, le m\'ecanisme
d'inflation lui-m\^eme peut donner naissance \`a des
cordes~\cite{pollock87,lyth87}, et le devenir de l'inflaton
peut \'egalement conduire \`a des transitions de phase les
g\'en\'erant. Sur la figure~\ref{figmix} est repr\'esent\'ee le
meilleur ajustement des donn\'ees actuelles par un m\'elange des
spectres de puissance issus des d\'efauts topologiques et de
l'inflation. Bien que la d\'eg\'en\'erescence dans les param\`etres
soit importante, il est int\'eressant de noter que les donn\'ees
observationnelles sont mieux ajust\'ees par un tel m\'elange que par
un pur mod\`ele d'inflation~\cite{bouchet02}.

\section{Conclusion}

L'utilisation de la physique des hautes \'energies en cosmologie
permet de r\'esoudre certains probl\`emes de taille du mod\`ele
standard de FLRW (probl\`eme de l'horizon, de la platitude
\dots). Cependant, cette physique pose de nouveaux probl\`emes cosmologiques
comme la formation de d\'efauts topologiques dont font parties les
cordes cosmiques. En plus de leur influence potentielle sur
l'\'evolution de l'univers, le statut de leur existence est d'un
int\'er\^et direct pour la physique des particules du fait des
contraintes qu'elle impose sur les sym\'etries pouvant \^etre
bris\'ees dans l'histoire de l'univers. Comme nous le verrons dans les
parties~\ref{partiedyn} et \ref{partieferm}, il est possible
d'\'etendre ces contraintes aux couplages que peut avoir le champ de
Higgs formant la corde aux autres particules. De telles cordes
s'habillent en effet de courants de charges modifiant leurs
propri\'et\'es microscopiques et par voie de cons\'equence leur
\'evolution cosmologique. La physique \`a l'\oe uvre dans ces cordes
est plut\^ot complexe, et il est d'usage d'utiliser un formalisme
macroscopique unifi\'e pour les d\'ecrire. Dans la
partie~\ref{partiedyn}, ce formalisme sera d\'etaill\'e et appliqu\'e
pour des cordes de Kibble avec et sans courants bosoniques. Il rend
possible la r\'esolution des \'equation du mouvement et l'\'etude
num\'erique de l'\'evolution des r\'eseaux de cordes. De telles
simulations seront pr\'esent\'ees pour des cordes sans courant, et
sont attendues pour statuer de mani\`ere rigoureuse sur la proportion
de tels d\'efauts pr\'esents dans les donn\'ees du CMBR. L'\'etude
cosmologique des cordes conductrices est pour sa part une t\^ache plus
ardue par la complexit\'e de la dynamique induite. Bien qu'encore
applicable pour les courants de bosons, le formalisme macroscopique
semble \^etre moins adapt\'e \`a la description des courants
fermioniques (voir partie~\ref{partieferm}). N\'eanmoins, nous verrons
que la seule
\'etude microscopique des cordes conductrices donne d\'ej\`a des
contraintes fortes sur la physique des hautes \'energies en jeu dans
l'univers primordial.

\part{Dynamique des cordes cosmiques}
\label{partiedyn}
\chapter{Formalismes macroscopiques}
\label{chapitreform}
\minitoc
\section{Introduction}

Dans la partie~\ref{partiegene}, la th\'eorie microscopique minimale
menant \`a la formation de cordes cosmiques par le m\'ecanisme de
Higgs introduisait deux champs bosoniques: le champ de Higgs $\Phi$ et
le champ vectoriel $B_\mu$ associ\'e \`a l'invariance de jauge
$U(1)$. Les solutions de corde aux \'equations de champs sont
g\'en\'eralement compliqu\'ees [voir Eqs.~(\ref{tildehiggs}) et
(\ref{tildegauge})] et il est souvent impossible de les d\'eterminer
autrement que par des m\'ethodes
num\'eriques~\cite{neutral,bps,ringeval}. Il est par cons\'equent
utile de se placer dans la limite de corde d'\'epaisseur nulle. Une
telle approximation revient donc \`a ne plus se soucier de la
structure interne de la corde, i.e. du profil transverse de ses champs
(cf. Fig.~\ref{figback}), en la d\'ecrivant comme une $2$-surface de
l'espace-temps. Lorsque le choix d'un r\'ef\'erentiel est
privil\'egi\'e, ce qui est g\'en\'eralement le cas lorsque l'on
s'int\'eresse au mouvement de la corde ou \`a ses couplages
gravitationnels, il est commode de choisir un syst\`eme de
coordonn\'ees internes $\xi^{a}$ en fonction duquel les grandeurs
physiques peuvent \^etre calcul\'ees: il s'agit du formalisme dit
traditionnel qui sera d\'etaill\'e dans la
section~\ref{sectionnambu}. N\'eanmoins, certaines propri\'et\'es dues
\`a la structure interne des cordes peuvent modifier compl\`etement
leur dynamique, comme par exemple l'existence de courants de
particules s'y propageant. Il est possible de tenir compte de ces
effets par le biais d'une \'equation d'\'etat reliant la densit\'e
d'\'energie $U$ \`a la tension de la corde $T$ dans un formalisme
covariant~\cite{carter89,carter89b,carter94b,carter97}. Ce dernier est
en effet particuli\`erement adapt\'e
\`a la description des propri\'et\'es intrins\`eques \`a chaque corde.
Il permet en outre de discuter, d'un point de vue classique, sur la
stabilit\'e de ces objets en pr\'esence ou non de courants.

\section{Cordes de Goto-Nambu}
\label{sectionnambu}
Le fait de n\'egliger compl\`etement la structure interne d'une corde
impose une invariance de Lorentz longitudinale et celle-ci peut
\^etre repr\'esent\'ee par une $2$-surface
d'univers\footnote{\emph{String worldsheet.}} \'evoluant dans
l'espace-temps dont le mouvement est d\'ecrit par la donn\'ee des ses
coordonn\'ees
\begin{equation}
x^\mu=x^\mu\left(\xi^a\right).
\end{equation}
Les param\`etres $\xi^a$ sont les coordonn\'ees internes \`a la corde,
avec $a \in \{0,1\}$, dont la composante $a=0$ est choisie de genre
temps et $a=1$ de genre espace\footnote{Dans toute la suite, les
indices $a$ et $b$ seront utilis\'es pour d\'esigner les coordonn\'ees
intrins\`eques \`a la $2$-surface.}. Ces coordonn\'ees permettent de
d\'efinir un lagrangien de surface caract\'erisant la dynamique propre
de la corde~\cite{book}. Ce lagrangien $\Lcb_\ugn$ doit \^etre
invariant sous les transformations de coordonn\'ees de l'espace-temps
$\delta x^\mu$ et de la $2$-surface $\delta \xi^a$. Il a de plus la
dimension d'une masse au carr\'e, et peut \^etre identifi\'e \`a la
densit\'e lin\'eique d'\'energie de la corde $U$. Sur la $2$-surface
d'univers courbe, il vient l'action de Goto-Nambu~\cite{goto,nambu}
\begin{equation}
\label{actiongn}
\Scb_\ugn = -U \int{\ud^2 \xi} \sqrt{-\gamma},
\end{equation}
avec $\gamma$ le d\'eterminant de la m\'etrique induite sur la corde
\begin{equation}
\label{metriqueinduite}
\gamma_{ab} = g_{\mu \nu} \frac{\partial x^\mu}{\partial \xi^a}
\frac{\partial x^\nu}{\partial \xi^b},
\end{equation}
telle que
\begin{equation}
\ud s^2 = g_{\mu \nu}\, \ud x^\mu \ud x^\nu = \gamma_{a b}\, \ud \xi^a \ud
\xi^b,
\end{equation}
pour deux points sur la corde s\'epar\'es par $\ud x^\mu$ ou $\ud
\xi^a$.  La normalisation du lagrangien de Goto-Nambu \`a $\Lcb_\ugn =
-U$ est en accord avec la d\'efinition du tenseur \'energie impulsion
obtenu par minimisation de l'action (\ref{actiongn}) par rapport \`a
la m\'etrique $g_{\mu \nu}$:
\begin{equation}
\label{tmunugn}
\sqrt{-g}\, T^{\mu \nu} = 2 \frac{\delta \Scb_\ugn}{\delta g_{\mu \nu}}
= U \int{\ud \xi^2} \sqrt{-\gamma} \, \gamma^{ab}
\frac{\partial x^\mu}{\partial \xi^a} \frac{\partial x^\nu}{\partial
\xi^b} \, \delta^4 \left[\bx - \bx(\xi) \right],
\end{equation}
o\`u la fonction $\delta$ localise la corde dans l'espace-temps de
r\'ef\'erence. Pour une corde droite \`a sym\'etrie cylindrique dans
un espace-temps de Minkowski, en choisissant les coordonn\'ees
internes $\xi^0=x^0=t$ et $\xi^1=x^3=z$, il vient, par int\'egration sur
les variables transverses,
\begin{equation}
\int{T^{tt} \, \ud x_\perp^2}= U = -\int{T^{zz} \, \ud x_\perp^2} = T.
\end{equation}
Les cordes de Goto-Nambu suivent donc une \'equation d'\'etat
identique \`a celle obtenue dans le mod\`ele de Higgs
ab\'elien\footnote{Le formalisme covariant montre que celle-ci est en
fait directement reli\'ee \`a l'invariance de Lorentz longitudinale.},
i.e. $U=T$. Les \'equations du mouvement donnant explicitement la
trajectoire de la corde dans l'espace de r\'ef\'erence, i.e. les
fonctions $x^\mu$, sont obtenues
\`a partir des \'equations d'Euler-Lagrange issues de l'action de
Goto-Nambu. On trouve
\begin{equation}
\label{mvtgngene}
\partial_a \left(\sqrt{-\gamma} \, \gamma^{ab}\, \partial_b x^\mu
\right) + \Gamma^\mu_{\nu \rho} \sqrt{-\gamma} \, \gamma^{ab}
\frac{\partial x^\nu}{\partial \xi^a} \frac{\partial x^\rho}{\partial
\xi^b} = 0,
\end{equation}
les $\Gamma^\mu_{\nu \rho}$ \'etant les symboles de Christoffel de la
m\'etrique $g_{\mu \nu}$. En espace-temps plat, celles-ci se
simplifient en
\begin{equation}
\label{gnmvtplat}
\partial_a \left(\sqrt{-\gamma} \, \gamma^{ab}\, \partial_b x^\mu
\right)=0.
\end{equation}
L'invariance sous les transformations de coordonn\'ees internes
requiert un choix de jauge suppl\'ementaire. On choisit
g\'en\'eralement les conditions de jauge conforme~\cite{book}:
\begin{equation}
\label{confgauge}
\dot{\bx} \cdot \bx' = 0 , \quad \dot{\bx}^{^2} + \bx'^{^2} = 0,
\end{equation}
o\`u le point d\'esigne la d\'eriv\'ee par rapport \`a $\xi^0$ et le
prime par rapport \`a $\xi^1$. Les \'equations du mouvement
(\ref{gnmvtplat}) se r\'eduisent alors \`a
\begin{equation}
\label{gnmvtconf}
\ddot{\bx} + \bx'' = 0.
\end{equation}
Ce choix de jauge n'est pas encore suffisant, et le degr\'e de
libert\'e restant doit \^etre fix\'e. Cela peut se faire, par example,
en se pla\c cant dans la jauge transverse, d\'efinie par
\begin{equation}
\label{transgauge}
\xi^0=x^0=t, \quad \xi^1=\xi.
\end{equation}
Les conditions de jauge conforme (\ref{confgauge}) deviennent alors
\begin{equation}
\label{gnvecgauge}
\dotvec{x}{} \cdot \primevec{x}{}=0, \quad \dotvec{x}{2} +
\primevec{x}{2} = 1.
\end{equation}
Le choix des coordonn\'ees (\ref{confgauge}) permet donc, d'apr\`es
(\ref{gnvecgauge}), de s'affranchir de la composante de vitesse
longitudinale qui est sans signification physique du fait de
l'invariance de Lorentz dans cette direction. La derni\`ere
\'egalit\'e dans l'\'equation (\ref{gnvecgauge}) est juste un choix de
normalisation fixant l'abscisse curviligne $\xi$. Dans ces
conditions, les \'equations du mouvement (\ref{gnmvtconf}) se
simplifient encore en
\begin{equation}
\label{gnmvtvec}
\ddotvec{x}{} - \pprimevec{x}{}=0,
\end{equation}
dont les solutions g\'en\'erales sont la propagation, \`a la vitesse
de la lumi\`ere, des d\'eformations de la corde
\begin{equation}
\label{gnpert}
\vec{x}(t,\xi) = \frac{1}{2} \left[\vec{p}(\xi-t) + \vec{q}(\xi
+t) \right],
\end{equation}
avec
\begin{equation}
\label{normgnpert}
\primevec{p}{2} = \primevec{q}{2} = 1.
\end{equation}
Cette derni\`ere \'equation r\'esulte de la normalisation de $\xi$
choisie dans la jauge conforme. La masse de la corde est donn\'ee,
d'apr\`es (\ref{tmunugn}), par
\begin{equation}
M = \int{\ud \vec{x}^{^3}} T^{tt} = U \int{\ud \xi},
\end{equation}
assurant sa proportionnalit\'e avec l'abscisse curviligne
$\xi$. D'autre part, la courbure de la corde \'etant donn\'ee par
$\ud^2\vec{x} / \ud\xi^{2}$, elle est d'apr\`es (\ref{gnmvtvec}),
proportionnelle \`a l'acc\'el\'eration $\ddotvec{x}{}$ dans le
r\'ef\'erentiel de repos o\`u la vitesse transverse
$\dotvec{x}{}=0$. Ainsi, le mouvement de la corde tend \`a chaque
instant \`a la rendre droite. La conservation de l'impulsion implique
que la corde va finalement osciller en chacun de ses points autour de
sa position d'\'equilibre rectiligne.

Jusqu'\`a pr\'esent, nous avons uniquement consid\'er\'e des cordes
infinies. Cependant, la description lagrangienne \'etant purement
locale, les \'equations du mouvements obtenues s'appliquent tout aussi
bien \`a des configurations de cordes ferm\'ees, ou
boucles\footnote{La d\'esignation de corde \emph{infinie} se
r\'ef\`ere \`a des cordes non ferm\'ees sur des distances
inf\'erieures
\`a l'horizon, celles-ci pouvant tout \`a fait \^etre des boucles sur
des \'echelles de distances sup\'erieures.}. Celles-ci repr\'esentent
en fait la forme dominante sous laquelle les cordes \'evoluent apr\`es
la transition de phase (cf. Chap.~\ref{chapitreevol}). La topologie
ferm\'ee impose cependant des conditions p\'eriodiques \`a la
propagation des perturbations, de la forme
\begin{equation}
\vec{x}(\xi + L) = \vec{x}(\xi),
\end{equation}
o\`u $L$ d\'esigne la longueur propre totale de la boucle,
i.e. $L=M/U$. Cette condition ne peut \^etre v\'erifi\'ee d'apr\`es
(\ref{gnpert}) que si
\begin{equation}
\label{condperiod}
\vec{p}(\xi+L) = \vec{p}(\xi), \quad \textrm{et} \quad \vec{q}(\xi+L)
= \vec{q}(\xi),
\end{equation}
autrement dit le mouvement de la boucle doit \'egalement \^etre
p\'eriodique dans le temps. Il est ainsi possible de d\'evelopper les
vecteurs $\vec{p}$ et $\vec{q}$ en s\'erie de Fourier dont les
coefficients doivent v\'erifier la condition de normalisation
(\ref{normgnpert}) montrant que les vecteurs $\primevec{p}{}$ et
$\primevec{q}{}$ \'evoluent sur une sph\`ere de rayon unit\'e. Si l'on
s'int\'eresse \`a la vitesse des perturbations de courbure, en
d\'erivant l'\'equation (\ref{gnpert}) il vient
\begin{equation}
\dotvec{x}{2} = \frac{1}{4} \left[\primevec{p}{}(\xi - t) -
\primevec{q}{}(\xi+t) \right]^2,
\end{equation}
et les points pour lesquels $\primevec{p}{}(\xi - t)=
-\primevec{q}{}(\xi+t)$ v\'erifient $\dotvec{x}{2}=1$. L'existence de
tels points est g\'en\'erique du fait des conditions de
p\'eriodicit\'e (\ref{condperiod}) qui impliquent
\begin{equation}
\int_0^L \primevec{p}{} \, \ud \xi = \int_0^L \primevec{q}{} \, \ud
\xi = 0.
\end{equation}
Autrement dit, les vecteurs $\primevec{p}{}$ et $\primevec{q}{}$ ne
restent g\'en\'eralement pas dans un seul h\'emisph\`ere de la
sph\`ere de rayon unit\'e (\ref{normgnpert}) assurant de ce fait leur
intersection en certains points. La vitesse de ces points sur la corde
est donc \'egale \`a celle de la lumi\`ere, et \`a partir du
d\'eveloppement de Fourier des vecteurs $\vec{p}$ et $\vec{q}$ en leur
voisinage, il est possible de montrer~\cite{kibble82,turok84} qu'ils
correspondent \`a des points de rebroussement dans la forme de la
corde\footnote{Cusps.}. De telles discontinuit\'es de courbure peuvent
\^etre \`a l'origine de l'\'emission de rayonnement, de particules si
la corde est travers\'ee de courant (cf. Chap.~\ref{chapitrecour}), et
g\'en\'eralement d'ondes gravitationnelles potentiellement
d\'etectables~\cite{damour01}. Le mouvement d'oscillation, et la
pr\'esence de points de rebroussement permettent aux boucles
d'\'evacuer de l'\'energie sous forme de rayonnement
gravitationnel. En premi\`ere approximation, en supposant l'\'energie
gravitationnelle $\Ec$ \'emise par le seul moment quadrupolaire $Q
\simeq M L^2$, il vient~\cite{garfinkle87}
\begin{equation}
\label{quadrupole}
\frac{\ud \Ec}{\ud t} \simeq \Gc \left( \frac{\delta^3 Q}{\delta t^3}
\right)^2 \simeq \Gc Q^2 \nu^6,
\end{equation}
o\`u $\nu$ est la fr\'equence caract\'eristique des oscillations de la
boucle, dont un ordre de grandeur est donn\'e par $\nu \simeq 1/L$. De
mani\`ere g\'en\'erale $\dot{\Ec}
\simeq \gamma_\ug \Gc U^2$ o\`u $\gamma_\ug$ est sans dimension et tient
compte des \'ecarts \`a l'\'equation (\ref{quadrupole}). Le temps de
vie d'une boucle est alors
\begin{equation}
\label{tempsvie}
\tau \simeq \frac{M}{\dot{\Ec}} \simeq \frac{1}{\gamma_\ug} \frac{L}{\Gc U}.
\end{equation}
Le coefficient de proportionnalit\'e peut \^etre estim\'e
num\'eriquement et est voisin de $\gamma_\ug \simeq
65$~\cite{allen92}. Comme nous le verrons dans le
chapitre~\ref{chapitreevol}, la d\'esint\'egration gravitationnelle
des boucles est d\'eterminante dans l'\'evolution et l'existence de
r\'eseaux de cordes cosmiques dans l'univers.

\section{Formalisme covariant}
\label{sectioncov}
La description pr\'ec\'edente est particuli\`erement adapt\'ee au
cordes cosmiques sans structure interne, de Goto-Nambu. Cependant,
comme nous le verrons dans les chapitres suivants, des courants
peuvent prendre naissance sur les cordes. Un formalisme covariant,
d\'evelopp\'e par
B.~Carter~\cite{carter89,carter89b,carter94b,carter97}, permet de
tenir compte de ces nouveaux param\`etres internes dans la dynamique
de la corde uniquement au travers d'une \'equation d'\'etat reliant la
densit\'e d'\'energie $U$ \`a la tension $T$.

\subsection{G\'eom\'etrie des $2$-surfaces}

L'\'evolution de la surface d'univers de la corde
$x^{\mu}\left(\xi^a\right)$ nous a permis, dans la section
pr\'ec\'edente, de d\'efinir la m\'etrique $\gamma_{a b}$ induite sur
la corde [voir. Eq.~(\ref{metriqueinduite})]. La corde peut, d'une
fa\c{c}on g\'en\'erale, \^etre consid\'er\'ee comme une vari\'et\'e de
dimension $p=2$ immerg\'ee dans un espace de dimension
sup\'erieure\footnote{On parle alors de $p$-\emph{brane}.} $n=4$. En
inversant la relation (\ref{metriqueinduite}) il est possible de
d\'efinir le premier tenseur fondamental de la corde
\begin{equation}
\label{premiertenseur}
\first^{\mu \nu} = \gamma^{ab} \partial_a x^\mu \partial_b x^\nu_,
\end{equation}
o\`u la d\'erivation partielle porte sur les coordonn\'ees internes
$\xi^a$. Par construction, ce tenseur vit sur la surface d'univers de
la corde et en est donc un projecteur d\'efini sur l'espace-temps de
m\'etrique $g_{\mu \nu}$. Le projecteur orthogonal s'obtient
directement \`a partir de (\ref{premiertenseur})
\begin{equation}
\label{orthotenseur}
\perp^\mu_{\ \nu} = g^\mu_{ \ \nu} - \first^{\mu}_{\ \nu},
\end{equation}
et il vient
\begin{eqnarray}
\label{orthorel}
\first^\mu_{\ \rho} \first^\rho_{\ \nu} = \first^\mu_{\ \nu}, \qquad
\perp^\mu_{\ \rho} \perp^\rho_{\ \nu} = \perp^\mu_{\ \nu}, \qquad
\first^\mu_{\ \rho} \perp^\rho_{\ \nu} =0.
\end{eqnarray}
Le premier tenseur fondamental $\first^{\mu \nu}$ permet de se
restreindre aux propri\'et\'es intrins\`eques de la $2$-surface tout
en conservant une approche covariante. Dans cet esprit, il est
\'egalement commode d'introduire une d\'eriv\'ee covariante
$\nablab_\mu$ n'agissant que sur la surface d'univers d\'ecrite par
la corde. Une mani\`ere naturelle de l'introduire est de la d\'efinir
comme la composante tangentielle de la d\'eriv\'ee covariante usuelle
sur la $2$-surface, i.e.
\begin{equation}
\label{deriveeinduite}
\nablab_\mu = \first^\nu_{\ \mu} \nabla_\nu.
\end{equation}
La structure g\'eom\'etrique de la corde peut alors \^etre
caract\'eris\'ee par les variations de son premier tenseur fondamental
avec les coordonn\'ees internes. La formulation covariante associ\'ee
met en jeu le deuxi\`eme tenseur fondamental
\begin{equation}
\label{deuxiemefond}
{K_{\mu \nu}}^{ \rho} = \first^\sigma_{\ \nu} \nablab_\mu
\first^\rho_{\ \sigma}.
\end{equation}
Plus pr\'ecis\'ement, s'il existe des quadrivecteurs $u^\mu$ et $v^\mu$
de genre temps et espace, respectivement, formant une base locale
orthonorm\'ee de la $2$-surface
\begin{equation}
\label{orthobase}
u^\mu u_\mu = - v^\mu v_\mu = 1, \quad u^\mu v_\mu =0,
\end{equation}
alors le premier tenseur fondamental devient, dans ce r\'ef\'erentiel
privil\'egi\'e,
\begin{equation}
\first^\mu_{\ \nu} = u^\mu u_\nu - v^\mu v_\nu,
\end{equation}
et selon l'\'equation (\ref{deuxiemefond}) le deuxi\`eme tenseur
fondamental se r\'eduit \`a
\begin{equation}
\label{deuxiemereduit}
{K_{\mu \nu}}^{ \rho} = \perp^\rho_{\ \sigma} \left(u_\nu \nablab_\mu
u^\sigma - v_\nu \nablab_\mu v^\sigma \right).
\end{equation}
La projection orthogonale \`a la corde de l'acc\'el\'eration du
quadrivecteur unit\'e $u^\mu$ s'obtient alors \`a partir de
(\ref{deuxiemereduit})
\begin{equation}
\perp^\rho_{\ \mu} \left(u^\nu \nablab_\nu u^\mu \right) = {K_{\mu
\nu}}^\rho u^\mu u^\nu.
\end{equation}
Le deuxi\`eme tenseur fondamental caract\'erise ainsi la forme
g\'eom\'etrique de la corde dans les dimensions transverses. Il
poss\`ede \'egalement des propri\'et\'es de
sym\'etrie\footnote{Propri\'et\'es de Weingarten.} n\'ecessaire
\`a la description de la corde comme une $2$-surface~\cite{carter92}
\begin{equation}
\label{weingarten}
{K_{[\mu \nu]}}^\rho = 0.
\end{equation}
Sa trace permet de plus de d\'efinir le vecteur de courbure
extrins\`eque
\begin{equation}
\label{vectex}
K^\rho = {{K^\mu}_\mu}^\rho = \nablab_\mu \first^{\mu \rho},
\end{equation}
qui s'exprime \`a partir de la m\'etrique induite
(\ref{metriqueinduite}) et des connexions de l'espace-temps via
\begin{equation}
\label{extrinseque}
K^\mu = \frac{1}{\sqrt{-\gamma}}\partial_a \left(\sqrt{-\gamma}
\gamma^{a b} \partial_b x^\mu \right) + \Gamma^\mu_{\nu \rho}
\gamma^{a b} \partial_a x^\nu \partial_b x^\rho.
\end{equation}
Par comparaison avec l'\'equation (\ref{mvtgngene}), on voit que les
\'equations du mouvement extrins\`eque de la corde se ram\`enent
simplement \`a $K^\mu = 0$ pour des cordes de Goto-Nambu. La
d\'etermination de celui-ci, dans le cas g\'en\'eral, n\'ecessite
n\'eanmoins la connaissance de la dynamique interne \`a la corde, ce
qui peut se faire au moyen du lagrangien effectif de surface.

\subsection{Lois de conservation}

Comme dans le cas de Goto-Nambu, la m\'etrique induite
(\ref{metriqueinduite}) permet de d\'efinir une action de surface
d\'ecrivant les propri\'et\'es physiques intrins\`eques \`a la
corde. Le choix d'un lagrangien effectif $\Lcb$ est en g\'en\'eral
motiv\'e par la th\'eorie des champs sous-jacente et les divers
couplages entre le champ de Higgs formant la corde et les champs
externes. En plus des \'eventuelles sym\'etries de jauge impos\'ees
par les divers couplages choisis, l'invariance de l'action vis-\`a-vis
des transformations infinit\'esimales de la m\'etrique permet de
d\'efinir le tenseur \'energie impulsion associ\'e. L'\'equation
(\ref{tmunulag}) devient sur la $2$-surface
\begin{equation}
\label{tmunuinduit}
\Tb^{\mu \nu} = 2 \frac{\delta \Lcb}{\delta g_{\mu
\nu}} + \Lcb \, \first^{\mu \nu}.
\end{equation}
La conservation du tenseur \'energie impulsion sur la surface
d'univers est assur\'ee par le th\'eor\`eme de Noether
et peut \^etre mis sous sa forme surfacique covariante
\begin{equation}
\label{consreduit}
\nablab_\mu \Tb^{\mu \nu} = \fb^\nu,
\end{equation}
avec la condition tangentielle ${\perp^\mu}_\nu \Tb^{\nu \rho} = 0$. La
densit\'e totale de force $\fb^\mu$ sur la $2$-surface tient compte de
l'effet des champs ext\'erieurs dans la th\'eorie effective, et
\'eventuellement de l'interaction de la corde avec d'autres
$p$-surfaces. Dans le cas d'une corde de Goto-Nambu, $\fb^\mu$ est
identiquement nul, alors que l'existence d'un champ vectoriel
ext\'erieur $A^\mu$, associ\'e \`a une invariance $U(1)$, conduirait
\`a une force de type \'electromagn\'etique  $\fb^\mu = F^{\mu \nu} 
\jb_\nu$, le courant surfacique \'etant naturellement
d\'efini par $\jb_\mu = \delta \Lcb/\delta A^\mu$. Lorsque l'on
s'int\'eresse au mouvement extrins\`eque de la corde, seule la
projection orthogonale de (\ref{consreduit}) est d\'eterminante. \`A
partir des \'equations (\ref{orthotenseur}) et (\ref{deriveeinduite}),
et de la d\'efinition du deuxi\`eme tenseur fondamental
(\ref{deuxiemefond}), il vient~\cite{carter92b}
\begin{equation}
\nablab_\mu \perp^\nu_{\ \rho} = -{K_{\mu \nu}}^\rho -
K_{\mu \, \, \nu}^{\,\rho},
\end{equation}
et la projection orthogonale de (\ref{consreduit}) est alors
\begin{equation}
\label{perpcons}
\Tb^{\mu \nu} {K_{\mu \nu}}^\rho = \perp^\rho_{\ \sigma} \fb^\sigma.
\end{equation}
La connaissance du tenseur \'energie impulsion $\Tb^{\mu \nu}$ et des
densit\'es de force $\fb^\mu$ permet ainsi de d\'eterminer \`a l'aide des
Eqs.~(\ref{vectex}) et (\ref{perpcons}), le vecteur de courbure
extrins\`eque, et par la relation (\ref{extrinseque}) les \'equations
du mouvement de la corde.

La forme g\'en\'erique du tenseur \'energie impulsion d'une
$2$-surface s'obtient en supposant que son \'energie est bien
d\'efinie, i.e. positive ou nulle, et que la causalit\'e n'est pas
viol\'ee, i.e. qu'il poss\`ede un vecteur propre tangentiel de genre
temps $u^\mu$ v\'erifiant $u^\mu u_\mu > 0$. La densit\'e d'\'energie
est alors la valeur propre du tenseur \'energie impulsion associ\'ee
\`a $u^\mu$. En le normalisant \`a l'unit\'e $u^\mu u_\mu=1$, il vient
\begin{equation}
\Tb^\mu_{\ \nu} u^\nu = U u^\mu,
\end{equation}
avec $U \ge 0$. Il est alors possible d'introduire un autre vecteur
tangentiel $v^\mu$ de genre espace v\'erifiant (\ref{orthobase}) pour
construire la base orthornorm\'e introduite dans la section
pr\'ec\'edente. En supposant, dans cette base, le tenseur \'energie
impulsion diagonal\footnote{Si ce n'est pas le cas, il est toujours
possible de s'y ramener par un changement de base lorsque les vecteurs
propres ne sont pas de genre lumi\`ere.}, la tension $T$ de la corde
est d\'efinie par
\begin{equation}
\Tb^{\mu \nu} = U u^\mu u^\nu - T v^\mu v^\nu,
\end{equation}
soit, en fonction du premier tenseur fondamental
(\ref{premiertenseur})
\begin{equation}
\label{tmunuperfect}
\Tb^{\mu \nu} = (U - T) u^\mu u^\nu + T \, \first^{\mu \nu}.
\end{equation}
La d\'enomination de tension se justifie donc par analogie avec la
forme hydrodynamique (\ref{tmunutout}) o\`u $T$ s'identifirait \`a
$-P$, comme intuitivement attendu sur un banal fil en m\'ecanique
classique. En reportant l'expression (\ref{tmunuperfect}) dans les
\'equations du mouvement extrins\`eque (\ref{perpcons}), il vient
\begin{equation}
\label{mvtextpriv}
\Tb^{\mu \nu} {K_{\mu \nu}}^\rho = \perp^\rho_{\ \sigma}
\left( U \dot{u}^\sigma - T v'^\sigma \right)=\perp^\rho_{\ \sigma}
\fb^\sigma,
\end{equation}
avec l'acc\'el\'eration $\dot{u}^\sigma =u^\mu \nablab_\mu u^\sigma$
et $v'^\sigma = v^\mu \nablab_\mu v^\sigma$. Dans le cas o\`u il n'y a
pas de champ de force ext\'erieur, $\fb^\mu = 0$ et le vecteur de
courbure extrins\`eque est alors donn\'e par l'\'equation
(\ref{vectex}), qui \`a l'aide de (\ref{mvtextpriv}), se simplifie en
\begin{equation}
\label{vecexparfait}
K^\rho = \perp^\rho_{\ \sigma} \left(1-\frac{U}{T} \right)
\dot{u}^\sigma.
\end{equation}
Les corde de Goto-Nambu v\'erifiant $U=T$ poss\`edent donc un vecteur
de courbure extrins\`eque nul $K^\rho=0$ redonnant \`a partir de
(\ref{extrinseque}) les \'equations du mouvements
(\ref{mvtgngene}). En fait, l'expression (\ref{tmunuperfect}) du
tenseur \'energie impulsion ne peut \^etre invariante sous les
transformation de Lorentz longitudinales que si le terme en $u^\mu
u^\nu$ s'annule, soit pour $U=T$. L'\'egalit\'e entre la densit\'e
d'\'energie et la tension est donc une cons\'equence directe de
l'absence de structure interne le long de la corde. R\'eciproquement,
en modifiant cette relation, il est possible de construire des
mod\`eles effectifs tenant compte des propri\'et\'es microscopiques
rendant la structure interne non triviale. L'\'equation reliant $U$
\`a $T$ est alors une \'equation d'\'etat pour la dynamique de la
corde. 

\subsection{\'Equation d'\'etat}
\label{sectionetatform}
La plus simple des \'equations d'\'etat autre que celle de Goto-Nambu
$U=T$ est celle r\'egissant la dynamique d'une corde parfaitement
\'elastique, ou barotropique, dont la densit\'e d'\'energie ne
d\'epend que de la tension\footnote{Comme nous le verrons dans la
partie~\ref{partieferm}, ce n'est pas la seule possible.} $U =
U(T)$. Puisqu'une telle corde n'est plus invariante de Lorentz selon
sa direction longitudinale, elle poss\`ede une structure interne qui
doit \^etre d\'ecrite par des param\`etres physiques. \`A partir de
l'\'equation d'\'etat $U=U(T)$, il est en effet possible de d\'efinir
le potentiel chimique $\mub$ et un nombre de densit\'e $\nub$ tels que
\begin{equation}
\label{defintparam}
\ln{\nub} = \int{\frac{\ud U}{U-T}}, \qquad \ln{\mub} = -\int{\frac{\ud
T}{U-T}}.
\end{equation}
Par int\'egration de l'\'equation (\ref{defintparam}), le potentiel
chimique et le nombre de densit\'e apparaissent transform\'es de
Legendre l'un de l'autre
\begin{equation}
\label{etatbarotrope}
U - T = \mub \, \nub. 
\end{equation}
\`A partir des \'equations (\ref{tmunuperfect}) et
(\ref{etatbarotrope}), le tenseur \'energie impulsion de la corde
barotropique devient
\begin{equation}
\label{tmunubarotrope}
\Tb^{\rho \sigma} = \nub^\rho \mub^\sigma + T \first^{\rho \sigma},
\end{equation}
o\`u le vecteur densit\'e de courant $\nub^\rho$ et le flux d'\'energie
$\mub^\rho$ sont d\'efinis par
\begin{equation}
\label{defflux}
\nub^\rho = \nub u^\rho, \qquad \mub^\rho = \mub u^\rho.
\end{equation}
La dynamique de la corde est alors obtenue \`a l'aide de l'\'equation
de conservation (\ref{consreduit}) avec, selon (\ref{tmunubarotrope})
\begin{equation}
\label{divtmunu}
\nablab_\rho \Tb^\rho_{\ \sigma} = \mub_\sigma \nablab_\rho \nub^\rho +
\nub^\rho \nablab_\rho \mub_\sigma - \nub^\rho \nablab_\sigma
\mub_\rho + T K_\sigma.
\end{equation}
Comme dans le cas de Goto-Nambu, les \'equations du mouvement
extrins\`eques sont obtenues en projetant (\ref{consreduit}) sur les
dimensions transverses \`a l'aide de $\perp^\rho_{\ \sigma}$, alors que la
dynamique interne est donn\'ee par la composante longitudinale. En
supposant la corde barotrope isol\'ee, i.e. $\fb^\rho = 0$ dans
l'\'equation (\ref{consreduit}), la divergence de $\Tb^\rho_{\
\sigma}$ est nulle, et par contraction de (\ref{divtmunu}) avec la
densit\'e de courant $\nub^\sigma$, il vient
\begin{equation}
\label{conscourant}
\nub^\sigma \nablab_\rho \Tb^\rho_{\ \sigma} =0 \quad \Rightarrow \quad
\nablab_\rho \nub^\rho = 0.
\end{equation}
Le quadrivecteur densit\'e de courant est donc conserv\'e. La relation
(\ref{etatbarotrope}) est caract\'eristique d'une corde cosmique
parcourue par un courant conserv\'e brisant l'invariance de Lorentz
longitudinale. La conservation du flux d''\'energie est
\'egalement assur\'ee par la projection longitudinale des \'equations
de conservation (\ref{consreduit}), toujours pour le tenseur
(\ref{tmunubarotrope}),
\begin{equation}
\label{consenergie}
\first^\sigma_{\ \alpha} \nablab_\rho \Tb^\rho_{\ \sigma} = 0 \quad
\Rightarrow \quad \first^\sigma_{\ \alpha} u^\rho \nablab_{\left[\rho
\right.} \mub_{\left. \sigma \right]} = 0.
\end{equation}
Le potentiel chimique $\mub$ peut alors \^etre interpr\'et\'e comme la
masse effective par particule \`a l'origine du courant conserv\'e
$\nub^\rho$ le long de la corde~\cite{carter92,carter92b}. Comme nous
le verrons dans le chapitre~\ref{chapitrecour}, ce mod\`ele est
particuli\`erement bien adapt\'e \`a la description de cordes
cosmiques poss\'edant un condens\^at de Bose \`a l'origine d'un
courant de particules scalaires~\cite{neutral,enon0}.

L'\'equation d'\'etat permettant de conna\^\i tre compl\`etement la
dynamique de la corde, il doit \^etre \'egalement possible d'en
d\'eriver le lagrangien de surface en fonction des param\`etres
internes. Dans le cas barotrope, $\mub$ et $\nub$ \'etant reli\'es par
l'\'equation d'\'etat, il n'y a qu'un seul param\`etre ind\'ependant
et ce lagrangien ne peut \^etre fonction que d'une seule variable $\varpib$,
\begin{equation}
\Lcb = \Lambdab(\varpib),
\end{equation}
o\`u $\Lambdab$ est g\'en\'eriquement non lin\'eaire. D'apr\`es
l'\'equation (\ref{consenergie}), le rotationnel du flux $\mub^\rho$
s'annulant, celui-ci d\'erive d'un potentiel surfacique $\varphi$ tel
que $\mub_\rho = \nablab_\rho \varphi$. Ceci sugg\`ere, dans la
repr\'esentation lagrangienne, de choisir un tel potentiel scalaire
$\varphi$ comme param\`etre interne ind\'ependant, et
d'imposer\footnote{Malgr\'e l'apparence arbitraire de ce choix, il
n'est qu'une reformulation des propri\'et\'es intrins\`eques \`a la
corde.} l'existence d'un courant conserv\'e $\pb^\rho$ tel que
\begin{equation}
\pb_\rho = \nablab_\rho \varphi.
\end{equation}
Il existe alors un autre courant trivialement conserv\'e $\cb^\sigma =
\varepsilon^{\sigma \rho} \pb_\rho$ permettant de d\'efinir la
variable $\varpib$
\begin{equation}
\varpib = \cb_\rho \cb^\rho = -\pb_\rho \pb^\rho = -\gamma^{a b}
\partial_a \varphi \partial_b \varphi.
\end{equation}
L'identification de ces param\`etres avec les variables physiques de
la corde s'effectue naturellement \`a l'aide du tenseur \'energie
impulsion en comparant les \'equations (\ref{consreduit}) et
(\ref{tmunubarotrope}). Pour cela, il est commode d'introduire les
quantit\'es \emph{duales}
\begin{equation}
\ctd^\rho = 2 \frac{\ud \Lambdab}{\ud \varpib}\, \pb^\rho, \quad \ptd_\rho
= 2 \frac{\ud \Lambdab}{\ud \varpib} \, \cb_\rho, \quad \varpitd =
\ctd^\rho
\ctd_\rho = -\ptd^\rho \ptd_\rho,
\end{equation}
ainsi que la fonction ma\^\i tresse duale
\begin{equation}
\Lambdatd = \Lambdab - 2 \varpib  \frac{\ud \Lambdab}{\ud \varpib}.
\end{equation}
Le tenseur \'energie impulsion (\ref{consreduit}) prend alors une
forme sym\'etrique
\begin{equation}
\Tb^{\rho \sigma} = \frac{\Lambdab}{\varpib} \, \cb^\rho \cb^\sigma +
\frac{\Lambdatd}{\varpitd} \, \ctd^\rho \ctd^\sigma,
\end{equation}
et d'apr\`es l'\'equation (\ref{tmunubarotrope}), selon que le courant
conserv\'e $\cb^\rho$ est de genre temps ou espace, les fonctions
ma\^\i tresse $\Lambdab$ et $\Lambdatd$ s'identifient \`a $-U$ et
$-T$, ou $-T$ et $-U$ respectivement, alors que le param\`etre
d'\'etat $\varpib$ est \'egal \`a $-\nub^2$ ou $\mub^2$
respectivement. La diff\'erenciation de ces deux r\'egimes,
respectivement d\'esign\'es \emph{\'electrique} et \emph{magn\'etique},
r\'esulte uniquement de la compatibilit\'e du choix des quadrivecteurs
de base $u^\rho$ et $v^\rho$ avec la d\'efinition du quadrivecteur
courant $\cb^\rho$. L'action de Goto-Nambu (\ref{actiongn}) est
trivialement retrouv\'ee dans la limite de courant nul $\varpib
\rightarrow 0$ o\`u $\Lcb = -U = -T$ ind\'ependamment du r\'egime.

\subsection{Stabilit\'e}
\label{sectionstabilite}
En plus de d\'ecrire de mani\`ere unifi\'ee les diff\'erentes
propri\'et\'es microscopiques des cordes au travers d'une \'equation
d'\'etat, le formalisme covariant permet \'egalement de statuer sur la
stabilit\'e de celles-ci vis-\`a-vis de la propagation de
perturbations transverses et longitudinales de leur g\'eom\'etrie. La
propagation de perturbations transverses concerne \'evidemment la
partie extrins\`eque des \'equations du mouvement. La premi\`ere
\'equation est donn\'ee par la condition d'int\'egrabilit\'e
(\ref{weingarten}) du tenseur de courbure extrins\`eque. En la
projetant orthogonalement
\`a la surface d'univers, il vient \`a l'aide de l'\'equation
(\ref{deuxiemereduit}),
\begin{equation}
\label{perttrans1}
\perp^\beta_{\ \rho} {K_{[\mu \nu]}}^\rho = 0 \quad \Leftrightarrow \quad 
\perp^\beta_{\ \rho} u^\alpha \nablab_\alpha v^\rho = \perp^\beta_{\ \rho}
v^\alpha \nablab_\alpha u^\rho.
\end{equation}
La deuxi\`eme \'equation est directement donn\'ee par les \'equations
du mouvement extrins\`eques (\ref{mvtextpriv}). Si maintenant on
s'int\'eresse \`a la propagation de perturbations transverses de la
g\'eom\'etrie de la forme
\begin{equation}
\label{defperttrans}
\delta u^\rho = \varepsilon(u^\rho) \, \ue^{i \chi}, \qquad
\delta v^\rho = \varepsilon(v^\rho) \, \ue^{i \chi},
\end{equation}
de vecteur d'onde $k^\sigma$ d\'efini par
\begin{equation}
\nablab_\sigma \chi = k_\sigma,
\end{equation}
les \'equations (\ref{mvtextpriv}) et (\ref{perttrans1}), pour les
infiniments petits du premier ordre $\varepsilon$, deviennent
\begin{equation}
\label{perttranstot}
\begin{array}{lcl}
\omega \perp^\beta_{\ \rho} \varepsilon(v^\rho) + k \perp^\beta_{\ \rho}
\varepsilon(u^\rho) & = & 0, \\ \\
\omega \, U \perp^\beta_{\ \rho} \varepsilon(u^\rho) + k \, T
\perp^\beta_{\ \rho} \varepsilon(v^\rho) & = & 0,
\end{array}
\end{equation}
o\`u l'\'energie et l'impulsion longitudinale des perturbations sont
respectivement d\'efinies par
\begin{equation}
\omega = k_\sigma u^\sigma, \qquad k = - k_\sigma v^\sigma.
\end{equation}
La condition d'existence de solutions aux \'equations
(\ref{perttranstot}) donne alors la valeur de la vitesse des perturbations
transverses
\begin{equation}
\label{ct2}
\ct^2 = \frac{\omega^2}{k^2} = \frac{T}{U}.
\end{equation}
Encore une fois, pour des cordes de Goto-Nambu, $\ct^2=1$ comme attendu
par la r\'esolution explicite des \'equations du mouvement donn\'ee
en (\ref{gnpert}). Dans le cas g\'en\'eral, la tension $T$ doit
\^etre toujours positive pour que la corde soit transversalement
stable. Si ce n'\'etait pas le cas, il pourrait exister des r\'egions
o\`u $\ct^2<0$, et les perturbations transverses (\ref{defperttrans})
seraient exponentiellement amplifi\'ees.

De la m\^eme mani\`ere l'existence d'une structure interne autorise la
propagation de perturbations longitudinales le long de la corde. Ces
ondes de type acoustique vont sur leur passage modifier les
variables internes, i.e. $\mub$ et $\nub$ dans le cas barotrope. En
plus des perturbations (\ref{defperttrans}), il convient donc
d'introduire
\begin{equation}
\delta\nub = \varepsilon(\nub) \, \ue^{i \chi}, \qquad
\delta\mub = \varepsilon(\mub) \, \ue^{i \chi},
\end{equation}
avec, comme pr\'ec\'edemment, $\varepsilon$ un infiniment petit du
premier ordre.  La projection longitudinale de la conservation du
courant (\ref{conscourant}) donne la premi\`ere \'equation
\begin{equation}
\label{pertlong1}
\first^\alpha_{\ \rho} k_\alpha \left[ \nub \, \varepsilon(u^\rho) +
u^\rho \,
\varepsilon(\nub) \right] = 0,
\end{equation}
alors que la deuxi\`eme r\'esulte de l'\'equation de conservation de
l'\'energie (\ref{consenergie}),
\begin{equation}
\label{pertlong2}
u^\beta k_\beta \, \mub \, v^\rho \, \varepsilon(u_\rho) - v^\beta k_\beta \,
\varepsilon(\mub) = 0.
\end{equation}
La vitesse de propagation des perturbations longitudinales se r\'eduit
donc, d'apr\`es (\ref{pertlong1}) et (\ref{pertlong2}), \`a
\begin{equation}
\cl^2 = \frac{\omega^2}{k^2} = \frac{\nub \, \varepsilon(\mub)}{\mub
\, \varepsilon(\nub)}.
\end{equation}
Les param\`etres internes $\mub$ et $\nub$ \'etant reli\'es par les
relations de d\'efinition (\ref{defintparam}) et l'\'equation d'\'etat
(\ref{etatbarotrope}), il vient finalement
\begin{equation}
\label{cl2}
\cl^2 = - \frac{\ud T}{\ud U},
\end{equation}
et la stabilit\'e de la corde vis-\`a-vis des perturbations
longitudinales requiert $\ud T/\ud U < 0$.

\begin{figure}
\begin{center}
\input{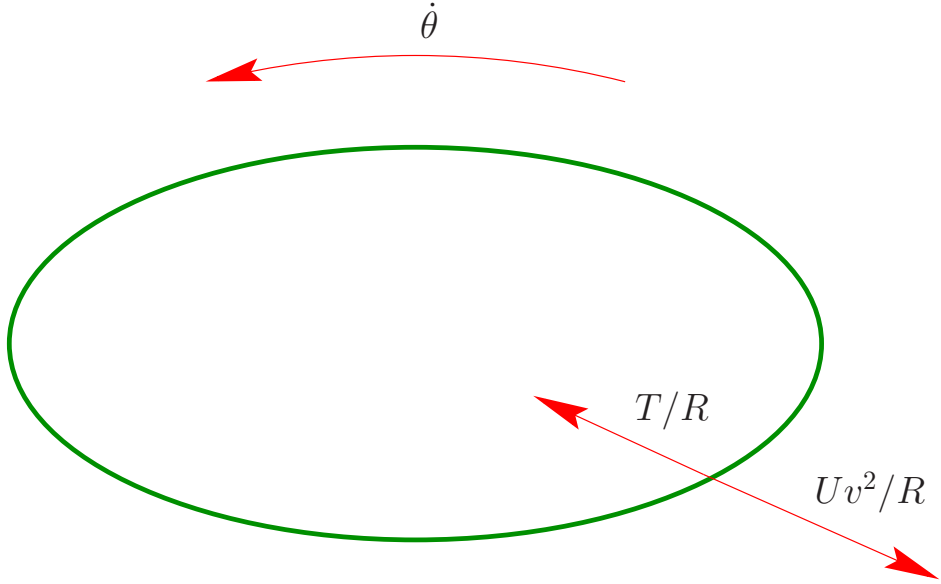}
\caption[Sch\'ematisation d'un vorton.]{La brisure de l'invariance de
Lorentz longitudinale par l'existence d'une structure interne permet
l'existence de boucles stables de corde cosmique, les vortons. La
tension est alors
\'equilibr\'ee par la force centrifuge lorsque la vitesse de rotation
\'egale celle de propagation des perturbations extrins\`eques $v^2 =
T/U$~\cite{carter89,carter89b}.}
\label{figvortons}
\end{center}
\end{figure}

\begin{figure}
\begin{center}
\input{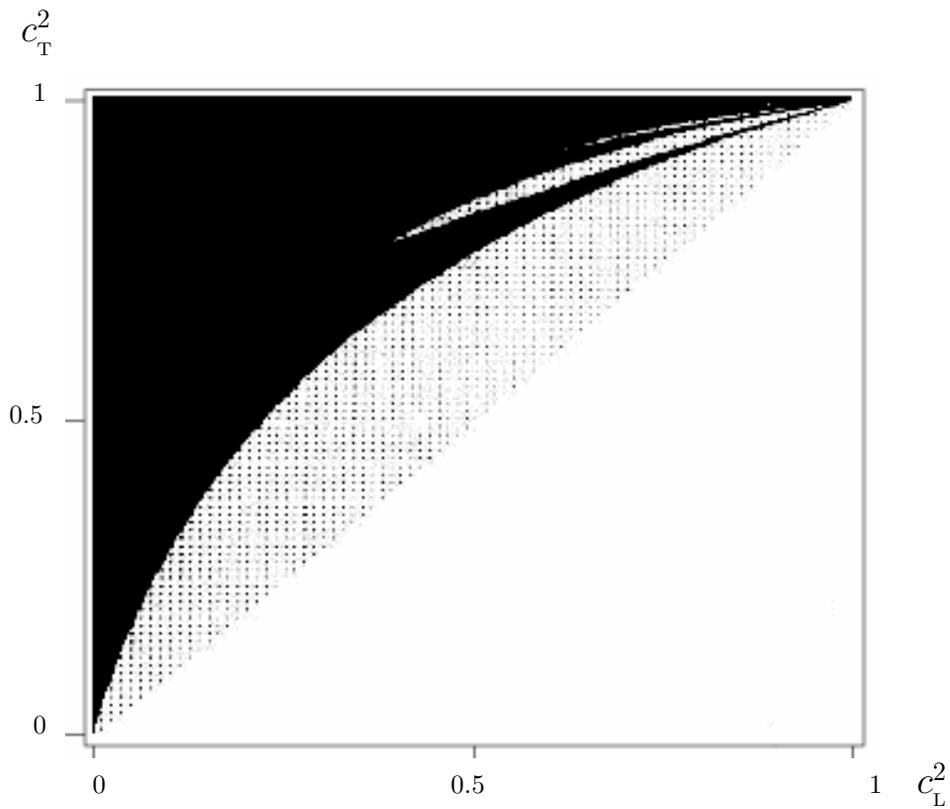}
\caption[Domaines d'instabilit\'e des vortons.]{Domaines
d'instabilit\'e classiques (en noir) des anneaux de corde cosmique en
fonction du carr\'e des vitesses de propagation des perturbations
transverses et longitudinales. Les boucles associ\'ees aux mod\`eles
de corde supersoniques $\ct^2 >\cl^2$ sont donc g\'en\'eriquement
instables~\cite{martin94}.}
\label{figclct}
\end{center}
\end{figure}

Le formalisme macroscopique permet donc, \`a partir de l'\'equation
d'\'etat, de conna\^\i tre la stabilit\'e locale des cordes par
rapport aux perturbations de leur mouvement, et de statuer sur leur
existence potentielle. Il a \'et\'e ainsi possible d'\'eliminer toute
une classe de cordes caract\'eris\'ees par une tension
n\'egative~\cite{nospring}. Ces crit\`eres de stabilit\'e bas\'es sur
l'existence de vitesses de propagation physiques des perturbations ne
sont cependant que n\'ecessaires dans le cas des boucles de corde. La
structure p\'eriodique induisant l'apparition d'une infinit\'e de
r\'esonances, la stabilit\'e de la boucle vis-\`a-vis des
perturbations ne sera assur\'ee que si la superposition de tous ces
modes d'oscillations reste finie. Ce probl\`eme est d'autant plus
important que la brisure de l'invariance de Lorentz longitudinale par
l'existence d'une structure interne peut conduire \`a l'apparition
d'anneaux stables de corde cosmique, appel\'es
\emph{vortons}~\cite{davisRL,brandi96}. Comme pour des boucles de
Goto-Nambu, la perte d'\'energie par rayonnement gravitationnel se
traduit par une diminution de la longueur propre des anneaux sous
l'effet de leur tension. Cependant, l'existence d'un moment
cin\'etique\footnote{Ce n'est que parce que l'invariance de Lorentz
longitudinale est bris\'ee qu'il est possible de d\'efinir un moment
cin\'etique.} impose dans le m\^eme temps l'augmentation de leur
vitesse angulaire jusqu'au moment o\`u l'\'equilibre entre la tension
et la force centrifuge est atteint (voir Fig.~\ref{figvortons}).
Si l'espace temps de r\'ef\'erence est plat, dans le cas le plus
simple d'un anneau circulaire de rayon $R$ et de vitesse angulaire
$\dot{\theta}$, les quadrivecteurs de base $u^\rho$ et $v^\rho$
[cf. Eq.~(\ref{orthobase})] peuvent se mettre, en coordonn\'ees
cylindriques, sous la forme~\cite{cartermeca}
\begin{equation}
\label{baserot}
u^\rho = \left(\gammal \, v,0,\gammal,0 \right), \qquad v^\rho=\left(
\gammal,0, \gammal \, v,0 \right),
\end{equation}
avec le facteur de Lorentz $\gammal = 1/\sqrt{1-v^2}$, et la vitesse de
rotation $v = R \dot{\theta}$. En reportant (\ref{baserot}) dans les
\'equations du mouvement extrins\`eque (\ref{mvtextpriv}) en
l'absence de force externe $\fb^\rho=0$, il vient
\begin{equation}
v^2 = \ct^2 = \frac{T}{U},
\end{equation}
comme intuitivement attendu par des consid\'erations classiques
(cf. Fig.~\ref{figvortons}). Il est alors possible de montrer que le
crit\`ere de stabilit\'e classique de tels anneaux vis-\`a-vis des
divers modes de propagation des perturbations est encore donn\'e par
les valeurs des vitesses $\ct^2$ et
$\cl^2$~\cite{martin95,mp2,cartermartin93}. Sur la
figure~\ref{figclct} sont repr\'esent\'es en noir les domaines du plan
$\left(\cl^2,\ct^2 \right)$ menant \`a des anneaux de corde
instables~\cite{martin94}. Ainsi, les cordes cosmiques ayant une
\'equation d'\'etat de type \emph{subsonique}, i.e. $\ct^2<\cl^2$,
pr\'edisent l'existence de vortons stables, alors que les mod\`eles
\emph{supersoniques} o\`u $\ct^2 > \cl^2$ m\`enent \`a des vortons
g\'en\'eriquement instables.

\section{Conclusion}

La limite d'\'epaisseur nulle des cordes cosmiques permet donc, par le
biais des formalismes macroscopiques, d'en d\'eduire les
caract\'eristiques essentielles de leur dynamique, ainsi que leur
stabilit\'e vis-\`a-vis des perturbations de leur forme. L'\'energie
par unit\'e de longueur et la tension apparaissent comme les
quantit\'es physiques d\'eterminantes dans la description de telles
$2$-surfaces: l'\'equation d'\'etat $U=U(T)$ fixe la dynamique propre
de la corde alors que les valeurs des vitesses de propagation des
perturbations $\ct^2 = T/U$ et $\cl^2 = -\ud T/\ud U$ en sont les
crit\`eres de stabilit\'e.

Cependant, les formalismes macroscopiques ne peuvent rendre compte des
propri\'et\'es dues \`a la structure transverses du vortex, et les
r\'esultats pr\'ec\'edents ne peuvent s'appliquer que si cette
structure n'est pas d\'eterminante dans les processus physiques \`a
l'\'etude. Dans les cas contraires, il est alors n\'ecessaire de
recourir \`a des mod\`eles microscopiques invoquant les champs formant
la corde, et n\'ecessitant le plus souvent l'emploi de m\'ethodes
num\'eriques.

Comme nous le verrons dans les chapitres suivants les deux approches
sont compl\'ementaires et autorisent l'\'etude des cordes cosmiques du
point de vue cosmologique tout en mettant en jeu les param\`etres des
th\'eories de physique des particules responsables de leur formation.

\chapter{\'Evolution cosmologique}
\label{chapitreevol}
\minitoc
\section{Introduction}

La microphysique des brisures de sym\'etrie par le m\'ecanisme de
Higgs (voir chapitre~\ref{chapitreaudela}) et les formalismes
macroscopiques (voir chapitre~\ref{chapitreform}) d\'ecrivent les
propri\'et\'es physiques des cordes cosmiques prises
individuellement. Cependant, dans l'univers primordial, les
transitions de phase menant \`a leur formation, cr\'eent un r\'eseau
de cordes dans un univers de FLRW en expansion. Afin d'en \'etudier
l'\'evolution cosmologique, il faut tout d'abord d\'eterminer ses
caract\'eristiques physiques initiales, puis les faire \'evoluer \`a
partir de la connaissance de la dynamique et des interactions de
chacune de ses cordes. Bien qu'il existe des mod\`eles effectifs
d\'ecrivant de tels r\'eseaux~\cite{kibble76, kibble80, vilenkin81},
leur mod\'elisation r\'ealiste passe par l'emploi de simulations
num\'eriques~\cite{albrecht85,bennett88,allen90,sakellariadou90}.
Initialement, ces simulations ont \'et\'e d\'evelopp\'ees dans les
ann\'ees 1980 pour calculer l'\'evolution de la densit\'e d'\'energie
associ\'ee aux r\'eseaux de cordes de Goto-Nambu, afin d'en
d\'eterminer leur impact sur la formation des grandes
structures. D'une part, elles ont montr\'e que ces r\'eseaux \'etaient
compatibles avec les contraintes cosmologiques \`a condition qu'une
certaine partie de leur \'energie soit \'evacu\'ee sous forme de
rayonnement, ce qui est effectivement r\'ealis\'e par le biais de la
d\'esint\'egration gravitationnelle des boucles (voir
Sect.~\ref{sectionnambu}). D'autres parts, il est apparu que les
r\'eseaux de cordes n'\'etaient pas l'effet dominant dans la formation
des grandes structures. Aujourd'hui, ces codes d'\'evolution sont \`a
nouveau \`a l'\'etude dans le but d'obtenir les signatures
observationnelles que pourrait avoir l'existence d'un
r\'eseau~\cite{wu02}, essentiellement dans les donn\'ees actuelles et
\`a venir concernant le CMBR~\cite{boomerang, boomerang2, maxima,
maxima2}. Dans cette optique, le plus pr\'ecis des codes de
l'\'epoque, d\'evelopp\'e par F.~Bouchet~\cite{bennett88,bouchet88}, a
\'et\'e repris et modernis\'e \`a l'aide des avanc\'ees technologiques
de la derni\`ere d\'ecennie, en particulier par l'utilisation de
m\'ethode de parall\'elisation (voir annexe~\ref{annexeomp}). Comme
nous verrons dans la section~\ref{sectionevol2cmb}, il est maintenant
possible d'extraire le spectre de puissance du CMBR d'un univers
poss\'edant un r\'eseau de cordes cosmiques.

\section{Intercommutation}
\label{sectionintercom}
\begin{figure}
\begin{center}
\epsfig{file=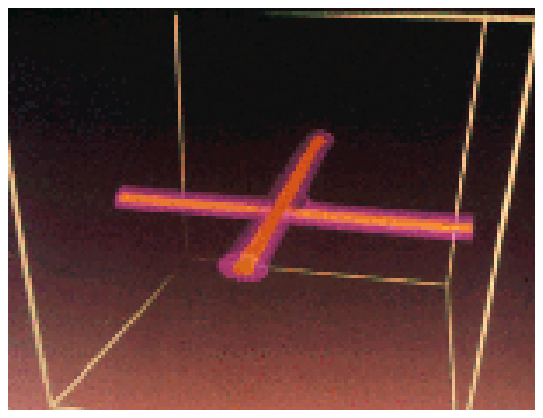,width=7.5cm,height=7.5cm}
\hspace{1.5mm}
\epsfig{file=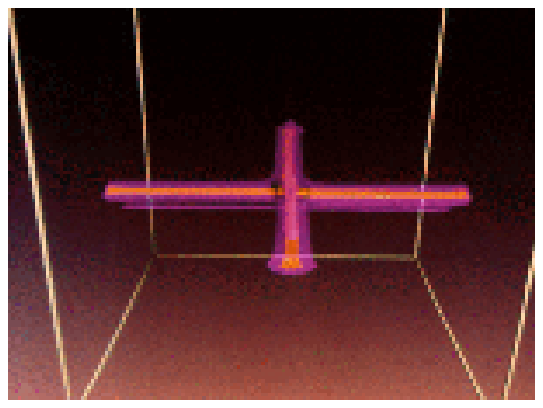,width=7.5cm,height=7.5cm}
\epsfig{file=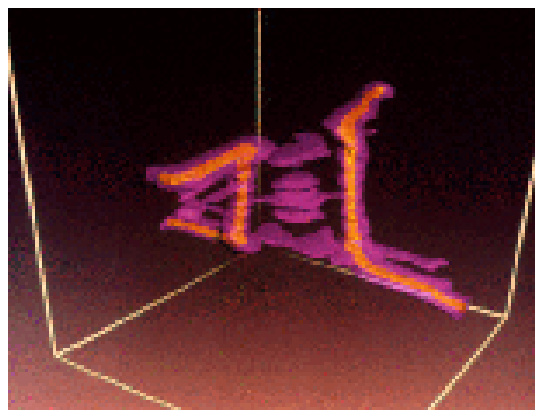,width=7.5cm,height=7.49cm}
\hspace{1.5mm}
\epsfig{file=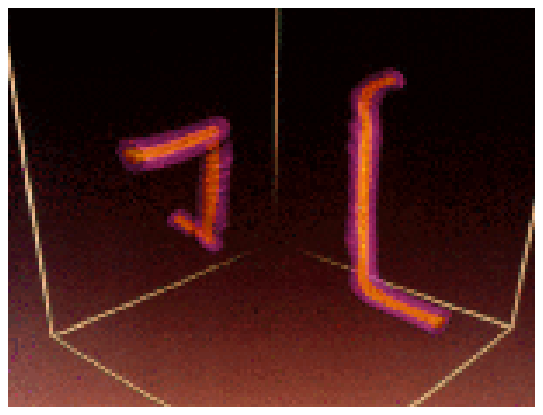,width=7.5cm,height=7.5cm}
\caption[Intercommutation de deux cordes cosmiques.]{Intercommutation
de deux cordes cosmiques dans le mod\`ele de Higgs
ab\'e\-lien~\cite{matzner88}. Les cordes sont repr\'esent\'ees par le
profil transverse des champ de Higgs (en orange) et de jauge (en
mauve), et apr\`es une phase interm\'ediaire d'excitation des divers
champs, elles \'echangent leurs segments respectifs. La probabilit\'e
d'\'echange estim\'ee par les diverses simulations num\'eriques est
$P_\ue \simeq 1$.}
\label{figvortexcross}
\end{center}
\end{figure}

En plus du mouvement propre de chaque corde, la dynamique d'un
r\'eseau fait que les cordes vont n\'ecessairement se croiser. Du
point de vue de la topologie du vide, il est impossible de statuer sur
l'issue d'une telle interaction\footnote{Ce n'est vrai que dans le cas
ab\'elien. Dans le cas de cordes cosmiques issues de la brisure d'un
groupe de sym\'etrie non-ab\'elien la topologie du vide impose
l'apparition d'une troisi\`eme corde les reliant.}: soit les cordes se
traversent simplement, soit elles \'echangent leur branches. C'est la
dynamique des champs formant le vortex qui va alors d\'eterminer ce
qui se passe effectivement. Les simulations num\'eriques effectu\'ees
pour r\'esoudre les \'equations de champs du type (\ref{tildehiggs})
et (\ref{tildegauge}) pour deux vortex en interaction, obtiennent
toutes que l'\'echange de partenaires est privil\'egi\'e, que ce soit
dans les intersections de cordes globales~\cite{shellard87} ou
locales~\cite{matzner88,moriarty88} (voir
Fig.~\ref{figvortexcross}). Au vue de ces r\'esultats, on estime que la
probabilit\'e d'intercommutation, i.e. d'\'echange des segments, est
suffisamment proche de l'unit\'e pour que l'on puisse consid\'erer que
cette reconnexion se produit syst\'ematiquement.

\begin{figure}
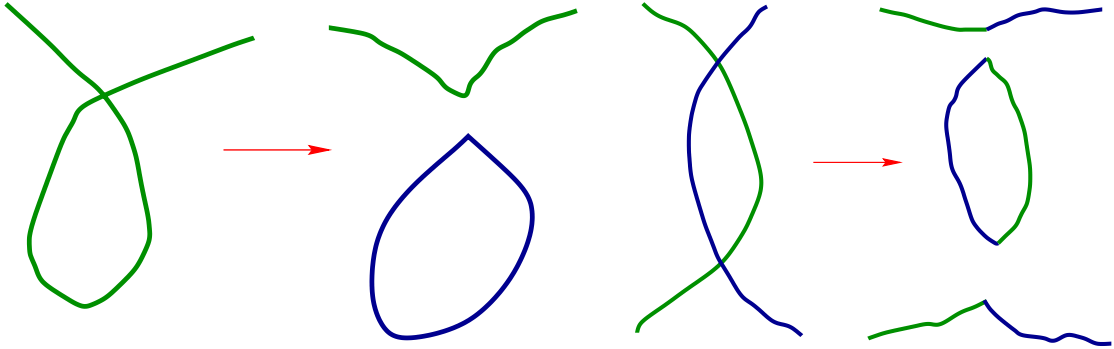

\begin{center}
\epsfig{file=autosection.eps,height=4.5cm}
\hspace{2cm}
\epsfig{file=collide.eps,height=4.5cm}
\caption[Formation de boucles de corde cosmique par
intercommutation.]{Formation de boucles de corde cosmique par
intercommutation d'une seule corde, ou intersection de deux cordes
diff\'erentes.}
\label{figboucle}
\end{center}
\end{figure}

Une cons\'equence majeure de ce m\'ecanisme d'intercommutation est la
formation incessante de boucles de cordes cosmiques. En effet, par ses
oscillations (voir Sect.~\ref{sectionnambu}), une corde peut
s'intercommuter avec elle m\^eme et \'echanger ses propres segments,
menant ainsi \`a une nouvelle corde infinie accompagn\'ee d'une boucle
(voir Fig.~\ref{figboucle}). De m\^eme, une collision entre deux
cordes diff\'erentes en deux points distincts conduit \'egalement \`a
la formation d'un anneau constitu\'e des segments de chaque corde
(voir Fig.~\ref{figboucle}).
Dans la section~\ref{sectionnambu}, nous avons vu que les mouvements
d'oscillation propres des boucles conduisent g\'en\'eralement \`a
l'apparition de discontinuit\'es dans la forme de la corde qui se
propagent \`a la vitesse de la lumi\`ere. Un autre type de
discontinuit\'e appara\^\i t \'egalement lors de l'intercommutation
des cordes, comme on peut le voir sur la
figure~\ref{figvortexcross}. La conservation de l'impulsion lors de
l'intercommutation impose aux points de la nouvelle corde issus des
deux cordes initiales (ou des deux zones diff\'erentes de la m\^eme
corde), d'avoir des vitesses voisines de celles qu'ils avaient avant
l'interaction. Autour du point d'\'echange, la nouvelle corde
pr\'esente donc des variations de vitesse et de forme tr\`es rapides
en fonction de l'abscisse curviligne $\xi$, qui, dans la limite
d'\'epaisseur nulle, apparaissent comme des discontinuit\'es dans les
d\'eriv\'ees $\dotvec{x}{}(\xi,t)$ et $\primevec{x}{}(\xi,t)$. Pour
des cordes de Goto-Nambu, d'apr\`es l'\'equation (\ref{gnpert}), ces
discontinuit\'es, appel\'ees \emph{kinks}, correspondent \'egalement
\`a des discontinuit\'es des vecteurs $\vec{p}(\xi-t)$ et
$\vec{q}(\xi+t)$, \`a $\xi-t$ et $\xi+t$ constants,
respectivement. Autrement dit, ces kinks se propagent \`a la vitesse
de la lumi\`ere dans chacune des directions de la corde \`a partir du
point d'intercommutation. La propagation de ces ondes de
discontinuit\'e donne alors une forme plus ou moins ``cisaill\'ee''
aux cordes ayant subi de multiples interactions. Comme les
points de rebroussement, les kinks sont des sources d'ondes
gravitationnelle intenses pouvant permettre leur \'eventuelle
d\'etection~\cite{damour01}.

L'intercommutation est un m\'ecanisme essentiel dans l'\'evolution
cosmologique d'un r\'eseau de cordes. Les cordes infinies vont avoir
tendance \`a se fragmenter en boucles, qui \`a leur tour peuvent se
raccrocher aux cordes infinies. Cependant, par la d\'esint\'egration
gravitationnelle des boucles, et leur plus faible probabilit\'e de
rencontre, le premier m\'ecanisme appara\^\i t favoris\'e. Afin de
pouvoir simuler num\'eriquement cette \'evolution, il faut de plus se
donner les conditions initiales, c'est-\`a-dire les propri\'et\'es
physiques de chaque corde nouvellement form\'ee lors de la transition
de phase, ainsi que le mouvement de chacune d'elles dans un univers de
FLRW.

\section{R\'eseau de cordes dans FLRW}

Les formalismes macroscopiques pr\'esent\'es dans le
chapitre~\ref{chapitreform} ne permettent pas d'obtenir de solution
analytique aux \'equations du mouvement des cordes dans un univers de
FLRW, m\^eme dans le cas le plus simple des cordes de Goto-Nambu [voir
Eq.~(\ref{mvtgngene})]. La prise en compte des intercommutations,
d\'ependant explicitement du mouvement et de la position de chaque
corde \`a chaque instant, motive \'egalement l'emploi de simulations
num\'eriques dans le but d'extraire les propri\'et\'es statistiques de
l'\'evolution cosmologique d'un r\'eseau de cordes.

\subsection{Approche qualitative}
\label{subsectionqual}
Deux propri\'et\'es essentielles r\'egissent, \emph{a priori},
l'\'evolution d'un r\'eseau de cordes: d'une part l'expansion de
l'univers, qui tend \`a en augmenter l'\'energie par l'entr\'ee dans
l'horizon de segments de corde infinie, et d'autres parts le
m\'ecanisme d'intercommutation, qui transf\`ere une part de cette
\'energie en rayonnement, par le biais de la d\'esint\'egration des
boucles.

Plus pr\'ecis\'ement du fait de l'expansion de l'univers, d'apr\`es
(\ref{metriqueFLRW}), la longueur propre $L$ d'une corde traversant
l'horizon, va se dilater avec le facteur d'\'echelle en $L \propto a$,
ainsi que son \'energie totale $M$,
\begin{equation}
\label{massgrow}
M = L U \propto a.
\end{equation}
En premi\`ere approximation, il est possible de ne consid\'erer qu'une
seule \'echelle de longueur caract\'erisant le
r\'eseau~\cite{kibble76,vilenkin81}. En effet, si la transition de
phase introduit uniquement la longueur de corr\'elation $\ell_\uc$
comme \'echelle de distance, la distance moyenne entre chaque corde
form\'ee, ainsi que leur rayon de courbure moyen, doivent \^etre de
cet ordre de grandeur. En appelant $L_\infty$ l'\'echelle de longueur
du r\'eseau, qui initialement s'identifie \`a $\ell_\uc$, le nombre
moyen de cordes dans un volume $V$ est voisin de
$V/L_\infty^3$. L'\'energie moyenne $E_\infty$ associ\'ee au r\'eseau
est donc
\begin{equation}
E_\infty \simeq \frac{V}{L_\infty^3} \times U L_\infty = U
\frac{V}{L_\infty^2},
\end{equation}
et la densit\'e d'\'energie correspondante
\begin{equation}
\label{densiteinfty}
\rho_\infty \simeq \frac{U}{L_\infty^2}.
\end{equation}
Du fait de l'expansion, $L_\infty \propto a$, et la densit\'e
d'\'energie associ\'e au r\'eseau varie donc en $\rho_\infty \propto
a^{-2}$, comme attendue par l'accroissement d'\'energie de chaque
corde (\ref{massgrow}) et la dilution par augmentation du volume en
$a^3$. Autrement dit, la densit\'e d'\'energie associ\'ee \`a un
r\'eseau de cordes infinies sans interaction finit toujours par
dominer l'univers, la densit\'e de mati\`ere \'evoluant en $1/a^3$ et
la radiation en $1/a^4$.

Si l'on tient compte des intercommutations, la partie d'\'energie
transf\'er\'ee du r\'eseau de cordes infinies vers les boucles est
certainement proportionnelle au nombre d'interactions. En premi\`ere
approximation, sur une \'echelle de distance de $L_\infty$, pour des
vitesses relativistes, le temps moyen entre deux intersections est de
$L_\infty$, soit un taux d'interaction par unit\'e de volume de
$1/L_\infty^4$. La perte d'\'energie par formation de boucle de
longueur voisine de $L_\infty$, pendant le temps $\delta t$, est
donc\footnote{Comme nous le verrons dans la section~\ref{sectionsimu},
cette hypoth\`ese n'est pas repr\'esentative des r\'esultats
num\'eriques, les boucles form\'ees sont de bien plus petites tailles,
mais leur nombre est \'egalement beaucoup plus grand assurant encore
une perte d'\'energie importante.}
\begin{equation}
\label{perteboucle}
\delta \rho_{\infty \rightarrow \ub} \simeq L_\infty^{-4} \delta t \,
U L_\infty.
\end{equation}
\`A l'aide de (\ref{hubbleparam}), (\ref{massgrow}),
(\ref{densiteinfty}) et (\ref{perteboucle}), l'\'equation
d'\'evolution de la densit\'e d'\'energie associ\'ee au r\'eseau de
cordes en interaction devient
\begin{equation}
\label{evolrhoinfty}
\frac{\ud \rho_\infty}{\ud t} \simeq -2 H \rho_\infty -
\frac{\rho_\infty}{L_\infty},
\end{equation}
o\`u $L_\infty$ est une fonction du temps. Si l'on pose
\begin{equation}
C(t) = \frac{L_\infty(t)}{t},
\end{equation}
et en exprimant le param\`etre de Hubble $H$ en fonction du temps
cosmique \`a l'aide de (\ref{hubbleparam}) et (\ref{evolechelle}),
l'\'equation d'\'evolution (\ref{evolrhoinfty}) devient
\begin{equation}
\label{autosiminfty}
\frac{1}{C} \frac{\ud C}{\ud t} \simeq -\frac{1}{2t}
\left(\frac{2+6w}{3+3w} - \frac{1}{C} \right).
\end{equation}
La solution constante $C(t)=C_\us=(3+3w)/(2+6w)$ appara\^\i t donc
comme un attracteur. Si initialement $C>C_\us$, d'apr\`es
(\ref{autosiminfty}), sa d\'eriv\'ee est n\'egative $\ud C/\ud t<0$,
inversement, pour des valeurs initiales plus faibles que la valeur
pivot, la d\'eriv\'ee est positive. Au cour de l'expansion, la
longueur caract\'eristique du r\'eseau de cordes en interaction tend
donc \`a varier comme la distance \`a l'horizon $d_\uH \propto t$. La
densit\'e d'\'energie, lorsque ce r\'egime stationnaire est atteint,
\'evolue en
\begin{equation}
\label{rhoscaling}
\rho_{\infty_\us} \propto \frac{U}{t^2} \propto \frac{U}{a^{3(1+w)}}.
\end{equation}
\`A la condition que l'\'energie associ\'ee aux boucles de corde soit
\'evacu\'ee sous forme de rayonnement, comme c'est effectivement le
cas pour les cordes de Goto-Nambu (voir section~\ref{sectionnambu}),
la densit\'e d'un tel r\'eseau en interaction finit par atteindre un
r\'egime stationnaire\footnote{\emph{Scaling.}} \'evitant sa
domination sur les autres formes d'\'energie. Sur la base de ce simple
mod\`ele, d'autres approches plus r\'ealistes ont \'et\'e
d\'evelopp\'ees~\cite{kibble85,bennett86,kibble91,copeland92}. Leur
conclusion est \'egalement l'existence d'une solution d'\'echelle de
type (\ref{rhoscaling}). Les simulations num\'eriques ont finalement
confirm\'e cette propri\'et\'e, en donnant de plus une valeur
num\'erique au coefficient de proportionnalit\'e dans l'\'equation
(\ref{rhoscaling}).

\subsection{Simulations num\'eriques}
\label{sectionsimu}

\subsubsection{Conditions initiales}

Lors de la transition de phase, les cordes cosmiques se forment par le
m\'ecanisme de Kibble~\cite{kibble76} lorsqu'il appara\^\i t des
configurations des phases du champ de Higgs qui ne sont pas
contractiles \`a un point (voir section~\ref{sectiondefauts}). Pour
une transition du second ordre, une fois les fluctuations de
temp\'erature suffisamment faibles pour que le champ de Higgs ne
puisse plus changer de phase~\cite{kibble76,book}, la distribution des
d\'efauts est directement donn\'ee par la distribution al\'eatoire des
phases du champ sur des distances sup\'erieures \`a la longueur de
corr\'elation $\ell_\uc$ de la transition. La distribution obtenue
peut \^etre raisonnablement mod\'elis\'ee par l'algorithme de
Vachaspati et Vilenkin~\cite{vachaspati84}: \`a chaque point d'une
grille cubique de pas de l'ordre de $\ell_\uc$ est attribu\'ee une
valeur al\'eatoire de la phase du champ de Higgs parmi trois valeurs
discr\`etes\footnote{Le choix de trois valeurs permet de statuer sur
l'enroulement de la phase
\`a partir des quatre points de chaque face. Pour pouvoir choisir plus
de trois points, il faudrait utiliser une grille non
cubique~\cite{kibble86}.} $0$, $2 \pi/3$ et $4 \pi/3$. Une corde
cosmique est alors suppos\'ee traverser chaque face des cubes
\'el\'ementaires si la phase du champ ``tourne'' de $2 \pi$ en se
d\'epla\c{c}ant le long des ar\^etes. La direction du vortex est
alors donn\'ee par le sens d'enroulement de la phase. On obtient donc
des cubes \'el\'ementaires poss\'edant des cordes entrantes et
sortantes. La conservation du flux permet ensuite de les connecter, et
lorsque le choix est ambigu, ces connections sont effectu\'ees de
mani\`ere al\'eatoire.

\begin{figure}
\begin{center}
\epsfig{file=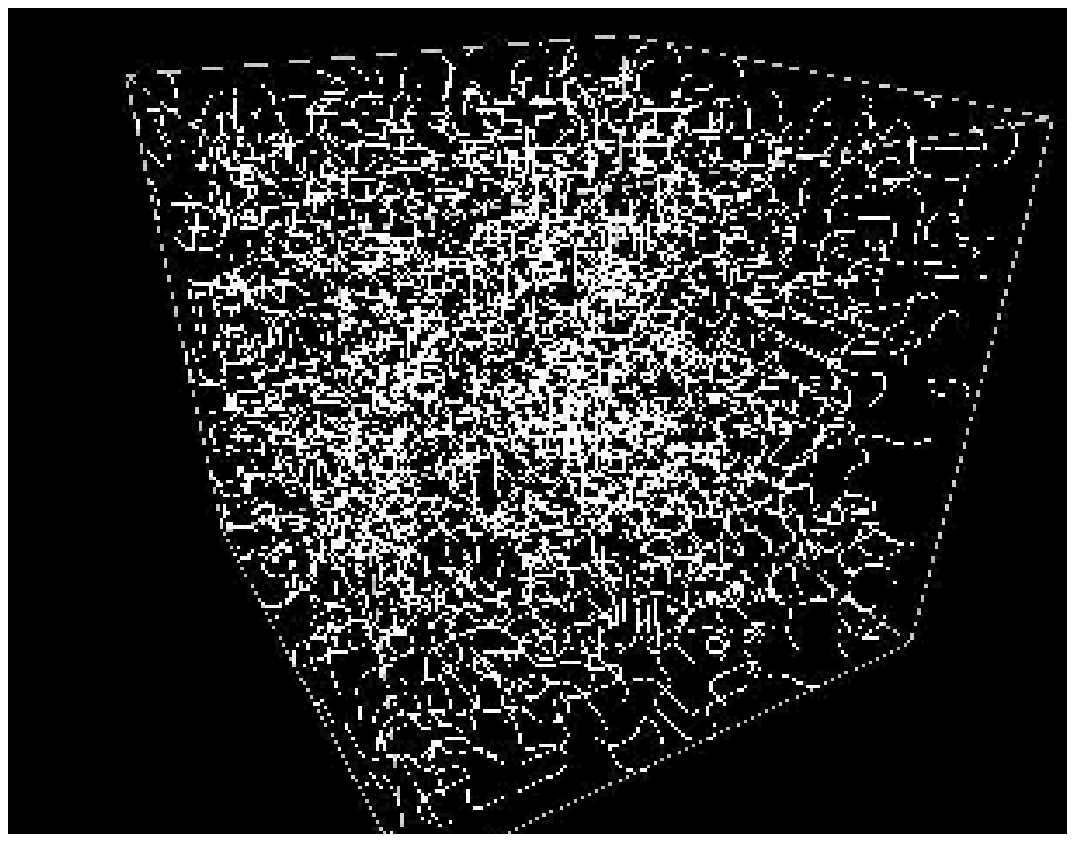,height=10cm}
\epsfig{file=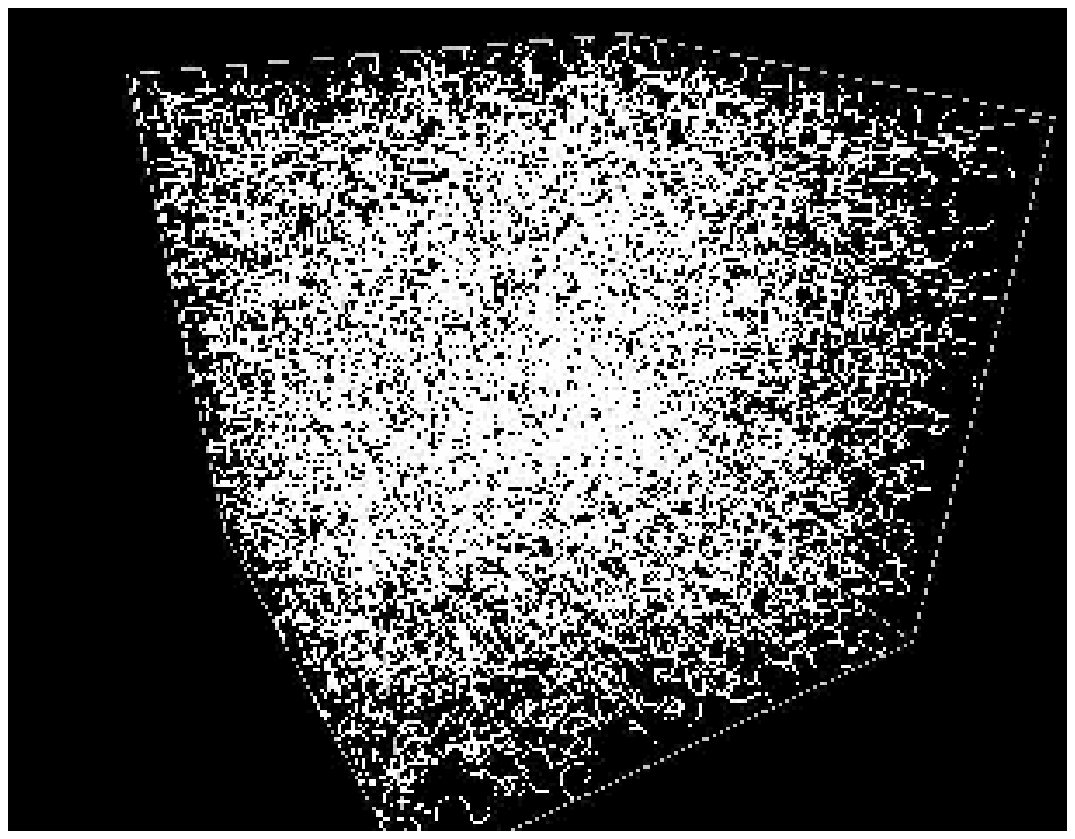,height=10cm}
\caption[Configurations d'un r\'eseau de cordes
cosmiques \`a sa formation.]{Configurations initiales d'un r\'eseau de
cordes, calcul\'ees par l'algorithme de Vachaspati et
Vilenkin~\cite{vachaspati84}, dans un volume comobile de c\^ot\'e $16
\ell_\uc$ et $26 \ell_\uc$, respectivement. La distribution de cordes
est constitu\'ee de $80 \%$ de cordes infinies contre $20 \%$ de
boucles, ind\'ependamment du volume de r\'ef\'erence.}
\label{figini}
\end{center}
\end{figure}

Sur la figure~\ref{figini} est repr\'esent\'ee la configuration de
cordes g\'en\'er\'ee par l'algorithme pr\'ec\'edent tel qu'il est
utilis\'e dans le code d'\'evolution de
F.~Bouchet~\cite{bouchet88}. Les param\`etres physiques ajustables
sont la densit\'e initiale de cordes, qui est fix\'ee par le rapport
de la longueur de corr\'elation $\ell_\uc$ \`a la distance \`a
l'horizon initiale, et l'amplitude d'une distribution al\'eatoire de
vitesse associ\'ee aux segments de corde de longueur $\ell_\uc$. Ce
dernier param\`etre est en fait un moyen d'introduire, \`a la main,
l'influence des interactions entre les cordes venant de na\^\i tre et
le fond intense de rayonnement existant lors de la transition de
phase~\cite{kibble80,vilenkin85}. Les configurations obtenues ne
peuvent cependant \^etre calcul\'ees, pour des raisons num\'eriques,
que dans un volume comobile de taille finie o\`u l'on impose des
conditions aux limites p\'eriodiques. Les cordes infinies sont alors
d\'efinies comme celle ne se bouclant pas \`a l'int\'erieur de ce
volume. Afin de pouvoir n\'egliger l'influence de ce volume, un
param\`etre purement num\'erique, sa longueur associ\'ee doit \^etre
toujours plus grande que les longueurs physiques du mod\`ele, i.e. la
longueur de corr\'elation $\ell_\uc$ et la distance \`a l'horizon
$d_\uH$. Cette derni\`ere augmentant proportionnellement au temps
cosmique [voir Eq.~(\ref{evolhorizon})], le volume de r\'ef\'erence
donne \'egalement une \'echelle de temps maximale \`a la simulation
num\'erique: celle pour laquelle l'horizon $d_\uH(t)$ y reste
confin\'ee. Au del\`a, les effets de bords ne peuvent en effet plus
\^etre n\'eglig\'es et deviennent rapidement dominant.

Les r\'esultats num\'eriques indiquent que le r\'eseau de cordes
form\'e par la transition de phase comprend $80 \%$ de cordes
infinies, dont la forme est typiquement celle d'une marche
al\'eatoire~\cite{vachaspati84}, et $20 \%$ de boucles de tailles
caract\'eristiques $R$, dont la densit\'e varie en $1/R^4$
ind\'ependamment des autres param\`etres physiques. Le calcul
num\'erique de l'\'evolution de ce r\'eseau n\'ecessite maintenant la
connaissance du mouvement de chaque corde dans FLRW, ainsi que la
prise en compte des intercommutations.

\subsubsection{\'Evolution dans FLRW}

L'espace-temps de FLRW privil\'egiant le choix d'un r\'ef\'erentiel
comobile, le formalisme traditionnel y est particuli\`erement adapt\'e
pour la description des cordes. Comme dans le cas de l'espace-temps de
Minkowski, l'action de Goto-Nambu permet d'obtenir les \'equations du
mouvement (\ref{mvtgngene}), et les degr\'es de libert\'e inh\'erents
au choix des syst\`emes de coordonn\'ees peuvent \^etre fix\'es par
des choix de jauge. Dans le cas de FLRW, il pourrait sembler commode
de se placer dans la jauge conforme (\ref{confgauge}), mais
malheureusement ce choix de jauge n'est plus compatible avec le choix
de jauge transverse (\ref{transgauge}), du fait des valeurs non nulles
des symboles de Christoffel. On choisit alors les conditions de jauge
mixtes
\begin{equation}
\label{mixgauge}
\xi^0 = \eta, \quad \xi^1=\xi, \quad \dotvec{x}{}.\primevec{x}{} = 0,
\end{equation}
avec $\eta$ le temps cosmique conforme [voir
Eq.~(\ref{metriqueFLRWconf})], et $\dotvec{x}{}$ la vitesse conforme
de la corde\footnote{Dans la suite de ce chapitre, le point
d\'esignera la d\'erivation par rapport \`a $\xi^0=\eta$.}, qui est
choisie purement transverse, comme cela est sugg\'er\'e par
l'invariance de Lorentz longitudinale (voir
Sect.~\ref{sectionnambu}). L'\'equation du mouvement (\ref{mvtgngene})
se simplifie alors en
\begin{equation}
\label{mvtflrw}
\ddotvec{x}{} + 2 \Hc \, \dotvec{x}{} \left(1 - \dotvec{x}{2} \right) =
\frac{1}{\epsilon} \left(\frac{\primevec{x}{}}{\epsilon} \right)',
\end{equation}
avec
\begin{equation}
\label{enerdef}
\epsilon = \sqrt{\frac{\primevec{x}{2}}{1 - \dotvec{x}{2}}}.
\end{equation}
Les condition de jauge (\ref{mixgauge}) et l'\'equation du mouvement
(\ref{mvtflrw}) imposent de plus
\begin{equation}
\label{enerevol}
\frac{\dot{\epsilon}}{\epsilon} = -2 \Hc \, \dotvec{x}{2},
\end{equation}
et \`a partir de (\ref{tmunugn}), l'\'energie totale d'une corde est
directement donn\'ee par $\epsilon$~\cite{turok84b}
\begin{equation}
\label{massflrw}
M = a U \int \epsilon \, \ud \xi .
\end{equation}
Ses variations avec le temps conforme s'obtiennent en reportant
(\ref{massflrw}) dans l'\'equation d'\'evolution (\ref{enerevol}). On
trouve
\begin{equation}
\label{massevol}
\frac{\dot{M}}{M} = \Hc \left( 1 - 2 \, \overline{v^2} \right),
\end{equation}
o\`u la vitesse quadratique moyenne $\overline{v^2}$ est d\'efinie par
\begin{equation}
\overline{v^2} = \frac{\displaystyle \int{\epsilon \,
\dotvec{x}{2}} \ud \xi}{\displaystyle \int{\epsilon}\, \ud \xi}.
\end{equation}
D'apr\`es l'\'equation (\ref{massevol}), l'\'energie totale de la
corde augmente donc proportionnellement au facteur d'\'echelle, comme
intuitivement attendu (voir Sect.~\ref{subsectionqual}). Cependant, le
terme en $\overline{v^2}$ tend \`a r\'eduire ce gain d'\'energie,
voire m\^eme \`a l'annuler pour $\overline{v^2} =0.5$. En fait, comme
le montre le terme de friction de l'\'equation (\ref{mvtflrw}),
l'expansion de l'univers tend \`a att\'enuer les oscillations propres
de la corde, et donc \`a r\'eduire $\overline{v^2}$ sur des \'echelles
de distance comparable \`a l'horizon. Inversement, aux petites
\'echelles de longueur, cette att\'enuation sera n\'egligeable et les
oscillations propres des boucles vont certainement dominer la
dynamique donnant $\dot{M} \simeq 0$, comme dans l'espace-temps de
Minkowski.

La r\'esolution num\'erique de l'\'equation (\ref{mvtflrw}) est un
probl\`eme d\'elicat \`a cause de la propagation des kinks
apparaissant lors de chaque intercommutation (voir
Sect.~\ref{sectionintercom}). Il est plus judicieux de transformer
(\ref{mvtflrw}) en deux \'equations du premier ordre portant sur des
variables ind\'ependantes et constantes pour chaque kink. Ceci est
r\'ealis\'e en posant~\cite{bennett90}
\begin{equation}
\vec{p}(\eta,u) = \frac{\primevec{x}{}}{\epsilon} - \dotvec{x}{}, \quad
\vec{q}(\eta,v) = \frac{\primevec{x}{}}{\epsilon} + \dotvec{x}{},
\end{equation}
avec $\vec{p}^{^{\,2}} = \vec{q}^{^{\,2}} =1$ et $(u,v)$ un nouveau choix de
coordonn\'ee d\'efini par
\begin{equation}
u \equiv \int{\epsilon} \, \ud \xi - \eta, \quad v \equiv
\int{\epsilon} \, \ud \xi + \eta.
\end{equation}
L'\'equation (\ref{mvtflrw}) devient, en fonction de ces nouveaux
param\`etres~\cite{bennett90},
\begin{eqnarray}
\label{mvtflrwsys}
\dotvec{p}{} & = &  \Hc \left[ \vec{q} - \left(\vec{p}.\vec{q} \, \right)
\vec{p} \, \right],\\ \nonumber \\
\dotvec{q}{} & = &  \Hc \left[ \vec{p} -
\left(\vec{p}.\vec{q} \, \right) \vec{p} \, \right],\\ \nonumber \\
\dot{\epsilon} & = & - \Hc \, \epsilon \left(1 - \vec{p}.\vec{q} \, \right).
\end{eqnarray}
Les kinks se propageant \`a la vitesse de la lumi\`ere le long des
deux directions de la corde, ils correspondent \`a des valeurs de $u$
et $v$ constantes. Ainsi, sur la grille $(u,v)$, les valeurs de
$\vec{p}$ et $\vec{q}$ correspondantes peuvent \^etre choisies
constantes le long de chaque segment, et seront plus ou moins
discontinues en chaque point par la pr\'esence des kinks. Cependant,
l'intercommutation n\'ecessite \'egalement la connaissance, \`a chaque
instant, des coordonn\'ees physiques $\vec{x}$ sur la grille initiale
$(\eta,\xi)$. Celles-ci sont obtenues en inversant les relations
pr\'ec\'edentes, et il vient
\begin{equation}
\label{grillephys}
\dotvec{x}{} = \frac{1}{2} \left(\vec{q} - \vec{p} \, \right), \quad
\primevec{x}{} = \frac{1}{2} \epsilon \left( \vec{p} + \vec{q} \,
\right).
\end{equation}
Une fois connue la position physique des points de chaque corde, la
d\'etection des intercommutations est essentiellement r\'ealis\'ee en
testant pour chaque segment de corde, si, dans une liste de segments
voisins, le volume du t\'etrah\`edre sous-tendu par les quatre points
des deux segments change de signe durant le pas de temps
\'el\'ementaire. Si c'est le cas, l'intercommutation est effectu\'ee.

\subsubsection{R\'esultats num\'eriques}

\`A partir des configurations initiales de la figure~\ref{figini}, il
est enfin possible de calculer leur \'evolution en r\'esolvant
num\'eriquement les \'equations (\ref{mvtflrwsys}) \`a
(\ref{grillephys}), et en d\'etectant toutes les intercommutations
survenant au cour du temps. Les r\'esultats pr\'esent\'es ici sont
issus d'un code num\'erique d\'evelopp\'e en \texttt{FORTRAN90} et
\texttt{OpenMP} fonctionnant sur des machines multiprocesseurs \`a
m\'emoire partag\'ee (voir annexe~\ref{annexeomp}). Il est construit
sur la base d'un des meilleurs codes des ann\'ees 1980, d\'evelopp\'e
par Bennett et Bouchet~\cite{bennett88,bennett90}, ayant
originellement confirm\'e l'existence de la solution d'\'echelle
(\ref{rhoscaling}).

\begin{figure}
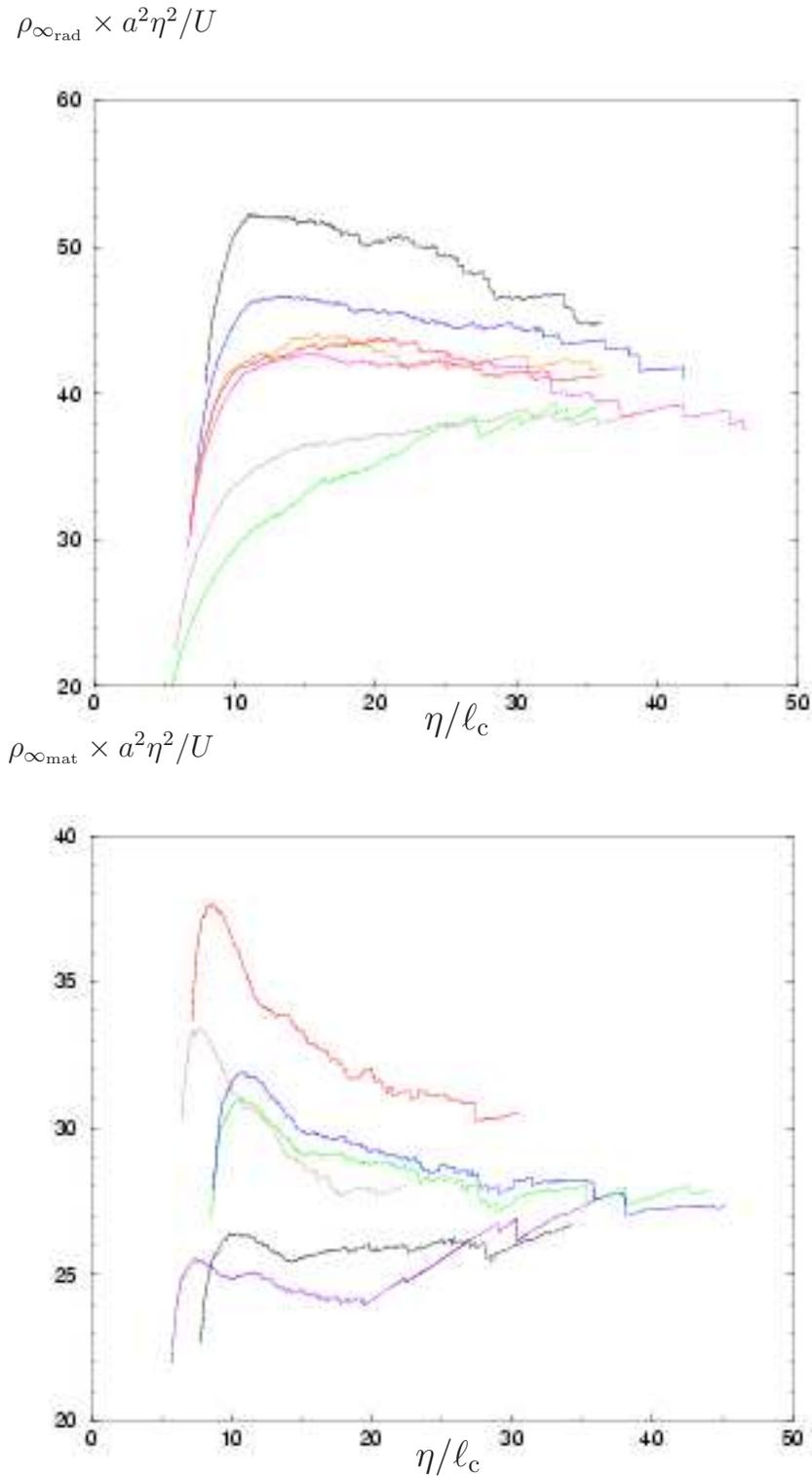

\begin{center}
\input{scalingrad.pstex_t}
\input{scalingmat.pstex_t}
\caption[\'Evolution cosmologique de la densit\'e d'\'energie d'un
r\'eseau de cordes cosmiques.]{\'Evolution de la densit\'e d'\'energie
totale associ\'ee aux cordes infinies, i.e. plus grandes que
l'horizon, en fonction du temps conforme, dans l'\`ere de radiation
(en haut) et de mati\`ere. Dans les deux cas, la loi d'\'echelle en
$1/t^2$ est un attracteur, quelques soient les conditions
initiales. La plus longue simulation est r\'ealis\'ee sur un volume
comobile de $(48 \ell_\uc)^3$ et met en jeu $2500000$ points.}
\label{figscaling}
\end{center}
\end{figure}

\begin{figure}
\begin{center}
\input{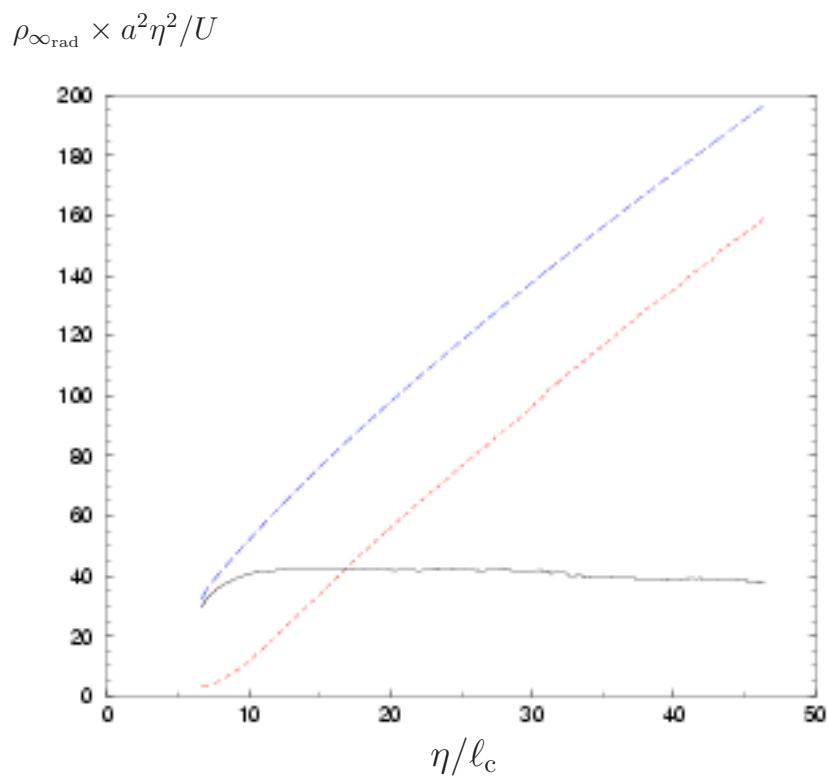}
\caption[\'Evolution cosmologique de la densit\'e d'\'energie
associ\'ee aux boucles de corde cosmique.]{\'Evolution de la densit\'e
d'\'energie totale des cordes infinies dans l'\`ere de radiation, en
trait plein noir, compar\'ee
\`a celle des boucles, en tirets rouges courts. La densit\'e
d'\'energie totale du r\'eseau, incluant les boucles et les cordes
plus grandes que l'horizon, est repr\'esent\'ee en tirets bleus
longs. La solution d'\'echelle en $1/t^2$ n'est donc rendue possible
que par la formation des boucles, et leur d\'esint\'egration en une
forme d'\'energie ne dominant pas la dynamique de l'univers (en
partique des ondes gravitationnelles).}
\label{figdom}
\end{center}
\end{figure}

\begin{figure}
\begin{center}
\epsfig{file=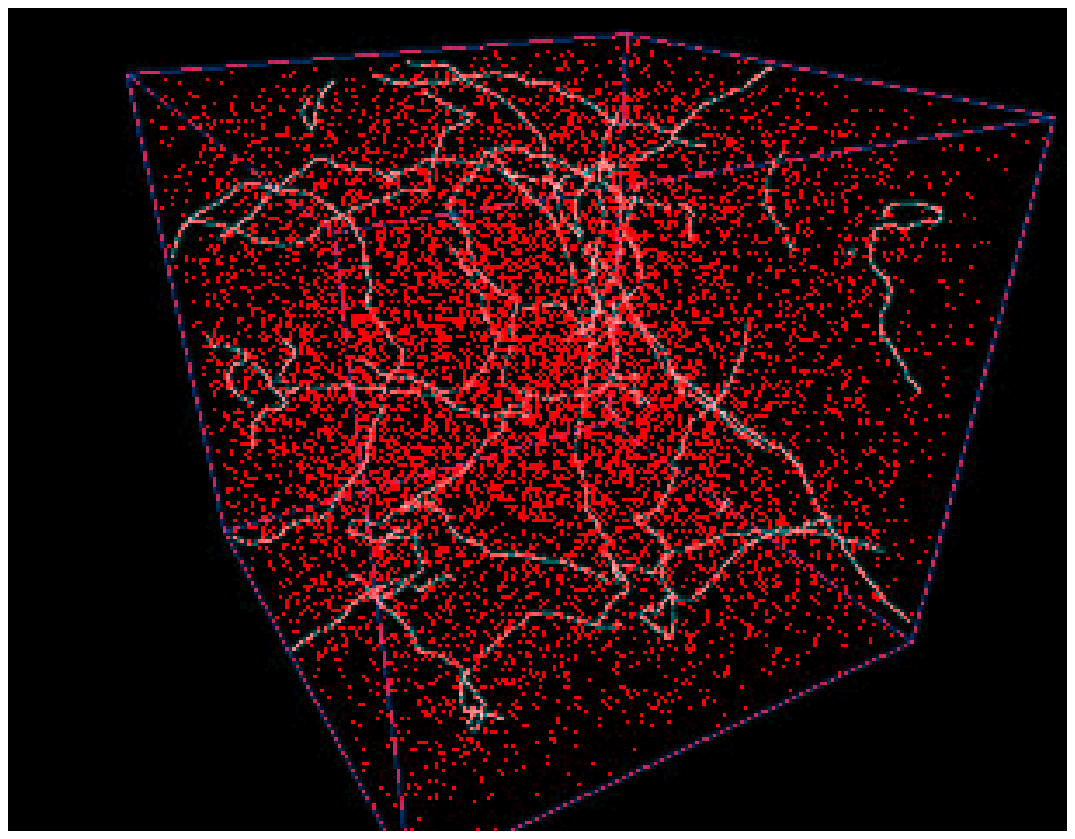,height=10cm}
\epsfig{file=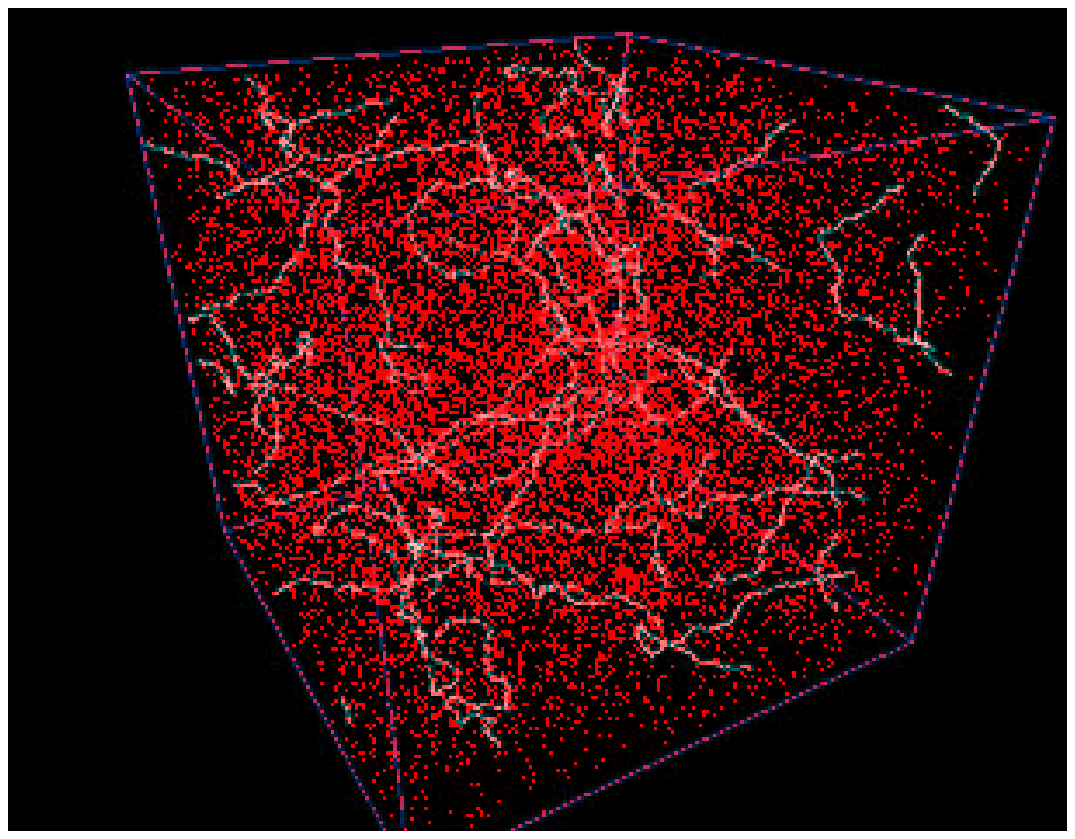,height=10cm}
\caption[Configuration cosmologique d'un r\'eseau de cordes cosmiques
dans l'\'ere de radiation et de mati\`ere.]{Configuration du r\'eseau
de cordes dans l'\`ere de mati\`ere (en haut) et de radiation, lorsque
le r\'egime stationnaire est atteint et lorsque le volume comobile
initial est compl\`etement contenu dans l'horizon. La structure \`a
petite \'echelle des cordes est clairement visible dans leur forme et
par la taille et le grand nombre de boucles (en rouge) g\'en\'er\'ees
au cour de l'\'evolution (la plupart sont en dessous de la
r\'esolution de l'image et apparaissent comme des points).}
\label{figend}
\end{center}
\end{figure}

Sur la figure~\ref{figscaling} sont repr\'esent\'ees les courbes
d'\'evolution de la densit\'e d'\'energie totale associ\'ee aux cordes
plus grandes que l'horizon en fonction du temps conforme. Que ce soit
dans l'\`ere de mati\`ere ou de radiation, celle-ci tend toujours,
ind\'ependamment des conditions initiales, vers la loi d'\'echelle
(\ref{rhoscaling}) en $1/t^2$. Dans l'\`ere de mati\`ere, et de
radiation, on obtient num\'eriquement, respectivement
\begin{equation}
\label{rhoscalingnum}
\rho_{\infty_\umat} \simeq 28 \, \frac{U}{a^2 \eta^2}, \quad
\rho_{\infty_\urad} \simeq 40 \, \frac{U}{a^2 \eta^2}.
\end{equation}
Les simulations num\'eriques confirment ainsi le m\'ecanisme
d'\'evacuation d'\'energie par formation des boucles. Sur la
figure~\ref{figdom} est repr\'esent\'ee l'\'evolution de la densit\'e
d'\'energie des cordes infinies (voir Fig.~\ref{figscaling})
compar\'ee \`a la densit\'e d'\'energie de toutes les cordes, incluant
les boucles. La loi d'\'echelle n'est clairement v\'erifi\'ee que pour
les cordes plus grandes que l'horizon, elle ne peut donc \^etre
applicable au r\'eseau que si l'\'energie associ\'ee aux boucles est
effectivement \'evacu\'ee, par rayonnement gravitationnel par
exemple. Cependant, comme on peut le voir sur la figure~\ref{figend},
la taille caract\'eristiques des boucles form\'ees est extr\^emement
plus faible que l'\'echelle de longueur typique du r\'eseau de cordes
infinies $L_\infty$. Cette observation~\cite{bennett88} montre donc
qu'il existe une autre \'echelle de longueur caract\'erisant le
r\'eseau \`a petite \'echelle et qu'elle est de plus d\'eterminante
dans la nature des intercommutations formant les boucles. Cette
longueur caract\'eristique est li\'ee \`a l'existence des kinks
apparaissant lors des intercommutations (voir
Sect.~\ref{sectionintercom}). En effet, la persistance de ces
discontinuit\'es dans la forme des cordes donne une structure \`a
petite \'echelle qui favorise de nouvelles intersections sur des
longueurs de plus en plus petites, et de fait la formation
privil\'egi\'ee de boucles de petites
tailles~\cite{bennett88,bouchet88,bennett90}. D'autre part, d'apr\`es
l'\'equation (\ref{mvtflrw}), le terme d'amortissement dans la
propagation des kinks d\'epend directement du taux d'expansion de
l'univers, et celui-ci \'etant plus important dans l'\`ere de
mati\`ere, le lissage de la structure induite par les kinks y est plus
efficace que dans l'\`ere de radiation. En cons\'equence, les boucles
produites dans l'\`ere de radiation devraient \^etre plus petites que
celles g\'en\'er\'ees dans l'\`ere de mati\`ere. Ceci appara\^\i t
clairement sur la figure~\ref{figend} si on compare la distribution
des boucles sur les deux prises de vue obtenues lorsque le r\'egime
stationnaire des grandes cordes est atteint. La plus petite taille des
boucles dans l'\`ere de radiation n\'ecessite \'egalement leur plus
grand nombre pour atteindre le r\'egime stationnaire (voir
Fig.~\ref{figend}).

La loi d'\'echelle atteinte par les cordes infinies sugg\`ere que la
densit\'e d'\'energie associ\'ee aux boucles, lorsque l'on consid\`ere
cette fois leur d\'esint\'egration par rayonnement gravitationnel,
doit \'egalement atteindre une loi d'\'echelle~\cite{bennett88}. La
longueur moyenne $L_\ub$ des boucles dans ce r\'egime stationnaire
reste alors une fraction constante de la distance \`a l'horizon
\begin{equation}
L_\ub = \alpha t.
\end{equation}
Les simulations num\'eriques ne donnent cependant qu'une valeur
maximale au coefficient de proportionnalit\'e, $\alpha < 10^{-3}$ dans
le cas du code utilis\'e ici~\cite{bennett88,bennett90}. Il existe
n\'eanmoins une limite inf\'erieure qui est donn\'ee par l'\'echelle
de distance associ\'ee au rayonnement gravitationnel $\alpha > \gamma_\ug
\Gc U$ [voir Eq.~(\ref{tempsvie})]. Il est possible de montrer,
\`a l'aide des mod\`eles effectifs~\cite{kibble85,bennett86,book}, que
la densit\'e d'\'energie totale des boucles rayonnantes suit la loi
d'\'echelle
\begin{equation}
\label{rhobouclerad}
\rho_{\ub_\urad} \simeq \sqrt{\frac{\alpha}{\gamma_\ug \Gc U}}
\rho_{\infty_\urad},
\end{equation}
dans l'\`ere de radiation, et
\begin{equation}
\label{rhobouclemat}
\rho_{\ub_\umat} \simeq 0.4 \sqrt{\alpha} \ln
\left(\frac{\alpha}{\gamma_\ug \Gc U} \right) \rho_{\infty_\umat},
\end{equation}
dans l'\`ere de mati\`ere. En supposant que la majeure partie de
l'\'energie gravitationnelle \'emise par le r\'eseau l'est par les
boucles, il est possible d'en estimer son intensit\'e. Il est d'usage
de consid\'erer le param\`etre de densit\'e d'\'energie
gravitationnelle par unit\'e logarithmique de
fr\'equence~\cite{vilenkin81d,hogan84}
\begin{equation}
\label{omegagrav}
\Omega_\ug = \frac{1}{\rho_\uc} \nu \frac{\ud \rho_\ug}{\ud \nu},
\end{equation}
o\`u $\rho_\ug$ est la densit\'e d'\'energie sous forme d'ondes
gravitationnelles de fr\'equence $\nu$, et $\rho_\uc$ la densit\'e
critique. Comme dans la section~\ref{sectionnambu}, ce param\`etre
peut grossi\`erement \^etre estim\'e en ne consid\'erant que la
fr\'equence caract\'eristique des boucles, i.e. $\nu \simeq
1/L_\ub$. L'\'energie gravitationnelle \'emise au temps $t_\ue $
s'identifie \`a la densit\'e totale d'\'energie des boucles qui se
sont d\'esint\'egr\'ees jusque l\`a, i.e. les boucles qui avaient une
longueur $L_\ub \simeq \alpha t_\ue$ aux temps ant\'erieurs [voir
Eq.~(\ref{tempsvie})]. Il vient alors, d'apr\`es les \'equations
(\ref{rhocrit}), (\ref{rhoscalingnum}), (\ref{rhobouclerad}) et
(\ref{omegagrav})
\begin{equation}
\label{omegagravtot}
\Omega_\ug \simeq 100 \sqrt{\frac{\alpha \Gc U}{\gamma_\ug}}.
\end{equation}
La connaissance des param\`etres $\alpha$ et $\gamma_\ug$ permet donc
d'estimer la contribution d'un r\'eseau de corde au fond d'ondes
gravitationnelles existant dans l'univers en fonction de l'\'echelle
de brisure de sym\'etrie l'ayant g\'en\'er\'ee $U$ [voir
Eq.~(\ref{omegagravtot})]. Ce fond d'ondes gravitationnelles est
cependant contraint par les pr\'edictions de la nucl\'eosynth\`ese
primordiale~\cite{nollett00,groom00,davisRL85,bennett91}, et les
mesures de r\'egularit\'e des temps de r\'eception des signaux radios
\'emis par les pulsars~\cite{bouchet90,stinebring90}. En effet, dans
le premier cas, il existe une densit\'e d'\'energie maximale des ondes
gravitationnelles au del\`a de laquelle l'abondance des \'el\'ements
l\'egers pr\'edite ne seraient plus compatible avec les observations
(voir Fig.~\ref{fignucleo}). Dans le deuxi\`eme cas, la modification
de la g\'eom\'etrie de l'espace-temps nous s\'eparant d'un pulsar, par
le passage d'ondes gravitationnelles, devrait induire des variations
dans les temps d'arriv\'ee de ses signaux. L'absence de ces
irr\'egularit\'es actuellement observ\'ee donne donc \'egalement une
limite sup\'erieure au fond d'ondes gravitationnelles. \`A l'aide des
simulations num\'eriques et de ces deux contraintes observationnelles,
on peut montrer que~\cite{bouchet90,bennett91}
\begin{equation}
\label{gravlim}
\Gc U \lesssim 10^{-6}.
\end{equation}
Un telle contrainte sur l'\'echelle de brisure de sym\'etrie n'est pas
en faveur de la formation d'un r\'eseau de cordes de Goto-Nambu aux
\'echelles d'\'energie de grande unification, et exclue quasiment leur
origine dans les m\'ecanismes de formation des grandes
structures~\cite{bouchet88,bennett91,albrecht97,
battye97,copeland99,albrecht99}.

\subsection{Influence sur le CMBR}
\label{sectionevol2cmb}

Les simulations num\'eriques permettent \'egalement de calculer les
d\'eformations subies par une onde plane \'electromagn\'etique
traversant le r\'eseau de cordes par l'effet Kaiser-Stebbins (voir
Sect.~\ref{sectioneffetgrav}). Appliqu\'es au rayonnement fossile, ces
r\'esultats permettent de dresser des cartes de temp\'erature du CMBR
o\`u les anisotropies sont g\'en\'er\'ees uniquement par les cordes
cosmiques~\cite{bouchet88} (voir Sect.~\ref{sectionqqcont}). Il est
alors possible d'en d\'eduire la valeur moyenne des fluctuations de
temp\'erature du CMBR cr\'e\'ees par ces cordes\cite{bouchet88}
\begin{equation}
\overline{\left(\frac{\delta \Theta}{\Theta}\right)}_2 \simeq 19 \,
\Gc U.
\end{equation}
L'observation de fluctuations de temp\'erature de l'ordre de $10^{-5}$
(voir Figs.~\ref{figfiras} et~\ref{figdata}) donne une condition sur
l'\'echelle de brisure de sym\'etrie formant le r\'eseau de corde du
m\^eme ordre de grandeur que (\ref{gravlim}). Cependant, dans la
section~\ref{sectionqqcont}, nous avons vu que le spectre de puissance
du CMBR actuellement observ\'e ne semblait pas \^etre uniquement
g\'en\'er\'e par des d\'efauts topologiques, et bien que donnant
toujours une limite sup\'erieure \`a $\Gc U$, les r\'esultats pr\'edis
par les cartes de temp\'erature sont difficilement comparables aux
observations.

Une autre approche consiste \`a ne s'int\'eresser qu'au spectre de
puissance du CMBR afin d'y rechercher la signature \'eventuelle d'un
r\'eseau de cordes cosmiques. Dans la section~\ref{sectionqqcont}, les
r\'esultats pr\'esent\'es sur la figure~\ref{figmix} ont \'et\'e
obtenus par superposition lin\'eaire des spectres de puissance
g\'en\'eriques associ\'es aux d\'efauts globaux seuls et \`a
l'inflation seule~\cite{bouchet02}. Cette hypoth\`ese est la plus
simple que l'on puisse faire mais ne tient pas compte des effets de
couplage entre les perturbations g\'en\'er\'ees pas des d\'efauts et
les autres esp\`eces pr\'esentes dans le plasma primordial. La
proportion de d\'efauts trouv\'ee avoisinant les $20 \%$ justifie une
investigation plus pouss\'ee par des m\'ethodes plus pr\'ecises.

Le calcul th\'eorique des anisotropies de temp\'erature du CMBR passe
par la d\'etermination de l'\'evolution cosmologique des perturbations
associ\'ees \`a la g\'eom\'etrie et \`a la mati\`ere qui sont
susceptibles d'en modifier les caract\'eristiques physiques. Celles-ci
sont g\'en\'eralement obtenues, dans l'approximation lin\'eaire, \`a
partir de la connaissance des perturbations $h_{\mu \nu}$ de la
m\'etrique
(\ref{metriqueFLRW})~\cite{lifchitz46,lifchitz63,riazueloT,uzanT}
\begin{equation}
\label{metriqueFLRWpert}
\ud s^2 = \left(g_{\mu \nu} + a^2 h_{\mu \nu} \right) \ud x^\mu \ud
x^\nu,
\end{equation}
et de la description, hydrodynamique ou particulaire, des
perturbations associ\'ees aux diff\'erents fluides pr\'esents dans
l'univers primordial. Les \'equations d'\'evolution sont alors
donn\'ees par les \'equations d'Einstein (\ref{einsteineq})
appliqu\'ees aux quantit\'es perturb\'ees
\begin{equation}
\label{einsteineqpert}
\widetilde{G}^{\mu \nu} = \kappa^2 \widetilde{T}_\utot^{\mu \nu},
\end{equation}
o\`u $\widetilde{G}^{\mu \nu}$ est le tenseur d'Einstein perturb\'e
obtenu \`a partir de la m\'etrique (\ref{metriqueFLRWpert}), et
$\widetilde{T}^{\mu \nu}_\utot$ le tenseur \'energie-impulsion total
associ\'e au diverses sources d'anisotropies. Celui-ci
d\'epend intrins\`equement des propri\'et\'es et des interactions des
divers fluides pr\'esents. L'influence de l'existence d'un r\'eseau de
cordes cosmiques peut \^etre prise en compte dans le terme source
$\widetilde{T}^{\mu \nu}_\utot$, une fois sa contribution $T^{\mu
\nu}_\us$ connu. En fait, puisque l'on s'int\'eresse uniquement \`a la
statistique des fluctuations au travers des fonctions de
corr\'elations (voir Sect.~\ref{sectionqqcont}), les quantit\'es \`a
d\'eterminer seront les corr\'elateurs \`a deux points du tenseur
\'energie-impulsion du r\'eseau
\begin{equation}
C^{\mu \nu \rho \sigma} = \langle T^{\mu \nu}_\us(\eta,\vec{x}\,) \,
T^{\rho \sigma}_\us(\eta',\primevec{x}{}) \rangle,
\end{equation}
o\`u la valeur moyenne porte sur l'ensemble des diff\'erentes
r\'ealisations de la transition de phase. La r\'esolution des
\'equations (\ref{einsteineqpert}), pour les corr\'elateurs, est
actuellement r\'ealis\'ee par diff\'erents codes
num\'eriques~\cite{riazueloT} n\'ecessitant en entr\'ee la
connaissance des corr\'elateurs des tenseurs
\'energie-impulsion des sources. L'int\'er\^et des simulations
num\'eriques de cordes est aujourd'hui de permettre une
d\'etermination de leur corr\'elateurs $C^{\mu \nu \rho \sigma}$ afin
d'inclure leurs effets au sein de la dynamique des perturbations
cosmologiques. \`A partir des \'equations (\ref{metriqueFLRW}),
(\ref{tmunugn}) et (\ref{mixgauge}), le tenseur \'energie-impulsion
d'une corde dans FLRW se r\'eduit \`a
\begin{equation}
\sqrt{-g} \, T^{\mu \nu}(\eta,\vec{r}\,) = U \int{\ud \xi} \left(\epsilon \,
\dot{x}^\mu \dot{x}^\nu - \frac{1}{\epsilon} x'^\mu x'^\nu
\right) \delta^{3}\left[\vec{r} - \vec{x}(\eta,\xi) \right],
\end{equation}
avec $\epsilon$ donn\'e par (\ref{enerdef}), dans la jauge mixte
(\ref{mixgauge}). Sa d\'etermination est particuli\`erement adapt\'ee
\`a la m\'ethode num\'erique du code utilis\'e ici, en effet, en
reportant l'\'equation (\ref{grillephys}) dans l'expression
pr\'ec\'edente, il vient
\begin{equation}
\sqrt{-g}\, T^{\mu \nu}(\eta,\vec{r}\,) = - U \int{\epsilon \, \ud \xi} \,
\frac{1}{2} \left(p^\mu q^\nu + q^\mu p^\nu \right)
\delta^{3}\left[\vec{r} - \vec{x}(\eta,\xi) \right].
\end{equation}
Le volume comobile de r\'ef\'erence (voir Fig.~\ref{figscaling}) peut
\^etre discr\'etis\'e afin de calculer, dans chaque cellule centr\'ee
sur $\vec{r}$, la valeur totale, au temps $\eta$, du tenseur
\'energie-impulsion des cordes qu'elle contient. On obtient ainsi un
ensemble de valeurs $T^{\mu
\nu}_\us(\eta, \vec{r}\,)$ pour le r\'eseau de corde, dont les
corr\'elateurs sont donn\'es, dans l'espace de Fourier, par
\begin{equation}
\label{corrcordes}
\widehat{C}^{\mu \nu \rho \sigma} (\eta,\eta',\vec{k}\,) =
\widehat{T}^{\mu \nu}_\us (\eta,-\vec{k}\,) \, \widehat{T}^{\rho
\sigma}_\us (\eta',\primevec{k}{}),
\end{equation}
o\`u la transform\'ee de Fourier est d\'efinie par
\begin{equation}
\widehat{C}(\vec{k}\,) = \int{\ud^3 \vec{r}} \, C(\vec{r}\,) \ue^{i
\vec{k}.\vec{r}},
\end{equation}
et o\`u l'on a utilis\'e, pour les fonctions r\'eelles $T^{\mu
\nu}_\us (\eta, \vec{r})$, la relation
\begin{equation}
\left[\widehat{T}^{\mu \nu}_\us (\eta, \vec{k}\,)\right]^\dag =
\widehat{T}^{\mu \nu}_\us (\eta, -\vec{k}\,).
\end{equation}
La moyenne statistique des corr\'elateurs (\ref{corrcordes}) peut
\^etre effectu\'ee sur diff\'erentes simulations num\'eriques obtenues
par diff\'erents choix de conditions initiales. En collaboration avec
A.~Riazuelo, auteur d'un code Boltzmann d'\'evolution de perturbations
cosmologiques~\cite{peeble70,riazueloT}, le calcul du spectre de puissance
correspondant est actuellement en cours. Nous pourrons donc, d'ici
peu, statuer pr\'ecis\'ement sur la proportion de cordes cosmiques
locales dans les observations du CMBR.

\section{Conclusion}

Les simulations num\'eriques montrent l'importance du m\'ecanisme de
formation de boucles dans l'\'evolution cosmologique des r\'eseaux de
cordes par la d\'esint\'egration n\'ecessaire de celles-ci en une
forme d'\'energie ne dominant pas l'univers (voir
Fig.~\ref{figdom}). Sous ces conditions, l'existence d'un r\'eseau est
compatible avec les observations pour des \'echelles d'\'energie de
brisure de sym\'etrie inf\'erieures \`a celles de grande unification,
i.e. $\Gc U < 10^{-6}$. Il est \'egalement possible qu'une
contribution non n\'egligeable de ces cordes soit actuellement
d\'etectable dans les donn\'ees du CMBR, et dont la signature pourra
\^etre d'ici peu v\'erifi\'ee ou infirm\'ee gr\^ace aux simulations
num\'eriques.

Cependant, comme on l'a vu par les formalismes macroscopiques (voir
Chap.~\ref{chapitreform}), il existe des types de corde pour
lesquelles les boucles form\'ees peuvent \^etre stables. Ce sont
celles poss\'edant une structure interne g\'en\'er\'ee par un courant
de particules. Il n'existe pas, \`a l'heure actuelle, de code
num\'erique permettant d'en calculer l'\'evolution cosmologique, mais
comme l'approche qualitative de la section~\ref{subsectionqual} le
sugg\`ere, la loi d'\'echelle atteinte par les r\'eseaux au cours de
leur \'evolution semble robuste, et le m\^eme m\'ecanisme appliqu\'e
\`a ces cordes devrait produire une grande proportion de boucles
conductrices potentiellement stables, aboutissant ainsi \`a leur
domination sur les autres formes d'\'energie. Un tel sc\'enario n'est
pas compatible avec les observations, et comme nous le verrons dans
les sections suivantes, la pr\'esence de courants sur les cordes
semble en \^etre une propri\'et\'e g\'en\'erique. La stabilit\'e des
boucles (voir Sect.~\ref{sectionstabilite}) devient donc un crit\`ere
cosmologique d'existence de ces cordes, et de ce fait, une contrainte sur
les sym\'etries effectivement bris\'ees dans l'univers primordial.

\chapter{G\'en\'eration de courants sur les cordes}
\label{chapitrecour}
\minitoc
\section{Introduction}

Les chapitres pr\'ec\'edents ont mis en \'evidence l'importance
cosmologique des boucles de corde cosmique. Leur conversion en une
forme d'\'energie ne dominant pas l'univers est une condition
n\'ecessaire \`a l'existence d'un r\'eseau dans l'univers. Cependant,
dans le chapitre~\ref{chapitreform}, le formalisme covariant nous a
permis de montrer que la brisure de l'invariance de Lorentz
longitudinale par un courant conserv\'e le long des cordes pouvait
stabiliser les boucles. Il est, de ce fait, l\'egitime de s'interroger
sur l'existence et la pertinence de m\'ecanismes microphysiques
pouvant \^etre \`a l'origine de ces courants. Dans le cadre du
mod\`ele de Higgs ab\'elien, il est possible de coupler la th\'eorie
microphysique, d\'ecrivant de purs vortex, \`a d'autres champs pouvant
\'eventuellement se propager le long de la corde. Le cas le plus
simple est celui d'un champ scalaire additionnel coupl\'e au champ de
Higgs. E.~Witten a en effet montr\'e qu'un condens\^at de ce champ
peut se former au centre du vortex et g\'en\'erer un courant de
particules scalaires~\cite{witten}. Les
\'equations du mouvement de ces cordes \emph{supraconductrices} sont
plus compliqu\'ees du fait de la pr\'esence de champs
suppl\'ementaires, il en r\'esulte que la description analytique de
leurs propri\'et\'es dynamiques est difficile. C'est pourquoi le
formalisme covariant est habituellement pr\'ef\'er\'e: la simple
connaissance de l'\'equation d'\'etat fixe la dynamique des cordes et
la stabilit\'e des boucles associ\'ees (voir
Chap.~\ref{chapitreform}). La correspondance entre l'approche
microscopique et le formalisme covariant s'effectue naturellement par
le passage \`a la limite d'\'epaisseur nulle de la th\'eorie
quadri-dimensionnelle. P.~Peter a obtenu, de cette mani\`ere,
l'\'equation d'\'etat des cordes poss\'edant un condens\^at de
Bose~\cite{neutral,enon0}. Ce lien est indispensable car il permet
d'une part, de justifier l'emploi de l'approche macroscopique pour des
cordes poss\'edant une structure interne, et d'autre part, d'exprimer
les grandeurs macroscopiques, telle l'\'energie par unit\'e de
longueur ou la tension, en fonction des param\`etres microscopiques de
la physique des particules sous-jacente. Ces param\`etres peuvent
ainsi \^etre fix\'es par les contraintes cosmologiques directement
d\'eduites du formalisme covariant.

\section{Le mod\`ele de Witten scalaire}
\label{sectionwitten}

\`A partir du mod\`ele de Higgs ab\'elien (\ref{laghiggs}) d\'ecrivant
la microphysique d'une corde de Goto-Nambu, il est possible de coupler
un champ scalaire $\Sigma$ suppl\'ementaire au champ de Higgs $\Phi$
formant la corde. Par analogie avec la supraconductivit\'e dans les
mat\'eriaux, l'id\'ee est de briser une sym\'etrie additionnelle
uniquement au centre du vortex afin que le champ $\Sigma$ puisse s'y
condenser. Le cas le plus simple est celui d'une invariance
$U_\ub(1)$, qui peut \^etre choisie locale ou globale selon que le
champ scalaire $\Sigma$ est charg\'e ou non, respectivement. La forme
g\'en\'erale du lagrangien associ\'e au champ $\Sigma$ s'\'ecrit, dans
le cas d'une sym\'etrie $U_\ub(1)$ locale, d'apr\`es
(\ref{emscalaire})
\begin{equation}
\label{lagsigma}
\Lc_\Sigma = \frac{1}{2} \left(\Dt_\mu \Sigma \right)^\dag
\left(\Dt^\mu \Sigma \right) - \frac{1}{4} N_{\mu \nu} N^{\mu \nu} -
\frac{\lambdat}{8} |\Sigma|^4 - \frac{1}{2} m_\sigma^2 |\Sigma|^2
\end{equation}
o\`u $N^{\mu \nu}$ est le tenseur de type Faraday associ\'e au champ de
jauge $C^\mu$ de la sym\'etrie $U_\ub(1)$, et
\begin{equation}
\Dt_\mu = \partial_\mu + i \, \gt \, C_\mu,
\end{equation}
la d\'eriv\'ee covariante associ\'ee. Cette sym\'etrie de jauge peut,
par ailleurs, \^etre identifi\'ee \`a l'\'electromagn\'etisme.
L'interaction de $\Sigma$ avec le champ de Higgs formant la corde
s'\'ecrit, de mani\`ere g\'en\'erale,
\begin{equation}
\label{lagintwitten}
\Lc_\uint = f \left(\eta^2 - |\Phi|^2 \right)|\Sigma|^2,
\end{equation}
o\`u $f$ est la constante de couplage entre les deux champs, et $f
\eta^2$ un terme de masse additionnel qui peut \^etre absorb\'e dans
$m_\sigma^2$. L'int\'er\^et d'avoir \'ecrit le lagrangien
d'interaction sous la forme (\ref{lagintwitten}) est que loin de la
corde, si le champ de Higgs $\Phi$ prend sa valeur moyenne dans le
vide $|\Phi|=\eta$, la masse physique du boson $\Sigma$ est donn\'ee
par $m_\sigma$. Le potentiel total associ\'e \`a $\Phi$ s'\'ecrivant,
d'apr\`es (\ref{laghiggs}), (\ref{pothiggs}) et (\ref{lagintwitten}),
\begin{equation}
V_\utot(\Phi) = \frac{\lambda}{8} \left(|\Phi|^2 -\eta^2 \right)^2 + f
|\Sigma|^2 |\Phi|^2,
\end{equation}
il est effectivement minimis\'e en $|\Phi|=\eta$ et
$|\Sigma|=0$. Inversement, au centre de la corde $\Phi=0$ et, pourvu
que $f \eta^2 > m_\sigma^2/2$, le champ $\Sigma$ semble acqu\'erir une
masse n\'egative caract\'eristique de la brisure de la sym\'etrie
$U_\ub(1)$ (voir Sect.~\ref{sectionhiggs}). En effet, le potentiel du
champ $\Sigma$ au centre de la corde est, d'apr\`es les \'equations
(\ref{lagsigma}) et (\ref{lagintwitten}),
\begin{equation}
\label{potsigmacore}
\left. \Vt(\Sigma)\right|_{\Phi=0} = \frac{\lambdat}{8} |\Sigma|^4 +
\frac{1}{2} \left(m_\sigma^2 - 2 f \eta^2 \right) |\Sigma|^2,
\end{equation}
Dans le cas o\`u $f \eta^2 < m_\sigma^2/2$, ce potentiel est
minimis\'e pour $|\Sigma|=0$ et la sym\'etrie $U_\ub(1)$ n'est pas
bris\'ee, alors que si $f \eta^2 > m_\sigma^2/2$, il est minimis\'e
pour
\begin{equation}
\label{condensatcore}
|\Sigma| = 2 \sqrt{\frac{2 f \eta^2 -m_\sigma^2}{\lambdat}}.
\end{equation}
On obtient, dans ce dernier cas, un condens\^at du champ au centre de
la corde o\`u la sym\'etrie $U_\ub(1)$ est bris\'ee. Par continuit\'e,
le champ $\Sigma$ va rejoindre son \'etat de vide $\Sigma=0$ loin de
la corde, et par sym\'etrie cylindrique, va \^etre uniquement une
fonction de la distance au vortex\footnote{La valeur du champ sur la
corde n'est en g\'en\'eral pas donn\'e par l'\'equation
(\ref{condensatcore}), mais d\'epend de la dynamique des champs dans
le vortex.} que nous noterons $\Sigma_\zero(r)$. Notons que le
lagrangien (\ref{lagsigma}) peut
\'egalement conduire \`a la formation de cordes cosmiques par la
brisure spontann\'ee de la sym\'etrie $U_\ub(1)$ lors du
refroidissement de l'univers. Ceci peut \^etre \'evit\'e en
choisissant les param\`etres physiques associ\'es \`a $\Sigma$ tels
que sa temp\'erature de transition de phase soit bien inf\'erieure \`a
celle de $\Phi$, i.e. d'apr\`es (\ref{tempcrit})
$m_\sigma/\lambdat^{1/2} < \eta$.

D'apr\`es (\ref{condensatcore}), l'ensemble des valeurs du champ
$\Sigma$ de la forme
\begin{equation}
\label{allsigmacore}
\Sigma = |\Sigma_\zero| \ue^{i \psi(\xi^a)},
\end{equation}
avec $\psi(\xi^a)$ une fonction r\'eelle des coordonn\'ees
longitudinales, minimisent \'egalement le potentiel dans le
vortex. Par le th\'eor\`eme de Noether, l'invariance $U_\ub(1)$ de la
th\'eorie impose au courant $j^\mu = \delta \Lc_\Sigma / \delta C_\mu$
d'\^etre conserv\'e. Ses composantes le long de la corde s'obtiennent
\`a partir de (\ref{lagsigma}) et (\ref{allsigmacore}), et on trouve
\begin{equation}
\label{courantsigma}
j^a = \gt \, |\Sigma_\zero|^2 \left(\partial^a \psi + \gt \, C^a
\right).
\end{equation}
La persistance de ce courant est assur\'ee pour des raisons
topologiques similaires \`a celles assurant l'existence des cordes: le
long d'une boucle de corde, la phase $\psi$ ne peut varier que d'un
multiple entier de $2 \pi$ permettant de d\'efinir un invariant
topologique~\cite{witten,book}
\begin{equation}
\label{chargetopo}
N = \frac{1}{2\pi} \oint{\ud \xi} \, \frac{\ud \psi}{\ud \xi},
\end{equation}
avec $\xi$ l'abscisse curviligne. Ce courant peut \^etre estim\'e en
calculant le potentiel vecteur $C^{a}$. Dans la jauge de Coulomb,
i.e. $\partial_i C^i=0$, la solution des \'equations (\ref{mvtem}) est
donn\'ee par
\begin{equation}
\label{biotsavart}
C^a(\vec{r}\,) = -\frac{1}{4\pi} \oint{\ud \xi} \,
\frac{j^a\left[\vec{x}(\xi)\right]}{\left| \vec{r} - \vec{x}(\xi)
\right|},
\end{equation}
avec $\vec{x}(\xi)$ la position de la corde dans un r\'ef\'erentiel
privil\'egi\'e. D'apr\`es (\ref{biotsavart}), le potentiel vecteur
diverge logarithmiquement avec la distance pr\`es du centre du
vortex. La largeur de la corde \'etant, en premi\`ere approximation,
donn\'ee par l'inverse de la masse du Higgs (voir
Sect.~\ref{sectiondefauts}), un ordre de grandeur de la partie
divergente de (\ref{biotsavart}) est $-\ln\left(m_\uh L_\ub \right)$,
avec $L_\ub$ la longueur de la boucle. Il vient finalement
\begin{equation}
\label{champtopo}
C^{a} \simeq - \ln\left(m_\uh L_\ub \right) j^{a},
\end{equation}
qui, combin\'e avec les \'equations (\ref{courantsigma}) et
(\ref{chargetopo}), donne la composante spatiale du courant le long de
la corde
\begin{equation}
\label{couranttopo}
j^z \simeq \frac{\gt \, |\Sigma_\zero|^2}{1 + \gt^2 |\Sigma_\zero|^2
\ln\left(m_\uh L_\ub\right)} \, \frac{N}{L_\ub} \simeq \frac{1}{\gt
\ln\left(m_\uh L_\ub\right) } \, \frac{N}{L_\ub}.
\end{equation}
Bien que son amplitude soit directement reli\'ee \`a la charge
topologique $N$, le courant $j^z$ est limit\'e physiquement par
l'existence du condens\^at sur la corde, c'est-\`a-dire qu'il est
n\'ecessaire que l'\'etat excit\'e du condens\^at reste l'\'etat
d'\'energie minimale. Or, d\`es que le courant est pr\'esent, pour $N
\neq 0$, le terme cin\'etique du lagrangien (\ref{lagsigma}) peut
se voir comme une contribution n\'egative \`a un potentiel effectif
pour $\Sigma$ dont la composante spatiale est de l'ordre de
\begin{equation}
\left(\Dt_z \Sigma\right)^\dag \left(\Dt_z \Sigma\right) =
-|\Sigma_\zero|^2 \left(N + \gt \, C_z \right)^2.
\end{equation}
La brisure de la sym\'etrie $U_\ub(1)$ n'est encore assur\'ee au
centre de la corde, d'apr\`es (\ref{lagsigma}) et
(\ref{potsigmacore}), que pour
\begin{equation}
\left(N + \gt \,C_z \right)^2 < f \eta^2- \frac{1}{2} m_\sigma^2,
\end{equation}
soit, \`a l'aide de (\ref{champtopo}) et (\ref{couranttopo}),
\begin{equation}
N^2 < \left(\frac{L_\ub}{L_\ub+1} \right)^2 \left(f \eta^2 -
\frac{1}{2} m_\sigma^2 \right).
\end{equation}
Il y a donc un courant maximal au del\`a duquel il est
\'energ\'etiquement pr\'ef\'erable aux particules se propageant le
long de la corde de rejoindre leur \'etat de vide usuel
$\Sigma=0$.

Le mod\`ele de Witten conduit, par le couplage le plus simple qu'un
champ scalaire additionnel puisse avoir avec le champ de Higgs, \`a la
naissance d'un courant conserv\'e le long de la corde. Il est clair
que la dynamique de celle-ci, ainsi que la stabilit\'e des boucles
associ\'ees, vont en \^etre modifi\'ees. Notons que le courant
(\ref{couranttopo}) introduit seulement un param\`etre physique
suppl\'ementaire, $\psi$, par rapport au cas de Goto-Nambu, lorsque
l'on passe \`a la limite d'\'epaisseur nulle. Il est donc raisonnable
d'esp\'erer d\'ecrire la dynamique macroscopique d'une telle corde par
le formalisme covariant introduit au chapitre~\ref{chapitreform}.

\section{Correspondance avec le formalisme covariant}
\label{sectioncorresp}
L'approche microscopique pr\'ec\'edente permet la d\'etermination, par
le passage \`a la limite d'\'epaisseur nulle, des quantit\'es
macroscopiques n\'ecessaires \`a la description d'une corde
conductrice dans le formalisme covariant, c'est-\`a-dire l'\'energie
par unit\'e de longueur et la tension, ainsi que le param\`etre
d'\'etat $\varpib$ tenant compte de l'influence du courant. Afin de ne
pas perdre compl\`etement l'information concernant la structure
transverse du vortex, la connaissance de la dynamique des champs en
interaction est indispensable avant de pouvoir la moyenner sur les
dimensions transverses \`a la corde. Comme c'\'etait le cas pour le
mod\`ele de Higgs ab\'elien (voir Sect.~\ref{sectionnielsen}), la
r\'esolution des \'equations du mouvement, dans le mod\`ele de Witten,
n\'ecessite l'emploi de m\'ethodes
num\'eriques~\cite{bps,neutral,enon0}. En utilisant des m\'ethodes de
relaxation~\cite{adler}, P.~Peter a, de cette mani\`ere, confirm\'e la
validit\'e du formalisme covariant pour ces types de corde tout en
d\'erivant la forme exacte de l'\'equation d'\'etat.

Les \'equations d'Euler-Lagrange des diff\'erents champs obtenues \`a
partir du lagrangien du mod\`ele de Witten scalaire,
\begin{equation}
\label{lagwittens}
\Lc_{\uw \ub} = \Lc_\uh + \Lc_\Sigma + \Lc_\uint,
\end{equation}
peuvent s'\'ecrire, sous leur forme adimensionn\'ee~\cite{enon0},
\begin{equation}
\label{systwitten}
\left\{
\begin{array}{lcl}
\displaystyle
\frac{\ud H}{\ud \varrho^2} + \frac{1}{\varrho} \frac{\ud H}{\ud
\varrho} & = & \displaystyle \frac{1}{\varrho^2} H Q^2 + \frac{1}{2} H
(H^2-1) + 2 \frac{f \eta^2}{m_\uh^2} \frac{m_\sigma^2}{\mcroix^2} H
S^2,\\ \\
\displaystyle
\frac{\ud S}{\ud \varrho^2} + \frac{1}{\varrho} \frac{\ud S}{\ud
\varrho} & = & \displaystyle w \frac{\mcroix^2}{m_\sigma^2} P^2 S + 2
\frac{f \eta^2}{m_\uh^2}(H^2 - 1) S + \frac{m_\sigma^2}{m_\uh^2}
S(S^2+1),\\ \\
\displaystyle
\frac{\ud Q}{\ud \varrho^2} - \frac{1}{\varrho} \frac{\ud Q}{\ud
\varrho} & = & \displaystyle \frac{m_\ub^2}{m_\uh^2} H^2 Q,\\ \\
\displaystyle
\frac{\ud P}{\ud \varrho^2} + \frac{1}{\varrho} \frac{\ud P}{\ud
\varrho} & = & \displaystyle \frac{m_\sigma^2}{m_\uh^2}
\frac{m_\uc^2}{\mcroix^2} S^2 P,
\end{array}
\right.
\end{equation}
o\`u l'on a introduit les param\`etres
\begin{equation}
\mcroix = \sqrt{\lambdat} \, \eta, \quad m_\uc= \gt \, \eta,
\end{equation}
par analogie avec les param\`etres donnant la masse du boson de Higgs $\Phi$
et du boson vecteur $B^\mu$ de la corde
\begin{equation}
m_\uh = \sqrt{\lambda} \, \eta, \quad m_\ub = g \, \eta.
\end{equation}
Le champ de Higgs adimensionn\'e $H$ et la composante orthoradiale du
champ de jauge de la corde $Q$, ainsi que la variable transverse
$\varrho$ sont donn\'es par les \'equations (\ref{defvarphi}) \`a
(\ref{defHQvarrho}). Le champ scalaire $S$ est d\'efini par
\begin{equation}
S= \frac{\mcroix}{m_\sigma} \frac{|\Sigma_\zero|}{\eta}.
\end{equation}
Quant au potentiel vecteur $C^\mu$, associ\'ee \`a la sym\'etrie
$U_\ub(1)$, et la phase $\psi$ \`a l'origine du courant topologique,
ils permettent de d\'efinir le bivecteur $P^a$ par la relation
\begin{equation}
P_a \equiv \partial_a \psi + \gt \, C_a.
\end{equation}
Seule la norme adimensionn\'ee $P$ de celui-ci intervient dans les
\'equations de champs (\ref{systwitten}). Elle permet en outre de
d\'efinir la quantit\'e $\varpib$ qui s'identifie au param\`etre
d'\'etat du formalisme covariant (voir Sect.~\ref{sectionetatform}):
\begin{equation}
P_t^2 - P_z^2 = \varpib P^2.
\end{equation}
Le param\`etre $\varpib$ appara\^\i t \'egalement dans les \'equations
du mouvement (\ref{systwitten}) sous sa forme adimensionn\'ee
\begin{equation}
w = \frac{m_\sigma^2}{m_\uh^2 \mcroix^2} \varpib.
\end{equation}
\begin{figure}
\begin{center}
\input{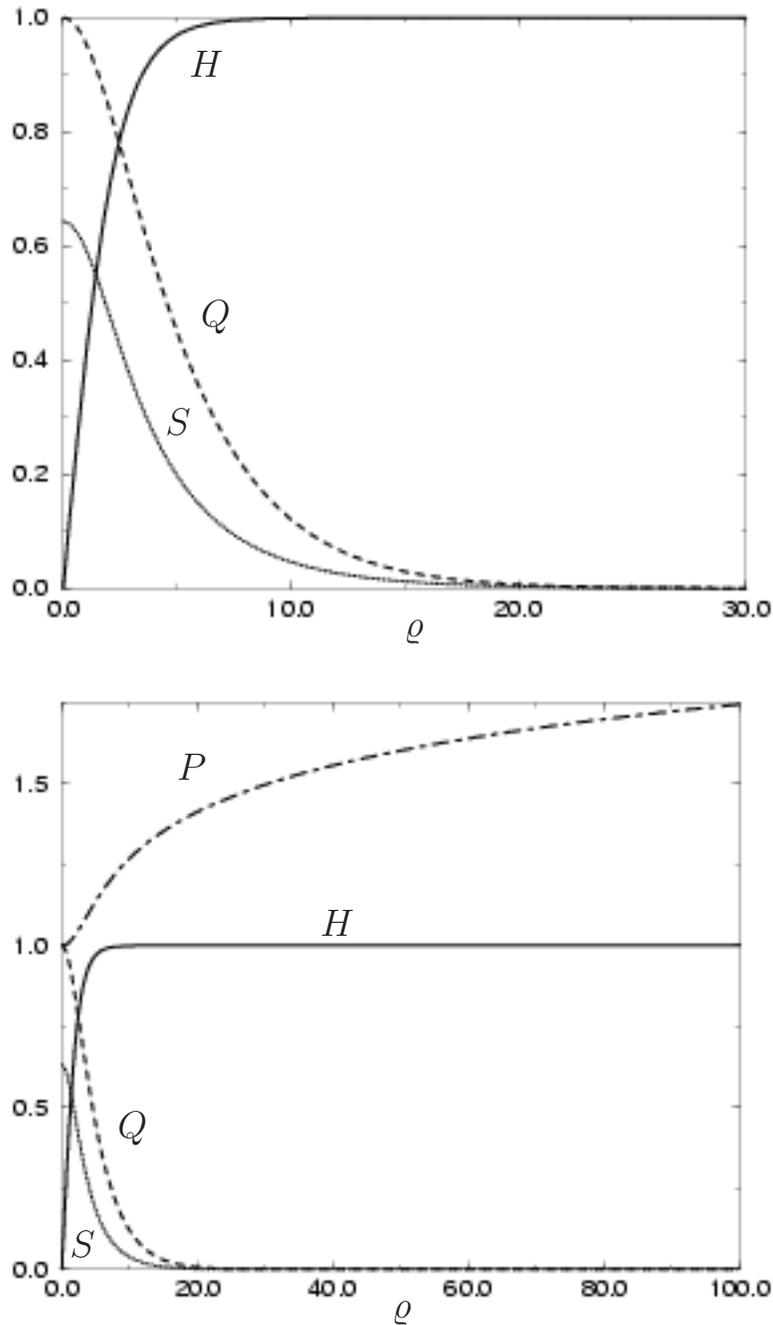}
\caption[Profils transverses des champs dans une corde cosmique
poss\'edant un condens\^at de Bose.]{Profils transverses des champs
dans une corde cosmique de Witten poss\'edant un condens\^at de Bose
en son c\oe ur~\cite{enon0}. Le champ de Higgs $H$ et le champ de
jauge formant la corde $Q$ ont les m\^emes caract\'eristiques que ceux
formant une corde de Goto-Nambu (voir Fig.~\ref{figback}). Le champ
scalaire $S$ se condense au centre du vortex, g\'en\'erant un courant
dont la divergence logarithmique du champ de jauge $P$ associ\'e est
caract\'eristique d'un fil conducteur.}
\label{figscalarcond}
\end{center}
\end{figure}

\`A partir de la r\'esolution des \'equations (\ref{systwitten}) (voir
Fig.~\ref{figscalarcond}) en fonction du param\`etre $w$, la
connaissance des variations transverses des champs permet de calculer
le tenseur \'energie impulsion $T^{ab}$, le long de la corde, \`a
l'aide de l'\'equation (\ref{tmunulag}) appliqu\'ee au lagrangien
(\ref{lagwittens}). Le courant topologique $j^a$ est pour sa part
directement donn\'e par (\ref{courantsigma}). Leur valeurs
bidimensionnelles, $\Tb^{ab}$ et $\jb^a$, dans la limite d'\'epaisseur
nulle, sont ensuite obtenues par int\'egration sur les directions
transverses~\cite{cartermeca,neutral,enon0} (voir
Sect.~\ref{sectionnielsen}). L'\'energie par unit\'e de longueur $U$
et la tension $T$ sont finalement donn\'ees, d'apr\`es la
section~\ref{sectioncov}, par
\begin{equation}
U=\Tb^{tt}, \quad T=-\Tb^{zz},
\end{equation}
et l'intensit\'e $\Cc$ du courant par
\begin{equation}
\Cc = \sqrt{\left| \jb_t^2 - \jb_z^2 \right|} = 2 \pi \eta^2
\frac{m_\sigma}{m_\uh \mcroix} \sqrt{|w|} \int{\varrho \, \ud
\varrho}\, S^2 P.
\end{equation}
On peut v\'erifier que ces grandeurs satisfont l'\'equation
barotropique attendue par le formalisme covariant
\begin{equation}
\label{eosbaroscalar}
U - T = \sqrt{|\varpib|} \, \Cc.
\end{equation}
Sur la figure~\ref{figscalarcour} sont repr\'esent\'ees les variations
de l'intensit\'e du courant $\Cc$ en fonction du param\`etre
$-w/\sqrt{|w|}$ dont le signe, positif ou n\'egatif, d\'epend
maintenant du type de r\'egime, magn\'etique ou \'electrique,
respectivement (voir Sect.~\ref{sectioncov}). Comme attendu par les
consid\'erations analytiques de la section pr\'ec\'edente, on observe
clairement une saturation du courant pour $|\varpib| \simeq
m_\sigma^2$, dans le r\'egime magn\'etique. Pour un param\`etre
d'\'etat sup\'erieur \`a cette valeur, l'intensit\'e du courant
diminue car l'\'etat de condensation le long de la corde n'est plus
\'energ\'etiquement favoris\'e, la fuite de particules scalaires vers
l'\'etat de vide usuel est privil\'egi\'ee. De la m\^eme mani\`ere,
dans le r\'egime \'electrique, $\Cc$ repr\'esente cette fois la
densit\'e de particules condens\'ees sur la corde, et, pour des
valeurs du param\`etre d'\'etat $|\varpib| \rightarrow m_\sigma^2$, on
observe une divergence de $\Cc$ appel\'ee \emph{seuil de
fr\'equence}. L'interpr\'etation physique en est similaire:
l'\'energie moyenne de chaque particule devient sup\'erieure \`a leur
masse $m_\sigma$ loin de la corde, et il n'existe plus de solution
stationnaire aux \'equations de champ (\ref{systwitten}), ce qui se
traduit par la divergence observ\'ee des solutions statiques sur la
figure~\ref{figscalarcour}.
\begin{figure}
\begin{center}
\input{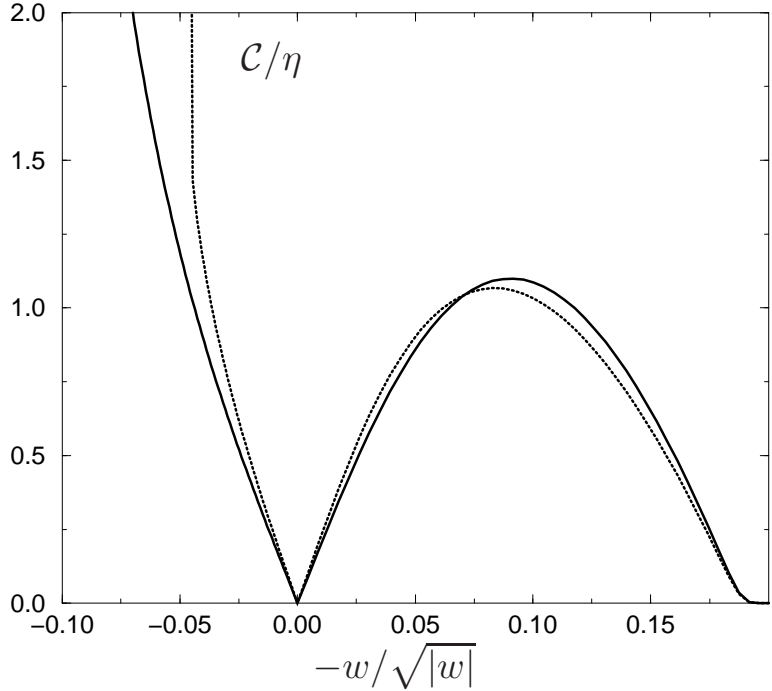}
\caption[Intensit\'es des courants de bosons sur une
corde cosmique.]{L'intensit\'e du courant $\Cc$ en fonction du
param\`etre adimensionn\'e $-w/\sqrt{|w|}$. Le condens\^at de Bose au
centre du vortex n'est \'energ\'etiquement privil\'egi\'e que pour
$|\varpib|<m_\sigma^2$, au del\`a, les particules scalaires rejoignent
leur \'etat de vide loin de la corde entra\^\i nant une saturation du
courant dans le r\'egime magn\'etique, et un seuil de fr\'equence,
caract\'eris\'e par la divergence de $\Cc$, dans le r\'egime
\'electrique~\cite{neutral}. La courbe en trait plein correspond \`a la
limite $\gt \rightarrow 0$ du mod\`ele de Witten (c'est la limite
neutre dans le cas o\`u la sym\'etrie $U_\ub(1)$ repr\'esente
l'\'electromagn\'etisme) et s'\'eloigne relativement peu de la courbe
incluant le couplage au champ de jauge $C^\mu$: les propri\'et\'es
physiques dominantes r\'esultent principalement des effets
m\'ecaniques du courant le long de la corde, comme attendu par le
formalisme covariant~\cite{cartermeca,carter97b}.}
\label{figscalarcour}
\end{center}
\end{figure}
\begin{figure}
\begin{center}
\input{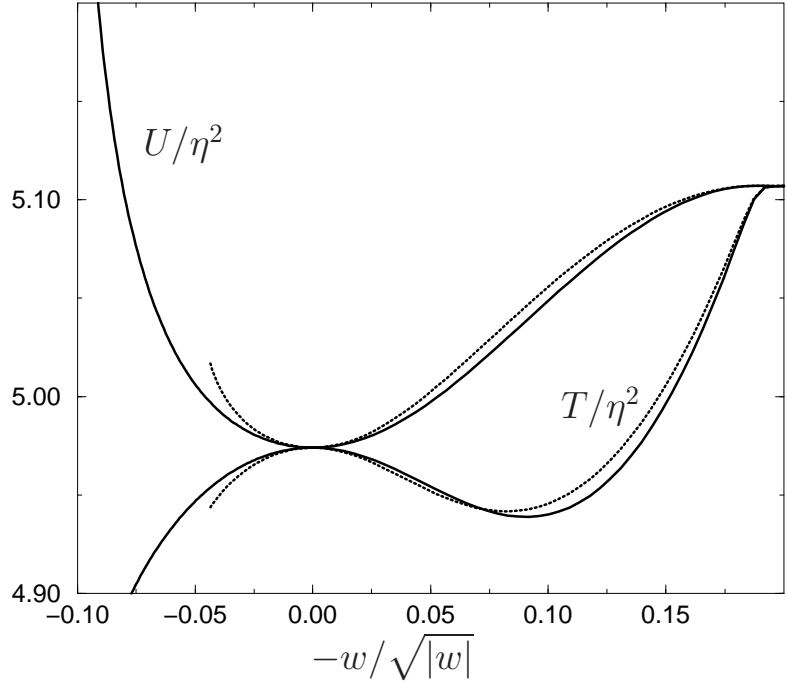}
\caption[\'Energie par unit\'e de longueur et tension d'une corde
cosmique parcourue par des bosons.]{L'\'energie par unit\'e de
longueur $U$ et la tension $T$ d'une corde cosmique, parcourue par un
courant de particules scalaires, en fonction du param\`etre
d'\'etat. Comme sur la figure~\ref{figscalarcour}, les courbes en
traits pleins et pointill\'es correspondent au condens\^at de Bose
obtenu par la brisure d'une sym\'etrie $U_\ub(1)$ globale et locale,
respectivement.  L'influence de la constante de couplage au champ de
jauge $C^\mu$ n'est pas d\'eterminante dans la dynamique. On retrouve
\'egalement le seuil de fr\'equence dans le r\'egime \'electrique,
o\`u ces deux quantit\'es divergent, et la cons\'equence de la
saturation du courant lorsque le minumum de tension, dans le r\'egime
magn\'etique, est atteint: la corde devient alors instable vis-\`a-vis
des perturbations longitudinales~\cite{enon0} (voir
Sect.~\ref{sectionstabilite}).}
\label{figscalareos}
\end{center}
\end{figure}

De la m\^eme mani\`ere, l'\'equation d'\'etat est repr\'esent\'ee sur
la figure~\ref{figscalareos} o\`u l'\'energie $U$ par unit\'e de
longueur et la tension $T$ sont trac\'ees en fonction du param\`etre
d'\'etat $\varpib$. Les ph\'enom\`enes de saturation et de seuil de
fr\'equence, li\'es \`a l'instabilit\'e du condens\^at, se retrouvent
\'egalement sur ces courbes. Comme pour le courant, le seuil de
fr\'equence appara\^\i t par la divergence de $U$ et $T$ dans le
r\'egime \'electrique, alors que la saturation de $\Cc$ correspond
maintenant au point d'inversion des variations de la tension, dans le
r\'egime magn\'etique, en $|\varpib|=m_\sigma^2$. D'apr\`es la
section~\ref{sectionstabilite}, la vitesse de propagation des
perturbations longitudinales \'etant donn\'ee par $\cl^2 = -\ud T/\ud
U$, la corde devient aussi instable au point $\varpib =
m_\sigma^2$. La valeur maximale atteinte par le courant est donc la
limite physique en de\c{c}a de laquelle ce type de corde conductrice
peut exister.

L'approche num\'erique permet de plus d'\'etudier l'influence de la
constante de couplage $\gt$ au champ de jauge $C^\mu$. Sur les
figures~\ref{figscalarcour} et~\ref{figscalareos}, les courbes en
traits pleins sont obtenues dans la limite neutre~\cite{neutral},
c'est-\`a-dire lorsque la sym\'etrie $U_\ub(1)$ est globale,
contrairement \`a celles en pointill\'es obtenues pour une sym\'etrie
locale avec $\gt^2 > 0.1$. Il est clair que les propri\'et\'es
physiques dominantes ne d\'ependent que tr\`es faiblement de $\gt$. En
plus de la relation (\ref{eosbaroscalar}), ce r\'esultat renforce
encore la validit\'e du formalisme covariant dans lequel seule la
brisure de l'invariance de Lorentz longitudinale, par le courant, est
d\'eterminante dans la dynamique de la corde.

\begin{figure}
\begin{center}
\input{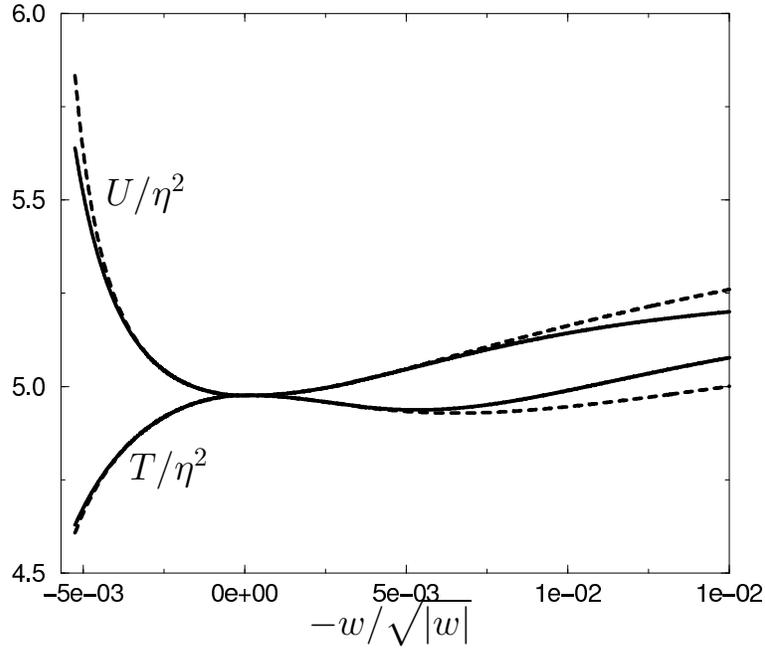}
\caption[Approximation analytique de l'\'equation d'\'etat d'une corde
poss\'edant un courant de bosons.]{Comparaison entre l'\'equation
d'\'etat issue de la r\'esolution num\'erique des \'equations de champ
(\ref{systwitten}), et son approximation analytique d\'eduite de la
fonction ma\^\i tresse duale d\'efinie par (\ref{dualscalarE}) et
(\ref{dualscalarM})~\cite{carter89}.}
\label{figanaeos}
\end{center}
\end{figure}

\begin{figure}
\begin{center}
\input{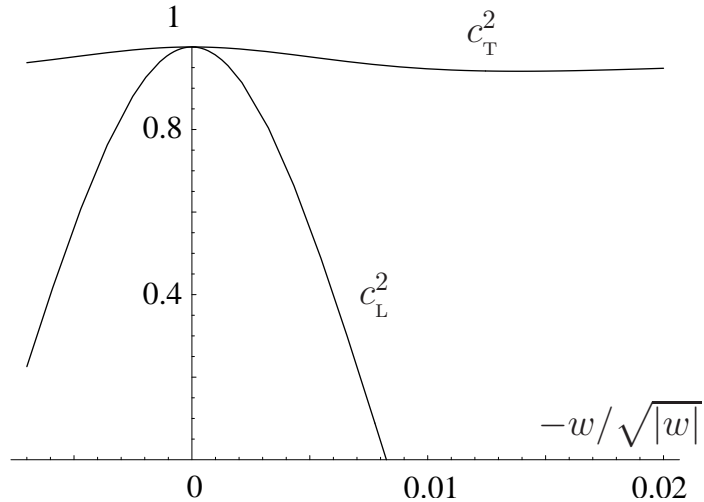}
\caption[Vitesses de propagation des perturbations d'une corde
cosmique parcourue par des bosons.]{Variation des vitesses de
propagation des perturbations transverses et longitudinales, pour une
corde poss\'edant un condens\^at de Bose, en fonction du param\`etre
d'\'etat~\cite{neutral}. Le r\'egime supersonique $\ct^2>\cl^2$ obtenu
est potentiellement porteur d'instabilit\'es pour ce type de boucles
(voir Sect.~\ref{sectionstabilite}), sauf peut \^etre autour de
$\varpib=0$~\cite{martinpeter,mp2}.}
\label{figclctscalar}
\end{center}
\end{figure}

Afin de pouvoir compl\`etement s'affranchir de l'approche
microscopique, il est d'usage d'approcher l'\'equation d'\'etat
calcul\'ee num\'eriquement (voir Fig.~\ref{figscalareos}) par une
forme analytique directement utilisable dans le formalisme
covariant. On montre que la fonction ma\^\i tresse duale (voir
Sect.~\ref{sectionetatform}) d\'efinie, dans le secteur magn\'etique
par
\begin{equation}
\label{dualscalarE}
\Lambdatd = m^2 + \frac{1}{2} \frac{\varpitd}{\displaystyle
\frac{\varpitd}{m_\star^2}-1},
\end{equation}
et dans le secteur \'electrique par,
\begin{equation}
\label{dualscalarM}
\Lambdatd = m^2 + \frac{m_\star^2}{2}
\ln\left(1-\frac{\varpib}{m_\star^2} \right),
\end{equation}
permet d'obtenir l'approximation voulue lorsque ses param\`etres sont
tels que $m \simeq m_\uh$ et $m_\star \simeq
m_\sigma$~\cite{larsen,carter97,gangui98}. L'\'equation d'\'etat analytique
obtenue est compar\'ee \`a la solution num\'erique pr\'ec\'edente sur
la figure~\ref{figanaeos}.
L'approximation est excellente tant que l'on reste dans les domaines
o\`u la corde peut exister physiquement, c'est-\`a-dire pour
$|\varpib|<m_\sigma^2$. D'apr\`es la section~\ref{sectionstabilite},
il est possible, par la connaissance des vitesses de propagations des
perturbations, de discuter la stabilit\'e des boucles de cordes
conductrices dans ce cadre. Sur la figure~\ref{figclctscalar}, on a
repr\'esent\'e l'\'evolution de $\cl^2$ et $\ct^2$ en fonction du
param\`etre d'\'etat. En plus de retrouver les crit\`eres de
stabilit\'e \'enonc\'es pr\'ec\'edemment, il est int\'eressant de
noter que la relation $\ct^2>\cl^2$ est toujours v\'erifi\'ee; seul le
point $\varpib=0$ implique $\ct^2=\cl^2$, comme attendu pour une corde
de Goto-Nambu. Les cordes de Witten parcourue par un courant de
scalaires sont donc g\'en\'eriquement de type \emph{supersonique},
autrement dit, d'apr\`es la section~\ref{sectionstabilite} les boucles
associ\'ees peuvent g\'en\'eriquement \^etre instables.

Le passage \`a la limite d'\'epaisseur nulle des solutions au mod\`ele
microscopique confirme la validit\'e du formalisme covariant dans la
description de cordes parcourues par un courant de scalaires. \`A
l'aide des approximations analytiques (\ref{dualscalarE}) et
(\ref{dualscalarM}), les param\`etres microscopiques peuvent
finalement \^etre reli\'es aux propri\'et\'es macroscopiques des
cordes, et en particulier permettre d'en \'etudier la
stabilit\'e. Bien que de r\'egime g\'en\'eriquement supersonique, les
boucles de corde de ce type peuvent n\'eanmoins \^etre stables sous
certaines conditions, en particulier au voisinage de $\varpib=0$ o\`u
$\ct^2 \simeq \cl^2$. Les cons\'equences cosmologique de l'existence
de ces vortons \'etant majeures, de nombreux travaux d\'epassant le
cadre de cette th\`ese ont \'et\'e r\'ealis\'es afin d'\'etudier plus
finement leurs crit\`eres de stabilit\'e~\cite{brandi96}. Quoiqu'il en
soit, l'existence de ces vortons dans l'univers ne peut \^etre
compatible avec les observations que pour des valeurs des param\`etres
$m_\uh$ et $m_\sigma$ bien en de\c{c}a des \'echelles de grande
unification. Sur la figure~\ref{figbcvortons} sont repr\'esent\'es les
domaines admissibles de ces param\`etres en accord avec les
observations cosmologiques actuelles: nucl\'eosynth\`ese et existence
de l'univers ($\Omega \simeq 1$)~\cite{brandi96}.
\begin{figure}
\begin{center}
\input{bcvortons.pstex_t}
\caption[Contraintes cosmologiques sur l'existence de
vortons.]{Contraintes cosmologiques sur les param\`etres $m_\uh$ et
$m_\sigma$ dans le cas o\`u les boucles de cordes conductrices sont
stables~\cite{brandi96}. L'\'echelle de brisure de sym\'etrie
privil\'egi\'ee est voisine de $10^{9} \GeV$.}
\label{figbcvortons}
\end{center}
\end{figure}

\section{Courants de fermions}

Bien que l'existence de champs scalaires dans l'univers primordial
soit motiv\'ee par les th\'eories de physique des particules (voir
Chap.~\ref{chapitreaudela}), le couplage d'un champ de Higgs avec des
fermions est, \`a nos \'echelles d'\'energie, pr\'edit par le mod\`ele
standard (voir Chap.~\ref{chapitrepp}). E.~Witten a \'egalement
montr\'e qu'un tel couplage entre des fermions et le champ de Higgs
formant la corde conduit \`a l'apparition de courants le long du
vortex~\cite{witten}.

\subsection{Le mod\`ele de Witten fermionique}

Le m\'ecanisme de Witten s'appuie sur un r\'esultat de Jackiw et
Rossi~\cite{jackiwrossi}, et Weinberg~\cite{weinberg81}, montrant que
l'op\'erateur de Dirac dans un vortex admet toujours $n$ \'etats
propres d'\'energie nulle, appel\'es \emph{modes z\'eros}, avec $n$ le
nombre d'enroulement du champ de Higgs\footnote{Indice de Pontryargin
de la corde.}. Si l'on consid\`ere deux fermions $\psi$ et $\chi$ de
chiralit\'e gauche\footnote{Il est d'usage de consid\'erer deux
fermions afin d'assurer l'invariance de jauge de la th\'eorie si
ceux-ci sont charg\'es par rapport \`a une sym\'etrie $U_\uf(1)$
additionnelle. Il suffit pour cela de choisir leurs charges
associ\'ees \'egales et oppos\'ees~\cite{witten}.}, ayant un couplage
au champ de Higgs formant la corde de type Yukawa (voir
Chap.~\ref{chapitrepp}), le lagrangien invariant de jauge le plus
g\'en\'eral dans le secteur fermionique s'\'ecrit
\begin{equation}
\label{lagwittenferm}
\Lc_{\uw \uf} = i \psib \gamma^\mu D_\mu \psi + i \chib \gamma^\mu
D_\mu \chi - \lambda_\uf \left(\Phi \, \varepsilon^{\alpha \beta}
\psi_\alpha \chi_\beta + \uc.\uc \right),
\end{equation}
o\`u les d\'eriv\'ees fermioniques sont covariantes sous la sym\'etrie
$U(1)$ du vortex [voir Eq.~(\ref{laghiggs})],
\begin{equation}
D_\mu = \partial_\mu + iq_\uf B_\mu,
\end{equation}
avec $q_\uf$ d\'esignant la charge du fermion consid\'er\'e,
c'est-\`a-dire $q_\psi$ ou $q_\chi$. L'invariance de jauge du terme de
Yukawa impose de plus la relation $q_\psi + q_\chi + g = 0$. En
supposant la corde align\'ee suivant l'axe $z$, les modes z\'eros de
l'op\'erateur $D_\mu$ assurent l'existence de solutions normalisables
dans le plan transverse du vortex, pour les deux fermions $\psi$ et
$\chi$. En notant $\psi_{_\uT}(x_\perp)$ et $\chi_{_\uT}(x_\perp)$ ces
solutions, il vient
\begin{equation}
\left(\gamma^{x_\perp} D_{x_\perp}\right) \psi_{_\uT} =
\left(\gamma^{x_\perp} D_{x_\perp}\right) \chi_{_\uT} = 0.
\end{equation}
On montre \'egalement que ces solutions sont \'etats propres de
l'op\'erateur $\gamma^0 \gamma^3$, de valeur propre
$-n$~\cite{jackiwrossi,witten}, c'est-\`a-dire pour une corde
d'enroulement $n=1$:
\begin{equation}
\label{etatpropre}
\gamma^0 \gamma^3 \psi_{_\uT} = -\psi_{_\uT}, \quad \gamma^0 \gamma^3
\chi_{_\uT}= -\chi_{_\uT}.
\end{equation}
Consid\`erons maintenant les champs fermioniques $\psi=\alpha(z,t)
\psi_{_\uT}$ et $\chi=\alpha(z,t) \chi_{_\uT}$, dont les \'equations
du mouvement, d\'eduites du lagrangien (\ref{lagwittenferm}), sont
\begin{eqnarray}
\left(\gamma^0 D_t + \gamma^3 D_z \right) \alpha(t,z) \,
\psi_{_\uT} = 0,\\ \nonumber \\
\left(\gamma^0 D_t + \gamma^3 D_z \right) \alpha(t,z) \,
\chi_{_\uT} = 0.
\end{eqnarray}
En remarquant que la seule composante non nulle du champ de jauge de
la corde est $B_\theta$ (voir Sect.~\ref{sectionnielsen}), et \`a
l'aide de l'\'equation (\ref{etatpropre}), il vient
\begin{equation}
\left(\partial_t - \partial_z \right) \alpha = 0,
\end{equation}
soit $\alpha=\alpha(t+z)$. Les fermions $\psi$ et $\chi$ se propagent
donc \`a la vitesse de la lumi\`ere le long de la corde, dans la
direction $-z$, et leur confinement dans le vortex est assur\'e par
l'existence des modes z\'eros normalisables dans le plan
transverse. De mani\`ere similaire, il est possible d'obtenir leur
propagation dans l'autre direction en les couplant \`a $\Phi^\dag$ au
lieu de $\Phi$ dans le lagrangien\footnote{La pr\'esence de ces
fermions est requise lorsqu'ils sont charg\'es sous une sym\'etrie
$U_\uf(1)$ additionnelle afin d'\'eliminer les anomalies de la
th\'eorie quantique associ\'ee. On peut montrer que ceci est
r\'ealis\'e pourvu que la somme des charges $U_\uf(1)$ des particules
se propageant dans une direction soit \'egale et oppos\'ee \`a la
somme des charges des particules se propageant dans l'autre
direction~\cite{witten}.} (\ref{lagwittenferm}). Dans ce cas, ils seront
\'etats propres de $\gamma^0 \gamma^3$ avec, cette fois, une valeur
propre \'egale \`a l'unit\'e donnant une d\'ependance longitudinale en
$\alpha = \alpha(t-z)$.

Plus intuitivement, le couplage des fermions au champ de Higgs formant
la corde dans (\ref{lagwittenferm}) leur donne une masse
proportionnelle \`a $|\Phi|$. Loin de la corde celui-ci prend sa
valeur moyenne dans le vide $\eta$, alors qu'au centre du vortex, il
est nul. L'\'energie de masse des fermions est donc minimale au centre
de la corde, et le profil transverse du champ de Higgs (voir
Fig.~\ref{figback}) peut \^etre vu comme un potentiel attractif. Les
fermions se retrouvant de masse nulle au centre de la corde, ils ne
peuvent que se propager \`a la vitesse de la lumi\`ere et cr\'eer un
courant.

\subsection{Bosonisation}

Comme pour les courants de bosons, il semblerait raisonnable de
d\'ecrire les cordes parcourues par des courants de fermions par le
formalisme macroscopique afin d'en \'etudier leur
stabilit\'e. Cependant, il ne serait pas correct de r\'esoudre les
\'equations de champs de mani\`ere purement classique, comme on l'a
fait dans la section~\ref{sectioncorresp}, du fait du principe de
Pauli. Il rend, en effet, impossible l'identification entre la valeur
classique des champs fermioniques et les \'etats d'excitation dans
lesquels se trouvent leurs particules associ\'ees. On peut n\'eanmoins
construire, sachant que ces courants existent, une action effective
dont la partie fermionique, associ\'ee \`a $\psi$ par exemple, est
donn\'ee par
\begin{equation}
\label{seffpsi}
\Scb_{\uf} = \int \, i \, \psib \gamma^a D_a \psi \,\ud z \, \ud t.
\end{equation}
En posant
\begin{equation}
\label{bosonisation}
\psib \gamma^a \psi  = \frac{1}{\sqrt{\pi}} \varepsilon^{ab} \partial_a
\Sigma,
\end{equation}
on peut montrer que l'action (\ref{seffpsi}) se r\'eduit
\`a~\cite{witten}
\begin{equation}
\label{seffsigma}
\Scb_{\uf} = \int \left[\frac{1}{2} \left(\partial_t
\Sigma\right)^2 - \frac{1}{2} \left(\partial_z \Sigma\right)^2 -
\frac{q_\psi}{\sqrt{\pi}} \Sigma \, \varepsilon^{ab} \partial_a B_b
\right] \ud z \, \ud t.
\end{equation}
Il est ainsi possible, \`a deux dimensions, de d\'ecrire le comportement
des fermions par celui d'un champ scalaire dont les \'equations du
mouvement sont, \`a partir de (\ref{seffsigma}),
\begin{equation}
\frac{\partial^2 \Sigma}{\partial t^2} = -\frac{q_\psi}{\sqrt{\pi}}
\varepsilon^{ab} \partial_a B_b,
\end{equation}
soit, \`a l'aide de (\ref{bosonisation})
\begin{equation}
\frac{\ud J}{\ud t} = \frac{q_\psi^2}{\pi} E,
\end{equation}
o\`u l'on a introduit l'intensit\'e $J$ du courant de fermions le long
de la corde, et $E$ celle du champ de type \'electrique associ\'e \`a
$B^a$, i.e
\begin{equation}
J=-q_\psi \psib \gamma^3 \psi, \quad E=\varepsilon^{ab} \partial_a
B_b.
\end{equation}
Autrement dit, s'il existe un champ de type \'electrique $E$, le
courant de fermions le long de la corde cro\^\i t proportionnellement
\`a $E$ et $t$. Ce courant appara\^\i t supraconducteur car lorsque $E
\rightarrow 0$, sa d\'eriv\'ee s'annule et il persiste au cours du
temps. La limite physique \`a ce courant est, comme dans le cas des
bosons, reli\'ee \`a la masse des fermions $m_\uf$ dans le vide loin
de la corde. Lorsque l'impulsion d'une particule de masse nulle sur la
corde d\'epasse sa masse dans le vide usuel, celle-ci aura tendance
\`a quitter la corde. La saturation du courant est finalement obtenue
pour~\cite{witten}
\begin{equation}
J_{\max} = \frac{q_\psi}{2 \pi} m_\uf.
\end{equation}

L'approche effective \`a deux dimensions semble montrer que les
courants de fermions conduisent \`a des effets comparables \`a ceux
g\'en\'er\'es par des courants de bosons. Le champ scalaire de
bosonisation $\Sigma$ n'introduisant finalement qu'un seul param\`etre
suppl\'ementaire par rapport au cas de Goto-Nambu, il semblerait
raisonnable de d\'ecrire les cordes parcourues par des courants de
fermions par l'\'equation d'\'etat barotropique scalaire
(\ref{eosbaroscalar}).

En fait, comme nous le verrons dans les chapitres suivants, il n'en
est rien. Bien que l'approche effective de Witten justifie la
g\'en\'eration du courant, elle n\'eglige totalement l'influence de la
structure transverse de la corde. Afin de tester le formalisme
covariant, il est plus raisonnable, comme cela a \'et\'e fait pour les
bosons (voir Sect.~\ref{sectioncorresp}), de ne passer \`a la limite
d'\'epaisseur nulle qu'une fois la structure du vortex comprise \`a
quatre dimensions. Pour cela, puisque des fermions sont mis en jeu, il
est n\'ecessaire de quantifier les champs fermioniques dans la corde
pour avoir une description particulaire rigoureuse de leurs
courants. On verra de plus que les modes z\'eros, jusqu'ici vue comme
les principaux \'etats li\'es fermioniques \`a l'origine des courants,
ne sont qu'un cas particulier instable des divers modes de propagation
dans une corde cosmique.

\section{Conclusion}

Le mod\`ele de Witten donne, du point de vue classique, une
justification en terme de physique des particules \`a l'existence de
courants le long des cordes cosmiques. Il semble m\^eme difficile
qu'il n'en soit pas autrement: que ce soit par le plus simple des
couplages entre le champ de Higgs et un champ scalaire
suppl\'ementaire, ou un champ fermionique, les particules qui y sont
associ\'ees peuvent toujours \^etre pi\'eg\'ees le long de la
corde. Les cordes sans structure interne semblerait donc devoir \^etre
produites dans des secteurs o\`u le champ de Higgs les formant devrait
\^etre isol\'e, ce qui semble peu naturel dans le cadre des
th\'eories unifi\'ees.

Concernant les bosons, la validit\'e et la pertinence de l'approche
macroscopique, au travers du formalisme covariant, a pu \^etre
d\'emontr\'ee. La correspondance ayant \'et\'e explicitement
r\'ealis\'ee, il est de plus possible d'obtenir, par le biais de
la cosmologie, des contraintes fortes sur l'\'echelle
d'\'energie de la transition formant la corde, et sur celle associ\'ee
au porteur de charge.

Du fait du principe d'exclusion, l'extraction des grandeurs physiques
macroscopiques dans le cas des cordes parcourues par des fermions
n\'ecessite une description quantique des champs spinoriels dans le
vortex. Il semble cependant difficile de construire une th\'eorie
quantique compl\`ete, \`a quatre dimensions, \`a partir du lagrangien
$\Lc = \Lc_\uh + \Lc_{\uw \uf}$. Mais puisque la notion de courant, du
point de vue macroscopique, ne concerne que les dimensions
longitudinales de la corde, on peut raisonnablement esp\'erer
restreindre la description quantique \`a deux dimensions
seulement. N\'eanmoins, afin de conserver l'information sur la
structure transverse, la description classique des champs doit \^etre
conserv\'ee dans les dimensions transverses. Une telle approche est
explicitement construite dans les chapitres suivants et nous permettra
de calculer l'\'energie par unit\'e de longueur et la tension de ces
cordes. Les r\'esultats trouv\'es diff\`erent essentiellement de ceux
obtenus dans le cas des scalaires, en particulier l'\'equation
d'\'etat barotropique n'est, en g\'en\'eral, plus v\'erifi\'ee.

\part{Courants fermioniques le long des cordes cosmiques}
\label{partieferm}

\chapter{Modes de masse nulle (article)}
\label{chapitrezero}
\minitoc
Dans ce chapitre, la quantification le long de la corde des modes
z\'eros est explicitement r\'ealis\'ee dans le but d'obtenir
l'\'equation d'\'etat correspondante. Contrairement au cas des bosons,
on montre que l'\'equation d'\'etat barotropique n'est plus
v\'erifi\'ee et que la quantification introduit naturellement autant
de param\`etres d'\'etat qu'il y a d'esp\`eces pi\'eg\'ees dans le
vortex. On peut toutefois v\'erifier que, ind\'ependamment du nombre
de param\`etres, une \'equation d'\'etat de type ``trace fix\'ee'' est
satisfaite, i.e. $\ud(U+T)=0$. Il en r\'esulte que le r\'egime
priv\'el\'egi\'e par les fermions est, contrairement aux bosons, de
type \emph{subsonique}, $\ct^2<\cl^2$, assurant de ce fait la
stabilit\'e classique des boucles de corde (voir
Sect.~\ref{sectionstabilite}). Enfin, l'existence m\^eme des modes
z\'eros est \'egalement discut\'ee lorsque l'on tient compte des
effets de r\'etroaction, c'est-\`a-dire de l'influence sur les
modes z\'eros des champs de jauge g\'en\'er\'es par le d\'eplacement
des fermions charg\'es le long de la corde. Ce chapitre est
pr\'esent\'e sous sa forme originale publi\'ee dans la revue
\journal{Physical Review} \numero{D}~\cite{ringeval}.

\newpage

\begin{center}
{\Large \textbf{
Equation of state of cosmic strings with fermionic current-carriers
}}
\end{center}
\vspace{5mm}
\begin{center}
Christophe Ringeval
\end{center}
\vspace{5mm}
\begin{center}
{\footnotesize{
D\'epartement d'Astrophysique Relativiste et de Cosmologie,\\
Observatoire de Paris-Meudon, UMR 8629, CNRS, 92195 Meudon, France,\\
Institut d'Astrophysique de Paris, 98bis boulevard Arago, 75014 Paris,
France.
}}
\end{center}
\vspace{5mm}
\begin{center}
\begin{minipage}[c]{14cm}
{\footnotesize \textbf{
The relevant characteristic features, including energy per unit length
and tension, of a cosmic string carrying massless fermionic currents
in the framework of the Witten model in the neutral limit are derived
through quantization of the spinor fields along the string. The
construction of a Fock space is performed by means of a separation
between longitudinal modes and the so-called transverse zero energy
solutions of the Dirac equation in the vortex. As a result,
quantization leads to a set of naturally defined state parameters
which are the number densities of particles and anti-particles trapped
in the cosmic string. It is seen that the usual one-parameter
formalism for describing the macroscopic dynamics of current-carrying
vortices is not sufficient in the case of fermionic carriers.
}}
\end{minipage}
\end{center}

\section{Introduction}

The mechanism of spontaneous symmetry breaking involved in early
universe phase transitions in some grand unified theories (GUT) might
lead to the formation of topological defects~\cite{kibble76,kibble80}.
Among them, only cosmic strings happen to be compatible with
observational cosmology if they form at the GUT scale. It was shown
however by Witten~\cite{witten} that, depending on the explicit
realization of the symmetry breaking scheme as well as on the various
particle couplings, a current could build along the strings, thereby
effectively turning them into superconducting wires. Such wires were
originally considered in the case the current couples to the
electromagnetic field so they may be responsible for a variety of new
effects, including an explosive scenario for large scale structure
formation for which an enormous energy release was realized in the
form of an expanding shell of non propagating photons in the
surrounding plasma~\cite{otw}.

The cosmology of strings has been the subject of intense work in the
past twenty years or so~\cite{peri95,bouchet88,allen97}, mainly based
on ordinary strings, global or local, aiming at deriving the large
scale structure properties stemming from their distribution as well as
their imprint in the microwave
background~\cite{avelino99,contaldi99,wu02}. It was even
shown~\cite{bouchet02} that the most recent data~\cite{boomerang} might
support a non-negligible contribution of such defects. As such a
result requires ordinary strings, it turns out to be of uttermost
importance to understand the influence of currents in the cosmological
context.

Indeed, it can be argued that currents might drastically modify
the cosmological evolution of a string network: The most clearly
defined consequence of the existence of a current flowing along a
vortex is the breaking of the boost invariance, since the current
itself defines a privileged frame. In other words, the energy per
unit length $U$ and the tension $T$ become two different numbers,
contrary to the ordinary (Goto-Nambu~\cite{goto,nambu}) case. As a
result, string loops become endowed with the capability of
rotation (the latter being meaningless for $U=T$), and the induced
centrifugal force permits equilibrium configurations, called
vortons~\cite{davisRL}. They would very rapidly reach a regime
where they would scale as ordinary non relativistic matter, until
they come to completely dominate the
Universe~\cite{brandi96}.

In the original Witten model~\cite{witten}, currents could form by
means of two different mechanisms. Scalar fields, directly coupled
with the string forming Higgs field, could feel a localized potential
into which they could accumulate in the form of bound states, while
fermions could be trapped along the string, propagating at the speed
of light, as zero energy solutions of the two dimensional Dirac
equation around the vortex. Other models were proposed where fermions
could also propagate in the string core at lower velocities in the
form of massive modes~\cite{davisRL87,hill97,hindmarsh88}, or
(possibly charged) vector fields could also
condense~\cite{everett88,ambjorn88}. All these models have essentially
made clear that the existence of currents in string is much more than
a mere possibility but rather an almost unavoidable fact in realistic
particle physics theories.

For scalar as well as vector carriers, the task of understanding the
microphysics is made simple thanks to their bosonic nature: all the
trapped particles go into the same lowest accessible energy state and
the field can be treated classically~\cite{bps,neutral}. Even the
surrounding electromagnetic~\cite{enon0} and
gravitational~\cite{garrigapeter,peterpuy93, peter94} fields
generated by the current can be treated this way and the back reaction
can be included easily~\cite{nospring}.

Meanwhile, a general formalism was set up by
Carter~\cite{carter89,carter89b,carter94b,carter97} to describe
current-carrying string dynamics. The formalism is based on a single
so-called state parameter, $w$ say, of which the energy per unit
length, the tension and the current itself are functions. Such a
formalism relied heavily on the fact that for a bosonic carrier, the
relevant quantity whose variation along the string leads to a current
is its phase, and the state parameter is essentially identifiable to
this phase gradient. Various equations of state relating the tension
to the energy per unit length were then derived~\cite{carpet1}, based on
numerical results and the existence of a phase frequency
threshold~\cite{neutral}.  It even includes the special case of a
chiral current~\cite{carpet2}, although the latter originates in principle
only for a purely fermionic current. Therefore, it was until now
implicitly assumed that such a formalism would be sufficient to
describe whatever current-carrying string configuration. It will be
shown in this paper that this is in fact not true and an extended
version, including more than one parameter, is needed~\cite{prep}.

The state parameter formalism, apart from being irrelevant
for fermionic current-carrying strings, can only provide a
purely classical description of their dynamics. This is
unfortunate since the most relevant prediction of superconducting
cosmic string models in cosmology is the existence of the vorton
states discussed above. These equilibrium configurations of
rotating loops are not necessarily stable, and in fact, this is
perhaps the most important question to be answered on this topic.
Indeed, any theory leading to the the existence of absolutely stable
vortons predicts a cosmological catastrophe~\cite{brandi96}
and must be ruled out.
One may therefore end up with a very stringent constraint
on particle physics extension of the standard model of electroweak
and strong interactions. To decide clearly on this point requires
to investigate both the classical and the quantum stability of
vortons.

Classical stability has already been established in the case of
bosonic carriers~\cite{martinpeter} for whatever equation of
state~\cite{carter93} on the basis of the one parameter
formalism. Yet it will also have to be addressed in the more
general context that will be discussed below. In the meantime
it was believed that a quantum treatment was necessary in order
to decide on the quantum stability: as one wants to compare the
characteristic lifetime of a vorton with the age of the
Universe, quantum effects can turn out to be relevant; hence the
following work in which the simplest of all fermionic Witten
models is detailed~\cite{witten} that can give rise to both
spacelike as well as timelike charge currents, generalizing
the usual point of view~\cite{davisS}.

Let us sketch the lines along which this work is made.

A two-dimensional quantization of the spinor fields involved
along a string is performed. Owing to anti-particle exitation
states, one can derive the conditions under which the current is
of arbitrary kind. Moreover, an equation of state giving the
energy per unit length and the tension is obtained that involves
four different state parameters which are found to be the number
densities of fermions, although three of them only happen to be
independent.

In Sec.~\ref{modele}, the model is presented and motivated, and
the equations of motion are derived. Then in section
\ref{modezero}, we obtain plane wave solutions along the string
by separating transverse and longitudinal dependencies of spinor
fields in the vortex. The zero mode transverse solutions are then
constraints to be normalizable in order to represent well defined
wave functions. The quantization restricted to massless
longitudinal modes is performed in Sect.~\ref{quantization}.
As a result, the classical conserved currents obtained from Noether
theorem, like energy-momentum tensor and fermionic currents, are
expressed in their quantum form.
All these quantum operators end up being
functions of the fermionic occupation numbers only. In the last
section (Sec.~\ref{etat}), the classical expressions for the
energy per unit length and the tension are derived and discussed
from computation of quantum observable values of the stress
tensor operator in the classical limit. Contrary to the bosonic
current-carrier case where there is only one state
parameter~\cite{neutral}, the classical limit of the model
involves four state parameters in order to fully determine the
energy per unit length and the tension. The cosmological
consequences of this new analysis are briefly discussed in the
concluding section.

\section{Equations of motion}
\label{modele}

We are going to be interested in the purely dynamical effects a
fermionic current flowing along a cosmic string may have. The
model we will be dealing with here is a simplified version of
that proposed by Witten~\cite{witten} which involves two kinds of
fermions, in the neutral limit. This limit, for which the
coupling between fermions and electromagnetic-like external fields
is made to vanish, permits an easy recognition of the dynamical
effects of the existence of an internal structure as in
Ref.~\cite{neutral}.

\subsection{Particle content}

The model we shall consider involves a complex scalar Higgs field,
$\Phi$ say, with conserved charge $qc_{\Phi}$ under a local $U(1)$
symmetry, together with the associated gauge vector field
$B^{\mu}$. In this simple Abelian Higgs model~\cite{higgs},
vortices can form after spontaneous breaking of the $U(1)$
symmetry. The minimal anomaly free model~\cite{witten} with spinor
fields requires two Dirac fermions denoted $\Psi$ and $\Chi$, with
opposite electromagnetic-like charges, getting their masses
from chiral coupling with the Higgs field and its complex
conjugate. They also have conserved gauge charges from invariance
under the broken $U(1)$ symmetry, $q\cpsir$, $q\cpsil$,
$q\cchir$, and $q\cchil$, for the right- and left-handed parts of
the two fermions respectively. The Lagrangian of the model
therefore reads
\begin{equation}
{\mathcal{L}}={\mathcal{L}}_\uh+{\mathcal{L}}_\ug+{\mathcal{L}}_\psi+
{\mathcal{L}}_\chi
\end{equation}
with ${\mathcal{L}}_\uh$, ${\mathcal{L}}_\ug$ and
${\mathcal{L}}_\psi$, ${\mathcal{L}}_\chi$, respectively the
Lagrangian in the Higgs, gauge, and fermionic sectors. In terms
of the underlying fields, they are
\begin{eqnarray}
{\mathcal{L}}_\uh & = & \frac{1}{2} (D_{\mu}\Phi)^{\dag}(D^{\mu}\Phi)
- V(\Phi),
\\
{\mathcal{L}}_\ug & = & -\frac{1}{4} H_{\mu \nu} H^{\mu \nu},
\\
\label{psilagrangian}
{\mathcal{L}}_\psi & = & \frac{i}{2} \left[\overline{\Psi}
\gamma^{\mu} D_{\mu} \Psi - (\overline{D_{\mu}\Psi}) \gamma^{\mu}
\Psi \right] -g \overline{\Psi} \frac{1+\gamma_5}{2} \Psi \Phi
-g\overline{\Psi} \frac{1-\gamma_5}{2} \Psi \Phi^\ast,
\\
\label{chilagrangian}
{\mathcal{L}}_\chi & = & \frac{i}{2} \left[\overline{{\Chi}}
\gamma^{\mu} D_{\mu} {\Chi} - (\overline{D_{\mu}{\Chi}})
\gamma^{\mu} {\Chi} \right] -g \overline{{\Chi}}
\frac{1+\gamma_5}{2} {\Chi} \Phi^{\ast} - g \overline{{\Chi}}
\frac{1-\gamma_5}{2} {\Chi} \Phi,
\end{eqnarray}
where we have used the notation
\begin{eqnarray}
D_{\mu}\Phi & = & (\nabla_{\mu} + i q c_{\phi}B_\mu) \Phi,
\\
D_{\mu}\Psi & = & (\nabla_{\mu} + i q \frac{\cpsir+\cpsil}{2}B_\mu
+i q \frac{\cpsir-\cpsil}{2} \gamma_5 B_\mu) \Psi,
\\
D_{\mu}{\Chi} & = & (\nabla_{\mu} + i q \frac{\cchir+
\cchil}{2}B_\mu +i q \frac{\cchir-\cchil}{2} \gamma_5 B_\mu)
{\Chi},
\\
H_{\mu \nu} & = & \nabla_\mu B_\nu - \nabla_\nu B_\mu,
\\
V(\Phi) & = & \frac{\lambda}{8} (|\Phi|^2 - \eta^2)^2.
\end{eqnarray}
The equivalence with the Witten model~\cite{witten} appears through a
separation into left- and right-handed spinors. Let us define
$\Psi_\uR$ and $\Psi_\uL$, respectively the right- and left-handed parts
of the Dirac spinor field $\Psi$ (and the same for ${\Chi}$),
eigenvectors of $\gamma_5$,
\begin{eqnarray}
\Psi_\uR  =  \frac{1+\gamma_5}{2} \Psi,
& \quad \hbox{\textrm{and}} \quad &
\Psi_\uL  =  \frac{1-\gamma_5}{2} \Psi.
\end{eqnarray}
The Lagrangian for the spinor field $\Psi$ now reads
\begin{eqnarray}
{\mathcal{L}}_\psi & = & \frac{i}{2} \left[\overline{\Psi}_\uR
\gamma^{\mu} D_{\mu} \Psi_\uR - (\overline{D_{\mu}\Psi}_\uR) \gamma^{\mu}
\Psi_\uR \right] + \frac{i}{2} \left[\overline{\Psi}_\uL \gamma^{\mu} D_{\mu}
\Psi_\uL - (\overline{D_{\mu}\Psi}_\uL) \gamma^{\mu} \Psi_\uL \right]
\nonumber \\
 & & 
-g \overline{\Psi}_\uL \Psi_\uR \Phi -g\overline{\Psi}_\uR \Psi_\uL \Phi^\ast,
\end{eqnarray}
with the associated covariant derivatives
\begin{equation}
D_{\mu}\Psi_{R(L)} = (\nabla_{\mu} + i qc_{\psi_{R(L)}}B_\mu) \Psi_{R(L)}.
\end{equation}
It is clear with the Lagrangian expressed in this way that the
invariance of the action under $U(1)$ transformations requires
\begin{equation}
\cpsil-\cpsir=c_\phi=\cchir-\cchil.
\end{equation}

\subsection{Equations of motion}

As we wish to deal with a cosmic string, the Higgs and gauge
fields can be set as a vortex-like Nielsen--Olesen solution and
they can be written in cylindrical coordinates as
follows~\cite{NO}:
\begin{equation}
\begin{array}{ccc}
\Phi  =  \varphi(r) \ue^{i \alpha(\theta)},
& \quad &
B_\mu  =  B(r) \delta_{\mu \theta}.
\end{array}
\end{equation}
In order for the Higgs field to be well defined by rotation around
the string, its phase has to be proportional to the orthoradial
coordinate, $\alpha(\theta)=n \theta$, where the integer $n$ is the
winding number. The new fields $\varphi$ and $\alpha$ are now real
scalar fields and are solutions of the equations of motion
\begin{eqnarray}
\label{higgsmvt}
\nabla_\mu \nabla^\mu \varphi & = & \varphi Q_\mu Q^\mu -
\frac{\ud V(\varphi)}{\ud \varphi}
- \overline{\Psi} \frac {\partial m_\psi}{\partial \varphi} \Psi
- \overline{{\Chi}} \frac {\partial m_\chi}{\partial \varphi}
{\Chi},
\\
\nabla_\mu \left(\varphi^2 Q^\mu \right) & = &
-\overline{\Psi} \frac {\partial m_\psi}{\partial \alpha} \Psi
- \overline{{\Chi}} \frac {\partial m_\chi}{\partial \alpha}
{\Chi},
\end{eqnarray}
where
\begin{eqnarray}
Q_\mu & = & \nabla_\mu \alpha + q c_\phi B_\mu,
\\
m_\psi & = & g \varphi \cos{\alpha} + i g \varphi \gamma_5 \sin{\alpha},
\\
m_\chi & = & g \varphi \cos{\alpha} - i g \varphi \gamma_5 \sin{\alpha}.
\end{eqnarray}
In the same way, the equations of motion for the gauge and spinor
fields are
\begin{eqnarray}
\label{gaugemvt}
\nabla_\mu H^{\mu \nu} & = & j^\nu_\psi + j^\nu_\chi - q c_\phi \varphi^2
Q^\nu,
\\
\label{psimvt}
i \gamma^\mu \nabla_\mu \Psi & = & \frac{\partial j^\mu_\psi}{\partial
\overline{\Psi}} B_\mu + m_\psi \Psi,
\\
\label{psibarmvt}
i \left(\nabla_\mu \overline{\Psi}\right) \gamma^\mu & = &
- \frac{\partial j^\mu_\psi}{\partial \Psi} B_\mu - \overline{\Psi} m_\psi,
\\
\label{chimvt}
i \gamma^\mu \nabla_\mu {\Chi} & = & \frac{\partial j^\mu_\chi}
{\partial \overline{{\Chi}}} B_\mu + m_\chi {\Chi},
\\
\label{chibarmvt}
i \left(\nabla_\mu \overline{{\Chi}} \right) \gamma^\mu & = &
- \frac{\partial j^\mu_\chi}{\partial {\Chi}} B_\mu
-\overline{{\Chi}} m_\chi.
\end{eqnarray}
The fermionic currents $j^\mu_{\psi(\chi)}$ have axial and
vectorial components due to the different coupling between left-
and right-handed spinors to the gauge field. The two kinds of
current are required to respect gauge invariance of the
Lagrangian. In terms of spinor fields, they read
\begin{equation}
j^\mu_{\psi(\chi)}=j^\mu_{\psi_\uV(\chi_\uV)}+j^\mu_{\psi_\uA(\chi_\uA)},
\end{equation}
with
\begin{equation}
\label{currents}
\begin{array}{ccccccclll}
j^\mu_{\psi_\uV} & = & q \displaystyle{\frac{\cpsir+ \cpsil}{2}}
\overline{\Psi} \gamma^\mu \Psi, & \quad & j^\mu_{\chi_\uV} & = & q
\displaystyle{\frac{\cchir+ \cchil}{2}} \overline{{\Chi}} \gamma^\mu {\Chi},
\\ \\
j^\mu_{\psi_\uA} & = & q \displaystyle{\frac{\cpsir- \cpsil}{2}}
\overline{\Psi} \gamma^\mu \gamma_5 \Psi, & \quad &
j^\mu_{\chi_\uA} & = & q \displaystyle{\frac{\cchir- \cchil}{2}}
\overline{{\Chi}} \gamma^\mu \gamma_5 {\Chi}.
\end{array}
\end{equation}

\section{Transverse solutions as zero modes}
\label{modezero}

\subsection{Plane wave solutions}

The study of the fermionic fields trapped along the string can be
performed by separating the longitudinal and transverse solutions
of the equations of motion. The plane wave solutions are
therefore expressed in the generic form
\begin{equation}
\label{planeansatz}
\begin{array}{lll}
\Psi_\uup^{(\pm)}  =  \ue^{\pm i(\omega t-kz)}
\left(\begin{array}{l}
\xi_1(r) \ue^{-im_1 \theta} \\
\xi_2(r) \ue^{-im_2 \theta} \\
\xi_3(r) \ue^{-im_3 \theta} \\
\xi_4(r) \ue^{-im_4 \theta}
\end{array} \right),
& \quad &
{\Chi}_\uup^{(\pm)}  =  \ue^{\pm i(\omega t-kz)}
\left(\begin{array}{l}
\zeta_1(r) \ue^{-il_1 \theta} \\
\zeta_2(r) \ue^{-il_2 \theta} \\
\zeta_3(r) \ue^{-il_3 \theta} \\
\zeta_4(r) \ue^{-il_4 \theta}
\end{array} \right).
\end{array}
\end{equation}
Inside the vortex, the numbers $m_i$ and $l_i$ have to be
integers in order to produce well defined spinors by rotation
around the string. In the following, the Dirac spinors will be
expressed in the chiral representation, and the metric is assumed
to have the signature $(+,-,-,-)$. Plugging the expression
(\ref{planeansatz}) into the equations of motion (\ref{psimvt})
and (\ref{chimvt}) yields the differential system
\begin{equation}
\label{syst}
\left\{
\begin{array}{lll}
\vspace{4pt}
\displaystyle
\left[\frac{\ud\xi_1}{\ud r} + \frac{1}{r}\left(-q \cpsir B + m_1\right)
\xi_1 \right] \ue^{-i(m_1-1)\theta} - i g \varphi \ue^{-i(m_4+n)\theta}
\xi_4 & = & \mp i (k+\omega) \xi_2 \ue^{-im_2\theta}, \\
\vspace{4pt}
\displaystyle
\left[\frac{\ud\xi_2}{\ud r} + \frac{1}{r}\left(q \cpsir B - m_2\right)
\xi_2 \right] \ue^{-i(m_2+1)\theta} - i g \varphi \ue^{-i(m_3+n)\theta}
\xi_3 & = & \pm i (k-\omega) \xi_1 \ue^{-im_1\theta}, \\
\vspace{4pt}
\displaystyle
\left[\frac{\ud\xi_3}{\ud r} + \frac{1}{r}\left(-q \cpsil B + m_3\right)
\xi_3 \right] \ue^{-i(m_3-1)\theta} + i g \varphi \ue^{-i(m_2-n)\theta}
\xi_2 & = & \mp i (k-\omega) \xi_4 \ue^{-im_4\theta}, \\
\displaystyle
\left[\frac{\ud\xi_4}{\ud r} + \frac{1}{r}\left(q \cpsil B - m_4\right)
\xi_4 \right] \ue^{-i(m_4+1)\theta} + i g \varphi \ue^{-i(m_1-n)\theta}
\xi_1 & = & \pm i (k+\omega) \xi_3 \ue^{-im_3\theta}.
\end{array}
\right.
\end{equation}
Similar equations are obtained for the field ${\Chi}$ with the
following transformations, $\xi \rightarrow \zeta$, $c_{\psi_{R(L)}}
\rightarrow c_{\chi_{R(L)}}$, and $n \rightarrow -n$, because of its
coupling to the anti-vortex instead of the vortex. Note that if,
instead of the vectorial phases ansatz (\ref{planeansatz}), we had chosen
a matricial phases ansatz in the form
\begin{equation}
\Psi_\uup^{(\pm)}  =  \ue^{\pm i(\omega t-kz)}
\left(\begin{array}{llll}
\xi_{11}(r) \ue^{-im_{11} \theta}+ \ldots + \xi_{14}(r) \ue^{-im_{14}
\theta} \\ 
\xi_{21}(r) \ue^{-im_{21} \theta}+ \ldots + \xi_{24}(r) \ue^{-im_{24}
\theta} \\ 
\xi_{31}(r) \ue^{-im_{31} \theta}+ \ldots + \xi_{34}(r) \ue^{-im_{34}
\theta} \\ 
\xi_{41}(r) \ue^{-im_{41} \theta}+ \ldots + \xi_{44}(r) \ue^{-im_{44}
\theta}
\end{array} \right),
\end{equation}
we would have found, from the requirement of having at most four independent
phases in the equations of motion, that the matrix $m_{ij}$ has to verify
for all $i$, $m_{ij}=m_{ik}$, for all $(j,k)$. Consequently, the vectorial
ansatz (\ref{planeansatz}) is the most general for solutions with separated
variables.

\subsection{Transverse solutions}

{}From the differential system (\ref{syst}), it is obvious that
the four phases $m_i$ cannot be independent parameters if the
spinors fields are not identically zero. It is also impossible to
find three independent phase parameters since each equation
involves precisely three different angular dependencies. The only
allowed angular separation requires two degrees of freedom in
$\theta$, and the only relevant relation for trapped modes in the
string reads from Eq.~(\ref{syst}),
\begin{eqnarray}
\label{angseparation}
m_1-1=m_4+n,
\end{eqnarray}
and,
\begin{eqnarray}
m_2+1=m_3+n.
\end{eqnarray}

\subsubsection{Zero modes}

Introducing the two integer parameters $p=m_1$ and
$q=m_3$, and using Eq.~(\ref{angseparation}), for $p\neq q+n$,
the system (\ref{syst}) reduces to the set
\begin{equation}
\label{zeromodes}
\left\{
\begin{array}{l}
\vspace{4pt}
\displaystyle
\frac{\ud \xi_2}{\ud r} + \frac{1}{r}\left(q \cpsir B - (q+n-1)\right)
\xi_2 - i g \varphi \xi_3  = 0, \\
\vspace{4pt}
\displaystyle
\frac{\ud \xi_3}{\ud r} + \frac{1}{r}\left(-q \cpsil B + q\right)
\xi_3 + i g \varphi \xi_2 = 0, \\
\vspace{4pt}
\displaystyle
(k+\omega) \xi_2 = 0, \\
\displaystyle
(k+\omega) \xi_3 = 0,
\end{array}
\right.
\end{equation}
\begin{equation}
\label{antizeromodes}
\left\{
\begin{array}{l}
\vspace{4pt}
\displaystyle
\frac{\ud \xi_1}{\ud r} + \frac{1}{r}\left(-q \cpsir B + p\right)
\xi_1 - i g \varphi \xi_4 = 0, \\
\vspace{4pt}
\displaystyle
\frac{\ud \xi_4}{\ud r} + \frac{1}{r}\left(q \cpsil B - (p-n-1)\right)
\xi_4 + i g \varphi \xi_1 = 0, \\
\vspace{4pt}
\displaystyle
(k-\omega) \xi_1 = 0, \\
\displaystyle
(k-\omega) \xi_4 = 0.
\end{array}
\right.
\end{equation}
There are two kinds of solutions which propagate along the two directions
of the string at the speed of light and which were originally found by Witten
\cite{witten}:
Either $k=\omega$ and $\xi_2 =\xi_3=0$, or $k=-\omega$ and $\xi_1=\xi_4=0$.
These zero modes must also be normalizable in the transverse plane of the
string in order to be acceptable as wave functions.

\subsubsection{Index theorem}

In the two cases $k=\omega$ and $k=-\omega$, we will call the
corresponding zero modes, $X_\psi(r,\theta)$ and
$Y_\psi(r,\theta)$, the solutions of the systems
(\ref{zeromodes}) and (\ref{antizeromodes}), respectively, i.e.,
\begin{equation}
\begin{array}{lll}
X_\psi = \left(
\begin{array}{l}
0 \\
\xi_2(r) \ue^{-i (q+n-1) \theta} \\
\xi_3(r) \ue^{-i q \theta} \\
0
\end{array}
\right),
& \quad &
Y_\psi = \left(
\begin{array}{l}
\xi_1(r) \ue^{-i p \theta} \\
0 \\
0 \\
\xi_4(r) \ue^{-i (p-n-1) \theta}
\end{array}
\right).
\end{array}
\end{equation}
These have to be normalizable in the sense that $\int{|X_\psi|^2
r\, \ud r \,\ud\theta}$ and $\int{|Y_\psi|^2 r\,\ud
r\,\ud\theta}$ must be finite. Thanks to the regularity of the
vortex background, the divergences in these integrals can only
arise close to the string core or asymptotically far away from it.
As a result, it is sufficient to study asymptotic behaviors of
the solutions to decide on their normalizability~\cite{jackiwrossi}. \\
Let us focus on the zero mode $X_\psi$ solution of
Eq.~(\ref{zeromodes}), keeping in mind that $Y_\psi$ can be dealt
with in the same way. The asymptotic behaviors at infinity are
easily found as solutions of the limit at infinity of the
differential system (\ref{zeromodes}). Note that the
identification between the equivalent solutions and the solutions
of the equivalent system is only allowed by the absence of
singular point for the system at infinity, and it will not be so
near the string since $r=0$ is a singular point, so that the
Cauchy theorem does no longer apply. From Eq.~(\ref{zeromodes}), the
eigensolutions of the equivalent system at infinity are in the form
$\exp{(\pm g \eta r)}$, and thus, there is only one normalizable
solution at infinity.

Near the string, i.e., where $r \rightarrow 0$, the system is no
longer well defined, the origin being a singular point.
Approximate solutions can however be found by looking at the
leading term of a power-law expansion of both system and
functions, as originally suggested by Jackiw and Rossi
\cite{jackiwrossi}. Because the Cauchy theorem does no longer
apply, many singular solutions might be found at the origin, and
among them, the two generic ones which match with the two
exponentials at infinity. Near the origin, the Higgs and gauge
fields are known to behave like~\cite{NO}
\begin{equation}
\label{equizero}
\begin{array}{lll}
\varphi(r) \sim \varphi_\zero r^{|n|}, & \quad & B(r) \propto r^2,
\end{array}
\end{equation}
so that the leading contribution near the string of the zero modes
can be found as
\begin{eqnarray}
\xi_2(r) & \sim & a_2 r^{\alpha_2}, \\
\xi_3(r) & \sim & a_3 r^{\alpha_3},
\end{eqnarray}
with $a_i$ and $\alpha_i$ real parameters to be determined.
The values of the exponents $\alpha_i$ are therefore given by the
leading order terms in system (\ref{zeromodes}), and one obtains
three solutions, the first one of which being singular,
\begin{equation}
\begin{array}{lllll}
\left(
\begin{array}{l}
\xi_2 \\
\xi_3
\end{array}
\right)_s
=
\left(
\begin{array}{l}
a_2 r^{q+n-1} \\
a_3 r^{-q}
\end{array}
\right),
& \quad & \textrm{provided} & \quad &
\displaystyle
-\frac{|n|+n}{2} < q <1+ \frac{|n|-n}{2}.
\end{array}
\end{equation}
The two other solutions are the generic ones which have to match with
the solutions at infinity
\begin{equation}
\begin{array}{lllllllll}
\left(
\begin{array}{l}
\xi_2 \\
\xi_3
\end{array}
\right)_{g_1}
& = &
\left(
\begin{array}{l}
a_2(a_3) r^{-q+|n|+1} \\
a_3 r^{-q}
\end{array}
\right),
& \quad  &
\left(
\begin{array}{l}
\xi_2 \\
\xi_3
\end{array}
\right)_{g_2}
& = &
\left(
\begin{array}{l}
a_2(a_3) r^{q+n-1} \\
a_3 r^{q+|n|+n}
\end{array}
\right),
\end{array}
\end{equation}
where the relationships between the parameters $a_i$ have not been written,
since they are clearly obtained from Eq.~(\ref{zeromodes}).
Normalizability near the origin requires that the integrals $\int{|\xi_2|^2\,
r\, \ud r}$ and $\int{|\xi_3|^2 \,r \,\ud r}$ converge, and this yields
\begin{equation}
\label{criterionzeromodes}
-n < q < 1.
\end{equation}
Analogous considerations for the system (\ref{antizeromodes}) show the
convergence criterion in this case to be
\begin{equation}
\label{criterionantizeromodes}
n < p < 1.
\end{equation}
The number of sets of parameters $p$ and $q$ satisfying the
previous inequalities is precisely the number of well defined
zero modes, respectively $Y_\psi$ and $X_\psi$, which are also
normalizable. In order to match with the single well behaved
solution at infinity, the two independent solutions near the
string have to be integrable. Therefore, for a vortex solution
with a positive winding number $n$, there are only $n$
normalizable zero modes, which are the $X_\psi$ ones. Similarly
in the case of an anti-vortex with negative winding number
$-|n|$, one finds  also $|n|$ zero modes $Y_\psi$. This is the
index theorem found by Jackiw and Rossi~\cite{jackiwrossi}.
Recall that the model involves two kinds of fermions, and all the
previous considerations apply as well for the field ${\Chi}$ with
the simple transformation $n \rightarrow -n$. Therefore, the
normalizable zero modes are swapped compared to those of the
field $\Psi$.

Finally, for a vortex with positive winding number
$n$, there are always $n$ massless plane wave solutions for both
spinor fields, which read
\begin{equation}
\label{psisolutions}
\Psi_\uup^{(\pm)} = \ue^{\pm i k(t+z)} \left(
\begin{array}{llll}
0 \\
\xi_2(r) \ue^{-i(q+n-1) \theta} \\
\xi_3(r) \ue^{-iq \theta}\\
0
\end{array}
\right)
=  \ue^{\pm i k(t+z)} X_\psi(r,\theta),
\end{equation}
\begin{equation}
\label{chisolutions}
{\Chi}_\uup^{(\pm)} = \ue^{\pm ik(t-z)} \left(
\begin{array}{llll}
\zeta_1(r) \ue^{-ip \theta} \\
0 \\
0 \\
\zeta_4(r) \ue^{-i(p+n-1) \theta}
\end{array}
\right)
=  \ue^{\pm ik(t-z)} Y_\chi(r,\theta),
\end{equation}
with now $q=m_3$ and $p=l_1$ which satisfy
\begin{equation}
\begin{array}{ccccc}
-n < q < 1, & \quad & \textrm{and} & \quad & -n < p < 1.
\end{array}
\end{equation}
Note that they are eigenvectors of the $\gamma^0 \gamma^3$ operator,
and they basically verify
\begin{equation}
\label{currentszeromodes}
\begin{array}{lclllcrllcl}
\overline{X_\psi} \gamma^0 X_\psi & = & |X_\psi|^2
&,  &
\overline{X_\psi}\gamma_5 \gamma^0 X_\psi & = & -|\xi_2|^2+|\xi_3|^2
&,  &
\overline{X_\psi}\gamma^3 X_\psi & = & -|X_\psi|^2,
\\
\\
\overline{X_\psi} \gamma^{1(2)} X_\psi & = & 0
&,  &
\overline{X_\psi} \gamma_5 \gamma^3 X_\psi & = & |\xi_2|^2-|\xi_3|^2
&,  &
\overline{X_\psi} \gamma_5 \gamma^{1(2)} X_\psi & = & 0,
\\
\\
\overline{Y_\psi} \gamma^0 Y_\psi & = & |Y_\psi|^2
&, &
\overline{Y_\psi} \gamma_5 \gamma^0 Y_\psi & = & -|\xi_1|^2+|\xi_4|^2
&, &
\overline{Y_\psi} \gamma^3 Y_\psi & = & |Y_\psi|^2,
\\
\\
\overline{Y_\psi} \gamma^{1(2)} Y_\psi & = & 0
&,  &
\overline{Y_\psi} \gamma_5 \gamma^3 Y_\psi & = & -|\xi_1|^2+|\xi_4|^2
&,  &
\overline{Y_\psi} \gamma_5 \gamma^{1(2)} Y_\psi & = & 0.
\end{array}
\end{equation}
The ${\Chi}$ zero modes $X_\chi$ and $Y_\chi$ verify the same
relationships with $\xi_i$ replaced by $\zeta_i$.

\subsubsection{Massive modes}

The case $m_1=m_3+n$ allows four-dimensional solutions of the
system (\ref{syst}). In particular, these solutions do no longer
require $\omega=\pm k$ and therefore represent massive modes. As
before, the interesting behaviors of these modes are found by
studying the solutions of the equivalent system asymptotically
and by looking for the leading term of a power-law expansion of
both system and solution near the string core.

At infinity, the system is well defined and there are two twice
degenerate eigensolutions $\exp{(\pm \Omega r)}$ out of which two
are normalizable, with
\begin{equation}
\Omega = \sqrt{g^2 \eta^2-(\omega^2-k^2)}.
\end{equation}
Near the origin, at $r=0$, the system is singular, and because of
its four dimensions there are much more singular solutions than
the previous two dimensional case, and among them the four
generic ones which match with the four ones at infinity. The
leading term in asymptotic expansion can be written in a standard
way
\begin{equation}
\xi_i(r) \sim a_i r^{\alpha_i}.
\end{equation}
Plugging these expressions in the system (\ref{syst}) with
Eq.~(\ref{equizero}), and keeping only leading terms at $r=0$ gives,
after some algebra, the four generic solutions
\begin{equation}
\label{onephasemodes}
\begin{array}{lllllll}
\left(
\begin{array}{l}
\xi_1 \\
\xi_2 \\
\xi_3 \\
\xi_4
\end{array}
\right)_{g_1}
 & = &
\left(
\begin{array}{l}
a_1 r^{-m} \\
a_2(a_1) r^{-m+1} \\
a_3(a_1) r^{-m+|n|+2} \\
a_4(a_1) r^{-m+|n|+1}
\end{array}
\right),
& \quad &
\left(
\begin{array}{l}
\xi_1 \\
\xi_2 \\
\xi_3 \\
\xi_4
\end{array}
\right)_{g_2}
& = &
\left(
\begin{array}{l}
a_1 r^{m+|n|-n} \\
a_2(a_1) r^{m+|n|-n+1} \\
a_3(a_1) r^{m-n} \\
a_4(a_1) r^{m-n-1}
\end{array}
\right),
\\ \\
\left(
\begin{array}{l}
\xi_1 \\
\xi_2 \\
\xi_3 \\
\xi_4
\end{array}
\right)_{g_3}
 & = &
\left(
\begin{array}{l}
a_1 r^{m} \\
a_2(a_1) r^{m-1} \\
a_3(a_1) r^{m+|n|} \\
a_4(a_1) r^{m+|n|+1}
\end{array}
\right),
& \quad &
\left(
\begin{array}{l}
\xi_1 \\
\xi_2 \\
\xi_3 \\
\xi_4
\end{array}
\right)_{g_4}
& = &
\left(
\begin{array}{l}
a_1 r^{-m+|n|+n+2} \\
a_2(a_1) r^{-m+|n|+n+1} \\
a_3(a_1) r^{-m+n} \\
a_4(a_1) r^{-m+n+1}
\end{array}
\right),
\end{array}
\end{equation}
with $m=m_1$, and where the relationships between the coefficients
$a_i$ have not been written as they are essentially given by a
linear system in $a_i$ given by Eq.~(\ref{syst}).
The solutions (\ref{onephasemodes}) will be normalizable near the
string if, for all $i$, $\int{|\xi_i|^2\, r\, \ud r}$ is finite.
Moreover there will be, at least, always one massive bound state
if there are at least three normalizable eigensolutions to match
with the well-defined ones at infinity. This is only allowed if
the parameter $m$ verifies simultaneously three of the following
conditions:
\begin{equation}
\sup{(0,n)} < m < \inf{(1,1+n)}.
\end{equation}
Because $m$ is necessary an integer, this condition cannot be
achieved. This criterion, originally derived and used by Jackiw
and Rossi in order to enumerate the number of zero modes in a
vortex-fermion system~\cite{jackiwrossi}, is only sufficient and
thus, normalizable massive bound states may exist, but are model
dependent since it is necessary that a particular combination of
the two normalizable eigenmodes near the string core match
exactly with a particular combination of the two well-defined
ones at infinity.

Such massive bound states depend therefore of the particular
values of the model parameters. Recently, it was shown
numerically~\cite{davisS} that the Abelian Higgs model with one
Weyl fermion admits always at least two massive bound states, as
a result, the present toy model also may have such states.
However, in order to simplify the quantization, we will only
consider the generic zero modes, and consequently, the following
results will be relevant for cosmic string only when the
occupancy of the massive bound states can be neglected compared
to the occupancy of the zero mode states. Such physical
situations are likely to occur far below the energy scale where
the string was formed, since the massive states are generally
expected to decay much more rapidly than the massless ones
\cite{davisS}.

The generic massless normalizable transverse solutions of the
fermionic equations of motion in the string with winding number
$n$ are the $n$ zero modes. For the spinor field $\Psi$ coupled
with the vortex, we find that the particles and the
anti-particles can only propagate at the speed of light in one
direction, ``$-z$'' say direction along the string, whereas the
spinor field ${\Chi}$ propagates in the opposite, ``$+z$''
direction. The existence of such plane waves allows us to
quantize the spinor fields along the string. The zero modes
themselves will therefore be transverse wave functions giving the
probability density for finding a trapped mode at a chosen
distance from the string core.

\section{Fock space along the string}
\label{quantization}

The spinor fields can be expanded on the basis of the plane wave
solutions computed above. A canonical quantization can then be
performed along the $z$-axis which provides analytical
expressions for these fields in two dimensions once the
transverse degrees of freedom have been integrated over. It is
therefore possible to compute the current operators as well as
their observable values given by their averages in a particular
Fock state. In the following we shall take a vortex with a unit
winding number $n=1$ and the subscript of the zero modes $X_\psi$
and $Y_\chi$ will be forgotten since there is not possible
confusion.

\subsection{Canonical quantization}

We shall first be looking for a physical expansion of the spinor
fields in plane waves, in the sense that creation and
annihilation operators are well defined. The Hamiltonian is then
calculable as a function of these and will be required to be
positive to yield a reasonable theory.

\subsubsection{Quantum fields}

As shown above, the spinor fields $\Psi$ and ${\Chi}$ propagate in
only one direction, therefore in expressions (\ref{psisolutions})
and (\ref{chisolutions}) the momentum $k$ can be chosen positive
definite. Let us, once again, focus on the spinor field $\Psi$.
The natural way to expand it in plane waves of positive and
negative energies is
\begin{equation}
\label{psiexpansion}
\Psi=\int_{0}^{\infty}{\frac{\ud k}{2\pi2k}\left[b^{\dag}(-k)
\ue^{ik(t+z)} +
\underline{b}(-k) \ue^{-ik(t+z)} \right] X}.
\end{equation}
The Fourier transform of $\Psi$ on positive and negative energies
has been written with similar notation $b^\dag$ and
$\underline{b}$ unlike in the free spinor case. Indeed, note
that, in the string, the zero modes are the same for both
positive and negative energy waves, so that the only way to
distinguish particles from anti-particles is in the sign of the
energy. The integration measure $\ud k/(2\pi2k)$ is the usual
Lorentz invariant measure in two dimensions. Note that $k$ is
chosen always positive in order to represent physical energy and
momentum actually carried by the field along the string; hence the
negative sign in $b^\dag(-k)$ and $\underline{b}(-k)$, which is a
reminder that the spinor field $\Psi$ propagates in the ``$-z$''
direction. In the same way, the field ${\Chi}$ is expanded as
\begin{equation}
\label{chiexpansion}
{\Chi}=\int_{0}^{\infty}{\frac{\ud k}{2\pi2k}\left[d^{\dag}(k)
 \ue^{ik(t-z)} + \underline{d}(k) \ue^{-ik(t-z)} \right] Y}.
\end{equation}
The Fourier transform will be written with the normalization
convention
\begin{equation}
\label{delta}
\int{\ud z\, \ue^{i(k-k^\prime)z}} = 2\pi \delta(k-k^\prime).
\end{equation}

\subsubsection{Creation and annihilation operators}

The Fourier coefficients can be expressed as functions of the
spinor field $\Psi$ or $\Psi^\dag$. With equation
(\ref{psiexpansion}) and (\ref{delta}), let us compute the
following integral
\begin{eqnarray}
\int{r \,\ud r\, \ud \theta \,\ud z\, \ue^{ik(t+z)} X^\dag \Psi} & = &
\int \frac{\ud k^\prime}{2k^\prime}
\left[b^\dag(-k^\prime)\delta(k+k^\prime) + \underline{b}(-k^\prime)
\delta(k-k^\prime) \right] \|X\|^2 \nonumber \\ & & =
\frac{\|X\|^2}{2k} \underline{b}(-k),
\end{eqnarray}
where we have defined
\begin{equation}
\|X\|^2 \equiv \int{r\, \ud r \,\ud\theta \, |X|^2}.
\end{equation}
Note that the separation between $b$ and $b^\dag$ only arises
from the chirality of the spinor field because the integration is
performed only over positive values of the momentum $k$; this is
why the $\delta(k+k^\prime)$ term vanishes. In the following we
will assume that the zero modes are normalized to unity,
$\|X\|^2=1$, and $\|Y\|^2=1$. Playing with similar integrals gives
us the other expansion coefficients
\begin{equation}
\begin{array}{lllllll}
\label{psioperators}
\underline{b}(-k) & = & 2k \displaystyle \int{r\,\ud r\,\ud \theta
\,\ud z\, \ue^{ik(t+z)} X^\dag \Psi}, & & \underline{b}^\dag(-k) & = &
2k \displaystyle \int{r\,\ud r\,\ud \theta \,\ud z\, \ue^{-ik(t+z)}
\Psi^\dag X}, \\ \\ b^\dag(-k) & = & 2k \displaystyle \int{r\,\ud
r\,\ud \theta \,\ud z\, \ue^{-ik(t+z)} X^\dag \Psi}, & & b(-k) & = &
2k \displaystyle \int{r\,\ud r\,\ud \theta \,\ud z\, \ue^{ik(t+z)}
\Psi^\dag X},
\end{array}
\end{equation}
and the corresponding relations for the spinor field ${\Chi}$
\begin{equation}
\begin{array}{lllllll}
\label{chioperators}
d^\dag(k) & = & 2k \displaystyle \int{r\,\ud r\,\ud \theta \,\ud z\,
\ue^{-ik(t-z)} Y^\dag {\Chi}},
&  &
d(k) & = & 2k \displaystyle \int{r\,\ud r\,\ud \theta \,\ud z\,
\ue^{ik(t-z)} {\Chi}^\dag Y},
\\ \\
\underline{d}(k) & = & 2k \displaystyle \int{r\,\ud r\,\ud \theta
\,\ud z\, \ue^{ik(t-z)} Y^\dag {\Chi}},
&  &
\underline{d}^\dag(k) & = & 2k \displaystyle \int{r\,\ud r\,\ud \theta
\,\ud z\, \ue^{-ik(t-z)} {\Chi}^\dag Y}.
\end{array}
\end{equation}
From these, one gets the necessary relations to define creation
and annihilation operators
\begin{equation}
\begin{array}{lll}
\underline{b}^\dag(-k) = \left[\underline{b}(-k)\right]^\dag,
& &
b^\dag(-k) = \left[b(-k)\right]^\dag,
\\ \\
d^\dag(k) = \left[d(k)\right]^\dag,
& &
\underline{d}^\dag(k) = \left[\underline{d}(k)\right]^\dag.
\end{array}
\end{equation}

\subsubsection{Commutation relations}

The canonical quantization is performed by the transformation of
Poisson brackets into anticommutators. Here, we want to quantize
the spinor fields only along the string, and therefore let us
postulate the anticommutation rules \emph{at equal times} for the
quantum fields
\begin{eqnarray}
\label{psiquantization}
\left\{\Psi_\alpha(t,\vec x), \Psi^{\dag \beta}
(t,\vec x^\prime) \right\} & = & \delta(z-z^\prime) X_{\alpha}(r,
\theta) X^{\dag \beta}(r^\prime, \theta^\prime),
\\
\label{chiquantization}
\left\{{\Chi}_\alpha(t,\vec x), {\Chi}^{\dag \beta}
(t,\vec x^\prime) \right\} & = & \delta(z-z^\prime) Y_{\alpha}(r,
\theta) Y^{\dag \beta}(r^\prime, \theta^\prime),
\end{eqnarray}
with $\alpha$ and $\beta$ the spinorial indices, and all the other
anticommutators vanishing. With Eq.~(\ref{psioperators}) and
these anticommutation rules, it follows immediately, for creation
and annihilation operators, that
\begin{equation}
\label{creatoranticom}
\begin{array}{ccccclll}
\left\{b(-k), b^\dag(-k^\prime) \right\} & = & \left\{\underline{b}(-k),
\underline{b}^\dag(-k^\prime)
\right\} & = & 2\pi2k2k^\prime \delta(k-k^\prime),
\\ \\
\left\{\underline{d}(k), \underline{d}^\dag(k^\prime)\right\} & = &
\left\{d(k), d^\dag(k^\prime) \right\} & = & 2\pi2k2k^\prime
\delta(k-k^\prime),
\end{array}
\end{equation}
with all other anticommutators vanishing. From the expressions
(\ref{psiexpansion}) and (\ref{chiexpansion}) and with the
anticommutation rules (\ref{creatoranticom}), it is possible to derive the
anticommutator between two quantum field operators \emph{at any time}. For
instance, the anticommutator between $\Psi$ and $\Psi^\dag$ reads
\begin{eqnarray}
\left\{\Psi_\alpha({\mathbf{x}}), \Psi^{\dag \beta} ({\mathbf{x}}^\prime) \right\}
& = &\int{\frac{\ud k\, \ud k^\prime}{(2\pi)^2 2k 2k^\prime}
\left\{b^\dag(-k)\ue^{ik(t+z)} + \underline{b}(-k)\ue^{-ik(t+z)} , b(-k^\prime)
\ue^{-ik^\prime (t^\prime+z^\prime)} \right. } + \nonumber \\
& + & \left. \underline{b}^\dag(-k^\prime) \ue^{ik^\prime (t^\prime+z^\prime)}
\right\} X_\alpha(r,\theta) X^{\dag \beta}(r',\theta').
\end{eqnarray}
Thanks to the delta function coming from the anticommutators
between the $b$ and $b^\dag$, this expression reduces to
\begin{eqnarray}
\left\{\Psi_\alpha({\mathbf{x}}), \Psi^{\dag \beta}
 ({\mathbf{x}}^\prime) \right\} & = & i \partial_t \Delta(t-t'+ z-z')
 X_\alpha(r,\theta) X^{\dag \beta}(r',\theta'),
\end{eqnarray}
with $\Delta({\mathbf{x}})$ the well known Pauli-Jordan function which
vanishes for spacelike separation, so the spinor fields indeed respect
micro-causality along the string.

\subsubsection{Fock states}

The Fock space can be built by application of the creation
operators on the vacuum state $|\emptyset\rangle$ which by
definition has to satisfy
\begin{equation}
\label{vide}
\underline{b}(-k)|\emptyset\rangle = b(-k)|\emptyset\rangle =
d(k)|\emptyset \rangle = \underline{d}(k)|\emptyset\rangle = 0,
\end{equation}
and is normalized to unity, i.e., $\langle \emptyset|\emptyset
\rangle = 1$. Each Fock state represents one possible combination
of the fields exitation levels. Let $|{\mathcal{P}} \rangle$ be a
Fock state representing $N_\psi$ particles labeled by $i$ and
$\overline{N}_\psi$ anti-particles labeled by $j$, of kind
$\Psi$, with respective momenta $k_i$ and $l_j$, and, $N_\chi$
particles labeled by $p$ and $\overline{N}_\chi$ anti-particles
labeled by $q$, of kind ${\Chi}$, with respective momenta $r_p$
and $s_q$. By construction the state is
\begin{eqnarray}
\label{fockstate}
|{\mathcal{P}} \rangle & = & b^\dag(-k_1) \ldots b^\dag(-k_i) \ldots
b^\dag(-k_{N_\psi}) \underline{b}^\dag(-l_1) \ldots
\underline{b}^\dag(-l_j) \ldots
\underline{b}^\dag(-l_{\overline{N}_\psi}) d^\dag(r_1) \ldots
\nonumber
\\
& \ldots & d^\dag(r_p) \ldots d^\dag(r_{N_\chi})
\underline{d}^\dag(s_1) \ldots \underline{d}^\dag(s_q) \ldots
\underline{d}^\dag(s_{\overline{N}_\chi}) |\emptyset\rangle.
\end{eqnarray}
Normalizing such a state is done thanks to the anticommutators
(\ref{creatoranticom}). For instance, for a one particle $\Psi$ state
with $k$ momentum, using Eq.~(\ref{vide}), the orthonormalization of
the corresponding states reads
\begin{equation}
\langle k'|k \rangle = 2\pi 2k2k'\delta(k-k').
\end{equation}
Obviously similar relations apply to all the other particle and
anti-particle states. Keeping in mind that the observable values of
quantum operators are their eigenvalues in a given quantum state, let
us compute the expectation value of the occupation number operator
involved in many quantum operators, as will be shown. For $\Psi$
particles it is
\begin{equation}
\label{twoaverage}
\frac{\langle {\mathcal{P}}|b^\dag(-k) b(-k')|{\mathcal{P}}\rangle}{\langle
{\mathcal{P}}|{\mathcal{P}} \rangle} = \frac{2\pi}{\delta(0)}2k2k' \sum_i
\delta(k-k_i)\delta(k'-k_i)
\end{equation}
and analogous relations for the other particle and anti-particle
states. By definition of the Fourier transform (\ref{delta}), the
infinite factor $\delta(0)$ is simply an artifact related to the
length of the string $L$ by
\begin{equation}
L = 2\pi \delta(0),
\end{equation}
in the limit where this string length $L\to \infty$. Note that
using periodic boundary conditions on $L$ allows to consider
large loops with negligible radius of curvature.

\subsection{Fermionic energy momentum tensor}

The simplest way to derive an energy momentum tensor already
symmetrized is basically from the variation of the action with
respect to the metric. Moreover, the Hamiltonian density of the
fermion $\Psi$,
\begin{equation}
{\mathcal{H}}_\psi=\partial_t \Pi_\psi \Psi + \partial_t \overline{\Psi}
\overline{\Pi}_{\psi} - {\mathcal{L}}_\psi,
\end{equation}
with $\Pi$ the conjugate field $\Pi=i \overline{\Psi}\gamma^0$, is also equal
to the $T_\psi^{tt}$ component of the stress tensor. In our case the metric is
cylindrical and we assume a flat Minkowski space-time background, thus in the
fermionic sector the stress tensor reads
\begin{eqnarray}
\label{tmunu}
T_\psi^{\mu \nu} = 2\frac{\delta{\mathcal{L}}_\psi}{\delta g_{\mu \nu}} -
g^{\mu \nu} {\mathcal{L}}_\psi
\quad & \textrm{and} & \quad
T_\chi^{\mu \nu} = 2\frac{\delta{\mathcal{L}}_\chi}{\delta g_{\mu \nu}} -
g^{\mu \nu} {\mathcal{L}}_\chi.
\end{eqnarray}
Once again, let us focus on $\Psi$. Plugging Eq.~(\ref{psilagrangian})
into Eq.~(\ref{tmunu}) gives
\begin{equation}
\label{psitensor}
T_\psi^{\mu \nu} = \frac{i}{2} \overline{\Psi} \gamma^{(\mu}
\partial^{\nu)}
\Psi - \frac{i}{2} \left(\partial^{(\mu}\overline{\Psi}\right)
\gamma^{\nu)} \Psi - B^{(\mu} j^{\nu)}_{_\psi}.
\end{equation}

\subsubsection{Symmetrized Hamiltonian}
\label{toyvac}
{}From Noether theorem, the Hamiltonian $P^t$ is also given by the
conserved charge associated with the time component of the
energy momentum tensor
\begin{equation}
T^{tt}_\psi=i \overline{\Psi}\gamma^0 \partial_t \Psi - i\left(\partial_t
\overline{\Psi} \right) \gamma^0 \Psi.
\end{equation}
Thanks to the expression of the quantum fields in equations
(\ref{psiexpansion}) and (\ref{chiexpansion}), and using the
properties of the zero modes from Eq.~(\ref{currentszeromodes}), the
quantum operator associated to $T_\psi^{tt}$ reads
\begin{eqnarray}
\label{psitensortt}
T_\psi^{tt} & = & \frac{1}{2} \int{\frac{\ud k\,\ud k'}{(2\pi)^22k}}
\left\{\left[-b(-k)b^\dag(-k') + \underline{b}^\dag(-k')\underline{b}(-k)
\right] \ue^{i(k'-k)(t+z)} \right.
\nonumber \\
& + &
\left.
\left[\underline{b}^\dag(-k)\underline{b}(-k') -b(-k') b^\dag(-k)
\right] \ue^{-i(k'-k)(t+z)} \right.
\nonumber \\
& + &
\left.
\left[b(-k)\underline{b}(-k')-b(-k')\underline{b}(-k)\right]
\ue^{-i(k'+k)(t+z)} \right.
\nonumber \\
& + &
\left.
\left[-\underline{b}^\dag(-k) b^\dag(-k')+ \underline{b}^\dag(-k')b^\dag(-k)
\right]\ue^{i(k'+k)(t+z)} \right\} |X|^2.
\end{eqnarray}
The Hamiltonian is given by spatial integration of the Hamiltonian
density, or similarly from Eq.~(\ref{psitensortt}),
\begin{equation}
\label{hamiltonian}
P^t_\psi= \int{\frac{\ud k}{2\pi2k} \left[-b(-k) b^\dag(-k)
+\underline{b}^\dag(-k) \underline{b}(-k) \right]}.
\end{equation}
Note, once again, that all the terms in the form $b^\dag b^\dag$
or $b{}b$ vanish as a consequence of the chiral nature of the
spinor fields which only allows $k>0$. The expectation value of this
Hamiltonian in the vacuum is not at all positive, but a simple
renormalization shift is sufficient to produce a reasonable
Hamiltonian provided one uses fermionic creation and annihilation
operators with the corresponding definition for the normal ordered
product (antisymmetric form). The normal ordered Hamiltonian is
therefore well behaved and reads
\begin{eqnarray}
\label{normorder}
:P^t_\psi: & = & \int{\frac{\ud k}{2\pi2k} \left[b^\dag(-k) b(-k)
+ \underline{b}^\dag(-k) \underline{b}(-k) \right]}.
\end{eqnarray}
However, note that such a normal ordering prescription overlooks the
differences between the vacuum energy of empty space, and that in the
presence of the string, for the massless fermions. Formally, from
Eq.~(\ref{psitensortt}), the normal ordered Hamiltonian is also obtained
by adding the operator $\int{\ud \vec x \, \uV_\psi}$ to the infinite
Hamiltonian in Eq.~(\ref{hamiltonian}), with
\begin{eqnarray}
\uV_\psi=:T^{tt}_\psi: \, - \, T^{tt}_\psi & = &
\frac{1}{2} \int{\frac{\ud k\,\ud k'}{(2\pi)^22k}}
\left[\left\{ b(-k),b^\dag(-k') \right\} \ue^{i(k'-k)(t+z)} \right.
\nonumber \\
& + &
\left.
\left\{b(-k'), b^\dag(-k) \right\} \ue^{-i(k'-k)(t+z)} \right]|X|^{2}.
\end{eqnarray}
Owing to the anticommutation rules in Eq.~(\ref{psiquantization}), this
expression reduces to
\begin{equation}
\label{inftystruct}
\uV_\psi=\int{\frac{\ud k}{\pi}\, k \,|X|^2}.
\end{equation}
This infinite renormalizing term of the vacuum associated with
the zero modes on the string comes from the contribution of the
infinite renormalization of the usual empty space together with a
finite term representing the difference between the two kinds of
vacua. The previous expression (\ref{inftystruct}) emphasizes the
structure of the divergence, and it can be conjectured that the
finite part is simply obtained by a cut-off $\Lambda$ in momentum
values. The finite vacuum contribution to the stress tensor can
therefore be represented from Eq.~(\ref{inftystruct}), up to the
sign, by the energy density
\begin{eqnarray}
\label{vacuum}
\frac{\Lambda^2}{2\pi}|X|^2=\frac{2\pi}{L_\uv^2}|X|^2,
& \quad \textrm{with} \quad &
L_\uv=\frac{2\pi}{\Lambda}.
\end{eqnarray}
The precise determination of the value of $L_\uv$ is outside the
scope of this simple model. It is well known however, that the
vacuum effects generally involve energies smaller than the first
quantum energy level and consequently it seems reasonable to
consider that $L_\uv > L$.
For a large loop, $L_\uv$ can be roughly estimated using the
discretization of the momentum values. With $k_n=2\pi n/L$,
$\uV_\psi$ therefore reads
\begin{equation}
\uV_\psi = \frac{4 \pi}{L^2}|X|^2\sum_{n=0}^{\infty}n.
\end{equation}
Assuming that the vacuum associated with the fermionic zero modes
on the string matches the Minkowski one associated with massless
fermionic modes, in the infinite string limit~\cite{fulling,kay},
$L_\uv$ can be obtained by substracting the two respective
values of $\uV_\psi$, once the transverse coordinates have been
integrated over.
The infinite sum over $n$ can be regularized by a cut-off factor
$\ue^{-\varepsilon k_n}$, letting $\varepsilon$ equal to zero at the
end of the calculation~\cite{birrell}. The regularized expression of
$\uV_\psi$ finally reads
\begin{equation}
\uV_\psi=\frac{4 \pi}{L^2} |X^2| \sum_{n=0}^{\infty} n
\ue^{-\varepsilon\frac{2\pi}{L}n},
\end{equation}
and expanded asymptotically around $\varepsilon=0$, it yields, once
the transverse coordinates have been integrated,
\begin{equation}
\label{devvac}
\int{r \, \ud r \, \ud \theta \, \uV_\psi} \sim \frac{1}{\pi
\varepsilon^2}- \frac{\pi}{3L^2}.
\end{equation}
As a result, the infinite renormalizing term relevant with the
usual vacuum associated with two dimensional chiral waves
is just $1/\pi \varepsilon^2$ whereas the relevant vacuum
associated with the zero modes along the string is exactly
renormalized by $\int{r \, \ud r \, \ud \theta \, \uV_\psi}$ given in
Eq.~(\ref{devvac}), and therefore involves the finite term $\pi/3L^2$
with a minus sign. As a result, the string zero mode vacuum appears
as an exited state in the Minkowski vacuum associated with two
dimensional chiral modes, with positive energy density
$\pi/3L^2$. In this case, the short distance cut-off therefore
reads
\begin{equation}
L_\uv^2=6L^2.
\end{equation}
It is therefore necessary to add the $2\pi L/L_\uv^2$ term to the
canonical normal ordered prescription, in Eq.~(\ref{normorder}), in
order to obtained an Hamiltonian with zero energy for the
Minkowski vacuum associated with two dimensional chiral modes.

It is important to note that, in order to be consistent, the
previous calculations can only involve the vacuum associated with
the corresponding quantized modes, i.e. in this case the zero modes.
Therefore this does not take care of the massive modes. In fact, even
for zero occupancy of the massive bound states, the physical vacuum
along the string must involve a similar dependence in the vacuum
associated with the massive modes. The influence of the two
dimensional massive vacuum on the equation of state will be more
discussed in Sec.~\ref{enertens}.

\subsubsection{Stress tensor}

All the other terms of the energy momentum tensor can be derived
from Eq.~(\ref{tmunu}). From the relationships verified by
the zero modes currents in Eq.~(\ref{currentszeromodes}), all the
transverse kinetic terms vanish. Moreover, the only non-vanishing
components of the axial and vectorial currents (\ref{currents})
are $j^t_\psi$ and $j^z_\psi$. Finally, the energy momentum
tensor reads
\begin{equation}
T^{\mu \nu}_\psi = \left(
\begin{array}{ccccllll}
T_\psi^{tt} & 0 & T^{t \theta}_\psi & T^{tz}_\psi \\
0 & 0 & 0 & 0 \\
T^{t \theta}_\psi & 0 & 0 & T^{z \theta}_\psi \\
T^{tz}_\psi & 0 & T^{z \theta}_\psi & T^{zz}_\psi
\end{array}
\right).
\end{equation}
Because the zero modes are eigenvectors of $\gamma^0 \gamma^3$, the
operators $\gamma^0 \partial^0$ and $\gamma^3 \partial^3$ are formally
identical for each spinor field, and therefore the diagonal terms of the
stress tensor are identical
\begin{equation}
\label{eigenstates}
T^{zz}_\psi=T^{tt}_\psi.
\end{equation}
{}From Eq.~(\ref{tmunu}), we find the transverse terms to be
$T^{t \theta}_\psi = - B^\theta j^t_\psi$, and $T^{z \theta}_\psi
= -B^\theta j^z_\psi$. In a Cartesian basis, these components
yield the transverse terms $T^{tx}_\psi$, $T^{ty}_\psi$,
$T^{zx}_\psi$, and $T^{zy}_\psi$, which vanish once the
transverse degrees of freedom have been integrated over. The only
non-vanishing non-diagonal part of the stress tensor comes from
the lightlike nature of each fermion current and reads
\begin{equation}
T^{tz}_\psi=-2i \overline{\Psi} \gamma^0 \partial_t \Psi,
\end{equation}
while the counterpart of the field $\Chi$ gets a minus sign because
of its propagation in the ``$+z$'' direction,
\begin{equation}
T^{tz}_\chi=2i \overline{\Chi} \gamma^0 \partial_t \Chi.
\end{equation}
In the above expressions, the back reaction is neglected, but the
fermionic current, $j^{\mu}=j^{\mu}_\psi+j^{\mu}_\chi$,
generates,  from Eq.~(\ref{gaugemvt}), new gauge field components,
$B_t$ and $B_z$, which have to be small, compared to the
orthoradial component, $B_\theta$, in order to avoid significant
change in the vortex background. However, up to first order, they
provide corrections to the energy momentum tensor whose effects
on energy per unit length and tension will be detailed, in the
classical limit, in Sec.~\ref{etat}. The back reaction
correction to the two-dimensional stress tensor then reads
\begin{equation}
\label{brcorr}
T^{\alpha \beta}_{\ub.\ur.}= \left(
\begin{array}{ccccllll}
\displaystyle{-2B_t j^t + \frac{1}{2} (\partial_r B_t)^2
+ \frac{1}{2} (\partial_r B_z)^2}
& \displaystyle{ B_z j^t-B_t j^z - (\partial_r B_t)(\partial_r B_z)} \\
\displaystyle{B_z j^t-B_t j^z - (\partial_r B_t)(\partial_r B_z)} &
\displaystyle{2B_z j^z+ \frac{1}{2} (\partial_r B_t)^2
+ \frac{1}{2} (\partial_r B_z)^2}
\end{array}
\right).
\end{equation}

\subsection{Axial-vector and vectorial currents}

From the expression of the spinor fields, the current operators
are immediately found in the Fock space. Moreover, it is
interesting to compute the electromagnetic-like fermionic
current, its scalar analogue being involved in the equation of
state for a cosmic string with bosonic current carriers
\cite{neutral}. From an additional \emph{global} $U(1)$
invariance of the Lagrangian, the electromagnetic-like current
takes similar form as the vectorial one coupled to the string
gauge field. It physically represents the neutral limit of the
full electromagnetic coupling.

\subsubsection{Vectorial currents}

Let $J^\mu$ be the electromagnetic current in the neutral limit.
{}From Noether theorem with global $U(1)$ invariance, this is
\begin{equation}
J^{\mu} = J^\mu_\psi+J^\mu_\chi=- \overline{\Psi} \gamma^{\mu} \Psi +
\overline{{\Chi}} \gamma^\mu {\Chi},
\end{equation}
The fermions $\Psi$ and $\Chi$ carry opposite
electromagnetic-like charges in order to cancel anomalies
\cite{witten}. Using Eq.~(\ref{psiexpansion}), their
components read
\begin{equation}
\begin{array}{cccccll}
\label{empsicurrent}
:J^t_\psi: & = & -:J^z_\psi: & = &  :{\mathcal{J}}_\psi: |X|^2,
\\
\label{emchicurrent}
:J^t_\chi: & = & :J^z_\chi: & = &  - :{\mathcal{J}}_\chi: |Y|^2,
\end{array}
\end{equation}
with the quantum operators defined as
\begin{eqnarray}
:{\mathcal{J}}_\psi: & = & \int{\frac{\ud k\,\ud k'}{(2\pi)^2 2k2k'}}
\left[-b^\dag(-k')b(-k) \ue^{i(k'-k)(t+z)} + \underline{b}^\dag(-k)
\underline{b}(-k') \ue^{-i(k'-k)(t+z)}+ \right.
\nonumber \\
& + & \left. b(-k)\underline{b}(-k') \ue^{-i(k+k')(t+z)} + \underline{b}^\dag(-k)
b^\dag(-k') \ue^{i(k+k')(t+z)} \right] |X|^2,
\\
:{\mathcal{J}}_\chi: & = & \int{\frac{\ud k\,\ud k'}{(2\pi)^2 2k2k'}}
\left[-d^\dag(k')d(k) \ue^{i(k'-k)(t-z)} + \underline{d}^\dag(k)
\underline{d}(k') \ue^{-i(k'-k)(t-z)} + \right.
\nonumber \\
& + & \left. d(k)\underline{d}(k') \ue^{-i(k+k')(t-z)} + \underline{d}^\dag(-k)
d^\dag(k') \ue^{i(k+k')(t-z)} \right] |Y|^2.
\end{eqnarray}
The conserved charges carried by these currents are basically derived from
spatial integration of the corresponding current densities, and are
\begin{eqnarray}
:Q_\psi: & = &  \int{\frac{\ud k}{2\pi (2k)^2} \left[-b^\dag(-k)b(-k)
+ \underline{b}^\dag(-k) \underline{b}(-k) \right]},
\\
:Q_\chi: & = & - \int{\frac{\ud k}{2\pi (2k)^2} \left[-d^\dag(k)d(k)
+ \underline{d}^\dag(k) \underline{d}(k) \right]}.
\end{eqnarray}
As expected at the quantum level, the anti-particles carry
charges that are opposite to that of the particles for both fields.
This is again because the opposite chirality of the fields makes
the $b{}b$ and $b^\dag b^\dag$ terms vanishing. The vectorial
gauge currents are easily obtained from the neutral limit ones
replacing the electromagnetic charge by the $U(1)$ gauge one,
namely
\begin{equation}
\label{vectorialcurrents}
\begin{array}{lllll}
\vspace{4pt} \displaystyle :j^t_{\psi_\uV}: & = & :-j^z_{\psi_\uV}: &
= & q \displaystyle{\frac{\cpsir+ \cpsil}{2}} :{\mathcal{J}}_\psi:
|X|^2,
\\
\displaystyle :j^t_{\chi_\uV}: & = & :j^z_{\chi_\uV}: & = & q
\displaystyle{\frac{\cchir+ \cchil}{2}} :{\mathcal{J}}_\chi:
|Y|^2.
\end{array}
\end{equation}

\subsubsection{Axial-vector currents}

In the same way, the axial currents are derived from their
classical expressions as function of the quantum fields, from
Eq.~(\ref{currentszeromodes}),
\begin{equation}
\label{axialcurrents}
\begin{array}{lllll}
\vspace{4pt} \displaystyle :j^t_{\psi_\uA}: & = & - :j^z_{\psi_\uA}:
& = & q \displaystyle{\frac{\cpsir- \cpsil}{2}}
:{\mathcal{J}}_\psi: \left(|\xi_2|^2-|\xi_3|^2 \right),
\\
\displaystyle :j^t_{\chi_\uA}: & = & :j^z_{\chi_\uA}: & = & q
\displaystyle{\frac{\cchir- \cchil}{2}} :{\mathcal{J}}_\chi:
\left(|\xi_1|^2-|\xi_4|^2 \right).
\end{array}
\end{equation}

Thanks to the normalizable zero modes in the transverse plane of the
string, it is possible to construct a Fock space along the string.
The chirality of each spinor field being well defined, anti-particle
states appear at quantum level as another mode propagating at the
speed of light in the same direction than the particle mode, but
carrying opposite gauge and electromagnetic-like charges.
The observable values of the quantum operators previously defined are
given by their expectation value in the corresponding Fock state. In
particular, the energy per unit length, the tension, and the current
per unit length, can now be derived from the previous expressions.

\section{Equation of state}
\label{etat}

In the case of a scalar condensate in a cosmic string, it was shown by
Peter~\cite{neutral} that the classical formalism of
Carter~\cite{carter89,carter89b,carter94b,carter97} with one single
state parameter could apply and an equation of state for the bosonic
cosmic string could be derived in the form
\begin{equation}
\label{scalareos}
U-T=\sqrt{|w|} {\mathcal{C}},
\end{equation}
where $U$ and $T$ are respectively the energy per unit length and
the tension of the string, ${\mathcal{C}}$ is the current density
along the string and $w$ a state parameter which appear as the
conjugate parameter of ${\mathcal{C}}$ by a Legendre
transformation. An analogous relation can be sought for our string
with fermionic current-carriers from the classical energy per
unit length, tension, and current density values.

Consider the fermionic cosmic string in the quantum state
(\ref{fockstate}). The energy per unit length and tension in this
state are basically given by the eigenvalues associated with
timelike and spacelike eigenvectors, respectively, of the expectation
value in $|{\mathcal{P}}\rangle$ of the energy momentum tensor,
once the transverse coordinates have been integrated over. The
stress tensor is obviously the total energy momentum tensor
\begin{equation}
T^{\mu \nu} = T^{\mu \nu}_\ug + T^{\mu \nu}_\uh +
:T^{\mu \nu}_\psi: + :T^{\mu \nu}_\chi:,
\end{equation}
where $ T^{\mu \nu}_\ug$ and $T^{\mu \nu}_\uh$ are the gauge and Higgs
contributions which describe the Goto-Nambu string and which
integrated over the transverse plane provides only two opposite
non-vanishing terms
\begin{equation}
\int{r \,\ud r\,\ud \theta \left(T^{tt}_\ug + T^{tt}_\uh \right)} =
-\int{r\,\ud r \,\ud \theta \left(T^{zz}_\ug + T^{zz}_\uh \right)}
\equiv M^2,
\end{equation}
thus defining the unit of mass $M$.

\subsection{Expectation values in the Fock state $|{\mathcal{P}}\rangle$}

\subsubsection{Two-dimensional energy momentum tensor}

Now, let us define the energy momentum tensor operator in two
dimensions, $\overline{T}^{\alpha \beta}$ say, once the transverse
coordinates have been integrated over, and where we have suppressed
the corresponding vanishing terms. Therefore, with $\alpha$ and
$\beta$ equal to $t$ or $z$, and neglecting for the moment the back
reaction, it reads
\begin{equation}
\overline{T}^{\alpha \beta}= \left(
\begin{array}{ccll}
\displaystyle
M^2 + \int{r \,\ud r\,\ud \theta \left(:T^{tt}_\psi: + :T^{tt}_\chi:
\right)} &
\displaystyle
\int{r \,\ud r\,\ud \theta \left(:T^{tz}_\psi: +
:T^{tz}_\chi: \right)} \\
\displaystyle
\int{r \,\ud r\,\ud \theta \left(:T^{tz}_\psi: + :T^{tz}_\chi:
\right)} &
\displaystyle
-M^2 + \int{r \,\ud r\,\ud \theta \left(:T^{tt}_\psi:
+ :T^{tt}_\chi: \right)}
\end{array}
\right).
\end{equation}
The expectation value in the Fock state $|{\mathcal{P}}\rangle$ of
$\overline{T}^{\alpha \beta}$, is immediately obtained from
equations (\ref{twoaverage}) and (\ref{psitensortt})
\begin{equation}
\label{twostress}
\langle \overline{T}^{\alpha \beta} \rangle_{{\mathcal{P}}} =
\frac{\langle {\mathcal{P}}| \overline{T}^{\alpha \beta}|{\mathcal{P}}
\rangle}{ \langle {\mathcal{P}}|{\mathcal{P}}\rangle} =
\left(
\begin{array}{ccll}
M^2 + E_\chip + E_\psip & E_\chip-E_\psip \\
E_\chip-E_\psip & -M^2+ E_\chip+E_\psip
\end{array}
\right),
\end{equation}
with the notations
\begin{eqnarray}
\label{enerchi}
E_\chip & = &\int{r \,\ud r\,\ud \theta \langle :T^{tt}_\chi:
\rangle_{{\mathcal{P}}}}
= \frac{2}{L} \left(\sum_{i=1}^{N_\psi} k_i + \sum_{j=1}^{
\overline{N}_\psi} l_j \right)_{\mathcal{P}},
\\
\label{enerpsi}
E_\psip & = &\int{r \,\ud r\,\ud \theta\langle :T^{tt}_\psi:
\rangle_{{\mathcal{P}}}}
=\frac{2}{L} \left(\sum_{p=1}^{N_\chi} r_p
+ \sum_{q=1}^{\overline{N}_\chi} s_q \right)_{\mathcal{P}}.
\end{eqnarray}
The summations are just over the momentum values taken in each
particle and anti-particle exitation states, so that $E_\chip$ and
$E_\psip$ depend on the quantum state $|{\mathcal{P}}\rangle$.
The $2 \pi \delta(0)$ factor has been replaced by the physical
length $L$ in order to deal only with finite quantities.
Moreover, from the integral expression of the normal ordering
prescription in Eq.~(\ref{vacuum}), the quantum zero mode vacuum
effects appear simply as a shift of the previous expressions,
and the corrected values of the parameters $E_\psip$ and
$E_\chip$ therefore read
\begin{eqnarray}
\label{fullener}
E_{\psip_\uv}=E_{\psip}+\frac{2\pi}{L^2_\uv},
& \quad \textrm{and} \quad &
E_{\chip_\uv}=E_{\chip}+\frac{2\pi}{L^2_\uv}.
\end{eqnarray}
Note that, from equations (\ref{enerchi}), (\ref{enerpsi}) and
(\ref{fullener}), if $L_\uv> L$, the vacuum contribution can be
neglected for non-zero exitation states.

\subsubsection{Current densities}

In the same way, the expectation value of current operators in the
Fock state $|{\mathcal{P}}\rangle$ are basically derived from the
average of the operators $:{\mathcal{J}}_\psi:$ and
$:{\mathcal{J}}_\chi:$
\begin{equation}
\begin{array}{ccc}
\langle :{\mathcal{J}}_\psi: \rangle_{\mathcal{P}} = -N_\psi +
\overline{N}_\psi,
& \quad &
\langle :{\mathcal{J}}_\chi: \rangle_{\mathcal{P}} = -N_\chi +
\overline{N}_\chi.
\end{array}
\end{equation}
Therefore, the electromagnetic-like current per unit length in the neutral
limit becomes, after transverse integration,
\begin{eqnarray}
\langle \overline{J}^t \rangle_{\mathcal{P}}& = & \frac{1}{L}
\left[\left(-N_\psi + \overline{N}_\psi \right) +
\left(N_\chi - \overline{N}_\chi \right) \right] ,
\\
\langle \overline{J}^z \rangle_{\mathcal{P}}& = & \frac{1}{L}
\left[\left(N_\psi - \overline{N}_\psi \right) +
\left(N_\chi - \overline{N}_\chi \right) \right] .
\end{eqnarray}
Averaging the vectorial and axial gauge currents in equations
(\ref{vectorialcurrents}) and (\ref{axialcurrents}) allows a
derivation of the total fermionic gauge current density
\begin{eqnarray}
\label{gaugecurrentt}
\langle j^t \rangle_{\mathcal{P}} & = & \frac{1}{L}
\left[ f_\psi(r)\left(-N_\psi + \overline{N}_\psi \right)
+ f_\chi(r) \left(-N_\chi + \overline{N}_\chi \right) \right],
\\
\label{gaugecurrentz}
\langle j^z \rangle_{\mathcal{P}} & = & \frac{1}{L}
\left[f_\psi(r) \left(N_\psi - \overline{N}_\psi \right)
+ f_\chi(r) \left(-N_\chi + \overline{N}_\chi \right) \right],
\end{eqnarray}
with the radial functions
\begin{eqnarray}
\label{fchi}
f_\chi(r) & = & q \cchir |\xi_1|^2 + q \cchil |\xi_4|^2,
\\\label{fpsi}
f_\psi(r) & = & q \cpsir |\xi_2|^2 + q \cpsil |\xi_3|^2.
\end{eqnarray}
Moreover, note that these currents can be lightlike, spacelike or
timelike according to the number of each particle species trapped in
the string. For instance, the square magnitude of the
electromagnetic-like line density current reads
\begin{equation}
\overline{{\mathcal{C}}}_{\mathcal{P}}^2 = \langle \overline{J}^t
\rangle_{\mathcal{P}}^2 - \langle \overline{J}^z \rangle_{\mathcal{P}}^2
= \left(\frac{2}{L}\right)^2 \left(N_\psi N_\chi +
\overline{N}_\psi \overline{N}_\chi - \overline{N}_\psi N_\chi
- \overline{N}_\chi N_\psi \right).
\end{equation}
As expected, if there is only one kind of fermion, $\Psi$ or
${\Chi}$, which respectively means $N_\chi=\overline{N}_\chi=0$
or $N_\psi=\overline{N}_\psi=0$, the current is lightlike.
However spacelike currents are also allowed from the existence of
anti-particles as they result from simultaneous exitations between
particles of one kind and anti-particles of the other kind (as for
instance $N_\psi \neq 0$ and $\overline{N}_\chi \neq 0$, or
$N_\chi \neq 0$ and $\overline{N}_\psi \neq 0$). Finally,
timelike currents are obtained from simultaneous exitation between
particles or anti-particles of both kind ($N_\chi \neq 0$ and
$N_\psi \neq 0$, or $\overline{N}_\chi \neq 0$ and
$\overline{N}_\psi \neq 0$).

\subsection{Energy per unit length and tension}
\label{enertens}
In the case of a string having a finite length $L$, periodic
boundary conditions on spinor fields impose the discretization of
the momentum exitation values
\begin{eqnarray}
k_i = \frac{2 \pi}{L} n_{\psi_i},  \qquad  l_j = \frac{2 \pi}{L}
\overline{n}_{\psi_j},  \qquad  r_p = \frac{2 \pi}{L} n_{\chi_p},
\qquad s_q = \frac{2 \pi}{L} \overline{n}_{\chi_q},
\end{eqnarray}
where $n_{\psi_i}$, $\overline{n}_{\psi_j}$, $n_{\chi_p}$ and
$\overline{n}_{\chi_q}$ are positive integers given by the particular
choice of a Fock state. From Eq.~(\ref{enerchi}) and
Eq.~(\ref{enerpsi}), the parameters $E_\chip$ and $E_\psip$ therefore
read
\begin{eqnarray}
\label{enerperio}
E_\chip = \frac{4\pi}{L^2} \left(\sum_{p=1}^{N_\chi} n_{\chi_p} +
\sum_{q=1}^{\overline{N}_\chi} \overline{n}_{\chi_q} \right),
& &
E_\psip=\frac{4\pi}{L^2} \left(\sum_{i=1}^{N_\psi} n_{\psi_i}
+ \sum_{j=1}^{\overline{N}_\psi} \overline{n}_{\psi_j} \right).
\end{eqnarray}
In the preferred frame where the two-dimensional energy momentum
tensor is diagonal, the energy per unit length and the tension
appear as the eigenvalues associated with the timelike and
spacelike eigenvectors, respectively. By means of equation
(\ref{twostress}), they read
\begin{eqnarray}
\label{ugene}
U_{\mathcal{P}}& = & M^2+2\sqrt{E_\chip E_\psip},
\\
\label{tgene}
T_{\mathcal{P}}& = & M^2-2\sqrt{E_\chip E_\psip}.
\end{eqnarray}
Note, first that the line energy density and the tension always
verify~\cite{prep}
\begin{equation}
\label{tracefix}
U+T=2M^2.
\end{equation}
Moreover the zero mode vacuum effects just modify the parameters
$E_\chip$ and $E_\psip$ as in Eq.~(\ref{fullener}), and therefore do
not modify this relationship. On the other hand, massive modes,
because they are not eigenstates of the $\gamma^0 \gamma^3$
operator, yield vacuum effects which certainly do not modify the
time and space part of the stress tensor in the same way, as was the
case for the zero modes [see Eq.~(\ref{eigenstates})]. As a result,
it is reasonable to assume that the massive mode vacuum effects
modify the Eq.~(\ref{tracefix}) by just shifting the right hand
side by a finite amount, of the order $1/L^2$.

\subsubsection{Classical limit for excited strings}

In order to derive classical values for the energy per unit length
and tension, we do not want to specify in what exitation quantum
states the system is. If the string is in thermal equilibrium
with the external medium, it is necessary to perform quantum
statistics. The number of accessible states in the string is
precisely the total number of combinations between the integer
$n_{\psi_i}$, $\overline{n}_{\psi_j}$, $n_{\chi_p}$, and
$\overline{n}_{\chi_q}$, which satisfies equations (\ref{enerchi})
and (\ref{enerpsi}) for fixed values of the stress tensor,
or, similarly, at given $E_\chi$ and $E_\psi$.
A possible representation of such equilibrium is naturally through
the microcanonical entropy
\begin{equation}
S = k_\ub \ln {\Omega},
\end{equation}
with $\Omega$ the number of accessible states and $k_\ub$ the
Boltzmann constant. Let $Q(p,N)$ be the well known partition
function $Q$ which gives the number of partitions of the integer
$p$ into exactly $N$ distinct non-zero integers. With
the following integers:
\begin{eqnarray}
K_\chi = \frac{L^2}{4\pi} E_\chi,
& \quad \textrm{and} \quad &
K_\psi = \frac{L^2}{4\pi} E_\psi,
\end{eqnarray}
the number of accessible states $\Omega$ reads
\begin{eqnarray}
\Omega & = & \sum_{n_{\chi}= \frac{N_\chi(N_\chi+1)}{2}}^{K_\chi
-\frac{\overline{N}_\chi( \overline{N}_\chi+1)}{2}} Q(n_\chi, N_\chi)
\sum_{\overline{n}_{\chi}= \frac{\overline{N}_\chi(\overline{N}_\chi
+1)}{2}}^{K_\chi-n_\chi} Q(\overline{n}_\chi, \overline{N}_\chi)
\nonumber \\
& \times &
\sum_{n_{\psi}= \frac{N_\psi(N_\psi+1)}{2}}^{K_\psi-
 \frac{\overline{N}_\psi(\overline{N}_\psi+1)}{2}} Q(n_\psi,N_\psi)
\sum_{\overline{n}_{\psi}= \frac{\overline{N}_\psi(
\overline{N}_\psi+1)}{2}}^{K_\psi-n_\psi} Q(\overline{n}_\psi,
\overline{N}_\psi).
\end{eqnarray}
The energy per unit length and tension of the string will therefore be
the values of $U$ and $T$ which maximize the entropy at given
$N_\psi$, $\overline{N}_\psi$, $N_\chi$ and $\overline{N}_\chi$. This
formalism might be useful whenever one wants to investigate the
dynamics of the string when the massless current forms, i.e., near the
phase transition at high temperatures. In what follows, we shall
assume that whatever the mechanism through which the fermions got
trapped in the string, they had enough time to reach an equilibrium
state with vanishing temperature. This can be due for instance by a
small effective coupling with the electromagnetic field opening the
possibility of radiative decay~\cite{prep}. If no such effect is
present, then one might argue that the string is frozen in an exited
state, the temperature of which possibly playing the role of a state
parameter~\cite{carter89,carter89b,carter94b,carter97} for a
macroscopic description~\cite{bcpc}.

Note that for a given distribution such as those we will be
considering later, the occupation numbers at zero temperature
must be such that, owing to Pauli exclusion principle, the
interaction terms implying for instance a $\Psi\bar\Psi$ decay
into a pair $\Chi\bar\Chi$ through Higgs bosons or $B_\mu$ exchange
are forbidden (vanishing cross-section due to lack of phase space).
In practice, this means that the following analysis is meaningful
at least up to one loop order.

\subsubsection{String at zero temperature}

\begin{figure}
\begin{center}
\epsfig{file=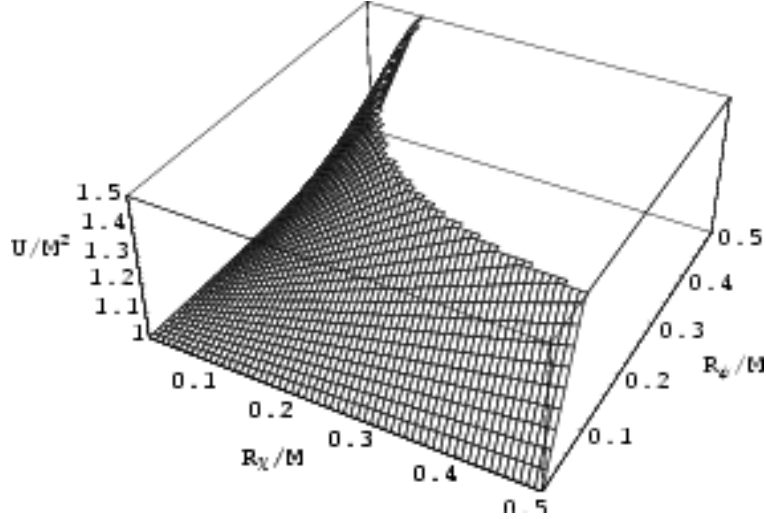,width=10cm}
\end{center}
\caption[\'Energie par unit\'e de longueur d'une corde
cosmique parcourue par des fermions de masse nulle.]{The energy per
unit length, in unit of $M^2$ and at zeroth order, as function of
$\Chi$ and $\Psi$ fermion densities plotted in unit of $M$, in the
infinite string limit.}
\label{figuzerorder}
\end{figure}

\begin{figure}
\begin{center}
\epsfig{file=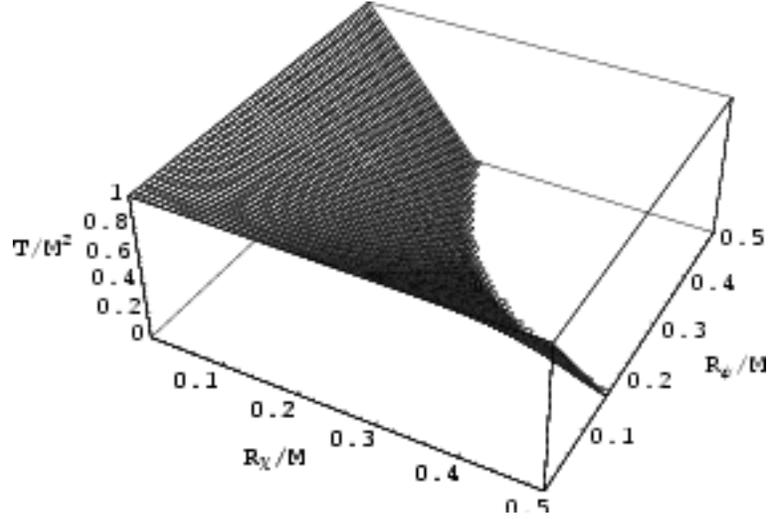,width=10cm}
\end{center}
\caption[Tension d'une corde cosmique parcourue par des fermions sans
masse.]{The tension, in unit of $M^2$ and at zeroth order, as function
of $\Chi$ and $\Psi$ fermion densities plotted in unit of $M$, in the
infinite string limit.}
\label{figtzerorder}
\end{figure}

\begin{figure}
\begin{center}
\epsfig{file=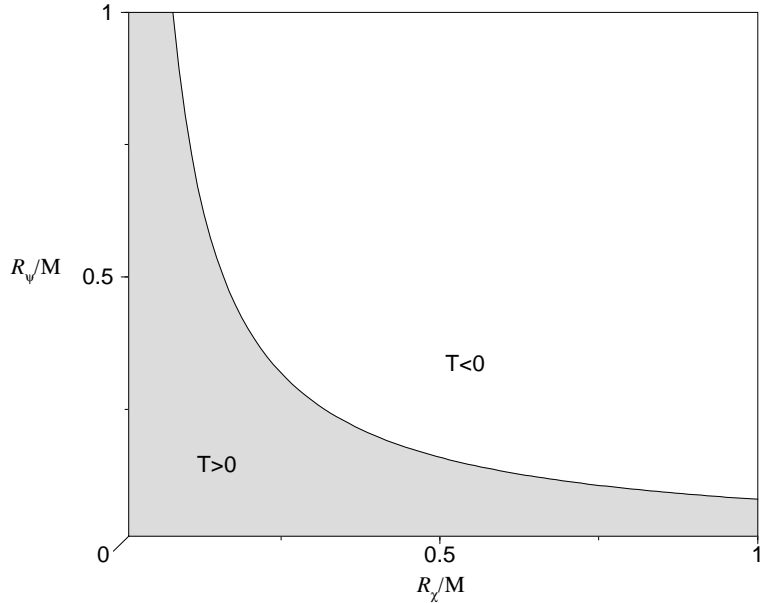,width=10cm}
\caption[Domaines de stabilit\'e d'une corde cosmique parcourue par des
fermions de masse nulle.]{Sign of the tension in the
$(\Rho_\chi/M,\Rho_\psi/M)$ plane, in the infinite string
limit. According to the macroscopic formalism, the string is unstable
with respect to transverse perturbations for $T<0$.}
\label{figtzerorderneg}
\end{center}
\end{figure}

For weak coupling between fermions trapped in the string and
external fields, as is to be expected far below the energy scale
where the string was formed, the set of particles is assumed to
fall in the ground state and because of anticommutation rules
(\ref{creatoranticom}) it obeys Fermi-Dirac statistic at zero
temperature. Consequently, the parameters $E_\chip$ and $E_\psip$
reads
\begin{eqnarray}
\label{energround}
E_\psi & = & \frac{2 \pi}{L} \left[\rho_\psi(L \rho_\psi+1) +
\overline{\rho}_\psi (L \overline{\rho}_\psi+ 1) \right],\\
E_\chi & = & \frac{2 \pi}{L} \left[ \rho_\chi(L \rho_\chi + 1)
+\overline{\rho}_\chi(L \overline{\rho}_\chi + 1) \right],
\end{eqnarray}
where the new parameters $\rho=N/L$ are the line number densities of
the corresponding particles and anti-particles trapped in the string.
Strictly speaking, these are the four independent state parameters
which fully determine the energy per unit length and the tension in
Eq.~(\ref{ugene}) and Eq.~(\ref{tgene}), and so cosmic strings with
fermionic current carriers do not verify the same equation of state
as the bosonic current-carrier case. This is all the more so manifest
with another more intuitive set of state parameters, $\Rho$ and
$\Theta$, defined for each fermion by
\begin{eqnarray}
\Rho^2  =  \left(\rho+\frac{1}{2L} \right)^2 + \left(\overline{\rho}
+\frac{1}{2L} \right)^2,
& &
\Theta = \arctan{\left(\frac{\displaystyle \overline{\rho} + 
\frac{1}{2L}}{\displaystyle \rho +\frac{1}{2L}}\right)}.
\end{eqnarray}
These are simply polar coordinates in the two-dimensional space
defined by the particle and anti-particle densities of each spinor
field. The parameter $\Rho$ physically represents the fermion density
trapped in the string regardless of the particle or anti-particle
nature of the current carriers. It is therefore the parameter that we
would expect to be relevant in a purely classical approach. On the
other hand, $\Theta$ quantifies the asymmetry between particles and
anti-particles since
\begin{eqnarray}
\rho=\Rho \cos{\Theta} - \frac{1}{2L},
& &
\overline{\rho}=\Rho \sin{\Theta} -\frac{1}{2L}.
\end{eqnarray}
The energy per unit length, tension and line density current now
read
\begin{eqnarray}
\label{uzerorder}
U & = & M^2 + 4\pi\sqrt{\left(\Rho_\chi^2-\frac{1}{2 L^2}\right)
\left(\Rho_\psi^2-\frac{1}{2L^2}\right)},
\\
\label{tzerorder}
T & = & M^2 - 4\pi\sqrt{\left(\Rho_\chi^2-\frac{1}{2 L^2}\right)
\left(\Rho_\psi^2-\frac{1}{2L^2}\right)},
\\
\overline{{\mathcal{C}}}^2 & = & 8 \Rho_\chi \Rho_\psi
\sin{\left(\Theta_\chi-\frac{\pi}{4}\right)}
\sin{\left(\Theta_\psi - \frac{\pi}{4} \right)}.
\end{eqnarray}
There are always four independent state parameters but only two,
$\Rho_\chi$ and $\Rho_\psi$, are relevant for line density energy
and tension. Compared to the scalar case where only one kind of charge
carrier propagates along the string, it is not surprising that we
found two degrees of freedom with two kinds of charge carriers. On the
other hand, the nature of the line density current is not relevant
because it only appears through $\Theta_\psi$ and $\Theta_\chi$, which
not modify $U$ and $T$, at least at the zeroth order. The energy per
unit length and tension relative to $M^2$ are represented in
Fig.~\ref{figuzerorder} and Fig.~\ref{figtzerorder} as function of
$\Rho_\chi/M$ and $\Rho_\psi/M$, in the infinite string limit.
As expected from their analytical expressions in the infinite
string limit, the energy per unit length is always positive and
grows with both fermion densities, $\Rho_\chi$ and $\Rho_\psi$,
whereas the tension always decreases and takes negatives values for
large fermion densities. Obviously, in the case $\Rho_\chi=\Rho_\psi=0$
there is no current along the string and we recover the Goto-Nambu
case, $U=T=M^2$. The chiral case, where the fermionic current is lightlike,
is obtained for $\Rho_\chi=0$, or $\Rho_\psi=0$, and also verifies
$U=T=M^2$ as in the chiral scalar current case~\cite{carpet2}.
From Eq.~(\ref{tzerorder}), and in the infinite string limit, the densities
for which the tension vanishes verify
\begin{equation}
\Rho_\chi\Rho_\psi = \frac{M^2}{4 \pi},
\end{equation}
This curve separates the plane $(\Rho_\chi,\Rho_\psi)$ in two regions
where $T$ is positive near the origin, and negative on the other side
(see Fig.~\ref{figtzerorderneg}). In the macroscopic formalism of
Carter~\cite{carter89,carter89b,carter94b,carter97}, the transverse
perturbations propagation speed is given by $\ct^2 = T/U$, and
therefore the domains where $T<0$ correspond to strings which are
always locally unstable with respect to transverse perturbations.
The tension of the string becomes negative only for carrier
densities close to the mass of the Goto-Nambu string
$M$. For such currents, it is necessary to derive the
back reaction in order to see how relevant it is for the energy
per unit length and tension. Moreover, in a renormalizable model,
the vacuum mass acquired by the fermions from their coupling to
the Higgs field is less than the Goto-Nambu string mass, and
thus, another quantum effects may take place before the negative
tension is reached, like tunneling into massive states. Besides,
note that $M$, the string unit of mass, arising from
non-perturbative effects, may well be much larger than the Higgs
boson mass, and so, it is expected that $\Rho_\chi, \Rho_\psi \ll M$.
Thus, the no-spring conjecture~\cite{nospring} proposed in the
case of bosonic carrier presumably apply to the fermionic carrier
case as well. Moreover, the zero mode vacuum effects on energy per
unit length and tension appear clearly from Eq.~(\ref{fullener}) as
additional string length. The corrected values of the equation of
state are therefore obtained by replacing the physical length of the
string, $L$, in Eq.~(\ref{uzerorder}) and Eq.~(\ref{tzerorder}), by
an equivalent length, $L_{\ue}$ say, which verifies
\begin{equation}
\label{eqlength}
\frac{1}{2L_\ue^2}=\frac{1}{2L^2}-\frac{1}{L_\uv^2}.
\end{equation}
In the particular case where $L_\uv^2=6L^2$, it reads $L_\ue^2=(3/2)
L^2$. On the other hand, the massive vacuum effects certainly shift
in a different way $U$ and $T$ by a finite amount as previously
discussed, but will not be considered in the following.
In the next section, the back reaction is derived in
the classical limit in order to find corrected values of energy
per unit length and tension. Moreover we shall take care of the
finite length of the string $L$, keeping in mind that its
value, and consequently the value of $L_\ue$, have to be larger than
$1/M$ since all physical values have been derived in the classical
vortex background, i.e., the quantum effects of the Higgs field have
been neglected.

\subsection{Back Reaction}

The existence of fermionic currents carrying gauge charge along
the string gives rise to new gauge field components, $B_t$ and
$B_z$, from the equations of motion (\ref{gaugemvt}). These, being
coupled with the corresponding currents, provide additional terms
in the energy momentum tensor (\ref{brcorr}). As a first step, the
new gauge field components are computed numerically from the zero
modes solutions of Eq.~(\ref{zeromodes}). The corrected equation of
state is then analytically derived, the numerical dependencies
having been isolated in model dependent coefficients.

\subsubsection{Back reacted gauge fields}

\begin{figure}
\begin{center}
\epsfig{file=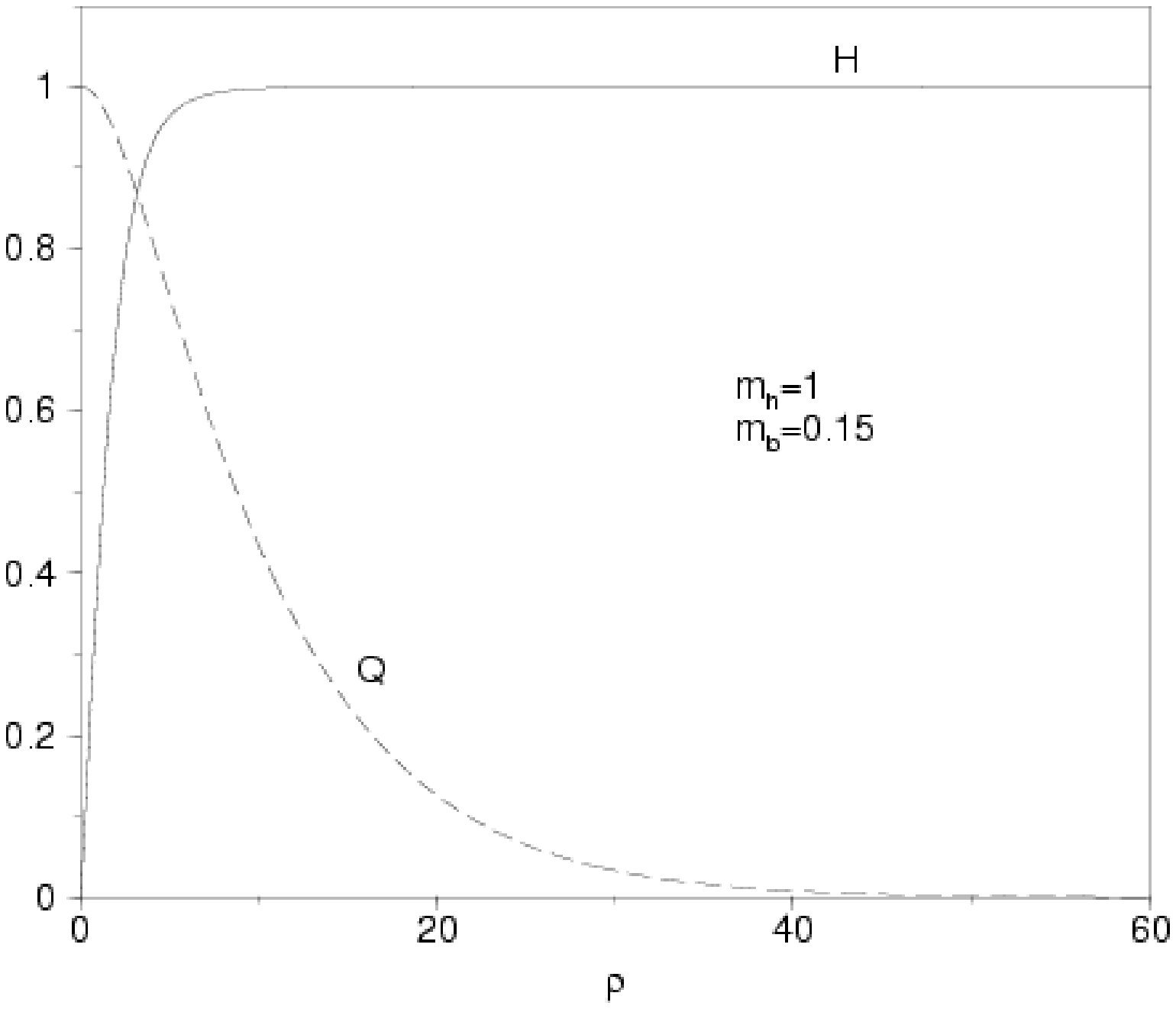,width=10cm}
\caption[Profils transverses des champs d'une corde cosmique.]{The
solutions of the field equations for the vortex background. The Higgs
field, $H$, takes its vacuum expectation value at infinity and the
gauge bosons condensate in the vortex.}
\label{figbackZM}
\end{center}
\end{figure}

\begin{figure}
\begin{center}
\epsfig{file=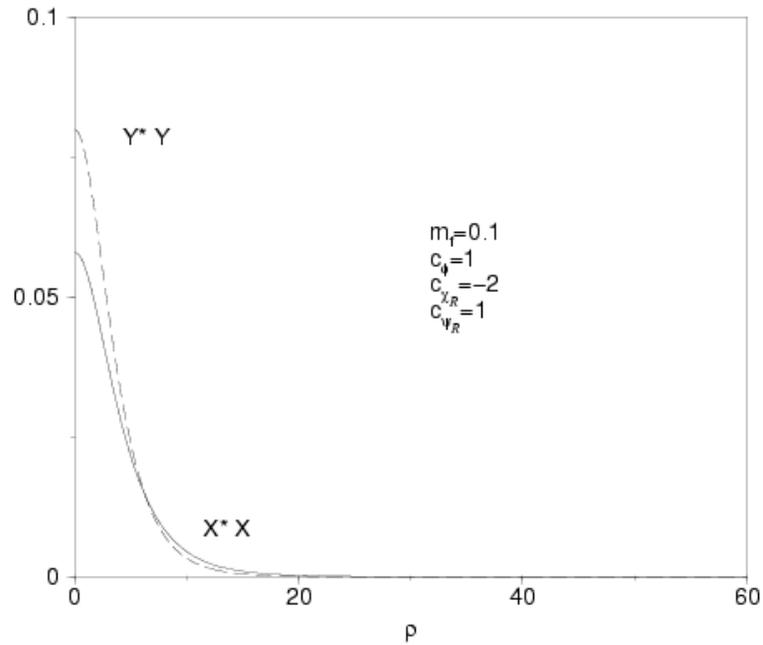,width=10cm} \caption[Densit\'es de
probabilit\'e de pr\'esence des fermions de masse nulle pi\'eg\'es sur
une corde cosmique.]{The normalized probability densities of the zero
modes, $|X|^2$ and $|Y|^2$. The rapid decay far from the string core
reflects the bound state nature of the condensates.} \label{figproba}
\end{center}
\end{figure}

\begin{figure}
\begin{center}
\epsfig{file=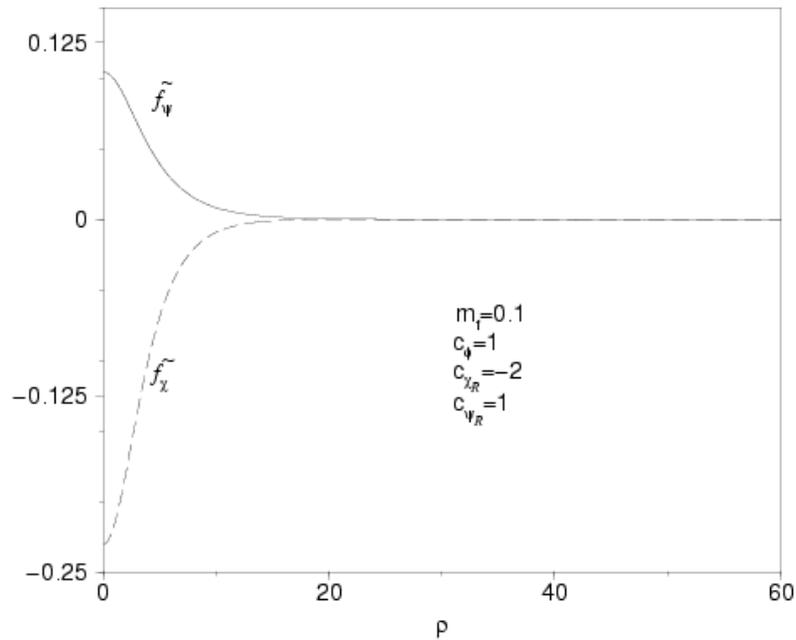,width=10.5cm}
\caption[Densit\'e transverse de charge associ\'ee aux modes z\'eros
fermioniques.]{The dimensionless radial current functions
$\tilde{f}_\chi$ and $\tilde{f}_\psi$. They can be viewed as the
effective transverse density charge carried by the fermion
currents. Their sign results in the initial choice of each conserved
fermion gauge charge.}
\label{figradfunc}
\end{center}
\end{figure}

In order to compute the $B_t$ and $B_z$ fields at first order, we
only need the zeroth order values of the zero modes and the
vortex background. Let us introduce the dimensionless scaled
fields and variables
\begin{eqnarray}
\varphi=\eta H,
\quad
Q_\theta=Q,
\quad
r=\frac{\varrho}{m_\uh},
\end{eqnarray}
with $m_\uh=\eta\sqrt{\lambda}$ the classical mass of the Higgs
field. From the equation of motion (\ref{higgsmvt}), the
orthoradial gauge field $Q$ and $H$ are solution of
\begin{eqnarray}
\label{tildehiggsZM}
\frac{\ud^2H}{\ud \varrho^2}+\frac{1}{\varrho} \frac{\ud H}{\ud
\varrho} & = & \frac{H Q^2}{\varrho^2}+\frac{1}{2}H(H^2-1), \\
\label{tildegaugeZM}
\frac{\ud^2 Q}{\ud \varrho^2} -\frac{1}{\varrho}\ \frac{\ud Q}{\ud
\varrho} & = & \frac{m_\ub^2}{m_\uh^2}H^2 Q,
\end{eqnarray}
where $m_\ub=qc_\phi \eta$ is the classical mass of the gauge boson.
The numerical solutions of these equations have been computed
earlier by many people~\cite{bps,neutral} using relaxation methods
\cite{adler}. They are presented in Fig.~\ref{figbackZM} for a
specific (assumed generic) set of parameters.
In the same way, deriving Eq.~(\ref{zeromodes}) with respect to
$\varrho$ yields the right component of the zero mode $Y$ as a
solution of
\begin{eqnarray}
\zeta_1'' - \left(\frac{H'}{H}+\frac{Q-n}{\varrho}\right)\zeta_1'
& + & \left[\frac{\cchir}{\cphi}\left(\frac{Q-n}{\varrho} \frac{H'}{H}
+ \frac{Q-n}{\varrho^2}-\frac{Q'}{\varrho}\right) \right.
\nonumber \\
& - &
\left.
\frac{\cchil \cchir}{\cphi^2} \left(\frac{Q-n}{\varrho}\right)^2
-\frac{m_\uf^2}{m_\uh^2} H^2 \right] \zeta_1 = 0,
\end{eqnarray}
while the left one satisfies
\begin{eqnarray}
i \frac{m_\uf}{m_\uh}H \zeta_4 & = &  \zeta_1' - \frac{\cchir}{\cphi}
\frac{Q-n}{\varrho} \zeta_1,
\end{eqnarray}
where a prime indicates a derivation with respect to the
dimensionless radial variable $\varrho$. The field $\Psi$
verifies similar equations with the transformation, $\zeta
\rightarrow \xi$ and $c_\chi \rightarrow c_\psi$. The numerical
integration has been performed with a relaxation method~\cite{adler}
and verified on the original system (\ref{zeromodes})
with a shooting method. As a result, the normalized probability
densities of the zero modes, $|X|^2$ and $|Y|^2$, are plotted in
Fig.~\ref{figproba}. The dimensionless radial functions
$\tilde{f}_\chi$ and $\tilde{f}_\psi$ defined from Eq.~(\ref{fchi})
and Eq.~(\ref{fpsi}) by
\begin{equation}
\tilde{f}_\chi=\frac{2 \pi}{m_\uh^2} f_\chi,
\quad
\tilde{f}_\psi=\frac{2 \pi}{m_\uh^2} f_\psi,
\end{equation}
are plotted in Fig.~\ref{figradfunc}. As expected, the fields
are confined in the string core, and so will the corresponding
fermionic currents.

\begin{figure}
\begin{center}
\epsfig{file=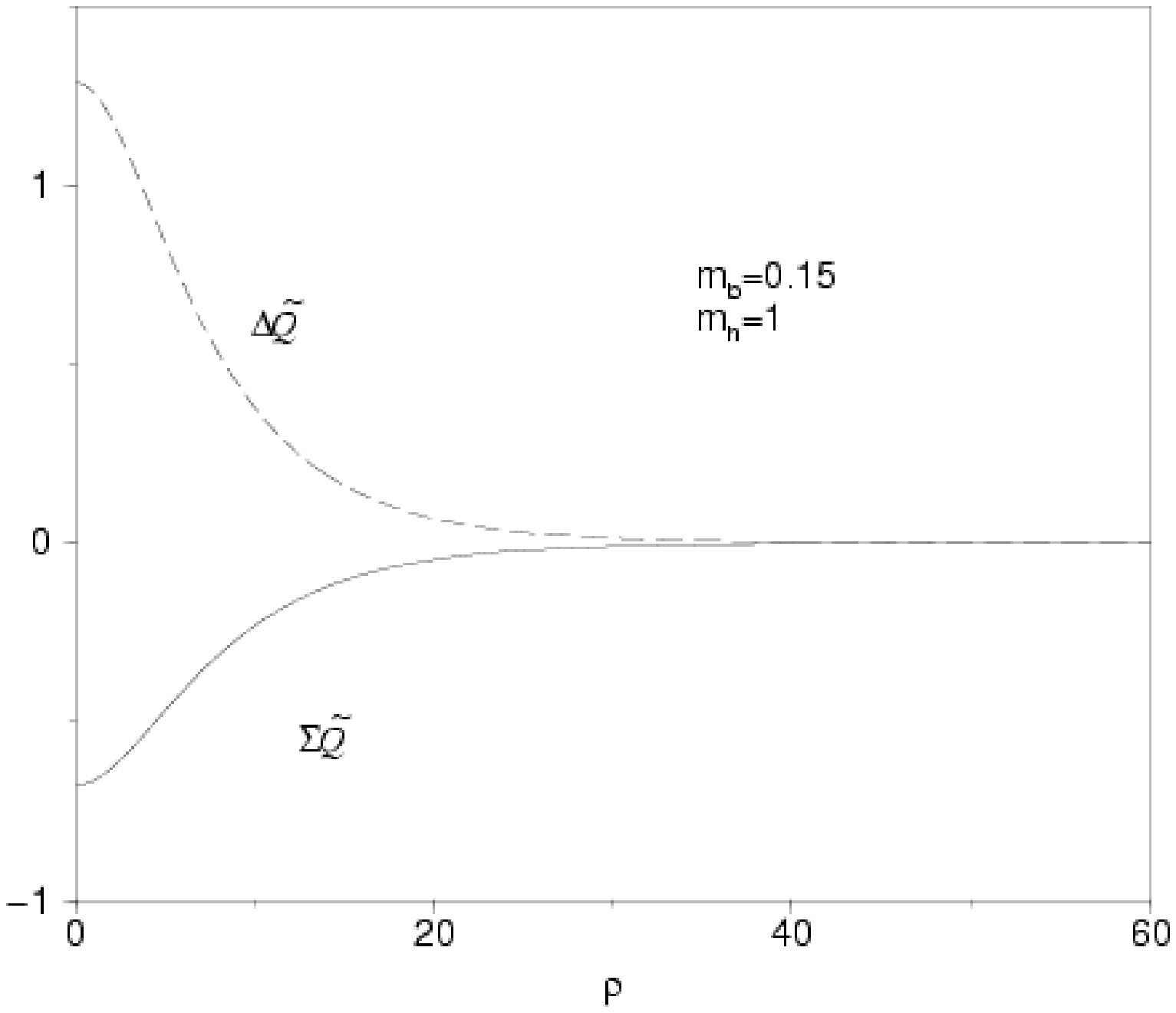,width=10cm}
\caption[Le champ de jauge g\'en\'er\'e par les modes z\'eros.]{The
dimensionless backreacted gauge fields $\Delta\tilde{Q}$ and
$\Sigma\tilde{Q}$, generated by the fermion currents. They do not lead
to new long-range effects since they acquire non-zero mass outside the
string core due to their coupling with the Higgs fields.}
\label{figgauge}
\end{center}
\end{figure}

Let us define the more relevant components of the backreacted
gauge field, $\Delta B=B_z-B_t$ and $\Sigma B=B_z+B_t$, with the
corresponding dimensionless scaled fields $\Delta \tilde{Q}$ and
$\Sigma \tilde{Q}$ defined by
\begin{eqnarray}
\label{newgaugetilde}
\Delta B  =  \left(\overline{\rho}_\chi - \rho_\chi \right)
\frac{m_\ub^2}{\pi \eta^2} \frac{\Delta \tilde{Q}}{q\cphi},
&  &
\Sigma B  =  -\left(\overline{\rho}_\psi - \rho_\psi \right)
\frac{m_\ub^2}{\pi \eta^2} \frac{\Sigma \tilde{Q}}{q\cphi}.
\end{eqnarray}
The equations of motion (\ref{gaugemvt}) in the classical limit now
reads
\begin{eqnarray}
\label{newgaugemvt}
\Delta \tilde{Q}''+\frac{1}{\rho}
\Delta\tilde{Q}'-\frac{m_\ub^2}{m_\uh^2}H^2 \Delta\tilde{Q} =
\frac{\tilde{f}_\chi}{q \cphi}, & & \Sigma
\tilde{Q}''+\frac{1}{\rho} \Sigma \tilde{Q}'-\frac{m_\ub^2}{m_\uh^2}H^2
\Sigma \tilde{Q} = \frac{\tilde{f}_\psi}{q \cphi}.
\end{eqnarray}
As for fermions, these new gauge fields get their masses from coupling
with the Higgs field, and therefore have non-zero mass outside the
string core. Moreover, they are generated by fermionic massless
currents confined in the core, therefore they also condense in and do
not lead to new long-range effects.  The solutions of these equations
(\ref{newgaugemvt}) have been obtained using, once again, a relaxation
method~\cite{adler} and are represented in Fig.~\ref{figgauge}. Note
that owing to the scaled field $\Delta \tilde{Q}$ and $\Sigma
\tilde{Q}$, we have separated the numerical dependence in the gauge
field and currents from the fermion densities content [see
Eq.~(\ref{newgaugetilde})].
Up to now, we have computed the fermionic gauge currents along the
string as well as the component $B_t$ and $B_z$, so that the
back reaction correction to the energy momentum tensor is
computable from Eq.~(\ref{brcorr}). As in the previous section,
the energy per unit length and tension can be derived in the
preferred frame where the stress tensor is diagonal, but now we
have to find the eigenvalues of the full two-dimensional energy
momentum tensor $\overline{T}^{\alpha \beta}_{\ub.\ur.}+\langle
\overline{T}^{\alpha \beta} \rangle$, with
\begin{equation}
\overline{T}^{\alpha \beta}_{\ub.\ur.}=\int{ r \, \ud r \, \ud \theta
\, T^{\alpha \beta}_{\ub.\ur.}}
\end{equation}

\subsubsection{Energy per unit length and tension with back reaction}

Using the dimensionless field, $\Delta \tilde{Q}$ and $\Sigma
\tilde{Q}$, with the expressions of the currents given in equations
(\ref{gaugecurrentt}) and (\ref{gaugecurrentz}), one gets, after
some algebra, the full expression of the stress tensor
with corresponding eigenvalues
\begin{eqnarray}
\label{ufull}
\widehat{U} & = & M^2 -\I(\Theta_\chi,\Theta_\psi)\Rhop_\chi \Rhop_\psi
+4 \pi \sqrt{\left(\Rhop_\chi^2 -\frac{1}{2L_\ue^2}\right)\left(\Rhop_\psi^2
-\frac{1}{2L_\ue^2}\right)},
\\
\label{tfull}
\widehat{T} & = & M^2 -\I(\Theta_\chi,\Theta_\psi)\Rhop_\chi \Rhop_\psi
-4 \pi \sqrt{\left(\Rhop_\chi^2 -\frac{1}{2L_\ue^2}\right)\left(\Rhop_\psi^2
-\frac{1}{2L_\ue^2}\right)},
\end{eqnarray}
with the scaled state parameters
\begin{eqnarray}
\label{dampdens}
\Rhop_\chi & = & \Rho_\chi \sqrt{1-\frac{m_\ub^2}{2 \pi^2 \eta^2} \left(
I_{\Delta \chi} - I_{\Delta 2} \right)
\sin^2{\left(\Theta_\chi-\frac{\pi}{4}\right)}},
\\
\Rhop_\psi & = & \Rho_\psi \sqrt{1-\frac{m_\ub^2}{2 \pi^2 \eta^2} \left(
I_{\Sigma \psi} - I_{\Sigma 2} \right)
\sin^2{\left(\Theta_\psi-\frac{\pi}{4}\right)}},
\end{eqnarray}
and the function $\I(\Theta_\chi,\Theta_\psi)$ defined by
\begin{eqnarray}
\label{ifonction}
\I(\Theta_\chi,\Theta_\psi) & = &\displaystyle \frac{m_\ub^2}{\pi \eta^2}
 \frac{\displaystyle \left(I_{\Sigma \chi}+I_{\Delta \psi} \right)
 \sin{\left(\Theta_\chi-\frac{\pi}{4} \right)}}{\sqrt{ \displaystyle
 1-\frac{m_\ub^2}{2 \pi^2 \eta^2} \left(I_{\Delta \chi}-I_{\Delta
 2}\right) \sin^2{\left(\Theta_\chi-\frac{\pi}{4}\right)}}}
 \nonumber \\ & & \times \frac{\sin{\left(\Theta_\psi-\frac{\pi}{4}
 \right)}}{\sqrt{\displaystyle 1-\frac{m_\ub^2}{2 \pi^2 \eta^2}
 \left(I_{\Sigma \psi}- I_{\Sigma 2}\right)
 \sin^2{\left(\Theta_\psi-\frac{\pi}{4}\right)}}}
\end{eqnarray}
The numerical integrations previously carried out appear through
pure numbers which depend only on the model parameters. The
coupling $B_\mu j^\mu$ leads to the following quantities
\begin{eqnarray}
\label{numintegspin}
I_{\Sigma \chi}=-2\int{\,\varrho \, \ud \varrho \, \Sigma
\tilde{Q}(\varrho) \frac{\tilde{f}_\chi(\varrho)}{q\cphi}}, & \quad &
I_{\Delta \psi}=-2\int{\,\varrho \, \ud \varrho \, \Delta
\tilde{Q}(\varrho) \frac{\tilde{f}_\psi(\varrho)}{q\cphi}}, \\
I_{\Sigma \psi}=-2\int{\,\varrho \, \ud \varrho \, \Sigma
\tilde{Q}(\varrho) \frac{\tilde{f}_\psi(\varrho)}{q\cphi}}, & \quad &
I_{\Delta \chi}=-2\int{\,\varrho \, \ud \varrho \, \Delta
\tilde{Q}(\varrho) \frac{\tilde{f}_\chi(\varrho)}{q\cphi}},
\end{eqnarray}
while the kinetic contribution of the new gauge fields appears through
\begin{eqnarray}
\label{numintegkin}
I_{\Delta 2}=\int{\, \varrho \, \ud \varrho \, \left[\partial_\varrho
\Delta \tilde{Q}(\varrho) \right]^2},
& \quad &
I_{\Sigma 2}=\int{\, \varrho \, \ud \varrho \, \left[\partial_\varrho
\Sigma \tilde{Q}(\varrho) \right]^2}.
\end{eqnarray}
By means of the equations of motion (\ref{newgaugemvt}) and the constant
sign of $\Delta \tilde{Q}$ and $\Sigma \tilde{Q}$,
$I_{\Delta \chi}$ and $I_{\Sigma \psi}$ are found to be always positive.
Intuitively, as in electromagnetism, the gauge field generated from charge
currents tends to resist to the currents which give birth to it. In
our case, the back reaction actually damps the weight of the charge
carriers in the energy per unit length and tension. In fact, the
relevant state parameters are now
$\Rhop$ instead of $\Rho$ with $\Rhop<\Rho$ since $I_{\Delta \chi}$ and
$I_{\Sigma \psi}$ are positive. Moreover, numerical calculations show
that the kinetic contribution numbers (\ref{numintegkin}) are always
one order of magnitude smaller than those resulting in the coupling
between gauge fields and currents (\ref{numintegspin}), as expected
for reasonable backreacted gauge field since they only involve the
square gradient of these fields [see Eq.~(\ref{numintegkin})].
However, there is an additional term involving new dependence in
the asymmetry between particles and anti-particles through the
$\I(\Theta_\chi,\Theta_\psi)$ function. In order to understand this
point physically, let us derive the magnitude of the gauge current
carried by the fermions. From Eq.~(\ref{gaugecurrentt}) and
Eq.~(\ref{gaugecurrentz}), once the transverse coordinates have been
integrated over, the dimensionless magnitude reads
\begin{equation}
\jbt^2=
\frac{\displaystyle 2 (2q\cphi)^2 \tilde{F}_\chi
\tilde{F_\psi}\sin{\left(\Theta_\chi-\frac{\pi}{4} 
\right)}
\sin{\left(\Theta_\psi-\frac{\pi}{4} \right)}\Rhop_\chi \Rhop_\psi}
{\sqrt{\displaystyle 1-\frac{m_\ub^2}{2 \pi^2
\eta^2} \left(I_{\Delta \chi}-I_{\Delta 2}\right)
\sin^2{\left(\Theta_\chi - \frac{\pi}{4}\right)}}
\sqrt{ \displaystyle 1-\frac{m_\ub^2}{2 \pi^2 \eta^2} \left(I_{\Sigma
\psi}-I_{\Sigma 2}\right)
\sin^2{\left(\Theta_\psi-\frac{\pi}{4}\right)}}},
\end{equation}
with the dimensionless constants are
\begin{eqnarray}
\tilde{F}_\chi = \int{\, \varrho \, \ud \varrho \,\frac{\tilde{f}_\chi
(\varrho)}{q \cphi}},
& &
\tilde{F}_\psi = \int{\, \varrho \, \ud \varrho \,\frac{\tilde{f}_\psi
(\varrho)}{q \cphi}}.
\end{eqnarray}
These numbers can be viewed as the effective charge carried by the
fermionic gauge currents since
\begin{equation}
\frac{\jbt^2}{(2q\cphi)^2\tilde{F_\chi}\tilde{F_\psi}}=\frac{\overline{
{\mathcal{C}}}^2}{4}.
\end{equation}
The function $\I(\Theta_\chi,\Theta_\psi)$ therefore verifies
\begin{equation}
\I(\Theta_\chi,\Theta_\psi)\Rhop_\chi \Rhop_\psi
=\frac{I_{\Sigma \chi}+I_{\Delta \psi}}{8 \pi \tilde{F}_\chi \tilde{F}_\psi}
\, \jbt^2,
\end{equation}
and, as before, according to Eq.~(\ref{newgaugemvt}),
$(I_{\Sigma \chi}+I_{\Delta \psi})/(\tilde{F}_\chi
\tilde{F}_\psi)$ is always positive, so the sign of $\I$ directly
reflects the spacelike or timelike nature of the current. Thus,
in addition to the back reaction damping effect, there is a
correction to the energy per unit length and tension directly
proportional to the magnitude of the fermionic current. Note that
this effect appears as a correction due to back reaction and not,
as it is the case for cosmic string with bosonic current
carriers, at the zeroth order~\cite{neutral}.

\subsubsection{Equation of state with back reaction}

Unfortunately, the corrected expressions of the line density
energy and tension involve four independent state parameters, and
consequently are not easily representable. However, they can be
studied as functions of the damped fermion densities $\Rhop$,
modified only by the function $\I(\Theta_\chi,\Theta_\psi)$ which
quantifies the efficiency of the fermionic currents in generating
backreacted gauge fields. In this way, the comparison with the
zeroth order case is all the more so easy.

The study of the surfaces defined by $\widehat{U}$ and
$\widehat{T}$ in the plane $(\Rhop_\chi,\Rhop_\psi)$ is less
canonical than at zeroth order. Three critical values of
the function $\I$ are found to modify the behaviors of the tension and
energy per unit length, namely, $-4\pi$, $0$, and $4 \pi$. However,
only small values of $\I$ are reasonable in this model as it is
discussed in the next section. This analysis is consequently
constraints to values of $|\I|<4 \pi$.

\paragraph{Energy per unit length.}

The line density energy follows different behaviors according
to the value of $\I$.

\begin{figure}
\begin{center}
\epsfig{file=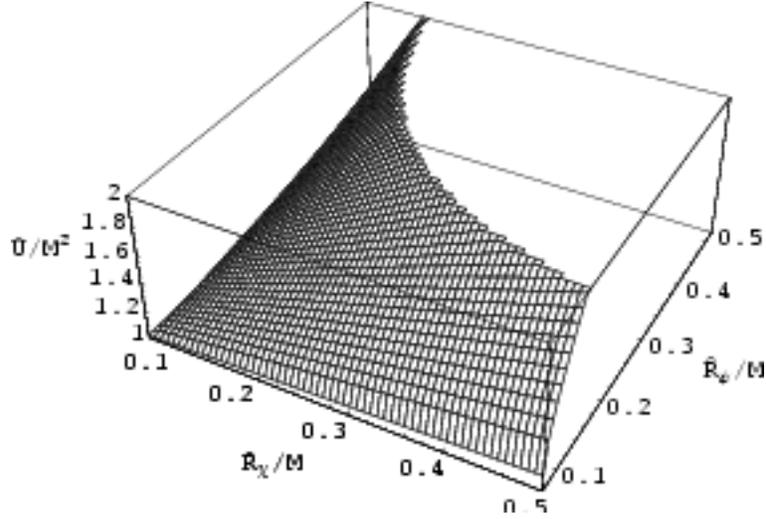,width=10cm}
\caption[Influence du champ de jauge g\'en\'er\'e par les modes z\'eros sur
la densit\'e d'\'energie de la corde.]{The energy per unit length in
unit of $M^2$ as function of $\Rhop_\chi/M$ and $\Rhop_\psi/M$, for
spacelike currents with $\I(\Theta_\chi,\Theta_\psi)<0$. The influence
of the finite length of the string just appears near the axes.}
\label{figuineg}
\end{center}
\end{figure}

The first and simplest case $\I<0$, obtained for spacelike
fermionic gauge currents, is very similar to the zeroth order
case, and the energy per unit length just grows a bit faster
with the damped fermion densities $\Rhop_\chi$ and $\Rhop_\psi$,
as on Fig.~\ref{figuineg}.

\begin{figure}
\begin{center}
\epsfig{file=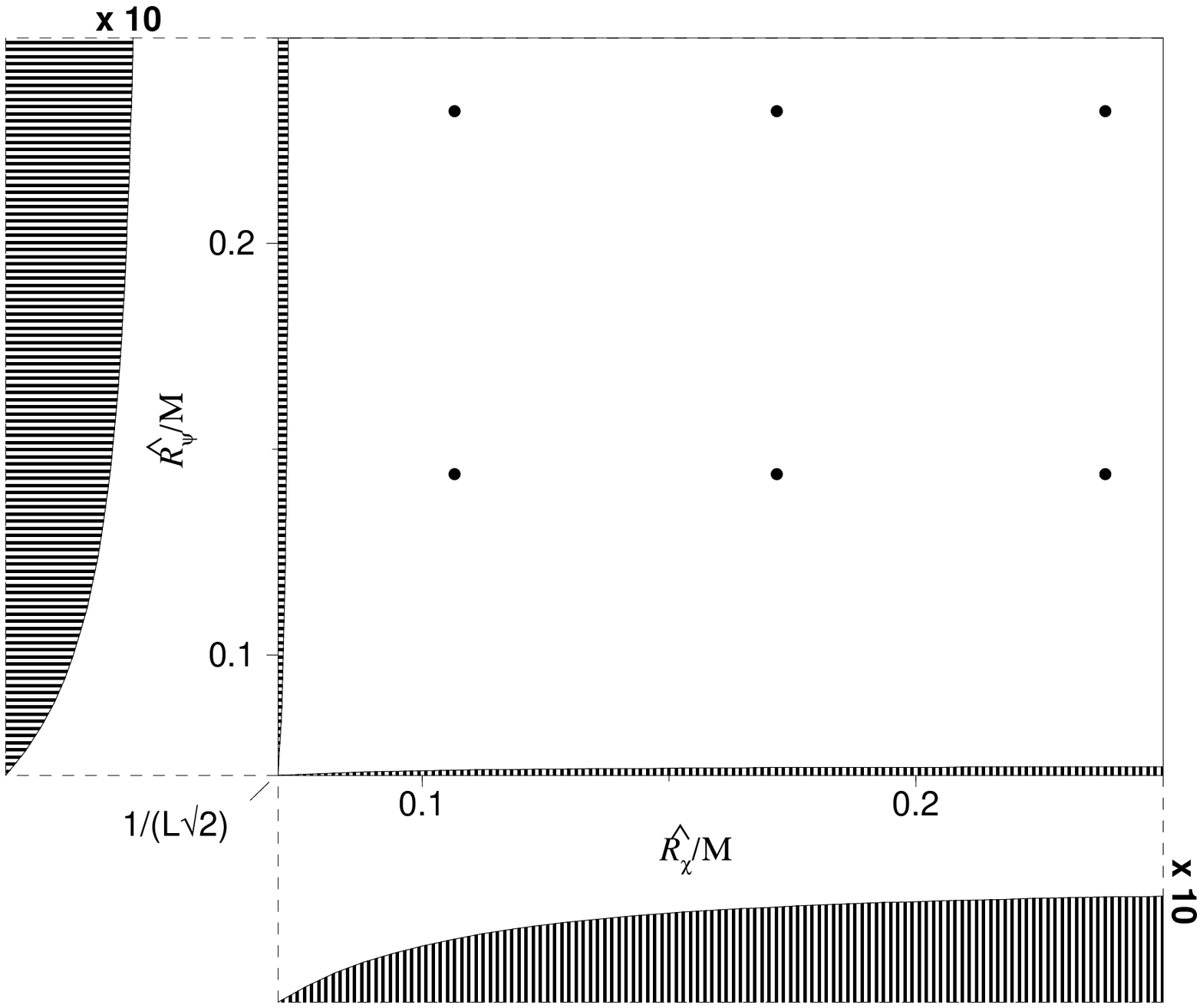,width=9cm}
\end{center}
\begin{center}
\epsfig{file=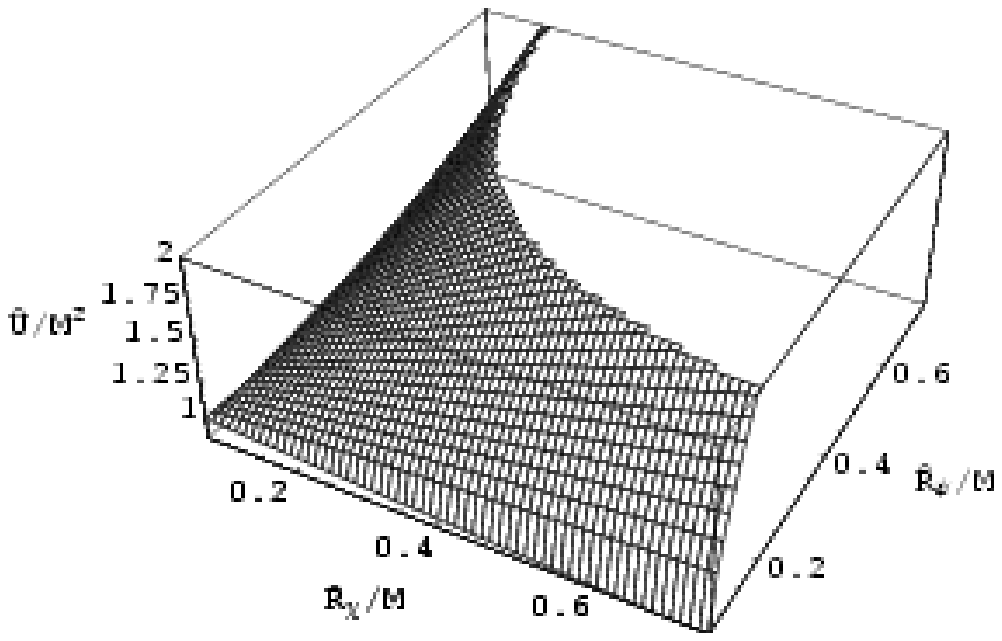,width=11cm}
\caption[Domaines de variation de la densit\'e d'\'energie d'une corde
cosmique parcourue par des fermions sans masse.]{Variation domains of
the energy per unit length, plotted in unit of $M^2$, in the
$(\Rhop_\chi/M,\Rhop_\psi/M)$ plane for timelike currents with
$0<\I<4\pi$. The vertical hatched regions define domains where the
line density energy could decrease with $\Rhop_\chi$ whereas the
horizontal ones are regions where the line density energy could
decrease with $\Rhop_\psi$. On one hand, these domains are
asymptotically limited by the curves $\Rhop=(2\pi/L_\ue)
\sqrt{2/(16\pi^2-\I^2)}$ and therefore coincide with the axis for
large values of $L_\ue$.  On the other hand, for reasonable values of
$\I \ll 4\pi$, there is no accessible quantum state inside, the first
one being shown as a dot. As a result the energy per unit length
always grows with the parameters $\Rhop$ and is always positive.}
\label{figuiposip}
\end{center}
\end{figure}

For timelike currents, $I>0$, we find that the back reaction damps
the growth of the density line energy with the fermion densities. As
a result the line density energy seems to decrease in some regions, and
the stationary curves of $\widehat{U}$ with respect to $\Rhop_\chi$
are given, from Eq.~(\ref{ufull}), by
\begin{eqnarray}
\label{energystat}
\frac{\partial{\widehat{U}}}{\partial \Rhop_\chi}=0 & \Leftrightarrow
& \Rhop_\psi=\frac{2 \pi}{L_\ue} \Rhop_\chi \sqrt{\displaystyle
\frac{2}{\displaystyle \frac{\I^2}{2L_\ue^2} + (16 \pi^2 -\I^2)
\Rhop_\chi^2}},
\end{eqnarray}
and thanks to the symmetry between $\Rhop_\chi$ and $\Rhop_\psi$,
similar equations are obtained for $\partial \widehat{U}/\partial
\Rhop_\psi=0$. Finally, the variation domains of the line density
energy are represented in Fig.~\ref{figuiposip} for $0<\I<4\pi$.
The first discrete values of the fermion densities (the length
of the string is finite) have been represented by dots in the
$(\Rhop_\chi$,$\Rhop_\psi)$-plane, and as can be seen in
Fig.~\ref{figuiposip}, for reasonable values of $\I$, there is no
available quantum state inside the tiny decreasing regions.
Consequently, the density line energy always grows with the fermions
densities and remains positive. Since the stationary curves of
$\widehat{U}$ are asymptotically proportional to $1/L_\ue < 1/L$ [see
Eq.~(\ref{energystat})], they coincide with the axis in the infinite
string limit. The surface describing $\widehat{U}(\Rhop_\chi,
\Rhop_\psi)$ has also been plotted in Fig.~\ref{figuiposip} in unit
normalized to $M^2$, and for minimal acceptable value of
$L_\ue=10/M$ just in order to show the influence of the finite length.


\paragraph{Tension.}

The study of the tension with respect to the fermion densities is
performed in the same way. As before the stationary curves
of $\widehat{T}$ with respect to $\Rhop_\chi$ or $\Rhop_\psi$ are
found from Eq.~(\ref{tfull}), and follow the same equation as those of
the energy per unit length in Eq.~(\ref{energystat}), although the
variation domains are not the same and have been plotted in
Fig.~\ref{figtinegip} for different values of the function $\I$.

\begin{figure}
\begin{center}
\epsfig{file=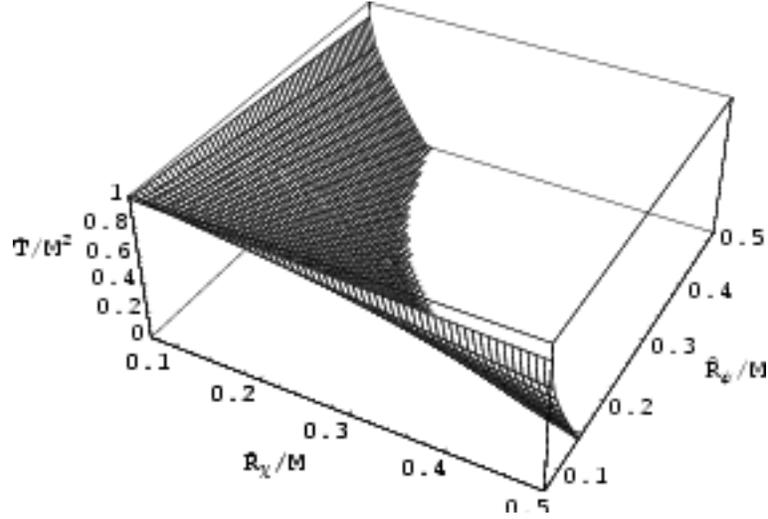,width=10cm}
\caption[Influence du champ de jauge g\'en\'er\'e par les modes z\'eros sur
la tension de la corde.]{The tension for timelike currents with
$\I>0$, plotted in unit of $M^2$ in the $(\Rhop_\chi/M,\Rhop_\psi/M)$
plane. Note the shift near the axes due to the finite length of the
string.}
\label{figtipos}
\end{center}
\end{figure}

\begin{figure}
\begin{center}
\epsfig{file=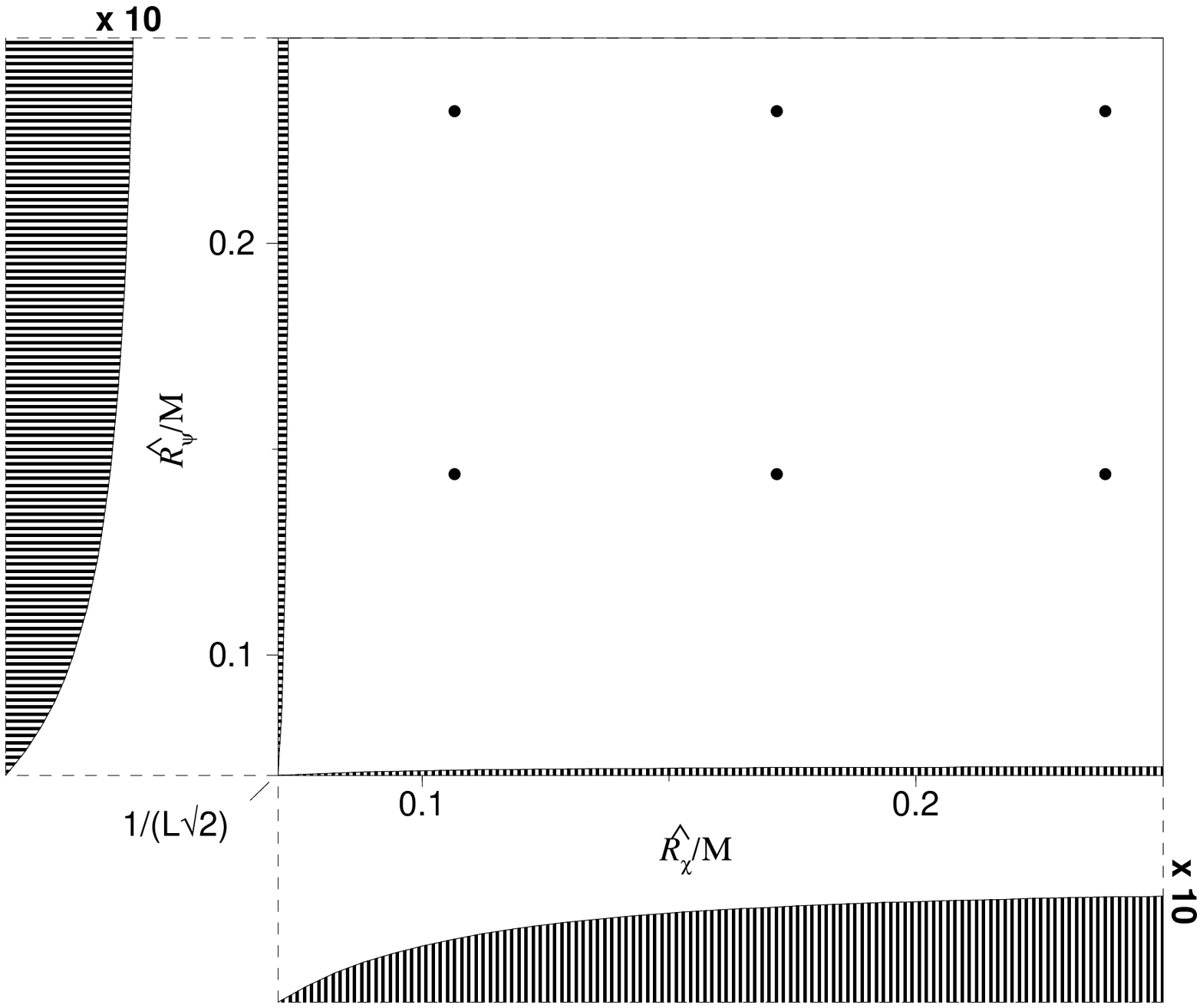,width=9cm}
\end{center}
\begin{center}
\epsfig{file=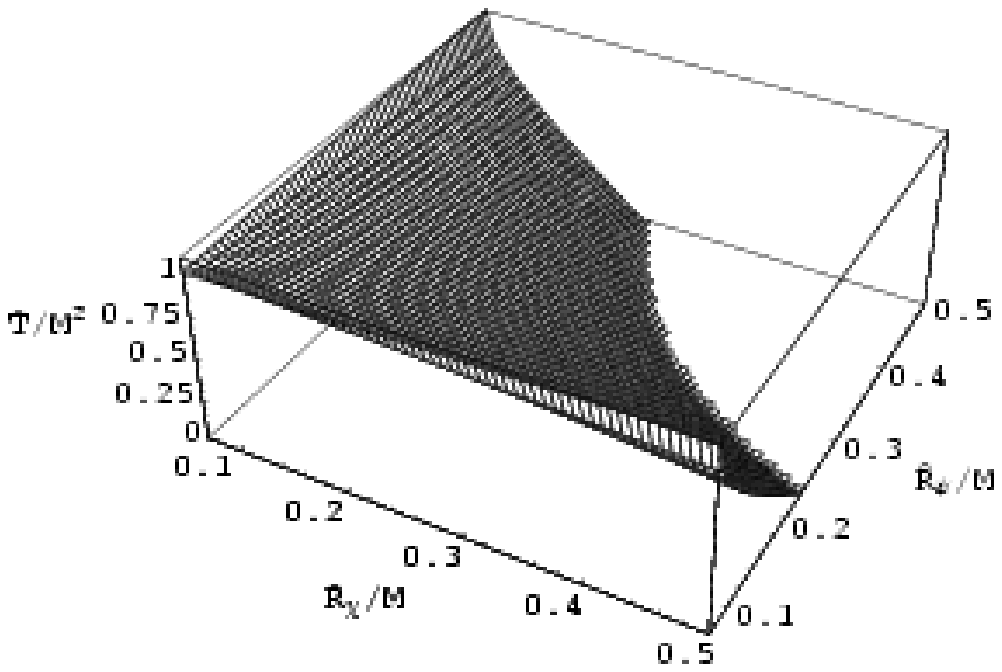,width=11cm}
\caption[Domaines de variation de la tension d'une corde
cosmique parcourue par des fermions de masse nulle.]{Variation domains
of the tension, plotted in unit of $M^2$, for spacelike currents with
$-4\pi<\I<0$, in the $(\Rhop_\chi/M,
\Rhop_\psi/M)$ plane. These regions have the same geometrical
properties as the line density energy ones in Fig.~\ref{figuiposip},
but this time, the zones with vertical hatches are domains where
$\widehat{T}$ could grow with respect to $\Rhop_\chi$, whereas the
horizontal ones correspond to growth with respect to $\Rhop_\psi$.
For reasonable values of $\I$, the discrete values of $\Rhop$,
represented by dots, are out of these regions, and the tension always
decrease with both fermion densities.}
\label{figtinegip}
\end{center}
\end{figure}

\begin{figure}
\begin{center}
\epsfig{file=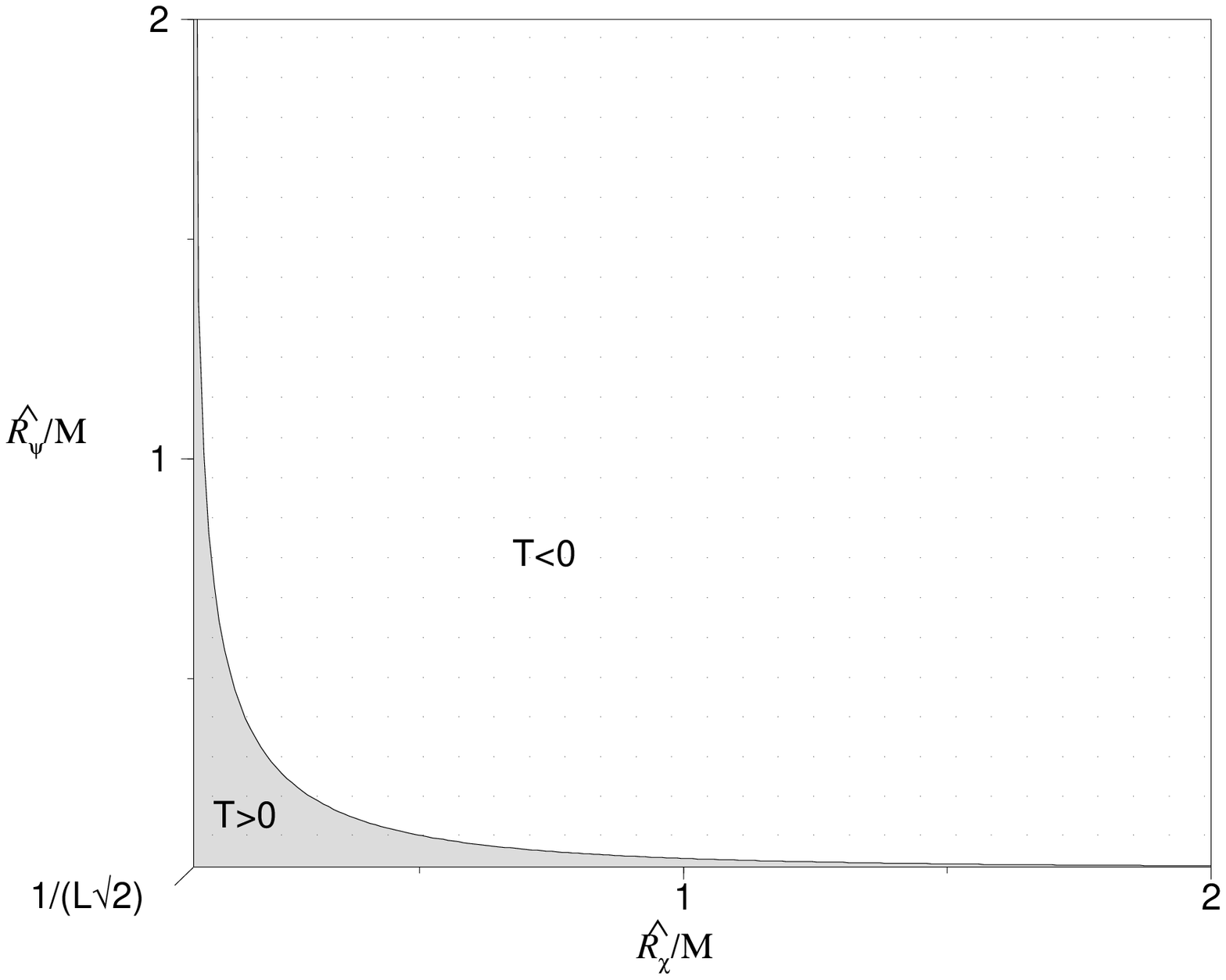,width=7cm}
\epsfig{file=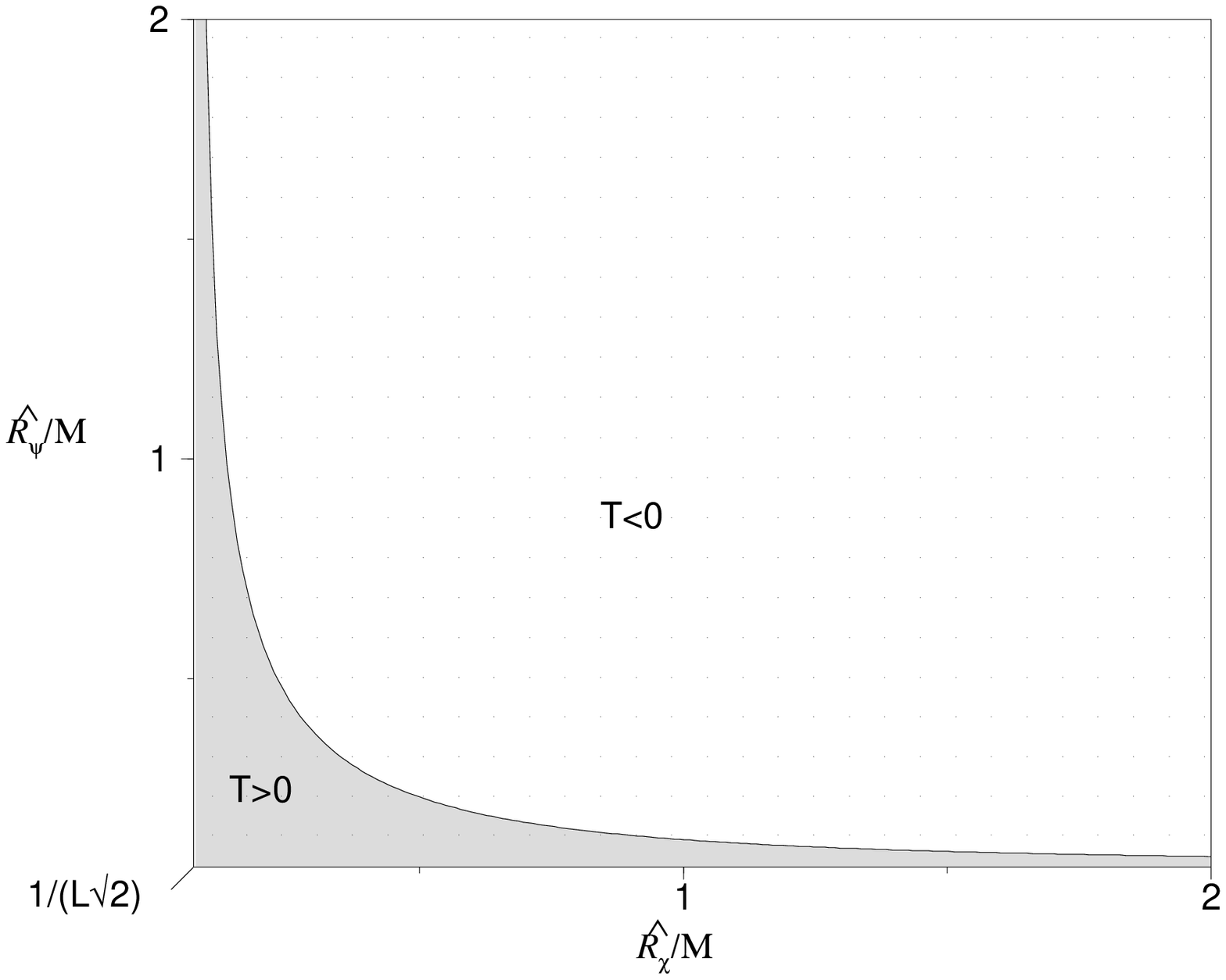,width=7cm}
\caption[Influence du champ de jauge g\'en\'er\'e par les modes z\'eros sur
la stabilit\'e de la corde.]{Sign of the tension for timelike currents
with $\I>0$, and spacelike currents with $-4\pi<\I<0$, the curves have
been plotted with $L_\ue=10/M$ in aim at clearly separating the
various regions. In the timelike case, the domain where the tension is
positive is closed, the null tension curves intersecting the axes at
$\Rhop=\sqrt{2}L_\ue M^2/\I$, whereas for spacelike currents, the null
tension curves tends asymptotically to
$\Rhop=(2\pi/L_\ue)\sqrt{2/(16\pi^2-\I^2)}$. Recall that the regions
where $T<0$ are unstable with respect to transverse string
perturbations according to the macroscopic formalism.}
\label{figtnull}
\end{center}
\end{figure}

For timelike fermionic gauge current, $\I>0$, the tension
decreases faster than in the zeroth order case, with the damped
fermionic densities $\Rhop_\chi$ and $\Rhop_\psi$ as on
Fig.~\ref{figtipos}, and reaches negative values at large densities
(see Fig.~\ref{figtnull}). The back reaction just increases the
slope of the surface, and thus, the negative values are reached
more rapidly. As for the energy per unit length, the equivalent length
was chosen equal to $L_\ue=10/M$ in the following figures.

For spacelike fermionic gauge currents, $-4\pi<\I<0$, the back reaction
damps the decrease of the tension with respect to the damped fermion
densities. There are also tiny regions near the axis, with areas
inversely proportional to $L_\ue$, and where $\widehat{T}$ could grow with
respect to one of the state parameters $\Rhop_\chi$ or $\Rhop_\psi$
(see Fig.~\ref{figtinegip}). As previously, for reasonable values of
$\I$, the first discrete values of the parameters are out of these
domains, and the tension always decreases with both fermion densities.
Finally, the tension reaches negative values at large damped
fermion densities (see Fig.~\ref{figtnull}).

\subsubsection{Relevant values of the parameters}
\label{relevant}

The previous derivation of the back reaction is built on the
classical vortex background and it is acceptable only if the
backreacted gauge fields do not perturb appreciably the Higgs and
orthoradial gauge fields profiles (see Fig.~\ref{figbackZM}). From
Eq.~(\ref{higgsmvt}), it will be the case only if $Q^t Q_t$ and $Q^z
Q_z$ can be neglected compared to $Q^\theta Q_\theta$. From
Eq.~(\ref{newgaugetilde}), and $Q_\theta Q^\theta \sim  m_\ub^2$, this
condition reads
\begin{eqnarray}
\frac{(q \cphi)^2 \Delta B \Sigma B}{Q_\theta Q^\theta} \sim
\frac{\Rho_\chi \Rho_\psi \sin{\left(\Theta_\chi-\frac{\pi}{4} \right)}
\sin{\left(\Theta_\psi -\frac{\pi}{4} \right)}}{\pi \eta^2}
\frac{m_\ub^2}{\pi \eta^2} \Delta \tilde{Q} \Sigma \tilde{Q} & \ll & 1,
\end{eqnarray}
or as function of $\I(\Theta_\psi,\Theta_\chi)$ and the damped fermion
densities,
\begin{eqnarray}
\label{domainok}
\I(\Theta_\chi,\Theta_\psi) \frac{\Rhop_\chi \Rhop_\psi}{\pi \eta^2}
& \ll & 1.
\end{eqnarray}
This condition is satisfied for damped fermion densities small
compared to the string energy scale, or for tiny values of the
function $\I(\Theta_\psi,\Theta_\chi)$.
Moreover, the backreacted gauge fields need to be small in
order to not significantly perturb the zero modes. From the equations of
motion (\ref{psimvt}) and (\ref{chimvt}), this condition leads to
$\Delta B, \Sigma B \ll \Rho$ and from Eq.~(\ref{newgaugetilde}) to
\begin{eqnarray}
\label{devlimok}
\frac{m_\ub^2}{\eta^2} \frac{c_{\psi(\chi)}}{\cphi} \Sigma (\Delta)
\tilde{Q}(0) \ll 1.
\end{eqnarray}
On the other hand, the maximum value of $\I$ in Eq.~(\ref{ifonction})
is clearly obtained when there are only particles or anti-particles
trapped in the string ($\Theta=0$ or $\Theta=\pi$), and deriving the
order of magnitude of the numerical integral in Eq.~(\ref{numintegspin}),
using equations (\ref{newgaugemvt}) and (\ref{fchi})-(\ref{fpsi}), one
shows that the large values of $\I$ (as $\I>4\pi$) can only be obtained
for model parameters which verify
\begin{equation}
\frac{m_\ub^2}{\eta^2} \frac{c_{\psi(\chi)}}{\cphi} \Sigma (\Delta)
\tilde{Q}(0) > 1.
\end{equation}
As a result, in order for the backreacted gauge fields not to modify
the equations of motion of the fermions at first order, the function
$\I$ has to be much smaller than $4\pi$. If it is not the case, then
the previous zero modes are no longer valid solutions and the
relevant equations of motion, in the case of the $\Psi$ fermions,
now read, from Eq.~(\ref{psimvt})
\begin{equation}
\label{pertzero}
\begin{array}{lll}
\vspace{4pt}
\displaystyle
\left[\frac{\ud\xi_1}{\ud r} + \frac{1}{r}\left(-q \cpsir B + m_1\right)
\xi_1 \right] - i g \varphi
\xi_4 & = & i [\mp(k+\omega) + q\cpsir \Delta B] \xi_2, \\
\vspace{4pt}
\displaystyle
\left[\frac{\ud\xi_2}{\ud r} + \frac{1}{r}\left(q \cpsir B - m_2\right)
\xi_2 \right] - i g \varphi
\xi_3 & = & i [\pm(k-\omega) - q\cpsir \Sigma B ] \xi_1, \\
\vspace{4pt}
\displaystyle
\left[\frac{\ud\xi_3}{\ud r} + \frac{1}{r}\left(-q \cpsil B + m_3\right)
\xi_3 \right] + i g \varphi 
\xi_2 & = & i [\mp(k-\omega) + q\cpsil \Sigma B] \xi_4, \\
\displaystyle
\left[\frac{\ud\xi_4}{\ud r} + \frac{1}{r}\left(q \cpsil B - m_4\right)
\xi_4 \right]  + i g \varphi
\xi_1 & = & i [\pm(k+\omega) - q\cpsil \Delta B] \xi_3,
\end{array}
\end{equation}
where the angular dependence has not been written owing to
Eq.~(\ref{angseparation}) and assuming $m_1=m_3+n$.
The zero modes seem to acquire an effective mass proportional to
$\Delta B$ or $\Sigma B$. More precisely, they are no longer
eigenstates of the $\gamma^0 \gamma^3$ operator since new spinor
components appear [$\xi_1$, $\xi_4$ here, see Eq.~(\ref{zeromodes})
and Eq.~(\ref{antizeromodes})].
It is clearly a second order effect since the gauge coupling constant
$q$ can be removed in the previous equations (\ref{pertzero}) using
Eq.~(\ref{newgaugetilde}) and assuming
\begin{equation}
\xi=X+q^2\delta\xi,
\end{equation}
with $X$ the zero mode solution $k=-\omega$ for $\Psi$ fermions
[see Eq.~(\ref{psisolutions})], and $q^2\delta\xi$ the perturbation
induced by the back reaction~\cite{ringeval3}. As a result, for strong
back reaction, the semi-classical approach can no longer be
used, since such second order effects appear as the semi-classical
manifestations of the one loop quantum corrections, and thus, only a
full quantum theory would be well defined.
However, if there is only one kind of fermion trapped in the
string, $\Psi$ say, the zero modes are not affected by the
back reaction since $\Delta B$ is only generated from the $\Chi$
current, and therefore vanishes [see Eq.~(\ref{newgaugemvt})], so
$\xi=X$ is always solution of the equations of motion
(\ref{pertzero}), and identically for the $\Chi$ zero modes
alone~\cite{moreno}.
Note, that there is no contradiction with the usual index
theorem since it is derived for Dirac operators, and thus without
backreacted fields. This just shows that the modes propagating in
the vortex with strong backreacted gauge fields are no longer well
described by the usual zero modes.
Physically, it might be the signature of a tunneling of the zero modes
to another states. The massive modes which have not been considered
here could be more relevant in such cases.

On the other hand, the shape of the string might allow the fermion
densities $\Rhop_\chi$ and $\Rhop_\psi$ to reach the tiny regions
where the energy decreases with one of them (see
Fig.~\ref{figuiposip}), by means of the zero mode vacuum quantum
effects. The present toy model does not involve the effect of the
radius of curvature $R$ of the string, and it is reasonable that the
contribution of the zero mode vacuum to the energy per unit length
involves $R$ through a redefinition of $L_\ue$. If $L_\ue$ becomes
smaller than $L$, the first discrete values of the fermion densities
could be inside the hatched regions in Fig.~\ref{figuiposip}, since
the discrete values of the fermion densities only depend on the
physical length of the string $L$. Note that it would therefore be
necessary that the zero mode vacuum energy is negative, which is not
the case without curvature in the simple framework of
Sec.~\ref{toyvac}. Such effects could be relevant for vorton
stability, as, for a small radius of curvature, the string could
become unstable to fermion condensation.

Finally, the model can be used only at the tree order, and the
conditions (\ref{domainok}) and (\ref{devlimok}) are the validity
criteria of the above derivations.

\section{Comparison with the scalar case}

Owing to the fermionic two-dimensional quantization along the
string, the energy per unit length and the tension of a string
carrying massless fermionic currents have been derived up to the
first order in back reaction corrections. The state of the string
is found to be well defined with four state parameters which are
the densities of each fermion trapped in the string, and asymmetry
angles between particles and anti-particles in each fermion
family. It seems quite different than the bosonic charge carriers
case, where the current magnitude is the only relevant state
parameter~\cite{neutral}, however, this is the result of the
allowed purely classical approach where the superposition of many
quantum states can be view as only one classical state owing to
the bosonic nature of the charge carriers. As a result, there is a
degeneracy between the number of bosons trapped in the string and
the charge current. The quantization introduced to deal with
fermions naturally leads to separate the charge current from the
particle current through the existence of anti-particle
exitations. Moreover, the magnitude of the current can only
modify the equation of state at non-zeroth order because the
chiral nature of fermions trapped in the string requires
simultaneous exitations between the two families to lead to
non-lightlike charge currents.

Nevertheless, some global comparisons can be made with the scalar
case. First, for reasonable values of $\I \ll 4\pi$, the energy per
unit length grows with the fermionic densities, whereas the tension
decreases with them.  However, note the relevant parameter for the
change in behaviors of the tension and line density energy is the
function $\I(\Theta_\chi,\Theta_\psi)$ instead of the current
magnitude in the scalar case. As it was said, $\I$ quantifies, through
the asymmetry between the number of particles and anti-particles trapped
in the string, the efficiency of the charge current per particle to be
timelike or spacelike. The more positive is $\I$, the more timelike the
fermionic charge current per particle will be, and conversely the more
negative $\I$ is, the more spacelike it will be. Once again, this
difference with the scalar case appears as a result of the degeneracy
breaking between particle current and charge current due to the
fermionic nature of the charge carriers.

The stability of the string with respect to transverse perturbations
is given from the macroscopic formalism by the sign of the tension
(for line density energy
positive)~\cite{carter89,carter89b,carter94b,carter97}, and we find
that instabilities always occur for densities roughly close to
$M/\sqrt{4\pi+\I}$, in finite domain for timelike current, in infinite
one for spacelike currents with $-4\pi<\I<0$.  Another new results are
obtained from the multi-dimensional properties of the equation of
state, in particular the problem of stability with respect to
longitudinal perturbations differs from the scalar barotropic case
where the longitudinal perturbations propagation speed is given by
$\cl^2= -dT/dU$~\cite{carter89,carter89b,carter94b,carter97}, and
therefore its two-dimensional form has to be derived to conclude on
these kinds of instabilities. Nevertheless, by analogy with the scalar
case, since, in the non-perturbed case and in the infinite string
limit, the equation of state verifies $U+T=2M^2$~\cite{prep}, the
longitudinal perturbation propagation speed might be close to the
speed of the light, even with small back reaction, and therefore, only
transverse stability would be relevant in macroscopic string stability
with massless fermionic currents.

\section{Conclusion}

The energy per unit length and the tension of a cosmic string
carrying fermionic massless currents were derived in the frame of
the Witten model in the neutral limit. Contrary to bosonic charge
carriers, the two-dimensional quantization required to deal with
fermions, leads to more than one state parameter in order to
yield a well-defined equation of state. They can be chosen, at
zeroth order, as fermion densities trapped in the string
regardless of charge conjugation. The minimal back reaction
correction appears through the fermionic charge current magnitude
which involves the asymmetry angles between the number of
particles and anti-particles trapped in the string, and which
might be identified with the baryonic number of the plasma in
which the string was formed during the phase transition. As a
result, it is shown that fermionic charge currents can be
lightlike, spacelike as well as timelike. Moreover the line
energy density and the tension evolve globally as in the bosonic
charge carriers case, but it was found that the tension can take
negative values in extreme regions where the fermion densities are
close to the string mass, and where the string is therefore
unstable with respect to transverse perturbations according to
the macroscopic formalism.

The present model has been built on the generic existence of
fermionic zero modes in the string and follows only a
semi-classical approach. It is no longer valid for higher
corrections in the back reaction when they modify notably the vortex
background and seem to give effective mass to the previous zero
modes. It may be conjectured, at this stage, that in a full
quantum theory, the quantum loop corrections give mass to the
zero modes for high currents and consequently might lead to their
decay by the mean of massive states. Only chiral charge currents could
be stable on cosmic string carrying large fermionic massless currents
in such a case. Another possible effect, relevant for vortons stability,
may be expected for loops with small radius of curvature, by means of
the vacuum effects which could render the loop unstable to fermion
condensation.

It will be interesting to quantify such modifications on the equation of
state in future works, as the effects of worldsheet curvature, and the
modification of the density line energy and tension by the massive
bound states. The field of validity of the model could therefore be
extended to higher energy scales which would be more relevant for vortons
and string formation.

\section*{Acknowledgments}

I would like to thank P. Peter for many fruitful discussions, and for
his help to clarify the presentation. I also wish to thank B. Carter
who helped me to enlighten some aspect of the subject.

\chapter{Approche macroscopique (article)}
\label{chapitrezeromacro}
\minitoc
Dans le chapitre pr\'ec\'edent, la quantification des modes z\'eros le
long de la corde a permis de montrer que, lorsque la r\'etroaction
\'etait n\'eglig\'ee, la dynamique de la corde \'etait r\'egie par une
\'equation d'\'etat de type trace fix\'ee $U+T = 2M^2$. Bien que la
quantification introduise naturellement plus d'un param\`etre
d'\'etat, il doit \^etre possible, par le formalisme covariant, de
red\'efinir un param\`etre effectif \`a partir de cette \'equation
d'\'etat.

Apr\`es avoir justifi\'e l'approximation de temp\'erature nulle
utilis\'ee dans la quantification (voir Chap.~\ref{chapitrezero}), la
relation $U+T=2M^2$ est retrouv\'ee par une approche purement
macroscopique. La description lagrangienne permet ensuite d'en
d\'efinir une fonction ma\^\i tresse de d\'ependant que d'un seul
param\`etre d'\'etat. Ce chapitre est essentiellement un compl\'ement
aux \emph{proceedings} de la conf\'erence des \journal{Journ\'ees
Relativistes 2001}~\cite{prep}.

\newpage

\begin{center}
{\Large \textbf{
Fermionic current-carrying cosmic strings: \\
zero-temperature limit and equation of state
}}
\end{center}
\vspace{5mm}
\begin{center}
Patrick Peter and Christophe Ringeval
\end{center}
\vspace{5mm}
\begin{center}
{\footnotesize{
Institut d'Astrophysique de Paris, 98bis boulevard
Arago, 75014 Paris, France.
}}
\end{center}
\vspace{5mm}
\begin{center}
\begin{minipage}[c]{14cm}
{\footnotesize \textbf{
The equation of state for a superconducting cosmic string whose
current is due to fermionic zero modes is derived analytically in the
case where the back reaction of the fermions to the background is
neglected. It is first shown that the zero mode fermions follow a zero
temperature distribution because of their interactions (or lack
thereof) with the string-forming Higgs and gauge fields. It is then
found that the energy per unit length $U$ and the tension $T$ are
related to the background string mass $m$ through the simple relation
$U+T= 2m^2$.  Cosmological consequences are briefly discussed.
}}
\end{minipage}
\end{center}

\section{Introduction}

Topological defects~\cite{NATO} have been considered in various
physical situations, e.g. in the context of condensed matter and
cosmology~\cite{kibble76,kibble80,book,book2}. In many cases of
interest in cosmology~\cite{bennett90,bouchet88,allen90,bouchet02}, they
can be approximated as structureless, the relevant dynamics being
often assumed not to depend on any specific choice of their internal
content. For cosmic defects, this internal content would correspond to
the particles that couple to the string-forming Higgs
field~\cite{witten}. However, in the latter example of cosmic strings,
it was shown that such a structure might lead to drastic modifications
not only of these object
dynamics~\cite{carter89,carter89b,carter94b,carter97,neutral}, which
could be seen as a mere academic situation given our present ignorance
on their very existence, but also, because of the appearance of new
accessible equilibrium states, of the cosmological setting, leading in
some instances to actual
catastrophes~\cite{davisRL88,carter91,brandi96,rdp}. To make a long
story short, let us just say that currents imply a breakdown of the
Lorentz-boost invariance along the string worldsheet, thereby allowing
loop configurations to rotate, the centrifugal force hereby induced
having the ability to sustain the loop tendency to shrink because of
the tension. The resulting states, called vortons, might be stable
even over cosmological timescales, scaling as matter and thus rapidly
coming to dominate the Universe evolution, in contradiction with the
observations. This leads to constraints on the particle physics
theories that predict them at energy scales that are believed to be
unreachable experimentally (in accelerators say) in the foreseeable
future.

Unfortunately, it appears that the string structure, contrary to
their counterparts as fundamental objects~\cite{superstring}, is
not determined by any consistency relation, and is therefore
somehow arbitrary, at least at the effective description
level~\cite{witten}. This means in practice that in order to be
able to tell anything relevant to (cosmic) string cosmology, one
needs to set up a complete underlying microscopic model, arising
say, from ones favorite Grand Unified Theory (GUT)~\cite{davisS}
or some low-energy approximation of some superstring-inspired
model~\cite{dine87,atick87,casas89,harvey89,binetruy98b}.

Some generic constructions can however be arranged, as was shown to be
the case whence a bosonic condensate gets frozen in the string
core~\cite{carter89,carter89b,carter94b,carter97,neutral}. In such a
situation, the boson field phase $\varphi$, thanks to a random
Kibble-like mechanism, may wind along the string itself, thereby
producing a current that turns out to be essentially a function of a
single state parameter $w$, thus expressible as a phase gradient as
\begin{equation} w \equiv \kappa_0 \gamma^{ab} \partial_a \varphi
\partial _b \varphi,\label{w}\end{equation}
with indices $a,b,...$ varying within the string worldsheet
coordinates defined by the relations
\begin{equation} x^\mu = X^\mu_{_{\uS}} (\xi^a), \ \ \ \ \ \
\xi^a \in \{\tau,\sigma\},\end{equation}
and $\gamma^{ab}$ the inverse of the induced metric defined with the
background metric $g_{\mu\nu}$ as
\begin{equation}\gamma_{ab} = g_{\mu\nu} {\partial
X^\mu_{_{\uS}} \over \partial \xi^a} {\partial X^\nu_{_{\uS}} \over
\partial \xi^b}.\label{gammaab}\end{equation} A straightforward
generalization of the Nambu-Goto action is then provided by the
$w-$weighted measure as
\begin{equation} {\cal S} = -m^2 \int \hbox{d}^2\xi \sqrt{-\gamma}
{\cal L} \{w\},\end{equation} with $m$ the typical mass scale of
symmetry breaking leading to string formation and $\gamma$ the
determinant of the induced metric~(\ref{gammaab}).  Reasonable
microscopic models~\cite{carpet1} then yield approximate forms for
the Lagrangian function ${\cal L}\{w\}$, out of which the
dynamical properties of the corresponding strings can be
derived~\cite{larsen,carter97,gangui98}.

Such a description remains however essentially classical even
though an alternative formalism, also proposed by
Carter~\cite{sigma}, in terms of a dilatonic model, appears more
suitable for quantization. This last formalism however, being
fully two-dimensional, cannot be used to derive interesting
quantities such as the relevant cross-sections for trapped
excitations to leave the string worldsheet. This is unfortunate
since this is precisely the information one would need for
cosmological applications~\cite{davisRL88,carter91,brandi96,davisS}.

It would therefore seem that by considering fermionic current carriers
instead of bosonic ones, one would, because of the intrinsically
quantum nature of fermions, obtain a more appropriate
description~\cite{davisS,ringeval}. Besides, fermions are trapped in
topological defects because of Yukawa couplings with the string
forming Higgs field in the form of zero modes~\cite{jackiwrossi}, so that
their dynamics is described by simple (although coupled) Dirac
equations, which are linear. In the bosonic case, the non-linear
(quartic) term is essential in order to ensure the dynamical stability
of the condensate so that a solitonic
treatment~\cite{christ75,tomboulis75,rajamaran82,hp} seems the only
way to deal with the underlying quantum physics. This fact
dramatically complicates matters and as a result, a complete
description yet fails to exist.

However, the fermionic case is not that simple either as here, one
faces another technical difficulty for the classical description: it
can be shown that there doesn't exist a simple state
parameter~\cite{ringeval}. An arbitrary spacelike or timelike current
can only be built out of at least two opposite chirality spinor fields
and will be given by the knowledge of four occupation numbers per unit
length. As this is true in particular at least in the zero-temperature
limit, we shall be concerned here first with this limit whose validity
was assumed to depend on the particular model under
consideration~\cite{davisS}. In the following section, we show that
setting the temperature to zero is always a good approximation because
of the couplings between the fermionic fields and the string-forming
Higgs and gauge fields\footnote{Otherwise, the temperature itself
could be taken as a state parameter, so that the usual formalism would
be applicable~\cite{vilenkin90,carter90b,carter94}}. These results
would seem to imply that the previously derived macroscopic formalism
is irrelevant to the fermionic situation (see Ref.~\cite{carter94b} for a
many-parameter formalism). In practice however, as a simple
relationship between the energy per unit length and the tension can be
found for fermionic currents, the single state parameter formalism can
be used that permits to draw some cosmological consequences.

\section{The zero-temperature limit}

In order to have an arbitrary current built upon fermionic fields,
one needs at least two Dirac fermions~\cite{witten} $\Psi$ and
$\chi$ say, coupled to the string-forming Higgs field $\Phi$
through Yukawa terms as well as to the associated gauge field
$B_\mu$, the later acquiring a mass from the vacuum expectation
value (VEV) of the Higgs field. Fermions may condense in the
string core in the form of zero modes, and by filling up the
accessible states, one forms a current which can be timelike,
spacelike, or lightlike. If one wants to give a classical
description of such a current-carrying string, one must be in a
configuration for which quantum effects are negligible. Such
quantum effects, as for instance tunneling outside the vortex,
will indeed be negligible provided most of the fermions are on
energy levels whose excitation energy is much below their vacuum
mass, the latter thus playing the role of a Fermi energy.

As a result, if a temperature may be defined for the fermions along
the vortex, quantum effects will be negligible if the temperature is
small compared to the vacuum mass of the fermions, which essentially
imply a zero temperature state. Note that the situation we are having
in mind is reached only whenever the background temperature is low
compared to that at which the string formed for otherwise interactions
with the surrounding plasma could populate the high energy levels. In
practice, this is what will happen at the time of string formation,
and if the fermions did not interact at all, or only through time
reversible interactions, one would be left with a frozen distribution
corresponding to a high temperature state~\cite{bcpc}.

That this is not the case can be seen through an exhaustive list
of all the possible fermion interactions. For that purpose, it
must be emphasized that fermionic condensates arise in the string
core in the form of zero modes, i.e. chiral states\footnote{We
do not consider here the possible massive bound states as those
are expected~\cite{davisS} to interact with each other because of
diagram $(c)$ of the figure and therefore move rapidly away from
the string core.}. One has essentially two coupling
possibilities, namely a coupling of the fermion with the Higgs
field or with the gauge field, illustrated on the figure. The
first case (diagram $a$) is seen to vanish identically in the
case of a chiral mode~\cite{ringeval} so we shall not consider it.
The second case is more interesting and comes from the second
diagram ($b$) of the figure. In this case, any trapped fermionic
zero mode is seen to be able to radiate a gauge vector boson.
These terms do not vanish identically, and in fact can be seen to
be the source for some back reacted components of the gauge field:
as all the vectors emitted this way will eventually condense into
a classical field, this diagram contributes to a small (and
indeed usually negligible~\cite{ringeval}) contribution in the
energy per unit length and tension.

\begin{figure}[h]
\begin{center}
\epsfig{file=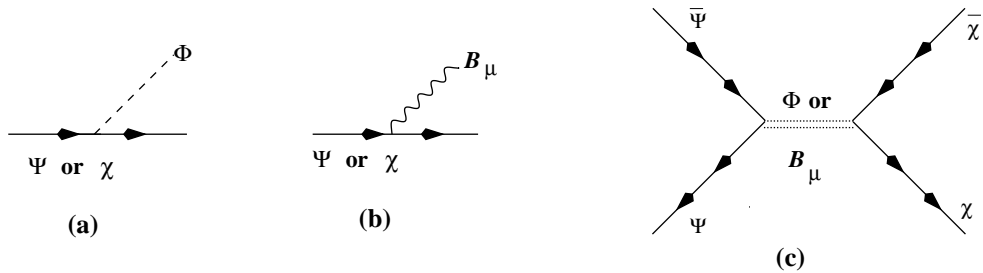,width=13cm}
\caption[Int\'eractions particulaires dans une corde cosmique
poss\'edant des fermions de masse nulle.]{Exhaustive list of all the
possible interaction terms between fermions and Higgs and gauge
fields. $(a)$ Higgs radiation by a fermion field denoted by $\psi$ or
$\chi$ (two fermion at least are necessary in order to generate an
arbitrary kind of current), $(b)$ gauge boson radiation, $(c)$
coupling between the various fermion fields.}
\end{center}
\end{figure}

Finally the third term, $(c)$, of the figure, represents a
would-be interaction term between the fermions that could be
responsible for an equilibrium configuration. When one considers
only zero modes in a usual model, such a term actually vanishes
because both fermions must have opposite chiralities in order for
the theory to be well defined. As a result, interactions between
fermions turn out to be negligible.

We are now in a position to understand the microphysics of what
might happen inside a fermionic current-carrying cosmic string.
First, when the fermions become trapped in the string core, they
do so on arbitrary high energy levels inside the string. Then,
they have the possibility to radiate most of their energy away in
the form of the vector field, thereby creating the back reacted
component. Note that the vector field itself can also interact
with the Higgs field thereby producing an effective lack of
symmetry between diagram $(b)$ and its time reversed counterpart.
This implies that the overall effect is indeed a radiative decay
and not an equilibrium. Thus, all the populated states end up
being the lowest reachable states. In practice, that means that
the effective temperature of the fermion gas is vanishing.
Moreover, such a configuration in turn is stable as no
interaction between the various fermion field can be present.

\section{Fermionic string equation of state}

As was discussed above, fermions that are coupled to the Higgs
field may be bound to cosmic strings in the form of zero
modes~\cite{witten,ringeval,jackiwrossi} and therefore the current they
generate arises from lightlike components. We recall briefly the
formalism necessary to handle this case and then move on to
derive the resulting stress-energy tensor as well as a simple
relationship relating its eigenvalues. This leads to a plausible
classical description in terms of a state parameter whose
validity is discussed.

\subsection{Stress-energy tensor and equation of state}

Chiral currents have special properties and need be studied on
their own. As was shown earlier, a phase gradient formalism hold
for them, similar to the $\sigma-$model~\cite{sigma} with
vanishing potential, namely~\cite{carpet2}
\begin{equation} {\cal S}_\sigma = - \int \hbox{d}^2\xi \sqrt{-\gamma}
\left( m^2 +{1\over 2} \psi^2 \gamma^{ab} \partial_a\varphi \partial_b
\varphi \right),\label{chiral}\end{equation}
where the normalization constant $\kappa_0$ of Eq.~(\ref{w}) in the
$w-$formalism has now been promoted to a dynamical field whose
variations lead to an ever-lightlike current.

Considering now a system of fermionic zero modes, and neglecting
back reaction, one may arrive at the conclusion that the relevant
action describing a general fermionic current-carrier cosmic string
will be given by
\begin{equation} {\cal S}_{_F} = \int \hbox{d}^2\xi \sqrt{-\gamma}
{\cal L} \end{equation}
where the Lagrangian function ${\cal L}$ is
\begin{equation}{\cal L} =-m^2 -{1\over 2} \sum_i^N \psi^2_{(i)} \gamma^{ab}
\partial_a\varphi_{(i)} \partial_b
\varphi_{(i)},\label{manych}\end{equation} $N$ being the number of
fermionic degrees of freedom (at least four~\cite{ringeval} if the
model is to describe arbitrary spacelike as well as timelike and
chiral currents). Having obtained the action, it is now a simple
matter to derive the corresponding dynamics by varying it with respect
to the various fields involved. However, as we show below, it turns
out not to be strictly necessary as many consequences, in particular
in cosmology, stem directly from the energy momentum tensor
eigenvalues $U$ and $T$, i.e. respectively the energy per unit length
and tension, for which a very simple relationship is now derived.

First of all, as all the fields $\psi_{(i)}$ are independent, it is
evident that variations of (\ref{manych}) lead to the chirality
condition on the various currents induced by the phase gradients,
namely
\begin{equation} {\delta {\cal S}_{F}\over \delta \psi_{(i)}} = 0 \ \
\Longrightarrow \ \ \ \gamma^{ab} \partial_a\varphi_{(i)}
\partial_b \varphi_{(i)} = 0 \ \ \ \forall i\in
[1,N],\label{dfdf}\end{equation} while the currents are obtained
through variations with respect to the phases themselves
\begin{equation} {\delta {\cal S}_{F}\over \delta \varphi_{(i)}} = 0 \ \
\Longrightarrow \ \ \ \nabla_a \left(\psi^2_{(i)} \gamma^{ab}
\partial_b \varphi_{(i)}\right) \equiv \nabla_a j^a_{(i)} =
0, \ \ \ \forall i\in [1,N].\label{curr}\end{equation}
Eq.~(\ref{dfdf}) can be seen to imply, in our two-dimensional
case, that each function $\varphi_{(i)}(\xi^a)$ separately is
harmonic, i.e.
\begin{equation} \gamma^{ab} \nabla_a \nabla_b \varphi_{(i)} = 0\ \ \
\forall i\in [1,N],\label{harm}\end{equation} so that the current
conservation equations (\ref{curr}) can be cast in the form
\begin{equation} \gamma^{ab} \nabla_a \psi_{(i)} \nabla_b
\varphi_{(i)} = 0\ \ \ \forall i\in [1,N],\end{equation} which
means that every $\psi_{(i)}$ is a function of $\varphi_{(i)}$
only, for any fixed value of $i$. As a result, the formalism
really describes only $N$ degrees of freedom, and not $2N$, and
one may interpret the phase gradients as occupation numbers per
unit length in a given underlying fermionic current-carrying
model.

The stress energy tensor is now obtained by the standard procedure of
variation with respect to the metric, i.e.,
\begin{equation} \overline T^{\mu\nu} = 2{\delta {\cal L}\over
\delta g_{\mu\nu}} +{\cal L} \eta^{\mu\nu},\end{equation} where
the first fundamental tensor of the
worldsheet~\cite{carter89,carter89b,carter94b,carter97}
\begin{equation} \eta^{\mu\nu} = \gamma^{ab} {\partial
X^\mu_{_{\uS}} \over \partial \xi^a} {\partial X^\nu_{_{\uS}} \over
\partial \xi^b}\label{eta}\end{equation} is definable in terms of the
eigenvectors $u^\mu$ and $v^\mu$, respectively timelike and
spacelike ($u_\mu u^\mu = - v_\mu v^\nu = -1$), of the stress
energy tensor for a non-chiral current:
\begin{equation} \eta^{\mu\nu} = v^\mu v^\nu - u^\mu u^\nu,
\end{equation}
and
\begin{equation} \overline T^{\mu\nu} = U u^\mu u^\nu - T v^\mu v^\nu
\label{UT}.\end{equation}

The stress-energy tensor now reads
\begin{equation} \overline T^{\mu\nu} = - m^2\eta^{\mu\nu}
+{1\over 2} \sum_i^N \psi_{(i)}^2 \left( X^\mu_{_{\uS},c}
X^\nu_{_{\uS},d} \varphi_{(i)} ^{,c}\varphi_{(i)} ^{,d} -
\eta^{\mu\nu} \varphi_{(i),a} \varphi_{(i)}
^{,a}\right)\label{stress}\end{equation} with a comma denoting
partial differentiation with respect to a worldsheet coordinate.
The last term identically vanishes because of the on-shell
relation (\ref{dfdf}), and we can now compute the eigenvalues by
projecting on the eigenvectors as
\begin{equation} U = \overline T^{\mu\nu} u_\mu u_\nu = m^2 + {1\over 2}
\left( \sum_i^N \psi_{(i)}^2 \varphi_{(i)}^{,a} \varphi_{(i)}^{,b}
\right) X^\mu_{_{\uS},a} X^\nu_{_{\uS},b} u_\mu u_\nu,
\end{equation}
and
\begin{equation} T = - \overline T^{\mu\nu} v_\mu v_\nu
= m^2 - {1\over 2} \left( \sum_i^N \psi_{(i)}^2
\varphi_{(i)}^{,a} \varphi_{(i)}^{,b} \right) X^\mu_{_{\uS},a}
X^\nu_{_{\uS},b} v_\mu v_\nu.
\end{equation}
By adding these two equations up, one gets that the last term is
proportional to the first fundamental tensor $\eta^{\mu\nu}$
projected onto the worldsheet coordinates, i.e. a term
proportional to the induced metric $\gamma_{ab}$. As each part of
the current is made of chiral fields, this last term eventually
cancels out and one is left with [a relation obtainable directly
by taking the trace of the stress tensor~(\ref{stress})]
\begin{equation} U+T = 2m^2,\label{UTm2}\end{equation}
which will be our final equation of state for a fermionic current
carrying cosmic string in the zero temperature limit. Note that
this relation holds in the specific case of the model discussed
in Ref.~\cite{ringeval} whenever one neglects the fermion
back reaction on the string fields.

\subsection{A macroscopic model}

Let us now discuss the various implications of this result. The
most important point related with cosmological models involving
current-carrying strings concerns vorton stability. As such a
model is exclusively classical in nature, we shall not examine the
quantum stability here, especially since this was already
discussed in Ref.~\cite{davisS}. Before however turning to this
physical point, we should like to stress a simple technical detail
concerning the equation of state itself.

As we have said, a fermionic current-carrying cosmic string does
not in general admit a classical description in terms of a single
state parameter. However, in the case where a functional
relationship exists between the energy per unit length and the
tension, as is indeed what happens in the situation under
consideration here, a state parameter can easily be derived as
follows.

Let us consider again the $w-$formalism. Performing the Legendre
transform
\begin{equation} \Lambda = {\cal L} - 2w{\hbox{d}{\cal
L}\over\hbox{d} w},\label{Legendre} \end{equation} it can be shown
that~\cite{carter89,carter89b,carter94b,carter97}, depending on the
timelike or spacelike character of the current, the energy per unit
length and tension can be identified, up to a sign, with ${\cal L}$
and $\Lambda$.  As a result, the knowledge of ${\cal L}$ as a function
of $\Lambda$ or, in other words that of $U(T)$, permits to integrate
Eq.~(\ref{Legendre}) to yield the state parameter through
\begin{equation}
\ln \left({w\over w_0}\right) = \int {\hbox{d}{\cal L}\over
2({\cal L} - \Lambda)},\label{state}
\end{equation} whose inversion, in turn, gives the functional
form of the Lagrangian ${\cal L}\{w\}$, up to a normalization
factor. Applied to our case, Eq.~(\ref{state}) implies immediately
\begin{equation} {\cal L}\{w\} = -m^2 -{w\over 2},
\label{linear}\end{equation} so we see that an arbitrary current
formed with many lightlike currents can be described by means of
a single state parameter with almost the simplest possible model;
in Eq.~(\ref{manych}), it suffices to replace the sum over the
many chiral models by the standard form of $w$, i.e.
Eq.~(\ref{w}), which can be viewed as the auxiliary field $\psi$
acquiring a fixed value ''on shell''. This is just the action of
Eq.~(\ref{chiral}) with $\psi^2 = \kappa_0$.

The model described by Eq.~(\ref{linear}) was however ruled out
as a valid description of a realistic Witten-like
current-carrying string in Ref.~\cite{carpet1}, so one may wonder
how it can be re-introduced here. There are two answers to that
question. First, it can be argued that most of the statements in
this reference applied to bosonic current-carriers, and have
therefore no reason to be true in the fermionic case, except that
bosons and fermions are known to be equivalent in two
dimensions~\cite{witten}. As a result, a classical description of
a vortex must somehow take into account the finite thickness
effects before averaging over the transverse degrees of freedom,
so that the string keeps a track of its $3+1$-dimensional nature.

The second, perhaps more important reason, why the model given by
Eq.~(\ref{linear}) was not considered seriously as a candidate to
describe a current-carrying string is the saturation effect. There
must indeed be a maximum current flowing along a string as individual
particles making the current are limited in energy because they are
bound states. In the case of bosons, thanks to Bose condensate, all
the particles are essentially in the same state and the saturation
effect stems from the non-linear (interaction) term between them. As
it turns out, even the interaction terms can be adequately treated
through a mean field approximation, so that a classical field
description is valid in this case. For fermions however, this effect
finds its origin in a completely different mechanism, related to the
exclusion principle: it is necessary, in order to increase the value
of the current, to add more particles on higher and higher energy
levels, up to the point where it becomes energetically favorable for
them to leave the worldsheet as massive modes. This is therefore a
purely quantum effect which cannot, of course, be properly taken into
account in the classical description developed here whose range of
validity is thus limited to small currents. Moreover, contrary to the
bosonic situation in which the boson mass enters explicitly as a
relevant dynamical parameter, fermionic zero modes exist independently
of the vacuum fermion mass, so there is no mass scale that could
determine the saturation regime in such a classical description.

The conclusion of the previous discussion is that the model described
by Eq.~(\ref{linear}) is indeed an accurate representation for
fermionic current-carrying cosmic strings provided the current is far
from the saturation regime. It should be emphasized that it will be
the case for most of the evolution of a network of such strings, so
that one is entitled, for cosmological application purposes
(e.g. numerical simulation), to derive the string dynamics with the
linear model.

\section{Consequences}

Let us now move to the consequences of such an equation of state.  We
shall assume for now on that the macroscopic formalism with the
Lagrangian given by Eq.~(\ref{linear}) is valid to describe a
fermionic carrier cosmic string, provided the string never leaves the
elastic regime. In other words, we shall assume that the string,
whatever its shape, has a curvature radius everywhere much larger than
its thickness and that the Fermi level is below the vacuum mass of the
fermion so that the quantum effects are negligible.

Given a functional relationship between the energy per unit length $U$
and the tension $T$, one can calculate the perturbation velocities
respectively as\cite{carter89,carter89b,carter94b,carter97}
\begin{equation} c^2_{_T} = {T\over U}\label{ct}
\end{equation}
for the transverse perturbations, and
\begin{equation} c^2_{_L} = -{\hbox{d}T\over \hbox{d}U}\label{cl}
\end{equation}
for the longitudinal ones. In the case at hand~(\ref{UTm2}), this
gives
\begin{equation} c^2_{_L} = 1, \ \ \ \ \ \ \ \ \ \ c^2_{_T} =
{2m^2\over U}-1< 1 \label{clct}
\end{equation}
since $m^2<U<2m^2$ by construction. On a plot $\ct^2$ versus $\cl^2$,
such an equation of state would therefore just be the line
$\cl^2=1$. For cosmological considerations, one may also consider for
instance the back reaction of the fermions on the background vortex
fields, or even the electromagnetic back reaction for charged
carriers. This means in practice, if one suppose that these will
indeed lead to corrections which in principle could not be properly
placed on such a diagram, that the corrected equation of state would
be a curve somewhere near the $\cl^2=1$ line.

The situation is exactly the opposite of what happens for a boson
field~\cite{martinpeter} for which it had been found that the equation
of state in this plot is a curve close to the $\ct^2=1$ line. One can
understand this result as a kind of duality between fermion and boson
condensates, the corresponding equations of state being roughly
symmetrical with respect to the line $\cl^2=\ct^2$. This adds further
insight on the fact that a purely 2-dimensional description is not
valid before the full field theory has been solved. It may be
conjectured at this point that a string carrying a current generated
by both fermions and bosons with an underlying supersymmetric
model~\cite{stern86,davisAC97} could produce an equation of state
exactly lying on the line $\cl^2=\ct^2$, i.e. the so-called fixed
determinant model (arising also from a Kaluza-Klein
projection~\cite{KK} or as a smoothed average description of the large
scale behavior of a simple Nambu-Goto model over the small scale
wiggles~\cite{martin95,vilenkin90,carter90b}) for which
$UT=m^4$. The advantage of this model, if the conjecture turned out to
be a reasonable approximation of a more realistic equation of state,
lies in its complete integrability~\cite{carter90b} in the case of
a flat background. Such a feature might be useful in network
simulations.

The last point that needs be mentioned here concerns vorton
stability. It was shown under rather general conditions that circular
loops reaching an equilibrium state thanks to a current may suffer
from classical instabilities, the fate of which presumably leading to
quantum effects~\cite{mp2}, provided the equation of state is in the
region above the $\cl^2=\ct^2$
line~\cite{cartermartin93,martin94}. Inclusion of the electromagnetic
corrections has also been achieved, showing that these can reduce the
number of vortons that can form during the loops
evolution~\cite{gangui98}, but that once they are formed, they are,
classically, more stable~\cite{vortelec}. In our case, if the
corrections do not change drastically the form of the equation of
state, the vortons would exist comfortably below the critical
line. Therefore, we expect them to be much more stable with respect to
classical perturbations. In fact, it is very hard to imagine anything,
except quantum background interaction~\cite{davisS}, that could
destabilize a vorton whose dynamics stems from the
Lagrangian~(\ref{linear}).

\section{Conclusion}

Fermionic zero modes trapped in cosmic strings are shown to
follow a vanishing temperature Fermi-Dirac distribution. This is
so because the chirality of the zero modes involved are such that
the only possible interaction of the fermions is through gauge
boson radiation, leading to an effective loss of energy (on
average). As a result, as strings are formed and fermions get
condensed along them in the form of zero modes, populating
arbitrary high energy levels, they have the possibility to decay
radiatively until they reach a zero temperature distribution.
Then, as all other interaction terms are identically vanishing,
they remain in this state which thus happens to be stable.

Assuming therefore such a vanishing temperature fermionic
current-carrying cosmic string, it turns out that the equation of
state relating the energy per unit length $U$ and the tension $T$
is of the self-dual~\cite{carpet1} fixed trace kind, namely
$U+T=2m^2$, with $m$ the characteristic string-forming Higgs
mass. Although fermionic carriers imply the need of more than one
state parameter, this implies that the simplest linear
Lagrangian~(\ref{linear}) provides a good approximation for a
classical description of such a vortex. This could in fact have been
anticipated as this is the only available equation of state that does
not involve any new dimensionfull constant.

Vortons formed with such currents are completely stable, at least at
the classical level (see however Ref.~\cite{davisS} for quantum
excitations). Assuming back reaction and electromagnetic corrections to
be small, one finds that the vorton excess
problem~\cite{davisRL88,carter91,brandi96} is therefore seriously
enhanced for fermionic current-carrier cosmic strings.

\section*{Acknowledgments}

It is a pleasure to thank Brandon Carter, Xavier Martin and Mairi
Sakellariadou for various enlightening discussions.

\chapter{Modes massifs (article)}
\label{chapitremassif}
\minitoc
Dans le chapitre~\ref{chapitrezero}, nous avons montr\'e que les
effets de r\'etroaction, induis par la propagation des modes z\'eros
charg\'es le long de la corde, modifient, au premier ordre dans les
charges fermioniques $q$, l'\'equation d'\'etat. Au deuxi\`eme ordre
$q^2$, nous avons \'egalement vu que les modes z\'eros n'\'etaient
plus solutions des \'equations du mouvement (voir
Sect.~\ref{relevant}). Il est donc l\'egitime de s'interroger sur la
pertinence physique de ces modes, d'autant plus que leur existence
pr\'evoie la stabilit\'e des boucles de corde associ\'ees (voir
Chap.~\ref{chapitrezeromacro}). Plus pr\'ecis\'ement, les modes
z\'eros poss\`edent la propri\'et\'e d'\^etre \'etats propres de
l'op\'erateur de conjugaison de charge $\gamma^0 \gamma^3$ (voir
Chap.~\ref{chapitrecour} et~\ref{chapitrezero}), et les \'equations du
mouvement perturb\'ees (\ref{pertzero}) montrent clairement que la
r\'etroaction d\'etruit cette caract\'eristique. Si il existe des
solutions de propagation dans le vortex n'\'etant pas \'etats propres,
\`a l'ordre le plus bas, de cet op\'erateur, la topologie de celles-ci
n'en sera certainement pas modifi\'ee aussi radicalement. Dans cet
article, publi\'e dans la revue
\journal{Physical Review} \numero{D}~\cite{ringeval2}, nous calculons
explicitement ces solutions et montrons qu'elles correspondent \`a des
modes massifs se propageant le long de la corde. L'\'equation d'\'etat
correspondante est calcul\'ee en g\'en\'eralisant la m\'ethode de
quantification introduite initialement pour les modes z\'eros (voir
Chap.~\ref{chapitrezero}). Le type de r\'egime de propagation des
perturbations transverses et longitudinales est finalement discut\'e,
et il appara\^\i t que les cordes cosmiques poss\'edant des courants
de fermions subissent de multiples transitions entre les r\'egimes
subsoniques et supersoniques.

\newpage

\begin{center}
{\Large \textbf{
Fermionic massive modes along cosmic strings
}}
\end{center}
\vspace{5mm}
\begin{center}
Christophe Ringeval
\end{center}
\vspace{5mm}
\begin{center}
{\footnotesize{
Institut d'Astrophysique
de Paris, 98bis boulevard Arago, 75014 Paris, France.
}}
\end{center}
\vspace{5mm}
\begin{center}
\begin{minipage}[c]{14cm}
{\footnotesize \textbf{
The influence on cosmic string dynamics of fermionic massive bound
states propagating in the vortex, and getting their mass only from
coupling to the string forming Higgs field, is studied. Such massive
fermionic currents are numerically found to exist for a wide range of
model parameters and seen to modify drastically the usual string
dynamics coming from the zero mode currents alone. In particular, by
means of a quantization procedure, a new equation of state describing
cosmic strings with any kind of fermionic current, massive or
massless, is derived and found to involve, at least, one state
parameter per trapped fermion species. This equation of state
exhibits transitions from subsonic to supersonic regimes while
the massive modes are filled.
}}
\end{minipage}
\end{center}

\section{Introduction}

Since it was realized that some early universe phase transitions might
lead to the formation of topological defects~\cite{kibble76,kibble80},
cosmic strings have been the subject of intense work within the
context of cosmology~\cite{peri95,bouchet88,allen97}. The large scale
structure generated by an ordinary string network in an expanding
universe, as well as its imprint on the microwave background, have
thus been derived~\cite{avelino99,contaldi99,wu02,bouchet02} in order to
state on their significance in the wide range of mechanisms in which
they had been originally involved~\cite{zeldov,vilenkin}. These
predictions, compared with the observations therefore constrain the
symmetry breaking schemes effectively realized in the early
Universe. These, associated with the most recent data for the
microwave background anisotropies~\cite{boomerang,maxima}, even seem
to show that such ordinary string networks could not have play the
dominant role in the Universe evolution, thereby all the more so
constraining the particle physics symmetries leading to their
formation. However, as was recently shown~\cite{bouchet02}, a
non-negligible fraction of such defects could have contributed to the
overall cosmic microwave background (CMB) anisotropies.

Meanwhile, it was shown by Witten~\cite{witten} that in realistic
physical models, involving various couplings of the string forming
Higgs field to other scalar or fermion fields, currents could build
along the strings, turning them into ``superconducting wires.''
Without even introducing couplings with the electromagnetic
fields~\cite{otw}, the breaking of Lorentz invariance along the vortex
induced by such currents may drastically modify the string properties,
and thus, the cosmological evolution of the associated networks. In
particular, cosmic string loops can reach centrifugally supported
equilibrium state, called vortons~\cite{davisRL}, that would
completely dominate the Universe~\cite{brandi96}. Theories
predicting stable vortons thus turn out to be incompatible with
observational cosmology, hence the particular interest focused on
``superconducting'' models.

Unfortunately, all the new properties and cosmological consequences
stemming from string conductivity have not yet been clearly
established, because of the complicated, and somehow arbitrary,
microphysics possible in these models. However, although the
microscopic properties induced by such currents depend on the explicit
underlying field
theory~\cite{davisRL87,hill97,hindmarsh88,everett88,ambjorn88}, a
macroscopic formalism was introduced by
Carter~\cite{carter89,carter89b,carter94b,carter97} which permits a
unified description of the string dynamics through the knowledge of
its energy per unit length $U$ and tension $T$. These ones end up
being functions of a so-called state parameter $w$, as the current
itself, through an equation of state. Such a formalism is, in
particular, well designed for scalar currents, as shown in, e.g.
Refs.~\cite{neutral,enon0}: due to their bosonic nature, all trapped
scalar particles go into the lowest accessible state, and thus can be
described through the classical values taken by the relevant scalar
fields~\cite{bps}. The induced gravitational
field~\cite{garrigapeter,peterpuy93,peter94} or the back reaction
effects~\cite{nospring} depend only on this state parameter. The
classical string stability~\cite{carter93,martinpeter} has already
been investigated for various equations of state relating $U$ and $T$,
on the basis of scalar and chiral currents
microphysics~\cite{carpet1,carpet2}. Moreover, it was also shown, through a
semiclassical approach, that fermionic current carrying cosmic
strings, even though in principle involving more than one state
parameter~\cite{ringeval}, can also be described by an equation of
state of the so-called ``fixed trace'' kind, i.e. $U+T=2M^2$.  Such a
relationship has the property of allowing stable loop configurations
to exist, at least at the classical
level~\cite{martinpeter}. Nevertheless, these results have been
derived for fermionic currents flowing along the string in the form of
zero modes only, as they were originally introduced by
Witten~\cite{witten}, although it was shown that the fermions may also
be trapped in the vortex with nonvanishing masses~\cite{davisS}: hence
the following work in which the influence of such massive modes is
studied for the simplest of all fermionic Witten model.

In this paper, after deriving numerically the relevant properties of
the trapped massive wave solutions of the Dirac equation in the
vortex, we show that the quantization procedure, originally performed
to deal with the fermionic zero modes~\cite{ringeval}, can be
generalized to include the massive ones, and leads to a new equation
of state with more than one state parameter. In particular, it is
found that the fixed trace equation of state, that holds for
massless fermionic currents alone, is no longer verified. Besides, the
massive modes are actually found to rapidly dominate the string
dynamics, thereby modifying the classical vorton stability induced by
the zero modes alone.

Let us sketch the lines along which this work is made. In
Sec.~\ref{modeleMM}, the model and the notations are set, while we
derive the equations of motion. Then, in Sec.~\ref{modemassif}, by
means of a separation between transverse and longitudinal degrees of
freedom of the spinor fields, the massive wave solutions along the
string are computed numerically for a wide range of fermion charges
and coupling constants. The constraint of transverse normalizability
is found to be satisfied only for particular values of the trapped
modes mass, $\miv$ say, whose dependence with the model parameters is
investigated. The two-dimensional quantization of the $\miv$ normalizable
massive modes is then performed in Sec.~\ref{quantizationMM}, using the
canonical procedure. In the way previously discussed in the
case of zero modes~\cite{ringeval}, the conserved quantities, i.e.
energy-momentum tensor and charge currents, are then expressed
in their quantum form. Their average
values, in the zero-temperature case, and infinite string limit, lead
to macroscopic expressions for the energy per unit length $U$ and
tension $T$ which end up being functions of the number densities of
fermions propagating along the string. Their derivation and extension
to any kind and number of fermionic carriers is performed in
Sec.~\ref{etatMM}, while the cosmological consequences of this new
analysis are briefly discussed in the concluding section.

\section{Model}
\label{modeleMM}

We shall consider here an Abelian Higgs model with scalar $\Phi$ and
gauge field $B_\mu$, coupled, following Witten~\cite{witten}, to two
spinor fields, $\Psi$ and $\Chi$ say. Since we are only interested in
the purely dynamical effects the current may induce on the strings,
we will not
consider any additional electromagneticlike coupling of the fermion
fields to an extra gauge field. Thus, we consider here the so-called
``neutral limit''~\cite{neutral}

\subsection{Microscopic Lagrangian}

The previous assumptions imply one needs one local $U(1)$ symmetry
which is spontaneously broken through the Higgs mechanism, yielding
vortices formation. The Higgs field is chosen as complex scalar field
with conserved charge $q \cphi$ under the local $U(1)$ symmetry,
associated with a gauge vector field $B_\mu$. The two spinor fields
acquire masses from a chiral coupling to the Higgs field, and have
opposite electromagnetic charges in order for the full
(four-dimensional) model to be anomaly free~\cite{witten}. Under the
broken symmetry they also have conserved charges $q \cpsir$, $q
\cpsil$ and $q \cchir$, $q \cchil$ for their right- and left-handed
parts, respectively. With ${\mathcal{L}}_{\mathrm{h}}$,
${\mathcal{L}}_{\mathrm{g}}$ and ${\mathcal{L}}_\psi$,
${\mathcal{L}}_\chi$, the Lagrangian in the Higgs, gauge, and
fermionic sectors, respectively, the theory reads
\begin{equation}
\label{lagrangien}
{\mathcal{L}}={\mathcal{L}}_{\mathrm{h}} + {\mathcal{L}}_{\mathrm{g}}
+ {\mathcal{L}}_\psi + {\mathcal{L}}_\chi,
\end{equation}
with
\begin{eqnarray}
{\mathcal{L}}_{\mathrm{h}} & = & \frac{1}{2}
(D_{\mu}\Phi)^{\dag}(D^{\mu}\Phi) - V(\Phi), \\
{\mathcal{L}}_{\mathrm{g}} & = & -\frac{1}{4} H_{\mu \nu} H^{\mu \nu},
\\
\label{psilagrangianMM}
{\mathcal{L}}_\psi & = & \frac{i}{2} \left[\overline{\Psi}
\gamma^{\mu} D_{\mu} \Psi - (\overline{D_{\mu}\Psi}) \gamma^{\mu}
\Psi \right] -g \overline{\Psi} \frac{1+\gamma_5}{2} \Psi \Phi
-g\overline{\Psi} \frac{1-\gamma_5}{2} \Psi \Phi^\ast,
\\
\label{chilagrangianMM}
{\mathcal{L}}_\chi & = & \frac{i}{2} \left[\overline{{\Chi}}
\gamma^{\mu} D_{\mu} {\Chi} - (\overline{D_{\mu}{\Chi}})
\gamma^{\mu} {\Chi} \right] -g \overline{{\Chi}}
\frac{1+\gamma_5}{2} {\Chi} \Phi^{\ast} - g \overline{{\Chi}}
\frac{1-\gamma_5}{2} {\Chi} \Phi,
\end{eqnarray}
where the $U(1)$ field strength tensor and the scalar potential
are
\begin{eqnarray}
H_{\mu \nu} & = & \nabla_\mu B_\nu - \nabla_\nu B_\mu,
\\
V(\Phi) & = & \frac{\lambda}{8} (|\Phi|^2 - \eta^2)^2,
\end{eqnarray}
while covariant derivatives involve the field charges through
\begin{eqnarray}
D_{\mu}\Phi & = & \left(\nabla_{\mu} + i q c_{\phi}B_\mu \right) \Phi,
\\
D_{\mu}\Psi & = & \left(\nabla_{\mu} + i q \frac{\cpsir+\cpsil}{2}B_\mu
+i q \frac{\cpsir-\cpsil}{2} \gamma_5 B_\mu \right) \Psi,
\\
D_{\mu}{\Chi} & = & \left(\nabla_{\mu} + i q \frac{\cchir+
\cchil}{2}B_\mu +i q \frac{\cchir-\cchil}{2} \gamma_5 B_\mu \right)
{\Chi},
\end{eqnarray}
and the relation
\begin{equation}
\cpsil-\cpsir=c_\phi=\cchir-\cchil
\end{equation}
should hold in order for the Yukawa terms in ${\mathcal{L}}_\psi$ and
${\mathcal{L}}_\chi$ to be gauge invariant.
\subsection{Basic equations}

This theory admits vortex solutions which are expected to form in the
early universe by means of the Kibble mechanism~\cite{kibble76,kibble80}. A
cosmic string configuration can be chosen to lie along the
$z$ axis, and we will use Nielsen-Olesen solutions of the field
equations~\cite{NO}. In cylindrical coordinates, the string forming
Higgs and gauge fields thus read
\begin{equation}
\begin{array}{ccc}
\Phi  =  \varphi(r) \ue^{i n\theta},
& \quad &
B_\mu  =  B(r) \delta_{\mu \theta},
\end{array}
\end{equation}
where the winding number $n$ is an integer, in order for the Higgs
field to be single valued under rotation around the string.
In such vortex background, the equations of motion in the fermionic
sector, for both spinor fields $\Ferm$ read (here and in various places
throughout this paper, we shall denote by $\Ferm$ an arbitrary
fermion, namely a spinor $\Psi$ or $\Chi$)
\begin{eqnarray}
\label{fermmvt}
i \gamma^\mu \nabla_\mu \Ferm & = & \frac{\partial j^\mu_\Ferm}{\partial
\overline{\Ferm}} B_\mu + M_\Ferm \Ferm
\end{eqnarray}
with the fermionic gauge currents 
\begin{eqnarray}
\label{currentsMM}
j^\mu_\Ferm & = & q \displaystyle{\frac{\cfr+ \cfl}{2}} \overline{\Ferm}
\gamma^\mu \Ferm + q \displaystyle{\frac{\cfr - \cfl}{2}}
\overline{\Ferm} \gamma^\mu \gamma_5 \Ferm,
\end{eqnarray}
and the mass terms
\begin{eqnarray}
\label{psihiggs}
M_\psi & = & g \varphi \cos{n \theta} + i g \varphi \gamma_5 \sin{n \theta},
\\
\label{chihiggs}
M_\chi & = & g \varphi \cos{n \theta} - i g \varphi \gamma_5 \sin{n \theta}.
\end{eqnarray}
Note the fermionic currents have an axial and vectorial component
because of the chiral coupling of the fermions to the Higgs field, as
can be seen through the mass terms $M_\Ferm$ in Eqs.~(\ref{psihiggs})
and (\ref{chihiggs}). Moreover, since the Higgs field vanishes in the
string core while taking nonzero vacuum expectation value, $\eta$
say, outside, the mass term acts as an attractive potential. As a
result, fermionic bound states, with energy between zero and $g \eta$,
are expected to exist and propagate in the string core.

\section{Fermionic bound states}
\label{modemassif}

\subsection{Trapped wave solutions}

Since the string is assumed axially symmetric, it is convenient to
look for trapped solutions of the fermionic equations of motion, by
separating longitudinal and transverse dependencies of the spinor
fields. Using the same notations as in Ref.~\cite{ringeval}, the
two-dimensional plane-wave solutions along the string, for both
fermions, read
\begin{equation}
\label{planeansatzMM}
\begin{array}{lll}
\Psi_{\mathrm{p}}^{(\varepsilon)}  =  \ue^{\varepsilon i(\omega t-kz)}
\left(\begin{array}{l}
\xi_1(r) \ue^{-im_1 \theta} \\
\xi_2(r) \ue^{-im_2 \theta} \\
\xi_3(r) \ue^{-im_3 \theta} \\
\xi_4(r) \ue^{-im_4 \theta}
\end{array} \right),
& \quad &
{\Chi}_{\mathrm{p}}^{(\varepsilon)}  =  \ue^{\varepsilon i(\omega t-kz)}
\left(\begin{array}{l}
\zeta_1(r) \ue^{-il_1 \theta} \\
\zeta_2(r) \ue^{-il_2 \theta} \\
\zeta_3(r) \ue^{-il_3 \theta} \\
\zeta_4(r) \ue^{-il_4 \theta}
\end{array} \right),
\end{array}
\end{equation}
where $\varepsilon=\pm 1$ labels the positive and negative energy
solutions. Similarly to the Higgs field case, the winding numbers of
the fermions, $m_i$ and $l_i$, are necessary integers.  In order to
simplify the notations, it is more convenient to work with
dimensionless scaled fields and coordinates. With
$m_{\mathrm{h}}=\eta\sqrt{\lambda}$ the mass of the Higgs boson, we
can write
\begin{eqnarray}
\varphi=\eta H,
\quad
Q=n + q \cphi \, B,
\quad \textrm{and} \quad
r=\frac{\varrho}{m_{\mathrm{h}}}.
\end{eqnarray}
In the same way, the spinorial components of the $\Psi$ field are
rescaled as
\begin{equation}
\begin{array}{ccccccc}
\vsep
\displaystyle
\xi_1(\varrho) & = & 
\displaystyle 
\frac{m_{\mathrm{h}}}{\sqrt{2\pi}}\sqrt{\omega+k}\,\atd_1(\varrho),
& \quad &
\displaystyle
\xi_2(\varrho) & = & 
\displaystyle
i \frac{m_{\mathrm{h}}}{\sqrt{2\pi}}\sqrt{\omega - k}\,\atd_2(\varrho), \\
\vsep
\displaystyle
\xi_3(\varrho) & = & 
\displaystyle
\frac{m_{\mathrm{h}}}{\sqrt{2\pi}}\sqrt{\omega-k}\,\atd_3(\varrho),
& \quad &
\displaystyle
\xi_4(\varrho) & = &
\displaystyle
i \frac{m_{\mathrm{h}}}{\sqrt{2\pi}}\sqrt{\omega+k}\,\atd_4(\varrho).
\end{array}
\end{equation}
In the chiral representation, and with the metric signature
$(+,-,-,-)$, in terms of these new variables, Eqs.~(\ref{fermmvt}) and
(\ref{planeansatzMM}) yield, for the $\Psi$ field,
\begin{equation}
\label{systadim}
\begin{array}{ccc}
\vsep \displaystyle \ue^{-i(m_1-1)\theta} \left[\frac{\ud \atd_1}{\ud
 \varrho} - \ft_1(\varrho) \atd_1(\varrho)\right] & = & \displaystyle
 \varepsilon \frac{\miv}{m_{\mathrm{h}}} \ue^{-im_2\theta}
 \atd_2(\varrho) - \frac{m_{\mathrm{f}}}{m_{\mathrm{h}}} H(\varrho)
 \ue^{-i(m_4+n) \theta} \atd_4(\varrho), \\ \vsep \displaystyle
 \ue^{-i(m_2+1)\theta} \left[\frac{\ud \atd_2}{\ud \varrho} -
 \ft_2(\varrho) \atd_2(\varrho)\right] & = & \displaystyle -
 \varepsilon \frac{\miv}{m_{\mathrm{h}}} \ue^{-im_1\theta}
 \atd_1(\varrho) + \frac{m_{\mathrm{f}}}{m_{\mathrm{h}}}H(\varrho)
 \ue^{-i(m_3+n)\theta} \atd_3(\varrho), \\ \vsep \displaystyle
 \ue^{-i(m_3-1)\theta} \left[\frac{\ud\atd_3}{\ud \varrho} -
 \ft_3(\varrho) \atd_3(\varrho)\right] & = & \displaystyle -
 \varepsilon \frac{\miv}{m_{\mathrm{h}}} \ue^{-i m_4 \theta}
 \atd_4(\varrho) + \frac{m_{\mathrm{f}}}{m_{\mathrm{h}}} H(\varrho)
 \ue^{-i(m_2-n)\theta} \atd_2(\varrho), \\ \vsep \displaystyle
 \ue^{-i(m_4+1)\theta} \left[\frac{\ud \atd_4}{\ud \varrho} -
 \ft_4(\varrho) \atd_4(\varrho) \right] & = & \displaystyle \varepsilon
 \frac{\miv}{m_{\mathrm{h}}} \ue^{-i m_3\theta} \atd_3(\varrho) -
 \frac{m_{\mathrm{f}}}{m_{\mathrm{h}}} H(\varrho)
 \ue^{-i(m_1-n)\theta} \atd_1(\varrho),
\end{array}
\end{equation}
where $m_{\mathrm{f}}=g \eta$ is the fermion mass in the vacuum in
which the Higgs field takes its vacuum expectation value $\eta$, and
$\miv=\sqrt{\omega^2-k^2}$ is the mass of the trapped mode. The
coupling to the gauge field $B_\mu$ appears through the purely radial
functions $\ft$:
\begin{equation}
\begin{array}{ccccccc}
\vsep
\displaystyle
\ft_1(\varrho) & = & 
\displaystyle 
\frac{\cpsir}{\cphi} \frac{Q-n}{\varrho} - \frac{m_1}{\varrho},
& \quad &
\displaystyle
\ft_2(\varrho) & = & 
\displaystyle
-\frac{\cpsir}{\cphi}\frac{Q-n}{\varrho} +\frac{m_2}{\varrho}, \\
\vsep
\displaystyle
\ft_3(\varrho) & = & 
\displaystyle
\frac{\cpsil}{\cphi}\frac{Q-n}{\varrho} - \frac{m_3}{\varrho},
& \quad &
\displaystyle
\ft_4(\varrho) & = &
\displaystyle
-\frac{\cpsil}{\cphi}\frac{Q-n}{\varrho} + \frac{m_4}{\varrho}.
\end{array}
\end{equation}
The spinor field ${\Chi}$ verifies the same equations apart from the
fact that, due to its coupling to $\Phi^\dag$ [see
Eq.~(\ref{chilagrangianMM})], it is necessary to transform $n
\rightarrow -n$.

As was originally found by Jackiw and Rossi~\cite{jackiwrossi} and
Witten~\cite{witten}, there are always $n$ normalizable zero energy
solutions of the Dirac operator in the vortex which allow fermions to
propagate at the speed of light in the ``$-z$'' and ``$+z$,'' say,
directions, for the $\Psi$ and $\Chi$ fields, respectively. These
solutions are found to be eigenvectors of the $\gamma^0 \gamma^3$
operator and are clearly obtained from the above equations by setting
the consistency angular relationships $m_1-1=m_4+n$ and $m_2+1=m_3+n$,
those leading to the zero mode dispersion relation $\miv=0
\Leftrightarrow \omega=\pm k$. Note that only one eigenstate of
$\gamma^0 \gamma^3$ end up being normalizable for each kind of chiral
coupling to the Higgs field, and thus the relevant dispersion
relations reduce to $\omega=-k$ and $\omega=k$, for the $\Psi$ and
$\Chi$ zero modes, respectively~\cite{ringeval}.

Such zero modes have a simple interpretation: since the Higgs field
vanishes in the string core, the mass term $M_\Ferm$ in
Eq.~(\ref{fermmvt}) vanishes too, and the fermions trapped in have
zero mass. As a result, they propagate at the speed of light and they
verify the dispersion relations $\omega= k$ or $\omega=-k$.

\subsection{Massive trapped waves}

However, it is also possible \emph{a priori}, for the trapped
fermions, to explore outer regions surrounding the string core where
the Higgs field takes nonexactly vanishing values. In practice, this
is achieved by means of a nonvanishing fermion angular momentum, which
will lead to a nonvanishing effective mass $\miv^2=\omega^2 -k^2 \neq
0$. For the $\Psi$ field, such massive solutions of the equations of
motion (\ref{systadim}) can only be obtained for four-dimensional
solutions, in order to ease the zero mode constraint $\omega=\pm
k$. The required angular consistency relations therefore read
\begin{equation}
\label{angconst}
m=m_1=m_2+1=m_3+n=m_4+n+1.
\end{equation}
Similarly, the angular dependence of $\Chi$ field has to verify
analogous conditions with the transformation $n \rightarrow -n$. It
was previously shown numerically that the Abelian Higgs model with one
Weyl spinor always admits such kind of normalizable solutions
\cite{davisS}. In the following, massive solutions for Dirac spinors
are numerically derived for our model and shown to exist for a wide
range of fermion charges and coupling constants.

\subsubsection{Analytical considerations}

Some interesting analytical asymptotic behaviors of these modes have
been previously studied~\cite{ringeval,davisS}. In particular, there
are only two degenerate \emph{normalizable} eigensolutions of
Eqs.~(\ref{systadim}) at infinity. Since the Higgs field goes to its
constant vacuum expectation value and the gauge coupling functions
vanish, we found the eigensolutions to scale as $\exp{(\pm \Omega
\varrho)}$, with
\begin{equation}
\Omega = \sqrt{\frac{m_{\mathrm{f}}^2-\miv^2}{m_{\mathrm{h}}^2}}.
\end{equation}
First, note that in order to have decreasing solutions at infinity,
the mass of the trapped modes $\miv$ has to be less than the fermion
vacuum mass $m_{\mathrm{f}}$, as intuitively expected (for
$\miv>m_{\mathrm{f}}$, one recovers the oscillating behaviour that is
typical of free particle solutions). Moreover, from Cauchy theorem,
two degrees of freedom can be set in order to keep only the two well
defined solution at infinity. On the other hand, by looking at the
power-law expansion of both system and solutions near the string
core~\cite{ringeval,jackiwrossi}, only two such solutions are also
found to be normalizable. More precisely, normalizability of each
eigensolution at $\varrho=0$ leads to one condition on the winding
numbers $m_i$ of each spinorial component $\xi_i$. Moreover, in order
for the fermion field to be well defined by rotation around the
string, each spinorial component $\xi_i$ with nonzero winding number
$m_i$ has to vanish in the string core, and so behaves like a
positive power of the radial distance to the core. The analytical
expression of the eigensolutions near $\varrho=0$
reads~\cite{ringeval}
\begin{equation}
\label{corepairs}
\begin{array}{ccccccc}
\left(
\begin{array}{l}
\xi_1 \\
\xi_2 \\
\xi_3 \\
\xi_4
\end{array}
\right)_{s_1}
 & \sim &
\left(
\begin{array}{l}
a_1 \varrho^{-m} \\
a_2(a_1) \varrho^{-m+1} \\
a_3(a_1) \varrho^{-m+|n|+2} \\
a_4(a_1) \varrho^{-m+|n|+1}
\end{array}
\right),
& \quad &
\left(
\begin{array}{l}
\xi_1 \\
\xi_2 \\
\xi_3 \\
\xi_4
\end{array}
\right)_{s_2}
& \sim &
\left(
\begin{array}{l}
a_1 \varrho^{m+|n|-n} \\
a_2(a_1) \varrho^{m+|n|-n+1} \\
a_3(a_1) \varrho^{m-n} \\
a_4(a_1) \varrho^{m-n-1}
\end{array}
\right),
\\ \\
\left(
\begin{array}{l}
\xi_1 \\
\xi_2 \\
\xi_3 \\
\xi_4
\end{array}
\right)_{s_3}
 & \sim &
\left(
\begin{array}{l}
a_1 \varrho^{m} \\
a_2(a_1) \varrho^{m-1} \\
a_3(a_1) \varrho^{m+|n|} \\
a_4(a_1) \varrho^{m+|n|+1}
\end{array}
\right),
& \quad &
\left(
\begin{array}{l}
\xi_1 \\
\xi_2 \\
\xi_3 \\
\xi_4
\end{array}
\right)_{s_4}
& \sim &
\left(
\begin{array}{l}
a_1 \varrho^{-m+|n|+n+2} \\
a_2(a_1) \varrho^{-m+|n|+n+1} \\
a_3(a_1) \varrho^{-m+n} \\
a_4(a_1) \varrho^{-m+n+1}
\end{array}
\right).
\end{array}
\end{equation}
The normalizability condition for the four eigensolutions can be
summarized by 
\begin{equation}
\sup{(0,n)} < m < \inf{(1,1+n)},
\end{equation}
and so, for any value of $m$ there are only two conditions
satisfied. However, from the consistency angular conditions on each
spinorial components, only three pairs of solutions are acceptable
near the string. Assuming $n>0$, if $m \le 0$ then only the pair
$(s_1,s_4)$ is both normalizable and well defined by rotation around
the vortex, similarly for $m \ge n+1$ the relevant solutions are
$(s_2,s_3)$, whereas for $1 \le m \le n$, they are $(s_3,s_4)$.  As a
result, the two remaining degrees of freedom can be set to get only
these pairs near the string core for a given value of $m$, but there
is no reason that they should match with the two normalizable
solutions at infinity. In order to realize this matching we have to
fine tune another parameter which turns out to be the mass of the
modes, $\miv$. As expected for bound states, this mass is therefore
necessarily quantized. Note at this point that $\miv=0$ is, in such a
procedure, nothing but a particular case of the general solution here
presented. The three different pairs of well defined solutions at the
origin suggest that there are three kinds of similar massive bound
states in the vortex, according to the values of the winding number
$m$. Intuitively, the more the field winds around the string, the
farther the particle explores regions surrounding the core due to the
higher values taken by its angular momentum, meaning the largest the
extension of its wave function is, the more it acquires mass from
coupling to a nonexactly vanishing Higgs field. As a result, the
lowest massive modes will certainly be obtained from values of $m$
which correspond to vanishing winding numbers $m_i$.

\subsubsection{Symmetries}
\label{symmetries}

In the following, the equations of motion (\ref{systadim}) will be
summarized in the form $\syse_i^j\atd_j=0$, with implicit summation
implied over repeated indices.

The first symmetry is obtained from the complex conjugation of the
equations of motion (\ref{systadim}). Since complex conjugation does
not modify Eqs.~(\ref{systadim}), once the angular consistency
relations (\ref{angconst}) are set, there is an arbitrary complex
phase in the choice of solutions, and it will be sufficient to look
for real rescaled spinorial components $\atd_i$.

There is another symmetry between the positive and negative energy
solutions of the equations of motion (\ref{systadim}) that may be
useful. With the label $\varepsilon=\pm $ for particle and
antiparticle states, respectively, one has
\begin{eqnarray}
\sysp_i^j \atd_{j_+}=0 &\quad \Rightarrow \quad & \sysm_i^j\atd_{j_-}=0,
\end{eqnarray}
provided
\begin{eqnarray}
\atd_{i_-} & = & \left(\gamma^0 \gamma^3\right)_i^j \atd_{j_+}.
\end{eqnarray}
As a result, the negative energy solutions are obtained from the
positive ones by the action of the $\gamma^0 \gamma^3$ operator, thereby
generalizing the properties of the zero modes which were precisely
found as eigenstates of this
operator~\cite{witten,ringeval,jackiwrossi}.

The last symmetry concerns the gauge coupling functions $\ft_i$. Under
the transformations
\begin{equation}
\label{symcphiwind}
\begin{array}{lllll}
\displaystyle
m & \rightarrow &\widehat{m} & = & n+1-m, \\
\displaystyle
\cpsil & \rightarrow & \widehat{c}_{\psi_{\mathrm{L}}} & = & -\cpsir, \\
\displaystyle
\cpsir & \rightarrow & \widehat{c}_{\psi_{\mathrm{R}}} & = & -\cpsil,
\end{array}
\end{equation}
the gauge functions $\ft_i$, in Eqs.~(\ref{systadim}), are simply
swapped according to $ \ft_1 \leftrightarrow \ft_4$ and $\ft_2
\leftrightarrow \ft_3$. As a result, for every $\atd$ solution found
at given $\cpsil$ and $m$, there is another solution $\ah$, with
charge $\widehat{c}_{\psi_{\mathrm{L}}}=\cphi-\cpsil$ and winding number
$\widehat{m}=n+1-m$, namely
\begin{equation}
\label{symcphi}
\begin{array}{ccccccc}
\vsep
\displaystyle
\ah_1(\varrho) & = &  \atd_4(\varrho),
& \quad &
\displaystyle
\ah_2(\varrho) & = & \atd_3(\varrho), \\
\vsep
\displaystyle
\ah_3(\varrho) & = &  \atd_2(\varrho),
& \quad &
\displaystyle
\ah_4(\varrho) & = & \atd_1(\varrho).
\end{array}
\end{equation}
Note that the particular case
$\cpsil=\widehat{c}_{\psi_{\mathrm{L}}}=\cphi/2$ appears as a frontier
separating two symmetric kinds of solutions with two differents
winding numbers lying on both sides of $m=(n+1)/2$. As a result, the
three different behaviors found above from normalization and angular
consistency conditions seem to reduce to only two, since the domains
where $m \le 0$ and $m \ge n+1$ are actually connected by charge
symmetry in relation to $\cphi/2$.

On the other hand, due to its coupling to the antivortex instead of
the vortex, the equations of motion of the $\Chi$ field are simply
obtained from Eqs.~(\ref{systadim}) by the transformations $\atd_i
\rightarrow \bt_j$, $c_{\psi_{\mathrm{L}(\mathrm{R})}} \rightarrow
c_{\chi_{\mathrm{L}(\mathrm{R})}}$, and $m_i \rightarrow l_i$. The
$l_i$ are the winding numbers of the scaled $\Chi$ spinorial
components, namely the $\bt_i$, and they verify the angular
consistency relations (\ref{angconst}) with $n$ replaced by $-n$ as
previously discussed.  Let us introduce one more transformation on the
$\Psi$ parameters,
\begin{equation}
\begin{array}{lllll}
\displaystyle
m & \rightarrow & \widehat{m} & = & l+n  , \\
\displaystyle
\cpsil & \rightarrow & \widehat{c}_{\psi_{\mathrm{L}}} & = & \cchir, \\
\displaystyle
\cpsir & \rightarrow & \widehat{c}_{\psi_{\mathrm{R}}} & = & \cchil.
\end{array}
\end{equation}
Naming $\gt_i$ the scaled gauge coupling functions of the $\Chi$ spinor,
the $\Psi$ ones are found to transform according to $\ft_1 \rightarrow
\gt_3$, $\ft_2 \rightarrow \gt_4$, $\ft_3 \rightarrow \gt_1$, and
$\ft_4 \rightarrow \gt_2$. Thus, if the $\atd$ are solutions of the
$\Psi$ equations of motion (\ref{systadim}), with $m$ winding number
and $\cpsil$ charge, then there exist $\bt$ solutions for the $\Chi$
field with same mass $\miv$, provided $l=m-n$ and
$\cchil=\cpsir=\cpsil-\cphi$, and they read
\begin{equation}
\label{sympsichi}
\begin{array}{ccccccc}
\vsep
\displaystyle
\bt_1(\varrho) & = & \atd_3(\varrho),
& \quad &
\displaystyle
\bt_2(\varrho) & = & -\atd_4(\varrho), \\
\vsep
\displaystyle
\bt_3(\varrho) & = & \atd_1(\varrho),
& \quad &
\displaystyle
\bt_4(\varrho) & = & -\atd_2(\varrho).
\end{array}
\end{equation}
Owing to these symmetries, it is sufficient to study the $\Psi$
equations of motion (\ref{systadim}), for various values of the
winding number $m$ and for left-handed part charges, namely $\cpsil$,
higher or equal than $\cphi/2$.

\subsubsection{Numerical methods}

\begin{figure}
\begin{center}
\epsfig{file=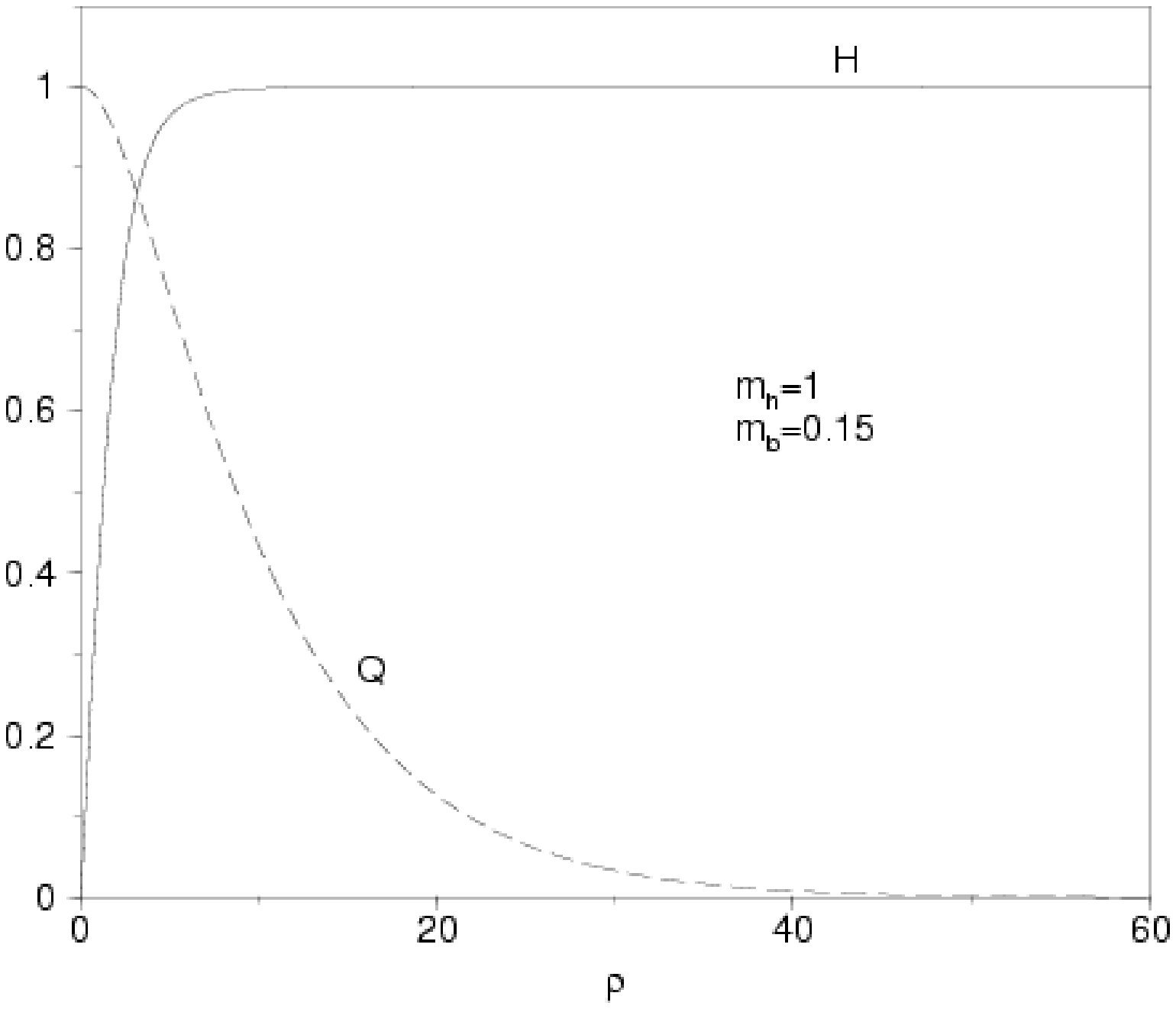,width=10cm}
\caption[Solutions des \'equations de champ d'une corde
cosmique.]{The solutions of the field equations for the vortex
background. The Higgs field, $H$, takes its vacuum expectation value
at infinity and the gauge bosons condensate in the vortex.}
\label{figbackMM}
\end{center}
\end{figure}

In order to compute the relevant massive wave solutions for the $\Psi$
fermions on the string, it is necessary to solve first the vortex
background. At zeroth order, neglecting the back reaction of the
fermionic currents, and in terms of the dimensionless fields and
parameters, the equations of motion for the string forming Higgs and
gauge fields read, from Eq.~(\ref{lagrangien}),
\begin{eqnarray}
\label{tildehiggsMM}
\frac{\ud^2 H}{\ud \varrho^2}+\frac{1}{\varrho} \frac{\ud H}{\ud
\varrho} & = & \frac{H Q^2}{\varrho^2}+\frac{1}{2}H(H^2-1), \\
\label{tildegaugeMM}
\frac{\ud^2 Q}{\ud \varrho^2} -\frac{1}{\varrho}\ \frac{\ud Q}{\ud
\varrho} & = & \frac{m_{\mathrm{b}}^2}{m_{\mathrm{h}}^2}H^2 Q,
\end{eqnarray}
where $m_{\mathrm{b}}=qc_\phi \eta$ is the classical mass of the gauge boson.
The solution of these equations is well known~\cite{neutral,bps,adler}
and shown in Fig.~\ref{figbackMM} for a specific (assumed generic) set
of parameters.

The system of Eqs.~(\ref{systadim}) being linear and involving only
first order derivatives of the spinor components, a Runge-Kutta
numerical method of integration has been used. However, as noted
above, since we are interested only in normalizable solutions, it is
more convenient to perform the resolution from an arbitrary cutoff at
infinity, toward the string core. Let us introduce $\varrho_\infty$,
the cutoff value on the dimensionless radial distance. From the
asymptotic form of Eqs.~(\ref{systadim}) at infinity, and in order to
suppress the exponential growth, the spinorial components $\atd_i$ have
to verify
\begin{eqnarray}
\atd_1(\varrho_\infty) & = & -\frac{\miv}{\Omega m_{\mathrm{h}}}
\atd_2(\varrho_\infty) + \frac{m_{\mathrm{f}}}{\Omega m_{\mathrm{h}}}
\atd_4(\varrho_\infty),
\\
\atd_3(\varrho_\infty) & = & - \frac{m_{\mathrm{f}}}{\Omega m_{\mathrm{h}}}
\atd_2(\varrho_\infty) + \frac{\miv}{\Omega m_{\mathrm{h}}} \atd_4(\varrho_\infty).
\end{eqnarray}
These conditions constrain two degrees of freedom, and another one is
fixed by normalization of the wave functions at $\varrho_\infty$. As a
result, only one free parameter can be used yet in order to achieve
the matching between these well defined solutions and the two
normalizable ones near the string core. It will be the case only for
particular values of the mass $\miv$. Numerically, the matching is
performed in two steps. First, by means of the last free parameter,
one of the usually divergent component near the string core is made to
vanish at $\varrho=0$. Obviously, this component is chosen among those
having a nonzero winding number since, in order to be single valued
by rotation around the vortex, it necessarily vanishes at the string
core. Once it is performed, the last divergent component at
$\varrho=0$ is regularized, its turn, by calculating the mass of the
mode $\miv$ leading to a convergent solution. For the range of model
parameters previously defined, the numerical computations thus lead to
the mass of the trapped wave solutions as well as their components as
function of the radial distance to the string core $\atd_i(\varrho)$.

\subsubsection{Numerical results}
\label{numresults}

In what follows, the Higgs winding number is assumed fixed to the
value $n=1$, and the range of $\cpsil$ restricted to $\cpsil \ge
\cphi/2$, the other case being derivable from the symmetric properties
discussed above.

\begin{figure}
\begin{center}
\epsfig{file=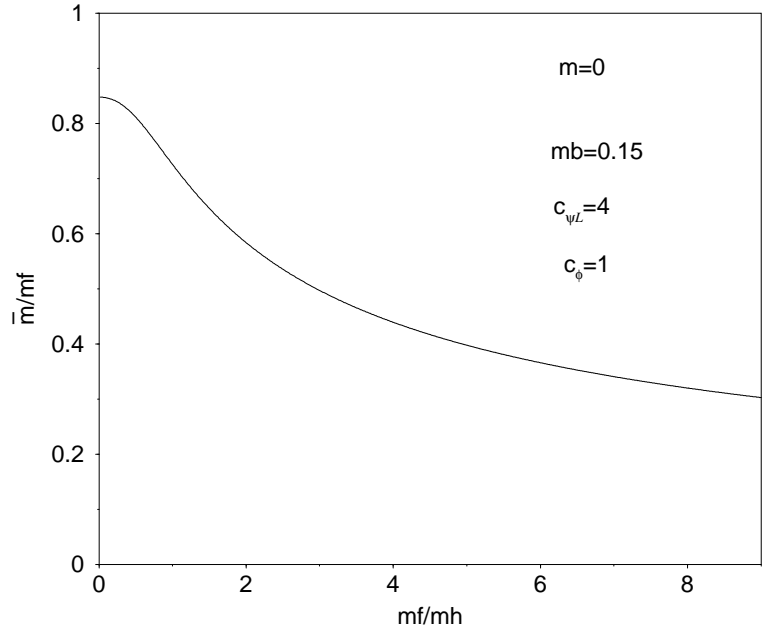,width=10cm}
\caption[Existence de fermions massifs dans une corde cosmique.]{The
mass of the lowest massive bound state, relative to the fermion vacuum
mass, plotted as function of the coupling constant to the Higgs field,
i.e. the fermion vacuum mass relative to the mass of the Higgs
boson.}
\label{figmfond}
\end{center}
\end{figure}

\begin{figure}
\begin{center}
\epsfig{file=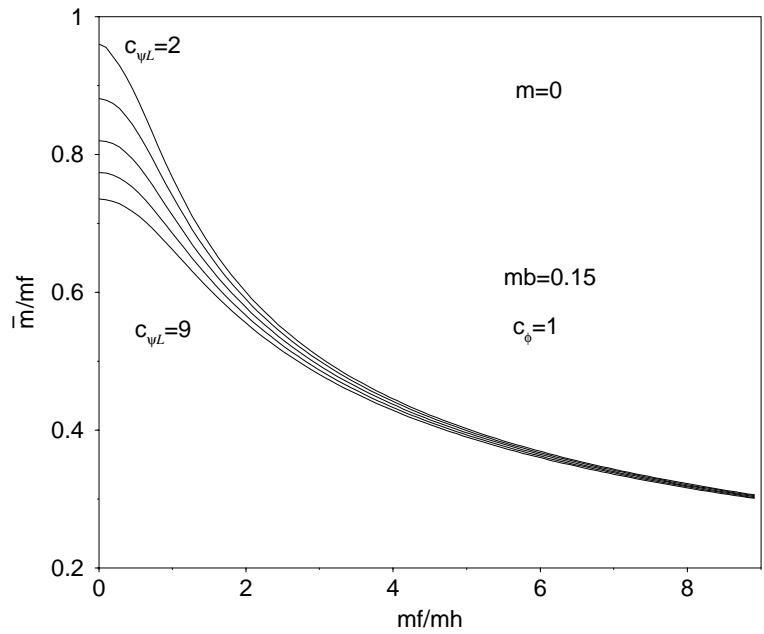,width=10cm}
\caption[Influence des constantes de couplage sur la masse des
fermions pi\'eg\'es dans une corde cosmique.]{The mass of the lowest
massive bound state, relative to the fermion vacuum mass, plotted as
function of the coupling constant to the Higgs field, for several
values of the fermionic charges. The closest $\cpsil$ is to $\cphi/2$,
the higher mode mass $\miv$ is. In the extreme case
$\cpsil\sim\cphi/2$, $\miv \sim m_{\mathrm{f}}$ there is no longer
normalizable massive bound state \emph{in the perturbative sector}.}
\label{figefond}
\end{center}
\end{figure}

\begin{figure}
\begin{center}
\epsfig{file=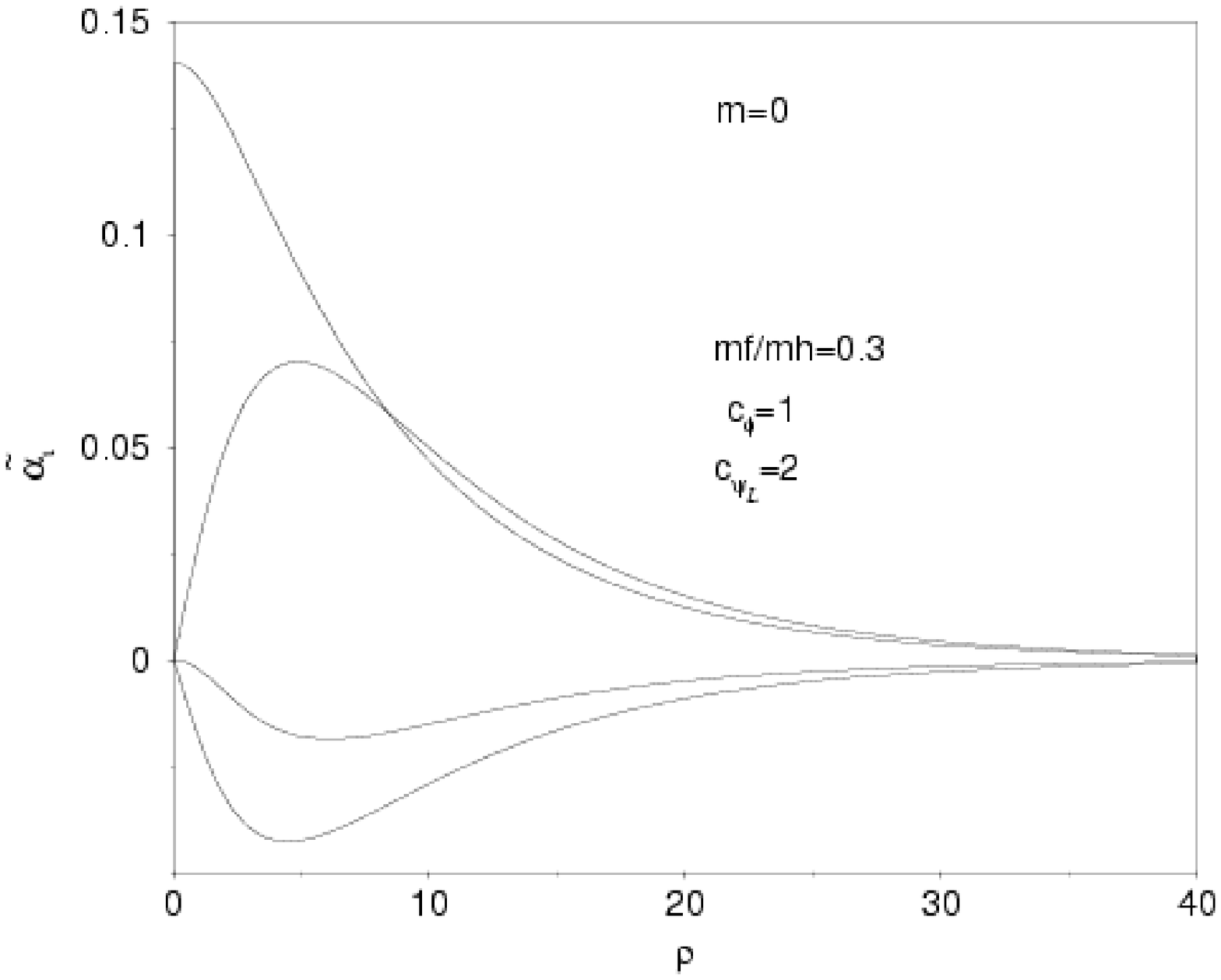,width=10cm}
\epsfig{file=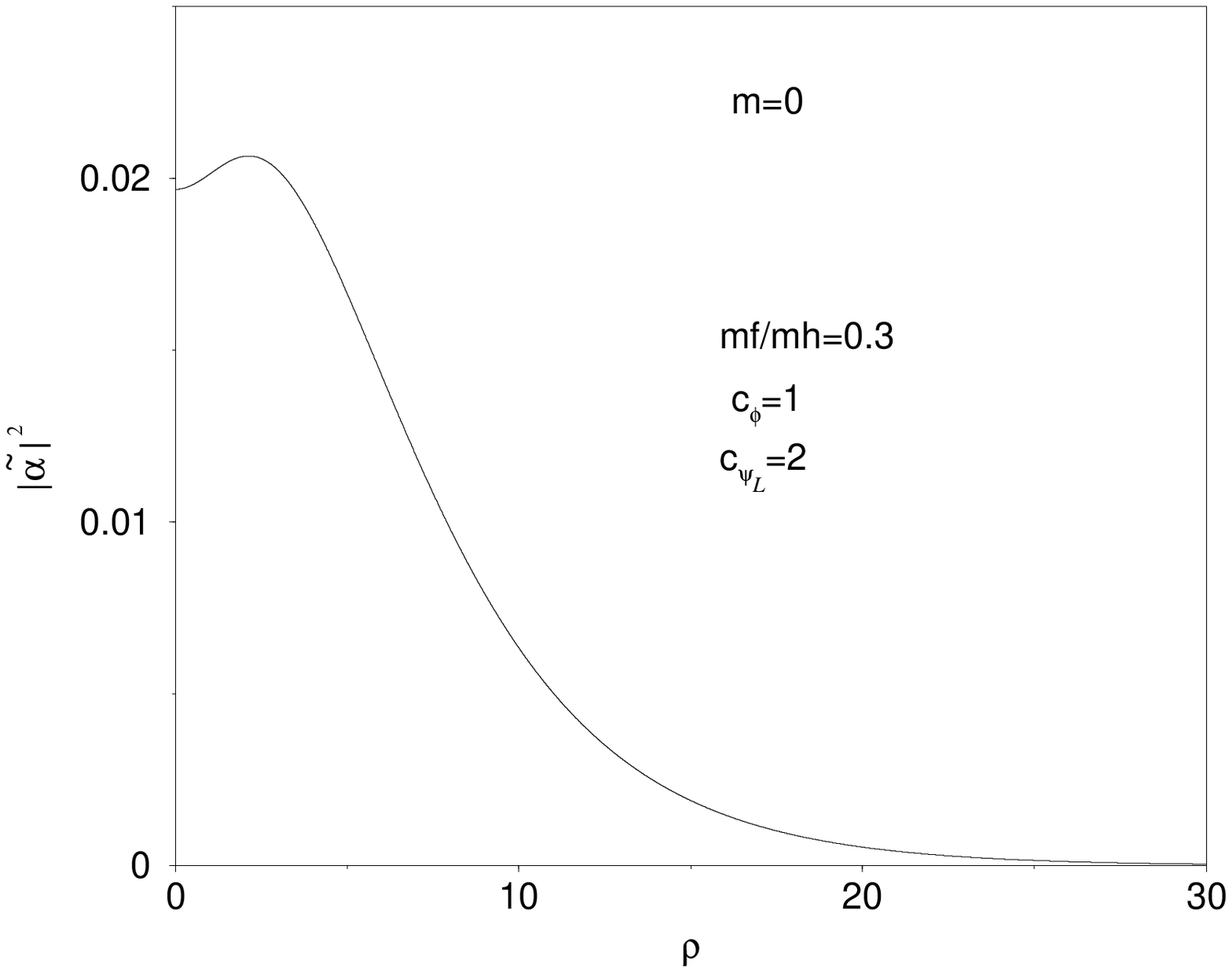,width=10cm}
\caption[Fonctions d'onde des fermions massifs dans la corde.]{The
transverse spinorial components of the $\Psi$ field, as functions of
the distance to the string core, for the lowest massive bound
state. The transverse normalized probability density is also plotted
and takes its maximum value nearby the string core, as expected for
massive bound states exploring the neighborhood of the core by means
of nonvanishing angular momentum. Note that for $m=0$, one spinorial
component behaves like a zero mode one, i.e. it condenses in the
string core contrary to the others.}
\label{figprobafond}
\end{center}
\end{figure}

The first results concern the ``perturbative sector'' where the
fermion vacuum mass verifies $m_{\mathrm{f}}<m_{\mathrm{h}}$, or
equivalently, for a smaller Yukawa coupling constant than the Higgs
self-coupling, i.e. $g<\sqrt{\lambda}$. In this case, for reasonable
values of the fermion charges, i.e. of the same order of magnitude
than the Higgs one $\cpsil \gtrsim \cphi/2$, only one normalizable
massive bound state is found with null winding number $m=0$. As a
result, by means of transformations~(\ref{symcphiwind}), there are
also symmetric modes for $\cpsil \le \cphi/2$, with winding number
$m=2$. The dependency of the mode mass $\miv$ with the fermion vacuum
mass and charges (i.e. the coupling constants to Higgs and gauge
fields) is plotted in Fig.~\ref{figmfond} and Fig.~\ref{figefond}. The
study has been also extended to the nonperturbative sector where this
massive mode thus appears as the lowest massive bound state. First, it
is found that the mass of the trapped mode always decreases with
respect to the coupling constant, i.e. with the fermion vacuum mass
$m_{\mathrm{f}}$. Moreover, for small values of
$m_{\mathrm{f}}/m_{\mathrm{h}}$, the derivative of the curve
$\miv(m_{\mathrm{f}}/m_{\mathrm{h}})$ vanishes near the origin (see
Fig.~\ref{figmfond}). As a result, the mass modes in the full
perturbative sector does not depend on the coupling constant to the
Higgs field, at first order. On the other hand, Fig.~\ref{figefond}
shows that the mass of the bound state hardy depends at all on its
coupling with the gauge field (i.e. on the charges $\cpsil$) in the
nonperturbative sector, where all the curves have the same asymptotic
behavior. Near the origin, the closest $\cpsil$ is to $\cphi/2$, the
higher mode mass $\miv$ is. In the particular limiting case $\cpsil
\sim \cphi/2$, there is no normalizable massive bound state, and as
can be seen in Fig.~\ref{figefond}, already for $\cpsil/\cphi=2$, the
mode mass is close to $m_{\mathrm{f}}$. It is not surprising since, as
it was above noted, $\cpsil=\cphi/2$ is a frontier between two kinds
of solutions with different winding numbers, and thus, at this point,
the ``normalizable'' winding numbers are not well defined. Note that
this is only true if $m\neq (n+1)/2$ as it is the case here in the
perturbative sector with $n=1$ and $m=0$.
The normalized scaled spinorial components $\atd(\varrho)$ have been
plotted in Fig.~\ref{figprobafond} for the lowest massive bound
state, with the normalization
\begin{equation}
\int{\varrho \,\ud \varrho\, \atd_i^{\dag} \atd_i} = 1.
\end{equation}
The corresponding transverse probability density has also been plotted
in Fig.~\ref{figprobafond}. Note that the massive mode wave function
is larger around the string rather on it, as expected for a nonvanishing
angular momentum.

\begin{figure}
\begin{center}
\epsfig{file=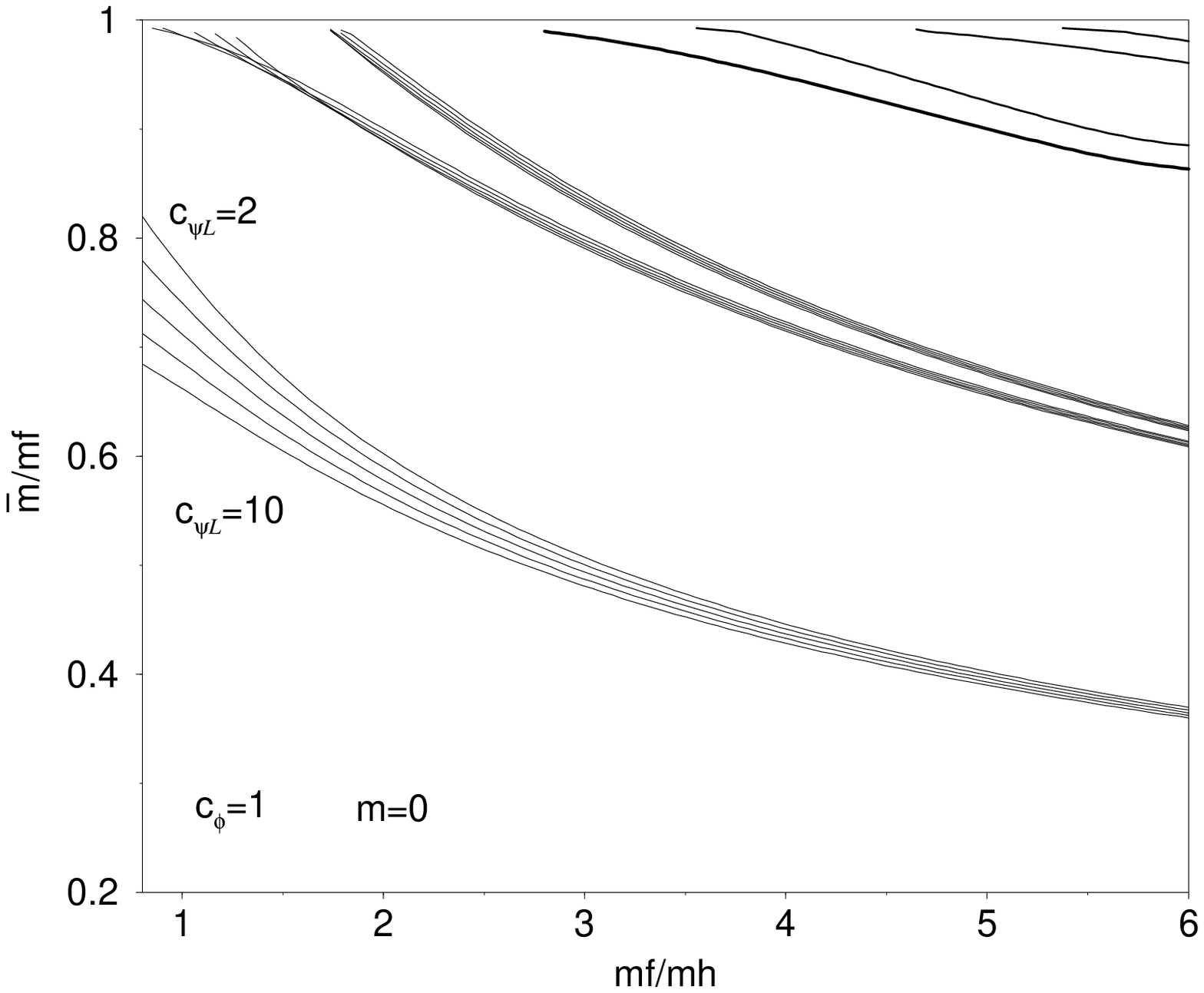,width=10cm}
\caption[Le spectre de masse des fermions dans une corde
cosmique.]{The evolution of the mass spectrum for $m=0$ winding
number, as function of the coupling constants. Each main branch
represents one massive mode whereas the substructures show its
evolution with respect to the fermion charge $\cpsil$. Five values of
the fermion charge have been plotted, from $\cpsil=2$ to $\cpsil=10$,
and the spectrum has been computed only in the nonperturbative sector,
since only the lowest mode exists for lower values of
$m_{\mathrm{f}}/m_{\mathrm{h}}$ (see Fig.~\ref{figefond}). As
expected, all the modes have mass $\miv$ decreasing with their
coupling constant to the Higgs field. Moreover, the substructures show
that, for sufficiently large values of
$m_{\mathrm{f}}/m_{\mathrm{h}}$, the mode mass is a decreasing
function of the charge $\cpsil$. However, note that this behavior can
be inverted for some modes close to their appearance region, as it is
the case for the second one.}
\label{figmasspect}
\end{center}
\end{figure}

\begin{figure}
\begin{center}
\epsfig{file=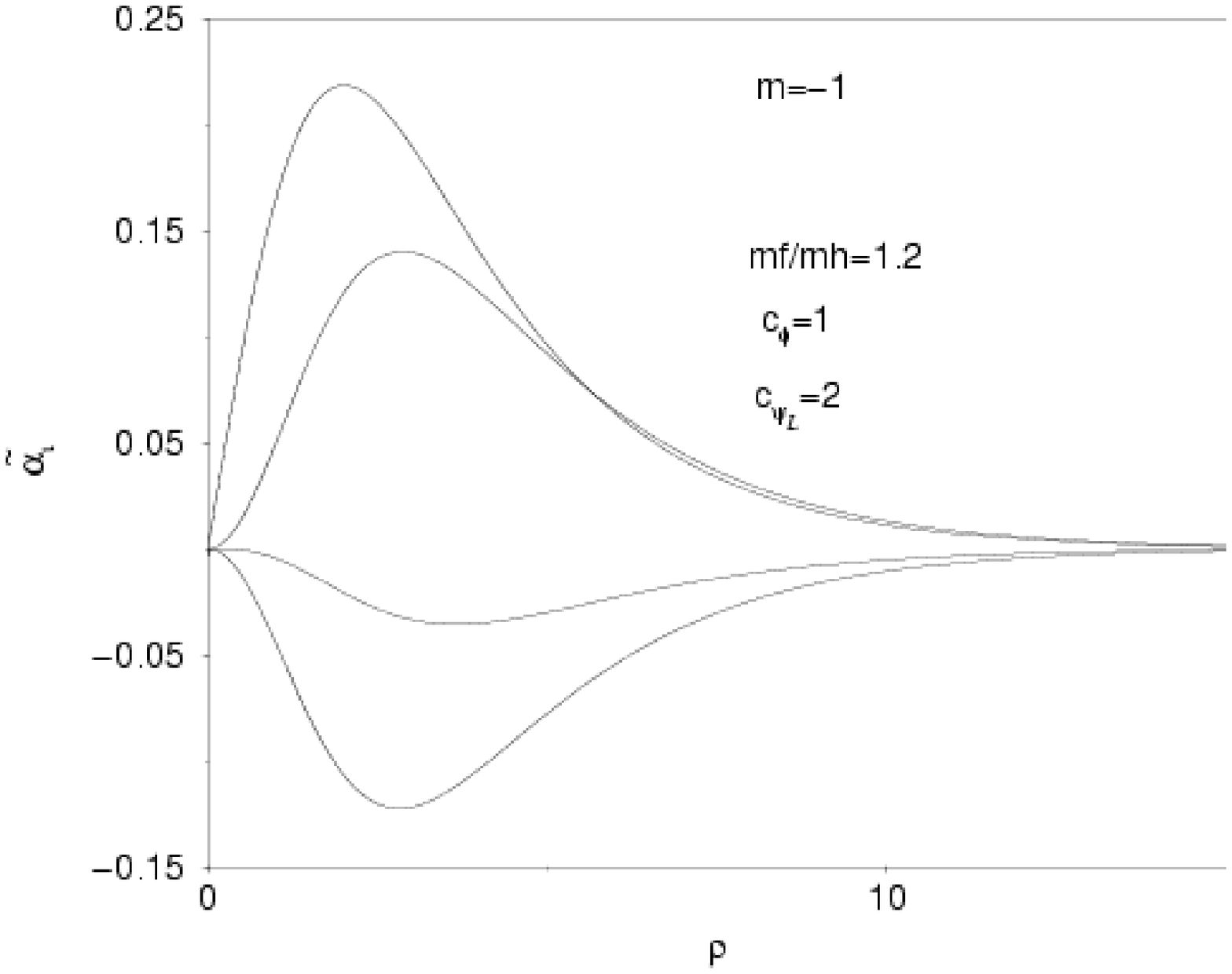,width=10cm}
\end{center}
\begin{center}
\hspace{7pt}
\epsfig{file=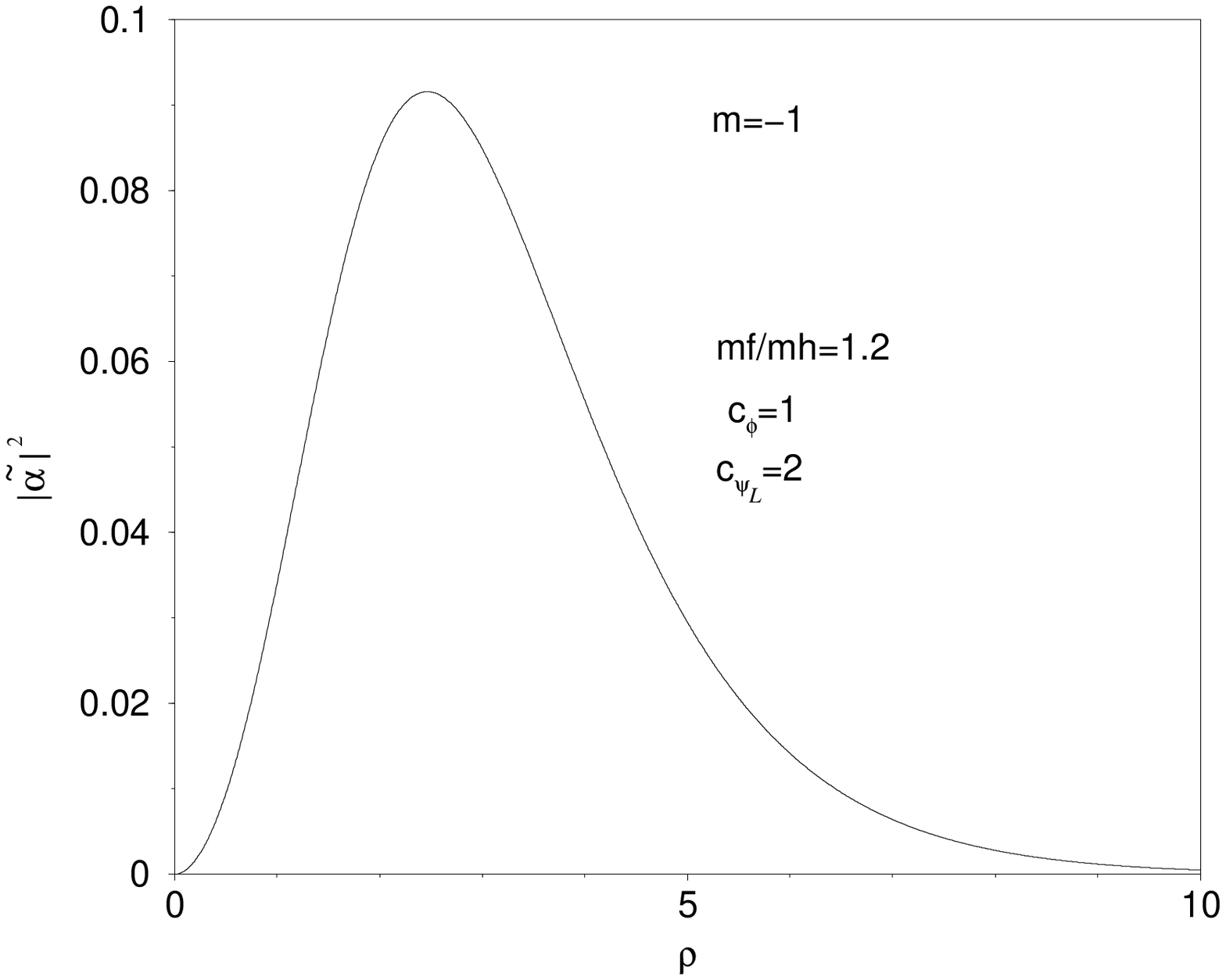,width=10cm}
\caption[Influence du moment angulaire des fermions sur leur
probabilit\'e de pr\'esence dans la corde.]{The transverse spinorial
components of the $\Psi$ field, plotted as functions of the distance
to the string core, for the $m=-1$ lowest massive bound state. The
transverse normalized probability density is also plotted and vanishes
in the string core. In this case, all components of the spinor wind
around the string and the corresponding mode is thus centrifugally
confined in a shell nearby the core, as expected for a nonzero angular
momentum eigenstate.}
\label{figprobafond1}
\end{center}
\end{figure}

\begin{figure}
\begin{center}
\epsfig{file=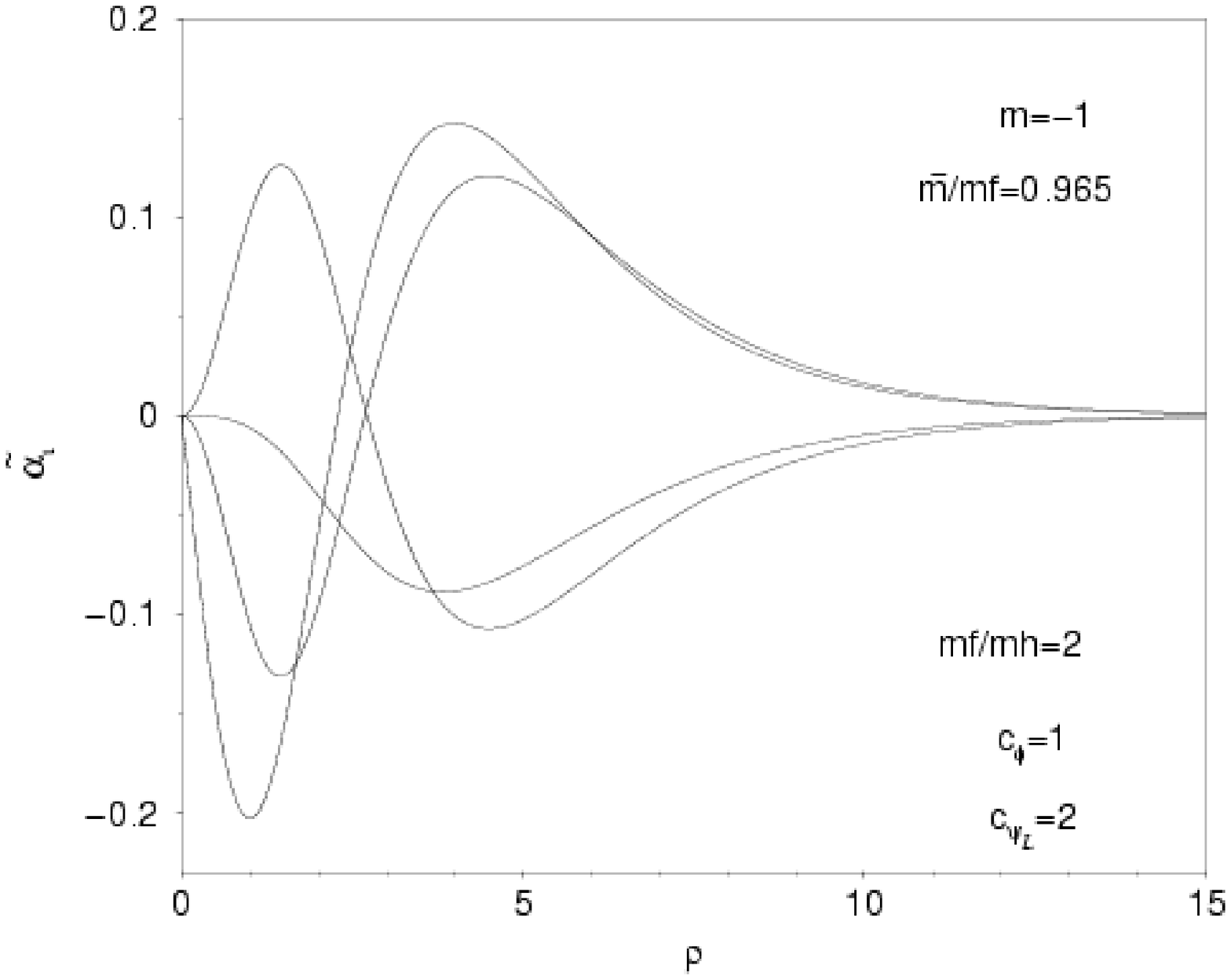,height=6cm}
\hspace{1pt}
\epsfig{file=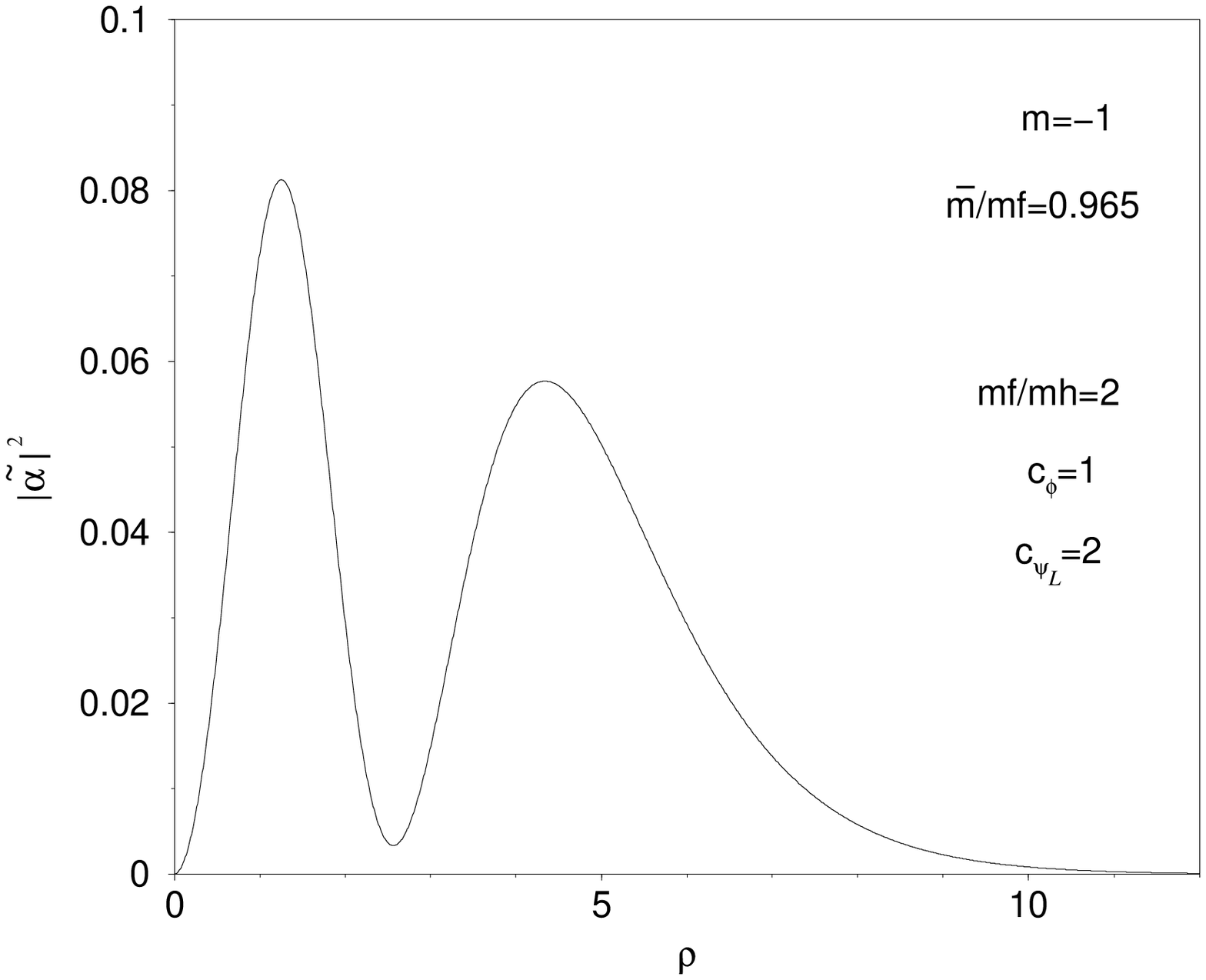,height=6cm}
\end{center}
\begin{center}
\epsfig{file=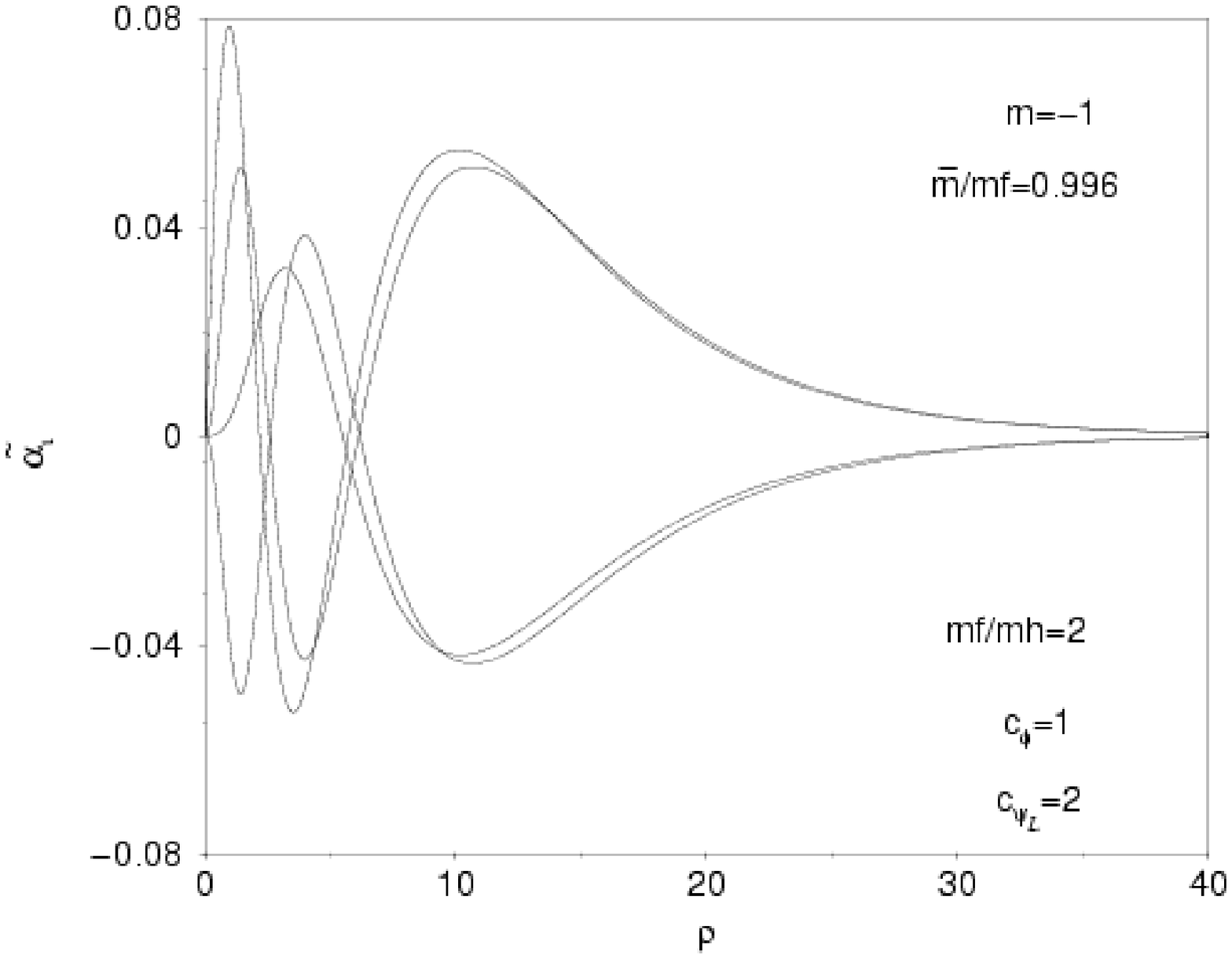,height=6cm}
\epsfig{file=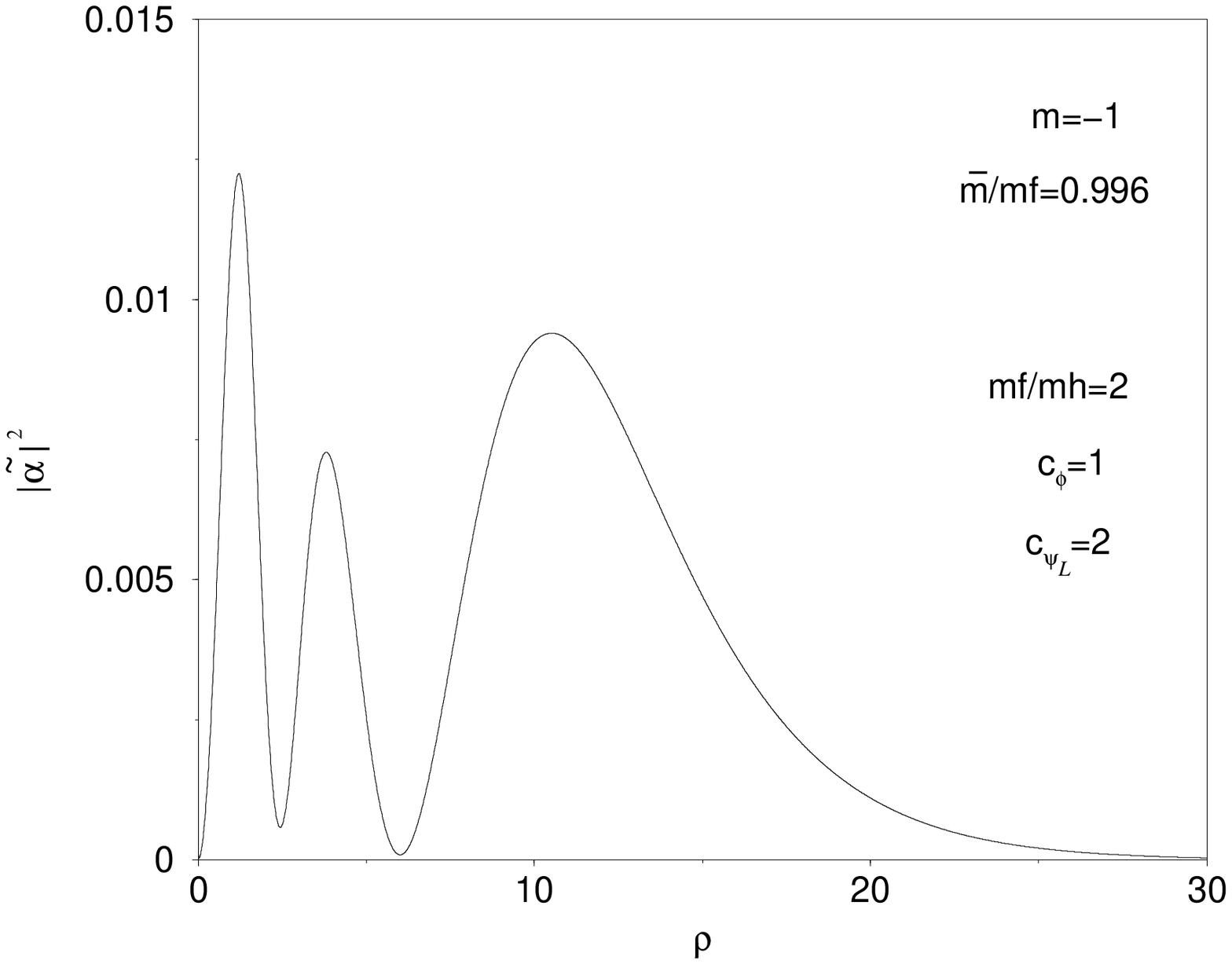,height=6cm}
\caption[Franges d'interf\'erences des fermions pi\'eg\'es dans une
corde cosmique.]{The components of the $\Psi$ field as function of the
radial distance, and the corresponding transverse probability
densities. The curves have been computed for $m=-1$ winding number,
and for two additional incoming modes in the nonperturbative
sector. As expected, interferences fringes appear from the nonzero
angular momentum eigenvalues of these modes.}
\label{figprobaex}
\end{center}
\end{figure}

\begin{figure}
\begin{center}
\epsfig{file=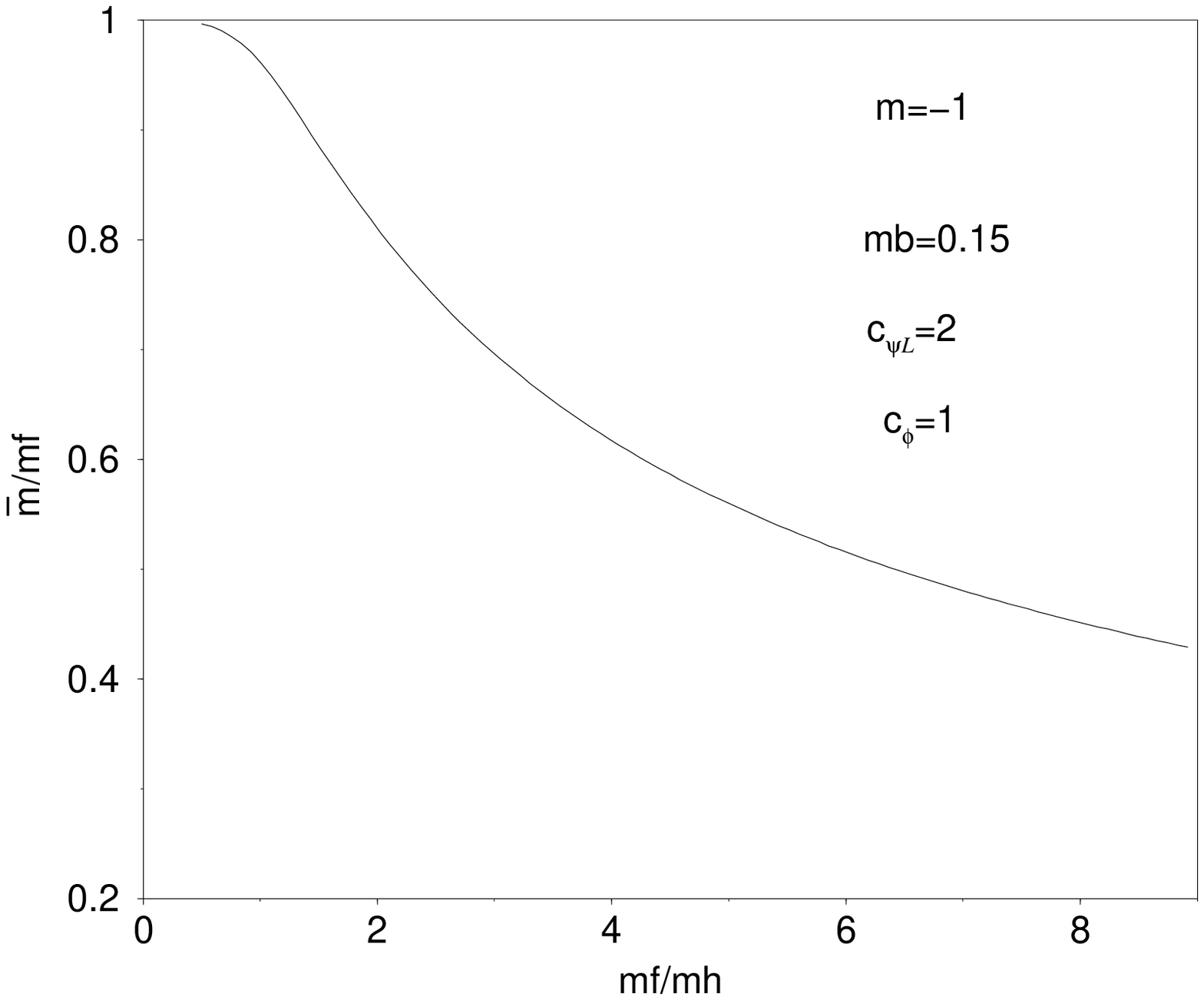,width=10cm}
\caption[Masse des fermions de moment angulaire non nul autour d'une
corde cosmique.]{The mass of the $m=-1$ lowest massive bound state,
relative to the fermion vacuum mass, with respect to the coupling
constant to the Higgs field. Note that the mode does not exist in the
perturbative sector.}
\label{figmassfond1}
\end{center}
\end{figure}

The nonperturbative cases with $m_{\mathrm{f}}>m_{\mathrm{h}}$,
involve much more massive bound states. First, another mode appears in
addition to the previous one, with the same winding number. Because of
the fact that $\miv$ decreases with $m_{\mathrm{f}}$ [see
Fig.~\ref{figmfond}], for higher values of $m_{\mathrm{f}}$, another
mode comes into the normalizable mass range. Since normalizability at
infinity requires $\miv<m_{\mathrm{f}}$, the number of massive modes
increases with the value of $m_{\mathrm{f}}$. Moreover, there are also
solutions involving all the other possible winding numbers. The
evolution of the mass spectrum, for winding number $m=0$, and with
respect to the coupling constant to the Higgs and gauge fields is
plotted in Fig.~\ref{figmasspect}. The behavior of each mass is the
same as that of the lowest mode previously studied, the new properties
resulting only in the appearance of new states for higher values of
the fermion vacuum mass $m_{\mathrm{f}}$, as found for two-dimensional
Weyl spinors in Ref.~\cite{davisS}.
Physically, the additional massive modes at a given winding number can
be interpreted as normalizable eigenstates of the angular momentum
operator in the vortex potential, with higher eigenvalues. From
Fig.~\ref{figprobafond} and Fig.~\ref{figprobafond1}, one can see
that for each value of $m$, the lowest massive state is confined
around the string with a transverse probability density showing only
one peak whereas the higher massive modes have transverse probability
density profiles with an increasing number of maxima, as can
be seen in Fig.~\ref{figprobaex}. In fact, as for the structure of
atomic spectra, the two spatial degrees of freedom of the attractive
potential certainly lead to two quantum numbers labeling the observable
eigenstates, one of them being clearly $m$, and the other appearing
through the number of zeros of the spinorial components, or, equivalently,
the number of maxima of the associated transverse probability density.
The massive modes with higher winding numbers behave in the same way.
However, they exist only for nonzero values of the coupling constant
$m_{\mathrm{f}}/m_{\mathrm{h}}$, this one increasing with the value of
the winding number $m$. The scaled spinorial components and the
transverse normalized probability density of the lowest massive bound
state with next $m=-1$ winding number are plotted in
Fig.~\ref{figprobafond1}. They are found to be normalizable for
coupling constant $m_{\mathrm{f}}/m_{\mathrm{h}} \gtrsim 0.5$ when
$\cpsil/\cphi=2$, as can be seen in Fig.~\ref{figmassfond1}. Contrary
to the $m=0$ lowest massive state, all spinorial components wind
around the string, and the transverse probability of finding such a
mode vanishes in the string core, as expected for a nonzero angular
momentum eigenstate. Obviously, this is also true for all higher
values of $m$, as for the $m=-2$ massive mode which appears to be
normalizable for $m_{\mathrm{f}}/m_{\mathrm{h}} \gtrsim 1.2$.
It is clear from the numerical results that the fermions can be
trapped in the string in the form of massive bound states, for a wide
range of model parameters. The only exception takes place for fermion
charges close to the particular value $\cpsil=\cphi/2$ where there is
no normalizable massive bound state in the perturbative sector. Note,
again, that all the previous results are also relevant for the
massive modes with symmetric winding numbers $\widehat{m}=n+1-m$, as
well as for the $\Chi$ spinor field and for the antiparticle states of
both $\Psi$ and $\Chi$ fermions.

\section{Fock space for massive modes}
\label{quantizationMM}
The existence of massive trapped waves requires that the quantum
state space~\cite{ringeval} be enlarged to include them. For each
normalizable mode with mass $\miv$, a two-dimensional Fock space can
be constructed by spinor field expansion over the relevant massive
plane waves. The full quantum theory can therefore be obtained from
tensorial product of the different Fock spaces belonging to their own
mass representation, together with the Fock space associated with the
zero modes~\cite{ringeval}. As a first step toward a full theory, we
will only consider the plane waves associated with one massive mode of
mass $\miv$.

\subsection{Quantum field operators}

Quantization is performed through the canonical procedure by
defining creation and annihilation operators satisfying
anticommutation rules. However, the particular structure of the
trapped massive waves yields relationships between longitudinal
quantum operators with nontrivial transverse dependencies.

\subsubsection{Fourier transform}

In the previous section, it was shown that the fermions could
propagate along the string direction with given mass $\miv$ belonging
to the spectrum. From Eq.~(\ref{planeansatzMM}), setting
\begin{eqnarray}
\Psi_{\mathrm{p}}^{(+)}  = u_\psi(k,r,\theta) \, \ue^{i(\omega t-kz)},
& &
\Psi_{\mathrm{p}}^{(-)}  = v_\psi(k,r,\theta) \, \ue^{-i(\omega t-kz)},
\end{eqnarray}
with
\begin{equation}
\label{dispmass}
\omega=\sqrt{\miv^2+k^2},
\end{equation}
and using the symmetry properties shown in Sec.~(\ref{symmetries}),
the transverse parts of the massive trapped waves for particle
and antiparticle states can be written as
\begin{equation}
\label{psiuv}
\begin{array}{ccccc}
u_\psi(k,\xper)
 & = &
\left(
\begin{array}{c}
\sqrt{\omega+k}\,\ab_1(r)\, \ue^{-im\theta}\\
i\sqrt{\omega-k}\,\ab_2(r)\, \ue^{-i(m-1)\theta}\\
\sqrt{\omega-k}\,\ab_3(r) \,\ue^{-i(m-n)\theta}\\
i\sqrt{\omega+k}\,\ab_4(r)\,\ue^{-i(m-n-1)\theta}
\end{array}
\right),
\\ \\
v_\psi(k,\xper)
& = &
\left(
\begin{array}{c}
\sqrt{\omega+k}\,\ab_1(r)\, \ue^{-im\theta}\\
-i\sqrt{\omega-k}\,\ab_2(r)\, \ue^{-i(m-1)\theta}\\
-\sqrt{\omega-k}\,\ab_3(r) \,\ue^{-i(m-n)\theta}\\
i\sqrt{\omega+k}\,\ab_4(r) \,\ue^{-i(m-n-1)\theta}
\end{array}
\right),
\end{array}
\end{equation}
with the notations
\begin{eqnarray}
\label{defalphabar}
\xper=(r,\theta), & &
\ab=\frac{m_{\mathrm{h}}}{\sqrt{2\pi}}\atd.
\end{eqnarray}
Contrary to the zero mode case, fermions can now propagate in both
directions of the string, so that the momentum $k$ of the massive
waves can take positive and negative values. As a result, the $\Psi$
field can be Fourier expanded over positive and negative energy states
as
\begin{equation}
\label{psiexpansionMM}
\Psi=\int{\frac{\ud k}{2\pi 2\omega} \left[b^{\dag}(k)\, u(k,\xper)\,
\ue^{i(\omega t-kz)} + d(k)\, v(k,\xper)\, \ue^{-i(\omega t-kz)} \right]},
\end{equation}
where the subscripts have been omitted. The normalization convention
of the Fourier transform is chosen as in the zero modes
case~\cite{ringeval}, i.e.
\begin{equation}
\label{deltaMM}
\int{\ud z\, \ue^{i(k-k^\prime)z}} = 2\pi \delta(k-k^\prime).
\end{equation}
Obviously, the $\Chi$ field verifies similar relations with the
transformation $n \rightarrow -n$, as was noted previously.

\subsubsection{Commutation relations}

In order to express the Fourier coefficients $b(k)$ and $d^\dag(k)$ as
function of the spinor field $\Psi$, let us introduce another unit
spinors
\begin{equation}
\label{psiuhvh}
\begin{array}{ccccc}
\uht(k,\xper)
 & = &
\left(
\begin{array}{c}
\sqrt{\omega+k}\,\ab_3(r)\, \ue^{-im\theta}\\
i\sqrt{\omega-k}\,\ab_4(r)\, \ue^{-i(m-1)\theta}\\
\sqrt{\omega-k}\,\ab_1(r) \,\ue^{-i(m-n)\theta}\\
i\sqrt{\omega+k}\,\ab_2(r)\,\ue^{-i(m-n-1)\theta}
\end{array}
\right),
\\ \\
\vht(k,\xper)
& = &
\left(
\begin{array}{c}
\sqrt{\omega+k}\,\ab_3(r)\, \ue^{-im\theta}\\
-i\sqrt{\omega-k}\,\ab_4(r)\, \ue^{-i(m-1)\theta}\\
-\sqrt{\omega-k}\,\ab_1(r) \,\ue^{-i(m-n)\theta}\\
i\sqrt{\omega+k}\,\ab_2(r) \,\ue^{-i(m-n-1)\theta}
\end{array}
\right).
\end{array}
\end{equation}
They clearly verify $\uht=\gamma^0\gamma^3 \vht$ and from
Eq.~(\ref{psiuv})
\begin{equation}
\label{orthonorm}
\begin{array}{ccccl}
\vsep \displaystyle
\uht^\dag(k)u(k) & = & \vht^\dag(k) v(k) & = &
2\omega \, \nub(r),
\\
\vsep \displaystyle
\uht(k)^\dag v(-k) & = & \vht^\dag(k) u(-k) & = & 0,
\end{array}
\end{equation}
where the dependency with respect to transverse coordinates have been
omitted in order to simplify the notation, and where we introduced the
function
\begin{equation}
\nub(r)=\ab_1(r) \ab_3(r) + \ab_2(r) \ab_4(r).
\end{equation}
From Eqs.~(\ref{psiexpansionMM}), (\ref{deltaMM}), and (\ref{orthonorm}),
the Fourier coefficients are found to be functions of the $\Psi$ field,
and read
\begin{equation}
\label{fouriercoeff}
\begin{array}{lcl}
\vsep
\displaystyle
b^\dag(k) & = & \displaystyle
\frac{1}{\norm} \int{r\,\ud r\,\ud \theta \,\ud z\,
\ue^{-i(\omega t-kz)} \uht^\dag(k,\xper) \Psi },
\\
\vsep
\displaystyle
d(k) & = & \displaystyle
\frac{1}{\norm} \int{r\,\ud r\,\ud
\theta \,\ud z\, \ue^{i(\omega t-kz)} \vht^\dag(k,\xper) \Psi},
\end{array}
\end{equation}
where we have defined the normalization factor
\begin{equation}
\norm=\int{r\,\ud r\,\ud\theta\, \nub(r)}=\int{\varrho\,\ud\varrho \,
\nut(\varrho)}
\end{equation}
Similarly, the expansion of the $\Psi^\dag$ field on the same positive
and negative energy solutions leads to the definition of its Fourier
coefficients, namely $b(k)$ and $d^{\dag}(k)$. From Eqs.~(\ref{deltaMM})
and (\ref{orthonorm}), they can also be expressed as functions of
$\Psi^\dag$, and verify
\begin{eqnarray}
\label{fourierconj}
b(k)=[b^\dag(k)]^\dag, &  &
d^\dag(k)=[d(k)]^\dag.
\end{eqnarray}

In order to perform a canonical quantization along the string
world sheet, let us postulate the anticommutation rules, \emph{at equal
times}, between the spinor fields
\begin{eqnarray}
\label{psiquantizationMM}
\left\{\Psi_i(t,\vec x), \Psi^{\dag j} (t,\vec x^\prime) \right\} & =
& \delta(z-z^\prime) \left(\Gamma^{0}\right)_i^j(\xper,\xper'),
\end{eqnarray}
where $\Gamma^0$ is a matrix with respect to spinor components whose
utility will be justified later, and which reads
\begin{eqnarray}
\Gamma^0(\xper,\xper') & = & \frac{1}{2 \miv^2} \left(\omega
{\mathrm{I}} - k \gamma^0 \gamma^3\right)\left[u(k,\xper)
u^\dag(k,\xper')+v(k,\xper) v^\dag(k,\xper') \right].
\end{eqnarray}
Note that $\Gamma^0$ does not depend on $\omega$ and $k$. Explicit
calculations show that the first terms involving $\omega$ and $k$ are
mixed with $u(k)$ and $v(k)$, and yield Lorentz invariant
quantities, such as $\miv$. Moreover, $\Gamma^0$ has the following
orthonormalization properties
\begin{equation}
\label{gammaprop}
\begin{array}{lcl}
\vsep
\displaystyle
\uht^\dag(k,\xper)\,\Gamma^0(\xper,\xper')\,\uht(k,\xper')
& = & 2\omega \,\nub(r) \nub(r')
\\
\vsep
\displaystyle
\uht^\dag(k,\xper)\,
\Gamma^0(\xper,\xper')\, \vht(-k,\xper') & = & 0,
\end{array}
\end{equation}
and similar relationships are obtained for $\vht$ by swapping $\uht$ and
$\vht$.

The anticommutation rules for the $\Psi$ Fourier coefficients are
immediately obtained from Eqs.~(\ref{deltaMM}), (\ref{fouriercoeff}),
(\ref{fourierconj}), and Eq.~(\ref{psiquantizationMM}), using the
properties of $\Gamma^0$ in Eq.~(\ref{gammaprop}), and read
\begin{equation}
\label{creatoranticomMM}
\begin{array}{ccccclll}
\left\{b(k), b^\dag(k^\prime) \right\} & = & \left\{d(k),
d^\dag(k^\prime)
\right\} & = & 2\pi2\omega \delta(k-k^\prime),
\end{array}
\end{equation}
with all the other anticommutators vanishing. As a result, the
Fourier coefficients $b^\dag$ and $d^\dag$ behave as well defined
creation operators, whereas their complex conjugates, $b$ and $d$, act
as annihilation operators of a particle and antiparticle massive
state, respectively.

In order to verify the microcausality of the theory and to justify,
\emph{a posteriori}, the ansatz of Eq.~(\ref{psiquantizationMM}), let us
derive the anticommutator between the quantum field operators $\Psi$
and $\Psi^\dag$, \emph{at any time}. The $\Psi$ expansion in
Eq.~(\ref{psiexpansionMM}) and its complex conjugate yield
\begin{eqnarray}
\label{fieldsanticom}
\left\{\Psi_i({\mathbf{x}}), \Psi^{\dag j} ({\mathbf{x}}^\prime) \right\}
& = &\int{\frac{\ud k\, \ud k^\prime}{(2\pi)^2 2\omega 2\omega'}}
\left[ \left\{b^\dag(k),b(k')\right\} u_i(k,\xper) u^{\dag
j}(k',\xper') \ue^{i[(\omega t -\omega' t')-(kz-k'z')]} \right.
\nonumber \\ & + & \left.  \left\{d(k),d^\dag(k')\right\} v_i(k,\xper)
v^{\dag j}(k',\xper')\ue^{-i[(\omega t -\omega'
t')-(kz-k'z')]}\right].
\end{eqnarray}
Using Eq.~(\ref{creatoranticomMM}), this equation simplifies to involve
tensorial products of unit spinors evaluated at the same momentum. It
is therefore convenient to introduce two additional matrices, namely
$\Gamma^3(\xper,\xper')$ and $\M(\xper,\xper')$, which verify
\begin{equation}
\label{tensorsplit}
\begin{array}{lcl}
\vsep \displaystyle u(k,\xper) u^{\dag}(k,\xper') & = & \displaystyle
\omega \Gamma^0(\xper,\xper')-k
\Gamma^3(\xper,\xper')-\M(\xper,\xper') \\ \vsep \displaystyle
v(k,\xper) v^{\dag}(k,\xper') & = & \displaystyle \omega
\Gamma^0(\xper,\xper')-k \Gamma^3(\xper,\xper')+\M(\xper,\xper').
\end{array}
\end{equation}
From Eq.~(\ref{psiuv}), these matrices are simply related to $\Gamma^0$
by
\begin{equation}
\begin{array}{lcl}
\label{gamma03m}
\vsep
\displaystyle
\Gamma^3(\xper,\xper') & = & \Gamma^0(\xper,\xper')\, \gamma^3
\gamma^0,\\
\vsep
\displaystyle
\M(\xper,\xper') & = &
\displaystyle
\Gamma^0(\xper,\xper')\,
\M_{\mathrm{d}}(\xper') \gamma^0,
\end{array}
\end{equation}
where $\M_{\mathrm{d}}(\xper)$ is the diagonal matrix
\begin{eqnarray}
\label{md}
\M_{\mathrm{d}}(\xper) & = & \miv \,{\mathrm{Diag}}
\left(\frac{\ab_3(r)}{\ab_1(r)}
\ue^{-in\theta},\frac{\ab_4(r)}{\ab_2(r)}
\ue^{-in\theta},\frac{\ab_1(r)}{\ab_3(r)}
\ue^{in\theta},\frac{\ab_2(r)}{\ab_4(r)} \ue^{in\theta}\right).
\end{eqnarray}
From Eqs.~(\ref{tensorsplit}) and (\ref{gamma03m}), the anticommutator
(\ref{fieldsanticom}) reduces to
\begin{eqnarray}\label{reducecom}
\left\{\Psi({\mathbf{x}}), \Psi^{\dag} ({\mathbf{x}}^\prime) \right\}
& = & \left[\Gamma^0(\xper,\xper')\,i\partial_0 +
\Gamma^3(\xper,\xper')\,
i \partial_3 + \M(\xper,\xper') \right] \nonumber \\
& \times & i \Delta(\xpar-\xpar'),
\end{eqnarray}
where $\xpar=(t,z)$, and $\Delta$ is the well-known Pauli Jordan
function which reads
\begin{eqnarray}
i\Delta(\xpar-\xpar') & = & \int{\frac{\ud k}{2\pi 2\omega}
\left[\ue^{-ik(\xpar-\xpar')} - \ue^{ik(\xpar-\xpar')}\right]},
\end{eqnarray}
and vanishes outside the light cone. As a result, the quantum fields
indeed respect microcausality along the string. The matrices
$\Gamma^\mu$ appear as the analogues of the matrices $\gamma^\mu$ for
the Dirac spinors living in the vortex. The two-dimensional
quantization along the string is thus not independent of the
transverse structure. It is all the more so manifest in the
anticommutator expression between $\Psi$ and $\overline{\Psi}$: from
Eq.~(\ref{reducecom}), and using Eq.~(\ref{gamma03m}), one gets
\begin{eqnarray}
\left\{\Psi({\mathbf{x}}), \overline{\Psi} ({\mathbf{x}}^\prime) \right\}
& = & \Gamma^0(\xper,\xper')\left[i\gamma^0 \partial_0 + i\gamma^3
\partial_3 + \M_{\mathrm{d}}(\xper')\right] \,  i \Delta(\xpar-\xpar').
\end{eqnarray}
The matrix $\Gamma^0$ now appears clearly as a local transverse
normalization of the longitudinal quantum field operators. Note that
the mass term also depends on the transverse coordinates due to the
nontrivial profile of the Higgs field around the string. Moreover,
setting $t=t'$ in Eq.~(\ref{reducecom}), leads to the postulated
anticommutator at equal times (\ref{psiquantizationMM}), and therefore
justifies the introduction of the $\Gamma^0$ term.

\subsubsection{Fock states}

In the following, $|\Pstate\rangle$ will design a Fock state
constructed by applying creation operators associated with a massive
mode $\miv$, on the relevant string vacuum. Such a state was similarly
defined for zero modes in Ref.~\cite{ringeval}. From the
anticommutators (\ref{creatoranticomMM}), a massive $\Psi$ state with
momentum $k$ is now normalized according to
\begin{equation}
\langle k'|k \rangle = 2\pi 2\omega \delta(k-k').
\end{equation}
Similarly, it will turn out to be convenient to derive the average of
the occupation number operator since it will appear in the derivation
of the equation of state. From Eq.~(\ref{creatoranticomMM}), and for a
$\Psi$ massive mode, it reads
\begin{equation}
\label{twoaverageMM}
\frac{\langle \Pstate |b^\dag(k) b(k')| \Pstate \rangle}{\langle
\Pstate|\Pstate \rangle} = \frac{2 \pi}{L} 2 \omega 2\pi \sum_i
\delta(k-k_i)\delta(k'-k_i),
\end{equation}
where the summation runs over all $\Psi$ massive particle states
present in the relevant $\miv$ Fock state $|\Pstate\rangle$, and $L$
is the physical string length, coming from the $\delta(0)$
regularization by means of Eq.~(\ref{deltaMM}).

\subsection{Stress tensor and Hamiltonian}

The classical stress tensor can immediately be derived from variation
of the full Lagrangian (\ref{lagrangien}) with respect to the metric,
and the $\Psi$ fermionic part thus reads~\cite{ringeval}
\begin{equation}
\label{psitensorMM}
T_\psi^{\mu \nu} = \frac{i}{2} \overline{\Psi} \gamma^{(\mu}
\partial^{\nu)}
\Psi - \frac{i}{2} \left[\partial^{(\mu}\overline{\Psi}\right]
\gamma^{\nu)} \Psi - B^{(\mu} j^{\nu)}_{_\psi}.
\end{equation}

\subsubsection{Hamiltonian}

The quantum operators associated with the classically conserved
charges can be obtained by replacing the classical fields by their
quantum forms involving creation and annihilation operators. In this
way, the Hamiltonian appears, from Noether theorem, as the charge
associated with the time component of the energy momentum tensor
\begin{equation}
T^{tt}_\psi=i \overline{\Psi}\gamma^0 \partial_t \Psi - i\left(\partial_t
\overline{\Psi} \right) \gamma^0 \Psi.
\end{equation}
Using Eqs.~(\ref{psiuv}) and (\ref{psiexpansionMM}) in the previous
equation, and performing a spatial integration, the Hamiltonian operator
reads, after some algebra,
\begin{equation}
\displaystyle
P^t_\psi= \int{\frac{\ud k}{2\pi2\omega}}\int{ \ud^2\xper}
\left[-b(k) b^\dag(k) +d^\dag(k) d(k) \right] \overline{u}(k,\xper)
\gamma^0 u(k,\xper).
\end{equation}
In order to simplify this expression, let us introduce the parameters
\begin{eqnarray}
\label{sigmadef}
\Sigmab_{\mathrm{X}}^2  =  \ab_2^2+\ab_3^2, & \quad \textrm{and} \quad &
\Sigmab_{\mathrm{Y}}^2  =  \ab_1^2+\ab_4^2.
\end{eqnarray}
From Eq.~(\ref{psiuv}), the Hamiltonian now reads
\begin{equation}
\label{hamiltonien}
\displaystyle P^t_\psi= \int{\frac{\ud k}{2\pi2\omega}} \left[-b(k)
b^\dag(k) +d^\dag(k) d(k) \right]\left[(\omega -
k)||\Sigma_{\mathrm{X}}^2||+(\omega+k)||\Sigma_{\mathrm{Y}}^2||\right],
\end{equation}
with
\begin{eqnarray}
||\Sigma^2|| = \int{r \, \ud r\, \ud \theta\,\Sigmab^2} = \int{\varrho
\,  \ud \varrho\,\widetilde{\Sigma}^2(\varrho)}.
\end{eqnarray}
Analogous relations also hold for the $\Chi$ field. It is interesting
to note that Eq.~(\ref{hamiltonien}) generalizes the expression
previously derived in the zero modes case~\cite{ringeval}, being found
again by setting $\omega=-k$ for the $\Psi$ zero modes, or $\omega=k$
for the $\Chi$ ones. The normal ordered Hamiltonian is obtained if one
uses the anticommuting normal ordered product for creation and
annihilation operators, i.e.
\begin{equation}
\label{hamiltonienorder}
\displaystyle :P^t_\psi:= \int{\frac{\ud k}{2\pi2\omega}}
\left[b^\dag(k)b(k) +d^\dag(k) d(k) \right]\left[(\omega -
k)||\Sigma_{\mathrm{X}}^2||+(\omega+k)||\Sigma_{\mathrm{Y}}^2||\right].
\end{equation}
Since $\omega \ge k$, this Hamiltonian is always positive definite and
is thus well defined. Note that, as in the zero mode case, such a
prescription overlooks the energy density difference between the
vacuum on the string and the usual one, but it can be shown to 
vary as $1/L^2$ and therefore goes to zero in the infinite string
limit~\cite{ringeval,fulling,kay}.

\subsubsection{Effective stress tensor}

In order to make contact with the macroscopic
formalism~\cite{carter89,carter89b,carter94b,carter97}, it is
necessary to express the classically observable quantities with no
explicit dependence in the microscopic structure. The relevant
two-dimensional fermionic energy momentum tensor can be identified
with the full one in Eq.~(\ref{psitensorMM}), once the transverse
coordinates have been integrated over. Due to the cylindrical symmetry
around the string direction, $z$ say, all nondiagonal components, in a
Cartesian basis, involving a transverse coordinate vanish after
integration. Moreover, since the fermion fields are normalizable in
the transverse plane, the diagonal terms, $T_\Ferm^{rr}$ and
$T_\Ferm^{\theta\theta}$, are also well defined, and by means of local
transverse stress tensor conservation, the integrated diagonal
components, $T_\Ferm^{xx}$ and $T_\Ferm^{yy}$, also
vanish~\cite{peterpuy93,peter94}. As expected, the only relevant
terms in the macroscopic formalism are thus $T_\Ferm^{\alpha \beta}$
with $\alpha,
\beta \in \{t,z\}$, i.e. the ones that live only in the string
world sheet. On the other hand, the macroscopic limit of the involved
quantum operators is simply obtained by taking their average over the
relevant Fock states.

Replacing the quantum fields in Eq.~(\ref{psitensorMM}) by their
expansion (\ref{psiexpansionMM}), and using Eq.~(\ref{psiuv}), one gets
the quantum expression of the energy momentum tensor. Averaging the
relevant components $T_\psi^{\alpha \beta}$ in the Fock state
$|\Pstate\rangle$, by means of Eqs.~(\ref{psiuv}) and
(\ref{twoaverageMM}), one obtains
\begin{eqnarray}
\label{ttpsitensor}
\langle :T_\psi^{tt}: \rangle_\Pstate & = & \frac{1}{L}
\left[\sum_i^{N_\psi} \overline{u}(k_i)\gamma^0 u(k_i) +
\sum_j^{\overline{N}_\psi} \overline{u}(k_j)\gamma^0 u(k_j)\right], \\
\label{zzpsitensor}
\langle :T_\psi^{zz}:\rangle_\Pstate & = & \frac{1}{L} \left[
\sum_i^{N_\psi} \frac{k_i}{\omega_i}\overline{u}(k_i)\gamma^3 u(k_i) +
\sum_j^{\overline{N}_\psi}
\frac{k_j}{\omega_j}\overline{u}(k_j)\gamma^3 u(k_j)\right], \\
\label{tzpsitensor}
\langle :T_\psi^{tz}:\rangle_\Pstate & = &
\frac{1}{2L}\left\{\sum_i^{N_\psi} \left[\overline{u}(k_i)\gamma^3 u(k_i) +
\frac{k_i}{\omega_i}\overline{u}(k_i)\gamma^0 u(k_i)\right] +
(i,N_\psi)\leftrightarrow (j,\overline{N}_\psi) \right\}, \qquad
\end{eqnarray}
where the  $i$ summations run over the $N_\psi$ particle states
with momentum $k_i$ involved in the Fock state $|\Pstate\rangle$,
while the  $j$ summations take care of the $\overline{N}_\psi$
antiparticle states with momentum $k_j$, all with mass $\miv$. In
order to simplify the notation, the transverse dependence of the unit
spinors have not been written, and the averaged operators stand for
\begin{equation}
\langle T^{\alpha \beta} \rangle_{{\mathcal{P}}} =
\frac{\langle {\mathcal{P}}|T^{\alpha \beta}|{\mathcal{P}}
\rangle}{ \langle {\mathcal{P}}|{\mathcal{P}}\rangle}.
\end{equation}
Similarly, the same relationships can be derived for the $\Chi$ field,
by replacing the $\Psi$ unit spinors by the $\Chi$ ones with the
correct angular dependence, and certainly, in another mass
representation $\miv_\chi$. Once the transverse coordinates have been
integrated over, Eqs.~(\ref{ttpsitensor}), (\ref{zzpsitensor}), and
(\ref{tzpsitensor}), lead to the two-dimensional $\Psi$ stress tensor
\begin{equation}
\label{twopsistress}
\begin{array}{l}
\displaystyle
\langle \overline{T}_\psi^{\alpha \beta} \rangle_\Pstate = \\ \\
\left(
\begin{array}{ccc}
\displaystyle E_\psip ||\Sigma_{\mathrm{Y}}^2|| + \overline{E}_\psip
||\Sigma_{\mathrm{X}}^2|| &  \displaystyle
\frac{E_\psip+P_\psip}{2}||\Sigma_{\mathrm{Y}}^2||-\frac{\overline{E}_\psip
- \overline{P}_\psip}{2} ||\Sigma_{\mathrm{X}}^2|| \\ \\ \displaystyle
\frac{E_\psip+P_\psip}{2}||\Sigma_{\mathrm{Y}}^2||-\frac{\overline{E}_\psip
- \overline{P}_\psip}{2} ||\Sigma_{\mathrm{X}}^2|| &
 \displaystyle P_\psip ||\Sigma_{\mathrm{Y}}^2|| -
\overline{P}_\psip ||\Sigma_{\mathrm{X}}^2||
\end{array}
\right),
\end{array}
\end{equation}
with the notations
\begin{equation}
\label{defparam}
\begin{array}{lcl}
\displaystyle
E_\psip & = &
\displaystyle\frac{1}{L}
\left[\sum_i^{N_\psi}\left(\omega_i+k_i\right) +
\sum_j^{\overline{N}_\psi} \left(\omega_j+k_j\right)\right], \\ \\
\displaystyle
\overline{E}_\psip & = &
\displaystyle
\frac{1}{L}\left[\sum_i^{N_\psi}\left(\omega_i-k_i\right) +
\sum_j^{\overline{N}_\psi} \left(\omega_j-k_j\right) \right],\\ \\
\displaystyle
\end{array}
\end{equation}
and
\begin{equation}
\begin{array}{lcl}
P_\psip & = & \displaystyle\frac{1}{L}
\left[\sum_i^{N_\psi}\frac{k_i}{\omega_i}\left(\omega_i+k_i\right) +
\sum_j^{\overline{N}_\psi}
\frac{k_j}{\omega_j}\left(\omega_j+k_j\right)\right], \\ \\
\displaystyle
\overline{P}_\psip & = & \displaystyle\frac{1}{L}
\left[\sum_i^{N_\psi}\frac{k_i}{\omega_i}\left(\omega_i-k_i\right) +
\sum_j^{\overline{N}_\psi}
\frac{k_j}{\omega_j}\left(\omega_j-k_j\right)\right].
\end{array}
\end{equation}
Recall that the full effective energy momentum tensor also involves
the Higgs and gauge fields of the vortex background. Since they
essentially describe a Goto-Nambu string~\cite{goto,nambu}, their
transverse integration yields a traceless diagonal tensor
\begin{equation}
\label{gotonambu}
\int{r \,\ud r\,\ud \theta \left(T^{tt}_{\mathrm{g}} +
T^{tt}_{\mathrm{h}} \right)} = -\int{r\,\ud r \,\ud \theta
\left(T^{zz}_{\mathrm{g}} + T^{zz}_{\mathrm{h}} \right)} \equiv M^2.
\end{equation}
Note that the full stress tensor may also involve several massive
$\Psi$ and $\Chi$ states, with different masses belonging to the
spectrum. In this case, there will be as many additional terms in the
form of Eq.~(\ref{twopsistress}), as different massive states there
are in the chosen Fock states.

\subsection{Fermionic currents}

The quantum current operators can be derived from their classical
expressions (\ref{currentsMM}) by using Eq.~(\ref{psiexpansionMM}), while
the corresponding conserved charges are obtained from their spatial
integration. By means of Eq.~(\ref{twoaverageMM}), the current operator,
averaged in the relevant Fock state $|\Pstate\rangle$, reads
\begin{eqnarray}
\label{currentaverage}
\langle:j^\alpha_\Ferm:\rangle_\Pstate & = & q \displaystyle{\frac{\cfr+
\cfl}{2}} \frac{1}{L}\left(-\sum_i^{N_\Ferm}
\frac{\overline{u}_{i}\gamma^\alpha u_{i}}{\omega_i} +
\sum_j^{\overline{N}_\Ferm} \frac{\overline{u}_{j}
\gamma^\alpha u_{j}}{\omega_j} \right)  \nonumber
\\
& + & q \displaystyle{\frac{\cfr-
\cfl}{2}} \frac{1}{L}\left(-\sum_i^{N_\Ferm}
\frac{\overline{u}_{i}\gamma^\alpha\gamma_5 u_{i}}{\omega_i} +
\sum_j^{\overline{N}_\Ferm} \frac{\overline{u}_{j}
\gamma^\alpha\gamma_5 u_{j}}{\omega_j} \right),
\end{eqnarray}
with $\alpha \in \{t,z\}$, and once again, the sums run over $\Psi$,
or $\Chi$, particle and antiparticle states. The $u_i$ are the unit
spinors associated with the field dealt with. Concerning the transverse
components, due to the properties of the unit spinors $u$ and $v$ in
Eq.~(\ref{psiuv}), only the orthoradial one does not vanish and reads
\begin{eqnarray}
\langle:j^\theta_\Ferm:\rangle_\Pstate & = &- q \displaystyle{\frac{\cfr+
\cfl}{2}} \frac{1}{L}\left(\sum_i^{N_\Ferm}
\frac{\overline{u}_{i}\gamma^\theta u_{i}}{\omega_i} +
\sum_j^{\overline{N}_\Ferm} \frac{\overline{u}_{j}
\gamma^\theta u_{j}}{\omega_j} \right)  \nonumber
\\
& - & q \displaystyle{\frac{\cfr-
\cfl}{2}} \frac{1}{L}\left(\sum_i^{N_\Ferm}
\frac{\overline{u}_{i}\gamma^\theta\gamma_5 u_{i}}{\omega_i} +
\sum_j^{\overline{N}_\Ferm} \frac{\overline{u}_{j}
\gamma^\theta\gamma_5 u_{j}}{\omega_j} \right),
\end{eqnarray}
whereas $\langle:j_\Ferm^r:\rangle=0$ due to the bound state nature of
the studied currents. As expected, the gauge charges carried by each
trapped fermion in the form of massive mode, generate only macroscopic
charge and current densities along the string, as was the case for the
zero modes~\cite{ringeval}. However, the nonvanishing orthoradial
component shows that the local charges also wind around the string
while propagating in the $z$ direction, as suggested by the above
numerical studies. However, this component will be no longer relevant
in the macroscopic formalism, since it vanishes in a Cartesian basis,
once the transverse coordinates have been integrated over.\\
Nevertheless, this nonzero angular momentum of the massive modes is
found to generate new properties for the longitudinal currents. Let us
focus on the vectorial gauge currents generated by one exitation
state, with energy $\omega$ and momentum $k$, of a $\Psi$
massive mode, $\miv$ say. From Eq.~(\ref{currentaverage}), using
Eqs.~(\ref{psiuv}) and (\ref{sigmadef}), the world sheet vectorial
charge current reads
\begin{eqnarray}
\langle:j^0_{\psi \mathrm{V}}:\rangle_\varepsilon & = &- q
\frac{\cpsir+\cpsil}{2} \frac{\varepsilon}{L}
\left[\left(1+\frac{k}{\omega}\right)\Sigmab_{\mathrm{Y}}^2 + \left(1
- \frac{k}{\omega} \right)\Sigmab_{\mathrm{X}}^2\right],\\
\label{anommoment}
\langle:j^3_{\psi \mathrm{V}}:\rangle_\varepsilon & = &- q
\frac{\cpsir+\cpsil}{2} \frac{\varepsilon}{L}
\left[\left(1+\frac{k}{\omega}\right)\Sigmab_{\mathrm{Y}}^2 - \left(1
- \frac{k}{\omega} \right)\Sigmab_{\mathrm{X}}^2\right],
\end{eqnarray}
where $\varepsilon=\pm 1$ stands a one particle or antiparticle
exitation state. Now, even setting $k=0$ in the previous equations
yields a nonvanishing spatial current. Physically, it can be simply
interpreted as an anomalous magnetic-like moment of the considered
massive mode in its rest frame. Examining Eq.~(\ref{anommoment}) shows
that it could be null only if
$\Sigmab_{\mathrm{Y}}^2(r)=\Sigmab_{\mathrm{X}}^2(r)$, which is
generally not satisfied due to the particular shapes of massive spinor
components trapped in the string (see Sec.~\ref{modemassif}). These
ones being associated with nonzero winding numbers, it is therefore
not surprising that, even for a massive stationary state along the
string, the nonvanishing charge angular momentum around the string
generates such additional magneticlike moment. Note that it does not
concern the zero modes, first because they precisely involve
vanishing winding numbers~\cite{ringeval}, and then because for them,
there is no defined rest frame due to their vanishing mass. Obviously,
this property can be generalized for the axial part of the current,
and thus is also valid for the total current of any massive spinor
field trapped in the string.

All the above construction of the Fock space and the derivation of the
quantum operators associated with the energy momentum tensor and gauge
currents remains valid for each $\Psi$ and $\Chi$ massive mode. More
precisely, the other masses belonging to the $\Psi$ spectrum verify
analogous relationships provided $\miv$ is replaced by the relevant
one, as for the unit spinors. In addition, the $\Chi$ massive states
require to transform $n\rightarrow-n$ in Eq.~(\ref{md}), due to their
coupling to the antivortex. At this stage, the averaged values of the
stress tensor and currents have been obtained, and therefore allow the
derivation of an equation of state, once the Fock states are known.

\section{Equation of state}
\label{etatMM}

The energy per unit length and tension in a given Fock state
$|\Pstate\rangle$ are basically the eigenvalues associated with
timelike and spacelike eigenvectors of the effective two-dimensional
full stress tensor. Obviously this one includes the classical
Goto-Nambu term resulting of the string forming Higgs and gauge fields
[see Eq.~(\ref{gotonambu})], with the fermionic part generated by the
massive currents [see Eq.~(\ref{twopsistress})]. Moreover, in order to
describe the string by an adequate macroscopic
formalism, it is necessary to choose a quantum
statistics for the relevant Fock states, and for energy scales far
below the ones where the string was formed, it is reasonable to
consider a Fermi-Dirac distribution at zero
temperature~\cite{ringeval,prep}. In the following, the equation of
state is first derived for the lowest massive modes associated with
the $\Psi$ and $\Chi$ field, and simplified in the zero-temperature
and infinite string limit. As a second step, these derivations are
generalized to any number and kind of trapped fermionic mode.

\subsection{Lowest massive modes}

\subsubsection{Energy per unit length and tension}

In this section, we will only consider the lowest massive modes
belonging to the $\Psi$ and $\Chi$ mass spectrums, with masses
$\miv_\psi$ and $\miv_\chi$, respectively. From
Eqs.~(\ref{twopsistress}) and (\ref{gotonambu}), the full effective
energy momentum tensor reads
\begin{equation}
\label{lowtwotensor}
\langle\overline{T}^{\alpha \beta}\rangle_\Pstate = \int{r \, \ud r \,
\ud \theta\,\left( T^{\alpha\beta}_{\mathrm{g}} + T^{\alpha
\beta}_{\mathrm{h}}\right)} + \langle\overline{T}^{\alpha
\beta}_\psi\rangle_\Pstate + \langle \overline{T}^{\alpha
\beta}_\chi\rangle_\Pstate,
\end{equation}
where $\overline{T}^{\alpha \beta}_\chi$ takes the same form as
$\overline{T}^{\alpha \beta}_\psi$ in Eqs.~(\ref{twopsistress}) and
(\ref{defparam}) once the $\Psi$ relevant parameters have been
replaced by the $\Chi$ ones. In the preferred frame where the stress
tensor is diagonal, we can identify its timelike and spacelike
eigenvalues with energy per unit length $U$ and tension $T$. Upon
using Eqs.~(\ref{twopsistress}), (\ref{gotonambu}), and
(\ref{lowtwotensor}), these read
\begin{eqnarray}
\label{upsichi}
U_\Pstate & = & M^2+ \sum_{\Ferm \in \{\Psi,\Chi\}}\left(
\frac{E_\Ferm-P_\Ferm}{2}||\Sigma_{\mathrm{Y} \Ferm}^2|| +
\frac{\overline{E}_\Ferm + \overline{P}_\Ferm}{2} ||\Sigma_{\mathrm{X}
\Ferm}^2||\right) \nonumber \\ \vsep & & + \quad \left(\sum_{\Ferm \in
\{\Psi,\Chi\}}\left(E_\Ferm + P_\Ferm\right)||\Sigma_{\mathrm{Y}
\Ferm}^2||\, \times\, \sum_{\Ferm \in
\{\Psi,\Chi\}}\left(\overline{E}_\Ferm - \overline{P}_\Ferm\right)
||\Sigma_{\mathrm{X} \Ferm}^2||\right)^{1/2}\,,
\end{eqnarray}
for the energy per unit length, and
\begin{eqnarray}
\label{tpsichi}
T_\Pstate & = & M^2+ \sum_{\Ferm \in \{\Psi,\Chi\}}\left(
\frac{E_\Ferm-P_\Ferm}{2}||\Sigma_{\mathrm{Y} \Ferm}^2|| +
\frac{\overline{E}_\Ferm + \overline{P}_\Ferm}{2} ||\Sigma_{\mathrm{X}
\Ferm}^2||\right) \nonumber \\ \vsep & & - \quad \left(\sum_{\Ferm \in
\{\Psi,\Chi\}}\left(E_\Ferm + P_\Ferm\right)||\Sigma_{\mathrm{Y}
\Ferm}^2||\,\times\, \sum_{\Ferm \in
\{\Psi,\Chi\}}\left(\overline{E}_\Ferm - \overline{P}_\Ferm\right)
||\Sigma_{\mathrm{X} \Ferm}^2||\right)^{1/2}\,,
\end{eqnarray}
for the tension. It is interesting to note first that
$U_\Pstate+T_\Pstate \neq 2M^2$, and thus the fixed trace equation of
state previously found for zero modes~\cite{ringeval,prep} is no
longer verified by massive modes, as expected since they are no longer
eigenstates of the $\gamma^0\gamma^3$ operator. Moreover, the
expression of energy density and tension does not seem to involve the
conserved charge current magnitude, which played the role of a state
parameter in the case of a scalar condensate in a cosmic
string~\cite{carter89,carter89b,carter94b,carter97,neutral}. In fact,
as it was the case at zeroth order for the zero modes~\cite{ringeval},
the charge currents are only involved in the stress tensor through
their coupling to the gauge field [see Eq.~(\ref{psitensorMM})]. At
zeroth order, when the back reaction is neglected, the only
nonvanishing component of the gauge field is $B_\theta$, and it
therefore couples only with $j_\Ferm^\theta$, which vanishes once the
transverse coordinates have been integrated over. As a result, it is
not surprising that the equation of state does not involve the
fermionic currents without back reaction. As a result, it is more
natural from quantization to define the occupation numbers of the
involved species as state parameters.

\subsubsection{Zero-temperature and infinite string limit}
\label{zerolimit}
\begin{figure}
\begin{center}
\epsfig{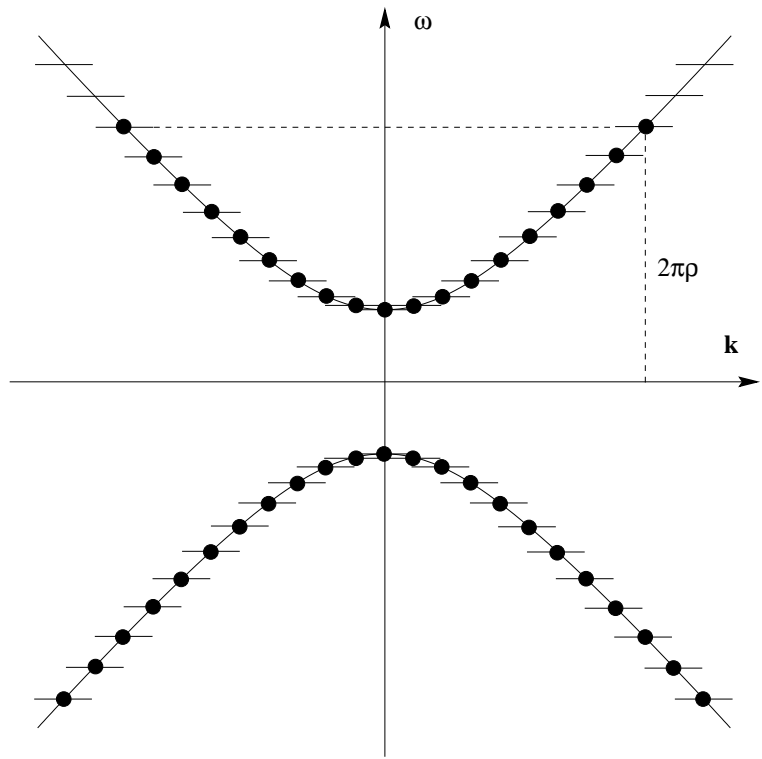}
\caption[L'occupation des \'etats massifs fermioniques.]{The filling
of massive states trapped in the string, as expected in the
zero-temperature limit, for a particular mass and for one species,
$\Psi$ or $\Chi$ say. All antiparticles have disappeared by
annihilation with particles during cooling, and the interactions
between particles moving in opposite directions, as their coupling to
the gauge field, lead to the energetically favored configuration with
same number of up and down movers. Obviously, the Fermi level is
necessarily below the vacuum mass of the relevant fermion.}
\label{figfillstates}
\end{center}
\end{figure}

\begin{figure}
\begin{center}
\epsfig{file=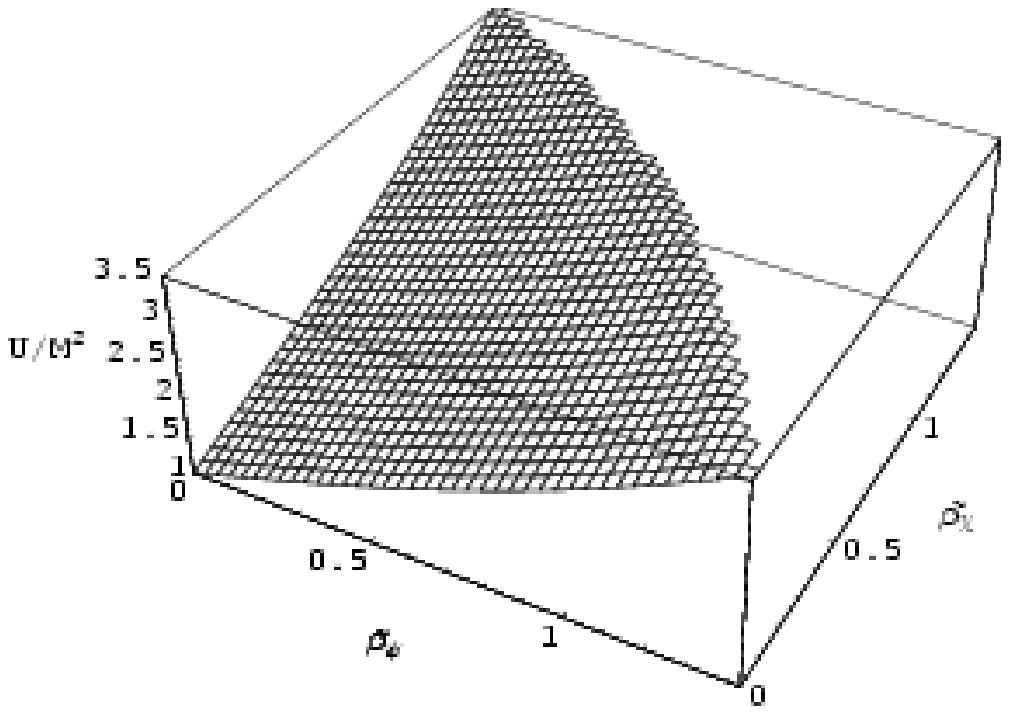,width=10cm}
\epsfig{file=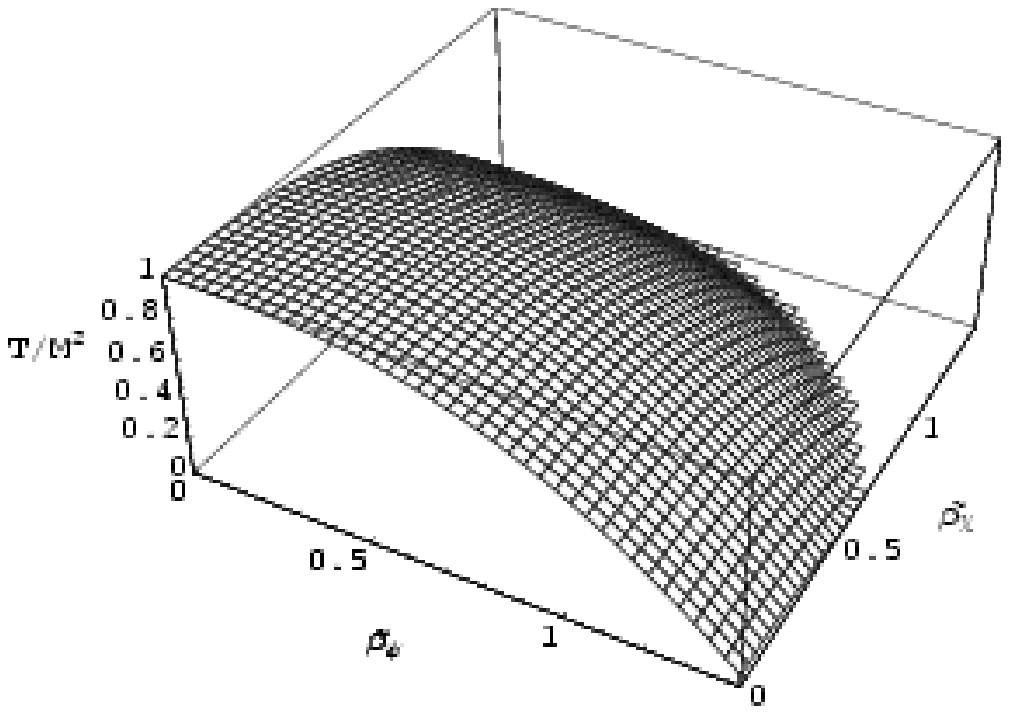,width=10cm}
\caption[Influence des modes massifs fermioniques sur la densit\'e
d'\'energie de la cor\-de.]{The energy per unit length and the tension,
in unit of $M^2$, for the lowest massive modes alone, plotted as
function of the dimensionless effective densities of the two fermion
fields, $\rhot_\psi$ and $\rhot_\chi$. The parameters have been chosen
in the nonperturbative sector, with $m_{\mathrm{f}}/m_{\mathrm{h}}\sim
3$ and $\miv_\chi \sim \miv_\psi \sim 0.6 m_{\mathrm{f}}$. Note the
linear variation of the energy density near the origin whereas the
tension varies quadratically. Moreover, in the allowed range for
fermion densities, i.e. less than the fermion vacuum mass, the
tension vanishes and the string becomes unstable with respect to
transverse perturbations.}
\label{figutm1}
\end{center}
\end{figure}

Assuming a Fermi-Dirac distribution at zero temperature for the
exitation states, the sums involved in Eq.~(\ref{defparam}) run over
the successive values of the allowed momentum $k_i$ until the Fermi
level of the considered species is reached. With periodic boundary
conditions on spinor fields, the allowed momentum exitation values are
discretized according to
\begin{equation}
k_n=\frac{2\pi}{L}n,
\end{equation}
where $n$ is an integer, playing the role of a quantum exitation
number. As a result, in the relevant $\miv$ representation of each
field, the exitation energies $\omega_i$ are also discrete according
to Eq.~(\ref{dispmass}), and for the $\Psi$ field, the parameters
$E_\psi$ and $P_\psi$ in Eq.~(\ref{defparam}) simplify to sums of
radical function of $n$, with $n$ running from the vacuum to the last
filled state. In order to express them as explicit functions of the
relevant Fermi level, it is convenient to consider the infinite
string limit $L\rightarrow\infty$. In this limit one gets
\begin{equation}
\label{thermlimit}
\lim_{L \to \infty}\frac{1}{L}\sum_{i=-N_\psi^-}^{N_\psi^+}f(k_i) =
\frac{1}{2\pi} \int_{-2 \pi \rho_\psi^-} ^{2\pi\rho_\psi^+} \ud k\,f(k),
\end{equation}
where $\rho_\psi^\pm=N_\psi^\pm/L$ are the $\Psi$ up and down
mover densities, $N_\psi^+$, $N_\psi^-$ standing for the number of
$\Psi$ particle moving in the $+z$ or $-z$ directions,
respectively. Note that the total number of particles of this
kind is thus $N_\psi=N_\psi^++N_\psi^-+1$ since there is the
additional rest state obtained for $k=0$. After some algebra in
Eq.~(\ref{defparam}), using Eq.~(\ref{thermlimit}), the parameters,
for the $\Psi$ field, read
\begin{eqnarray}
\label{limitparam}
E_\psi & = &
\frac{\miv^2}{4\pi}\left\{\rhot_\psi^{+^2}-\rhot_\psi^{-^2} +
\left(\rhot_\psi^+ \sqrt{1+\rhot_\psi^{+^2}} + \rhot_\psi^- \sqrt{1 +
\rhot_\psi^{-^2}}\right) \right. \nonumber \\
& & +  \left. \ln{\left[\left(\sqrt{1 + \rhot_\psi^{+^2}}
+ \rhot_\psi^{+}\right)\left(\sqrt{1 + \rhot_\psi^{-^2}} +
\rhot_\psi^{-}\right)\right]} \right\}  +
\left(\rhot_\psi^\pm \leftrightarrow \overline{\rhot}_\psi^\pm\right),
\\ P_\psi & = &
\frac{\miv^2}{4\pi}\left\{\rhot_\psi^{+^2}-\rhot_\psi^{-^2} +
\left(\rhot_\psi^+ \sqrt{1+\rhot_\psi^{+^2}} + \rhot_\psi^- \sqrt{1 +
\rhot_\psi^{-^2}}\right) \right. \nonumber \\
& & \left.  - \ln{\left[\left(\sqrt{1 + \rhot_\psi^{+^2}}
+ \rhot_\psi^{+}\right)\left(\sqrt{1 + \rhot_\psi^{-^2}} +
\rhot_\psi^{-}\right)\right]} \right\} +
\left(\rhot_\psi^\pm \leftrightarrow \overline{\rhot}_\psi^\pm\right),
\end{eqnarray}
where $\rhot_\psi$ stands for the dimensionless $\Psi$ mover density
\begin{equation}
\label{adimdens}
\rhot_\psi=\frac{2\pi}{\miv}\rho_\psi,
\end{equation}
while $\overline{\rhot}_\psi$ is defined in the same way for the
$\Psi$ antiparticle states. Similarly, the two other parameters
$\overline{E}_\psi$ and $\overline{P}_\psi$ read
\begin{eqnarray}
\overline{E}_\psi & = &
\frac{\miv^2}{4\pi}\left\{-\rhot_\psi^{+^2}+\rhot_\psi^{-^2} +
\left(\rhot_\psi^+ \sqrt{1+\rhot_\psi^{+^2}} + \rhot_\psi^- \sqrt{1 +
\rhot_\psi^{-^2}}\right) \right.  \nonumber \\ 
& & \left. + \ln{\left[\left(\sqrt{1 + \rhot_\psi^{+^2}}
+ \rhot_\psi^{+}\right)\left(\sqrt{1 + \rhot_\psi^{-^2}} +
\rhot_\psi^{-}\right)\right]} \right\} +
\left(\rhot_\psi^\pm \leftrightarrow \overline{\rhot}_\psi^\pm\right),
\\
\label{limitparambar}
\overline{P}_\psi & = &
\frac{\miv^2}{4\pi}\left\{\rhot_\psi^{+^2}-\rhot_\psi^{-^2} -
\left(\rhot_\psi^+ \sqrt{1+\rhot_\psi^{+^2}} + \rhot_\psi^- \sqrt{1 +
\rhot_\psi^{-^2}}\right] \right. \nonumber \\
& & \left. + \ln{\left[\left(\sqrt{1 + \rhot_\psi^{+^2}}
+ \rhot_\psi^{+}\right)\left(\sqrt{1 + \rhot_\psi^{-^2}} +
\rhot_\psi^{-}\right)\right]} \right\}  +
\left(\rhot_\psi^\pm \leftrightarrow \overline{\rhot}_\psi^\pm\right).
\end{eqnarray}
Note that these parameters depend differently on the up and
down mover densities as expected for chiral coupling of the
fermions to the string forming Higgs field. Recall that in the
massless case the zero modes associated with the $\Psi$ and $\Chi$
fields can only propagate in the $-z$ and $+z$ direction
respectively~\cite{witten,ringeval,jackiwrossi}. The same
relationships also hold for the $\Chi$ field by using the relevant
dimensionless mover densities $\rhot^\pm_\chi$ and
$\overline{\rhot}_\chi^\pm$. Although the equation of state can be
derived as a function of these four parameters for each fermion field
$\Ferm$, it is convenient at this stage to perform some physical
simplifications. Contrary to the zero mode case, the coupling between
massive particles and antiparticles of the same species $\Ferm$ does
not vanish along the string. As a result, it is reasonable to consider
that the only kind surviving at zero temperature corresponds to the
one which was in excess in the plasma in which the string was formed
during the phase transition. On the other hand, the energetically
favored distribution at zero temperature involves necessarily the same
number of $\Ferm$ ``up'' and ``down'' movers, each filling the
accessible states living on each branch of the mass hyperbola (see
Fig.~\ref{figfillstates}). As a result, in the considered energy
scale, it seems reasonable to consider only one state parameter per
mass instead of the four initially introduced by quantization, namely
$\rhot_\Ferm=\rhot_\Ferm^+=\rhot_\Ferm^-$, for a plasma dominated by
particles, say.
Setting these simplifications in Eqs.~(\ref{limitparam}) to
(\ref{limitparambar}), by means of Eqs.~(\ref{upsichi}) and
(\ref{tpsichi}) the energy density and the tension associated with the
lowest massive modes now read
\begin{eqnarray}
\label{loweqstate}
U & = & M^2+ \frac{1}{2\pi}\sum_{\Ferm} \miv_\Ferm^2 \ln{\left(\sqrt{1
+ \rhot_\Ferm^2} + \rhot_\Ferm\right)} \nonumber \\ & & +
\frac{1}{\pi}\,\left(\sum_\Ferm \miv_\Ferm^2 \, ||\Sigma_{\mathrm{Y}
\Ferm}^2||\, \rhot_\Ferm \sqrt{1+\rhot_\Ferm^2}\,\times\, \sum_\Ferm
\miv_\Ferm^2 \, ||\Sigma_{\mathrm{X} \Ferm}^2||\, \rhot_\Ferm
\sqrt{1+\rhot_\Ferm^2}\right)^{1/2}\,,\\ T & = & M^2+
\frac{1}{2\pi}\sum_{\Ferm} \miv_\Ferm^2 \ln{\left(\sqrt{1 +
\rhot_\Ferm^2} + \rhot_\Ferm \right)} \nonumber \\ & & -
\frac{1}{\pi}\,\left(\sum_\Ferm \miv_\Ferm^2 \, ||\Sigma_{\mathrm{Y}
\Ferm}^2||\, \rhot_\Ferm \sqrt{1+\rhot_\Ferm^2}\,\times\, \sum_\Ferm
\miv_\Ferm^2 \, ||\Sigma_{\mathrm{X} \Ferm}^2||\, \rhot_\Ferm
\sqrt{1+\rhot_\Ferm^2}\right)^{1/2}\,.
\end{eqnarray}
The sum runs over the two lowest massive bound states, each one being
associated to the two fermion fields trapped in the vortex, namely
$\Psi$ and $\Chi$, and have $\miv_\psi$ and $\miv_\chi$ masses,
respectively. As a result, the equation of state involves two
independent parameters, $\rhot_\psi$ and $\rhot_\chi$, in the
zero-temperature and infinite string limit. The energy per unit length
and the tension have been plotted in Fig.~\ref{figutm1}, for the
lowest massive modes in the nonperturbative sector. The curves are
essentially the same in the perturbative sector, but the variations
around the Goto-Nambu case $U=T=M^2$ are much more damped. For
reasonable values of the transverse normalizations, e.g.
$||\Sigma_{\mathrm{X}}^2|| \sim ||\Sigma_{\mathrm{Y}}^2|| \sim 0.5$,
and for small values of the dimensionless parameters $\rhot_\psi$ and
$\rhot_\chi$, the energy density is found to grow linearly with
$\rhot_\psi$ and $\rhot_\chi$, whereas the tension varies
quadratically. As can be seen in Eq.~(\ref{loweqstate}), due to the
minus sign in $T$, all linear terms in $\rhot$ vanish near origin,
whereas it is not the case for the energy density. However, for higher
values of the densities, the quadratic terms dominate and both energy
density and tension end up being quadratic functions of $\rhot$. On
the other hand, according to the macroscopic
formalism~\cite{carter89,carter89b,carter94b,carter97}, the string
becomes unstable with respect to transverse perturbations when the
tension takes on negative values, as in Fig.~\ref{figutm1} for high
densities. Moreover, the decrease of the tension is more damped in the
perturbative sector, and the negative values cannot actually be
reached for acceptable values of $\rhot$, i.e. $\rhot <
m_{\mathrm{f}}/\miv$. As a result, the rapid decrease of the tension
with respect to the fermion densities constrains the nonperturbative
sector where the string is able to carry massive fermionic
currents. For each mass, the higher acceptable value of the $\rhot$
ensuring transverse normalizability is roughly $m_{\mathrm{f}}/\miv$,
and from Eq.~(\ref{loweqstate}), the tension becomes negative at this
density for $m_{\mathrm{f}}^2 \sim 4\pi M^2$. Much higher values of
$m_{\mathrm{f}}$ will thus yield to empty massive states.
As previously noted, the energy density and tension for massive modes
no longer verify the fixed trace equation of state found with the zero
modes alone. As a result, the longitudinal perturbation propagation
speed $c_{\mathrm{L}}^2=-dT/dU$ is no longer equal to the speed of
light, and it is even no longer well defined since the equation of
state involves more than one state parameter. A necessary condition
for longitudinal stability can nevertheless be stated by verifying
that all the perturbation propagation speeds $-(\partial
T/\partial\rhot)/(\partial U/\partial\rhot)$ obtained from variation
of only one state parameter are positive and less than the speed of
light. The longitudinal and transverse perturbation propagation
speeds, $c^2_{\mathrm{L}}$ and $c^2_{\mathrm{T}}=T/U$, respectively,
have been plotted in Fig.~\ref{figclctm1} in the case where there is
only one species trapped as massive mode, $\Psi$ or $\Chi$ say. It is
interesting to note that there is a transition between a supersonic
regime obtained at low fermion density, and a subsonic at high fermion
density. Moreover, the transition density between the two regimes is
all the more so high as the coupling constant
$m_{\mathrm{f}}/m_{\mathrm{h}}$ is weak. It is not surprising to
recover such zero-mode-like subsonic behavior~\cite{ringeval,prep} for
densities much higher than the rest mass, since in these cases the
ultrarelativistic limit applies. On the other hand, since the mass of
the massive mode decreases with the coupling constant as in
Fig.~{\ref{figmasspect}, the transition will occur earlier in the
nonperturbative sector, as can be seen in
Fig.~\ref{figrhotcross}. Note that the subsonic region is also limited
by the maximum allowed values of the massive fermion densities, i.e,
$\sim m_{\mathrm{f}}/\miv$, and the regions of transverse
instabilities where $c^2_{\mathrm{T}}$ becomes negative. Inclusion of
the other species does not change significantly these behaviors, the
main effect being to lower $c_{\mathrm{T}}^2$ with respect to the
other fermion density, as can be seen in Fig.~\ref{figutm1}.
\begin{figure}
\begin{center}
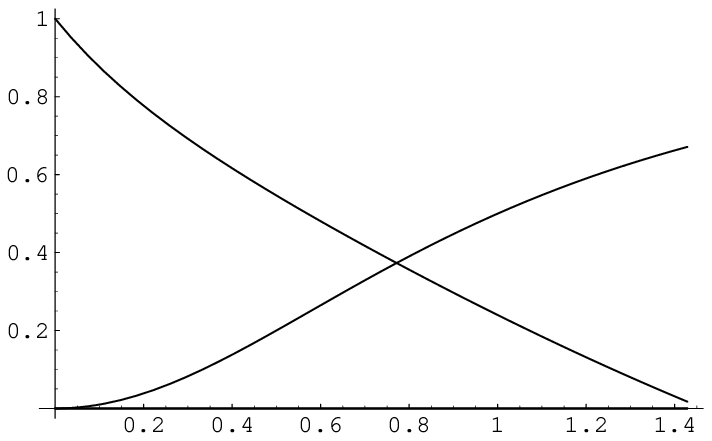
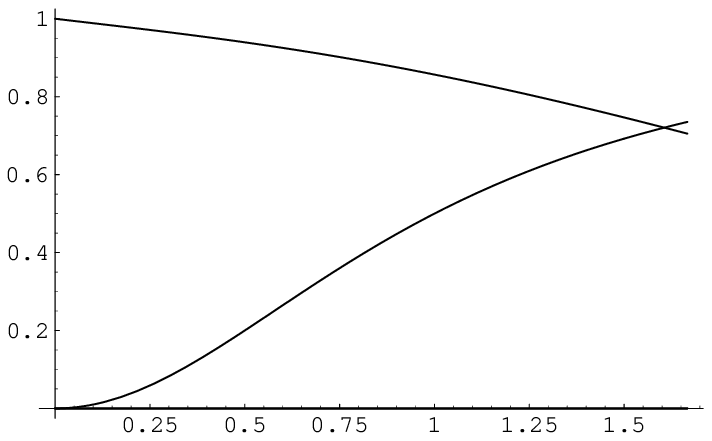
\caption[Influence des modes massifs sur les vitesses de propagations
des perturbations transverses et longitudinales.]{The squared
longitudinal and transverse perturbations propagation speeds for one
massive species only, plotted as functions of the dimensionless
fermion density $\rhot$, for two values of the coupling constant
$m_{\mathrm{f}}/m_{\mathrm{h}}$. Note the transition between subsonic
and supersonic behaviors takes place at a cross density,
$\rhot_\times$ say, which decreases with the coupling constant (see
Fig.~\ref{figrhotcross}).}
\label{figclctm1}
\end{center}
\end{figure}
\begin{figure}
\begin{center}
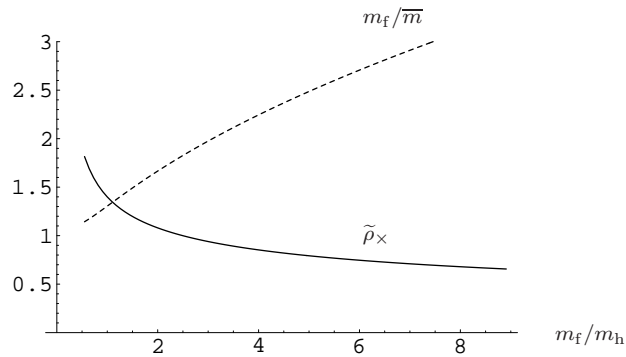
\caption[Densit\'e des fermions en r\'egime transonique.]{The cross
dimensionless density $\rhot_\times$, i.e. the dimensionless fermion
density for which the transverse and longitudinal perturbation
propagation speeds are equal, plotted as function of the coupling
constant $m_{\mathrm{f}}/m_{\mathrm{h}}$, for one massive species
only. The dashed curve shows the maximum allowed values
$m_{\mathrm{f}}/\miv$ of the massive fermion density ensuring
transverse normalizability. The transition from supersonic regime to
subsonic can thus occurs only in the nonperturbative sector, below
this frontier.}
\label{figrhotcross}
\end{center}
\end{figure}
The current magnitude can also be derived from the averaged current
operators in the zero temperature and infinite string limit. From
Eq.~(\ref{currentaverage}), using Eqs.~(\ref{psiuv}) and
(\ref{twoaverageMM}), once the transverse coordinates have been
integrated over, the world sheet components read
\begin{equation}
\label{gaugecurrents}
\begin{array}{lcl}
\vsep \displaystyle \langle\overline{j}^0\rangle & = & \displaystyle -
\sum_\Ferm \frac{\miv_\Ferm}{\pi} \left(q \cfr ||\ab_1^2+\ab_2^2|| + q
\cfl ||\ab_4^2 + \ab_3^2|| \right) \,\rhot_\Ferm, \\ \vsep
\displaystyle \langle\overline{j}^3\rangle & = & \displaystyle
-\sum_\Ferm \frac{\miv_\Ferm}{\pi} \left(q \cfr ||\ab_1^2-\ab_2^2|| +
q \cfl ||\ab_4^2 - \ab_3^2|| \right)\,\rhot_\Ferm.
\end{array}
\end{equation}
The current magnitude
${\mathcal{C}}^2 = \langle \overline{j}^0\rangle^2 -
\langle\overline{j}^3 \rangle^2$
therefore reads
\begin{eqnarray}
\label{currentmag}
{\mathcal{C}}^2 & = & 4 \left(\sum_\Ferm \frac{\miv_\Ferm}{\pi}
F_{\mathrm{Y} \Ferm}\, \rhot_\Ferm\right) \left(\sum_\Ferm
\frac{\miv_\Ferm}{\pi} F_{\mathrm{X} \Ferm}\, \rhot_\Ferm\right),
\end{eqnarray}
where $F_{\mathrm{X} \Ferm}$ and $F_{\mathrm{Y} \Ferm}$ denote the
transverse effective charges
\begin{equation}
\begin{array}{lcl}
\vsep \displaystyle F_{\mathrm{Y} \Ferm} & = & \displaystyle q \cfr
||\ab_1^2|| + q \cfl ||\ab_4^2|| \\ \vsep \displaystyle F_{\mathrm{X}
\Ferm} & = & \displaystyle q \cfr ||\ab_2^2|| + q \cfl ||\ab_3^2||,
\end{array}
\end{equation}
already introduced for the zero mode currents~\cite{ringeval}.
In the case of one massive species only, the charge current magnitude
simplifies to
\begin{equation}
{\mathcal{C}}^2=4 \frac{\miv^2}{\pi^2} F_{\mathrm{X}} F_{\mathrm{Y}}
\rhot^2,
\end{equation}
and thus the sign of ${\mathcal{C}}^2$ is only given by the sign of
$F_{\mathrm{X}} F_{\mathrm{Y}}$, which is generally positive for
reasonable values of the transverse normalizations,
e.g. $||\Sigma_{\mathrm{X}}^2|| \sim ||\Sigma_{\mathrm{Y}}^2|| \sim
0.5$. As a result, the charge current generated by only one massive
species is always timelike~\cite{davisS}, contrary to the zero mode
charge current which was found to be possibly timelike, but also
spacelike~\cite{ringeval}, owing to the allowed exitations of
antiparticle zero mode states. As noted above, the antiparticle states
cannot exist for massive modes due to the nonvanishing cross section
along the string between massive particles and antiparticles.
Moreover, and as it was the case for the zero modes, unless there is
only one massive species trapped in the string, the magnitude of the
charge current is not a sufficient state parameter, contrary to the
bosonic current-carrier case~\cite{neutral}.

\subsection{General case}

From the numerical approach in Sec.~(\ref{numresults}), as soon as
the nonperturbative sectors are considered, additional massive bound
states become relevant, and it is reasonable to consider that, in the
zero-temperature limit, all these accessible massive states will be
also be filled. Moreover, the complete description of the string state
also requires the inclusion of the zero modes in addition to the
massive ones.

\subsubsection{Full stress tensor}

The effective two-dimensional energy momentum tensor, involving all
trapped modes in the cosmic string, can be obtained from
Eq.~(\ref{lowtwotensor}) by replacing the sum over the two lowest
massive modes with the sum over all the accessible masses, plus the
zero mode terms previously derived in Ref.~\cite{ringeval}. In the
preferred frame where the stress tensor is diagonal, after some
algebra, the energy density and tension therefore read
\begin{eqnarray}
\label{ueqstate}
U & = & M^2+ \frac{1}{2\pi}\sum_{\Ferm, \ell} \miv_{\Ferm_\ell}^2
\ln \left(\sqrt{1 + \rhot_{\Ferm_\ell}^2} + \rhot_{\Ferm_\ell}\right)
\nonumber \\ 
& + & \frac{1}{\pi}\,\left[\left(4\pi^2 \Rho_{\chi}^2 +
\sum_{\Ferm, \ell} \miv_{\Ferm_\ell}^2 \, ||\Sigma_{\mathrm{Y}
\Ferm_\ell}^2||\, \rhot_{\Ferm_\ell}
\sqrt{1+\rhot_{\Ferm_\ell}^2}\right) \right. \nonumber \\
& & \times \left. \left(4\pi^2 \Rho_{\psi}^2 +
\sum_{\Ferm, \ell} \miv_{\Ferm_\ell}^2 \, ||\Sigma_{\mathrm{X}
\Ferm_\ell}^2||\, \rhot_{\Ferm_\ell}
\sqrt{1+\rhot_{\Ferm_\ell}^2}\right)\right]^{1/2}\,,\\
\label{teqstate}
T & = & M^2+ \frac{1}{2\pi}\sum_{\Ferm, \ell} \miv_{\Ferm_\ell}^2
\ln\left(\sqrt{1 + \rhot_{\Ferm_\ell}^2} + \rhot_\Ferm \right)
\nonumber \\
& - & \frac{1}{\pi}\,\left[\left(4\pi^2 \Rho_{\chi}^2 +
\sum_{\Ferm, \ell} \miv_{\Ferm_\ell}^2 \, ||\Sigma_{\mathrm{Y}
\Ferm_\ell}^2||\, \rhot_{\Ferm_\ell}
\sqrt{1+\rhot_{\Ferm_\ell}^2}\right) \right. \nonumber \\
& & \times \left. \left(4\pi^2 \Rho_{\psi}^2 +
\sum_{\Ferm, \ell} \miv_{\Ferm_\ell}^2 \, ||\Sigma_{\mathrm{X}
\Ferm_\ell}^2||\, \rhot_{\Ferm_\ell}
\sqrt{1+\rhot_{\Ferm_\ell}^2}\right)\right]^{1/2}\,.
\end{eqnarray}
The sums run over all accessible massive bound states $\ell$
with masses $\miv_{\Ferm_\ell}$ of each fermion $\Ferm$, i.e. $\Psi$
and $\Chi$. The additional parameters $\Rho_\chi$ and $\Rho_\psi$ are
the particle densities trapped in the string in the form of zero modes,
for the $\Chi$ and $\Psi$ field, respectively, with same notation as
in Ref.~\cite{ringeval}. Note that the zero mode contribution can also be
obtained from the null mass limit in Eq.~(\ref{loweqstate}). As a
result, the full expression of energy per unit length and tension
seems to involve as many state parameters as trapped modes in the
string.

\subsubsection{Equation of state}

As for the lowest massive modes, it is convenient to perform some
approximations owing to the energetically favored filling of the
involved states, in the zero-temperature limit. In particular, it is
reasonable to consider that the nonvanishing cross sections between
massive modes, and between zero modes and massive modes, lead to the
filling of all the accessible states with energy lower than a Fermi
energy, $\Eferm_\Ferm$ say, for each fermion field $\Ferm$. As a
result, the energetically favored filling takes place by successive
jumps from the lower masses to the highest ones, until the last mass
hyperbola with $\miv_{\Ferm_\ell} \sim \Eferm_\Ferm$ is
reached. Obviously, this filling begins with the zero modes, next with
the lowest massive modes and so on. On the other hand, only the
particle states are assumed to be relevant because of the assumed
annihilation of the antiparticle states, as discussed in
Sec.~(\ref{zerolimit}). As a result, the Fermi levels, $\nu_\Ferm$
say, can be defined through the zero modes filling only, as the line
densities of zero mode exitations trapped in the string (see
Fig.~\ref{figallfilled}), and thus play the role of state
parameters.
\begin{figure}
\begin{center}
\epsfig{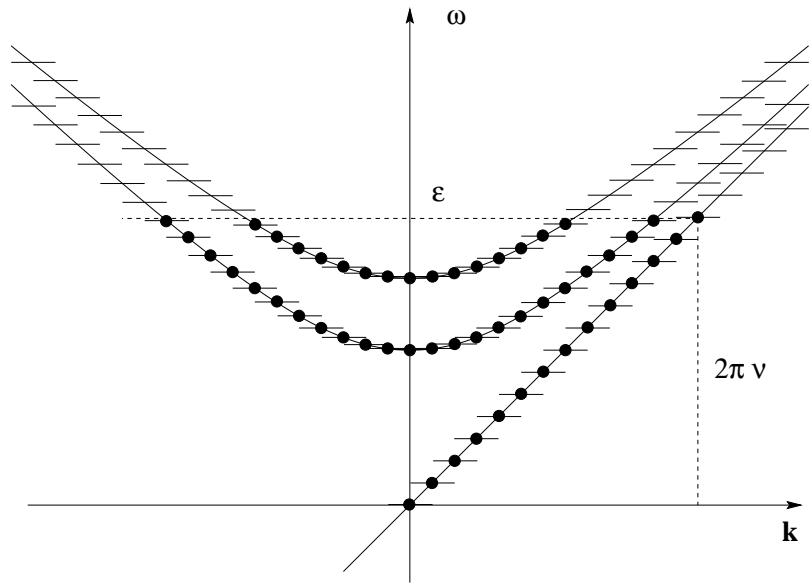}
\caption[Spectre des \'etats accessibles fermioniques dans une corde
cosmique.]{The accessible states, for the $\Chi$ fermions, in the
zero-temperature limit. The zero modes are represented by the chiral
line $\omega=k$, while the massive modes appear as mass
hyperbolae. The Fermi levels are therefore dependent of the considered
energy scale $\Eferm$, since the filling is performed by successive
jumps from the zero modes to the massive ones, with $\miv_{\ell} \sim
\Eferm$. As a result, under these approximations, each trapped species
leads to only one state parameter which can be identified with the
Fermi level of the zero mode exitations, namely
$\nu=\Eferm/2\pi$. Note that the antiparticle states have not been
represented due to their assumed annihilation.}
\label{figallfilled}
\end{center}
\end{figure}
According to the so-defined state parameters, the massive
exitation densities $\rho_{\Ferm_\ell}$, in Eqs.~(\ref{ueqstate}) and
(\ref{teqstate}), reduce to
\begin{eqnarray}
\label{massdens}
\rho_{\Ferm_\ell} & = & 
\left(\nu_\Ferm - \frac{\miv_{\Ferm_\ell}}{2\pi} \right) \Theta \! \!
\left(\nu_\Ferm - \frac{\miv_{\Ferm_\ell}}{2\pi}\right),
\end{eqnarray}
with $\Theta$ function is the Heavyside step function, as expected for
energy scales less than the rest mass of the considered massive mode.
The zero mode density simply reads
\begin{eqnarray}
\label{zerodens}
\Rho_\Ferm & = & \nu_\Ferm,
\end{eqnarray}
for zero mode particle states alone. From Eqs.~(\ref{massdens}) and
(\ref{zerodens}), and the definition of the dimensionless densities in
Eq.~(\ref{adimdens}),
\begin{equation}
\rhot_{\Ferm_\ell}=\frac{2 \pi}{\miv_{\Ferm_\ell}} \rho_{\Ferm_\ell},
\end{equation}
the energy per unit length and the tension now depend explicitly of
the two state parameters only, namely $\nu_\psi$ and $\nu_\chi$. By
means of Eq.~(\ref{ueqstate}), the energy density reads
\begin{eqnarray}
\label{fullenergy}
U & = & \displaystyle M^2+ \frac{1}{2\pi}\sum_{\miv_{\Ferm_\ell} \le 2
\pi \nu_\Ferm} \miv_{\Ferm_\ell}^2 \ln{\left[\sqrt{1 +
\left(\frac{2\pi}{\miv_{\Ferm_\ell}}\nu_\Ferm-1\right)^2} +
\frac{2\pi}{\miv_{\Ferm_\ell}}\nu_\Ferm-1 \right]} \nonumber \\ & + &
\, \frac{1}{\pi}\,\left[(2 \pi \nu_\chi)^2 + \sum_{\miv_{\Ferm_\ell}
\le 2 \pi \nu_\Ferm} \miv_{\Ferm_\ell}^2 \, ||\Sigma_{\mathrm{Y}
\Ferm_\ell}^2||\,\left(\frac{2\pi}{\miv_{\Ferm_\ell}}\nu_\Ferm-1
\right) \sqrt{1+\left(\frac{2\pi}{\miv_{\Ferm_\ell}} \nu_\Ferm- 1
\right)^2}\right]^{1/2} \nonumber \\ & \times & \, \left[(2 \pi
\nu_\psi)^2 + \sum_{\miv_{\Ferm_\ell} \le 2 \pi \nu_\Ferm}
\miv_{\Ferm_\ell}^2 \, ||\Sigma_{\mathrm{X}
\Ferm_\ell}^2||\,\left(\frac{2\pi}{\miv_{\Ferm_\ell}}\nu_\Ferm-1
\right) \sqrt{1+\left(\frac{2\pi}{\miv_{\Ferm_\ell}} \nu_\Ferm- 1
\right)^2}\right]^{1/2}\,,
\end{eqnarray}
while the tension is obtained from Eq.~(\ref{teqstate}),
\begin{eqnarray}
\label{fulltension}
T & = & \displaystyle M^2+ \frac{1}{2\pi}\sum_{\miv_{\Ferm_\ell} \le 2
\pi \nu_\Ferm} \miv_{\Ferm_\ell}^2 \ln{\left[\sqrt{1 +
\left(\frac{2\pi}{\miv_{\Ferm_\ell}}\nu_\Ferm-1\right)^2} +
\frac{2\pi}{\miv_{\Ferm_\ell}}\nu_\Ferm-1 \right]} \nonumber \\ & - & \,
\frac{1}{\pi}\,\left[(2 \pi \nu_\chi)^2 + \sum_{\miv_{\Ferm_\ell} \le 2
\pi \nu_\Ferm} \miv_{\Ferm_\ell}^2 \, ||\Sigma_{\mathrm{Y}
\Ferm_\ell}^2||\,\left(\frac{2\pi}{\miv_{\Ferm_\ell}}\nu_\Ferm-1
\right) \sqrt{1+\left(\frac{2\pi}{\miv_{\Ferm_\ell}} \nu_\Ferm- 1
\right)^2}\right]^{1/2}  \nonumber \\ & \times & \,
\left[(2 \pi \nu_\psi)^2 + \sum_{\miv_{\Ferm_\ell} \le 2
\pi \nu_\Ferm} \miv_{\Ferm_\ell}^2 \, ||\Sigma_{\mathrm{X}
\Ferm_\ell}^2||\,\left(\frac{2\pi}{\miv_{\Ferm_\ell}}\nu_\Ferm-1
\right) \sqrt{1+\left(\frac{2\pi}{\miv_{\Ferm_\ell}} \nu_\Ferm- 1
\right)^2}\right]^{1/2}\,.
\end{eqnarray}
The full energy per unit length and tension have been plotted in
Fig.~{\ref{figuttwomass} for a configuration including two massive
bound states, in addition to the zero mode ones. Due to the
zero-temperature limit, for densities smaller than the first
accessible mass, the Heavyside functions in Eq.~(\ref{massdens})
vanish, as a result, from Eqs.~(\ref{fullenergy}) and
(\ref{fulltension}) the fixed trace equation of state is
recovered~\cite{ringeval} with
\begin{equation}
\begin{array}{lclclcl}
\displaystyle
U & = &
\displaystyle
M^2 + 4 \pi \nu_\chi \nu_\psi,
& \quad &
\displaystyle
T & = &
\displaystyle
M^2 - 4 \pi \nu_\chi \nu_\psi.
\end{array}
\end{equation}
Once the first mass hyperbola is reached, the behaviors of the energy
per unit length and tension are clearly modified and become very
rapidly dominated by the mass terms, and, as found for the lowest
massive modes alone, the energy density begins to grow linearly with
respect to the state parameters, whereas the tension decreases
quadratically. Actually, the plotted curves in
Fig.~\ref{figuttwomass} show slope discontinuities each time the
phase space is enlarged due to the income of accessible massive bound
states (see Fig.~\ref{figallfilled}).
\begin{figure}
\begin{center}
\epsfig{file=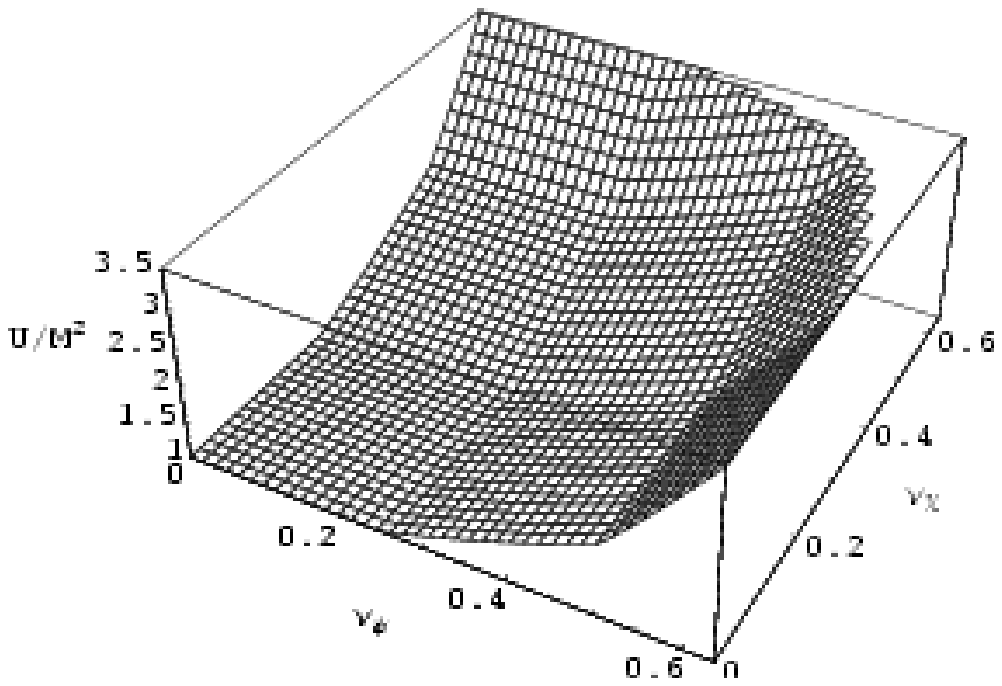,width=10cm}
\epsfig{file=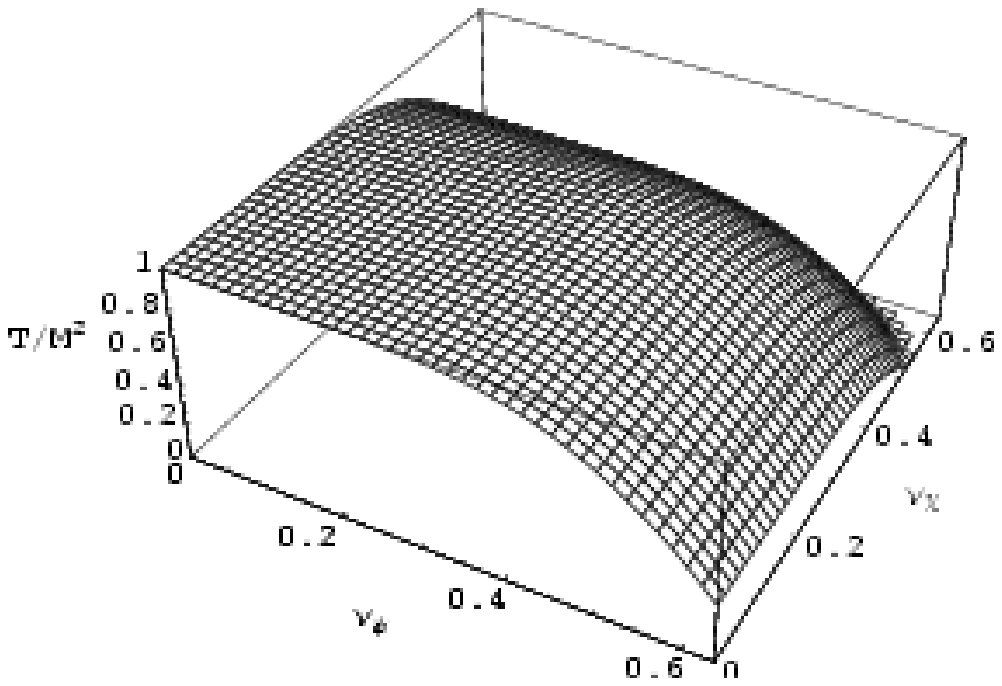,width=10cm}
\caption[L'\'equation d'\'etat d'une corde cosmique parcourue par des
fermions.]{The energy per unit length and the tension plotted as
function of the two state parameters, i.e. the zero mode densities,
in the nonperturbative sector $m_{\mathrm{f}}/m_{\mathrm{h}} \sim
4$. Two additional massive bound states have been considered with
respective masses $\miv/m_{\mathrm{f}} \sim 0.4$ and
$\miv/m_{\mathrm{f}} \sim 0.6$. In the zero-temperature limit, the
filling of the accessible states is performed by successive jumps as
soon as the Fermi level reaches one mass hyperbola (see
Fig.~\ref{figallfilled}). As a result, for the lowest values of the
state parameters, only the zero modes are relevant and the fixed trace
equation of state, $U+T=2M^2$, is verified, then the first and second
massive modes are successively reached and become rapidly dominant. As
can be seen near the origin, the smooth variations induced by the zero
modes appear completely negligible compared to the massive ones. In
the perturbative sectors, these behaviors are essentially the same,
but the induced variations of the density energy and the tension are
all the more small.}
\label{figuttwomass}
\end{center}
\end{figure}
On the other hand, it is reasonable to expect competition between the
subsonic regimes induced by zero mode currents, or ultrarelativistic
massive modes, and the supersonic ones coming from massive currents.
In all cases, when the state parameters remain small, only the chiral
massless states are accessible and the regime is obviously subsonic,
as can be seen in Fig.~\ref{figtwomassclct}. However, the massive mode
filling modifies radically this behavior, and as found for the lowest
massive modes alone, as soon as a mass hyperbola is reached, the
longitudinal perturbations propagation speed falls drastically and
ends up being less than the transverse perturbation velocity. There is
a rapid transition from the subsonic to the supersonic regime. For
higher densities $\nu$, the behavior depends on the coupling
constant. More precisely, in the nonperturbative sector, the
ultrarelativistic limit can be applied before the energy scales reach
the fermion vacuum masses, and thus the subsonic regime is recovered,
whereas it is not the case in the perturbative sector, as can be seen
in Fig.~\ref{figtwomassclct}.
\begin{figure}
\begin{center}
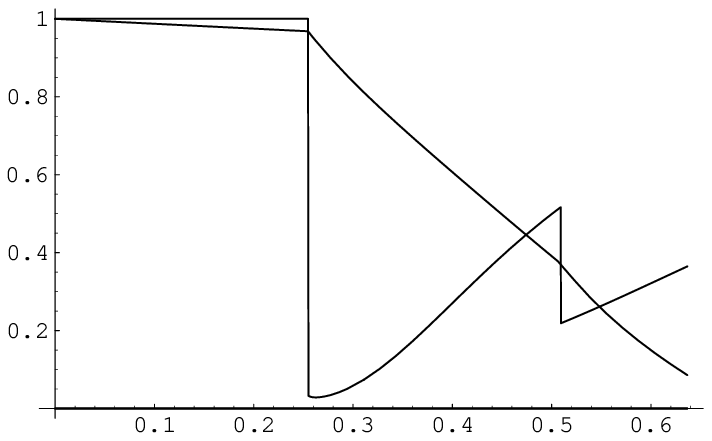
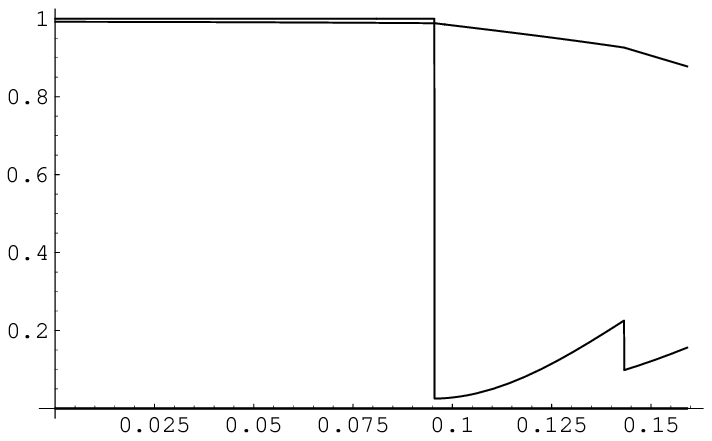
\caption[R\'egimes subsoniques et supersoniques d'une corde cosmique
parcourue par des fermions.]{The squared longitudinal and transverse
perturbation propagation speeds for a spectrum involving two massive
states in addition to the zero mode, plotted as functions of the state
parameter $\nu_\Ferm$ of one species, the other being fixed to a
particular value. The curves have been plotted for two values of the
coupling constant to the Higgs field, $m_{\mathrm{f}}/m_{\mathrm{h}}
\sim 1$ and $m_{\mathrm{f}}/m_{\mathrm{h}} \sim 4$. Note the
successive transitions between subsonic and supersonic behaviors
according to the allowed jumps to the mass hyperbolae. However, the
fermion vacuum mass limit does not allow the ultrarelativistic limit
to take place in the perturbative sector, as it was the case for the
lowest massive modes alone. In this case, the string dynamics follows
a supersonic regime as soon as the first massive bound state is
filled.}
\label{figtwomassclct}
\end{center}
\end{figure}

\subsubsection{Discussion}

All these results have been derived without considering the back
reaction effects induced by the trapped charge currents along the
string [see Eq.~(\ref{gaugecurrents})]. As was already discussed for
the zero modes in Ref.~\cite{ringeval}, these currents yield back reacted
gauge field, $B_t$ and $B_z$, which might modify the vortex background
and the fermionic equations of motion. However, such perturbations of
the Higgs and orthoradial gauge field profiles (see
Fig.~\ref{figbackMM}) can be neglected for $B_t B^t$ and $B_z B^z$ small
compared to the string forming gauge field $B_\theta B^\theta \sim
m_{\mathrm{b}}^2$. Using Eqs.~(\ref{defalphabar}) and
(\ref{gaugecurrents}), the dimensionless charge currents associated
with one massive bound state, and generating the dimensionless gauge
fields $B_\alpha/m_{\mathrm{b}}$, roughly read
\begin{equation}
\label{roughcurrent}
\widetilde{j}^\alpha \sim \frac{c_\Ferm}{\cphi} \rhot_\Ferm
\frac{\miv_\Ferm}{2 \pi^2 \eta},
\end{equation}
with $c_\Ferm$ the fermion charges, i.e. $\cfr$ or $\cfl$. As a
result, the back reaction on the vortex background is negligible as
long as $\miv < 2 \pi^2 \eta$, which is clearly satisfied in the full
perturbative sector. Moreover, since $m_{\mathrm{h}}=\eta
\sqrt{\lambda}$, the previous derivations of the equation of state are
also valid in the nonperturbative sector provided $\miv < 2 \pi^2
m_{\mathrm{h}}/\sqrt{\lambda}$, and thus depend on the values of
self-coupling constant of the Higgs field $\lambda$, but also on the
mass spectrum. As can be seen in Fig.~\ref{figmasspect}, the ratio
$\miv/m_{\mathrm{f}}$ decreases with the fermion vacuum mass
$m_{\mathrm{f}}$, as a result $\miv/m_{\mathrm{h}}$ increases all the
more slowly, which allows to have both $m_{\mathrm{f}}>m_{\mathrm{h}}$
and $\miv < 2 \pi^2 m_{\mathrm{h}}/\sqrt{\lambda}$.

Moreover, in order for the new gauge fields $B_t$ and $B_z$ to
not significantly modify the fermionic equations of motion, from
Eq.~(\ref{fermmvt}), they have to verify $q \cphi B_\alpha < \omega
\sim \miv_\Ferm$. As a result, Eq.~(\ref{roughcurrent}), and
$B_\alpha/m_{\mathrm{b}} \sim \widetilde{j}^\alpha$ yield the
condition
\begin{equation}
\frac{m_{\mathrm{b}}^2}{\eta^2} \frac{c_\Ferm}{c_\phi} < 1.
\end{equation}
As expected, it is essentially the same condition as the one
previously derived for the zero modes alone~\cite{ringeval}. On the
other hand, although the back reaction on the fermionic equations of
motion can deeply modify the zero mode currents~\cite{ringeval3}, since
the massive bound states are no longer eigenstates of the $\gamma^0
\gamma^3$ operator, it is reasonable to assume that rather than modify
their nature and existence significantly, the back reaction gauge
fields may only modify their mass spectrum. In this sense, back
reaction would indeed be a correction.

\section{Conclusion}

The relevant characteristic features of Dirac fermions trapped in a
cosmic string in the form of massive bound states have been study
numerically, in the framework of the Witten model, and in the neutral
limit. By means of a two-dimensional quantization of the associated
spinor fields along the string world sheet, the energy per unit length
and the tension of a cosmic string carrying any kind of fermionic
current, massive or massless, have been computed, and found to involve
as many state parameters as different trapped modes. However, in the
zero-temperature limit, only two have been found to be relevant and
they can be defined as the density numbers of the chiral zero mode
exitations associated with the two fermions $\Psi$ and $\Chi$ coupled
to the Higgs field.

As a result, it was shown that the fixed trace equation of state no
longer applies, as soon as massive states are filled, i.e. for energy
scales larger than the lowest massive mode belonging to the mass
spectrum. Moreover, the filling of massive states leads to a rapid
transition from the subsonic regime, relevant with massless, or
ultrarelativistic currents, to supersonic. Such properties could be
relevant in vorton evolution since it has been shown that supersonic
regimes generally lead to their classical
instabilities~\cite{carter93}. As a result, in the perturbative
sectors for which $m_{\mathrm{f}} < m_{\mathrm{h}}$, the protovortons
could be essentially produced at energy scales necessarily smaller
than the lower mass of the spectrum, where the fermionic currents
consist essentially in zero modes. In this way, vortons with fermionic
currents could be included in the more general two energy scale
models~\cite{brandi96}. However, the present conclusions are
restricted to parameter domains of the model where the back reaction
can be neglected. Although it is reasonable to consider that the back
reaction effects may simply modify the massive bound states through
their mass values, their influence on zero modes are expected to be
much more significant. In particular, the modified zero modes cannot
be any longer eigenstates of the $\gamma^0 \gamma^3$
operator~\cite{ringeval}, so one may conjecture that they acquire an
effective mass, leading to massive states potentially instable for
cosmic string loops

\section*{Acknowledgments}

It is a pleasure to thank particularly P. Peter for
many fruitful and helpfull discussions. I also wish to acknowledge
M. Lemoine, J. Martin, and O. Poujade for advice and various
enlightening discussions.

\part{Cosmologie en dimensions suppl\'ementaires}

\chapter{Le mod\`ele de Randall--Sundrum}
\label{chapitrers}
\minitoc
\section{Introduction}

Les mod\`eles standards de cosmologie et de physique des particules
supposent que notre espace-temps poss\`ede quatre dimensions. La
validit\'e de leurs pr\'edictions (voir chapitres~\ref{chapitrecosmo}
et~\ref{chapitrepp}), ainsi que notre perception du monde, semblent
donner le statut ``d'\'evidence'' \`a cette hypoth\`ese. Cependant,
l'id\'ee qu'il puisse exister des dimensions suppl\'ementaires n'est
pas nouvelle. Elle a \'et\'e initialement motiv\'ee par les travaux de
Weyl qui tentait d'unifier, de mani\`ere g\'eom\'etrique, la
gravitation d'Einstein et l'\'electromagn\'etisme par l'adjonction de
degr\'es de libert\'e suppl\'ementaires \`a l'espace-temps, au travers
d'une ``covariance conforme''~\cite{weyl18}. Dans les ann\'ees 1920,
Kaluza et Klein reprirent cette id\'ee en associant ces degr\'es de
libert\'e \`a une dimension spatiale additionnelle compactifi\'ee sur
une \'echelle de longueur $R$~\cite{kaluza21,klein26}. L'existence
physique d'une telle dimension induit cependant des r\'esonances des
divers champs pouvant s'y propager, et dont les masses se retrouvent
\^etre des multiples entiers de $1/R$. L'absence de ces particules,
dites de Kaluza Klein\footnote{KK.}, a finalement repouss\'e les
valeurs admissibles de $R$ aux petites \'echelles de longueur ($R \ll
\TeV^{-1}$).

Ce sont les th\'eories de cordes fondamentales qui ont ensuite
motiv\'e l'existence de dimensions suppl\'ementaires par le fait
qu'elles ne peuvent \^etre quantifi\'ees, de mani\`ere coh\'erente,
que dans un espace-temps \`a plus de quatre
dimensions~\cite{superstring}. Bien que les \'echelles de longueur
typiques des th\'eories quantiques de gravitation soient voisines de
celle de Planck, certaines th\'eories de cordes peuvent n\'eanmoins
\^etre \'etendues jusqu'\`a des \'echelles de longueur, $R$, pouvant
atteindre le
millim\`etre~\cite{horava96,horava96b,arkani98,antoniadis98}. Un tel
ordre de grandeur permet d'envisager des tests exp\'erimentaux sur
l'influence que pourraient avoir ces dimensions cach\'ees, et de ce
fait, de contraindre ces th\'eories~\cite{hoyle01,long99}. Du point de
vue de la physique des particules, ces th\'eories ont l'avantage de
r\'esoudre les probl\`emes de hi\'erarchie des masses. On montre en
effet que la compactification des dimensions suppl\'ementaires m\`ene
\`a une \'echelle de masse effective, dans l'espace-temps usuel,
d\'ependant \`a la fois de l'\'echelle de masse typique de la
th\'eorie et du volume des
\emph{extra-dimensions}~\cite{arkani98,arkani99,RS}. Le rapport de
seize ordres de grandeur existant entre la masse de Planck et
l'\'echelle de brisure \'electrofaible peut ainsi \^etre, du moins en
partie, justifi\'e par la pr\'esence de ce volume.

Afin de pouvoir \'etudier les cons\'equences physiques de l'existence
de dimensions suppl\'ementaires, il est plus commode de construire des
mod\`eles ph\'enom\'enologiques \`a partir de la th\'eorie des
champs. Pour retrouver les propri\'et\'es des mod\`eles standards, les
champs observ\'es doivent \^etre confin\'es sur l'hypersurface \`a
trois dimensions dans laquelle nous vivons\footnote{La \emph{brane}.}
qui est suppos\'ee \^etre immerg\'ee dans un espace de dimension
sup\'erieure\footnote{Le \emph{bulk}.}. L'autoconsistence de ces
mod\`eles requiert n\'eanmoins de trouver des m\'ecanismes de
localisation des divers champs. En particulier, la gravit\'e peut
\^etre localis\'ee, comme dans les th\'eories de Kaluza-Klein, par
compactification des dimensions
suppl\'ementaires~\cite{arkani98,arkani99}, mais
\'egalement par la courbure de l'espace-temps du bulk. Ce dernier cas,
introduit par Randall et Sundrum~\cite{RS}, permet d'avoir des
dimensions suppl\'ementaires non compactes, et comme nous le verrons
dans ce chapitre, on peut y calculer explicitement les \'ecarts
induits par les extra-dimensions \`a la gravit\'e d'Einstein
usuelle~\cite{RS,garriga00}. Cette gravit\'e modifi\'ee autorise, par
ailleurs, l'\'etude de la viabilit\'e de ces mod\`eles dans le
contexte cosmologique~\cite{binetruy00,binetruy00b}.

\section{Gravit\'e lin\'earis\'ee sur la brane}

Le mod\`ele de Randall et Sundrum~\cite{RS} suppose un espace-temps \`a
cinq dimensions dont la m\'etrique de fond est
\begin{equation}
\label{metriqueRS}
\ud s^2 = g_{AB} \,  \ud x^A \ud x^B = -\ud y^2 + \ue^{-2\sigma(y)}
\eta_{\mu \nu} \ud x^\mu \ud x^\nu,
\end{equation}
o\`u la fonction $\sigma$ est donn\'ee par
\begin{equation}
\label{defsigma}
\sigma(y) = \frac{|y|}{R},
\end{equation}
$y$ \'etant la coordonn\'ee dans la cinqui\`eme dimension, et $R$ une
\'echelle de distance caract\'eristique\footnote{Les indices latins
majuscules sont suppos\'es varier de $0$ \`a $4$.}. La th\'eorie de la
gravit\'e est de plus suppos\'ee \^etre une g\'en\'eralisation de
celle d'Einstein \`a cinq dimensions, c'est-\`a-dire d'action
\begin{equation}
\label{actionEinsteinRS}
\Sc_{\ug} = \int \frac{1}{2 \kappabu^2} \left(R - 2 \Lambda \right)
\sqrt{g} \, \ud^5 x,
\end{equation}
avec $g$ le d\'eterminant de la m\'etrique (\ref{metriqueRS}) et
$\kappabu^2$ la constante de couplage gravitationnelle \`a cinq
dimensions. Elle permet de d\'efinir la constante de Newton $\Gbu$ et la
masse de Planck $\mbu$, dans le bulk, par
\begin{equation}
\kappabu^2 = 6 \pi^2 \Gbu = \frac{6 \pi^2}{\mbu^3}.
\end{equation}
Comme nous le verrons dans le chapitre suivant, la m\'etrique
(\ref{metriqueRS}) est solution des \'equations d'Einstein d\'eduites
de l'action (\ref{actionEinsteinRS}),
\begin{equation}
\label{eqeinsteinRS}
G^{AB} + \Lambda g^{A B} = \kappabu^2 T_\infty^{A B},
\end{equation}
lorsque la source est une brane dont le tenseur \'energie-impulsion,
dans la limite d'\'epaisseur nulle, est donn\'e par~\cite{garriga00}
\begin{equation}
T_\infty^{\mu \nu} = T_\infty g^{\mu \nu} \delta(y).
\end{equation}
L'existence de la solution statique (\ref{metriqueRS}) n'est cependant
assur\'ee que lorsque la tension $T_\infty$ de la brane, et la
constante cosmologique $\Lambda$ dans le bulk v\'erifient les
relations
\begin{equation}
T_\infty = \frac{1}{\pi^2 R \Gbu}= \frac{\mbu^3}{\pi^2 R}, \quad \Lambda =
-\frac{6}{R^2}.
\end{equation}

La mati\`ere existant sur la brane va modifier cette solution, et en
particulier le tenseur m\'etrique $g_{AB}$. En notant $h_{AB}$ ces
perturbations, et \`a partir de l'\'equation (\ref{metriqueRS}), on
\'ecrit donc la m\'etrique perturb\'ee sous la forme
\begin{equation}
\label{metriquepertRS}
\ud s^2 = -\ud y^2 + \ue^{-2\sigma(y)} \ud x^\mu \ud x^\nu + h_{A B}
\, \ud
x^A \ud x^B.
\end{equation}
Comme dans le cas des \'equations du mouvement des cordes cosmiques
(voir Chap.~\ref{chapitreform}), l'invariance sous les transformations
de de coordonn\'ees requiert un choix de jauge. Il est ici commode de
se placer dans la jauge dite transverse et sans trace\footnote{TT,
pour \emph{traceless transverse}.} pour $h$~\cite{RS}, d\'efinie par
\begin{equation}
\label{gaugeTT}
\partial_\mu h^{\mu \nu} = \eta_{\mu \nu} h^{\mu \nu} = 0.
\end{equation}
Ce choix de jauge n'est pas encore suffisant, et les degr\'es de
libert\'e restant doivent \^etre fix\'es. Cela peut se faire en se
pla\c{c}ant dans la jauge gaussienne normale\footnote{GN.} d\'efinie
par
\begin{equation}
\label{gaugeGN}
h_{\mu 4} = h_{44} = 0.
\end{equation}
En reportant (\ref{metriquepertRS}) dans les \'equations d'Einstein
(\ref{eqeinsteinRS}), et en utilisant les conditions de jauge
(\ref{gaugeTT}) et (\ref{gaugeGN}), on trouve les \'equations du
mouvement~\cite{garriga00},
\begin{equation}
\label{mvtpertRS}
\left(\ue^{2 \sigma} \partial_\mu^2 + \partial_y^2 -\frac{4}{R^2}
\right) h_{\mu \nu} = 0.
\end{equation}
On peut montrer que les conditions de jauge ainsi choisies fixent
\'egalement la position de la brane en $y=-f^4(x^\mu)$, o\`u $f^4$ est
une certaine fonction des coordonn\'ees internes \`a la
brane. Cependant, la discontinuit\'e induite par la brane,
d'\'epaisseur nulle, impose des conditions de
jonction~\cite{garriga00} de la m\'etrique sur celle-ci qu'il est plus
commode de fixer en $y=0$. Pour cela, la condition (\ref{gaugeTT})
doit \^etre relax\'ee, et il vient, en \'egalant les parties
distributionnelles dans les \'equations d'Einstein
(\ref{eqeinsteinRS}), toujours pour la m\'etrique
(\ref{metriquepertRS}),
\begin{equation}
\label{jonction}
\left(\partial_{y} + \frac{2}{R} \right) \hb_{\mu \nu} = -
\kappabu^2 \left(T_{\mu \nu} - \frac{1}{3} \ue^{-2\sigma} \eta_{\mu \nu}
{T^\alpha}_{\alpha} \right),
\end{equation}
o\`u les perturbations surlign\'ees sont exprim\'ees dans la jauge GN,
avec $y=0^+$. Afin de pouvoir utiliser cette relation dans les
\'equations du mouvement (\ref{mvtpertRS}), il convient de pouvoir
passer d'une jauge \`a l'autre. D'apr\`es l'\'equation
(\ref{gaugeGN}), la jauge gaussienne normale ne reste v\'erifi\'ee que
pour des changements de coordonn\'ees engendr\'es par des vecteur
$v^A$ de la forme
\begin{equation}
v^4 =f^4(x^\nu), \quad v^\mu= f^\mu(x^\nu) - \frac{R}{2} \ue^{-2
\sigma} \partial^\mu f^4(x^\nu),
\end{equation}
avec $f^\mu$ des fonctions des coordonn\'ees internes \`a la
brane. Les perturbations de la m\'etrique entre les deux jauges sont
donc reli\'ees par une transformation \'egalement engendr\'ee par les
vecteurs $v^A$ donn\'ee par~\cite{garriga00}
\begin{equation}
\label{chgtpert}
h_{\mu \nu}= \hb_{\mu \nu} - R \partial_\mu \partial_\nu f^4
-\frac{2}{R} \ue^{-2 \sigma} \eta_{\mu \nu} f^4 + \ue^{-2 \sigma}
\partial_{(\mu} f_{\nu)}.
\end{equation}
Cette \'equation permet d'obtenir la condition de jonction
(\ref{jonction}) pour les perturbations $h$, c'est-\`a-dire,
\begin{equation}
\label{jonctionRS}
\left(\partial_y +\frac{2}{R} \right) h_{\mu \nu} = -\kappabu^5
\left(T_{\mu \nu} - \frac{1}{3} \ue^{-2\sigma} \eta_{\mu \nu}
{T^\alpha}_\alpha \right) - 2 \partial_\mu \partial_\nu f^4.
\end{equation}
En combinant les \'equations (\ref{mvtpertRS}) et (\ref{jonctionRS}),
les perturbations de la m\'etrique sont finalement solutions de
\begin{equation}
\label{mvtperttotRS}
\left[\ue^{2 \sigma} \partial_\mu^2 + \partial_y^2 -\frac{4}{R^2} +
\frac{4}{R} \delta(y)
\right] h_{\mu \nu} = -2 \kappabu^5 S_{\mu \nu} \delta(y),
\end{equation}
avec
\begin{equation}
\label{sourceRS}
S_{\mu \nu} = \left(T_{\mu \nu} - \frac{1}{3} \ue^{-2\sigma} \eta_{\mu
\nu} {T^\alpha}_\alpha \right) + \frac{2}{\kappabu^2}
\partial_\mu \partial_\nu f^4.
\end{equation}
\`A l'aide de la fonction de Green retard\'ee $G(\bx,y,\bx',y')$
associ\'ee au membre de gauche de l'\'equation (\ref{mvtperttotRS}),
les solutions g\'en\'erales sont donn\'ees par
\begin{equation}
\label{solutionpert}
h_{\mu \nu} = -2 \kappabu^5 \int G_\zero(\bx,\bx') S_{\mu
\nu}(\bx') \, \ud^4 \bx',
\end{equation}
avec $G_\zero$ la fonction de Green $G$ prise sur la brane. Les
conditions de jauge transverse (\ref{gaugeTT}) peuvent maintenant
\^etre fix\'ees au travers du terme source $S_{\mu \nu}$. En
particulier, l'annulation de sa trace donne l'expression de la
coordonn\'ee transverse $f^4$. D'apr\`es (\ref{sourceRS}), il
vient\footnote{$\Box$ est le d'Alembertien quadri-dimensionnel.}
\begin{equation}
\label{TTcond}
\Box f^4 = \frac{\kappabu^2}{6} {T^\alpha}_\alpha.
\end{equation}
Le calcul explicite de la fonction de Green $G$ montre qu'elle se
d\'ecompose en deux parties: l'une faisant intervenir des modes de
masse nulle, similaires aux gravitons, et l'autre un continuum de modes
massifs~\cite{garriga00},
\begin{equation}
\label{green}
G(\bx,y,\bx',y') = \int \frac{\ud^4 \bk}{(2 \pi)^4} \ue^{i \bk(\bx -
\bx')} \left[\frac{\ue^{-2 \sigma(y)} \ue^{-2 \sigma(y')}/R}{(\omega +
i \varepsilon)^2-\vec{k}^{^2}} + \int_0^\infty \frac{u_m(y)
u_m(y')}{(\omega + i \varepsilon)^2 - \vec{k}^{^2} - m^2} \ud m
\right],
\end{equation}
o\`u les fonctions propres $u_m$ sont donn\'ees par
\begin{equation}
u_m(y) = \sqrt{\frac{m R}{2}} \, \frac{J_1(mR) Y_2(\ue^{\sigma} m R)-
Y_1(mR) J_2(\ue^{\sigma} m R)}{\sqrt{J_1^2(mR) + Y_1^2(mR)}},
\end{equation}
avec $J_n$ et $Y_n$ les fonctions de Bessel de premi\`ere et deuxi\`eme
esp\`eces, d'ordre $n$.

En choisissant de mani\`ere appropri\'e les fonctions $f^\mu$,
l'\'equation (\ref{chgtpert}) peut \^etre simplifi\'ee, \`a l'aide de
(\ref{sourceRS}) et (\ref{solutionpert}), en
\begin{equation}
\hb_{\mu \nu} = -2 \kappabu^2 \int G_\zero(\bx,\bx')
\left(T_{\mu \nu} - \frac{1}{3} \ue^{-2\sigma} \eta_{\mu \nu}
{T^\alpha}_\alpha \right) \ud^4 \bx' + \frac{2}{R} \ue^{-2\sigma}
\eta_{\mu \nu} f^4,
\end{equation}
o\`u la fonction $f^4$ est solution de l'\'equation (\ref{TTcond}). La
gravit\'e d'Einstein usuelle lin\'earis\'ee est finalement retrouv\'ee
en tronquant la fonction de Green (\ref{green}) \`a son mode z\'ero,
et il vient
\begin{equation}
\hb_{\mu \nu} = -\frac{2 \kappabu^2}{R} \frac{1}{\nabla^2_\rho} \left(T_{\mu
\nu} - \frac{1}{3} \eta_{\mu \nu} {T^\alpha}_\alpha
\right).
\end{equation}
Le premier terme dans le membre de droite de cette \'equation
s'identifie naturellement au terme de couplage quadri-dimensionnel en
$-16 \pi \Gc$, et la constante de gravitation universelle $\Gc$ se
trouve reli\'ee \`a la constante de couplage gravitationnelle dans le
bulk par
\begin{equation}
\label{relationconst}
\kappabu^2 = 6 \pi^2 \Gbu = 8 \pi \Gc R.
\end{equation}
Afin d'\'etudier l'influence du continuum de modes massifs dans la
fonction de Green [voir Eq.~(\ref{green})], on peut calculer le
potentiel gravitationnel statique dans la brane g\'en\'er\'e par une
masse plac\'ee sur celle-ci. En choisissant le tenseur
\'energie-impulsion de la mati\`ere de la forme
\begin{equation}
\label{tmunupertRS}
T_{\mu \nu}= \rho \, u_\mu u_\nu,
\end{equation}
les $u^\mu$ \'etant les quadrivecteurs vitesses associ\'es, \`a l'aide
des \'equations (\ref{TTcond}), (\ref{green}) et (\ref{tmunupertRS}),
le potentiel newtonnien \`a l'ext\'erieur de la source, en supposant la
sym\'etrie sph\'erique, est
\begin{equation}
\frac{1}{2}\hb_{00} \simeq \frac{\Gc M}{r} \left(1 + \frac{2}{3}
\frac{R^2}{r^2} \right),
\end{equation}
avec $r=|\vec{x}-\primevec{x}{}|$ et $M$ la masse totale $ M = \int{
\rho} \, \ud^3 \vec{x}$.  Les modes massifs induisent donc des
corrections en $R^2/r^2$ au potentiel gravitationnel statique,
potentiellement d\'etectables aux petites distances telles que
$r\simeq R$. Les tests de gravit\'e aux \'echelles inf\'erieures au
millim\`etre permettent ainsi de fixer une limite sup\'erieure \`a
l'\'echelle de distance $R$ caract\'eristique de la cinqui\`eme
dimension. Notons
\'egalement que le mod\`ele de Randall-Sundrum permet \'egalement de
resituer le probl\`eme de la hi\'erarchie des masses. En effet,
d'apr\`es l'\'equation (\ref{relationconst}), la masse de Planck dans
la brane est donn\'ee par
\begin{equation}
m_\uPl^2 = \frac{1}{8 \pi \Gc} = \frac{R}{6 \pi^2} \mbu^3.
\end{equation}
Pour $R$ voisin du millim\`etre, la masse de Planck dans le bulk
$\mbu$ est abaiss\'ee \`a $10^8 \GeV$ pour obtenir $m_\uPl \simeq
10^{19} \GeV$.

Le mod\`ele de Randall-Sundrum donne un cadre ph\'enom\'enologique,
par la th\'eorie des champs, \`a l'existence d'une dimension
suppl\'ementaire dont les effets pourraient \^etre d\'etect\'es aux
petites \'echelles de longueur. Cependant, bien que le confinement des
gravitons de masse nulle soit implicitement assur\'e, les
m\'ecanismes de confinement de la mati\`ere ne sont \emph{a priori}
pas inclus dans ce mod\`ele.

\section{Le confinement des fermions}

De la m\^eme mani\`ere que les gravitons sont pi\'eg\'es sur la brane
par la forme de la m\'etrique (\ref{metriqueRS}), on peut chercher \`a
confiner des fermions du bulk par la
gravitation~\cite{rubakov01,bajc}. Comme nous le verrons dans le
chapitre suivant, il est possible de d\'ecrire des fermions de masse
nulle, dans un espace-temps \`a cinq dimensions, par des spineurs \`a
quatre composantes. Leur lagrangien, en espace-temps courbe de
m\'etrique (\ref{metriqueRS}), est donn\'e par~\cite{birrell}
\begin{equation}
\label{lagzerofermRS}
\Lc = i \sqrt{g} \, \Psib \Gamma^A \left[\partial_A -\frac{1}{2}
\sigma'(y) \ue^{-\sigma} \Gamma_\mu \Gamma^4 \right] \Psi,
\end{equation}
avec $\Gamma^A$ les matrices de Dirac en cinq dimensions. On peut
maintenant chercher des solutions d\'ecoupl\'ees de la forme
\begin{equation}
\Psi(\bx,y)=\Uc(y) \psi(\bx),
\end{equation}
en imposant
\begin{equation}
\gamma^\mu \partial_\mu \psi = 0,
\end{equation}
pour retrouver un comportement dans la brane similaire \`a celui d'un
fermion quadri-dimensionnel de masse nulle. Dans ce cas, les
\'equations du mouvement d\'eduites du lagrangien
(\ref{lagzerofermRS}) donnent
\begin{equation}
\left[ \partial_y - 2 \sigma'(y) \right] \Uc = 0,
\end{equation}
dont la solution est
\begin{equation}
\Uc \propto \ue^{2 \sigma}.
\end{equation}
D'apr\`es l'\'equation (\ref{defsigma}), ces fermions sont
compl\`etement d\'elocalis\'es de la brane. Il est \'egalement facile
de se convaincre que l'ajout d'un terme de masse dans le lagrangien
(\ref{lagzerofermRS}) du type $m \Psib \Psi$ conduit \'egalement \`a
des solutions qui ne sont pas normalisables dans la cinqui\`eme
dimension. Le mod\`ele de Randall Sundrum ne permet donc pas le
confinement des fermions sur la brane.

\section{Conclusion}

La gravit\'e usuelle dans l'espace-temps quadri-dimensionnel peut
s'accommoder de l'existence d'une dimension suppl\'ementaire non
compacte pourvu que la g\'eom\'etrie dans cette dimension soit de type
anti-de Sitter\footnote{AdS$_5$.} sur une \'echelle de longueur
typiquement inf\'erieure au millim\`etre. Concernant le secteur de la
gravit\'e, de nombreux travaux ont g\'en\'eralis\'e le mod\`ele de
Randall Sundrum \`a un nombre de dimensions additionnelles sup\'erieur
\`a l'unit\'e~\cite{kanti01,gherghetta00,giovannini01}. Il est
\'egalement possible de reproduire dans la brane un espace-temps de
type FLRW pr\'esentant, comme dans le cas de Minkowski [(voir
Eq.~\ref{metriqueRS})], des corrections qui pourraient
\^etre observables en cosmologie~\cite{binetruy00,binetruy00b}. D'un
autre c\^ot\'e, l'autoconsistence ph\'enom\'enologique de ce mod\`ele
est alt\'er\'ee par l'impossibilit\'e de pi\'eger des fermions dans la
brane. Une solution consiste \`a remarquer que, comme les gravitons,
il est possible de pi\'eger le mode z\'ero d'un champ scalaire
additionnel~\cite{bajc}, et de coupler celui-ci \`a des
fermions. Comme nous le verrons dans le chapitre suivant, si le
couplage entre ce champ et les fermions du bulk est suffisamment fort,
des fermions de masse nulle peuvent exister sur la
brane~\cite{bajc,rubakov01}. Le probl\`eme est encore de leur donner
une masse: ceci peut \^etre finalement obtenu en supposant qu'il
existe un autre champ scalaire sur la brane, de type Higgs, se
couplant aux fermions initialement de masse nulle~\cite{bajc}. Ces
m\'ecanismes semblent cependant peu naturels vu le nombre de degr\'es
de libert\'e qu'ils introduisent par rapport au simple mod\`ele
initial.

Une autre approche consiste \`a g\'en\'eraliser les r\'esultats
obtenus dans le chapitre~\ref{chapitremassif} concernant les fermions
pi\'eg\'es le long des cordes cosmiques. Nous avons vu qu'il existe
tout un spectre de fermions, de masses quantifi\'ees, se propageant le
long de la corde pourvu que le champ de Higgs auquel ils sont
coupl\'es forme un d\'efaut topologique. L'id\'ee est de coupler, \`a
cinq dimensions, des fermions \`a un unique champ scalaire ayant la
propri\'et\'e de former un mur de domaine qui s'identifie \`a la
brane. Dans le chapitre suivant, publi\'e dans la revue
\journal{Physical Review} \numero{D}, nous montrons que la gravit\'e
g\'en\'er\'ee par un tel mur de domaine redonne asymptotiquement le
mod\`ele de Randall Sundrum, assurant ainsi le confinement des
gravitons de masse nulle, et que le couplage des fermions \`a l'unique
champ scalaire conduit \`a l'apparition de fermions pi\'eg\'es dans la
brane ayant un spectre de masse discret~\cite{rpu}.

\chapter{Mur de domaine quadri-dimensionnel (article)}
\label{chapitremur}
\minitoc
\begin{center}
{\Large \textbf{
Localisation of massive fermions on the brane
}}
\end{center}
\vspace{5mm}
\begin{center}
Christophe Ringeval$^1$, Patrick Peter$^1$ and Jean--Philippe Uzan$^2$
\end{center}
\vspace{5mm}
\begin{center}
{\footnotesize{
(1) Institut d'Astrophysique de Paris,
             98bis Boulevard Arago, F--75014 Paris (France).\\
         \vskip0.25cm
(2) Laboratoire de Physique Th\'eorique, CNRS-UMR 8627,
             B\^at. 210,\\ Universit\'e Paris XI,
             F--91405 Orsay Cedex (France).
}}
\end{center}
\vspace{5mm}
\begin{center}
\begin{minipage}[c]{14cm}
{\footnotesize \textbf{
We construct an explicit model to describe fermions confined on a four
dimensional brane embedded in a five dimensional anti-de~Sitter
spacetime. We extend previous works to accommodate massive bound
states on the brane and exhibit the transverse structure of the
fermionic fields.  We estimate analytically and calculate numerically
the fermion mass spectrum on the brane, which we show to be
discrete. The confinement life-time of the bound states is evaluated,
and it is shown that existing constraints can be made compatible with
the existence of massive fermions trapped on the brane for durations
much longer than the age of the Universe.
}}
\end{minipage}
\end{center}

\section{Introduction}\label{sec_intro}

The idea that our universe may be a hypermembrane in a five
dimensional spacetime has received some attention in the last few
years after it was realized that gravity could be localized on a
three-brane embedded in an anti-de~Sitter spacetime~\cite{RS}. Since
then, much work has been done in a cosmological
context~\cite{binetruy00, csaki99,cline99,binetruy00b,
kraus99,shiromizu00, flanagan00,maartens02,rubakov01} and there is
hope that a consistent (i.e. mathematically self-contained and
observationally satisfying) high dimensional model might soon be
formulated.  For instance, it has been proposed that a
model~\cite{khoury01,donagi01} based on such ideas could present
itself as an alternative to the inflationary paradigm, although for
the time being the controversy as to whether or not such a model might
have anything in common with our Universe is still going
on~\cite{kallosh01,kallosh01b}.

The idea is not new however, but has evolved from the standard
Kaluza-Klein approach to that of particle localization on a higher
dimensional defect~\cite{akama82,rubakov83,visser85,gibbons}. In
particular, it has been shown that massless bulk scalars and gravitons
share the property to have a zero mode localized on the
brane~\cite{bajc} in the Randall-Sundrum model. Various
mechanisms~\cite{dvali97,dvali97b,dvali01, dimopoulos01,duff01,
oda,ghoruku, akhmedov01} have been invoked according to which it would
be possible to confine massless gauge bosons on a brane, so that there
is hope to achieve a reasonable model including all the known
interactions in a purely four dimensional effective model.

A mechanism permitting localization of massless fermions on a domain
wall was described in
Refs.~\cite{akama82,rubakov83,visser85,jackiw}. However, although
appealing this mechanism might be, it should be emphasized that actual
fermions, as seen on an everyday basis in whatever particle physics
experiment, are massive, so that a realistic fermionic matter model on
the brane must accommodate for such a mass. The question of
localization of massive fermions on the brane thus arises naturally,
and it is the purpose of this work to provide the transverse brane and
fermionic structure that leads to this localization. Up to now,
fermions have been confined under the restricting hypothesis that the
brane self gravity was negligible~\cite{dubovski}, or that it was
embedded in a Minkowski spacetime with one~\cite{hisano} or
two~\cite{ringeval2} transverse dimensions (see also~\cite{neronov}
for the localization of fermion on a string-like defect in five
dimension).

Our goal is to transpose the original work of Ref.~\cite{ringeval2} to
the brane context. For that purpose, we realize the brane as a domain
wall. Such domain wall configurations in anti-de~Sitter space have
already been studied~\cite{lee,charmousis}. We will assume that five
dimensional fermions are Yukawa-coupled to the domain wall forming
Higgs field, as in the usual case of cosmic strings. In this respect,
our work somehow extends Ref.~\cite{dubovski}, where the
mass term was put by hand, and Ref.~\cite{hisano} where the gravity of
the wall was neglected.

We start, in the following section, by recalling the domain wall
configuration of a Higgs field in a five dimensional anti-de~Sitter
spacetime and discuss briefly its properties. In
section~\ref{sec_dirac}, we describe the dynamical equations of
fermions coupled to this domain wall in order to show that they obey a
Schr\"odinger-like equation with an effective potential which can trap
massive modes on the wall. The asymptotic structure, i.e. deep in the
bulk (far from the brane), is not Minkowski space, so that the
effective potential felt by the fermions possesses a local minimum at
the brane location, but no global minimum, as first pointed out in
Ref.~\cite{dubovski}. As a consequence, the bound states are
metastable and fermions can tunnel to the bulk.

We then provide an analytical approximation of the effective
potential, thanks to which we compute analytically, in
section~\ref{sec_analy}, the mass spectrum of the fermions trapped on
the brane. We obtain the mass of the heaviest fermion that can live on
the brane and estimate its tunneling rate. This result is compared to
a full numerical integration, performed in section~\ref{sec_num}. In a
last section, we investigate the parameter space and, after having
compared our results to previous ones, we conclude that there exists a
wide region in the parameter space for which the fermion masses can be
made arbitrary low, i.e. comparable to the observed small values
(with respect to the brane characteristic energy scale), while their
confinement life-time can be made much larger than the age of the
Universe. Such models can therefore be made viable as describing
realistic matter on the brane.

\section{Membrane configuration in AdS$_5$}\label{sec_wall}

We consider the action for a real scalar field $\Phi$ coupled to
gravity in a five dimensional spacetime
\begin{equation}\label{action}
S=\int\left[\frac{1}{2\kappa_{_5}^2}(R-2\Lambda)
+\frac{1}{2}g^{AB}\partial_A\Phi\partial_B\Phi
-V(\Phi)\right]\sqrt{g}\,\dd^5 x \equiv \int \sqrt{g}\,\dd^5 x [ {\cal
L}_{_{\rm grav}}+{\cal L}_\Lambda+{\cal L}_\Phi ]
\end{equation}
where $g_{AB}$ is the five dimensional metric with signature
$(+,-,-,-,-)$, $R$ its Ricci scalar, $\Lambda$ the five dimensional
cosmological constant and $\kappa_{_5}^2\equiv6\pi^2\Gbu$, $\Gbu$
being the five dimensional gravity constant. Capital Latin indices
$A,B\ldots$ run from 0 to 4. The potential of the scalar field $\Phi$
is chosen to allow for topological membrane (domain wall like)
configurations,
\begin{equation}\label{V}
V(\Phi)=\frac{\lambda}{8}\left(\Phi^2-\eta^2\right)^2,
\end{equation}
where $\lambda$ is a coupling constant and $\eta=\langle
|\Phi|\rangle$ is the magnitude of the scalar field vacuum expectation
values (VEV)\footnote{Note, that, because of the unusual number of
spacetime dimensions, the fields have dimensions given by $[R]=M^2$,
$[\Phi]=M^{3/2}$, $[\Lambda]=M^2$, $[\lambda]=M^{-1}$,
$[\eta]=M^{3/2}$ and $[\kappa_5]=M^{-3/2}$ ($M$ being a unit of
mass).}.

Motivated by the brane picture, we choose the metric of the bulk
spacetime to be of the warped static form
\begin{equation}\label{metric}
\dd s^2=g_{AB} \dd x^A \dd x^B = -\dd y^2 +
\hbox{e}^{-2\sigma(y)}\eta_{\mu\nu} \dd x^\mu\dd x^\nu = -\dd y^2 +
g_{\mu\nu} \dd x^\mu\dd x^\nu,
\end{equation}
where $\eta_{\mu\nu}$ is the four dimensional Minkowski metric of
signature ($+,-,-,-$), and $y$ the coordinate along the
extra-dimension. Greek indices $\mu,\nu\ldots$ run from 0 to 3.

With this metric ansatz, the Einstein tensor components reduce to
\begin{equation} G_{\mu\nu} = -g_{\mu\nu} (6\sigma'^2 - 3\sigma''), \
\ \ G_{yy} = -6\sigma'^2,\end{equation} where a prime denotes
differentiation with respect to $y$. The non-vanishing components of
the matter stress-energy tensor
\begin{equation}T_{AB} \equiv 2{\delta {\cal L}_\Phi \over \delta
g^{AB}} - g_{AB} {\cal L}_\Phi
\end{equation}
are given by
\begin{equation} T_{\mu\nu} = {1\over 2} g_{\mu\nu} (\Phi'^2 + 2 V), \
\ \ \ T_{yy} = {1\over 2} (\Phi'^2 - 2 V).\label{tmunuBF}\end{equation}
It follows that the five dimensional Einstein equations
\begin{equation} G_{AB} + \Lambda g_{AB} = \kappa_5^2 T_{AB}
\end{equation}
can be cast in the form
\begin{eqnarray}
\frac{3}{\kappa_{_5}^2}\sigma'' & = &
\Phi^{\prime2}\label{einstein1}\\ 6\sigma^{\prime 2}& =
&\frac{\kappa_{_5}^2}{2}\left(\Phi^{\prime2}-2V\right) - \Lambda ,
\label{einstein2}
\end{eqnarray}
while the Klein--Gordon equation takes the form
\begin{equation}\label{kg}
\Phi''-4\sigma'\Phi'=\frac{\dd V}{\dd\Phi} .
\end{equation}
Eqs.~(\ref{einstein1}-\ref{kg}) is a set of three differential
equations for two independent variables ($\Phi$ and $\sigma$). Indeed,
as can easily be checked, the Klein-Gordon equation stems from the
Einstein equations provided $\Phi'\neq 0$. To study the domain wall
configuration, we choose to solve the first Einstein equation
(\ref{einstein1}) together with the Klein-Gordon equation (\ref{kg}).

This system of equations must be supplemented with boundary
conditions. By definition of the topological defect like
configuration, we require that the Higgs field vanishes on the
membrane itself, i.e. $\Phi = 0$ for $y=0$, while it recovers its VEV
in the bulk, so that $\lim_{y\to\pm\infty} \Phi = \pm \eta$. Note that
the sign choice made here is arbitrary and corresponds to the
so-called kink solution~; the opposite choice
(i.e. $\lim_{y\to\pm\infty} \Phi = \mp \eta$) would lead to an
anti-kink whose physical properties, as far as we are concerned, are
exactly equivalent. As for the metric function $\sigma$, it stems from
the requirement that one wants to recover anti-de~Sitter
asymptotically, so that one demands that $\sigma'$ tends to a constant
for $y\to\pm\infty$. This constant can be determined using
Eq.~(\ref{einstein2}), so that $\lim_{y\to\pm\infty} \sigma' = \pm
\sqrt{-\Lambda/6}$. Note that as $y$ changes sign at the brane
location, there is no choice for the sign of the function $\sigma$ in
this case. Note also that, as is well known, the static hypothesis
implies that the bulk cosmological constant $\Lambda$ must be
negative, and therefore the five dimensional spacetime to be
anti-de~Sitter.

With the convenient dimensionless rescaled variables
\begin{equation} \varrho\equiv  y\sqrt{|\Lambda|},\ \ \ \ H\equiv
\frac{\Phi}{\eta},\ \ \ \ \ S\equiv\frac{\dd\sigma}{\dd \varrho},
\end{equation}
the dynamical equations read
\begin{equation} \dot S = \frac{\alpha}{3} \dot
H^2,\label{eq1}\end{equation}
\begin{equation} \ddot H -4S\dot H = 4\beta H (H^2 -1),\label{eq2}
\end{equation}
where a dot refers to a derivative with respect to $\varrho$ and the
two dimensionless (positive) parameters $\alpha$ and $\beta$ are
defined by
\begin{equation} \alpha \equiv \kappa_{_5}^2 \eta^2,\ \ \ 
\beta \equiv \frac{\lambda \eta^2}{8 |\Lambda|}.
\end{equation}

These parameters are not independent since, for an arbitrary value of
$\beta$ say, there is only one value of $\alpha$ for which the
boundary condition $S (0)= 0$, or equivalently $\lim_{\varrho\to
-\infty} S(\varrho) = -1/\sqrt{6}$, is satisfied. This stems from the
fact that Eq.~(\ref{eq1}) is a first order equation in $S$, so that
only one boundary condition is freely adjustable, and we choose it to
be at $\varrho\to+\infty$.  Once this choice is made, the value of $S$
on the brane is completely determined, and unless the parameters are
given the correct values, it does not vanish. As the solution must be
symmetric with respect to the extra dimension coordinate $y$, one must
tune the parameters in order to have a meaningfull solution (i.e. for
which the metric and its first derivative are continuous at
$\varrho=0$). This is reminiscent of the relation that should hold
between the brane and bulk cosmological constants~\cite{charmousis}.
\begin{figure}
\begin{center}
\epsfig{file=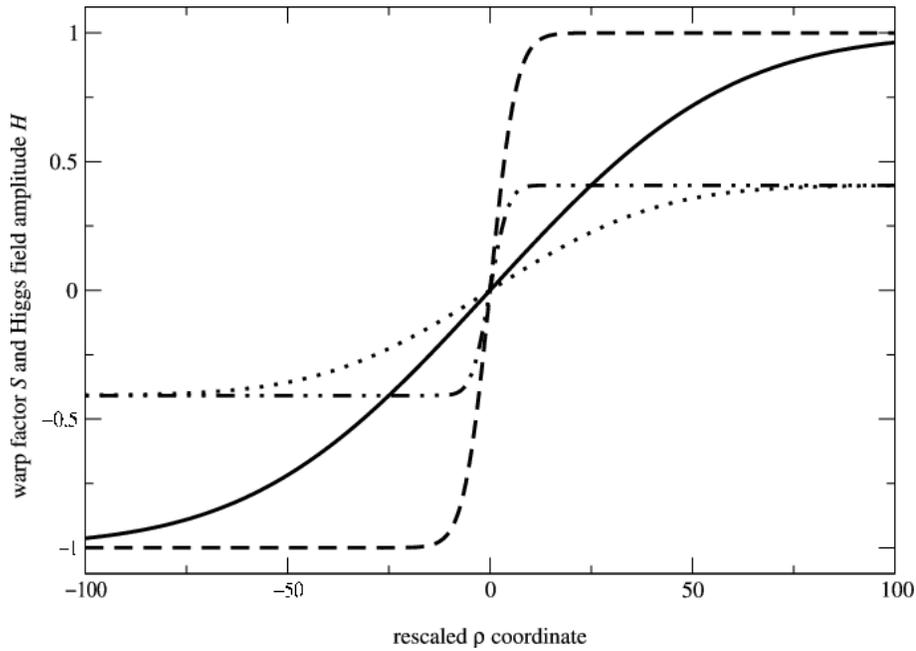,angle=270,width=12cm}
\caption[Profils des champs formant la brane dans la dimension
suppl\'ementaire.]{The rescaled Higgs field amplitude $H$ (full line
for $\beta = 0.01$ and dashed line for $\beta =0.1$) and warp factor
derivative $S$ (dotted line for $\beta=0.01$ and dot-dashed line for
$\beta=0.1$) as functions of the rescaled extra dimension coordinate
$\varrho$.}
\label{fig3}
\end{center}
\end{figure}
Eqs.~(\ref{eq1}~-\ref{eq2}) have been solved numerically with the
relevant boundary conditions. The field profiles are depicted on
Fig.~\ref{fig3} for two arbitrary values of the parameter $\beta$.
\begin{figure}
\begin{center}
\epsfig{file=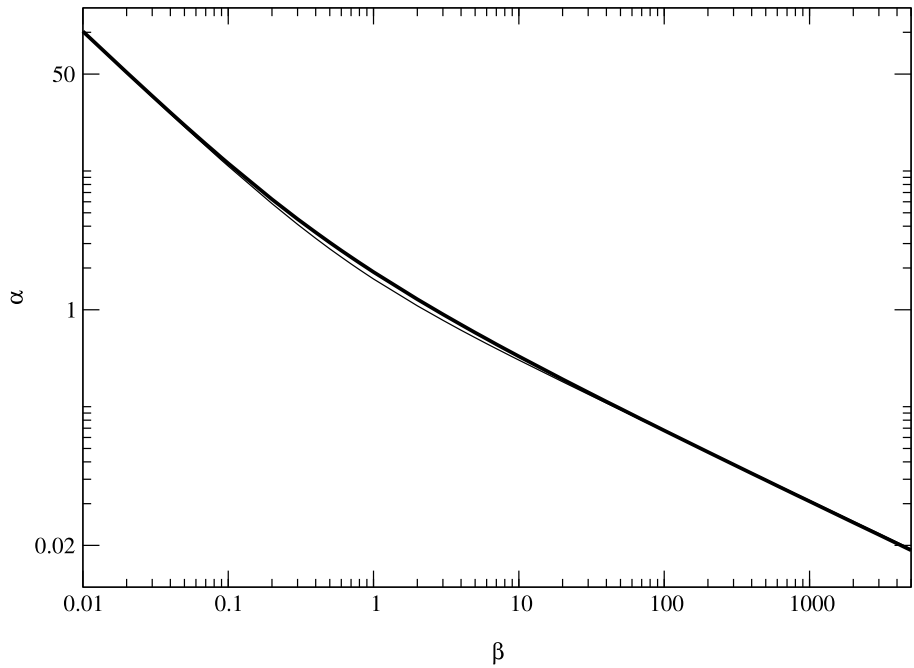,angle=270,width=12cm}
\caption[Contraintes impos\'ees par le confinement de gravitons de
masse nulle sur la brane.]{Relation between the dimensionless
constants $\alpha$ and $\beta$. The thick curve represent the result
of the numerical integration, while the (hardly distinguishable) thin
curve is the best analytical fit.}
\label{RPU2}
\end{center}
\end{figure}
The relation between the parameters $\alpha$ and $\beta$ such that the
metric is regular at the brane location is depicted on
Fig.~\ref{RPU2}.  As can be seen on the figure, it consists
essentially in two power laws. For small values of $\beta$, one finds
roughly $\alpha\sim 1/\beta$, which becomes exact in the limit
$\beta\to 0$, while for large values of $\beta$, one gets $\alpha\sim
{4\over 3} \beta^{-1/2}$, which is, again, exact in the limit
$\beta\to\infty$.  We were able to find the best fit
\begin{equation}
\alpha^2 = {1\over \beta} \left[ {1\over \beta} + \left( {4\over 3}
\right)^2 \right],
\label{relab}
\end{equation}
which, as can be seen on the figure, is almost exact everywhere. This
translates into the relation 
\begin{equation}
\label{rellambda}
|\Lambda | = {1\over 9}\lambda\eta \left[
\sqrt{1+\left( {9\kappa_5^2\eta^2\over
8}\right)^2}-1\right]
\end{equation}
between the five dimensional cosmological constant and the microscopic
parameters, which corresponds to the usual relations between brane and
bulk cosmological constants.

\section{Localization of fermions on the wall}\label{sec_dirac}

This section is devoted to the description of the Dirac equation in
five dimensions with the domain wall configuration obtained in the
previous section.

The minimal representation of spinors in five dimensions can be chosen
to be four dimensional~\cite{polchinsky}. The five dimensional
Clifford algebra can then be constructed from the usual four
dimensional one by adding the $\gamma_5$ matrix to close the
algebra. If $\gamma^\mu$ design the usual four dimensional Dirac
matrices in Minkowski space in the chiral representation, the Dirac
matrices in five dimensional Minkowski space, $\Gamma^A$, are
\begin{eqnarray}\label{def_dirac}
\Gamma^\mu =  \gamma^\mu, \qquad \Gamma^4 = -i \gamma_5
\end{eqnarray}
and they satisfied the usual Clifford algebra
\begin{equation}\label{clifford}
\{\Gamma^A,\Gamma^B\}=2 \eta^{A B}
\end{equation}
where $\eta^{A B}$ stands for the five dimensional Minkowski
metric. Since in this representation the Dirac matrices satisfy
\begin{equation}\label{Gamma_prop}
\Gamma^5=i\Gamma^0\dots \Gamma^4=\mathrm{Id}
\end{equation}
the five dimensional spinors have neither Weyl nor Majorana
represention. It follows that the Dirac Lagrangian in five dimension
for fermions coupled to the Higgs domain wall is necessary of the
form
\begin{equation}\label{lagBF}
{\mathcal{L}}_\psi= \sqrt{g} \left( i \overline\Psi \Gamma^A \DD_A \Psi
- g_{_{\rm F}} {\overline\Psi} \Phi \Psi \right) = 
\sqrt{g}\, {\overline\Psi}\left[i
\hbox{e}^{\sigma(y)} \Gamma^\mu \DD_\mu + i \Gamma^4
\partial_y  -g_{_{\rm F}}\Phi\right]\Psi
\end{equation}
where the Lorentz covariant derivative with  spin connection
is~\cite{birrell}
\begin{equation}\label{derivative}
\DD_\mu\equiv\partial_\mu-\frac{1}{2} \sigma'(y) \hbox{e}^{-\sigma(y)}
\Gamma_\mu \Gamma^4.
\end{equation}
We emphasize that the sign of the coupling $g_{_{\rm F}}$ of the
spinor $\Psi$ to the Higgs field is arbitrary and represents a
coupling either to kink or to anti-kink domain wall. For definiteness,
we shall consider in what follows only the case of a kink coupling,
and thus assume without lack of generality\footnote{Note also that the
dimensions are given by $[\Psi ]=M^2$ and $[g_{_{\rm F}} ]=
M^{-1/2}$.} that $g_{_{\rm F}}>0$.

The variation of the Lagrangian (\ref{lagBF}) leads to the
equation of motion of the spinor field, namely the Dirac equation in
five dimensional anti-de~Sitter space for a fermionic field coupled to
a Higgs field,
\begin{equation}\label{fermmvtBF}
\left\{i \Gamma^4\left[\partial_y- 2\sigma'(y)\right] +
i \hbox{e}^{\sigma(y)} \Gamma^\mu\partial_\mu
 - g_{_{\rm F}} \Phi \right\} \Psi=0.
\end{equation}
This equation involves the matrix $\gamma_5$ (through $\Gamma_4$)
and it is thus convenient to split the four dimensional right-
and left-handed components of the five dimensional spinor and to
separate the variables as
\begin{equation}\label{sepansatz}
\Psi(x^\mu,y)= \left[ \frac{1+\gamma_5}{2}\U_{\rm
R}(y)+\frac{1-\gamma_5}{2}\U_{\rm L}(y)\right] \psi(x^\mu),
\end{equation}
where $\psi(x^\mu)$ is a four dimensional Dirac spinor, while $\U_{\rm
R}(y)$ and $\U_{\rm L}(y)$ are yet undetermined functions of $y$.
In what follows, we want the five dimensional Dirac equation to yield
an effective four dimensional massive Dirac equation, with an
effective mass $m$ (energy eigenvalue of the bound state).
Such a requirement implies that
\begin{equation}
i \gamma^\mu \partial_\mu \psi =m \psi,
\end{equation}
or, equivalently, in terms of the right- and left- handed
components
\begin{equation}\label{diracmass}
i\gamma^\mu \partial_\mu \psi_{\rm R} =  m \psi_{\rm L},\qquad
i\gamma^\mu \partial_\mu \psi_{\rm L} = m \psi_{\rm R},
\end{equation}
where the right- and left-handed components of the four
dimensional spinor are defined as
\begin{equation}\label{leftright4d}
\psi_{\rm R}\equiv\frac{1+\gamma_5}{2}\psi, \quad
\psi_{\rm L}\equiv\frac{1-\gamma_5}{2}\psi.
\end{equation}
Contrary to the case studied in Ref.~\cite{dubovski}, the mass $m$ is
not an arbitrary parameter and will be determined later.

Choosing $\psi_{\rm R}$ and $\psi_{\rm L}$ as the independent
variables instead of $\Psi$ and $\overline\Psi$, and inserting
equation (\ref{diracmass}) into the equation of motion (\ref{fermmvtBF})
while using the splitting ansatz (\ref{sepansatz}) yields the
differential system for the two functions $\U_{{\rm R}/{\rm L}}(y)$,
\begin{eqnarray}\label{systurul}
\left[\partial_y-2 \sigma'(y)-g_{_{\rm F}} \Phi\right]\U_{\rm R}(y)&=&
-m\de^{\sigma(y)}\U_{\rm L}(y), \label{systurul1} \\
\left[\partial_y-2 \sigma'(y)+g_{_{\rm F}} \Phi\right]\U_{\rm L}(y)&=&
m\de^{\sigma(y)}\U_{\rm R}(y). \label{systurul2}
\end{eqnarray}
To simplify the notations, it is convenient to introduce the
dimensionless rescaled bulk components of the fermions
\begin{equation}\label{defut}
\Ut(\varrho)\equiv\de^{-\frac{3}{2}\sigma(\varrho)}
\frac{\U(\varrho)}{|\Lambda|^{1/4}},
\end{equation}
in terms of which the system (\ref{systurul1}-\ref{systurul2})
takes the form
\begin{eqnarray}
\left[\partial_\varrho - \left(\Mt H + \frac{1}{2}
S \right)\right]\, \Ut_{\rm R} &=& - \mt\, \de^{\sigma}\Ut_{\rm L},
\label{syst1}\\
\left[\partial_\varrho + \left(\Mt H - \frac{1}{2}
S\right)\right]\, \Ut_{\rm L} &=& \mt\, \de^{\sigma}\Ut_{\rm R},
\label{syst2}
\end{eqnarray}
with the dimensionless rescaled mass and coupling constant
\begin{eqnarray}\label{reduce_para}
\mt\equiv\frac{m}{\sqrt{|\Lambda|}}, \quad
\Mt\equiv\frac{g_{_{\rm F}} \eta}{\sqrt{|\Lambda|}}.
\end{eqnarray}
Let us first concentrate on the special case $\mt=0$. The system
(\ref{syst1}-\ref{syst2}) then consists in two decoupled differential
equations and the zero mode states~\cite{randjbar} are
recovered. Asymptotically, these functions behave as
\begin{eqnarray}
\Ut_{\rm R}(\varrho\rightarrow\pm\infty) & \sim & \de^{\left(\Mt +
\frac{1}{2\sqrt{6}}\right)|\varrho|},\\ \Ut_{\rm
L}(\varrho\rightarrow\pm\infty) & \sim & \de^{-\left(\Mt -
\frac{1}{2\sqrt{6}} \right)|\varrho|}.
\end{eqnarray}
Thus, only the left-handed solution $\Ut_{\rm L}$ may remain
bounded~\cite{bajc,jackiw}, and yet provided
\begin{equation}\label{cond_bounded}
\Mt > \frac{1}{2\sqrt{6}}.
\end{equation}
Indeed, the right-handed zero modes could have been obtained by
considering the coupling of fermions to the anti-kink Higgs
profile\footnote{The coupling with and anti-kink for which $g_{_{\rm
F}}<0$ would have yield right-handed solution with the constraint $\Mt
< -\frac{1}{2\sqrt{6}}.$}. We thus recover the well-known fact that
massless fermions {\it must} be single-handed in a brane model,
contrary to the ordinary four dimensional field theory in which they
simply {\it can}.

Let us now focus on the more interesting massive case for which
$\mt\not=0$. Then the system (\ref{syst1}-\ref{syst2}) can be
decoupled by eliminating $\Ut_{\rm R}$ say. For that purpose, we
differentiate (\ref{syst2}) with respect to $\varrho$ and express
$\partial_\varrho\Ut_{\rm R}$ using equation (\ref{syst1}) and
$\Ut_{\rm R}$ using equation (\ref{syst2}) again to get
\begin{eqnarray}
\label{ordertwoul}
\left[\partial^2_\varrho - 2 S \,\partial_\varrho + \left(\mt^2
\de^{2 \sigma} +\frac{3}{4} S^2 - \Mt^2 H^2 - \Mt S H
- \frac{1}{2} \partial_\varrho S +\Mt  \partial_\varrho H
\right)\right] \Ut_{\rm L} = 0,\\ 
\label{orderzerour}
\Ut_{\rm R} - \frac{\de^{-\sigma}}{\mt}\left[\partial_\varrho +
\left(\Mt H - \frac{1}{2} S \right)\right]\Ut_{\rm L} = 0.
\end{eqnarray}
This system is strictly equivalent to the initial system
(\ref{syst1}-\ref{syst2}) since differentiating
Eq.~(\ref{orderzerour}) and then using Eq.~(\ref{ordertwoul}) gives
back Eq.~(\ref{syst1}). It is thus important to keep both
equations. Note that the integration of the first equation
(\ref{ordertwoul}) will require two initial conditions but that
$\Ut_{\rm R}$ will then be completely determined and thus requires no
extra constant of integration. As a consequence, it is sufficient to
solve the second order equation (\ref{ordertwoul}) for $\Ut_{\rm L}$
in order to fully determine the left- and right-handed bulk fermion
profiles.

Eq.~(\ref{ordertwoul}) can be recast into a Schr\"odinger-like second
order differential equation
\begin{equation}
\label{schrodlike}
\partial^2_\varrho \,\Uh_{\rm L} + \omega^2(\varrho)\, \Uh_{\rm L} = 0,
\end{equation}
where the function $\omega$ is defined by
\begin{equation}\label{omega2}
\omega^2(\varrho) \equiv \mt^2 \de^{2 \sigma(\varrho)} +
\partial_\varrho\left(\Mt H + \frac{1}{2} S \right) - 
\left(\Mt H+\frac{1}{2} S\right)^2
\end{equation}
with the new function
\begin{equation}\label{defuh}
\Uh(\varrho)\equiv\de^{-\sigma(\varrho)} \Ut(\varrho).
\end{equation}
Our aim will now be to find the zero modes of this new equation; as
previously discussed, they will be equivalent to the massive bound
states we are looking for on the brane.

In order for the fermions to be confined on the brane, the minimum of
$\omega^2$ needs to be negative to imply an exponential decrease of
$\Uh_{\rm L}$ in the bulk. This is essentially equivalent to the
condition (\ref{cond_bounded}) that was obtained for the case of zero
modes. We shall assume henceforth that this condition also holds for
massive modes, i.e. that the value of $\Mt$ necessary to bind massive
fermions on the brane is at least that to bind massless ones. Indeed,
Eq.~(\ref{omega2}) shows that the minimum of $\omega^2$ can only be
negative for large values of the parameter $\Mt$. However,
since the first term of (\ref{omega2}) increases exponentially at
large distance from the brane, then, if $\mt\not=0$, $\omega^2$ will
necessarily become positive.  This will yield asymptotic radiative
behaviors of the spinor bulk components. Physically, it can be
interpreted as a tunneling of the fermions from the brane to the
bulk~\cite{dubovski}. On the other hand, on the brane, the Higgs field
$H$ and the derivative of the warp factor $S$ vanish, so that
$\omega^2(0)$ is positive. As a result, the fermions can freely
propagate in a tiny region around the brane, but certainly only for
particular values of $\omega^2$ (and thus of $\mt$) satisfying the
boundary conditions with the surrounding exponential decreasing
regions. The effective potential $V_{_{\rm eff}}=-\omega^2$, depicted
in Fig.~\ref{figveff}, exhibits a local minimum on the brane and
minima at infinity.  The modes trapped on the brane are thus expected
to have discrete masses $\mt$ on the brane and non-zero probability of
tunneling into the bulk.
\begin{figure}
\begin{center}
\epsfig{file=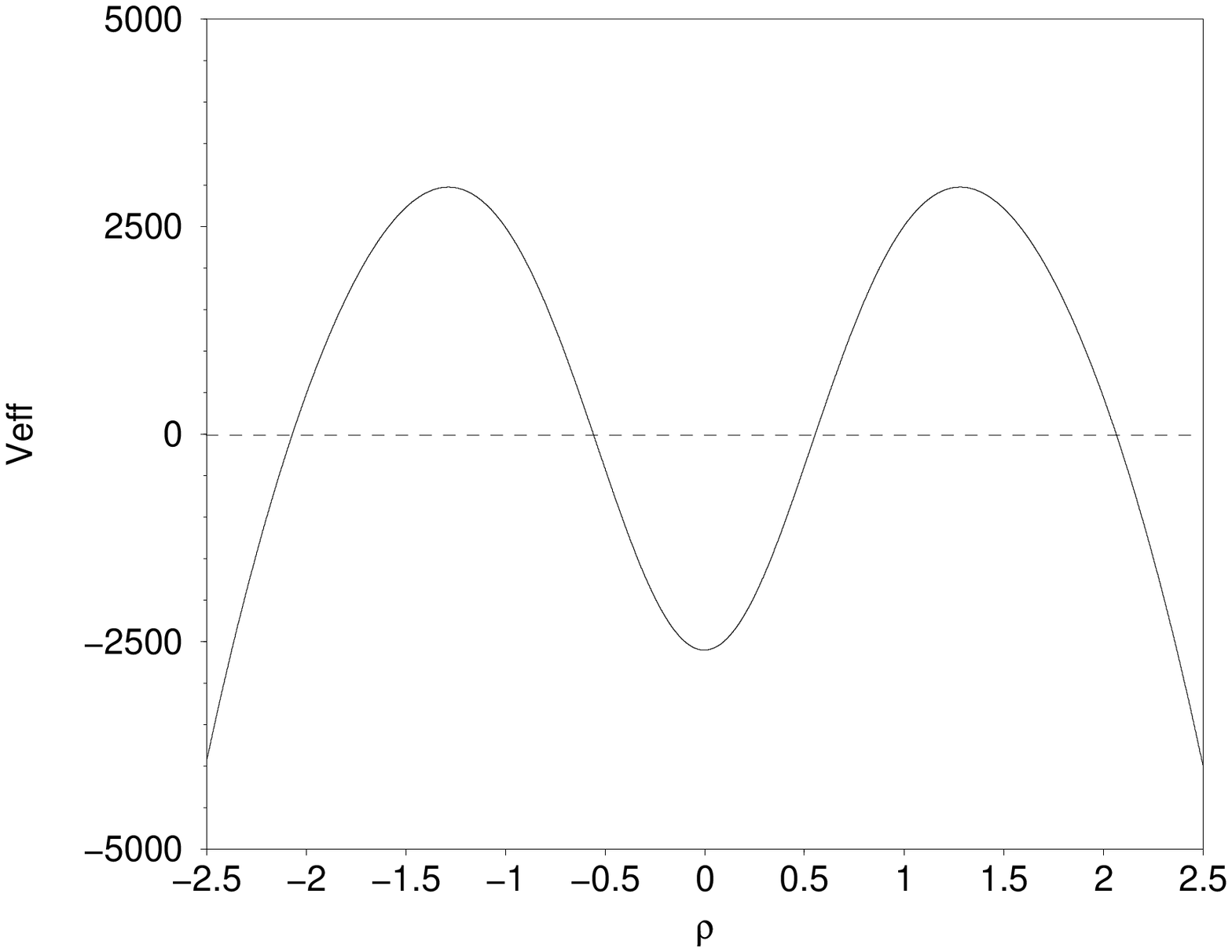,width=10cm}
\caption[Le potentiel des fermions dans la cinqui\`eme dimension.]{The
effective potential $V_{_{\rm eff}}=-\omega^2$ felt by $\Uh_{\rm
L}$. We chose the parameters $\Mt= 100$ and $\mt = 0.5 \Mt$ (an
illustrative value, not necessarily leading to a bound state) in the
Higgs and gravity background similar to those of Fig.~\ref{fig3}, and
we have assumed, for definiteness, the values $\alpha =1.63$,
$\beta=1.23$. The fermions are trapped on the brane where $V_{_{\rm
eff}}$ is negative and have a non-zero probability of tunneling into
the bulk due to combined effects of Higgs and gravity which produce a
finite potential barrier.}\label{figveff}
\end{center}
\end{figure}

\section{Mass spectrum and tunneling rate}
\label{sec_analy}
Since the Higgs and warp factor profiles are not known analytically,
it is a priori impossible to solve Eq.~(\ref{schrodlike})
analytically. Nevertheless, since the fermions are expected to be
trapped in a neighborhood of the brane, we can look for series
solutions in $\varrho$.

In \S~\ref{subsec_approx}, we give an approximation of the effective
potential $V_{_{\rm eff}}$ which will then be used to determine the
bound states and the mass spectrum in \S~\ref{subsec_bound}. We end
this section by determining in \S~\ref{subsec_tunel} the tunneling
rate in this approximation. The validity of this approach is difficult
to assess and it will be justified {\sl a posteriori} on the ground of
a full numerical integration of the system in the following section.
In the whole section, it is assumed that $\mt>0$.

\subsection{Approximation of the effective potential}\label{subsec_approx}

In a neighborhood of the brane we can expand
the Higgs and warp factor profiles as
\begin{eqnarray}
H(\varrho) & = & H_1\varrho + H_3\varrho^3 +
{\mathcal{O}}(\varrho^4),\label{fieldexpand1}\\
S(\varrho) & = & S_1\varrho + S_3 \varrho^3 + {\mathcal{O}}(\varrho^4),
\label{fieldexpand2}
\end{eqnarray}
where both the constant and quadratic terms vanish for symmetry
reasons. To simplify the analysis, we shall make use of the equations
of motion in the form
\begin{equation} \dot S  = {2\over 3} F(S,H),\label{dotS}
\end{equation}
and
\begin{equation} \dot H  = \pm \sqrt{{2\over \alpha} F(S,H)},\label{dotH}
\end{equation}
where the function $F(S,H)$ is defined as
\begin{equation}F(S,H) \equiv 6 S^2 +\alpha\beta (H^2-1)^2 -1, \label{F}
\end{equation}
(recall that $\Lambda < 0$).

Plugging the expansions (\ref{fieldexpand1}) and (\ref{fieldexpand2})
into Eqs.~(\ref{dotS}) and (\ref{dotH}), it follows that the function
$F(S,H)$ can be expanded up to third order as
\begin{equation}\label{Fexpand}
F(S,H) \sim \alpha \beta - 1 + \left(6 S_1^2 - 2 \alpha \beta H_1^2
\right) \varrho^2 + {\mathcal{O}}(\varrho^4).
\end{equation}
Inserting this expression back into the equations of motion yields the
three coefficients
\begin{eqnarray}
S_1 & = &\frac{\alpha}{3}H_1^2,\label{coeffs1}\\
S_3 & = & \frac{4}{9} \alpha H_1^2
\left[\frac{\alpha}{3} H_1^2 - \beta \right],\label{coeffs2}\\
H_3 & = & \frac{2}{3} H_1 \left[\frac{\alpha}{3} H_1^2 -
\beta \right]\label{coeffs3}
\end{eqnarray}
in terms of the coefficient $H_1$
\begin{equation}\label{def_H1}
H_1\equiv\left.\partial_\varrho H\right\vert_{\varrho=0}=
\sqrt{\frac{2}{\alpha}(\alpha\beta-1)}.
\end{equation}
Then, the frequency $\omega^2$ can be expanded as an harmonic
oscillator potential
\begin{equation}\label{omega2near}
\omega^2(\varrho)=\omega_{\rm b}^2(\varrho)+{\mathcal{O}}(\varrho^4)\qquad
\hbox{with}\qquad
\omega^2_{\rm b}(\varrho)\equiv\omega_0^2 - \Omega \,\varrho^2
\end{equation}
where $\omega_0^2$ and $\Omega$ are given by
\begin{eqnarray}
\omega_0^2 &\equiv& \mt^2 + \Mt H_1\left(1 + \frac{\alpha}{6 \Mt} H_1\right)
\label{omega2brane}\\
\Omega &\equiv& H_1 \Mt^2 \left[H_1 + \frac{2}{\Mt} \left(\beta -
\frac{\alpha}{6} H_1^2 \right) + \frac{\alpha}{3\Mt^2}H_1 \left(2\beta
- \mt^2 - \frac{7\alpha}{12} H_1^2 \right) \right].
\label{omega2shape}
\end{eqnarray}

The function $\omega^2$ is well approximated by $\omega_{\rm b}^2$
only near the brane and the expansion (\ref{omega2near}) is no longer
valid at large distance where the exponential term dominates [see
Eq.~(\ref{omega2})]. Once the fixed asymptotic values of the
Higgs and warp factor are reached, the frequency (\ref{omega2})
behaves as $\omega^2 \sim \omega^2_\infty(\varrho)$, with
\begin{equation}
\label{omega2infty}
\omega^2_\infty = \mt^2 \de^{2|\varrho|/\sqrt{6}} - \left(\Mt
+ \frac{1}{2\sqrt{6}} \right)^2.
\end{equation}
The analytical estimate of the function $\omega^2$ is thus obtained by
matching the two limiting asymptotic behaviors (\ref{omega2near}) and
(\ref{omega2infty}), respectively $\omega_{\rm b}^2$ closes to the
brane and $\omega^2_\infty$ far from it, as
\begin{equation}\label{omega2estim}
\omega^2(\rho)=\left\lbrace
\begin{array}{cc}
\omega_{\rm b}^2, & |\rho|<\rho_{_{\rm m}}\\
\omega_{\infty}^2, & |\rho|>\rho_{_{\rm m}}
\end{array}\right. .
\end{equation}
The dimensionless matching distance $\varrho_{_{\rm m}}$ has to be
solution of
\begin{equation}
\omega^2_{_{\rm b}}(\varrho_{_{\rm m}})=\omega^2_\infty(\varrho_{_{\rm m}}),
\end{equation}
in order to get a continuous function. Note that, because of the
symmetry on both sides of the wall, we can assume $\varrho_{_{\rm
m}}>0$ without lack of generality. For large values of $\Mt$, and
using Eq.~(\ref{omega2shape}), we get
\begin{equation}
\label{shapeapp}
\sqrt{\Omega} \sim \Mt H_1,
\end{equation}
leading to
\begin{equation}\label{rhomatch}
\varrho_{_{\rm m}} \sim \frac{1}{H_1}.
\end{equation}
As it turns out, the faster the asymptotic solution is reached, the
better the approximation works. This is the case in particular for
$H_1>1$,

The exact (numerically integrated) effective potential and its
approximation are compared in Fig.~\ref{figapprox}. The global shapes
are effectively the same, and in spite of uncertainties at
intermediate regions due to this crude approximation, it is reasonable
to expect the same fermion physical behaviors in both potentials. Note
that for (cosmologically favored) higher value of $\alpha \beta$, the
Higgs field and the warp factor reach more rapidly their asymptotic
values leading thus to a better agreement between the two potentials,
as can be seen on Fig.~\ref{figapprox}.

\subsection{Determination of the bound states}\label{subsec_bound}

Given the approximate frequency (\ref{omega2estim}), the equation of
motion for the left-handed bulk spinor component $\Uh_{\rm L}$ reduces
to
\begin{eqnarray}
\varrho>\varrho_{_{\mathrm{m}}} & \Rightarrow &
\left[\partial_\varrho^2 + \mt^2 \de^{2 |\varrho|/\sqrt{6}} -
\left(\Mt + \frac{1}{2\sqrt{6}}
\right)^2\right]\Uh_{\mathrm{L}}=0,\label{approxfar}\\
\varrho<\varrho_{_{\mathrm{m}}} & \Rightarrow &
\left(\partial_\varrho^2+ \omega_0^2-\Omega \varrho^2\right)\Uh_{\rm
L}=0,\label{approxnear}
\end{eqnarray}
with the requirement that $\Uh_{\rm L}$ and its derivative are
continuous at $\varrho_{_{\rm m}}$. Note also that we consider only
the case $\varrho>0$, physics on both sides of the brane being
completely symmetric under the transformation $\varrho\to
-\varrho$. Let us consider the solutions in each region separately.
\begin{figure}
\begin{center}
\epsfig{file=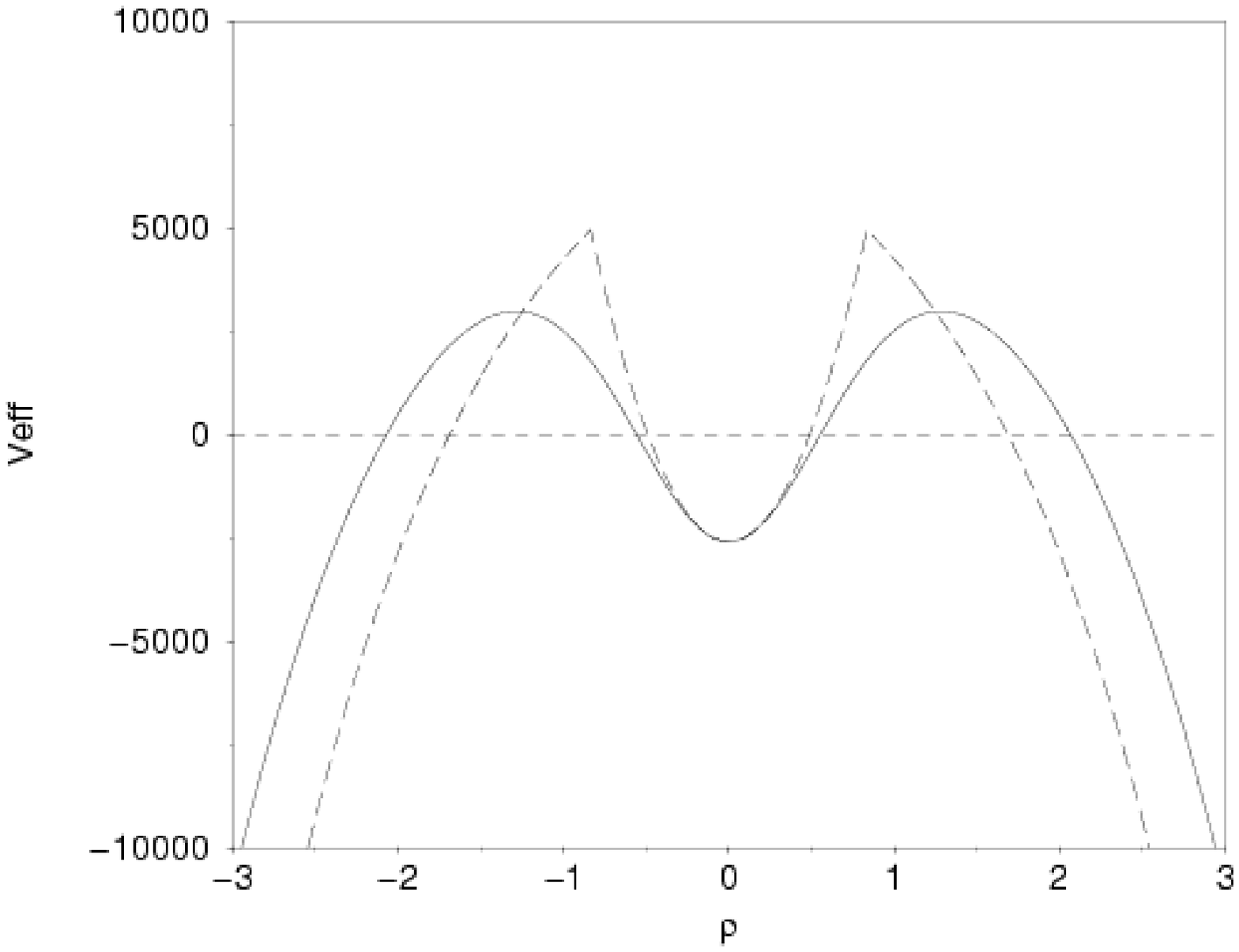,width=10cm}
\epsfig{file=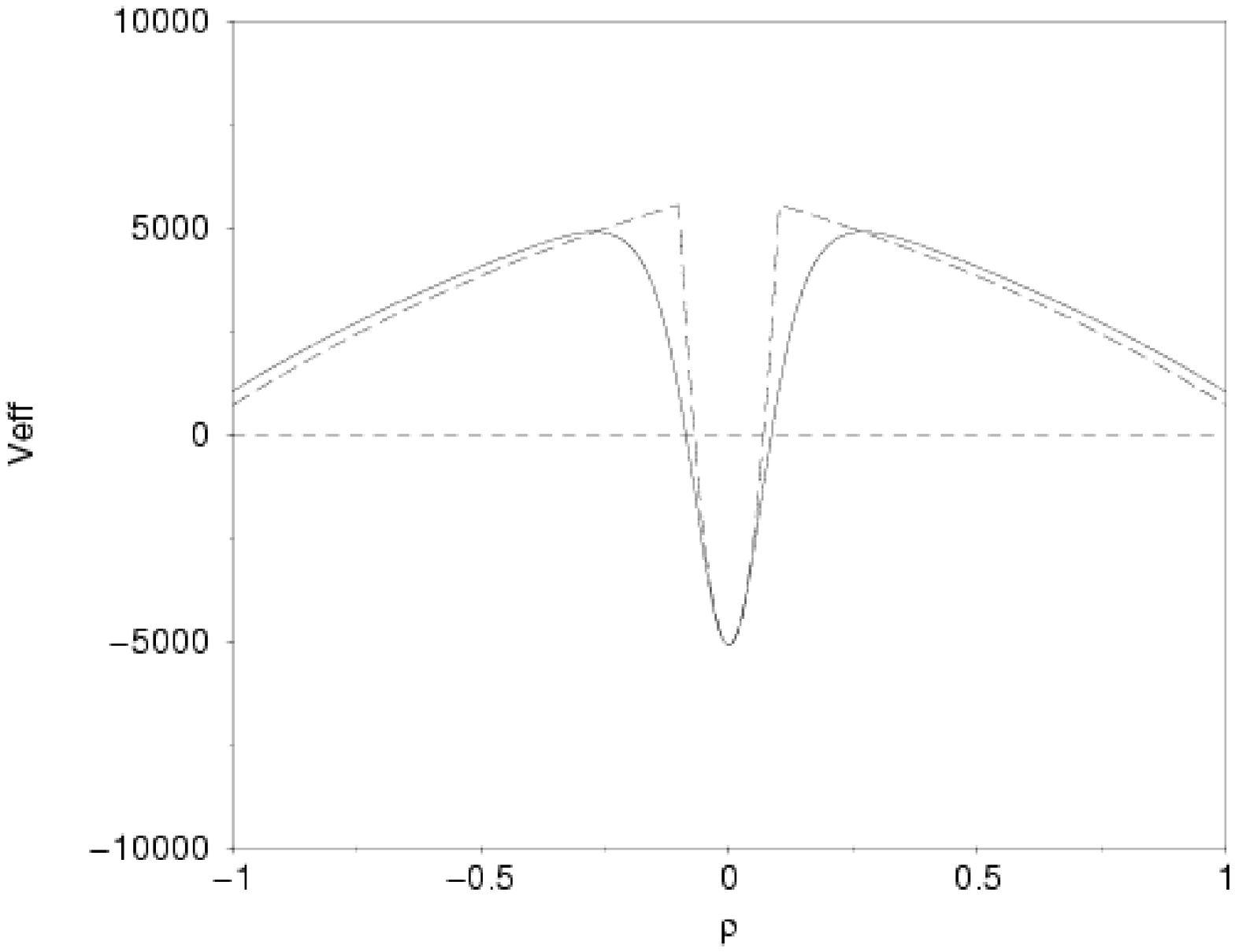,width=10cm}
\caption[Approximation analytique du potentiel des fermions dans la
dimension suppl\'ementaire.]{The effective potential $V_{_{\rm
eff}}=-\omega^2$ and the approximate analytic effective potential
(dashed curve) obtained from the matching of its two asymptotic
expansions near the brane and at infinity. On the first panel, the
parameters are the same as on Fig.~\ref{figveff}, i.e. $\Mt= 100$ and
$\mt = 0.5 \Mt$. As the asymptotic solution is not reached very close
to the brane, the approximation is rather poor and the two solutions
are not in good agreement at intermediate regions. On the panel, we
use the new parameter values $\Mt= 100$ and $\mt = 0.6 \Mt$, obtained
for $\alpha = 0.1865$ and $\beta = 53.62$, a much better approximation
is obtained around the potential barrier provided the Higgs field and
warp factor reach their vacuum value rapidely, i.e. for larger value
of $\alpha \beta$. As on Fig.~\ref{figveff}, the parameters are chosen
to illustrate the point and do not necessarily correspond to existing
bound states.}
\label{figapprox}
\end{center}
\end{figure}

\paragraph{$\varrho>\varrho_{_{\rm m}}$:} we introduce the new variable
\begin{equation}\label{defz}
z\equiv\sqrt{6}\mt\de^{|\varrho|/\sqrt{6}},
\end{equation}
in terms of which Eq.~(\ref{approxfar}) reduces to a standard Bessel
differential equation
\begin{equation}\label{bessel}
\left[\frac{\dd^2}{\dd z^2} + \frac{1}{z}\frac{\dd}{\dd z} + \left(1 -
\frac{\mut^2}{z^2}\right)\right]\, \Uh_{\rm L} = 0,
\end{equation}
the order of which being given by
\begin{equation}
\label{defmut}
\mut=\sqrt{6}\Mt+\frac{1}{2}.
\end{equation}
Since $\omega^2$ is positive at infinity, the asymptotic form of the
solution is necessary radiative, as was already pointed out in the
previous section.  The most general solution of the Bessel equation
(\ref{bessel}) is a linear superposition of Hankel functions. Since we
are interested only in ingoing waves in order to study a tunneling
process, the most general solution takes the form
\begin{equation}
\Uh_{\rm L}(z)=B \, H^{(1)}_{\mut}(z),
\end{equation}
where $H^{(1)}_{\mut}(z)$ is the Hankel function of the first kind,
propagating towards the brane at infinity~\cite{abramo}
\begin{equation}
H^{(1)}_{\mut}(z\rightarrow\infty) \sim \sqrt{\frac{2}{\pi z}}
\,\de^{i(z-\mut\pi/2-\pi/4)}.
\end{equation}
and $B$ is an arbitrary complex constant.
\paragraph{$\varrho<\varrho_{_{\rm m}}$:} performing operations similar to
those of the previous case, we cast Eq.~(\ref{approxnear}) on the form
\begin{equation}\label{pcf}
\left[\frac{\dd^2}{\dd x^2} - \left(\frac{1}{4} x^2 + a \right)
\right] \Uh_{\rm L}=0,
\end{equation}
in which we have introduced the new variable and parameter
\begin{equation} x\equiv \left(4 \Omega\right)^{\frac{1}{4}} \varrho,\
\ \ \ \ a \equiv - \frac{\omega_0^2}{2 \sqrt{\Omega}}.
\label{defa}\end{equation}
The general solutions of Eq.~(\ref{pcf}) are the parabolic cylinder
functions, namely $U(a,x)$ and $V(a,x)$, of which $\Uh_{\rm L}(x)$ can
be expressed as linear superpositions. In the limit $x \gg |a|$, these
solutions scale as~\cite{miller}
\begin{eqnarray}\label{asymppcf}
U(a,x) \sim \de^{-\frac{1}{4}x^2} x^{-a-\frac{1}{2}}, & & V(a,x) \sim
\sqrt{\frac{2}{\pi}} \,\de^{\frac{1}{4}x^2} x^{a-\frac{1}{2}}.
\end{eqnarray}
Since we are interested in confined fermion states on the brane, only
the exponentially decreasing function is relevant, so that the general
solution near the brane reads
\begin{equation}
\Uh_{\rm L}(x) = A \, U(a,x),
\end{equation}
where $A$ is a complex integration constant.

The general solution $\Uh_{\rm L}(\varrho)$ for all $\varrho$ is
obtained by matching the two different solutions at
$\varrho=\varrho_{_{\rm m}}$. Since $\varrho_{_{\rm m}}$ corresponds
to the maximum positive value of the effective potential [see
Fig.~\ref{figapprox}], it is reasonable to consider that the Hankel
function at that point can be expanded around small values of their
argument with respect to their order~\cite{abramo}, i.e.
\begin{equation}
\label{hankelmatch}
H^{(1)}_{\mut}(z_{_{\rm m}}) \sim -\frac{2^{\mut}}{\pi} \Gamma(\mut)
z_{_{\rm m}}^{-\mut},
\end{equation}
while the parabolic cylinder functions can be taken in their large
argument asymptotic limit (\ref{asymppcf}). This is the same kind of
approximation as that made to derive the effective
potential. Physically, the initial conditions on the brane, i.e.
$\Uh_{\rm L}(0)$ and $\partial_\varrho\Uh_{\rm L}|_0$, are chosen in
such a way that the asymptotic exponentially growing function $V(a,x)$
contribution is everywhere negligible. Once these initial conditions
are fixed, they fully determine the solution on the other side of the
brane, i.e. for $x<0$. The asymptotic expansion (\ref{asymppcf}) can
be analytically extended to $-|x|=|x|\de^{i\pi}$ and
yields~\cite{gradsh}
\begin{equation}\label{pcfneg}
U(a,-|x|) \sim \de^{-\frac{1}{4}|x|^2} |x|^{-a-\frac{1}{2}}
\de^{-i\pi\left(a+\frac{1}{2}\right)}.
\end{equation}
Thus, once $\Uh_{\rm L}(0)$ and $\partial_\varrho\Uh_{\rm L}|_0$ are
fixed, the matchings between $H^{(1)}_{\mut}(z_{_{\rm m}})$ and
$U(a,-|x_{_{\rm m}}|)$ on one side, and $U(a,|x_{_{\rm m}}|)$ on the
other side fully determine the bulk component $\Uh_{\rm L}$ for all
$\varrho$.

The last constraint comes from Eq.~(\ref{orderzerour}) determining the
right-handed spinor bulk function. It is well defined if and only if
both $\Uh_{\rm L}$ and $\partial_\varrho \Uh_{\rm L}$ are not
singular. In fact, the derivative of the parabolic cylinder function
$U(a,x)$ is generally discontinuous at $x=0$. With the help of the
Wronskian of $U(a,x)$ and $U(a,-x)$~\cite{miller}
\begin{equation}
U(a,x) \frac{\dd U(a,-x)}{\dd x}- U(a,-x) \frac{\dd
U(a,x)}{\dd x}=\frac{2\pi}{\Gamma\left(\frac{1}{2}+a\right)},
\end{equation}
we can construct the derivative discontinuity at $x=0$. This is
\begin{equation}\label{jump}
\frac{\dd U(a,0^-)}{\dd x} - \frac{\dd U(a,0^+)}{\dd x}=2^{\frac{a}{2}
+ \frac{3}{4}} \frac{\Gamma\left(\frac{3}{4} +
\frac{a}{2}\right)}{\Gamma\left(\frac{1}{2} + a\right)},
\end{equation}
where we have used the particular value~\cite{miller}
\begin{equation}
\label{uzero}
U(a,0)=\frac{\sqrt{\pi}}{2^{\frac{a}{2} + \frac{1}{4}}
\Gamma \left(\frac{3}{4} + \frac{a}{2}\right)}.
\end{equation}
Imposing that the derivative of $\Uh_{\rm L}$ is continuous at $x=0$
results in imposing that the jump (\ref{jump}) vanishes.  This is the
case if and only if $a$ is solution of
\begin{equation}\label{nulwronsk}
\frac{\Gamma \left(\frac{3}{4} +
\frac{a}{2}\right)}{\Gamma\left(\frac{1}{2} + a\right)} = 0.
\end{equation}
Since $\Gamma$ is singular for negative integer arguments, this
condition is satisfied only for
\begin{equation}
\label{quantizationBF}
-a-\frac{1}{2}=2n,
\end{equation}
where $n$ is a positive integer. Note that $-a-1/2$ cannot be odd
since then the numerator of the Wronskian~(\ref{nulwronsk}) will also
be singular resulting in a finite derivative jump at $x=0$. The
condition (\ref{quantizationBF}) shows that the trapped fermions on the
brane have necessarily discrete masses $\mt_n$ which read, using the
values of the parameters~(\ref{omega2brane}), (\ref{omega2shape}) and
(\ref{defa}),
\begin{eqnarray}
\mt^2_n & = & \Mt H_1 \left[(4n+1) \sqrt{1 + \frac{6\beta}{3\Mt H_1} +
\frac{24 \alpha \beta + \alpha^2 H_1^2
\left[(4n+1)^2-5\right]}{36\Mt^2}} \right.  \nonumber \\ & & -
\left. \left(1+\frac{\alpha H_1}{6\Mt} \left[(4n + 1)^2 + 1 \right]
\right) \right].
\end{eqnarray}
This mass spectrum is valid for $n>0$ since our derivation assumed
that $\mt>0$. In the limiting case where $\Mt\gg 1$, it reduces to the
much simpler form for the lowest masses
\begin{equation}\label{appspectrum}
\mt_n \sim  2 \sqrt{n} \sqrt{\Mt H_1}.
\end{equation}
On the other hand, $\mt^2$ cannot reach very large values since it is
necessary to have a potential barrier in order to have bound
states. From the expression of the effective potential
(\ref{omega2infty}), the barrier is found to disappear when
\begin{equation}
\omega_\infty^2(\varrho_{_{\rm m}},\mt_{\max}) \sim 0.
\end{equation}
Again in the limit where $\Mt\gg 1$, using the value (\ref{rhomatch})
of $\varrho_{_{\rm m}}$, one gets the maximum accessible reduced mass
$\mt_{\max}$ for $\mt$ as
\begin{equation}\label{defmtmax}
\mt_{\max} \sim \Mt \de^{-1/\sqrt{6} H_1}.
\end{equation}
The maximum number of distinct massive states trapped on the brane
can thus be estimated to be
\begin{equation}
\label{defnmax}
n_{\max} \sim {\mathrm{Int}}\left[\Mt \frac{1}{4 H_1}\de^{-2
/\sqrt{6}H_1} \right].
\end{equation}
For the parameters chosen in Fig.~\ref{figveff}, one obtains
$\mt_{\max} \sim 0.68 \Mt$ and there are $n_{\max}=11$ massive modes
trapped on the brane.

\subsection{Fermion tunneling rate}\label{subsec_tunel}

Since the effective potential becomes negative at infinity, the
massive modes trapped on the brane are subject only to a finite
potential barrier. They are in a metastable state and can tunnel from
the brane to the bulk. In this section we use our previous analytic
solution to estimate the tunneling rate. Would this rate be too high,
one would observe an effective violation of energy-momentum
conservation on the brane, i.e. in four dimension, thereby
contradicting observation.

Our starting point is the analytic solution for the left-handed
bulk function that was derived in the previous section
\begin{eqnarray}
\varrho<\varrho_{_{\rm m}} & \Rightarrow & \Uh_{\rm L}(\varrho)=A\,
U\left(a,\sqrt{2}\Omega^{1/4} \varrho\right), \\
\varrho>\varrho_{_{\rm m}} & \Rightarrow & \Uh_{\rm L}(\varrho)=B \,
H^{(1)}_{\mut}\left(\sqrt{6}\mt\de^{|\varrho|/\sqrt{6}}\right).
\end{eqnarray}
The transmission factor can easily be derived from the matching
conditions of the left-handed bulk fermion component at
$\varrho=\varrho_{_{\rm m}}$. First of all, $\Uh_{\rm L}$ has to be
continuous. Using the expansions~(\ref{asymppcf}) and
(\ref{hankelmatch}) and the value (\ref{quantizationBF}) of the
parameter $a$, permits to find the relation
\begin{equation}\label{amppos}
A \, \de^{-\frac{1}{2}\sqrt{\Omega}\varrho_{_{\rm m}}^2}
\left(2\sqrt{\Omega}\right)^n \varrho_{_{\rm m}}^{2n} =
-\frac{i}{\pi}B \, \Gamma(\mut)
\left(\sqrt{\frac{3}{2}}\mt_n\right)^{-\mut} \de^{-\mut
\varrho_{_{\rm m}}/\sqrt{6}}
\end{equation}
between the coefficients $A$ and $B$. Making use of the
expression~(\ref{rhomatch}) for $\varrho_{_{\rm m}}$ yields
\begin{equation}\label{baba}
\frac{B}{A}= \frac{i \pi}{\Gamma(\mut)}
\left(\sqrt{\frac{3}{2}}\mt_n\right)^{\mut}
\left(\frac{\mut^2}{3\sqrt{\Omega}}\right)^n
\exp{\left(\frac{\mut^2}{12\sqrt{\Omega}}\right)}.
\end{equation}

Assuming, as above, that $\Mt \gg 1$ (so that $\mut\gg1$) we can
expand $\Gamma(\mut)$ as
\begin{equation}
\Gamma(\mut) \sim \mut^{\mut-1/2} \de^{-\mut} \sqrt{2 \pi},
\end{equation}
so that, using expressions~(\ref{shapeapp}), (\ref{rhomatch}) and
(\ref{defmut}) respectively of $\Omega$, $\varrho_{_{\rm m}}$, and
$\mut$, we get that $B/A$ is approximated by
\begin{equation}\label{rappamp}
\frac{B}{A} \sim i \sqrt{\frac{\pi}{2}} \left(\frac{2}{H_1}\right)^n
6^{1/4} \Mt^{n+1/2} \exp{\left[- \sqrt{6}\Mt
\left(\ln{\frac{2\mt_{\max}}{\mt}} - 1 + \frac{1}{2 \sqrt{6}H_1}\right)
\right]}.
\end{equation}

The transmission coefficient from the brane to the bulk associated
with $\Uh_{\rm L}$ can thus be defined by
\begin{equation}\label{th}
\Th \equiv \frac{\Uh_{\rm L}(\mut)}{\Uh_{\rm L}(0)},
\end{equation}
with $\Uh_{\rm L}(\mut)=B H_\mut^{(1)}(\mut)$, is evaluated at the
turning point $\varrho=\mut$ where the spinor bulk component begins to
propagate freely. Using the behaviour (\ref{uzero}) of the function at
the origin, the ratio (\ref{rappamp}) and the properties~\cite{abramo}
of the Hankel function
\begin{equation}
H^{(1)}_{\mut}(\mut)=\left(\frac{4}{3}\right)^{2/3}
\frac{\de^{-i\pi/3}}{\Gamma(2/3)}
\, \mut^{-1/3},
\end{equation}
the transmission coefficient (\ref{th}) reduces to
\begin{eqnarray}\label{transmh}
\Th \sim \frac{2^{4/3}\Gamma(1/2-n)}{3^{7/12} \Gamma(2/3)} \de^{i\pi/6}
H_1^{-n} \Mt^{n+1/6}
\exp{\left[- \sqrt{6}\Mt \left(\ln{\frac{2\mt_{\max}}{\mt}} - 1 +
\frac{1}{2\sqrt{6} H_1}\right) \right]}.
\end{eqnarray}
It follows, using the definition~(\ref{defuh}), that the probability
for a trapped particle on the brane to tunnel to the bulk is given by
\begin{equation}
\label{probtunnel}
\Pb\equiv|\Tt|^2=\de^{2\mut/\sqrt{6}} \, |\Th|^2.
\end{equation}

The characteristic time for a fermionic mode trapped
on the brane can be roughly estimated by
\begin{equation}\label{timecar}
\tau_{_{\mathrm{b}}} \sim y_{_{\rm b}}=\frac{\varrho_{_{\rm
b}}}{\sqrt{|\Lambda|}},
\end{equation}
where $y_{_{\rm b}}$ represents the typical length, in the fifth
dimension, felt by a particle on the brane. As can be seen on
Fig.~\ref{figbulkradiate}, the spinor bulk components are
exponentially damped as soon as the effective potential becomes
positive. Thus $\varrho_{_{\rm b}}$ can be estimated by the solution
of $\omega_{_{\rm b}}^2(\varrho_{_{\rm b}})=0$ so that, keeping in
mind that $\mt<\Mt$,
\begin{equation}
\label{varrhob}
{\varrho_{_{\rm b}}}^{-1} \sim \sqrt{H_1 \Mt}.
\end{equation}
The life-time $\tau_n$ of a fermionic bound state on the brane
labelled by $n$
\begin{equation}
\tau_n=\frac{\tau_{_{\rm b}}}{\Pb}
\end{equation}
can be estimated by
\begin{eqnarray}\label{lifetime}
\tau_n & \sim &\frac{3^{7/6} \, \Gamma^2(2/3)}{2^{11/6} \,
\Gamma^2(1/2-n)}H_1^{2n -1/2} \frac{\Mt^{-2n - 5/6}}{\sqrt{|\Lambda|}}
\, \nonumber \\ & & \times \exp{\left[2\sqrt{6}\Mt\left(
\ln{\frac{2\mt_{\max}}{\mt_n}} - 1 - {1\over \sqrt{6}} +
\frac{1}{2\sqrt{6}H_1}\right)\right]}.
\end{eqnarray}

We recall that, due to the approximations performed in the previous
derivation, this estimate is valid only for
$\mt\ll\mt_{\max}$. Nevertheless, the argument in the exponential
amplifies
the transition from bound states to tunneling ones for masses $\mt
\sim \mt_{\max}$, as intuitively expected. An order of magnitude of
the minimal coupling constant $\Mt$ leading to stable bound states can
thus be estimated by requiring that the lowest massive state does not
tunnel,
\begin{equation}
\mt_1 <2 \mt_{\max} \exp{\left(-1 - {1\over \sqrt{6}} +
\frac{1}{2\sqrt{6}H_1}\right)}.
\end{equation}
Using the two values~(\ref{appspectrum}) and (\ref{defmtmax}),
this implies that
\begin{equation}
\Mt \gtrsim H_1 \exp{\left(2 + {2\over \sqrt{6}} +
\frac{1}{\sqrt{6}H_1} \right)}.
\label{frontier}\end{equation}
As a numerical application, for the Higgs and gravity parameters used
in Fig.~\ref{figveff}, we get $\Mt \gtrsim 25$.

\section{Numerical investigation}
\label{sec_num}

Numerically, it is simpler and more convenient to solve the first
order differential system (\ref{syst1}-\ref{syst2}). A Runge-Kutta
integration method was used, on both sides of the brane. In order to
suppress the exponential growth, we integrate from the turning point,
$\varrho=\mut$, where the solution begins to propagate freely, toward
the brane. In this way, we get only $U(a,x)$ near the brane. The
radiative solution for $\varrho>\mut$, is simply obtained by
integrating from the turning point toward infinity, with initial
conditions determined by the matching with the exponential decreasing
solution near the wall.  The same method is used on the other side of
the brane, but this time, by means of the last free parameter, we
impose the continuity of one bulk spinor component on the brane
($\Ut_{\rm L}$ say). Generally, the other bulk spinor component will
be discontinuous at $\varrho=0$, as expected from the analytical study
since $U'(a,0)$ is generally discontinuous. The mass spectrum is thus
obtained by requiring the continuity of $\Ut_{\rm R}$ on the brane.

The bulk spinor components computed this scheme have been plotted for
the first massive modes trapped on the brane in
Fig.~\ref{figbulkcomp}, for $\Mt=100$. The lowest mass is numerically
found to be $\mt_1\sim 0.209 \Mt$ and was estimated analytically,
from~(\ref{appspectrum}), to be $\mt_1 \sim 0.210 \Mt$ leading to a
precision of 0.5\% for the analytical estimate. The second mass is
numerically found to be $\mt_2 \sim 0.291\Mt$, which has to be
compared to its analytical estimate $\mt_2 \sim 0.295 \Mt$. Again, the
precision of this estimate is of about 1\%. As predicted
from~(\ref{defnmax}), there are $n_{\max}=11$ massive bound states the
lightest masses of which are summed up in table~\ref{table}. On
Fig.~\ref{figbulkradiate}, we plot the last $n=11$ trapped mode; it
has a tiny radiative component, as expected for a tunneling mode.

\begin{figure}
\begin{center}
\epsfig{file=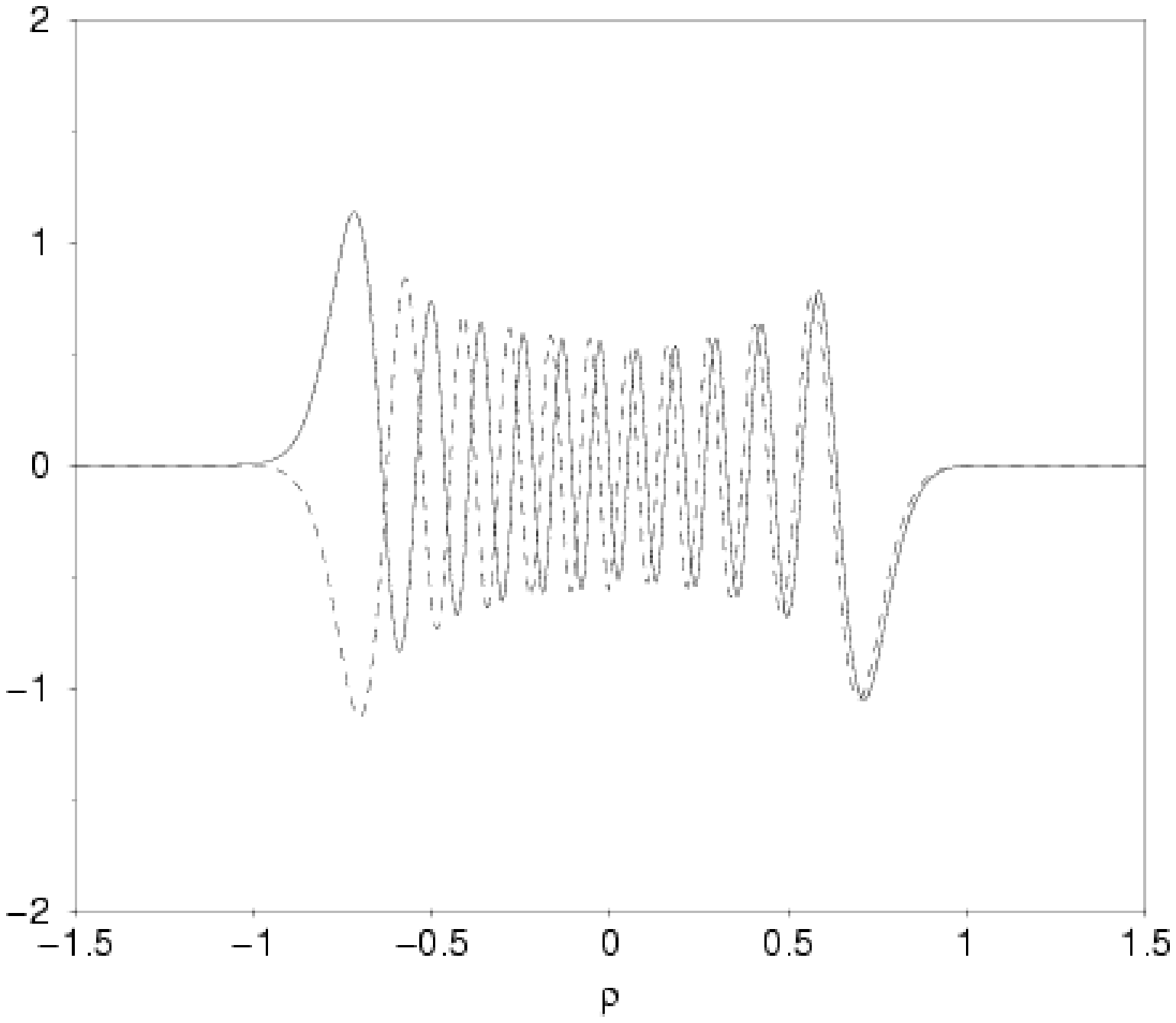,width=7cm}
\epsfig{file=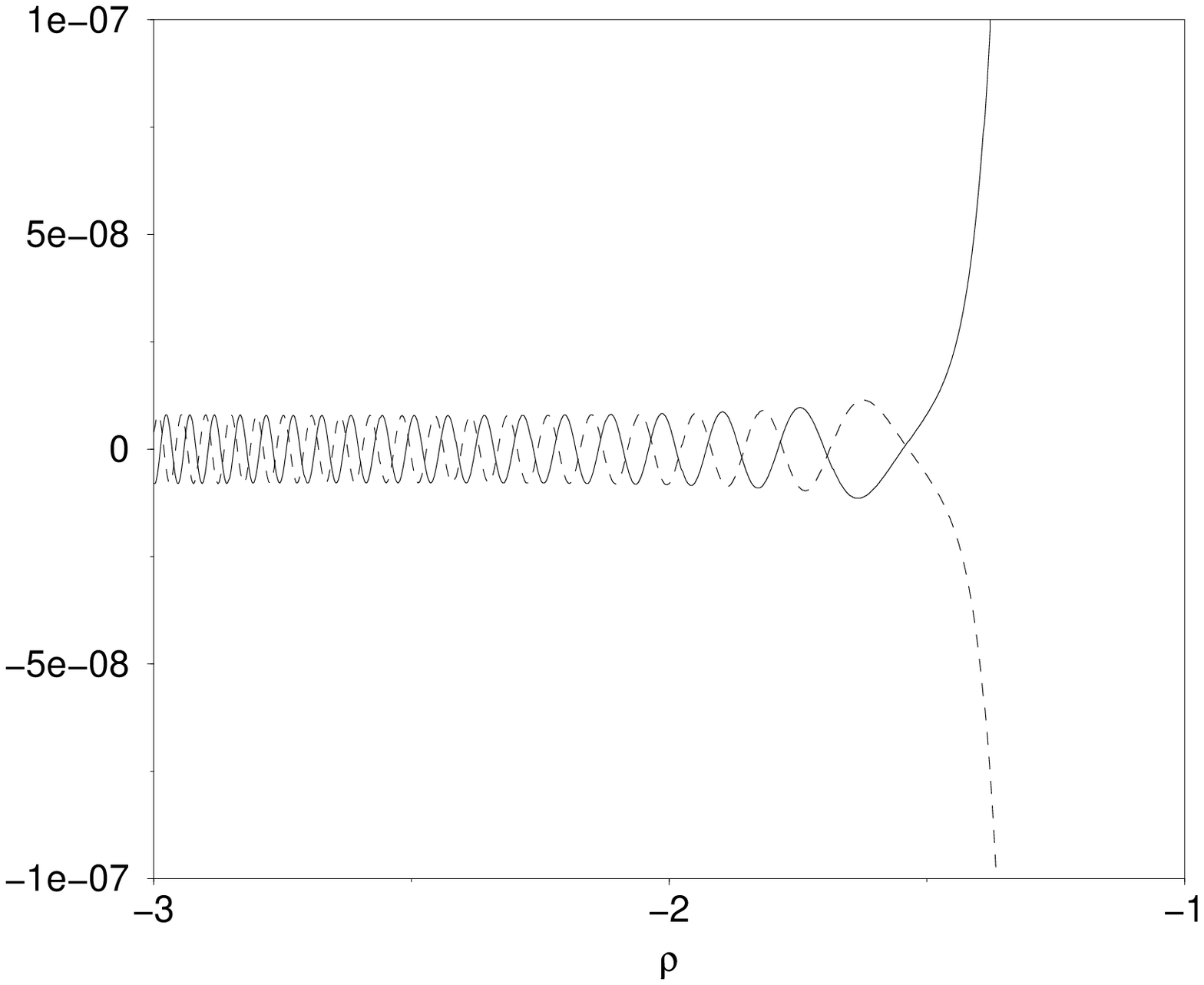,width=7.6cm}
\caption[Fonctions d'onde des fermions pi\'eg\'es sur la
brane.]{[Left] The right (dashed line) and left (solid line) bulk
spinor components $\Ut_{\rm R}$ and $\Ut_{\rm L}$, as functions of the
dimensionless distance $\varrho$ to the brane, for the heaviest
massive state ($n=11$) and $\Mt=100$. [Right] Zoom near the turning
point and transition to the radiative behaviour which takes place at a
finite distance to the brane due to the tunneling of this mode from
the brane to the bulk.}
\label{figbulkradiate}
\end{center}
\end{figure}

\begin{table}
\begin{center}
\begin{tabular}{c|c|c|c|c|c|c|c c}
  $n$                 &   1   &   2   &   3   &  4    &   5   & 6 &
  \dots & 11 \\
  $\mt_n$ (numerical) & 0.209 & 0.291 & 0.353 & 0.402 & 0.444 & 0.480
  & \dots & 0.593  \\
  $\mt_n$ (estimates)   & 0.210 & 0.295 & 0.359 & 0.412 & 0.458 & 0.499
  & \dots & 0.657  \\
  precision (\%)      &  0.5  &  1.3  &  1.7  & 2.4   &   3   &  3.8 &
  \dots & 9.7\\
\end{tabular}
\caption[Validit\'e de l'approximation analytique du potentiel des
fermions.]{Comparison of the numerical values and analytical estimates
of the first six bound states reduced mass $\mt$, together with the
heaviest mode, computed for $\Mt=100$.}\label{table}
\end{center}
\end{table}

In conclusion, the numerics confirm that the approximations of the
previous section and our estimates are accurate up to 1\%-10\% (see
table~\ref{table}).

\begin{figure}
\begin{center}
\epsfig{file=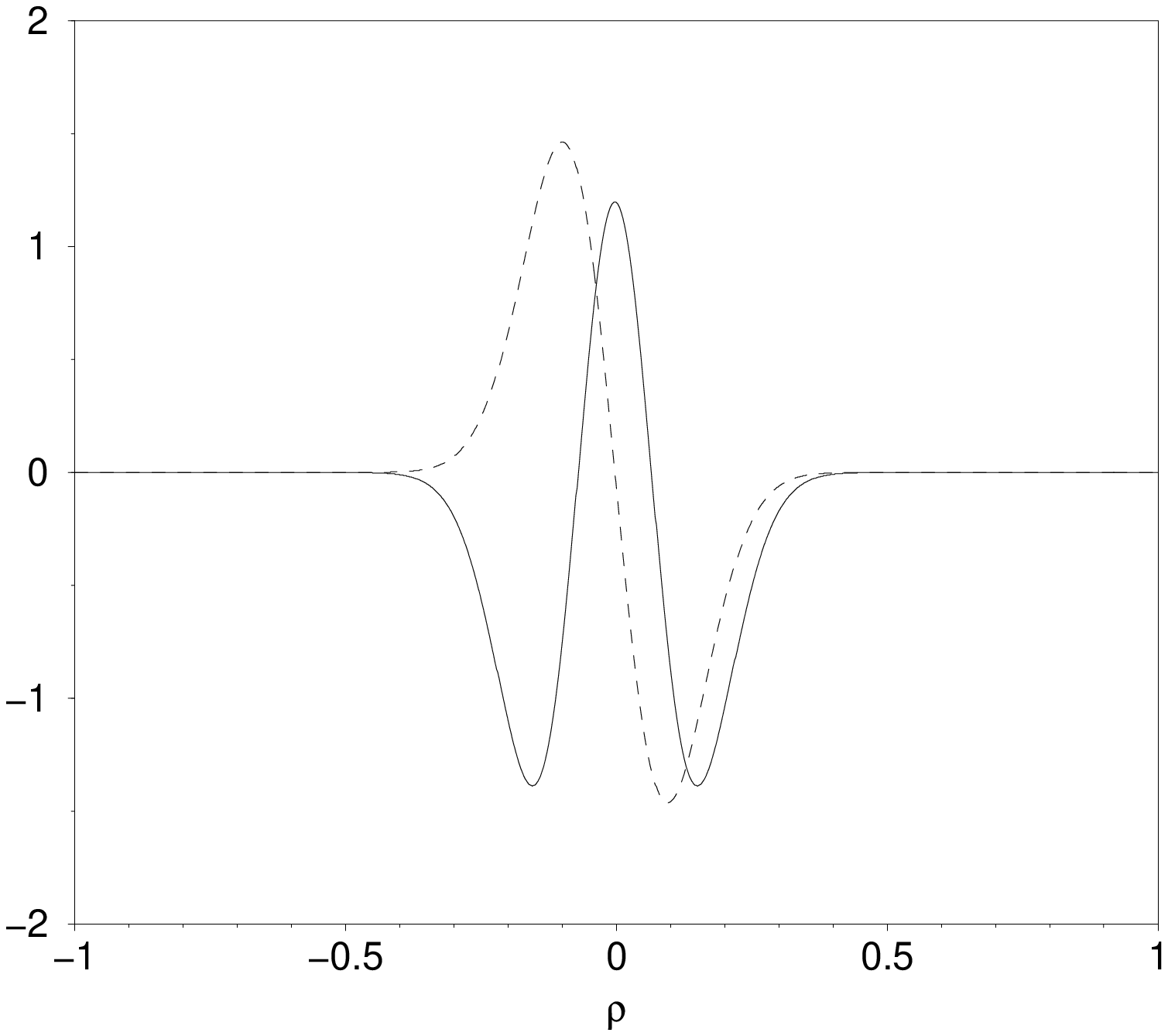,width=7cm}
\epsfig{file=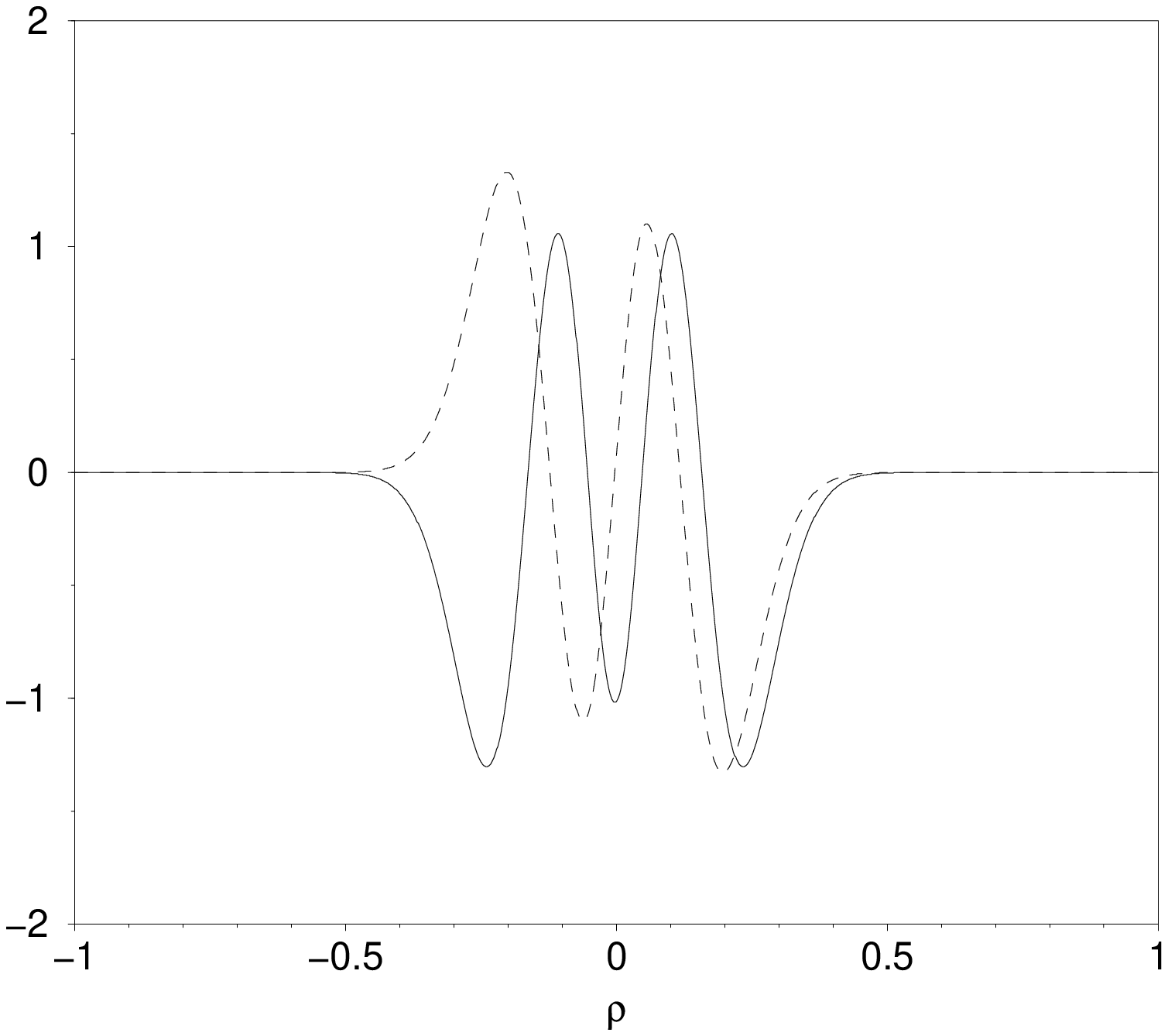,width=7cm}
\epsfig{file=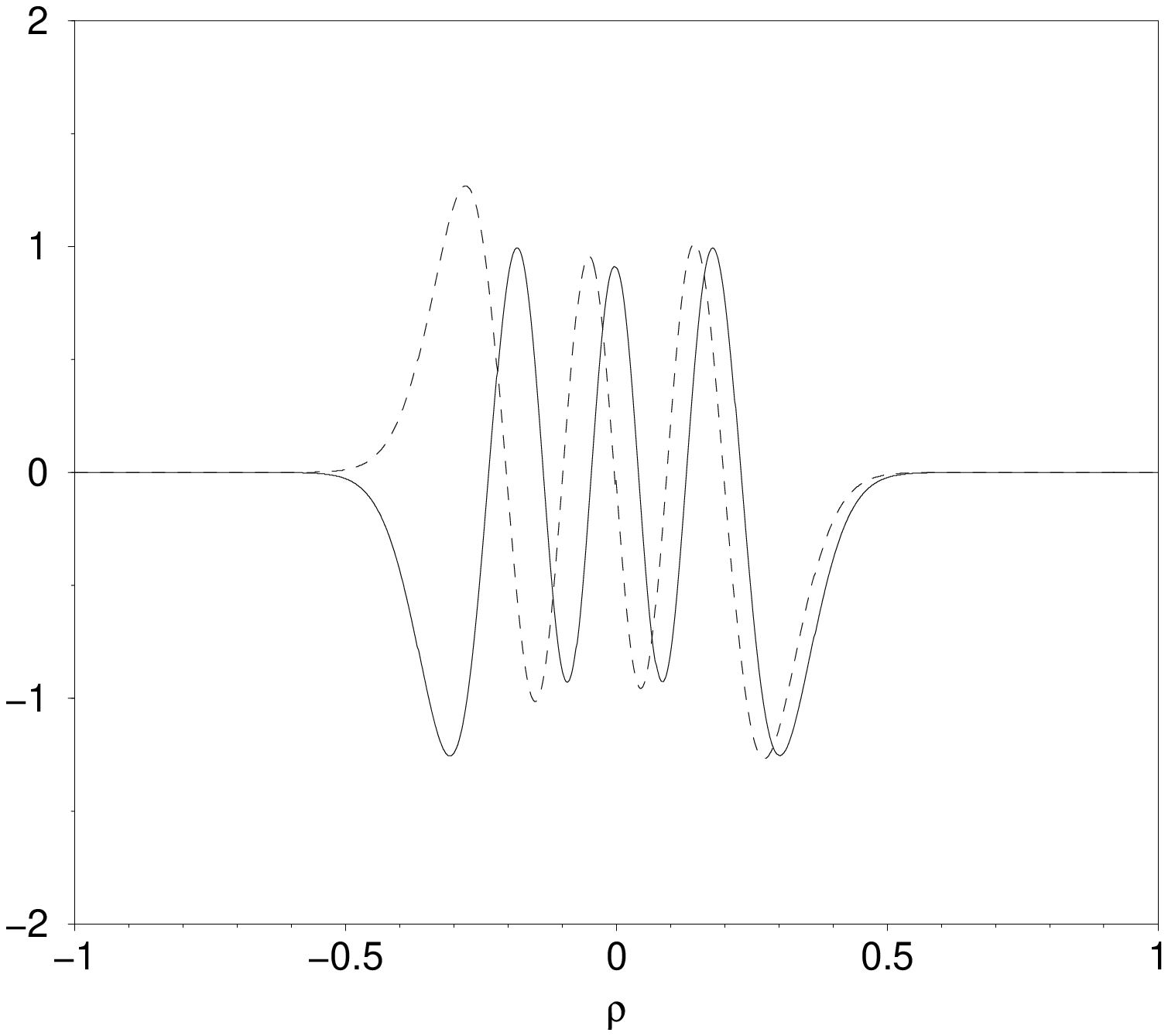,width=7cm}
\epsfig{file=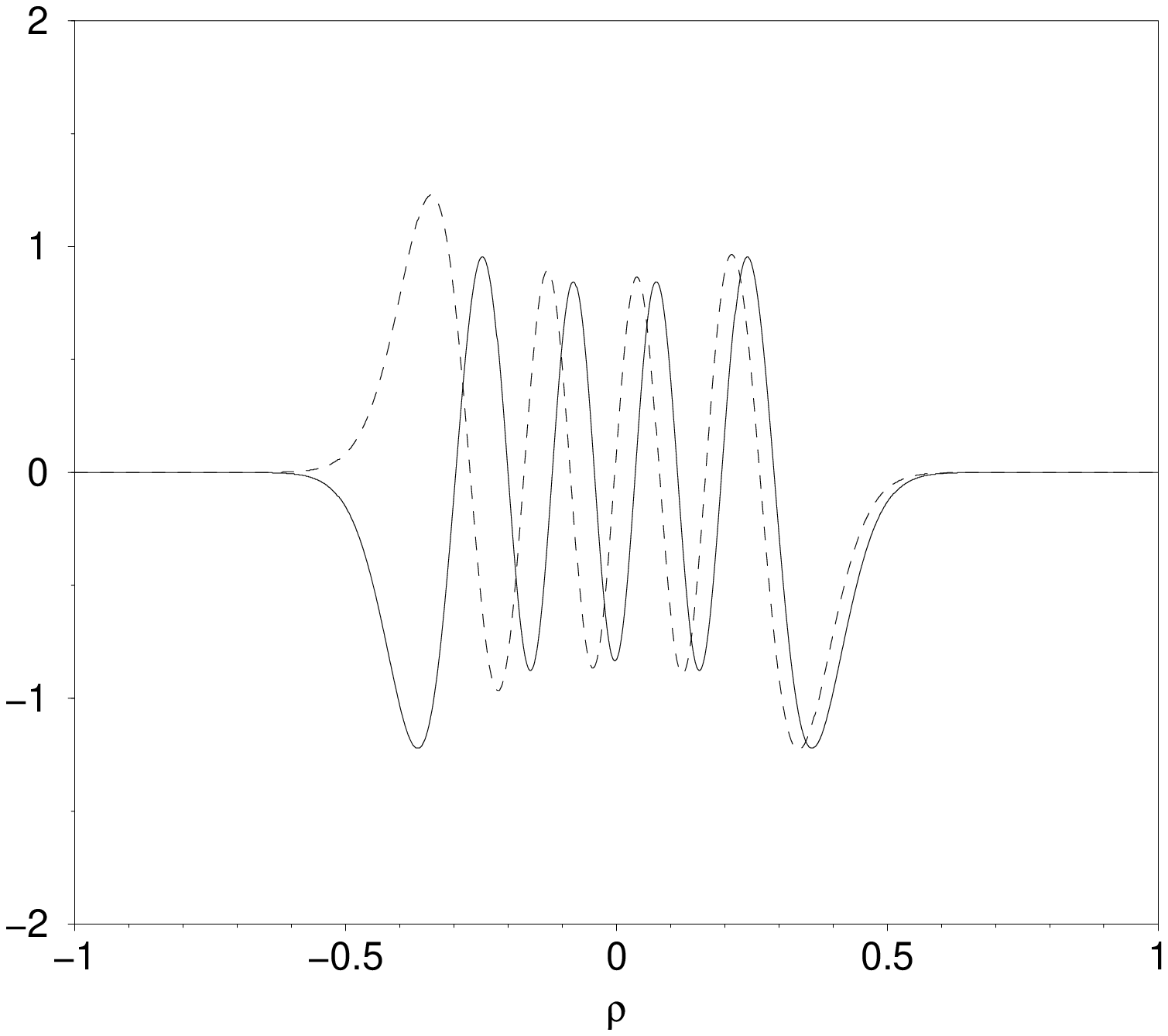,width=7cm}
\epsfig{file=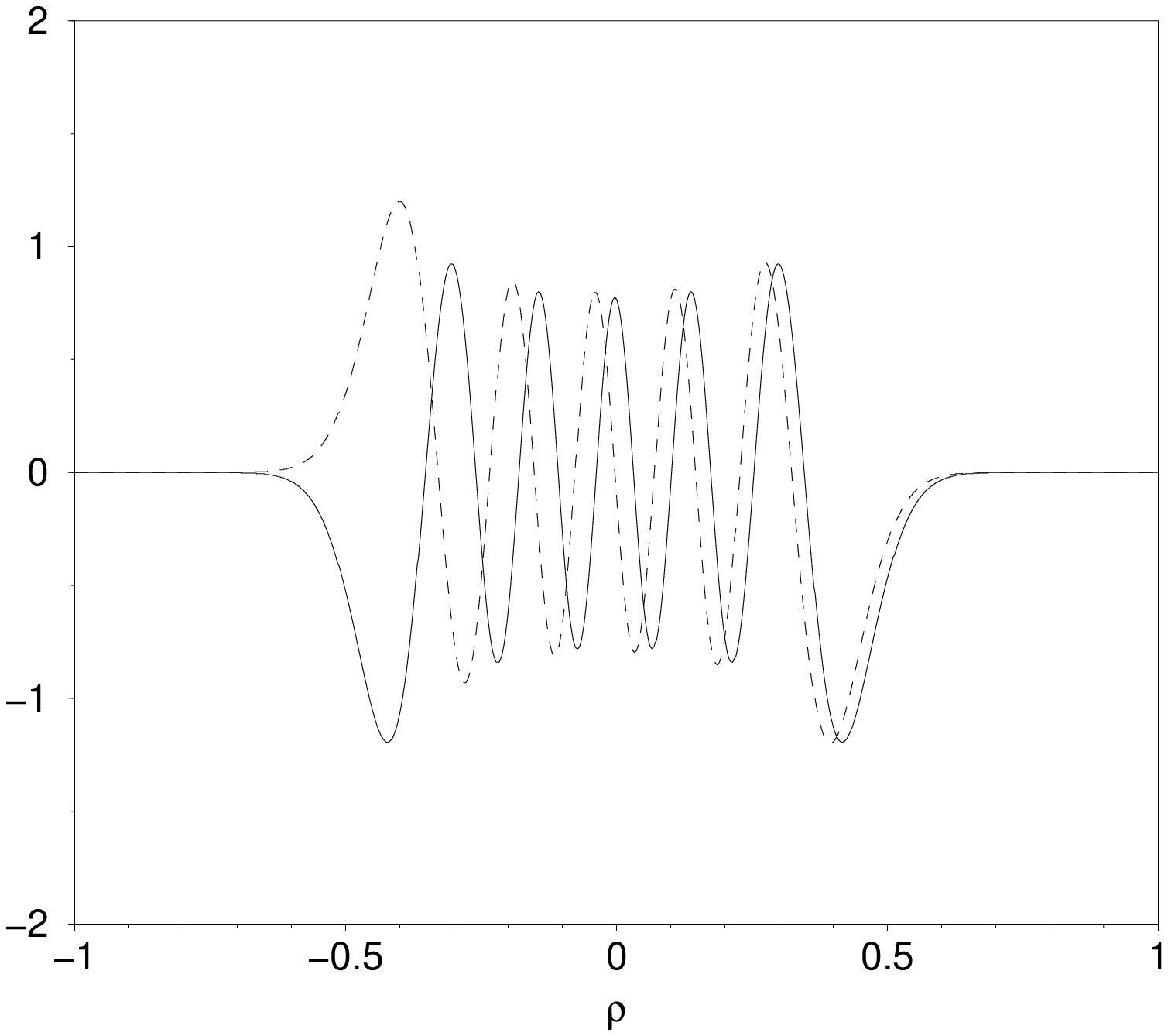,width=7cm}
\epsfig{file=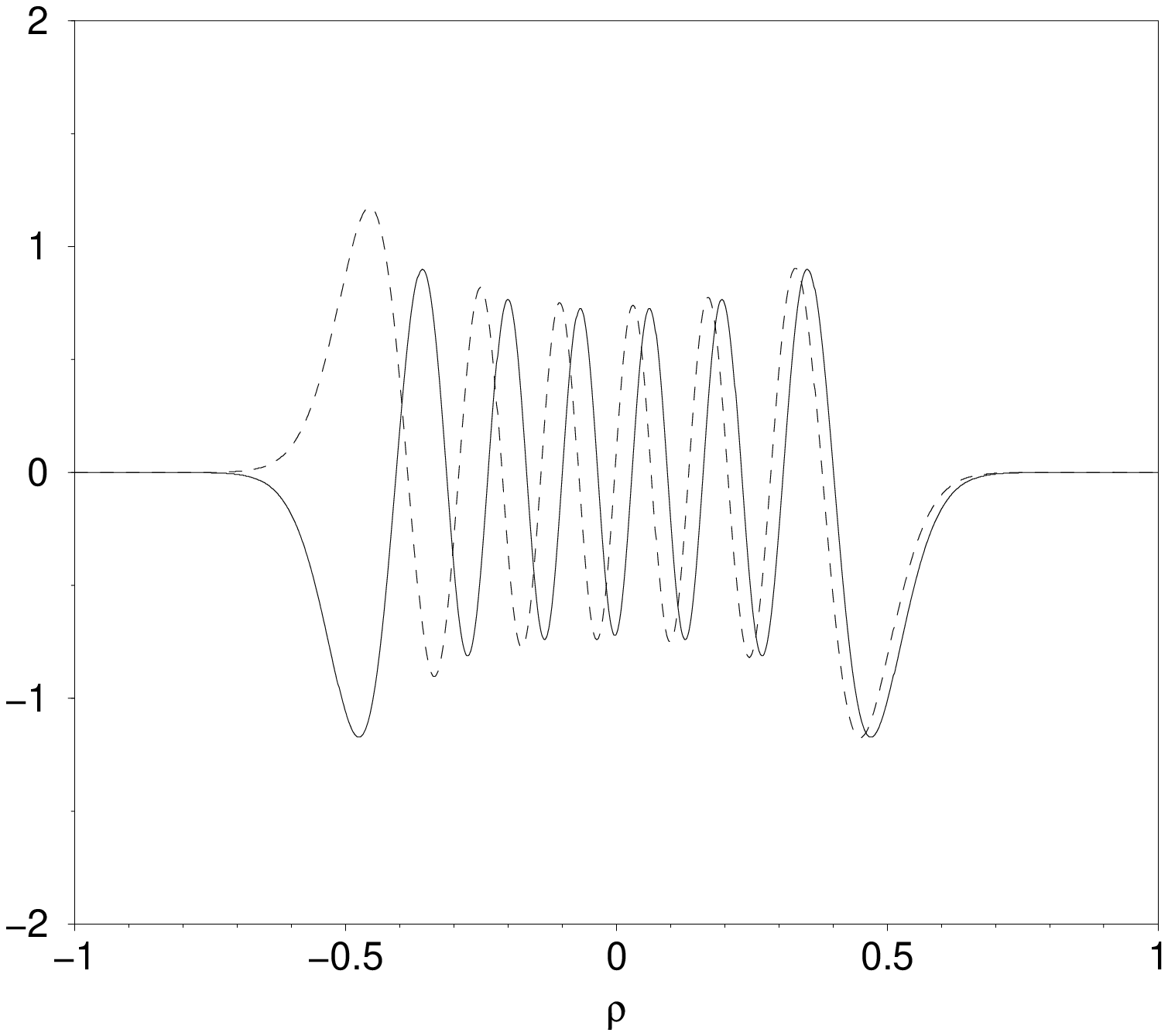,width=7cm}
\caption[Probabilit\'e de pr\'esence, dans la cinqui\`eme dimension, des
fermions les plus l\'egers pi\'eg\'es sur la brane.]{The right (dashed
curve) and left bulk spinor components, $\Ut_{\rm R}$ and $\Ut_{\rm
L}$, as functions of the dimensionless distance $\varrho$ from the
brane for the lightest massive bound states. They have been computed
for $\Mt=100$ and in the Higgs and warp factor profiles obtained with
parameter of Fig.~\ref{figveff}. The numerical values of the
corresponding reduced masses are reported on table~\ref{table}.}
\label{figbulkcomp}
\end{center}
\end{figure}

\section{Discussion and conclusions}\label{sec_concl}

We shall now discuss the cosmological constraints existing on the kind
of model we have been considering here. Most of these constraints come
from brane models in which the wall structure is replaced by an
infinitely thin four dimensional layer. As discussed in the previous
sections, such an approximation is equivalent, within our framework,
to asking that the combination $\alpha\beta$ be much larger than
unity. In this limit, equivalent to the large $\beta$ limit since,
from Eq.~(\ref{relab}), $\alpha\beta\sim {4\over 3} \beta^{1/2}$, we can
replace the stress-energy tensor~(\ref{tmunuBF}) by the effective four
dimensional surface distribution
\begin{equation} T^{^{\rm eff}}_{\mu\nu} = T_\infty g_{\mu\nu} \delta
(y),\end{equation} whose isotropic tension $T_\infty$ is obtained by
integration in the transverse direction to yield
\begin{equation} T_\infty = \sqrt{|\Lambda|} \eta^2 \int \dd\varrho\,
\de^{-6\sigma(\varrho)} \left[ {1\over 2} \Phi'^2 + 2 V \right] \equiv
\sqrt{|\Lambda|} \eta^2 \xi (\beta) \end{equation}
where the function $\xi(\beta)$ can be expressed as
\begin{equation} \xi = \int \dd\varrho\,
\de^{-6\sigma(\varrho)} \left[ {6S^2-1\over \alpha} + 2 \beta (H^2-1)^2
\right]. \end{equation}

\begin{figure}
\begin{center}
\epsfig{file=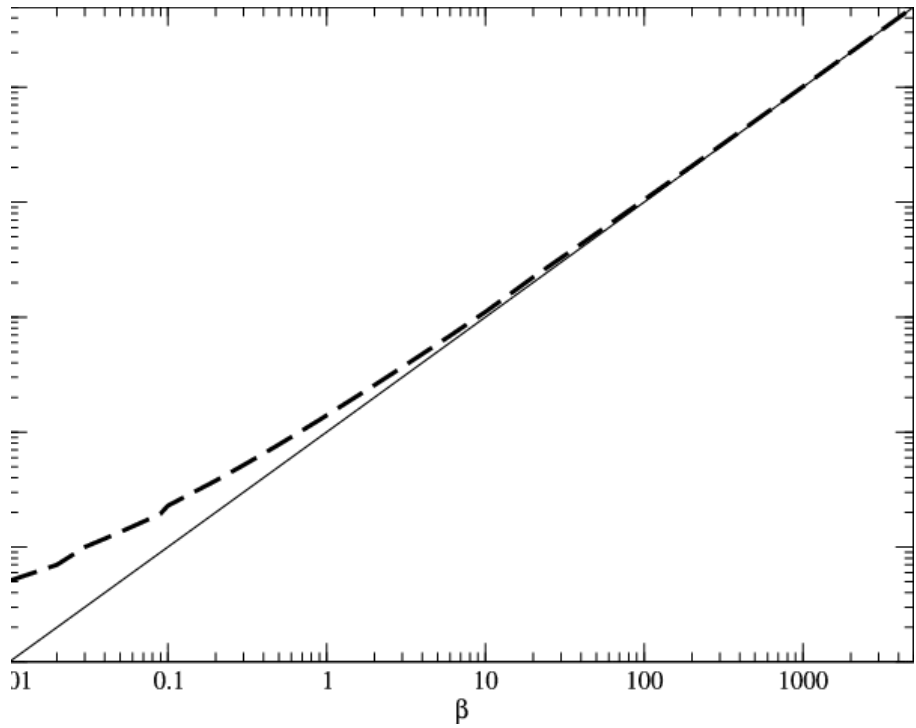,width=12cm}
\caption[Tension effective de la brane en fonction des constantes de
couplage.]{Effective four dimensional brane tension, in units of
$\sqrt{|\Lambda|} \eta^2$, of the domain wall as a function of $\beta$
(dashed line). It is clear that in the thin brane limit,
$\xi\simeq\beta$ (thin line).}
\label{Tinf}
\end{center}
\end{figure}

Fig.~\ref{Tinf} shows the variation of $\xi$ as a function of
$\beta$. In the limit $\beta\gg 1$, it is clear that
$\xi\simeq\beta$, so that the brane tension behaves as
\begin{equation} T_\infty \simeq {\lambda\eta^4\over 8\sqrt{|\Lambda|}},
\label{tension} \end{equation}
which will be used to derive the relevant cosmological
constraints. Note also that the discrepancy between the above formula
and the actual value of $T_\infty$ becomes important (more than 100~\%
error say) for $\beta \lesssim 0.1$, which is already rather far from the
thin brane limit usually considered.

\subsection{Investigation of the parameter space}

The model described in this article depends of five parameters, four
describing the spacetime and scalar field dynamics
$(\Gbu,\Lambda,\eta,\lambda)$ and one concerning the fermions
$(g_{_{\rm F}}$). With the domain wall structure assumed, only four of
these parameters are independent [see Eq.~(\ref{rellambda})].  It is
convenient to replace this set of parameters by the three mass scales
\begin{equation}\label{mass_scale}
m_{_5}\equiv \Gbu^{-1/3},\qquad
m_{_\Lambda}\equiv \frac{\sqrt{-\Lambda}}{6}, \qquad
m_{_\infty}\equiv T_{_\infty}^{1/4},
\end{equation}
and the dimensionless parameter $\gamma_{_{\rm F}}$. These parameters
are subject to a number of constraints, namely:
\paragraph{}The four dimensional gravitational constant must agree with its
observed value $G_{_4}\equiv m_{_4}^{-2}$ with $m_{_4}\sim
10^{19}\,$GeV.  Using the expression of the four dimensional Planck
mass in terms of the five dimensional analog and of the brane tension
gives~\cite{binetruy00, csaki99,cline99,binetruy00b,
kraus99,shiromizu00, flanagan00,maartens02,rubakov01,carter,shiro}
\begin{equation}\label{rel4}
m_{_5}^3 \sim m_{_4}m_{_\infty}^2.
\end{equation}
\paragraph{}The brane cosmological constant~\cite{binetruy00, csaki99,cline99,binetruy00b,
kraus99,shiromizu00, flanagan00,maartens02,rubakov01,carter,shiro}
\begin{equation}
2\Lambda_{_4}=\Lambda+6\pi^4\Gbu^2T_{_\infty}^2,
\end{equation}
must also agree with the standard observational bound
$m_{\Lambda_{4}}<10^{-60}m_{_4}$. This implies
\begin{eqnarray}\label{rel2}
m_{_\Lambda}=\frac{\pi^2}{\sqrt{6}}\frac{m_{_\infty}^4}{m_{_5}^3} &
\Rightarrow & m_\infty^4 \sim m_\Lambda m_5^3.
\end{eqnarray}
Note that in the limit $\beta\gg 1$, this relation is equivalent to
Eq.~(\ref{tension}). This means that this condition is readily
satisfied in the thin brane limit. At this point, it is worth
emphasizing that this is precisely the limit in which the analytic
approximation for fermion masses are the most accurate.
\paragraph{}There must not be any deviation of the law of gravity on the
brane with respect to the inverse square Newton law above 1
millimeter~\cite{limit}. This implies~\cite{chung}
\begin{equation}\label{rel5}
m_{_\Lambda}\gsim 10^{-3}\,\mathrm{eV}.
\end{equation}
\paragraph{}Finally, we require the fermion stress-energy tensor to be
negligible with the brane stress-energy, so that we impose that the
mass of the heaviest fermion is smaller than the brane mass scale. By
means of Eq.~(\ref{defmtmax}), this condition reads
\begin{equation}
m_{\max} \sim m_\Lambda \Mt \de^{-1/(\sqrt{6}H_1)} < m_\infty,
\label{mmax}\end{equation}
where $H_1$ ends up being function of $\beta$ only by means of
Eqs~(\ref{relab}) and (\ref{def_H1}),
\begin{equation}
\label{h1beta}
H_1^2 = 2\beta - \frac{6 \beta}{\sqrt{16 \beta + 9}}.
\end{equation}
In the limit $\beta\gg 1$, Eqs.~(\ref{mmax}) and (\ref{h1beta})
combine to give the constraint on the coupling constant
\begin{equation} \Mt < {m_\infty\over m_\Lambda}
\de^{1/\sqrt{12\beta}} \simeq {m_\infty\over
m_\Lambda}.\label{gammaf}
\end{equation}
It follows from the relations (\ref{rel4}) and (\ref{rel2}) that
$m_\infty^2 \sim m_\Lambda m_4$ so that the three mass scales must
satisfy
\begin{equation}\label{numval}
m_{_\Lambda} \gsim 10^{-3}\,{\rm eV},\qquad
m_{_\infty} \gsim 1 \,{\rm TeV},\qquad
m_{_5} \gsim 10^6\,{\rm TeV},
\end{equation}
which, together with Eq.~(\ref{gammaf}) and (\ref{cond_bounded})
yields
\begin{equation} {1\over 2\sqrt{6}} < \Mt \lesssim
10^{15}\left( {m_\infty\over 1\,\hbox{TeV}}\right)
,\label{cstgammaf}\end{equation} where, as discussed below
Eq.~(\ref{defuh}), the lower bound is very conservative.

Now, let us examine the stability of the fermion confinement on the
brane and the restriction on their life-time imposed by the previous
conditions. We require, at least, one massive bound state to have a
life-time longer than the age of the Universe, i.e.
\begin{equation}\label{t_u}
\tau_1 > \tau_{\mathrm{univ}} \sim 10^{17} \mathrm{s}.
\end{equation}
Using Eq.~(\ref{lifetime}), one roughly gets
\begin{equation}
m_\Lambda\tau_1 \sim \Upsilon(\beta,\Mt)\equiv H_1^{3/2} \Mt^{-17/6}
\exp{\left[2\sqrt{6}\Mt \left(\frac{1}{2} \ln{\frac{\Mt}{H_1}} - 1 -
1/\sqrt{6} - \frac{1}{2\sqrt{6}H_1}\right)\right]}.
\end{equation}
The condition (\ref{t_u}) together with the former constraint
(\ref{numval}) can be written as
\begin{equation}
\Upsilon(\beta,\Mt)\gtrsim 10^{30}.
\end{equation} 
\begin{figure}
\begin{center}
\epsfig{file=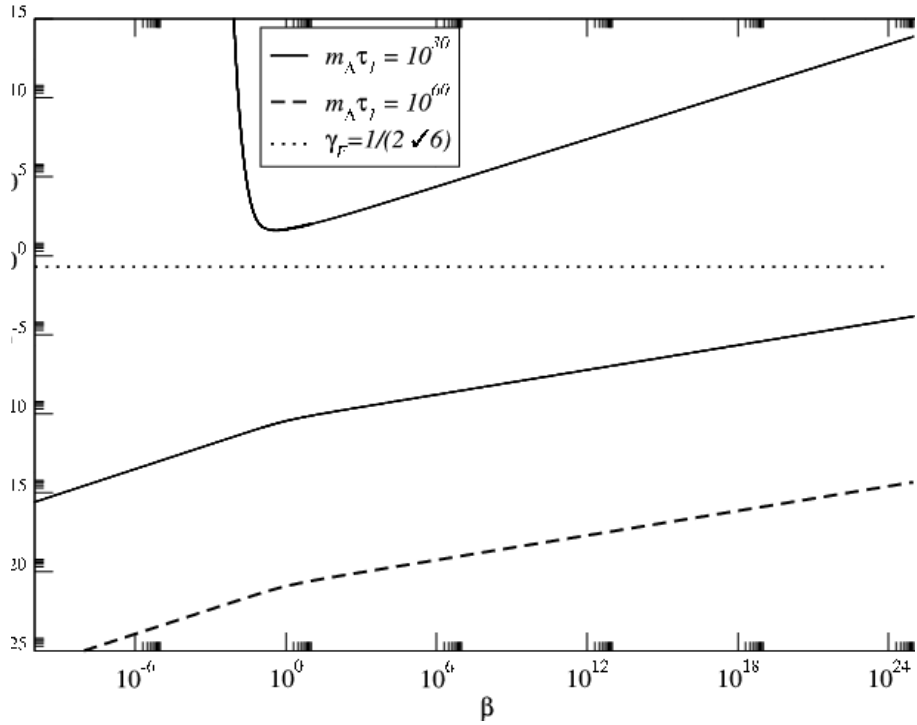,width=12cm}
\caption[Domaines de stabilit\'e des fermions pi\'eg\'es sur la
brane.]{Contour plot of the function $\Upsilon (\beta,\Mt)$ for
$\Upsilon =10^{30}$ [solid lines] and $10^{60}$ [dashed line], i.e.
respectively a particle life-time of the order of the age of the
Universe, and of the proton life-time lower limit. Note that both top
curves are indistinguishable due to the exponential behavior of
$\Upsilon$ in $\Mt$. They turn out to be equivalent to the analytical
requirement given by Eq.~(\ref{frontier}). Points above the highest
curve and below the lowest curves satisfy the contraint. We also
superimpose the conservative constraint (\ref{cond_bounded}) necessary
for the existence of a massless bound state, the allowed region being
above the dotted line.}
\label{figlifetime}
\end{center}
\end{figure}

On Fig.~\ref{figlifetime}, we present the contour plot of the
dimensionless function $\Upsilon(\beta,\Mt)$, for $\Upsilon =10^{30}$
and $10^{60}$, which correspond respectively to a particle life-time
of the order of the age of the Universe, and of the proton life-time
lower limit. For $\beta \gtrsim 1$, there are in principle two allowed
regions, corresponding to strong and weak coupling limits,
i.e. $\Mt\gg 1$ and $\Mt\ll 1$. However, the lower bound on $\Mt$,
which comes from the requirement that fermions are actually trapped on
the brane, pushes the weak coupling allowed region to very high values
of $\beta$, in practice $\beta \gtrsim 7.5\times 10^{57}$ for $\Upsilon
=10^{30}$. Note also that this already rather extreme value is based
on the conservative estimate given by Eq.~(\ref{cond_bounded}).

For $\beta\lesssim 1$, the weak coupling region completely disappears,
while the strong coupling allowed region shrinks rapidely: for
$\beta\lesssim 8.5\times 10^{-3}$, the life-time cannot exceed the age of
the Universe because $\Mt\gtrsim 10^{15}$. Considering $\beta\gtrsim 1$
therefore turns out to be the relevant limit if one wishes to have
fermionic bound states living on the brane.

\subsection{Conclusion}

In this article, we have considered fermions coupled to a Higgs field
with a domain wall structure in a five dimensional anti-de~Sitter
spacetime. This domain wall can be thought of as a realization of a
brane universe.

After, studying the domain wall configuration, we solved the Dirac
equation and showed that there exists massive fermionic bound states
trapped on the wall. We develop both analytic approximation to compute
the mass spectrum and the tunneling time. This was compared to a full
numerical integration of the dynamical equations that revealed the
accuracy of our approximation scheme.

We recover the fact that massive fermions tunnel to the
bulk~\cite{dubovski}. Investigation in the parameter space shows that,
for models satisfying the cosmological constraints, the relevant
confinement life-time can be much greater that either the age of the
Universe or the proton life-time. This was made possible by the
derivation of the analytic estimate.

One of our central result is the derivation of an analytic mass
spectrum for fermions trapped on a brane-like four dimensional
spacetime. In particular, as could have been
anticipated~\cite{neronov}, it was shown that the allowed masses are
quantized, with a spectrum varrying, in the strong coupling limit, as
$\sqrt{n}$. Such a spectrum is indeed in contradiction with
experimental measurements of particle masses~\cite{groom00}, which is
not surprising given the simplicity of the model. It however opens the
possibility to build more realistic theories in which mass
quantization would stem naturally from extra dimensions.

\section*{Acknowledgments}
We thank J\'er\^ome Martin for numerous enlightening discussion, and in
particular for pointing to us Ref.~\cite{miller} on parabolic cylinder
functions. JPU thanks l'Institut d'Astrophysique de Paris for
hospitality while this work was carried out.

\SpecialSection{Conclusion}

Les principaux r\'esultats de cette th\`ese concernent le confinement
de fermions sur des objets \'etendus de type cordes cosmiques et
branes en dimensions suppl\'ementaires.

Du point de vue de la physique des particules, il y est montr\'e qu'un
couplage de Yukawa entre des champs fermioniques et un champ de Higgs
inhomog\`ene, de type d\'efaut topologique, conduisait, de mani\`ere
g\'en\'erique, \`a l'existence d'\'etats li\'es dont le spectre de
masse a \'et\'e \'etudi\'e en d\'etail. La prise en compte de l'effet
des courants g\'en\'er\'es par le remplissage de ces \'etats sur le
spectre lui-m\^eme semble montrer que l'\'etat fondamental ne peut pas
\^etre de masse nulle, invalidant de ce fait l'existence physique de
modes z\'eros dans ces mod\`eles. Ce r\'esultat remet en cause les
\'etudes pr\'ec\'edentes sur les courants de fermions dans les cordes,
tant du point de vue microscopique que cosmologique.

D'un point de vue cosmologique, ces r\'esultats nous ont permis, dans
un premier temps, d'\'etudier la dynamique des cordes cosmiques
parcourues par des courants de fermions, et d'en extraire une
\'equation d'\'etat. Contrairement aux cas connus des cordes
poss\'edant des courants de bosons, cette \'equation d'\'etat met en
jeu plusieurs param\`etres internes qui s'identifient aux densit\'es
des divers fermions pr\'esents dans la corde. Ce r\'esultat fait qu'il
n'est pas toujours possible d'appliquer le formalisme covariant \`a un
param\`etre pour les courants de fermions. Ces courants se distinguent
\'egalement de ceux de bosons, dans la dynamique des cordes, par le
fait qu'ils privil\'egient \`a la fois des r\'egimes subsoniques,
lorsque les courants sont ultrarelativistes, et des r\'egimes
supersoniques, lorsque les modes massifs sont pr\'edominants. Les
vortons associ\'es \`a ces deux types de dynamique sont alors
classiquement respectivement stables et instables. Leur persistance
au cours de l'\'evolution de l'univers pouvant conduire \`a une
catastrophe cosmologique, les cons\'equences des transitions entre les
r\'egimes de stabilit\'e et d'instabilit\'e de ces vortons pourraient
\^etre importantes.

En ce qui concerne la cosmologie en dimensions suppl\'ementaires, le
m\'ecanisme de confinement \'etudi\'e sur les cordes a permis de
proposer une solution simple au probl\`eme de la localisation des
fermions massifs sur une brane repr\'esentant notre univers. Il est
ainsi possible de donner une explication \`a la quantification de la
masse, en g\'en\'eral, par l'existence d'un spectre de masse
naturellement discret des \'etats li\'es fermioniques. La prise en
compte des effets gravitationnels a cependant montr\'e que ces modes
pouvaient s'\'echapper dans la dimension suppl\'ementaire, et, en
imposant que la dur\'ee de vie de ces \'etats soit plus grande que
l'\^age de l'univers, des contraintes num\'eriques sur les
param\`etres du mod\`ele ont pu \^etre trouv\'ees.

Le prolongement naturel de ces travaux porte \`a la fois sur l'aspect
cosmologique et microscopique. Concernant le premier, la r\'ealisation
de simulations num\'eriques d'\'evolution de cordes conductrices
permettrait de fixer pr\'ecis\'ement les contraintes cosmologiques
qu'elles doivent v\'erifier, et, par le biais de l'\'equation d'\'etat,
de contraindre la physique des particules sous-jacente. Par analogie
avec des cordes sans structure interne, l'\'evolution de ces r\'eseaux
doit \^etre \'etroitement li\'ee \`a la formation et \`a la
persistance des vortons qu'ils produisent. Bien que classiquement
stables pour les r\'egimes subsoniques, des effets quantiques
pourraient toutefois les d\'estabiliser. Il serait donc int\'eressant de
pouvoir g\'en\'eraliser l'approche introduite dans cette th\`ese aux
cas de cordes courb\'ees, et d'en \'etudier les cons\'equences sur les
\'etats li\'es fermioniques.

En prolongement de l'\'etude du confinement des fermions massifs sur
un mur de domaine quadri-dimensionnel, leur localisation sur une corde
quadri-dimensionnelle plong\'ee dans un espace-temps \`a six dimensions
permettrait d'introduire leur charge. En effet, la corde \'etant
g\'en\'er\'ee par la brisure d'une sym\'etrie locale, elle met en jeu
un champ de jauge auquel les fermions sont coupl\'es. Le spectre de
masse des fermions charg\'es pourrait ainsi \^etre compar\'e aux
observations afin d'\'etudier la viabilit\'e de ces mod\`eles. De
plus, comme pour les cordes, l'influence de la courbure de la brane
sur le spectre de masse reste \`a \'etudier.

\SpecialSection{Annexes}

\Annex{Parall\'elisation num\'erique}
\label{annexeomp}
\minitoc
\section{Introduction}

Dans le chapitre~\ref{chapitreevol}, nous avons pr\'esent\'e des
r\'esultats issus de simulations num\'eriques faisant \'evoluer un
r\'eseau de cordes cosmiques de Goto-Nambu dans un univers de FLRW,
\`a partir de diverses configurations initiales. La physique
r\'egissant l'\'evolution d'un tel r\'eseau ayant \'et\'e d\'ecrite
dans les chapitres~\ref{chapitreform} et~\ref{chapitreevol}, cette
annexe est consacr\'ee aux aspects purement num\'eriques du calcul. La
r\'esolution des \'equations du mouvements (\ref{mvtflrwsys}), ainsi
que la d\'etection et la r\'ealisation des intercommutations entre les
cordes, sont les deux t\^aches essentielles qui guident le calcul
num\'erique. Cependant, celles-ci doivent \^etre en permanence
accompagn\'ees de multiples v\'erifications et corrections afin
d'\'eviter la croissance d'erreurs purement num\'eriques, dues \`a la
pr\'ecision finie des machines. Dans le code utilis\'e ici, ce travail
a \'et\'e effectu\'e par F.~Bouchet et D.~Bennett dans les ann\'ees
1980~\cite{bennett88,bennett90}, \`a l'aide du langage de
programmation \texttt{Fortran77}. Cependant, l'utilisation actuelle
de ce code, pour le calcul du spectre de fluctuation du CMBR induit par
les cordes, requiert une grande dynamique temporelle de
simulation. Or, la taille finie du volume dans lequel le r\'eseau
\'evolue fixe \'egalement la dur\'ee de la simulation, l'horizon
devant y \^etre toujours confin\'e pour \'eviter des effets de bord
parasites (voir Chap.~\ref{chapitreevol}). Typiquement, les volumes,
et donc les dur\'ees de simulation maximales de l'\'epoque, \'etaient
de l'ordre de $(26 \ell_\uc)^3$, permettant d'\'etudier l'\'evolution
cosmologique du r\'eseau pour un accroissement du facteur d'\'echelle
d'un facteur quinze dans l'\`ere de mati\`ere, contre un facteur six
dans l'\`ere de radiation. Le nombre initial de points, par unit\'e de
longueur, utilis\'e pour d\'ecrire chaque corde est \'egalement un
param\`etre significatif. Il fixe la taille physique admissible des
plus petites boucles qui ne peuvent \^etre repr\'esent\'ees que par
trois points minimum. Ce \emph{cutoff} est \'egalement exprim\'e en
fonction de la longueur de corr\'elation de la transition de phase
$\ell_\uc$ (voir Chap.~\ref{chapitreevol}), et sa valeur typique est
d'environ $0.3
\ell_\uc$, soit un nombre initial de points par longueur de
corr\'elation de $10/\ell_\uc$. Si l'on souhaite augmenter la
r\'esolution des simulations sur la taille des petites boucles, ce
param\`etre est \'egalement \`a augmenter.

Hormis l'accroissement des performances des machines actuelles, la
mise en place de simulations num\'eriques de hautes r\'esolutions peut
\^etre op\'er\'ee par le biais de m\'ethodes de
parall\'elisation. Celles-ci consistent essentiellement \`a
reprogrammer le code afin que ses calculs, initialement op\'er\'es
successivement, soient partag\'es entre diff\'erents processeurs, ou
machines, dans le but d'augmenter la vitesse effective
d'ex\'ecution. Il n'est cependant pas possible de
\emph{parall\'eliser} tout un code, certaines parties n\'ecessitant
toujours une ex\'ecution s\'equentielle. Les sections suivantes
illustrent les am\'eliorations effectu\'ees, ainsi que les
difficult\'es rencontr\'ees, lorsque ces m\'ethodes ont \'et\'e
d\'evelopp\'ees sur le code de F.~Bouchet, et dont les r\'esultats
provisoires ont \'et\'e pr\'esent\'es dans le
chapitre~\ref{chapitreevol}.

\section{Directives de compilation \texttt{OpenMP}}

Le langage qui a \'et\'e utilis\'e pour parall\'eliser le code
correspond au standard \texttt{OpenMP}\footnote{Open
Multi-Processing.} apparu en 1997~\cite{omp}, et remis \`a jour en
novembre 2000~\cite{omp}. Bien que la parall\'elisation multit\^aches
existe depuis quelques ann\'ees, chaque constructeur poss\`ede son
propre jeu de directives, et le standard \texttt{OpenMP} est
actuellement le seul langage unifi\'e et r\'epandu. Sa jeunesse est
cependant \'egalement responsable de son instabilit\'e, et son
fonctionnement correct \`a requis certaines reprogrammations du code
\`a l'aide du langage \texttt{Fortran90}.

\subsection{Parall\'elisation des autocommutations}

L'identification des parties du programme pouvant \^etre
parall\'elis\'ees est souvent directement donn\'ee par la physique
qu'elles mod\'elisent. Le crit\`ere d\'eterminant est que les sections
parall\`eles doivent \^etre ind\'ependantes, c'est-\`a-dire que le
r\'esultat donn\'e par une section parall\`ele doit \^etre
ind\'ependant du r\'esultat donn\'e par l'autre section. Lorsque ce
n'est pas le cas, la structure et l'ex\'ecution s\'equentielle du
programme doivent \^etre pr\'eserv\'ees pour conserver la validit\'e
des r\'esultats.

Dans le code d'\'evolution de cordes cosmiques, l'\'etape demandant le
plus de temps de calcul est la d\'etection et la r\'ealisation des
autocommutations. Comme on peut le voir sur la
figure~\ref{figbench1p}, le sous programme nomm\'e
``\texttt{iniselfc}'' occupe pr\`es de $80 \%$ du temps
d'ex\'ecution. Sa fonction dans le code est d'initialiser des
variables relatives aux \emph{autocommutations} de chaque boucle de
corde en construisant une liste de points, sur la boucle en question,
susceptibles de se croiser. Il est appel\'e pour chaque point de
chaque boucle de corde et ne concerne que les points de la dite
boucle. Il est clair qu'une telle op\'eration peut \^etre ex\'ecut\'ee
en parall\`ele car la d\'etection d'un croisement est propre \`a
chaque boucle. En fait, dans le but d'augmenter au maximum
``l'\'epaisseur'' des sections parall\`eles, ceci sugg\`ere de
s\'eparer le calcul de l'\'evolution de chaque corde sur les
diff\'erents processeurs accessibles. Cette op\'eration est cependant
limit\'ee par le fait qu'il existe des intercommutations entre les
diff\'erentes cordes. Dans ce cas, il faut revenir \`a une ex\'ecution
en s\'erie.

\begin{figure}
\begin{center}
\epsfig{file=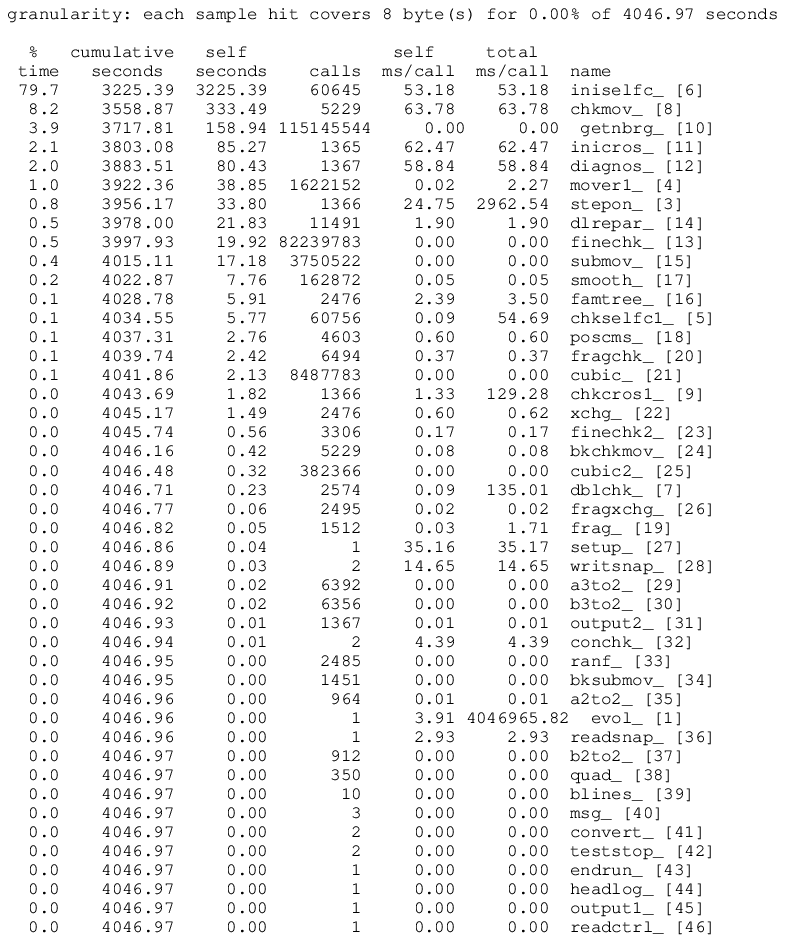,width=14cm}\
\caption[Chronom\'etrages du code d'\'evolution de cordes
cosmiques.]{Temps de calcul pass\'e dans les plus lents sous
programmes du code d'\'evolution de cordes cosmiques lorsqu'il est
ex\'ecut\'e en s\'erie. La d\'etection des autocommutation occupe
pr\`es de $80 \%$ du temps total.}
\label{figbench1p}
\end{center}
\end{figure}

La s\'eparation des t\^aches entre les divers processeurs peut
simplement \^etre mise en place \`a l'aide de directives de
compilation, i.e. de commandes ins\'er\'ees dans le code donnant les
instructions n\'ecessaires au compilateur. Dans le langage
\texttt{OpenMP}, la parall\'elisation sur chaque corde s'est
effectu\'ee par la directive \texttt{!{\$}OMP PARALLEL DO} appliqu\'ee
\`a une boucle de calcul \texttt{DO} s\'eparant le traitement des
diff\'erentes cordes (voir Fig.~\ref{figompdo}).
\begin{figure}
\begin{center}
\epsfig{file=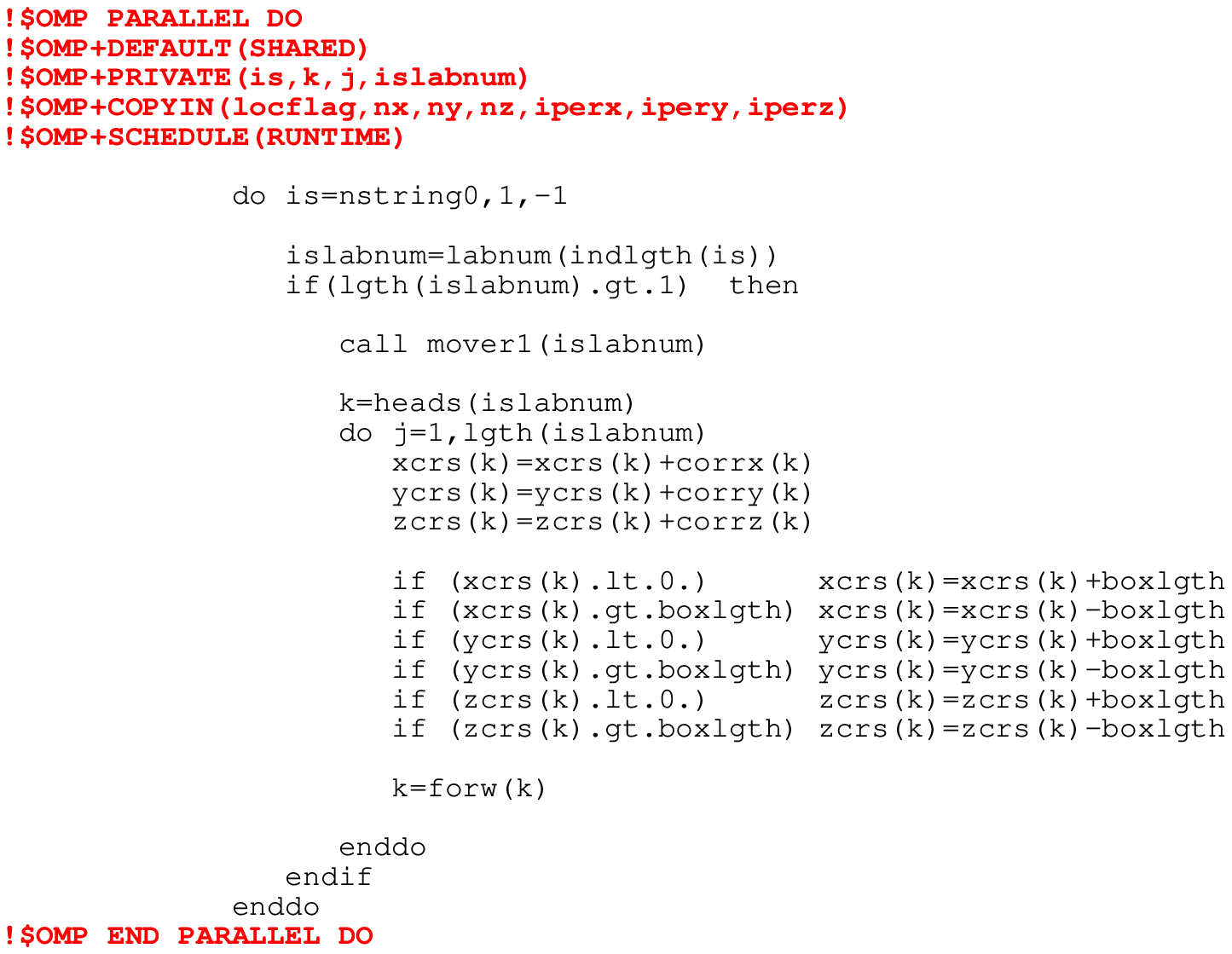,width=14cm}
\caption[La parall\'elisation d'une boucle de calcul.]{Directive de
compilation \texttt{!{\$}OMP PARALLEL DO} permettant l'ex\'ecution en
parall\`ele de la boucle de calcul, selon son indice, ici
\texttt{is}. L'ensemble des jeux d'instruction int\'erieurs \`a la
boucle de calcul \texttt{DO} est ex\'ecut\'e par un processeur
diff\'erent pour chaque valeur de \texttt{is}, modulo le nombre total
de processeurs.}
\label{figompdo}
\end{center}
\end{figure}
L'effet de cette directive est de faire calculer chaque bloc de
commandes, situ\'e entre \texttt{!{\$}OMP PARALLEL DO} et
\texttt{!{\$}OMP END PARALLEL DO}, par chaque processeur accessible,
le num\'ero du processeur \'etant reli\'e \`a l'indice \texttt{is},
modulo le nombre total de processeurs. Il est de plus possible
d'am\'eliorer la dynamique de r\'epartition des t\^aches par une
directive nomm\'ee \texttt{!{\$}OMP SCHEDULE} (voir
Fig.~\ref{figompdo}). Celle-ci sp\'ecifie dans quel ordre les blocs de
calcul index\'es par \texttt{is} sont r\'epartis sur les
processeurs. Le choix le plus efficace est obtenu par l'option
\texttt{DYNAMIC} qui affecte une t\^ache \texttt{is} quelconque au
premier processeur libre ayant termin\'e le calcul d'un bloc
ant\'erieur. On \'evite de cette mani\`ere que des blocs de calcul
plus long que les autres ne ralentissent l'ex\'ecution en
parall\`ele.

Les autres directives de compilation apparaissant dans tout le reste
du code (voir Fig.~\ref{figcode}) g\`erent le partage de la m\'emoire
entre les diff\'erentes r\'egions parall\`eles. 
\begin{figure}
\begin{center}
\epsfig{file=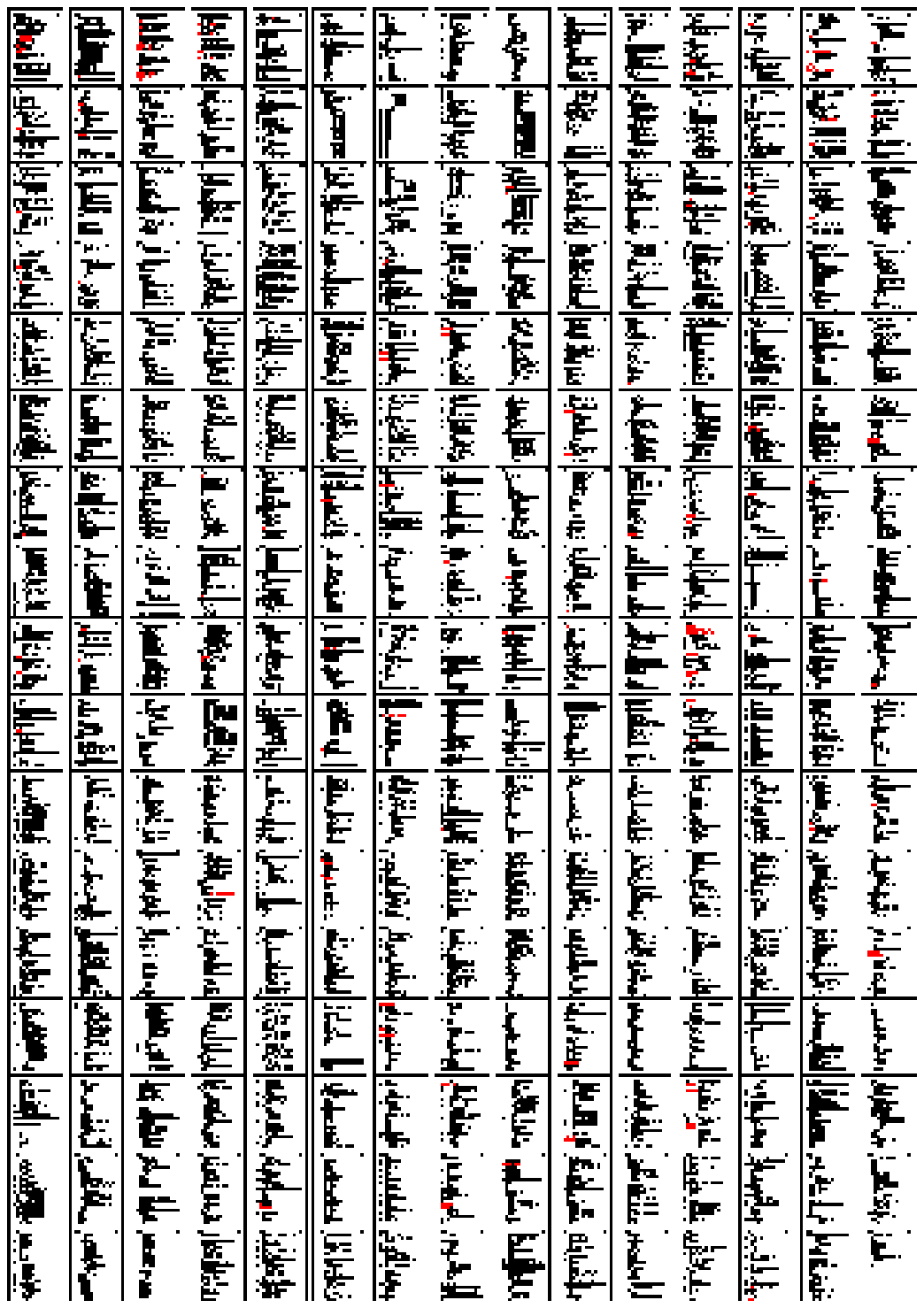,width=14cm}
\caption[Vue d'ensemble du code d'\'evolution de cordes cosmiques
apr\`es paral\-l\'eli\-sa\-tion.]{Le code d'\'evolution de cordes cosmiques
de F.~Bouchet et D.~Bennett~\cite{bennett88,bennett90}. Environ deux
cents directives de compilation \texttt{OpenMP} ont \'et\'e
introduites pour sa parall\'elisation et sont repr\'esent\'ees en
rouge.}
\label{figcode}
\end{center}
\end{figure}

\subsection{Les variables partag\'ees}

La principale difficult\'e lorsqu'une partie du code est
parall\'elis\'ee est de d\'efinir exactement quelles sont les
variables qui vont \^etre communes, ou non, \`a chaque section
parall\`ele, i.e. \`a chaque corde dans le cas qui nous concerne. Par
exemple, la plus \'evidente des variables priv\'ees est l'indice
\texttt{is}: sa valeur est n\'ecessairement diff\'erente pour chaque
corde, par contre, le nombre total de cordes \texttt{nstring0} est une
variable partag\'ee (voir Fig.~\ref{figompdo}). Les sp\'ecifications
de partage des variables sont ex\'ecut\'ees par les clauses
\texttt{SHARED} ou \texttt{PRIVATE} sur la figure~\ref{figompdo}, pour
d\'eclarer des variables communes ou priv\'ees, respectivement.

Dans le code utilis\'e ici, pr\`es de deux cents directives de ce type
ont \'et\'e introduites afin de correctement d\'eclarer le type de
chaque variable (voir fig.~\ref{figcode}). La deuxi\`eme difficult\'e
appara\^\i t lorsqu'une section parall\`ele modifie une variable
partag\'ee entre tous les processeurs. C'est par exemple le cas de
\texttt{nstring0}: lors des autocommutations de chaque corde, de
nombreuses petites boucles sont form\'ees (voir
Chap.~\ref{chapitreevol}) et dans toutes les sections parall\`eles il
y a une instruction du type
\begin{equation}
\label{sharedupdate}
\mathtt{nstring0}=\mathtt{nstring0}+1.
\end{equation}
Pour que la valeur finale soit correcte, il ne faut absolument pas que
deux processeurs acc\`edent en m\^eme temps \`a
\texttt{nstring0}. Ceci peut \^etre sp\'ecifi\'e par les commandes
\texttt{!{\$}OMP CRITICAL} et \texttt{!{\$}OMP END CRITICAL} qui
assurent que le bloc d'instruction compris entre ces deux directives
n'est jamais effectu\'e simultan\'ement par deux sections
parall\`eles. L'\'ecriture correcte de l'instruction
(\ref{sharedupdate}) devient finalement\footnote{Lorsqu'une seule
instruction n\'ecessite cette pr\'ecaution, on utilise \'egalement la
directive \texttt{!{\$}OMP ATOMIC}.}
\begin{equation}
\begin{array}{lll}
\mathtt{!{\$}OMP \ CRITICAL} & & \\
&\mathtt{nstring0}=\mathtt{nstring0}+1 & \\
\mathtt{!{\$}OMP \ END \ CRITICAL}& &
\end{array}
\end{equation}
Enfin, une fois les variables correctement d\'eclar\'ees et
calcul\'ees dans les sections parall\`eles, il faut s'assurer qu'elles
sont correctement reli\'ees \`a leur analogue dans les parties en
s\'erie du programme, et inversement. Pour cela, on utilise la clause
\texttt{!{\$}OMP COPYIN}}. Sur la figure~\ref{figompdo}, elle permet de
copier la valeur des variables utilis\'ees dans la section en s\'erie
vers les variables priv\'ees des sections parall\`eles.

Le test suffisant, mais non n\'ecessaire, permettant de v\'erifier si
ces diverses \'etapes de parall\'elisation ont \'et\'e correctement
programm\'ees, est ensuite de comparer les r\'esultats entre le
programme initial et le programme parall\'elis\'e\footnote{La
parall\'elisation peut \'egalement servir \`a d\'etecter
d'\'eventuelles erreurs de programmation dans le code initial,
celles-ci \'etant souvent la causes d'un \'echec d'ex\'ecution
parall\`ele, sans pour autant l'\^etre en s\'erie\dots}. Les valeurs
des diverses quantit\'es physiques ont ainsi \'et\'e retrouv\'ees, \`a
la pr\'ecision de la machine pr\`es, entre les deux types
d'ex\'ecution.  Sur la figure~\ref{figcharge}, sont repr\'esent\'es
les courbes de vitesse d'ex\'ecution instantan\'ee du programme
parall\'elis\'e lorsqu'il fonctionne sur deux, quatre, six ou huit
processeurs. On voit clairement que l'augmentation du nombre de
processeurs est de moins en moins efficace: le gain de temps entre six
et huit processeurs est relativement minime. Cette limite d\'epend de
plusieurs crit\`eres qui seront \'evoqu\'es dans la section suivante.
\begin{figure}
\begin{center}
\epsfig{file=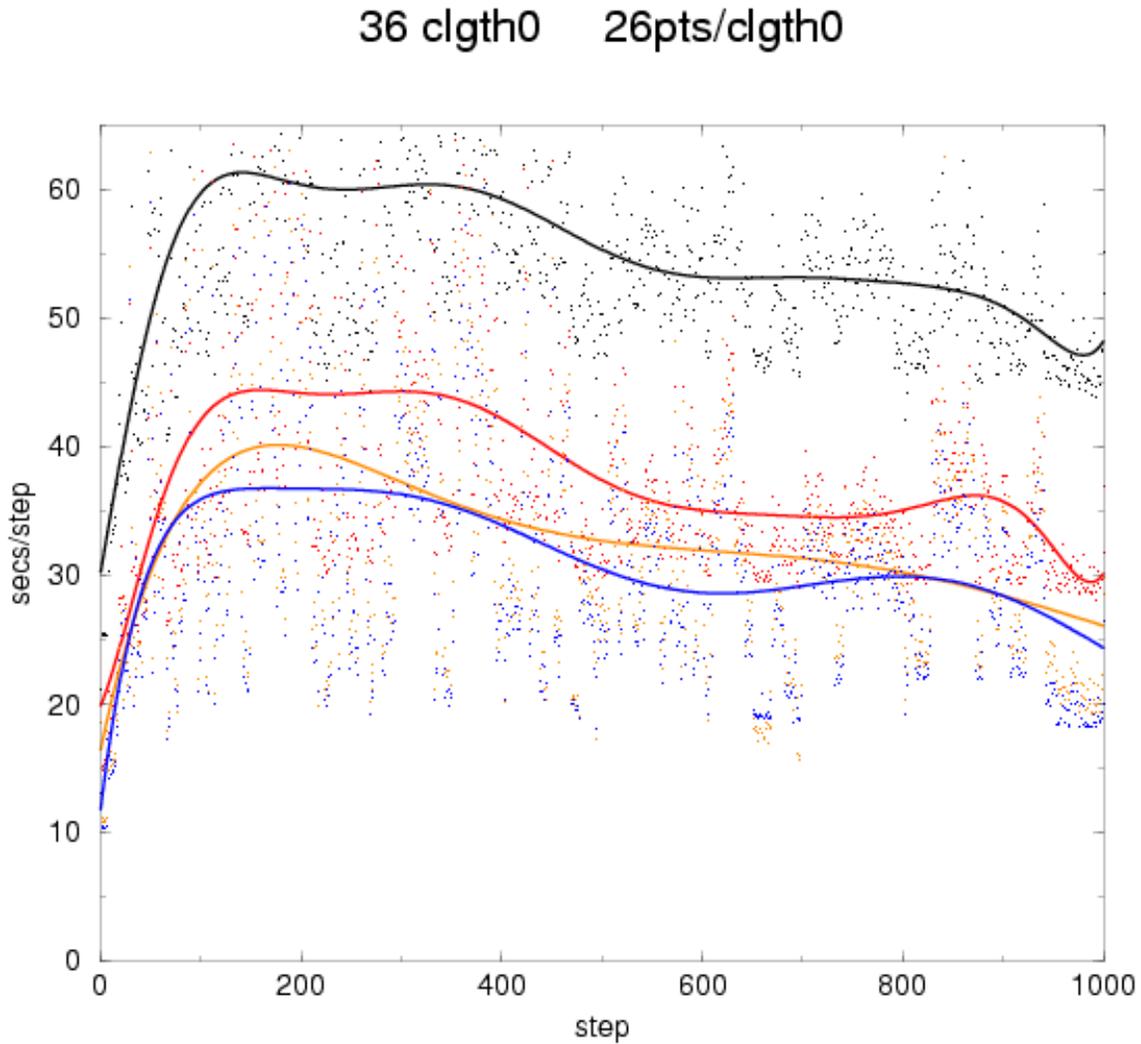,width=14cm,angle=270}
\caption[Temps d'ex\'ecution dynamiques en fonction du nombre de
processeurs.]{Temps d'ex\'ecution, par pas de temps, du code
d'\'evolution de cordes lorsqu'il fonctionne en parall\`ele sur deux
(noir), quatre (rouge), six (orange) ou huit (bleu) processeurs. Les
courbes sont des ajustement des points de mesure par un polyn\^ome du
dixi\`eme degr\'es. L'augmentation du nombre de processeurs n'est plus
efficace au del\`a de six, pour les param\`etres de la simulation
test\'ee, i.e un volume initial de $(36
\ell_\uc)^3$ pour $26/\ell_\uc$ points par longueur de corr\'elation.}
\label{figcharge}
\end{center}
\end{figure}

\section{Limites physiques}

La contrainte la plus naturelle limitant la rapidit\'e du programme
parall\'elis\'e est l'existence de sections en s\'erie pour traiter
les intercommutations. Celles-ci sont en effet ex\'ecut\'ees sur un
seul processeur, et lorsque le temps d'ex\'ecution des sections
parall\`eles devient n\'egligeable, les parties ex\'ecut\'ees
s\'equentiellement fixent la vitesse d'ex\'ecution du programme qui ne
d\'epend plus du nombre de processeurs. En fait, on peut montrer que
ce n'est pas ce probl\`eme qui sature le temps d'ex\'ecution sur la
figure~\ref{figcharge}, ses effets ne deviennent apparents que pour de
tr\`es longues simulations, typiquement $(50 \ell_\uc)^3$. Le facteur
limitant est ici le
\emph{load balancing}.

Celui-ci appara\^\i t dans les sections parall\`eles lorsque un
processeur re\c coit un jeu d'instructions beaucoup plus long que les
autres. Si il n'a pas finit de le traiter \`a la fin de l'ex\'ecution
parall\`ele, les autres processeurs vont l'attendre avant de pouvoir
continuer l'ex\'ecution du programme en s\'erie. Sur la
figure~\ref{figcharge}, ce ph\'enom\`ene correspond aux groupes de
points formant des pics s'\'ecartant sensiblement des courbes
moyennes. Dans les cas extr\^emes, le temps d'ex\'ecution pour les
pics est d\'ej\`a satur\'e pour quatre processeurs, les points rouges
se m\'elangeant aux points noirs sur la
figure~\ref{figcharge}. Physiquement, on montre que ces pics
correspondent aux longues cordes. Celles-ci n\'ecessitent beaucoup
plus de d\'etections de voisins, lors de leur autocommutation, qu'il en
est n\'ecessaire pour des cordes de plus petites tailles. La dynamique
de parall\'elisation, par la directive \texttt{!{\$}OMP SCHEDULE},
attribuant de mani\`ere al\'eatoire les cordes sur les processeurs
disponibles, peut donner une longue corde \`a un processeur \`a la fin
d'une section parall\`ele, et on se retrouve pr\'ecis\'ement dans le
cas probl\'ematique. La solution adopt\'ee consiste finalement \`a
classer les cordes selon leur longueur et \`a les attribuer
successivement, par ordre d\'ecroissant de longueur, aux processeurs
libres.

Un autre facteur limitant l'accroissement du nombre de processeurs
r\'esulte des limites physiques des machines utilis\'ees. Celles-ci
sont de type multiprocesseurs \`a m\'emoire partag\'ee, c'est-\`a-dire
que les diff\'erents processeurs stockent les variables dans des zones
m\'emoires communes. Les variables d\'eclar\'ees priv\'ees sont
copi\'ees autant de fois qu'il y a de processeurs utilis\'es, alors
que les variables communes ne le sont qu'une fois. La taille de
m\'emoire physique occup\'ee par le programme augmente ainsi
proportionnellement au nombre de processeurs utilis\'es. Typiquement,
pour huit processeurs, des simulations de $(48 \ell_\uc)^3$ avec une
densit\'e de point de $26/\ell_\uc$, occupent pr\`es de $6 \,
\mathrm{Go}$ de m\'emoire. De telles tailles ralentissent de mani\`ere
significative les temps d'acc\`es aux variables, et de ce fait
l'ex\'ecution du programme. On peut cependant montrer que cette limite
n'\'evolue plus au del\`a d'une certaine taille m\'emoire, ce
qui signifie que l'efficacit\'e de la parall\'elisation augmente
encore avec le nombre de processeurs pour des simulations de plus
grande r\'esolution.

Finalement, la parall\'elisation du code d'\'evolution de corde a
permis de gagner pr\`es d'un ordre de grandeur en vitesse
d'ex\'ecution par rapport \`a sa version originale. Les r\'esultats
pr\'esent\'es dans le chapitre~\ref{chapitreevol} en sont le fruit, et
il est actuellement envisageable de r\'ealiser des simulations de
volume $(100 \ell_\uc)^3$ pour le calcul du spectre de fluctuation du
CMBR.

\section{Conclusion}

La parall\'elisation du code d'\'evolution de cordes cosmiques permet
de r\'ealiser des simulations num\'eriques de hautes r\'esolutions
n\'ecessaires au calcul du spectre des fluctuations qu'elles
engendrent sur le CMBR. Son efficacit\'e pourrait cependant \^etre
encore am\'elior\'ee. Bien que le classement des cordes selon leur
longueur soit indispensable, la limite de \emph{load balancing}
pourrait \^etre encore abaiss\'ee en coupant virtuellement les grandes
cordes pour pouvoir r\'epartir le calcul de leur autocommutations sur
diff\'erents processeurs. De plus, quelques sections encore en s\'erie
pourraient \^etre parall\'elis\'ees moyennant des modifications du
code. La programmation parall\`ele am\'eliore de mani\`ere
significative la vitesse d'ex\'ecution du code sans recourir
\`a des machines extr\^emement puissantes. Il existe \'egalement des
directives de compilation, nomm\'ees \texttt{MPI} et similaires \`a
\texttt{OpenMP}, permettant de parall\'eliser un code sur diff\'erentes
machines, \`a m\'emoire s\'epar\'ee. Il devient ainsi possible
d'utiliser un r\'eseau d'ordinateurs de bureau comme un
super-calculateur.

\Annex{Effondrement d'un polytrope homog\`ene (article)}
\label{annexeaa}
\minitoc
Dans cette annexe, le m\'ecanisme d'instabilit\'e gravitationnelle de
Jeans~\cite{jeans}, donnant naissance aux grandes structures de
l'univers (voir Chap.~\ref{chapitrecosmo}), est illustr\'e sur
l'exemple tr\`es simple de l'effondrement gravitationnel d'une
\'etoile mod\'elis\'ee par un gaz polytropique homog\`ene. On y
\'etudie l'\'evolution newtonienne des perturbations de densit\'e par
une approche analytique. Cet article a \'et\'e publi\'e dans la revue
\journal{Astronomy and Astrophysics}~\cite{ringevalSN}.

\vspace{1cm}

\begin{center}
{\Large \textbf{ Dynamical stability for the gravitational evolution
of a homogeneous polytrope
}}
\end{center}
\vspace{5mm}
\begin{center}
Christophe Ringeval and Serge Bouquet
\end{center}
\vspace{5mm}
\begin{center}
{\footnotesize{ Commissariat \`a l'\'Energie Atomique, Centre
d'\'etude de Bruy\`eres-le-Ch\^atel, DAM/DIF,\\ D\'epartement de
Physique Th\'eorique et Appliqu\'ee, Service de Physique des Plasmas
et \'Electromagnetisme,\\ 91680 Bruy\`eres-le-Ch\^atel, France.
}}
\end{center}
\vspace{5mm}
\begin{center}
\begin{minipage}[c]{14cm}
{\footnotesize \textbf{The dynamic stability of the spherical
gravitational evolution (collapse or expansion) for a homogeneous
polytropic gas with any exponent $\gamma$, is studied using the
lagrangian formalism. We obtain the analytical expression for density
perturbations at the first order. In the case $\gamma=4/3$, the
Jeans' criterion is easily generalized to a self-similar expanding
background. The collapsing case is found to be always unstable.  The
stability of density modes obtained for $\gamma \neq 4/3$ does not
introduce any conditions on the wavelength perturbation, but only a
criterion on the polytropic index. As a result, stability is obtained
for an expanding gas provided $\gamma < 4/3$, and for a collapsing
one, for $\gamma>5/3$.
}}
\end{minipage}
\end{center}

\section{Introduction}
Within the framework of high energy laser experiments, the study of
dynamic stability for a gas in a microtarget under an external field
becomes experimentally possible (Kane et al. \cite{kane97a},
\cite{kane97b}; \cite{kane99}; Remington et al. \cite{remington}).
The extrapolation of the results to large self-gravitating masses
opens the way to the ``laboratory astrophysics''. In particular,
instabilities in giant molecular hydrogen clouds can be considered as
initial seeds to the gravitational collapse and, consequently, to the
birth of stars. Due to simple models, it is therefore conceivable to
find conditions on protostellar configurations which do, or do not
lead to their own gravitational collapse.  A first method for dealing
with this process is the analysis of non-linear equations by eulerian
self-similar techniques (Blottiau et al. \cite{blottiau88}; Bouquet et
al.  \cite{bouquet85a}; Shu \cite{shu}; Yahil \cite{yahil}). The
lagrangian way, often prefered in numerical studies, has also been
used by Blottiau (\cite{blottiau88}). However, whereas the numerical
results seem to agree with theoretical stability obtained from
self-similarity analysis, (Blottiau et al. \cite{blottiau88}),
analytical lagrangian approaches remain in discrepancy (Bonnor
\cite{bonnor}; Buff \& Gerola \cite{buff}).

In this study we use widely and intensively the
analytical lagrangian approach to check and to compare our results
with the ones previously found by eulerian self-similar ways.  The
``predilection'' model is still the one describing the evolution of a
homogeneous polytropic spherical mass. The stability is discussed from
the study of the time evolution of density perturbations at the first
order (Bonnor \cite{bonnor}; Bouquet \cite{bouquet99}). From the
simplicity of the assumptions, it is obvious that such treatment
cannot describe thoroughly stellar explosions or collapses.  However,
it can provide relevant conditions and results for the starting
processes leading to the dynamic evolution.  On the other hand,
laboratory experiments will allow us to delimit the domain of validity
of such ``simple'' models, but which are almost the only ones fully
computable analytically.

In Sect.~\ref{premierchapitre}, similar results of Blottiau et
al. (\cite{blottiau88}) and Bouquet (\cite{bouquet99}) are refered to
and used to generalize the Jeans' criterion in the case of an
expanding homogeneous polytropic gas with $\gamma=4/3$.

Sect.~\ref{deuxiemechapitre} deals with the lagrangian
description. The system which consists in the hydrodynamical equations
for the density perturbations, at first order, has been solved
analytically. The stability criteria are obtained from the study of
the asymptotic behaviour of these solutions for any value of
$\gamma$. The analytical expression is obtained from an infinite
summation over eigenmodes satisfying the appropriate boundary
conditions.  It turns out that the results confirm and extend those
presented in Sect.~\ref{premierchapitre}.  The conclusion is given in
Sect.~\ref{dernierchapitre}.
\section{Eulerian collapse}
\label{premierchapitre}
\subsection{Previous results}
The study of self-gravitating configurations can be made with the use
of scaling transformations (Bouquet et al. \cite{bouquet85a}; Blottiau
et al. \cite{blottiau88}; Chi\`eze et al. \cite{chieze}; Hanawa \&
Nakayama \cite{hanawa97}; Saigo \& Hanawa \cite{saigo}; Nakamura et
al. \cite{nakamura}; Hanawa et al. \cite{hanawa99}). The equations
governing the evolution of the gravitational system, written in the
new space (rescaled space) of transformed physical quantities
(rescaled quantities) are often easier to solve and to understand than
in the physical one.  In particular, the dynamic stability problem may
reduce to a static one. The Euler self-similar approach of Blottiau et
al. (\cite{blottiau88}) deals with a homogeneous self-gravitating
infinite mass which follows a polytropic equation of state
\begin{equation}
\label{equetat}
P=K\rho^{\gamma},
\end{equation}
with an exponent $\gamma=4/3$ and where $P$ and $\rho$ are
respectively the pressure and the density of the medium. The case
$\gamma \neq 4/3$ was also studied but only in a numerical way. In the
present paper, we first recall the Euler analytical study for
$\gamma=4/3$, and we recast it into the lagrangian frame. Second, we
extend this approach, analytically, to any value of the polytropic
exponent $\gamma$.

The evolution of the system is governed by the Euler, Poisson and
continuity equations which read respectively
\begin{eqnarray}
\label{eulerdepart}
\frac{\partial v}{\partial t}+v\frac{\partial v}{\partial r} & = &
-\frac{1}{\rho} \frac{\partial P}{\partial r}+g, \\
\label{poissondepart}
\frac{1}{r^{2}}\frac{\partial}{\partial r}\left(r^{2} g\right) & = &
-4\pi \rho \Gc, \\
\label{continuitedepart}
\frac{\partial \rho}{\partial t} & = & -
\frac{1}{r^{2}}\frac{\partial}{\partial r}\left(r^{2} \rho v\right),
\end{eqnarray}
where $r$, $t$, $v(r,t)$ and $g(r,t)$ are respectively the radial
position, the time, the eulerian velocity field and the value of the
gravitational field at the event $(r,t)$. A newtonian self-similar
solution for these equations is a parabolic homogeneous collapse,
therefore without any velocity at infinity (Blottiau et
al. \cite{blottiau88}; Henriksen \& Wesson \cite{henriksen})
\begin{eqnarray}
\label{rzero}
r_\zero(m,t) & = & \hat{r}_\zero (1+\Omega t)^{\frac{2}{3}}, \\
\label{rhozero}
\rho_\zero(t) & = & \hat{\rho}_\zero(1+\Omega t)^{-2}, \\ \hat{r}_\zero & = &
\left(\frac{9 \Gc m}{2 \Omega^{2}}\right)^{\frac{1}{3}}, \\
\label{rhozerochapo}
\hat{\rho}_\zero & = & \frac{\Omega^2}{6 \pi \Gc},
\end{eqnarray}
where $r_\zero(m,t)$ and $\rho_\zero(t)$ are, respectively, the
position of the shell whose interior mass is $m$, and the uniform
density, both of them being taken at time $t$.  The quantities
$\hat{r}_\zero$ and $\hat{\rho}_\zero$ represent, respectively, the
position of the shell (labelled by $m$) and the uniform density at the
initial time $t=0$. The parameter $\Omega$ (Blottiau et
al. \cite{blottiau88}; Bouquet et al. \cite{bouquet85a}) is an
integrating constant which reflects the freedom in the choice of the
time-origin.  It is just proportional to the \emph{initial} Jeans
frequency, $\hat{\Omega}_{\mathrm{J}}$, given by
\begin{equation}
\hat{\Omega}_{\mathrm{J}}=\sqrt{4 \pi \hat{\rho}_\zero \Gc},
\end{equation}
and the relationship between $\Omega$ and $\hat{\Omega}_{\mathrm{J}}$
is just [Eq.~(\ref{rhozerochapo})]
\begin{equation}
\label{omegajeans}
\Omega^{2}=\frac{3}{2} \hat{\Omega}_{\mathrm{J}}^{2}.
\end{equation}
This parameter may seen redundant with the Jeans frequency. However,
we are going to explain how relevant it can be. Usually, one works
with the variable $t$ which varies from $-\infty$ up to
$+\infty$. But, generally, a singularity arises at $t=0$ which, in our
opinion, is not so easy to be managed. For instance, the spatial
extension of the configuration must be zero.

In opposition, the introduction of the parameter $\Omega$ allows us to
leave $t=0$ as the initial time in any case. In order to describe
expansions, we take the positive solution in Eq.~(\ref{omegajeans}),
$\Omega=+\sqrt{3/2}~\hat{\Omega}_{\mathrm{J}}$ and $t$ elapses from
$0$ up to $+\infty$. In contrast, collapses will be obtained for
$\Omega=-\sqrt{3/2}~\hat {\Omega}_{\mathrm{J}}$ [negative solution in
Eq.~(\ref{omegajeans})] and the final gravitational singularity will
arise for $1+\Omega t_{\mathrm{sing}}=0$, i.e. at
$t_{\mathrm{sing}}=-1/\Omega$ (which, of course, is a positive value
since $\Omega$ is negative). It should be noted that this very simple
remark provides, in a very straightforward and easy way, the free
fall-time for a homogeneous gravitational system
\begin{displaymath}
t_{\mathrm{ff}}=\sqrt{\frac{1}{6 \pi \hat{\rho}_\zero \Gc}}.
\end{displaymath}
In addition, and for any situation, $t$ remains positive and its
initial value is finite, always $t=0$. Moreover, since at $t=0$ no
singularity arises, the extension of the configuration can be not (and
is not) zero whereas removing the parameter $\Omega$ could give rise
to expansions beginning at the singularity ($r=0$ and $t=0$) which, in
our mind, does not make sense. Thanks to the parameter $\Omega$, we
may specify any spatial profile (for the density, for the velocity,
etc.) at the initial time and study its influence on the further
evolution of the system. These properties are very convenient from a
physical viewpoint.  This parameter is not only useful in
astrophysical studies but it can also be used very fruitfully in
evolution problems plasma physics (Bouquet et al. \cite{bouquet85b};
Burgan et al. \cite{burgan78}, \cite{burgan83}), nonlinear evolution
equations and dynamic systems (Bouquet \cite{bouquet95}; Cair\'o \&
Feix \cite{cairo98}; Cair\'o et al.  \cite{cairo99}) and other
interesting domains.

We are going to see that by means of scaling transformations, the time
dependence of the solutions [Eq.~(\ref{rzero}) and
Eq.~(\ref{rhozero})] can be removed. The dynamic problem of stability
reduces, therefore, to a static one. The new physical quantities in
this rescalled space are written with a hat ``$\hat{\quad}$'' and,
according to Blottiau et al. (\cite{blottiau88}) and Bouquet et
al. (\cite{bouquet85a}), we have
\begin{eqnarray}
\label{space}
\hat{r} & = & r(1+\Omega t)^{\gamma-2},\\
\label{echelle}
\hat{t} & = & \frac{1}{\Omega} \ln{(1+\Omega t)},\\
\hat{\rho}(r,t) & = & (1+\Omega t)^{2}\rho(r,t),\\
\hat{P}(r,t) & = & (1+\Omega t)^{2\gamma}P(r,t),\\
\hat{g}(r,t) & = & (1+\Omega t)^{\frac{4}{3}}g(r,t),
\end{eqnarray}
where we set $\gamma = 4/3$ in the following. From these equations,
it is clear that at the initial time $t=\hat{t}=0$, the quantities
with and without ``$\hat{\quad}$'' coincide (the rescaled space and
the physical one are initially identical). Moreover, for $\Omega>0$
($t$ and $\hat{t}$ go from $0$ to $+\infty$), the transformation
describes an expansion, while for $\Omega<0$, the configuration
collapses up to the central singularity in a finite time given by
$t=-1/\Omega$. It must be noted that for this case ($\Omega<0$), the
times $t$ and $\hat{t}$ vary respectively, in the ranges
$[0,-1/\Omega[$ and $[0,+\infty[$. It can be easily shown (Blottiau
et al. \cite{blottiau88}; Bouquet et al. \cite{bouquet85a}; Bouquet
\cite{bouquet99}) that the system formed by equations
(\ref{eulerdepart}) to (\ref{continuitedepart}), becomes stationary in
the new space without any explicit dependence upon
$\hat{t}$. Moreover, assuming that
\begin{eqnarray}
\hat{\rho}(\hat{r},\hat{t}) & = & \hat{\rho}_\zero+\delta
\hat{\rho}(\hat{r},\hat{t}),\\
\label{deltarhoeul}
\delta \hat{\rho}(\hat{r},\hat{t}) & = & A(\hat{t})
\sin(\hat{k}\hat{r})/(\hat{k}\hat{r}),
\end{eqnarray}
where $\hat{k}$ is the wave number in the rescaled space and where
\begin{eqnarray}
\label{aeul}
A(\hat{t}) & = & A_\zero \exp(\omega \hat{t}).
\end{eqnarray}
$A_\zero$ and $\omega$ are two constants. The study of the evolution of
the perturbations for the various transformed quantities, at the first
order, provides a dispersion equation for the density modes. This
dispersion relationship is (Blottiau et al. \cite{blottiau88})
\begin{equation}
\label{dispeuler}
\omega^{2}+\frac{\Omega}{3}\omega+\hat{k}^{2}\hat{c}^{2} -
\hat{\Omega}_{\mathrm{J}}^{2}=0.
\end{equation}
where $\hat{c}$ is the initial sound velocity given by
\begin{equation}
\hat{c}^2=\gamma K \hat{\rho}_\zero^{\gamma-1},
\end{equation}
and where $\hat{\Omega}_{\mathrm{J}}$ is related to $\Omega$ from
Eq.~(\ref{omegajeans}).  Their physical values at time $t$ are
obtained from the inverse scale transformation (Blottiau et
al. \cite{blottiau88})
\begin{eqnarray}
\Omega_{\mathrm{J}}(t) & = & \hat{\Omega}_{\mathrm{J}}(1+\Omega
t)^{-1}, \\
c(t) & = & \hat{c}(1+\Omega t)^{-\frac{1}{3}}.
\end{eqnarray}
Coming back to Eq.~(\ref{dispeuler}), the eigenmodes are obtained by
the resolution of the dispersion equation, quadratic in $\omega$ with
the discriminant
\begin{equation}
\Delta_{\hat{k}}=\frac{25 \hat{\Omega}_{\mathrm{J}}^{2}-24 \hat{k}^{2}
\hat{c}^{2}}{6}.
\end{equation}
According to the sign of $\Delta_{\hat{k}}$, we obtain, therefore, the
two solutions for $\omega$
\begin{eqnarray}
\label{omegaeulre}
\hat{k}<\hat{k}_{\mathrm{trans}} \Rightarrow \omega_{\mathrm{r}\pm} &
= & -\frac{\Omega}{6} \pm \frac{\sqrt{\Delta_{\hat{k}}}}{2},\\
\label{omegaeulim}
\hat{k}>\hat{k}_{\mathrm{trans}} \Rightarrow \omega_{\mathrm{i}\pm} &
= & -\frac{\Omega}{6} \pm i\frac{\sqrt{-\Delta_{\hat{k}}}}{2},\\
\textrm{with} \quad
\label{ktrans}
\hat{k}_{\mathrm{trans}} & = & \sqrt{\frac{25}{24}}
\frac{\hat{\Omega}_{\mathrm{J}}}{\hat{c}}.
\end{eqnarray}
The imaginary values, $\omega_{\mathrm{i}\pm}$, for $\omega$ are
obtained for $\hat{k}>\hat{k}_{\mathrm{trans}}$.  These solutions give
rise to evanescent modes. This is the stability criterion, in the
rescaled space, found by Blottiau et al. (\cite{blottiau88}).  In the
next section, we are going to show that it is equivalent to the Jeans'
criterion in the physical space.
\subsection{Equivalence to the Jeans' criterion}
\label{sectionjeanseuler}
The time dependence of the density perturbations, in the physical
space, is deduced from Eq.~(\ref{echelle}), Eq.~(\ref{deltarhoeul})
and Eq.~(\ref{aeul}). We obtain,
\begin{equation}
\label{aeulphy}
A(t)=A_\zero(1+\Omega t)^{\frac{\omega}{\Omega}}.
\end{equation}
Moreover, since $\rho$ and $\delta\rho$ rescale in the same way, we
have, at the first order
\begin{equation}
\frac{\delta \rho}{\rho}=\frac{\delta \hat{\rho}}{\hat{\rho}} \sim
\frac{\delta\hat{\rho}}{\hat{\rho}_\zero}.
\end{equation}
The asymptotic time evolution of $\delta \rho / \rho$ is, therefore,
directly given by the real part sign of the exponent $\omega /
\Omega$.\\ First, for $\hat{k}<\hat{k}_{\mathrm{trans}}$, $\omega$ is real but
a critical value of $\hat{k}$, $\hat{k}_{\mathrm{crit}}$, makes changing the
sign of $\omega_{\mathrm{r}+}/\Omega$. With Eq.~(\ref{omegaeulre}) it comes
\begin{eqnarray}
\hat{k} < \hat{k}_{\mathrm{crit}}< \hat{k}_{\mathrm{trans}} &
\Rightarrow & \frac{\omega_{\mathrm{r}+}} {\Omega}>0,\\
\hat{k}_{\mathrm{crit}}< \hat{k} < \hat{k}_{\mathrm{trans}} &
\Rightarrow & \frac{\omega_{\mathrm{r}+}}{\Omega}<0,
\end{eqnarray}
where
\begin{equation}
\label{kcrit}
\hat{k}_{\mathrm{crit}} = \frac{\hat{\Omega}_{\mathrm{J}}}{\hat{c}}.
\end{equation}
We notice that $\hat{k}_{\mathrm{crit}}$ corresponds to the value given by
Jeans (\cite{jeans}).  In addition, keeping in mind the permanent
negative sign of $\omega_{\mathrm{r}-}/\Omega$, and since the solution is
written as the linear superposition of the two modes, the asymptotic
behaviour is given by the leading term. We get
\begin{eqnarray}
\label{stabinfkcrit}
\hat{k}<\hat{k}_{\mathrm{crit}}< \hat{k}_{\mathrm{trans}} \Rightarrow
& \left\{
\begin{array}{lll}
\forall \Omega>0 & \displaystyle \lim_{t \rightarrow
\infty}\frac{\delta \rho}{\rho} = \infty, \\ \\ \forall \Omega<0 &
\displaystyle\lim_{t \rightarrow -\frac{1}{\Omega}}\frac{\delta
\rho}{\rho} = \infty,
\end{array} \right.\\
\label{stabsupkcrit}
\hat{k}_{\mathrm{crit}} < \hat{k} < \hat{k}_{\mathrm{trans}}
\Rightarrow & \left\{
\begin{array}{lll}
\forall \Omega>0 & \displaystyle \lim_{t \rightarrow\infty}\frac{\delta
\rho}{\rho} = 0, \\ \\ \forall \Omega<0 & \displaystyle \lim_{t \rightarrow
-\frac{1}{\Omega}}\frac{\delta \rho}{\rho} = \infty.
\end{array} \right.
\end{eqnarray}
Second, for $\hat{k}>\hat{k}_{\mathrm{trans}}$, the imaginary part of
$\omega$ [given by Eq.~(\ref{omegaeulim})] produces an oscillating
contribution to $A(t)$. In contrast, the real part gives a time-power
evolution with a negative exponent $-1/6$. Consequently, one gets
\begin{eqnarray}
\label{stabsupktrans}
\hat{k}>\hat{k}_{\mathrm{trans}} \Rightarrow & \left\{
\begin{array}{lll}
\forall \Omega>0 & \displaystyle \lim_{t \rightarrow
\infty}\frac{\delta \rho}{\rho} = 0, \\ \\ \forall \Omega<0 &
\displaystyle \lim_{t \rightarrow -\frac{1}{\Omega}}\frac{\delta
\rho}{\rho} = \infty.
\end{array} \right.
\end{eqnarray}
Equations (\ref{stabinfkcrit}) to (\ref{stabsupktrans}) emphasize that
the asymptotic behaviour of the density perturbations depends strongly
on the value of the wave number $\hat{k}$, which is connected to the
value of $k$ in the physical space by the inverse transformation of
Eq.~(\ref{space}) (Blottiau et al. \cite{blottiau88})
\begin{equation}
k=\hat{k}(1+\Omega t)^{-\frac{2}{3}}.
\end{equation}
As a result, for explosions ($\Omega>0$), the density perturbations
are unstable as soon as the instantaneous wave number, $k(t)$,
satisfies $k(t)<k_{\mathrm{crit}}(t)$ with $k_{\mathrm{crit}}(t)=\Omega_{\mathrm{J}}(t)/c(t)$ is
the instantaneous Jeans wave number.  However, since $k$ and
$k_{\mathrm{crit}}$ have the same time-dependence, if the criteria is satisfied
at $t=0$, it is satisfied for any time. Consequently, the result
obtained by Jeans (\cite{jeans}) for a static background is also valid
for an expanding one provided $\gamma=4/3$. This is closely akin to
Bonnor's results (\cite{bonnor}).  On the other hand, in the implosion
case ($\Omega<0$), we always have instabilities: every density
perturbation is amplified during the collapse.  Finally, note that the
value $\hat{k}_{\mathrm{trans}}$ is not relevant to stability, but indicates
only changes in behaviour with wave number: beyond this value, the
perturbation oscillates and increases, and below, it explodes as a
time-power dependence.
\subsection{Conclusion}
The stability criterion for an eulerian self-similar evolution does
not agree with the one given by Buff \& Gerola (\cite{buff}). Instead
of Eq.~(\ref{kcrit}), they find a Jeans wave number equal to
$\sqrt{3/2}~k_{\mathrm{crit}}$. As a matter of fact, their dispersion equation,
derived in the physical space, for a fixed mass collapse is
\begin{equation}
\label{dispersbuff}
\omega^{2}-\frac{2}{3} \Omega_{\mathrm{J}}^{2}+k^{2}c^{2}=0.
\end{equation}
Buff \& Gerola (\cite{buff}) have chosen a density perturbation, at
first order, under the form $A(t)\sin(kr)/(kr)$ with $A(t)=A_\zero
\exp(\omega t)$. In our opinion, this ansatz is not possible. The
reason for this is that, in opposition to our approach in which we
obtain a second order automous differential equation for the density
perturbations, they get a linearized equation with time-varying
coefficients. But, in that case, it is well known that the exponential
solution, $\exp(\omega t)$, is no longer valid. Consequently, the
meaning of Eq.~(\ref{dispersbuff}) is not clear and one would have to
assume that $\omega$ be an explicit function of time. However, under
this assumption, additional terms proportional to $\ud \omega/\ud t$
should appear and Eq.~(\ref{dispersbuff}) would be modified.  In the
next section, an analytical lagrangian calculation is performed. It is
shown that obtaining a dispersion relation is not necessary and we are
going to recover and to extend the results found by Blottiau et
al. (\cite{blottiau88}) and by Bonnor (\cite{bonnor}).
\section{Lagrangian collapse}
\label{deuxiemechapitre}
Let $M$ be the mass of a spherical homogeneous configuration with
initial radius $\hat{R}$ submitted to its own gravitational field and
initially at rest. In the following, the physical quantities will be
expressed, either as a function of the lagrangian variable $m$ (where
$m$ is the internal mass of a shell), or in terms of $\hat{r}_\zero$ (with
$\hat{r}_\zero$ being the initial radius of the shell labelled by $m$),
plus the time, $t$, in both cases. The stability is again studied via
the time-evolution of density perturbations at the first order,
$\delta \rho$.  All parameters with the subscript ``$\zero$'' are associated
with the non-perturbated solution.  Finally, it must be pointed out
that this study is performed analytically for any arbitrary value of
the polytropic exponent.
\subsection{Equation of evolution}
The evolution of the non-perturbed system obeys the hydrodynamical
equations (\ref{equetat}) to (\ref{continuitedepart}) with the
solution given by (\ref{rzero}) and (\ref{rhozero}). The perturbation
is then written in the form
\begin{eqnarray}
\label{pert}
\rho(m,t) & = & \rho_\zero(t)+\delta \rho(m,t),\\
r(m,t) & = & r_\zero(m,t)+\delta r(m,t).
\end{eqnarray}
The solution will no longer be homogeneous and we have to keep the
pressure gradient term in the Euler equation (\ref{eulerdepart}). This
gradient is expressed as a function of the density according to the
polytropic equation of state (\ref{equetat}). After elimination of the
zero order terms from equations (\ref{rzero}) and (\ref{rhozero}), the
Euler equation reads
\begin{equation}
\label{eulerpert}
\frac{\partial^{2} \delta r}{\partial t^{2}}=-4 \pi r_\zero^{2} c^{2}
\frac{\partial \delta \rho}{\partial m}+\frac{8 \pi}{3} \Gc
\rho_\zero \delta r.
\end{equation}
The time-dependent sound velocity, $c$, is written at the zero order
\begin{eqnarray}
\label{son}
c^{2}(t) & \simeq & c_\zero^{2}(t)= \frac{\gamma
P}{\rho}=\hat{c}_\zero^{2}(1+\Omega t)^{2(1-\gamma)},
\end{eqnarray}
with
\begin{equation}
\hat{c}_\zero^{2} =  \gamma K \hat{\rho}_\zero^{\gamma - 1}.
\end{equation}
In addition, the conservation of mass from the non-perturbed to the
perturbed configuration provides the second equation
\begin{equation}
\ud m = 4 \pi r^{2} \rho \, \ud r = 4 \pi r_\zero^{2} \rho_\zero \, \ud
r_\zero,
\end{equation}
which becomes,
\begin{equation}
\label{massepert}
(3 m)^{\frac{2}{3}} \frac{\partial \delta r}{\partial m} + 2 (3
m)^{-\frac{1}{3}} \delta r = - \frac{\delta \rho}{\left(4 \pi
\rho_\zero^{4}\right)^{\frac{1}{3}}}.
\end{equation}
The differential system formed by (\ref{eulerpert}) and
(\ref{massepert}) can be solved by direct integration with the
physical assumption that there is no perturbation at the center of the
configuration ($\delta r << r_\zero $ in $r_\zero=0$ gives $\delta
r|_\zero=0$), which is a zero mass point. The solution of
Eq.~(\ref{massepert}) is, therefore
\begin{equation}
\label{deltar}
\delta r(m,t)=-\frac{1}{\left(36 \pi \rho_\zero^{4} m^{2}
\right)^{\frac{1}{3}}} \int_0^m \delta \rho(\mu,t) \, \ud \mu.
\end{equation}
Plugging this solution into Eq.~(\ref{eulerpert}), the evolution
equation for the density perturbation writes
\begin{eqnarray}
\label{deltarho}
\lefteqn{m^{\frac{4}{3}}\frac{\partial ^{2} \delta \rho}{\partial
m^{2}} +\frac{4}{3} m^{\frac{1}{3}} \frac{\partial \delta
\rho}{\partial m} } \nonumber\\ & {} = \displaystyle \frac{1}{(36 \pi
\rho_\zero)^{\frac{2}{3}} c_\zero^{2}} \left(\frac{\partial^{2} \delta
\rho}{\partial t^{2}} + 8 \frac{|\Omega|}{\Omega} \sqrt{\frac{8 \pi
\Gc \rho_\zero}{3}} \frac{\partial \delta \rho}{\partial t} + 24 \pi
\Gc \rho_\zero \delta \rho \right).
\end{eqnarray}
As expected, Eq.~(\ref{deltarho}) is linear but with a partial
differentiation with respect to the independant variables $m$ and
$t$. The eigenmodes may be found by the technique of separation of
variables. Then, the general solution will be the superposition of all
modes with the constraint that the boundary conditions must be
satisfied.
\subsection{Density eigenmodes}
Introducing the separation of variables for $\delta \rho(m,t)$ under
the form
\begin{equation}
\label{separation}
\delta \rho= \delta T(t) \delta R(m)
\end{equation}
the equation for the mass dependence becomes
\begin{equation}
\label{massique}
\frac{\ud^{2} \delta R}{\ud l^{2}} + \frac{2}{l} \frac{\ud \delta
R}{\ud l} - 9 \, \varepsilon \, \eta_k^{2} \, \delta R = 0,
\end{equation}
where the independant variable, $l$, is given by
\begin{equation}
l=m^{1/3}.
\end{equation}
In Eq.~(\ref{massique}), we have decided to write the separation
constant as $\varepsilon \, \eta_k^{2}$ with $\eta_k \ge 0$ which has the
dimension $[M]^{-1/3}$.  The parameter $\varepsilon=\pm 1$ has been
introduced for choosing the sign.  From $\eta_k$, let us introduce,
now, the wave number, $k$, labelling each density eigenmode, and a
dimensionless number, lets say $N_k$, which will help to separate the
various stability regimes
\begin{eqnarray}
k & = & (36 \pi)^{\frac{1}{3}} \hat{\rho}_\zero^{\frac{1}{3}} \eta_k, \\
\label{nombre}
N_k & = & \frac{k \hat{c}_\zero}{|\Omega|}.
\end{eqnarray}
The quantity $\eta_k$ is equivalent to a ``massic pulsation'' since we
have
\begin{equation}
3 \eta_k m^{\frac{1}{3}} = k \hat{r}_\zero.
\end{equation}
On the other hand, it comes from Eq.~(\ref{deltarho}) that the time
differential equation, for $\gamma \ne 4/3$, is
\begin{equation}
\label{temporelled}
z^{2}\frac{\ud^{2}\delta S}{\ud z^{2}} +z\frac{\ud \delta S}{\ud z} -
(\varepsilon z^{2}+n^{2}) \delta S=0,
\end{equation}
where the new variable, $z$, and function, $\delta S(z)$, are given
by
\begin{eqnarray}
\label{argument}
z & = & \frac{N_k (1+\Omega t)^{\mu}}{|\mu|},\\
\label{fonction}
\delta T(t) & = & N_k^{-\frac{13}{6 \mu}} (1+\Omega t)^{-\frac{13}{6}}
\delta S(z),\\
\label{changementdeb}
\mu & = & \frac{4}{3}-\gamma,\\
n & = & \frac{5}{6 |\mu|}.
\end{eqnarray}
The special case $\gamma = 4/3$ gives, from Eq.~(\ref{deltarho}), the
second order differential equation
\begin{equation}
\label{temporellee}
\frac{\ud^{2} \delta T}{\ud y^{2}}+\frac{13}{3} \frac{\ud \delta T}{\ud
y}+(4- \varepsilon N_k^{2}) \delta T = 0,
\end{equation}
with the new independant variable
\begin{equation}
y=\ln(1+\Omega t).
\end{equation}
It turns out that equations (\ref{massique}) and (\ref{temporelled})
are the so-called classical and modified Bessel equations according to
the value of $\varepsilon$. On the other hand, for $\gamma=4/3$,
Eq.~(\ref{temporellee}) is a linear homogeneous differential equation
with constant coefficients. It is therefore readily integrable in
terms of the exponential functions. This separation naturally leads us
to distinguish between the eigenmodes for $\gamma=4/3$ from the ones
for $\gamma \ne 4/3$ (see Sect.~\ref{deptemps}).

\subsubsection{Mass dependence of the solution}

The requirement of having a finite density perturbation at the center
of the configuration restricts the solutions of Eq.~(\ref{massique})
to
\begin{eqnarray}
\label{masstrigo}
\delta R_{\varepsilon=-1}^k(m) & \propto & \frac{\sin(3 \eta_k
m^{\frac{1}{3}})}{3 \eta_k m^{\frac{1}{3}}}, \\
\label{masshyper}
\delta R_{\varepsilon=1}^k(m) & \propto & \frac{\sinh(3 \eta_k
m^{\frac{1}{3}})}{3 \eta_k m^{\frac{1}{3}}}.
\end{eqnarray}
Note that the hyperbolic sine appears in the case $\varepsilon=1$.
\subsubsection{Time dependence of the solution}
\label{deptemps}
\paragraph{Case $\gamma=4/3$.}
The roots, $s_{\pm}^{\varepsilon}(k)$, of the characteristic equation
associated with Eq.~(\ref{temporellee}) are
\begin{eqnarray}
\label{racines}
s_{\pm}^{\varepsilon}(k) & = & -\frac{13}{6} \pm 
\frac{\sqrt{\Delta_k^{\varepsilon}}}{2},
\end{eqnarray}
where the discriminant is
\begin{eqnarray}
\Delta_k^{\varepsilon} = \frac{25+36 \, \varepsilon \, N_k^{2}}{9} & =
& \frac{25 \hat{\Omega}_{\mathrm{J}}^{2} +24 \, \varepsilon \, k^{2}
\hat{c}_\zero^{2}}{9 \hat{\Omega}_{\mathrm{J}}^{2}}.
\end{eqnarray}
Then, for the hyperbolic modes ($\varepsilon=1$), we have
\begin{equation}
\label{solutiontempplus}
\delta T_{+}^k(t)= \beta_k (1+\Omega t)^{s_+^{+}(k)}+
\gamma_k (1+\Omega t)^{s_-^{+}(k)}
\end{equation}
where $\beta_k$ and $\gamma_k$ are two arbitrary real constants and
where the superscript sign in the exponents is just the sign of the
parameter $\varepsilon$.\\ The trigonometric modes (with
$\varepsilon=-1$) introduce roots with imaginary part provided $N_k <
5/6$, i.e. $k<\sqrt{25/24}~\hat{\Omega}_{\mathrm{J}}/\hat{c}_\zero$. We have,
therefore, two kinds of solution:
\begin{eqnarray}
\label{modetempplus}
k<\sqrt{\frac{25}{24}}\frac{\hat{\Omega}_{\mathrm{J}}}{\hat{c}_\zero}
\Rightarrow \delta T_{-}^k(t) = \beta_k (1+\Omega t)^{s_+^{-}(k)} {} +
\nonumber\\ {} + \gamma_k (1+\Omega t)^{s_-^{-}(k)}, \\
\label{modetempmoins}
k>\sqrt{\frac{25}{24}}\frac{\hat{\Omega}_{\mathrm{J}}}{\hat{c}_\zero}
\Rightarrow \delta T_{-}^k(t) = \alpha_k (1+\Omega t)^{-\frac{13}{6}}
{} \times \nonumber\\ {} \times \cos
\frac{\sqrt{|\Delta_k^{-}|}\ln(1+\Omega t)}{2}, & &
\end{eqnarray}
where $\alpha_k$ is an arbitrary real constant and where the
unimportant phase in the cosine have been dropped. Let us notice the
transition at the same value of the wave number than in the eulerian
self-similar case, $k_{\mathrm{trans}}$, which according to
Eq.~(\ref{ktrans}), is given by
\begin{equation}
\label{ktranslag}
k_{\mathrm{trans}} = \sqrt{\frac{25}{24}}
\frac{\hat{\Omega}_{\mathrm{J}}}{\hat{c}_\zero}.
\end{equation}
It must be noted that, now, the transition wave number is no longer
defined in a rescaled space, but it applies directly in the physical
one.
\paragraph{Case $\gamma \ne 4/3$.}
From the inversion of the independent variables [$z$ and $t$ in
Eq.~(\ref{argument})] and the dependent ones [$\delta S$ and $\delta
T$ in Eq.~(\ref{fonction})], it turns out that for all $k$ and for
$\varepsilon=1$, we have
\begin{eqnarray}
\label{modetemporelpos}
\lefteqn{\delta T_{+}^k(t) = (1+\Omega t)^{-\frac{13}{6}} {} \times }
\nonumber \\ & {} \displaystyle \times \left\{\alpha_k
I_n\left[\frac{N_k(1+\Omega t)^{\mu}}{|\mu|} \right] + \beta_k
K_n\left[\frac{N_k(1+\Omega t)^{\mu}}{|\mu|} \right] \right\},
\end{eqnarray}
where $I_n$ and $K_n$ are respectively, the modified Bessel functions
of first and second kind of order $n$.\\ The other case is
$\varepsilon=-1$, and we get
\begin{eqnarray}
\label{modetemporelneg}
\lefteqn{\delta T_{-}^k(t) = (1+\Omega t)^{-\frac{13}{6}} {} \times }
\nonumber \\ & {} \displaystyle \times
\left\{\alpha_k J_n\left[\frac{N_k(1+\Omega
t)^{\mu}}{|\mu|} \right] + \beta_k Y_n\left[\frac{N_k(1+\Omega
t)^{\mu}}{|\mu|} \right] \right\}.
\end{eqnarray}
The functions $J_n$ and $Y_n$ are respectively the first and second,
classical Bessel functions.

\subsection{Stability of eigenmodes}
\label{stabilitedesmodes}
From the analytical expressions of the eigenmodes, it is possible to
derive their asymptotic behaviour. In the case where $\Omega > 0$
(expanding background), the time elapses from $t=0$ to $t \rightarrow
+\infty$. On the other hand, in a collapsing background ($\Omega<0$),
the initial time is again $t=0$, while the final one is defined when
the singularity at $r_\zero=0$ arises, i.e. $t \rightarrow
-1/\Omega$.

\subsubsection{Stability of the  eigenmodes for $\gamma=4/3$}

According to Eq.~(\ref{modetempplus}) and Eq.~(\ref{modetempmoins}),
the asymptotic time behaviour of the perturbation is given by the
value of the limit of $(1+\Omega t)^q$ where the exponent $q$ is
either $s^\varepsilon_\pm$ or $-13/6$, according to the studied case.
This value depends upon the sign of $q$ and $\Omega$.\\ Now, the
relevant quantity is the density contrast $\delta \rho
/\rho_\zero$. Keeping in mind the square contained into
Eq.~(\ref{rhozero}), the asymptotic variations of the hyperbolic modes
($\varepsilon=1$) are readily obtained. Since from
Eq.~(\ref{racines}), we have the inegality $s_+^+(k)>s_-^+(k)$, it is
clear that in equation (\ref{solutiontempplus}) the asymptotic leading
behaviour for a collapse (resp. for an expansion) is given by the
variation of the second term (resp. the first term) of the right hand
side of Eq.~(\ref{solutiontempplus}), i.e.
\begin{eqnarray}
\label{epsplus}
\Omega<0 & \Rightarrow & \frac{\delta T_+^k}{\rho_\zero} \propto \lim_{t
\rightarrow -\frac{1}{\Omega}} (1+\Omega t)^{s_-^+(k)+2} = \infty, \\
\Omega>0 & \Rightarrow & \frac{\delta T_+^k}{\rho_\zero} \propto \lim_{t
\rightarrow \infty} (\Omega t)^{s_+^+(k)+2} = \infty.
\end{eqnarray}
As a consequence, all hyperbolic modes are unstable for any value of
the wave number.\\ The behaviour of trigonometric modes
($\varepsilon=-1$) introduces the Jeans wave number through the
exponant sign, like in the eulerian derivation. A transition beetween
oscillating and non-oscillating modes is also obtained for the
``pivot'' value, $k_{\mathrm{trans}}$, given by Eq.~(\ref{ktranslag}). For the
implosions ($\Omega<0$), it becomes
\begin{eqnarray}
\label{epsmoinsommoins}
k<k_{\mathrm{trans}} & \Rightarrow & \frac{\delta T_-^k}{\rho_\zero}
\propto \lim_{t \rightarrow -\frac{1}{\Omega}}(1+\Omega
t)^{s_-^-(k)+2} = \infty, \\ k>k_{\mathrm{trans}} & \Rightarrow &
\frac{\delta T_-^k}{\rho_\zero} \propto \lim_{t \rightarrow
-\frac{1}{\Omega}} \frac{\cos \left(\ln(1+\Omega t) \right)}{
(1+\Omega t)^{\frac{1}{6}}}= \infty,
\end{eqnarray}
and for the explosion ($\Omega>0$),
\begin{eqnarray}
\label{epsmoinsomplus}
0<k<k_{\mathrm{J}} & \Rightarrow & \frac{\delta T_-^k}{\rho_\zero}
\propto \lim_{t\rightarrow \infty} (\Omega t)^{s_+^-(k)+2} = \infty,
\\ k_{\mathrm{J}}<k<k_{\mathrm{trans}} & \Rightarrow & \frac{\delta
T_-^k}{\rho_\zero} \propto \lim_{t\rightarrow \infty}(\Omega
t)^{s_+^-(k)+2} = 0, \\
\label{phaseocillante}
k>k_{\mathrm{trans}} & \Rightarrow & \frac{\delta T_-^k}{\rho_\zero} \propto
\lim_{t\rightarrow \infty} \frac{\cos \left(\ln(\Omega t) \right)}{
(\Omega t)^{\frac{1}{6}}} = 0,
\end{eqnarray}
where $k_{\mathrm{J}}$ is the Jeans wave number given by
\begin{equation}
\label{kjeans}
k_{\mathrm{J}}  =  \frac{\hat{\Omega}_{\mathrm{J}}}{\hat{c}_\zero}.
\end{equation}
Similarly to $k_{\mathrm{trans}}$, the wave number $k_{\mathrm{J}}$ is
now significant in the physical space.  Here, again, we have kept the
leading order term in $\delta T_-^k(t)$. The trigonometric modes are
found to be unstable only for $k<k_{\mathrm{J}}$ in a expanding
background and for all $k$ in a collapsing one. This wave number is
identical to $k_{\mathrm{crit}}$ defined in
Sect.~\ref{sectionjeanseuler}. We fit closely, therefore, with the
self-similar approach. However, an oscillating phase occurs before the
final divergence for $k>k_{\mathrm{trans}}$ for collapses. For
expansions, an oscillating phase arises for the stable case too [see
Eq.~(\ref{phaseocillante})].

\subsubsection{Stability of eigenmodes with $\gamma \ne 4/3$}

We use the same method as in the previous section. However the
eigenmodes are now expressed in terms of the Bessel functions. It is,
therefore, necessary to know their asymptotic form when their argument
goes to zero or to infinity. We have (Abramowitz \& Stegun
\cite{abramo})
\begin{eqnarray}
x \rightarrow  0 & \Rightarrow & \left\{ \begin{array}{lll}
\label{zerolimitEP}
J_n(x) &\sim& \displaystyle \frac{1}{\Gamma(n+1)} \left(\frac{x}{2}
\right)^n, \\ \\ Y_n(x) &\sim& \displaystyle -\frac{\Gamma(n)}{\pi}
\left(\frac{2}{x} \right)^n, \\ \\ I_n(x) &\sim& \displaystyle
\frac{1}{\Gamma(n+1)} \left(\frac{x}{2} \right)^n, \\ \\ K_{n}(x)
&\sim& \displaystyle \frac{\Gamma(n)}{2} \left(\frac{2}{x} \right)^n,
\end{array} \right.
\end{eqnarray}
and
\begin{eqnarray}
x \rightarrow \infty & \Rightarrow & \left\{ \begin{array}{lll}
\label{inftylimit}
J_n(x) &\sim& \displaystyle \sqrt{\frac{2}{\pi x}}
\cos\left(x-\frac{n\pi}{2}-\frac{\pi}{4}\right),\\ \\ Y_n(x) &\sim&
\displaystyle \sqrt{\frac{2}{\pi x}}
\sin\left(x-\frac{n\pi}{2}-\frac{\pi}{4}\right),\\ \\I_n(x) &\sim&
\displaystyle \sqrt{\frac{1}{2 \pi x}} \exp{(x)},\\ \\ K_n(x) &\sim&
\displaystyle \sqrt{\frac{\pi}{2 x}} \exp{(-x)}.
\end{array} \right.
\end{eqnarray}
Moreover, in equations (\ref{modetemporelpos}) and
(\ref{modetemporelneg}), the exponent $\mu$, i.e. the quantity
$(\gamma - 4/3)$ appears and, consequently, the asymptotic behaviour
will be dependent on whether $\gamma<4/3$ or $\gamma>4/3$.
\paragraph{Hyperbolic modes ($\varepsilon=1$).}
From equations (\ref{modetemporelpos}), (\ref{zerolimitEP}) and
(\ref{inftylimit}), the explosion case ($\Omega>0$) behaves according
to
\begin{eqnarray}
\label{exponentinf}
\gamma<\frac{4}{3} & \Rightarrow & \frac{\delta T_+^k}{\rho_\zero} \propto
\lim_{t \rightarrow \infty} \frac{\exp{(\Omega
t)^{\frac{4}{3}-\gamma}}}{(\Omega t)^{\frac{5/3 - \gamma}{2}}} =
\infty, \\
\label{divsup}
\gamma>\frac{4}{3} & \Rightarrow & \frac{\delta T_+^k}{\rho_\zero} \propto
\lim_{t \rightarrow \infty} (\Omega t)^{\frac{2}{3}} = \infty.
\end{eqnarray}
and for $\Omega<0$ (implosions), we have
\begin{eqnarray}
\label{divinf}
\gamma<\frac{4}{3} & \Rightarrow & \frac{\delta T_+^k}{\rho_\zero} \propto
\lim_{t \rightarrow -\frac{1}{\Omega}} (1+\Omega t)^{-1} = \infty, \\
\label{exponentsup}
\gamma>\frac{4}{3} & \Rightarrow & \frac{\delta T_+^k}{\rho_\zero}
\propto \lim_{t \rightarrow -\frac{1}{\Omega}} \frac{\exp{(1+\Omega
t)^{\frac{4}{3}-\gamma}}}{ (1+\Omega t)^{\frac{5/3-\gamma}{2}}} =
\infty.
\end{eqnarray}
We conclude that all hyperbolic modes are unstable for any values of
both the wave number $k$ and the polytropic exponent $\gamma$. It is
important to notice the exponential rise of the perturbation in
equations (\ref{exponentinf}) and (\ref{exponentsup}). Moreover, it is
quite surprising to see that expansions and collapses behave exactly
in the same way for $\gamma<4/3$ and $\gamma>4/3$, respectively. The
dependence upon the value of $\gamma$ is very sensitive and it is
interesting that, in addition to the ``critical value'' $\gamma= 4/3$,
the difference $\gamma-5/3$ arises quite naturally. This was not
expected from the beginning of the study. On the other hand, we see
that for $\gamma>4/3$ (resp.  $\gamma<4/3$), the leading time
evolution of the instability for explosions (resp.  implosions) does
not depend any more upon the value of $\gamma$.
\paragraph{Trigonometric modes ($\varepsilon=-1$).}
In the same way, from the asymptotic form of the classical Bessel
functions (Abramowitz \& Stegun \cite{abramo}), we have in the
explosive case ($\Omega>0$)
\begin{eqnarray}
\gamma<\frac{4}{3} & \Rightarrow & \frac{\delta T_-^k}{\rho_\zero}
\propto \lim_{t \rightarrow \infty} (\Omega t)^{\frac{\gamma -
5/3}{2}} \cos{\left[(\Omega t)^{\frac{4}{3}-\gamma} \right]} = 0, \\
\label{omplusgammasup43}
\gamma>\frac{4}{3} & \Rightarrow & \frac{\delta T_-^k}{\rho_\zero} \propto
\lim_{t \rightarrow \infty} (\Omega t)^{\frac{2}{3}} = \infty.
\end{eqnarray}
and for implosions ($\Omega<0$)
\begin{eqnarray}
\label{omneggammainf43}
\gamma<\frac{4}{3} & \Rightarrow & \frac{\delta T_-^k}{\rho_\zero} \propto
\lim_{t \rightarrow -\frac{1}{\Omega}} (1+\Omega t)^{-1} = \infty, \\
\frac{4}{3}<\gamma<\frac{5}{3} & \Rightarrow & \frac{\delta
T_-^k}{\rho_\zero} \propto \lim_{t \rightarrow -\frac{1}{\Omega}}
\frac{\cos{ \left[ (1+\Omega t)^{\frac{4}{3}-\gamma}
\right]}}{(1+\Omega t)^{\frac{5/3-\gamma}{2}}} = \infty, \\
\gamma>\frac{5}{3} & \Rightarrow & \frac{\delta T_-^k}{\rho_\zero} \propto
\lim_{t \rightarrow -\frac{1}{\Omega}} \frac{\cos{\left[(1+\Omega
t)^{\frac{4}{3}-\gamma} \right]} }{(1+\Omega
t)^{\frac{5/3-\gamma}{2}}} = 0.
\end{eqnarray}
It turns out that, for an expanding background, all modes with
polytropic exponent $\gamma<4/3$ are stable. This property is valid
for any value of the wave number $k$. For the collapsing case, only
the modes with $\gamma>5/3$ vanish. In particular, and in the frame of
this simple model, the core of a supernova, which can be described by
a polytrope with $\gamma \simeq 2 $, is stable during the implosion
regarding the evolution of density perturbations.\\ On the other hand,
from Eq.~(\ref{zerolimitEP}) and Eq.~(\ref{inftylimit}), the bessel
function of the second kind, $Y_n$, behaves near the origin, like a
power divergent function, whereas it is oscillating for arguments
greater than the first zero. It is, therefore, possible that the
initial value of the argument $z$ be greater than the first zero of
$Y_n$. In addition, if $z$ decreases with time, we may have transient
oscillating modes.\\ Let $z_n^\zero$ be the first zero of $Y_n$. From
equation (\ref{argument}), the argument $z(0)$ of $Y_n$ at $t=0$ is
\begin{equation}
z(0)=\frac{N_k}{|\mu|}=\sqrt{6}\frac{k
\hat{c}_\zero}{\hat{\Omega}_{\mathrm{J}}|4-3\gamma|}.
\end{equation}
Consequently, for a given $\gamma$, the value of this argument can be
greater than $z_n^\zero$ if the wave number is large enough, and
satisfies
\begin{equation}
k>\frac{|4-3\gamma|}{\sqrt{6}} z_n^\zero k_{\mathrm{J}} =
k_{\mathrm{J},\gamma}.
\end{equation}
In this derivation, we have used equation (\ref{kjeans}). At time $t$,
the argument of $Y_n$ is written from equation (\ref{argument})
\begin{equation}
z(t)=z(0)(1+\Omega t)^{\frac{4}{3}-\gamma},
\end{equation}
and provided the condition $\Omega (\gamma-4/3)>0$ is satisfied, the
variable $z$ will decrease to zero as the time will elapse. This is
the proof of our claim that, if $z(0)>z_n^\zero$, the first zero will be
crossed over in that case. The consequence for the evolution is that
the expanding (resp. collapsing) configuration will oscillate for
$\gamma>4/3$ (resp. $\gamma<4/3$) before the final divergence at $t
\rightarrow \infty$ (resp. $t \rightarrow -1/\Omega$). The amplitude
of such oscillating modes increases with time and as soon as
$z<z_n^\zero$, the mode grows as a power of time according to
Eq.~(\ref{omplusgammasup43}) and Eq.~(\ref{omneggammainf43}). Note
that this behaviour is observed only for eigenmodes with
$k>k_{J,\gamma}$. The other ones grow immediatly as a power of
time. This is also a result found by Bouquet (\cite{bouquet99}) and he
calls it the ``dynamic Jeans criterion''. In fact, this is just a
change in the behaviour, but it might lead to a true criterion in an
improved model. Thus, we have
\begin{eqnarray}
k< k_{J,\gamma} & \Rightarrow & \frac{\delta T_-^k}{\rho_\zero}
\rightarrow \infty \quad \textrm{no oscillation},\\ k>
k_{\mathrm{J},\gamma} & \Rightarrow & \frac{\delta T_-^k}{\rho_\zero}
\rightarrow \infty \quad \textrm{transient oscillations}.
\end{eqnarray}

\subsection{Perturbations in a finite medium}

Each eigenmode, of wave number $k$, can be written as
(\ref{separation})
\begin{equation}
\delta \rho_{\varepsilon}^k(m,t)=\delta R_{\varepsilon}^k(m) \delta
T_{\varepsilon}^k(t).
\end{equation}
Moreover, we have to distinguish between the two cases $\gamma=4/3$
and $\gamma \neq 4/3$.  The evolution of the radius, $R$, of a
configuration with total mass, $M$, is given by Eq.~(\ref{rzero}) and
the special form of the density, Eq.~(\ref{rhozero}), means that the
mass is preserved during the evolution. Considering that the
configuration is embedded into the interstellar medium, we consider
that the pressure remains constant at the surface $r=R(t)$. The
equation of state (\ref{equetat}) and the continuity of the pressure
through the surface make the density perturbation zero at
$r=R(t)$. Thus, we must have $\delta \rho(M,t)=0$ at all times. Since
each eigenmode has its own time variation, this condition should be
applied to each of them. Equations (\ref{masstrigo}) and
(\ref{masshyper}) provide respectively, for all $k$
\begin{eqnarray}
\delta \rho_{\varepsilon=-1}^k(M,t)  =  0 & \Rightarrow &
\frac{\sin(k \hat{R})}{k\hat{R}}=0, \\
\delta \rho_{\varepsilon=1}^k(M,t)  =  0 & \Rightarrow &
\frac{\sinh(k \hat{R})}{k\hat{R}}=0,
\end{eqnarray}
where $\hat{R}$ is the initial radius of the configuration. The second
condition leads to $k=0$, and, thus, all hyperbolic modes are
zero. The first one (trigonometric modes, $\varepsilon=-1$) gives a
quantification for the values of the wave number $k$
\begin{equation}
\label{discretisation}
k_q=\frac{\pi q}{\hat{R}}=\left(\frac{\hat{\Omega}_{\mathrm{J}}^{2}}{3 \Gc}
\right)^{\frac{1}{3}} \frac{\pi q}{M^{\frac{1}{3}}}, \qquad
\textrm{with} \quad q\in \mathbf{N^*}.
\end{equation}
As we can see in this equation, the lagrangian representation with
$(\hat{R},t)$ is more useful than the $(M,t)$ one. However, both of
them are strictly equivalent and the connection between the
$(\hat{r}_\zero,t)$ and the $(m,t)$ coordinates is obtained from the
conservation of mass
\begin{equation}
\label{partielle}
\frac{\partial}{\partial m}=\frac{1}{4 \pi r_\zero^{3} \rho_\zero}
\frac{\partial}{\partial r_\zero} =\frac{1}{4 \pi \hat{r}_\zero^{3}
\hat{\rho}_\zero} \frac{\partial}{\partial \hat{r}_\zero}.
\end{equation}
The most general expression for the density perturbation is a discrete
sum over the eigenmodes satisfaying Eq.~(\ref{discretisation})
\begin{equation}
\label{sommation}
\delta \rho(\hat{r}_\zero,t)=\sum_{q=1}^\infty \delta
\rho_{\varepsilon=-1}^{k_q}(\hat{r}_\zero,t).
\end{equation}
\subsubsection{Case $\gamma \ne 4/3$.}
Plugging Eq.~(\ref{discretisation}) into Eq.~(\ref{modetemporelneg})
and using Eq.~(\ref{separation}) the density perturbation is,
therefore
\begin{eqnarray}
\label{perturbneq}
\lefteqn{\delta \rho(\hat{r}_\zero,t)=(1+\Omega t)^{-\frac{13}{6}}
\sum_{q=1}^\infty \frac{\sin(k_q \hat{r}_\zero)}{k_q \hat{r}_\zero} {}
} \nonumber \\ & {}\displaystyle \times \left\{ \alpha_q
J_n\left[\frac{N_{q}(1+\Omega t)^{\mu}}{|\mu|}\right] + \beta_q
Y_n\left[\frac{N_{q}(1+\Omega t)^{\mu}}{|\mu|} \right] \right\},
\end{eqnarray}
where constant factors and integration constants have been absorbed in
the coefficients $\alpha_q$ and $\beta_q$. The orthonormalization of
trigonometric functions allow us to find their expressions (Abramowitz
\& Stegun \cite{abramo}). For the sake of simplicity, let us
introduce the quantities
\begin{eqnarray}
I_{\rho}^p & = & \int_0^{2\hat{R}} \delta \rho(\hat{r}_\zero,0)
\hat{r}_\zero \sin(k_p \hat{r}_\zero) \, \ud \hat{r}_\zero, \\
\dot{I}_{\rho}^p & = &\int_0^{2\hat{R}} \frac{\partial \delta
\rho}{\partial t} (\hat{r}_\zero,0) \hat{r}_\zero \sin(k_p
\hat{r}_\zero) \, \ud \hat{r}_\zero
\end{eqnarray}
\begin{eqnarray}
J_{n}^{p} = J_n\left(\frac{k_p \hat{c}_\zero}{|\mu \Omega|}\right), & &
J_{n}^{p'} =\frac{\ud J_n (x)}{\ud x}\left|_{x=\frac{k_p \hat{c}_\zero}{|\mu
\Omega|}}\right.,\\ Y_{n}^{p} = Y_n\left(\frac{k_p \hat{c}_\zero}{|\mu
\Omega|}\right), & & Y_{n}^{p'} =\frac{\dd Y_n (x)}{\dd x}
\left|_{x=\frac{k_p \hat{c}_\zero}{|\mu \Omega|}}\right..
\end{eqnarray}
After some algebra, one gets,
\begin{eqnarray}
\label{alphabeta}
\alpha_p = \frac{\displaystyle \frac{|\mu \Omega|}{ \mu \Omega}
\frac{Y_n^p}{\hat{c}_\zero}\dot{I}_\rho^p - \left( k_p
Y_n^{p'}-\frac{13}{6} \frac{|\mu \Omega|}{\mu \hat{c}_\zero} \right)
I_\rho^p}{{\hat{R} (J_{n-1}^p Y_n^p - J_n^p Y_{n-1}^p)}},\\ \beta_p =
\frac{\displaystyle \frac{|\mu \Omega|}{ \mu \Omega}
\frac{J_n^p}{\hat{c}_\zero}\dot{I}_\rho^p -\left( k_p
J_n^{p'}-\frac{13}{6} \frac{|\mu \Omega|}{\mu \hat{c}_\zero} \right)
I_\rho^p}{{\hat{R} (J_n^p Y_{n-1}^p - J_{n-1}^p Y_n^p)}}.
\end{eqnarray}
\subsubsection{Case $\gamma=4/3$.}
The discretization of wave numbers obeys
Eq.~(\ref{discretisation}). However, because of the critical value,
$k_{\mathrm{trans}}$ [see Eq.~(\ref{modetempplus}) to
Eq.~(\ref{ktranslag})], of the wave number, $k$, we are obliged to
separate the sum in two parts. With $p_{\mathrm{trans}} \in
\mathbf{N}$ defined as
\begin{equation}
p_{\mathrm{trans}}=\mathrm{Int} \left(\frac{5}{6}\frac{|\Omega|}{\pi
\hat{c}_\zero} \hat{R}\right),
\end{equation}
the perturbation is expressed as
\begin{eqnarray}
\label{perturbeq}
\lefteqn{\delta \rho(\hat{r}_\zero,t) =
\sum_{q=1}^{p_{\mathrm{trans}}-1} \frac{\sin(k_q \hat{r}_\zero)}{k_q
\hat{r}_\zero} \left[\beta_q (1+\Omega t)^{s_+^{-}(k_q)} \right. {}}
\nonumber\\ {} & & + \left. \gamma_q(1+\Omega t)^{s_-^{-}(k_q)}\right]
+ \sum_{q=p_{\mathrm{trans}}}^\infty \frac{\sin(k_q
\hat{r}_\zero)}{k_q \hat{r}_\zero} (1+\Omega t)^{-\frac{13}{6}}
{} \nonumber\\ {} & & \times \alpha_q
\cos\frac{\sqrt{-\Delta_{k_q}^{-}}\ln{(1+\Omega t)}}{2}.
\end{eqnarray}
As in the case $\gamma \neq 4/3$, the initial conditions, $\delta
\rho(\hat{r}_\zero,0)$ and $\delta \dot{\rho}(\hat{r}_\zero,0)$, completely
define the parameters $\alpha_q$, $\beta_q$ and $\gamma_q$. From the
orthonormalization conditions, it comes
\begin{eqnarray}
\alpha_p & = & \frac{\pi p}{\hat{R}^{2}} I_{\rho}^p,\\ \beta_p & = &
\frac{\hat{\Omega}_{\mathrm{J}} k_p}{\sqrt{\Delta_{k_p}^{-}}}
\left[\dot{I_\rho^p}- s_-^-(k_p)I_\rho^p \right], \\ \gamma_p & = &
\frac{\hat{\Omega}_{\mathrm{J}} k_p}{\sqrt{\Delta_{k_p}^{-}}}
\left[-\dot{I_\rho^p} -s_+^-(k_p)I_\rho^p \right].
\end{eqnarray}
\section{Conclusion}
\label{dernierchapitre}
In this paper, we have studied the dynamic stability of a homogeneous
collapsing or expanding spherical polytropic configuration. In
opposition to the usual studies performed up to now, we have used the
lagrangian formalism instead of the eulerian one. It turns out that
the polytrope must be split in the two cases $\gamma=4/3$ and $\gamma
\neq 4/3$. This is not really surprising because the
$\gamma=4/3$--polytrope is highly self-similar: $\partial v / \partial
t \propto v\partial v/\partial r \propto (1/\rho)\partial p/\partial r
\propto g \propto (1+\Omega t)^{-4/3}$ (Blottiau et
al. \cite{blottiau88}). However the langrangian approach makes the
study more difficult than the eulerian one because of the lack of
dispersion relation. Nevertheless, we have been able to come to a
conclusion about the gravitational stability and, unexpectedly, it
comes out that the polytrope $\gamma=5/3$ also plays a special role.\\
Let us come back to the particular $\gamma=4/3$--polytrope in more
detail. In spite of the decelerated (or accelerated) motion of the
expanding background, part of the stability criterion is still given
by the Jeans' result derived for a static configuration (Jeans
\cite{jeans}). We recover the classical threshold for the wave number,
$k_{\mathrm{J}}=\hat{\Omega}_{\mathrm{J}}/\hat{c}_\zero$, but, in
addition, a second pivot value,
$k_{\mathrm{trans}}=\sqrt{25/24}~k_{\mathrm{J}}$, separates oscillating
solutions ($k>k_{\mathrm{trans}}$) from monotonic ones ($k<k_{\mathrm{trans}}$), and
both of them are stable provided $k>k_{\mathrm{J}}$.  It is really
amazing that the macroscopic expanding motion of the background does
not alter the Jeans' criterion. In our opinion, this is due to the
beautiful property of ``sharp'' self-similarity. Collapses behave in a
quite different way: although, the pivot value, $k_{\mathrm{trans}}$, plays
exactly the same role as in expansions, we find that any disturbance
is instable.\\ Now, let us examine the case $\gamma \neq 4/3$. As
written above, the lagrangian treatment does not lead to a dispersion
relation.  The condition derived by Buff \& Gerola (\cite{buff}) does
not agree with our results. According to us, the time variation of the
coefficients arising in their linearized dispersion equation has not
been taken into account.  In opposition to the case $\gamma=4/3$, we
find that the stability does not depend any longer on the value of the
wave number (excepted for the apparition of transcient oscillating
phases). The critical parameter for the stability is just the value of
the polytropic exponent $\gamma$.  For expansions, $\gamma=4/3$ is a
threshold and stability (resp. unstability) is obtained for
$\gamma<4/3$ (resp. $\gamma>4/3$). Collapses are more complicated
since two critical values arise, i.e. $\gamma=4/3$ and
$\gamma=5/3$. For $\gamma<5/3$, unstable collapse occurs with
monotonic (resp. oscillating) behaviour for $\gamma<4/3$ (resp. for
$\gamma>4/3$). On the other hand, for $\gamma>5/3$, collapses are
always stable. To our knowledge, this is the first time that the
$\gamma=5/3$ has been derived as a threshold for gravitational
stability. The case $\gamma=4/3$ is not surprising, it corresponds to
a perfect gas of photons plus matter and it is very relevant in
astrophysics (Chandrasekhar \cite{chandrasekhar}). The value
$\gamma=5/3$ corresponds to the monoatomic perfect gas but, up to now,
we have not been able to associate this value with a specially
important phenomena in astrophysics.\\ Finally, it would be very
interesting to check numerically these theoritical predictions. This
will be the next step in further studies.

\section*{acknowledgements}
The authors thank Prof. R.N. Henriksen for his pertinent comments and
for providing relevant references dealing with this problem.

\WriteThisInToc
\listoffigures

\CenterTitles

\AbstractsOnEvenPage
\NumberAbstractPages

{\footnotesize

\begin{ThesisAbstract}
\begin{FrenchAbstract}
Cette th\`ese est une \'etude d\'etaill\'ee de la structure interne
d'objets cosmologiques, de type d\'efaut topologique et membrane en
dimensions suppl\'ementaires, poss\'edant des courants de fermions.

La premi\`ere partie pr\'esente le cadre g\'en\'eral de la cosmologie
primordiale dans lequel ces objets peuvent se former lors des brisures
de sym\'etrie associ\'ees aux th\'eories de physique des particules.
Dans la deuxi\`eme partie, la dynamique des cordes cosmiques, une
classe privil\'egi\'ee de d\'efauts, est d\'ecrite \`a l'aide d'un
formalisme macroscopique covariant. Qu'elles soient ou non parcourues
par des courants, ce formalisme offre une description unifi\'ee des
cordes permettant de pr\'edire leur \'evolution cosmologique par le
biais de simulations num\'eriques. Apr\`es avoir justifi\'e la
validit\'e de l'approche macroscopique pour des cordes poss\'edant un
courant de particules scalaires, le cas des courants fermioniques est
d\'etaill\'e dans la troisi\`eme partie. Dans un premier temps, le
spectre de masse des fermions pi\'eg\'es dans une corde est
d\'etermin\'e, et sugg\`ere que leurs modes de propagation
privil\'egi\'es sont de masse non nulle. Dans un deuxi\`eme temps,
l'\'equation d'\'etat d'une telle corde est obtenue \`a l'aide d'une
quantification des modes de propagation des fermions le long de la
corde. Il appara\^\i t que l'approche macroscopique usuelle \`a un
param\`etre n'est pas toujours suffisante pour d\'ecrire les
fermions. D'autres parts, contrairement aux cordes parcourues par des
bosons, les courants de fermions engendrent des transitions, dans la
dynamique des cordes, entre des r\'egimes subsoniques et supersoniques
dont les cons\'equences cosmologiques pourraient \^etre
importantes. La quatri\`eme partie g\'en\'eralise ces r\'esultats au
cas d'un mur de domaine quadri-dimensionnel, mod\'elisant notre
univers, et plong\'e dans un espace-temps \`a cinq dimensions. Dans le
cadre du mod\`ele de Randall-Sundrum, il est ainsi possible de
pr\'edire la masse des fermions stables pi\'eg\'es sur la membrane.

\KeyWords{Cosmologie, d\'efauts topologiques, dimensions
suppl\'ementaires, fermions}
\end{FrenchAbstract}

\begin{EnglishAbstract}
This PhD thesis discusses the internal structure of topological
defects, and branes in extra-dimensions, carrying fermionic currents.

The general framework in which these objects may appear, from
spontaneous symmetry breaking, is at the frontier between particle
physics and cosmology, and is presented in the first part. The second
part is devoted to the dynamic of cosmic strings, a class of
topological defects of uttermost importance to modern cosmology, as it
can be obtained from a macroscopic covariant formalism. This formalism
offers a unified description of cosmic strings, including the case for
which they carry internal currents, and allows the study of their
cosmological evolution, and implications, by means of numerical
simulations. Its validity has already been confirmed for cosmic string
carrying bosonic currents, and the third part provides new results
concerning the fermionic currents case. First, the fermion mass
spectrum in a cosmic string is computed, and suggests that fermionic
currents are usually built on massive propagation modes. As a result,
a new equation of state describing such cosmic strings is derived by
means of a quantization of the relevant spinor fields along the
string. This highlights that the usual one parameter macroscopic
formalism is not always sufficient in case of fermionic
currents. Moreover, contrary to the bosonic case, the dynamics
stemming from this equation of state exhibits transitions between the
subsonic and supersonic regimes whose consequences in cosmology could
be important. The last part is an extension of these results to brane
cosmology for which our universe is viewed as a four dimensional
domain wall embedded in a five dimensional space-time. In the
framework of the Randall-Sundrum model, the masses of the trapped
fermions on the brane can be predicted.

\KeyWords{Cosmology, topological defects, extra-dimensions, fermions}
\end{EnglishAbstract}
\end{ThesisAbstract}

}


\begin{thebibliography}{399}
{\footnotesize


\bibitem{abbott82}
L.~Abbott, E.~Farhi \ET M.~Wise, \journal{Phys. Lett.} \numero{B117},
\page{29} \annee{1982}.



\bibitem{abramo}
M.~Abramowitz \ET C.A.~Stegun, \journal{Handbook of Mathematical
Functions with Formulas, Graph, and Mathematical Tables},
\editeur{Dover Publications} \annee{1972}.



\bibitem{abrikosov57}
A.~A.~Abrikosov, \journal{Sov. Phys. JETP} \numero{5}, \page{1174}
\annee{1957}.



\bibitem{abrikosov59}
A.~A.~Abrikosov, L.~P.~Gorkov \ET I.~E.~Dzyaloshinski,
\journal{Sov. Phys. JETP} \numero{9}, \page{636} \annee{1959}.



\bibitem{achucarro92}
A.~Achucarro \ETAL, \journal{Nucl. Phys.} \numero{B388}, \page{435}
\annee{1992}.



\bibitem{adler}
S.~L.~Adler \ET T.~Piran, \journal{Rev. Mod. Phys.} \numero{56},
\page{1} \annee{1984}.



\bibitem{akama82}
K.~Akama, \journal{Lect. Notes Phys.} \numero{176}, \page{267}
\annee{1982}.



\bibitem{akhmedov01}
E.~K.~Akhmedov, \journal{Phys.Lett.} \numero{B521}, \page{79}
\annee{2001}, \eprint{hep-th/0107223}.



\bibitem{albrecht85}
A.~Albrecht \ET N.~Turok, \journal{Phys. Rev. Lett.} \numero{54},
\page{1868} \annee{1985}.



\bibitem{albrecht82}
A.~Albrecht \ETAL, \journal{Phys. Rev. Lett.} \numero{48},
\page{1437} \annee{1982}.



\bibitem{albrecht97}
A.~Albrecht, R.~A.~Battye \ET J.~Robinson, \journal{Phys. Rev. Lett.}
\numero{79}, \page{4736} \annee{1997}, \eprint{astro-ph/9707129}.



\bibitem{albrecht99}
J.~Albrecht, R.~A.~Battye \ET J.~Robinson, \journal{Phys. Rev.}
\numero{D59}, \page{023508} \annee{1999}, \eprint{astro-ph/9711121}.



\bibitem{allen90}
B.~Allen \ET E.~P.~S.~Shellard, \journal{Phys. Rev. Lett.} \numero{64},
\page{119} \annee{1990}.



\bibitem{allen92}
B.~Allen \ET E.~P.~S.~Shellard, \journal{Phys. Rev.} \numero{D45},
\page{1898} \annee{1992}.



\bibitem{allen97}
B.~Allen \ETAL, \journal{Phys. Rev. Lett.} \numero{79}, \page{2624}
\annee{1997}, \eprint{astro-ph/9704160}.



\bibitem{ambjorn88}
J.~Ambj{\o}rn, N.~K.~Nielsen \ET P.~Olesen, \journal{Nucl. Phys.}
\numero{B310}, \page{625} \annee{1988}.



\bibitem{antoniadis98}
I.~Antoniadis, S.~Dimopoulos \ET G.~Dvali, \journal{Nucl. Phys.}
\numero{B516}, \page{70} \annee{1998}, \eprint{hep-ph/9710204}.



\bibitem{arkani98}
N.~Arkani-Hamed, S.~Dimopoulos \ET G.~Dvali, \journal{Phys. Lett.}
\numero{B429}, \page{263} \annee{1998}, \eprint{hep-ph/9803315}.



\bibitem{arkani99}
N.~Arkani-Hamed, S.~Dimopoulos \ET G.~Dvali, \journal{Phys. Rev.}
\numero{D59}, \page{086004} \annee{1999}, \eprint{hep-ph/9807344}.



\bibitem{atick87}
J.~Atick, L.~Dixon \ET A.~Sen, \journal{Nucl. Phys.} \numero{B292},
\page{109} \annee{1987}.



\bibitem{avelino99}
P.~P.~Avelino \ETAL, \journal{Phys. Rev.} \numero{D60}, \page{023511}
\annee{1999}.



\bibitem{baade}
W.~Baade, \journal{Pub. Astron. Soc. Pacific} \numero{68}, \page{5}
\annee{1956}.



\bibitem{bps} 
A.~Babul, T.~Piran \ET D.~N.~Spergel, \journal{Phys. Lett.}
\numero{B202}, \page{307} \annee{1988}.



\bibitem{bailinlove}
D.~Bailin \ET A.~Love, \journal{Supersymmetric Gauge Field Theory and
String Theory}, \editeur{Inst. Phys. Pub., F.~Brewer} \annee{1994}.



\bibitem{bajc}
B.~Bajc \ET G.~Gabadadz\'e, \journal{Phys. Lett.}
\numero{B474}, \page{282} \annee{2000}, \eprint{hep-th/9912232}.



\bibitem{barbieri82}
R.~Barbieri, S.~Ferrara \ET C.~A.~Savoy, \journal{Phys. Lett.}
\numero{B119}, \page{343} \annee{1982}



\bibitem{battye97}
R.~A.~Battye, J.~Robinson \ET A.~Albrecht, \journal{Phys. Rev. Lett.}
\numero{80}, \page{4847} \annee{1998}, \eprint{astro-ph/9711336}.



\bibitem{shiro}
R.~Battye \ETAL, \journal{Phys. Rev.} \numero{D64}, \page{124007}
\annee{2001}, \eprint{hep-th/0105091}.



\bibitem{bennett91}
D.~P.~Bennett \ET F.~R.~Bouchet, \journal{Phys. Rev.} \numero{D43},
\page{2733} \annee{1991}.



\bibitem{bennett89}
D.~P.~Bennett \ET F.~R.~Bouchet, \journal{Phys. Rev. Lett.}
\numero{63}, \page{2776} \annee{1989}



\bibitem{bennett90}
D.~P.~Bennett \ET F.~R.~Bouchet, \journal{Phys. Rev.} \numero{D41},
\page{2408} \annee{1990}.



\bibitem{bennett86}
D.~P.~Bennett, \journal{Phys. Rev.} \numero{D34}, \page{3592}
\annee{1986}.



\bibitem{bennett88}
D.~P.~Bennett \ET F.~R.~Bouchet, \journal{Phys. Rev. Lett.}
\numero{60}, \page{257} \annee{1988}.



\bibitem{bernardeau00}
F.~Bernardeau \ET J.-P.~Uzan, \journal{Phys. Rev.} \numero{D23}, \page{023005} \annee{2000}, \eprint{astro-ph/0004102}.



\bibitem{boomerang2}
P.~de~Bernardis \ETAL, \journal{Nature} \numero{404}, \page{955}
\annee{2000}, \eprint{astro-ph/0105296}.



\bibitem{binetruy98b}
P.~Bin\'etruy, C.~Deffayet \ET P.~Peter, \journal{Phys. Lett.}
\numero{B441}, \page{52} \annee{1998}, \eprint{hep-ph/9807233}.



\bibitem{binetruy98}
P.~Bin\'etruy, \journal{Phys. Rev.} \numero{D60}, \page{063502}
\annee{1999}, \eprint{hep-ph/9810553}.



\bibitem{binetruy00b}
P.~Bin{\'e}truy \ETAL, \journal{Phys. Lett.} \numero{B477},
\page{285} \annee{2000}, \eprint{hep-th/9910219}.



\bibitem{binetruy00}
P.~Bin{\'e}truy, C.~Deffayet \ET D.~Langlois, \journal{Nucl. Phys.}
\numero{B565}, \page{269} \annee{2000}, \eprint{hep-th/9905012}.



\bibitem{birrell}
N.~D.~Birrell \ET P.~C.~W.~Davies,
\journal{Quantum fields in curved space}, \editeur{Cambridge
University Press, Cambridge, England} \annee{1994}.



\bibitem{bjorken69}
J.~D.~Bjorken \ET E.~A.~Paschos, \journal{Phys. Rev.} \numero{185},
\page{1975} \annee{1969}.



\bibitem{blottiau88}
P.~Blottiau, S.~Bouquet \ET J.-P.~Chi\`eze, \journal{Astron. Astrophys.}
\numero{207}, \page{24} \annee{1988}.



\bibitem{bondi48}
H.~Bondi \ET T.~Gold, \journal{M.N.R.A.S} \numero{108}, \page{252}
\annee{1948}.



\bibitem{charmousis}
F.~Bonjour, C.~Charmousis \ET R.~Gregory, \journal{Class. Quant. Grav.}
\numero{16}, \page{2427} \annee{1999}, \eprint{gr-qc/9902081}.



\bibitem{bonnor}
M.~Bonnor, \journal{Mont. Not. Roy. Astron. Soc.} \numero{117},
\page{104}
\annee{1957}.



\bibitem{bouchet90}
F.~R.~Bouchet \ET D.~P.~Bennett, \journal{Phys. Rev.} \numero{D41},
\page{720} \annee{1990}.



\bibitem{bouchet88}
F.~R.~Bouchet, D.~P.~Bennett \ET A.~Stebbins, \journal{Nature}
\numero{335}, \page{410}
\annee{1988}.



\bibitem{bouchet02}
F.~R.~Bouchet \ETAL, \journal{Phys. Rev.} \numero{D65},
\page{021301} \annee{2002}, \eprint{astro-ph/0005022}.



\bibitem{bouquet99}
S.~Bouquet, \journal{Dynamical Systems, Plasmas and Gravitation,
Lectures Notes in Physics} \numero{518}, \editeur{P.~G.~L.~Leach,
S.~Bouquet, J.~L.~Rouet, E.~Fijalkow, Springer--Verlag} \annee{1999}



\bibitem{bouquet95}
S.~Bouquet, \journal{J. Math. Phys.} \numero{36}, \page{1242}
\annee{1995}.



\bibitem{bouquet85b}
S.~Bouquet, L.~Cair\'o \ET M.~R.~Feix, \journal{J. Plas. Phys.}
\numero{34}, \page{127} \annee{1985}.



\bibitem{bouquet85a}
S.~Bouquet \ETAL, \journal{Astrophys. J.} \numero{293},
\page{494} \annee{1985}.



\bibitem{brandi85}
R.~Brandenberger, \journal{Rev. Mod. Phys.} \numero{57}, \page{1}
\annee{1985}.



\bibitem{brandi96}
R.~Brandenberger \ETAL, \journal{Phys. Rev.} \numero{D54},
\page{6059} \annee{1996}, \eprint{hep-ph/9605382}.



\bibitem{brax99}
P.~Brax \ET J.~Martin, \journal{Phys. Lett.} \numero{B468}, \page{40}
\annee{1999}, \eprint{astro-ph/9905040}.



\bibitem{brax01}
P.~Brax, J.~Martin \ET A.~Riazuelo, \journal{Phys. Rev.} \numero{D64},
\page{083505} \annee{2001}, \eprint{hep-ph/0104240}.



\bibitem{buff}
J.~Buff \ET H.~Gerola, \journal{Astrophys. J.} \numero{230}, \page{839}
\annee{1979}.



\bibitem{burgan83}
J.~R.~Burgan \ETAL, \journal{J. Plas. Phys.} \numero{29}, \page{139}
\annee{1983}.



\bibitem{burgan78}
J.~R.~Burgan \ETAL, \journal{Strongly Coupled Plasmas}, \editeur{G.~Kalman,
P.~Carini, Plenum Publishing Company, New-York} \annee{1978}.



\bibitem{cairo99}
L.~Cair\'o \ET M.~R.~Feix, \journal{J. Math. Phys.} \numero{40},
\page{2074} \annee{1999}.



\bibitem{cairo98}
L.~Cair\'o \ET M.~R.~Feix, \journal{Extr. Math.} \numero{1}, \page{1}
\annee{1998}.



\bibitem{sigma}
B.~Carter, \journal{Ann. Phys.} \numero{9},
\page{247} \annee{2000}, \eprint{hep-th/0002162}.



\bibitem{carter97}
B.~Carter, P.~Peter \ET A.~Gangui, \journal{Phys. Rev.} \numero{D55},
\page{4647} \annee{1997}, \eprint{hep-ph/9609401}.



\bibitem{carter93}
B.~Carter \ET X.~Martin, \journal{Ann. Phys. (N.Y.)} \numero{227},
\page{151} \annee{1993}.



\bibitem{carter00}
B.~Carter \ET A.-C.~Davis, \journal{Phys. Rev.} \numero{D61},
\page{123501} \annee{2000}, \eprint{hep-ph/9910560}.



\bibitem{carter92}
B.~Carter, \journal{J. Geom. Phys.} \numero{8}, \page{53}
\annee{1992}.



\bibitem{carter91}
B.~Carter, \journal{Ann. (N.Y.)
Acad.Sci.} \numero{647}, \page{758} \annee{1991}.



\bibitem{carter90}
B.~Carter, \journal{Phys. Lett.} \numero{B238}, \page{166}
\annee{1990}.



\bibitem{carter97b}
B.~Carter \annee{1997}, \eprint{hep-th/9705172}.



\bibitem{carter94b}
B.~Carter, \journal{Nucl. Phys.} \numero{B412}, \page{345}
\annee{1994}.



\bibitem{carter89b}
B.~Carter, \journal{Phys. Lett.} \numero{B228}, \page{466} \annee{1989}.



\bibitem{carter90b}
B.~Carter, \journal{Phys. Rev.} \numero{D41}, \page{3869}
\annee{1990}.



\bibitem{carter92b}
B.~Carter, \journal{Class. Quant. Grav.} \numero{9}, \page{19}
\annee{1992}.



\bibitem{carter94}
B.~Carter, M.~Sakellariadou \ET X.~Martin, \journal{Phys. Rev.}
\numero{D50}, \page{682} \annee{1994}.



\bibitem{carter95}
B.~Carter, \journal{Proceedings of the XXXth Rencontres de Moriond,
Villard--sur--Ollon}, \editeur{B.~Guiderdoni and J.~Tran Thanh V\^an,
Editions Fronti\`eres, Gif--sur--Yvette} \annee{1995}.



\bibitem{carpet1}
B.~Carter \ET P.~Peter, \journal{Phys. Rev.} \numero{D52}, \page{1744}
\annee{1995}, \eprint{hep-ph/9411425}.



\bibitem{carter}
B.~Carter \ET J.-P.~Uzan, \journal{Nucl. Phys.} \numero{B606}, \page{45}
\annee{2001}, \eprint{gr-qc/0101010}.



\bibitem{bcpc}
B.~Carter, private communication.



\bibitem{cartermeca}
B.~Carter, \journal{The Formation and Evolution of Cosmic Strings},
\editeur{G.~Gibbons, S.~Hawking \ET T.~Vachaspati, Cambridge},
\page{143} \annee{1990}.



\bibitem{carter89}
B.~Carter, \journal{Phys. Lett.} \numero{B224}, \page{61}
\annee{1989}.



\bibitem{cartermartin93}
B.~Carter \ET X.~Martin, \journal{Ann. Phys.} \numero{227}, \page{151}
\annee{1993}.



\bibitem{carpet2}
B.~Carter \ET P.~Peter, \journal{Phys. Lett.} \numero{B466}, \page{41}
\annee{1999}, \eprint{hep-th/9905025}.



\bibitem{casas89}
A.~Casas \ET C.~Mu\~{n}oz, \journal{Phys. Lett.} \numero{B216},
\page{37} \annee{1989}.



\bibitem{casas89b}
A.~Casas \ETAL, \journal{Nucl.  Phys.} \numero{B328}, \page{272}
\annee{1989}.



\bibitem{chandrasekhar}
S.~Chandrasekhar, \journal{The Study of
Stellar Structure}, \editeur{Dover Publication, New-York} \annee{1967}.



\bibitem{chieze}
J.-P.~Chi\`eze, R.~Teyssier \ET J.-M.~Alimi, \journal{Astrophys. J.}
\numero{484}, \page{40} \annee{1997}.



\bibitem{christ75}
N.~H.~Christ \ET T.~D.~Lee, \journal{Phys. Rev.}
\numero{D12}, \page{1606} \annee{1975}.



\bibitem{chung}
D.~J.~H.~Chung, H.~Davoudiasl \ET L.~Everett, \journal{Phys. Rev.}
\numero{D64}, \page{065002} \annee{2001}, \eprint{hep-ph/0010103}.



\bibitem{cline78}
D.~B.~Cline \ET F.~E.~Mills, \journal{Unification of Elementary Forces
and Gauge Theories}, \editeur{Harwood Academic Pub., London}
\annee{1978}.



\bibitem{cline99}
J.~M.~Cline, C.~Grojean \ET G.~Servant, \journal{Phys. Rev. Lett.}
\numero{83} \page{4245} \annee{1999}, \eprint{hep-ph/9906523}.



\bibitem{coleman73}
S.~Coleman \ET D.~J.~Gross, \journal{Phys. Rev.} \numero{D8},
\page{4383} \annee{1973}.



\bibitem{coleman67}
S.~Coleman \ET J.~Mandula, \journal{Phys. Rev.} \numero{159},
\page{1251} \annee{1967}.



\bibitem{L301}
L3 Collaboration, \journal{Phys. Lett.} \numero{B517}, \page{319}
\annee{2001}, \eprint{hep-ex/0107054}.



\bibitem{collins}
C.~A.~Collins \ETAL, \journal{Nature} \numero{320}, \page{506}
\annee{1986}.



\bibitem{contaldi99}
C.~Contaldi, M.~Hindmarsh \ET J.~Magueijo, \journal{Phys. Rev. Lett.}
\numero{82}, \page{679} \annee{1999}, \eprint{astro-ph/9809053}.



\bibitem{copeland86}
E.~J.~Copeland \ETAL, \journal{Nucl. Phys.} \numero{B298}, \page{445}
\annee{1986}.



\bibitem{copeland92}
E.~Copeland, T.~W.~B.~Kibble \ET D.~Austen, \journal{Phys. Rev.}
\numero{D45}, \page{1000} \annee{1992}.



\bibitem{copeland99}
E.~J.~Copeland, J.~Magueijo \ET D.~A.~Steer, \journal{Phys. Rev.}
\numero{D61}, \page{063505} \annee{1999}, \eprint{astro-ph/9903174}.



\bibitem{cremmer83}
E.~Cremmer \ETAL, \journal{Nucl. Phys.} \numero{B212}, \page{413}
\annee{1983}.



\bibitem{csaki99}
C.~Cs\'aki \ETAL, \journal{Phys. Lett.} \numero{B462}, \page{34}
\annee{1999}, \eprint{hep-ph/9906513}.



\bibitem{damour01}
T.~Damour \ET A.~Vilenkin, \journal{Phys. Rev.} \numero{D64},
\page{064008} \annee{2001}, \eprint{gr-qc/0104026}.



\bibitem{davisAC97}
A.-C.~Davis \ET S.~C.~Davis, \journal{Phys. Rev.} \numero{D55},
\page{1879} \annee{1997}, \eprint{hep-ph/9608206}.



\bibitem{davisRL88}
R.~L.~Davis \ET E.~P.~S.~Shellard, \journal{Phys. Rev.}
\numero{D38}, \page{4722} \annee{1988}.



\bibitem{davisRL85}
R.~L.~Davis, \journal{Phys. Lett.} \numero{B161}, \page{285}
\annee{1985}.



\bibitem{davisRL}
R.~L.~Davis, \journal{Phys. Rev.} \numero{D38}, \page{3722}
\annee{1988}.



\bibitem{davisRL89}
R.~L.~Davis \ET E.~P.~S.~Shellard, \journal{Nucl. Phys.} \numero{B323},
\page{209} \annee{1989}.



\bibitem{davisRL87}
R.~L.~Davis, \journal{Phys. Rev.} \numero{D36}, \page{2267}
\annee{1987}.



\bibitem{davisS}
S.~C.~Davis, W.~B.~Perkins \ET A.-C.~Davis, \journal{Phys. Rev.}
\numero{D62}, \page{043503} \annee{2000}, \eprint{hep-ph/9912356}.



\bibitem{NATO}
\journal{Formation and Interactions of
Topological Defects, NATO ASI} \numero{B349},
\editeur{R.~Brandenberger
\& A.-C.~Davis, Plenum, New York} \annee{1995}.



\bibitem{dicke70}
R.~H.~Dicke, \journal{Gravitation and the Universe}, \editeur{American
Philosophical Society} \annee{1970}.



\bibitem{dimopoulos01}
P.~Dimopoulos \ETAL, \journal{Nucl.Phys.} \numero{B617}, \page{237}
\annee{2001}, \eprint{hep-th/0007079}.



\bibitem{dine87}
M.~Dine, N.~Seiberg \ET E.~Witten,
\journal{Nucl. Phys.} \numero{B289}, \page{589} \annee{1987}.



\bibitem{dine88}
M.~Dine, I.~Ichinose \ET N.~Seiberg,
\journal{Nucl. Phys.} \numero{B293}, \page{253} \annee{1988}.



\bibitem{dirac33}
P.~A.~M.~Dirac, \journal{Physikalische Zeitschrift der Sowjetunion}
\numero{3}, \page{64} \annee{1933}.



\bibitem{dirac27}
P.~A.~M.~Dirac, \journal{Proc. Roy. Soc. London} \numero{114},
\page{243} \annee{1927}.



\bibitem{dolan74}
L.~Dolan \ET R.~Jackiw, \journal{Phys. Rev.} \numero{D9}, \page{3320}
\annee{1974}.



\bibitem{dolgov82}
A.~Dolgov \ET A.~Linde, \journal{Phys. Lett.} \numero{B116},
\page{329} \annee{1982}.



\bibitem{donagi01}
R.~Y.~Donagi \ETAL, \journal{J.H.E.P.} \numero{0111}, \page{041}
\annee{2001}, \eprint{hep-th/0105199}.



\bibitem{drell71}
S.~D.~Drell \ET T.~M.~Yan, \journal{Ann. Phys. (N.Y.)} \numero{66},
\page{578} \annee{1971}.



\bibitem{dressler}
A.~Dressler \ETAL, \journal{Astrophys. J.} \numero{313}, \page{L37}
\annee{1987}.



\bibitem{dubovski}
S.~L.~Dubovsky, V.~A.~Rubakov \ET P.~G.~Tinyakov,
\journal{Phys. Rev.} \numero{D62}, \page{105011} \annee{2000},
\eprint{hep-th/0006046}.



\bibitem{duff01}
M.~J.~Duff, J.~T.~Liu \ET W.~A.~Sabra, \journal{Nucl. Phys.}
\numero{B605}, \page{234} \annee{2001}, \eprint{hep-th/0009212}.



\bibitem{durrer97}
R.~Durrer \ET M.~Sakellariadou, \journal{Phys. Rev.} \numero{D56},
\page{4480} \annee{1997}, \eprint{astro-ph/9702028}.



\bibitem{durrer99}
R.~Durrer, M.~Kunz \ET A.~Melchiorri, \journal{Phys. Rev.}
\numero{D59}, \page{123005} \annee{1999}, \eprint{astro-ph/9811174}.



\bibitem{dvali97}
G.~Dvali \ET M.~Shifman, \journal{Phys. Lett.} \numero{B396},
\page{64} \annee{1997}, \eprint{hep-th/9612128}.



\bibitem{dvali97b}
G.~Dvali \ET M.~Shifman, \journal{Phys. Lett.} \numero{B407},
\page{452} \annee{1997}, \eprint{hep-th/9612128}.



\bibitem{dvali01}
 G.~Dvali, G.~Gabadadz\'e \ET M.~Shifman, \journal{Phys. Lett.}
\numero{B497}, \page{271} \annee{2001}, \eprint{hep-th/0010071}.


\bibitem{dyson49}
F.~J.~Dyson, \journal{Phys. Rev.} \numero{75}, \page{486}
\annee{1949}.



\bibitem{einstein17}
A.~Einstein, \journal{Preussische Akademie der Wissenschaften,
Sitzungsberichte}, \page{142} \annee{1917}.



\bibitem{einstein15}
A.~Einstein, \journal{Preussische Akademie der Wissenschaften,
Sitzungsberichte}, \page{778} \annee{1915};
\page{799} \annee{1915}; \page{844} \annee{1915}.



\bibitem{englert64}
F.~Englert \ET R.~Brout, \journal{Phys. Rev. Lett.} \numero{13},
\page{321} \annee{1964}.



\bibitem{everett88}
A.~E.~Everett, \journal{Phys. Rev. Lett.} \numero{61}, \page{1807}
\annee{1988}.



\bibitem{faber79}
S.~M.~Faber \ET J.~S.~Gallagher, \journal{Ann. Rev. Astron. Ap.}
\numero{17}, \page{135} \annee{1979}.



\bibitem{fadeev67}
L.~D.~Fadeev \ET V.~N.~Popov, \journal{Phys. Lett.} \numero{B25},
\page{29} \annee{1967}.



\bibitem{feynman58}
R.~P.~Feynman \ET M.~Gell-Mann, \journal{Phys. Rev.} \numero{109},
\page{193} \annee{1958}.



\bibitem{feynmann49}
R.~P.~Feynman, \journal{Phys. Rev.} \numero{76}, \page{769}
\annee{1949}.



\bibitem{feynman69}
R.~P.~Feynman, \journal{Phys. Rev. Lett.} \numero{23}, \page{1415}
\annee{1969}.



\bibitem{feynman48}
R.~P.~Feynman, \journal{Rev. Mod. Phys.} \numero{20}, \page{367}
\annee{1948}.



\bibitem{feynman63}
R.~P.~Feynman, \journal{Acta Phys. Polonica} \numero{24}, \page{697}
\annee{1963}.



\bibitem{flanagan00}
E.~Flanagan, S.~Tye \ET I.~Wasserman, \journal{Phys. Rev.} \numero{D62},
\page{044039} \annee{2000}, \eprint{hep-ph/9910498}



\bibitem{freedman76}
D.~Z.~Freedman, P.~van~Nieuwenhuizen \ET S.~Ferrera,
\journal{Phys. Rev.} \numero{D13}, \page{335} \annee{1976}.



\bibitem{friedmann22}
A.~Friedmann, \journal{Zeitschrift f\"ur Physik} \numero{10},
\page{377} \annee{1922}.



\bibitem{fulling}
S.~A.~Fulling, \journal{Phys. Rev.} \numero{D7}, \page{2850}
\annee{1973}.



\bibitem{gangui98}
A.~Gangui, P.~Peter \ET C.~Boehm, \journal{Phys. Rev.} \numero{D57},
\page{2580},\annee{1998}, \eprint{hep-ph/9705204}.



\bibitem{garfinkle87}
D.~Garfinkle \ET T.~Vachaspati, \journal{Phys. Rev.} \numero{D36},
\page{2229} \annee{1987}.



\bibitem{garriga00}
J.~Garriga \ET T.~Tanaka, \journal{Phys. Rev. Lett.} \numero{84},
\page{2778} \annee{2000}, \eprint{hep-th/9911055}.



\bibitem{garrigapeter}
J.~Garriga \ET P.~Peter, \journal{Class. Quant. Grav.}
\numero{11}, \page{1743} \annee{1994}, \eprint{gr-qc/9403025}.



\bibitem{gellmann63}
M.~Gell-Mann, \journal{Phys. Lett} \numero{8}, \page{214}
\annee{1963}.



\bibitem{georgi74}
H.~Georgi \ET S.~L.~Glashow, \journal{Phys. Rev. Lett.} \numero{32},
\page{438} \annee{1974}.



\bibitem{gherghetta00}
T.~Gherghetta, E.~Roessl \ET M.~Shaposhnikov, \journal{Phys. Lett.}
\numero{B491}, \page{353} \annee{2000}, \eprint{hep-th/0006251}.



\bibitem{ghoruku}
K.~Ghoroku \ET A.~Nakamura, \eprint{hep-th/0106145}.



\bibitem{gibbons}
G.~W.~Gibbons \ET D.~L.~Wiltshire,
\journal{Nucl. Phys.} \numero{B287}, \page{717} \annee{1987},
\eprint{hep-th/0109093}.



\bibitem{ginzburg60}
V.~L.~Ginzburg, \journal{Sov. Phys. Sol. Stat.} \numero{2},
\page{1824} \annee{1960}.



\bibitem{giovannini01}
M.~Giovannini, H.~Meyer \ET M.~Shaposhnikov, \journal{Nucl. Phys.}
\numero{B619}, \page{615} \annee{2001}, \eprint{hep-th/0104118}.



\bibitem{glashow61}
S.~L.~Glashow, \journal{Nucl. Phys.} \numero{22}, \page{579}
\annee{1961}.



\bibitem{golfand71}
Y.~A.~Gol'Fand \ET E.~P.~Likhtman, \journal{Sov. Phys. JETP}
\numero{13}, \page{452} \annee{1971}.



\bibitem{goldstone62}
J.~Goldstone, A.~Salam \ET S.~Weinberg, \journal{Phys. Rev.}
\numero{127}, \page{965} \annee{1962}.



\bibitem{goldstone61}
J.~Goldstone, \journal{Nuev. Cim.} \numero{19}, \page{154}
\annee{1961}.



\bibitem{goliath99}
M.~Goliath \ET G.~F.~R.~Ellis, \journal{Phys. Rev.} \numero{D60},
\page{023502}, \annee{1999}, \eprint{gr-qc/9811068}.



\bibitem{goliath01}
M.~Goliath \ETAL, \eprint{astro-ph/0104009}.



\bibitem{goto}
T.~Goto, \journal{Prog. Theor. Phys.} \numero{46}, \page{1560}
\annee{1971}



\bibitem{gott85}
J.~R.~Gott, \journal{Astrophys. J.} \numero{288}, \page{422}
\annee{1985}.



\bibitem{gradsh}
I.~S.~Gradshteyn \ET I.~M.~Ryzhik, \journal{Table of Integrals, Series
and Products}, \editeur{Academic Press} \annee{1980}.



\bibitem{groom00}
D.~E.~Groom \ETAL, \journal{Eur. Phys. J.} \numero{C15} \annee{2000},
\eprint{http://pdg.lbl.gov/}.



\bibitem{guralnik64}
G.~S.~Guralnik, C.~R.~Hagen \ET T.W.~B.~Kibble,
\journal{Phys. Rev. Lett.} \numero{13}, \page{585}
\annee{1964}.



\bibitem{guth81}
A.~Guth, \journal{Phys. Rev.} \numero{23}, \page{347} \annee{1981}.



\bibitem{guth82}
A.~Guth \ET S.~Y.~Pi, \journal{Phys. Rev. Lett.} \numero{49},
\page{1110} \annee{1982}.



\bibitem{haag75}
R.~Haag, J.~Lopuszanski \ET M.~Sohnius, \journal{Nucl. Phys.}
\numero{B88}, \page{257} \annee{1975}.



\bibitem{hanawa97}
T.~Hanawa \ET K.~Nakayama, \journal{Astrophys. J.} \numero{448},
\page{238} \annee{1997}.



\bibitem{hanawa99}
T.~Hanawa \ET T.~Matsumoto, \journal{Publ. Astron. Soc. Japan}
\numero{1}, \page{1} \annee{1999}.



\bibitem{harvey89}
J.~A.~Harvey \ET S.~G.~Naculich, \journal{Phys. Lett.}
\numero{B217}, \page{231} \annee{1989}.



\bibitem{henriksen}
R.~N.~Henriksen \ET P.~S.~Wesson, \journal{Astrophys. and Space Sci.}
\numero{53}, \page{429} \annee{1978}.



\bibitem{higgs64}
P.~W.~Higgs, \journal{Phys. Lett.} \numero{12}, \page{132}
\annee{1964}.



\bibitem{higgs}
P.~W.~Higgs, \journal{Phys. Rev.} \numero{145},
\page{1156} \annee{1966}.



\bibitem{hilbert15}
D.~Hilbert, \journal{K\"onigliche Gesellschaft der Wissenschaft zu
G\"ottigen, Mathematische-physikalische Klasse, Nachrichten},
\page{395} \annee{1915}.



\bibitem{hill97}
C.~T.~Hill \ET L.~M.~Widrow, \journal{Phys. Lett.}
\numero{B189}, \page{17} \annee{1987}.



\bibitem{hindmarsh88}
M.~Hindmarsh, \journal{Phys. Lett.} \numero{B200}, \page{429}
\annee{1988}.



\bibitem{book2}
M.~B.~Hindmarsh \ET T.~W.~B.~Kibble,
\journal{Rep. Prog. Phys.} \numero{58}, \page{477} \annee{1995},
\eprint{hep-ph/9411342}.



\bibitem{hindmarsh92}
M.~B.~Hindmarsh, \journal{Phys. Rev. Lett.} \numero{68}, \page{1263}
\annee{1992}.



\bibitem{hisano}
J.~Hisano \ET N.~Okada, \journal{Phys. Rev.} \numero{D61},
\page{106003} \annee{2000}, \eprint{hep-ph/9909555}.



\bibitem{horava96}
P.~Ho\v{r}ava \ET E.~Witten, \journal{Nucl. Phys.} \numero{B460},
\page{506} \annee{1996}, \eprint{hep-th/9510209}.



\bibitem{horava96b}
P.~Ho\v{r}ava \ET E.~Witten, \journal{Nucl. Phys.} \numero{B475},
\page{94} \annee{1996}, \eprint{hep-th/9603142}.



\bibitem{hogan84}
C.~J.~Hogan \ET M.~J.~Rees, \journal{Nature} \numero{311}, \page{109}
\annee{1984}.



\bibitem{holman92}
R.~Holman, T.~W.~B.~Kibble \ET S.-J.~Rey, \journal{Phys. Rev. Lett.}
\numero{69}, \page{241} \annee{1992}, \eprint{hep-ph/9203209}.



\bibitem{thooft71}
G.~'t Hooft, \journal{Nucl. Phys.} \numero{B33},
\page{173} \annee{1971}.



\bibitem{thooft71b}
G.~'t Hooft, \journal{Phys. Rev.} \numero{B35},
\page{167} \annee{1971}.



\bibitem{thooft72}
G.~'t Hooft \ET M.~Veltman, \journal{Nucl. Phys.} \numero{B44},
\page{189} \annee{1972}.



\bibitem{hoyle48}
F.~Hoyle, \journal{M. N. R. A. S.} \numero{108}, \page{372}
\annee{1948}.



\bibitem{hoyle01}
C.~D.~Hoyle \ETAL, \journal{Phys. Rev. Lett.} \numero{86},
\page{1418} \annee{2001}, \eprint{hep-ph/0011014}.



\bibitem{hubble25}
E.~Hubble, \journal{Astrophys. J.} \numero{62}, \page{409}
\annee{1925}.



\bibitem{hubble29}
E.~Hubble, \journal{Proceedings of The National Academy of Science
(U.S.A)} \numero{15}, \page{168} \annee{1929}.



\bibitem{hubble26}
E.~Hubble, \journal{Astrophys. J.} \numero{64}, \page{321}
\annee{1926}.



\bibitem{hp}
E.~Huguet \ET P.~Peter, \journal{Astropart. Phys.} \numero{12},
\page{277} \annee{2000}, \eprint{hep-ph/9901370}.



\bibitem{jackiw}
R.~Jackiw \ET C.~Rebbi, \journal{Phys. Rev.} \numero{D13},
\page{3398} \annee{1976}.



\bibitem{jackiwrossi}
R.~Jackiw \ET P.~Rossi, \journal{Nucl. Phys.} \numero{B190},
\page{681} \annee{1981}.



\bibitem{jeans}
J.~H.~Jeans, \journal{Astronomy and Cosmogony}, \editeur{Dover
Publication, New York} \annee{1961}.



\bibitem{kaiser84}
N.~Kaiser \ET A.~Stebbins, \journal{Nature} \numero{310}, \page{391}
\annee{1984}.



\bibitem{kallosh01}
R.~Kallosh, L.~Kofman \ET A.~Linde, \journal{Phys. Rev.} \numero{D64},
\page{123523} \annee{2001}, \eprint{hep-th/0104073}.



\bibitem{kallosh01b}
R.~Kallosh \ETAL, \journal{Phys. Rev.} \numero{D64}, \page{123524}
\annee{2001}, \eprint{hep-th/0106241}.



\bibitem{kaluza21}
T.~Kaluza, \journal{Preussische Akademie der Wissenschaften,
Sitzungsberichte}, \page{966} \annee{1921}.



\bibitem{kane97a}
J.~Kane \ETAL, \journal{Proceedings of Second Oak Ridge Symposium on
Atomic \& Nuclear Astrophysics}, \editeur{Oak Ridge, Tenessee}
\annee{1997}.



\bibitem{kane97b}
J.~Kane \ETAL, \journal{Astrophys. J.} \numero{478}, \page{L75}
\annee{1997}.



\bibitem{kane99}
J.~Kane \ETAL, \journal{Phys. Plas.} \numero{6} \annee{1999}



\bibitem{kanti01}
P.~Kanti, R.~Madden \ET K.~A.~Olive, \journal{Phys. Rev.} \numero{D64},
\page{044021} \annee{2001}, \eprint{hep-th/0104177}.



\bibitem{kay}
B.~S.~Kay, \journal{Phys. Rev.} \numero{D20}, \page{3052}
\annee{1979}.



\bibitem{khoury01}
J.~Khoury \ETAL, \journal{Phys. Rev.} \numero{D64}, \page{123522}
\annee{2001}, \eprint{hep-th/0103239}.



\bibitem{kibble91}
T.~W.~B.~Kibble \ET E.~Copeland, \journal{Phys. Script.} \numero{T36},
\page{153} \annee{1991}.



\bibitem{kibble82}
T.~W.~B.~Kibble \ET N.~G.~Turok, \journal{Phys. Lett.} \numero{B116},
\page{141} \annee{1982}.



\bibitem{kibble85}
T.~W.~B.~Kibble, \journal{Nucl. Phys.} \numero{B252}, \page{227}
\annee{1985}.



\bibitem{kibble80}
T.~W.~B.~Kibble, \journal{Phys. Rep.} \numero{67}, \page{183}
\annee{1980}.



\bibitem{kibble76}
T.~W.~B.~Kibble, \journal{J. Math. Phys.} \numero{A9}, \page{1387}
\annee{1976}.



\bibitem{kibble86}
T.~W.~B.~Kibble, \journal{Phys. Lett.} \numero{B166}, \page{311}
\annee{1986}.



\bibitem{kibble67}
T.~W.~B.~Kibble, \journal{Phys. Rev.} \numero{155}, \page{1554}
\annee{1967}.



\bibitem{kirzhnits74}
D.~A.~Kirzhnits \ET A.~Linde, \journal{Sov. Phys. JETP} \numero{40},
\page{628} \annee{1974}.



\bibitem{kirzhnits76}
D.~A.~Kirzhnits \ET A.~Linde, \journal{Ann. Phys.} \numero{101},
\page{195} \annee{1976}.



\bibitem{klein71}
O.~Klein, \journal{Science} \numero{171}, \page{339} \annee{1971}.



\bibitem{klein26}
O.~Klein, \journal{Nature} \numero{118}, \page{516} \annee{1926}.



\bibitem{kolbturner}
E.~W.~Kolb \ET M.~S.~Turner, \journal{The Early Universe},
\editeur{Frontier in Physics} \annee{1990}.



\bibitem{kraus99}
P.~Kraus, \journal{J.H.E.P.} \numero{9912}, \page{11} \annee{1999}.



\bibitem{la89}
D.~La \ET P.~J.~Steinhard, \journal{Phys. Rev. Lett.} \numero{62},
\page{376} \annee{1989}.



\bibitem{lachieze95}
M.~Lachi\`eze-Rey \ET J.-P.~Luminet, \journal{Physics Report},
\numero{254}, \page{135} \annee{1995}, \eprint{gr-qc/9605010}.



\bibitem{langacker80}
P.~Langacker \ET S.-Y.~Pi, \journal{Phys. Rev. Lett.} \numero{45},
\page{1} \annee{1980}.



\bibitem{larsen97}
A.~L.~Larsen \ET M.~Axenides, \journal{Class.  Quant. Grav.} \numero{14},
\page{443} \annee{1997}, \eprint{hep-th/9604135}.



\bibitem{larsen}
A.~L.~Larsen, \journal{Class. Quant. Grav.} \numero{10}, 1541 (1993),
\eprint{hep-th/9304086}.



\bibitem{leavitt}
H.~S.~Leavitt, \journal{Harvard College Obs. Circ.} \numero{173},
\page{3} \annee{1912}.



\bibitem{maxima}
A.~T.~Lee \ETAL, \journal{Astrophys. J.} \numero{561}, \page{L1}
\annee{2001}, \eprint{astro-ph/0104459}



\bibitem{lee}
H.~Lee \ET W.~S.~l'Yi \annee{2000}, \eprint{hep-th/0011144}.



\bibitem{lehmann55}
H.~Lehmann, K.~Symanzik \ET W.~Zimmermann, \journal{Nuev. Cim.}
\numero{1}, \page{205} \annee{1955}



\bibitem{lemaitre27}
G.~Lema\^itre, \journal{Ann. Soc. Sc. (Bruxelles)} \numero{A47},
\page{49} \annee{1927}.



\bibitem{lifchitz46}
E.~M.~Lifchitz, \journal{J. Exp. Theor. Phys.} \numero{16},
\page{576} \annee{1946}.



\bibitem{lifchitz63}
E.~M.~Lifchitz \ET I.~M.~Khalatnikov, \journal{Adv. Phys.}
\numero{12}, \page{185} \annee{1963}.



\bibitem{linde83}
A.~Linde, \journal{Phys. Lett.} \numero{B129}, \page{177}
\annee{1983}.



\bibitem{linde94}
A.~Linde, \journal{Phys. Rev.} \numero{D49}, \page{748}
\annee{1994}, \eprint{astro-ph/9307002}.



\bibitem{linde82}
A.~Linde, \journal{Phys. Lett.} \numero{B108}, \page{389}
\annee{1982}.



\bibitem{linet90}
B.~Linet, \journal{Class. Quant. Grav.} \numero{7}, \page{75}
\annee{1990}.



\bibitem{linet85}
B.~Linet, \journal{Gen. Rel. Grav.} \numero{17}, \page{1109}
\annee{1985}.



\bibitem{long99}
J.~C.~Long, H.~W.~Chang \ET J.~C.~Price, \journal{Nucl. Phys.}
\numero{B529}, \page{23} \annee{1999}, \eprint{hep-ph/9805217}.



\bibitem{limit}
J.~C.~Long, A.~B.~Churnside \ET J.~C.~Price, \journal{Proceedings of
the IXth Marcel Grossmann Meeting} \annee{2000},
\eprint{hep-ph/0009062}.



\bibitem{lopez96}
J.~Lopez, \journal{Reports on Progress in Physics} \numero{59},
\page{819} \annee{1996}, \eprint{hep-ph/9601208}.



\bibitem{lyth87}
D.~H.~Lyth, \journal{Phys. Lett.} \numero{B196}, \page{126}
\annee{1987}.



\bibitem{maartens02}
R.~Maartens \ETAL, \journal{Phys. Rev.} \numero{D62}, \page{041301}
\annee{2002}, \eprint{hep-ph/9912464}.



\bibitem{magueijo96}
J.~Magueijo \ETAL, \journal{Phys. Rev.} \numero{D56}, \page{3727}
\annee{1996}, \eprint{astro-ph/9605047}.



\bibitem{vortelec}
X.~Martin, \journal{Int. J. Mod. Phys.} \numero{A15}, \page{1031}
\annee{2000}.



\bibitem{martin59}
P.~C.~Martin \ET J.~Schwinger, \journal{Phys. Rev.} \numero{115},
\page{1342} \annee{1959}.



\bibitem{martinpeter}
X.~Martin \ET P.~Peter, \journal{Phys. Rev.} \numero{D51},
\page{4092} \annee{1995}, \eprint{hep-ph/9405220}.



\bibitem{martin94}
X.~Martin, \journal{Phys. Rev.} \numero{D50}, \page{7479}
\annee{1994}.



\bibitem{mp2}
X.~Martin \ET P.~Peter, \journal{Phys. Rev.} \numero{D61},
\page{043510} \annee{2000}, \eprint{hep-ph/9808222}.



\bibitem{martin95}
X.~Martin, \journal{Phys. Rev. Lett.} \numero{74}, \page{3102}
\annee{1995}.



\bibitem{mather90}
J.~C.~Mather \ETAL, \journal{Astrophys. J.} \numero{354},
\page{L37} \annee{1990}.



\bibitem{mather93}
J.~C.~Mather \ETAL, \journal{Astrophys. J.} \numero{432},
\page{L15} \annee{1993}.



\bibitem{matsubara55}
T.~Matsubara, \journal{Prog. Theor. Phys.} \numero{14}, \page{351}
\annee{1955}.



\bibitem{matzner88}
R.~A.~Matzner \ET J.~McCracken \journal{Cosmic Strings: The Current
Status}, F.~S.~Accetta \ET L.~M.~Krauss, \editeur{World Scientific,
Singapore} \annee{1988}.



\bibitem{maxwell}
J.~C.~Maxwell, \journal{An Elementary Treatise on Electricity},
\editeur{Oxford, Clarendon Press} \annee{1888}.



\bibitem{miller}
J.~C.~P.~Miller, \journal{Tables of Weber Parabolic Cylinder
Functions}, \editeur{Her majesty's stationery office, London}
\annee{1955}.



\bibitem{milne34}
E.~A.~Milne, \journal{Q. J. Math.} \numero{5}, \page{64}
\annee{1934}.



\bibitem{moreno}
J.~M.~Moreno, D.~H.~Oaknin \ET M.~Quir\'os,
\journal{Phys. Lett.} \numero{B347}, \page{332} \annee{1995},
\eprint{hep-ph/9411411}.



\bibitem{moriarty88}
K.~Moriarty \ETAL, \journal{Phys. Lett.} \numero{B207}, \page{411}
\annee{1988}.



\bibitem{nakahara90}
M.~Nakahara, \journal{Geometry, Topology and Physics}, \editeur{Adam
Hilger, Bristol, (N.Y.)} \annee{1990}.



\bibitem{nakamura}
F.~Nakamura \ETAL, \journal{Astrophys. J.} \numero{510}, \page{274}
\annee{1999}.



\bibitem{nambu61}
Y.~Nambu \ET G.~Jona-Lasinio, \journal{Phys. Rev.} \numero{122},
\page{345} \annee{1961}.



\bibitem{nambu}
Y.~Nambu, \journal{Phys. Rev.} \numero{D10}, \page{4262}
\annee{1974}.



\bibitem{neronov}
A.~Neronov, \journal{Phys.Rev.} \numero{D65}, \page{044004}
\annee{2002}, \eprint{gr-qc/0106092}.



\bibitem{boomerang}
C.~B.~Netterfield \ETAL \annee{2001}, \eprint{astro-ph/0104460}.



\bibitem{nielsen73}
N.~K.~Nielsen \ET P.~Olesen, \journal{Nucl. Phys.} \numero{B61},
\page{45} \annee{1973}.



\bibitem{NO}
N.~K.~Nielsen \ET P.~Olesen, \journal{Nucl. Phys.} \numero{B291},
\page{829} \annee{1987}.



\bibitem{KK}
N.~K.~Nielsen, \journal{Nucl. Phys.} \numero{B167}, \page{248}
\annee{1980}.



\bibitem{vannieuw81}
P.~van~Nieuwenhuizen, \journal{Phys. Rep.} \numero{68}, \page{189}
\annee{1981}.



\bibitem{nilles84}
H.~P.~Nilles, \journal{Phys. Rep.} \numero{110}, \page{1}
\annee{1984}.



\bibitem{nollett00}
K.~M.~Nollett \ET S.~Burles, \journal{Phys. Rev.} \numero{D61},
\page{123505} \annee{2000}, \eprint{astro-ph/0001440}.



\bibitem{oda}
I.~Oda \annee{2001}, \eprint{hep-th/0103052}.



\bibitem{otw}
J.~P.~Ostriker, C.~Thompson \ET E.~Witten,
\journal{Phys. Lett.} \numero{B180}, \page{231} \annee{1986}.



\bibitem{peebles}
P.~J.~E.~Peebles, \journal{Principles of Physical Cosmology},
\editeur{Princeton University Press} \annee{1980}.



\bibitem{peeble70}
P.~J.~E.~Peebles \ET J.~T.~Yu, \journal{Astrophys. J.} \numero{162},
\page{815} \annee{1970}.



\bibitem{peebles88}
P.~J.~E.~Peebles \ET B.~Ratra, \journal{Astrophys. J.} \numero{325},
\page{L17} \annee{1988}.



\bibitem{penzias65}
A.~A.~Penzias \ET R.~W.~Wilson, \journal{Astrophys. J.} \numero{142},
\page{419} \annee{1965}.



\bibitem{peri95}
L.~Perivolaropoulos, \journal{Astrophys. J.} \numero{451}, \page{429}
\annee{1995}, \eprint{astro-ph/9402024}.



\bibitem{perlmutter99}
S.~Perlmutter \ETAL, \journal{Astrophys. J.} \numero{517},
\page{565} \annee{1999}, \eprint{astro-ph/9812133}.



\bibitem{nospring}
P.~Peter, \journal{Phys. Rev.} \numero{D47}, \page{3169}
\annee{1993}.



\bibitem{peter94}
P.~Peter, \journal{Class. Quant. Grav.} \numero{11}, \page{131}
\annee{1994}.



\bibitem{peterpuy93}
P.~Peter \ET D.~Puy, \journal{Phys. Rev.} \numero{D48}, \page{5546}
\annee{1993}



\bibitem{enon0}
P.~Peter, \journal{Phys. Rev.} \numero{D46}, \page{3335}
\annee{1992}.



\bibitem{neutral}
P.~Peter, \journal{Phys. Rev.} \numero{D45}, \page{1091}
\annee{1992}.



\bibitem{prep}
P.~Peter \ET C.~Ringeval, \journal{Proceedings des Journ\'ees
Relativistes, Dublin} \annee{2001},
\eprint{hep-ph/0011308}.



\bibitem{polarski92}
D.~Polarski \ET A.~A.~Starobinsky, \journal{Nucl. Phys.} \numero{B385},
\page{623} \annee{1992}.



\bibitem{superstring}
J.~Polchinski, \journal{String Theory}, \numero{1},
\editeur{Cambridge University Press, Cambridge} \annee{1998}.



\bibitem{polchinsky}
J.~Polchinsky, \journal{String Theory}, \numero{2}, \editeur{Cambridge
University Press, Cambridge} \annee{1998}.



\bibitem{pollock87}
M.~Pollock, \journal{Phys. Lett.} \numero{B185}, \page{34}
\annee{1987}.



\bibitem{polonsky95}
N.~Polonsky, \journal{Phys. Rev.} \numero{D52}, \page{3081}
\annee{1995}, \eprint{hep-ph/9503214}.



\bibitem{preskill79}
J.~Preskill, \journal{Phys. Rev. Lett.} \numero{43}, \page{1365}
\annee{1979}.



\bibitem{rajamaran82}
R.~Rajamaran, \journal{Solitons and Instantons},
\editeur{North-Holland, Amsterdam} \annee{1982}.



\bibitem{RS}
L. Randall \ET R. Sundrum, \journal{Phys. Rev. Lett.} \numero{83},
\page{4690} \annee{1999}, \eprint{hep-th/9906064}.



\bibitem{randjbar}
S.~Randjbar-Daemi \ET M.~Shaposhnikov, \journal{Phys. Lett.}
\numero{B492}, \page{361} \annee{2000}, \eprint{hep-th/0008079}.



\bibitem{rarita41}
W.~Rarita \ET J.~Schwinger, \journal{Phys. Rev.} \numero{60},
\page{61} \annee{1941}.



\bibitem{ratra88}
B.~Ratra \ET P.~J.~E.~Peebles, \journal{Phys. Rev.} \numero{D37},
\page{3406} \annee{1988}.



\bibitem{remington}
B.~A.~Remington \ETAL, \journal{Phys. Plasmas} \numero{4}
\annee{1994}.



\bibitem{omp}
Open Multiprocessing Ressource, \eprint{http://www.openmp.org}



\bibitem{rdp}
A.~Riazuelo, N.~Deruelle \ET P.~Peter, \journal{Phys. Rev.}
\numero{D61}, 123504 (2000), \eprint{astro-ph/9910290}.



\bibitem{riazueloT}
A.~Riazuelo, \journal{Signature de divers mod\`eles d'Univers
primordial dans les anisotropies du rayonnement fossile},
\editeur{th\`ese de doctorat en Astrophysique, Universit\'e Paris 11}
\annee{2000}.



\bibitem{rpu}
C.~Ringeval, P.~Peter \ET J-P.~Uzan, \journal{Phys. Rev.} \numero{D65},
\page{044016} \annee{2002}, \eprint{hep-th/0109194}.



\bibitem{ringeval2}
C.~Ringeval, \journal{Phys. Rev.} \numero{D64}, \page{123505}
\annee{2001}, \eprint{hep-ph/0106179}.



\bibitem{ringevalSN}
C.~Ringeval \ET S.~Bouquet, \journal{Astron. Astrophys.} \numero{355},
\page{564} \annee{2000}.



\bibitem{ringeval3}
C.~Ringeval, (in preparation).



\bibitem{ringeval} 
C.~Ringeval, \journal{Phys. Rev.} \numero{D63},
\page{063508} \annee{2001}, \eprint{hep-ph/0007015}



\bibitem{robertson28}
H.~P.~Robertson, \journal{Phil. Mag.} \numero{5}, \page{62} \annee{1928}.



\bibitem{rubakov83}
V.~A.~Rubakov \ET M.~E.~Shaposhnikov, \journal{Phys. Lett.} \numero{B125},
\page{136} \annee{1983}.



\bibitem{rubakov01}
V.~A.~Rubakov, \journal{Phys. Usp.} \numero{44}, \page{871}
\annee{2001}, \eprint{hep-ph/0104152}.



\bibitem{saigo}
K.~Saigo \ET T.~Hanawa, \journal{Astrophys. J.} \numero{493}, \page{342}
\annee{1998}.



\bibitem{sakellariadou90}
M.~Sakellariadou \ET A.~Vilenkin, \journal{Phys. Rev.} \numero{D42},
\page{349} \annee{1990}.



\bibitem{sakurai58}
J.~J.~Sakurai, \journal{Nuev. Cim.} \numero{7}, \page{649}
\annee{1958}.



\bibitem{salam64}
A.~Salam \ET J.~C.~Ward, \journal{Phys. Lett.}, \numero{13},
\page{168} \annee{1964}.



\bibitem{salam74}
A.~Salam \ET J.~Stathdee, \journal{Nucl. Phys.} \numero{B76},
\page{477} \annee{1974}.



\bibitem{schwinger51}
J.~Schwinger, \journal{Phys. Rev.} \numero{82}, \page{664}
\annee{1951}.



\bibitem{schwinger59}
J.~Schwinger, \journal{Phys. Rev. Lett.} \numero{3}, \page{296}
\annee{1959}.



\bibitem{shellard87}
E.~P.~S.~Shellard, \journal{Nucl. Phys.} \numero{B283}, \page{624}
\annee{1987}.



\bibitem{shiromizu00}
T.~Shiromizu, K.~Maeda \ET M.~Sasaki, \journal{Phys. Rev.} \numero{D62},
\page{024012} \annee{2000}, \eprint{gr-qc/9910076}.



\bibitem{shu}
F.~H.~Shu, \journal{Astrophys. J.} \numero{214}, \page{488}
\annee{1977}.



\bibitem{skyrme61}
T.~H.~R.~Skyrme, \journal{Proc. Roy. Soc.} \numero{A262}, \page{233}
\annee{1961}.



\bibitem{slipher15}
V.~M.~Slipher, \journal{Popular Astronomy} \numero{23}, \page{21}
\annee{1915}.



\bibitem{smoot92}
G.~F.~Smoot \ETAL, \journal{Astrophys. J. Letters} \numero{396},
\page{L1} \annee{1992}.



\bibitem{stebbins89}
A.~Stebbins \ET M.~S.~Turner, \journal{Astrophys. J.} \numero{339},
\page{L13} \annee{1989}.



\bibitem{stern86}
A.~Stern \ET U.~A.~Yajnik, \journal{Nucl. Phys.} \numero{B267},
\page{158} \annee{1986}.



\bibitem{stinebring90}
D.~R.~Stinebring \ETAL, \journal{Phys. Rev. Lett.} \numero{65},
\page{285} \annee{1990}.



\bibitem{maxima2}
R.~Stompor \ETAL, \journal{Astrophys. J.} \numero{561}, \page{L7}
\annee{2001}, \eprint{astro-ph/0105062}.



\bibitem{strauss92}
M.~A.~Strauss \ETAL, \journal{Astrophys. J.} \numero{361}, \page{49}
\annee{1992}.



\bibitem{sudarshan58}
E.~C.~G.~Sudarshan \ET R.~E.~Marshak, \journal{Phys. Rev.}
\numero{109}, \page{1860} \annee{1958}.



\bibitem{tomboulis76}
E.~Tomboulis, G.~Woo, \journal{Nucl. Phys.} \numero{B107},
\page{221} \annee{1976}.



\bibitem{tomboulis75}
E.~Tomboulis, \journal{Phys. Rev.} \numero{D12}, \page{1678} \annee
{1975}.



\bibitem{turner82}
M.~S.~Turner, \journal{Phys. Lett.} \numero{B115}, \page{95}
\annee{1982}.



\bibitem{turok84b}
N.~Turok \ET P.~Bhattacharjee, \journal{Phys. Rev.} \numero{D29},
\page{1557} \annee{1984}.



\bibitem{turok84}
N.~G.~Turok, \journal{Nucl. Phys.} \numero{B242}, \page{520}
\annee{1984}.



\bibitem{uzan00}
J.-P.~Uzan, R.~Lehoucq \ET J.-P.~Luminet,
\journal{$\mathrm{XIX}^{\mathrm{th}}$ Texas Symposium on Relativistic
Astrophysics and Cosmology, Nucl. Phys. Proc. Suppl.} \numero{80},
\editeur{Elsevier Science, Amsterdam} \annee{2000},
\eprint{gr-qc/0005128}.



\bibitem{uzan00b}
J.-P.~Uzan \ET F.~Bernardeau, \journal{Phys. Rev.} \numero{D64},
\page{083004} \annee{2001}, \eprint{hep-ph/0012011}.



\bibitem{uzanT}
J.-P.~Uzan, \journal{D\'efauts topologiques et conditions aux limites
en cosmologie primordiale}, \editeur{th\`ese de doctorat en Physique
Th\'eorique, Universit\'e Paris 11} \annee{1998}.



\bibitem{uzan01}
J.-P.~Uzan \ET R.~Lehoucq, \journal{Eur. J. Phys.} \numero{22},
\page{371} \annee{2001}, \eprint{physics/0108066}.



\bibitem{vachaspati84}
T.~Vachaspati \ET A.~Vilenkin, \journal{Phys. Rev.} \numero{D30},
\page{2036} \annee{1984}.



\bibitem{vachaspati86}
T.~Vachaspati, \journal{Nucl. Phys.} \numero{B277}, \page{593}
\annee{1986}.



\bibitem{vachaspati91}
T.~Vachaspati \ET A.~Achucarro, \journal{Phys. Rev.} \numero{D44},
\page{3067} \annee{1991}.



\bibitem{vachaspati93}
T.~Vachaspati, \journal{Nucl. Phys.} \numero{B397}, \page{648}
\annee{1993}.



\bibitem{vilenkin81d}
A.~Vilenkin, \journal{Phys. Lett.} \numero{B107}, \page{47}
\annee{1981}.



\bibitem{vilenkin85}
A.~Vilenkin, \journal{Phys. Rep.} \numero{121}, \page{263}
\annee{1985}.



\bibitem{book}
A.~Vilenkin \ET E.~P.~S.~Shellard, \journal{Cosmic strings and other
topological defects}, \editeur{Cambridge University Press}
\annee{1994}.



\bibitem{vilenkin84}
A.~Vilenkin, \journal{Astrophys. J.} \numero{L51}, \page{282}
\annee{1984}.



\bibitem{vilenkin90}
A.~Vilenkin, \journal{Phys. Rev.} \numero{D41},
\page{3038} \annee{1990}.



\bibitem{vilenkin}
A.~Vilenkin, \journal{Phys. Rev. Lett.} \numero{46}, \page{1169}
\annee{1980}.



\bibitem{vilenkin81b}
A.~Vilenkin, \journal{Phys. Rev.} \numero{D23}, \page{852}
\annee{1981}.



\bibitem{vilenkin81}
A.~Vilenkin, \journal{Phys. Rev.} \numero{D24}, \page{2082}
\annee{1981}.



\bibitem{vincent97}
G.~R.~Vincent, M.~B.~Hindmarsh \ET M.~Sakellariadou,
\journal{Phys. Rev.} \numero{D55}, \page{573} \annee{1997},
\eprint{astro-ph/9606137}.



\bibitem{visser85}
M.~Visser, \journal{Phys. Lett.} \numero{B159}, \page{22}
\annee{1985}.



\bibitem{waldbook}
R.~M.~Wald, \journal{General Relativity}, \editeur{The University of
Chicago Press} \annee{1984}.



\bibitem{walker36}
A.~G.~Walker, \journal{Proceeding of the London Mathematical Society}
\numero{42}, \page{90} \annee{1936}.



\bibitem{wang00}
L.~Wang \ETAL, \journal{Astrophys. J.} \numero{530}, \page{17}
\annee{2000}.



\bibitem{weinberg81}
E.~J.~Weinberg, \journal{Phys. Rev.} \numero{D24}, \page{2669}
\annee{1981}.



\bibitem{weinberg74}
S.~Weinberg, \journal{Phys. Rev.} \numero{D9}, \page{3357}
\annee{1974}.



\bibitem{weinberg89}
S.~Weinberg, \journal{Rev. Mod. Phys.} \numero{61}, \page{1}
\annee{1989}.



\bibitem{weinberg67}
S.~Weinberg, \journal{Phys. Rev. Lett.} \numero{19}, \page{1264}
\annee{1967}.



\bibitem{weinbergbook}
S.~Weinberg, \journal{Gravitation and Cosmology}, \editeur{John Wiley \& Sons, (N.Y.)} \annee{1972}.



\bibitem{weinbergbook2}
S.~Weinberg, \journal{The Quantum Theory of Fields}, \editeur{Cambridge University Press} \annee{1996}.



\bibitem{wess74}
J.~Wess \ET B.~Zumino, \journal{Nucl. Phys.} \numero{B70}, \page{39}
\annee{1974}.



\bibitem{wetterich95}
C.~Wetterich, \journal{Astron. Astrophys.} \numero{301}, \page{321}
\annee{1995}, \eprint{hep-th/9408025}.



\bibitem{weyl18}
H.~Weyl, \journal{Akademie der Wissenschaften, Sitzungsberichte},
\page{465} \annee{1918}.



\bibitem{dewitt67}
B.~S.~De~Witt, \journal{Phys. Rev.} \numero{162}, \page{1195}
\annee{1967}.



\bibitem{witten}
E.~Witten, \journal{Nucl. Phys.} \numero{B249}, \page{557}
\annee{1985}.



\bibitem{witten00}
E.~Witten \annee{2000}, \eprint{hep-ph/0002297}.



\bibitem{wolf67}
J.~A.~Wolf, \journal{Spaces of constant curvature},
\editeur{Pub. Per. Inc., (U.S.A.)} \annee{1967}.



\bibitem{wu02}
J.~P.~Wu \ETAL, \journal{Int. J. Mod. Phys.} \numero{D11}, \page{61}
\annee{2002}, \eprint{astro-ph/9812156}.



\bibitem{yahil}
A.~Yahil, \journal{Astrophys. J.} \numero{265}, \page{1047}
\annee{1983}.


\bibitem{yang54}
C.~N.~Yang \ET R.~L.~Mills, \journal{Phys. Rev.} \numero{96},
\page{191} \annee{1954}.



\bibitem{yukawa35}
H.~Yukawa, \journal{Proc. Phys. Math. Soc. (Japan)} \numero{17},
\page{48} \annee{1935}.



\bibitem{zeldov75}
Y.~B.~Zel'dovich, I.~Y.~Kobzarev \ET L.~B.~Okun,
\journal{Sov. Phys. JETP} \numero{40}, \page{1} \annee{1975}.



\bibitem{zeldov}
Y.~B.~Zel'dovich, \journal{M.N.R.A.S.} \numero{192},
\page{663} \annee{1980}.



\bibitem{zeldov78}
Y.~B.~Zel'dovich, \ET M.~Y.~Khlopov, \journal{Phys. Lett.}
\numero{B79}, \page{239} \annee{1978}.



\bibitem{zimmermann73}
W.~Zimmermann, \journal{Ann. Phys. (N.Y.)} \numero{77}, \page{536}
\annee{1973}. 



\bibitem{zweig63}
G.~Zweig, \journal{C.E.R.N. report} \annee{1963}.

}
\end{thebibliography}
\end{document}